\pdfoutput=1
\documentclass[fleqn,usenatbib]{mnras}

\usepackage{mathptmx}
\usepackage{lscape}

\usepackage[T1]{fontenc}
\usepackage{ae,aecompl}

\usepackage{graphicx}	
\usepackage{amsmath}	
\usepackage{amssymb}	
\usepackage{textcomp}
\usepackage{color}
\setcitestyle{notesep={  }}
\title[The role of kinematics in molecular gas build]{VALES V:  A kinematic analysis of the molecular gas content in $H$-ATLAS galaxies at $z\sim0.03-0.35$ using ALMA}

\author[J. Molina et al.]{J.~Molina,$^{1}$\thanks{E-mail: jumolina@das.uchile.cl}
Edo~Ibar,$^{2}$
V.~Villanueva,$^{3}$
A.~Escala,$^{1}$
C.~Cheng,$^{2,4}$
M.~Baes,$^{5}$
\newauthor{H.~Messias,$^{6,7}$
C.~Yang,$^{7}$
F.~E.~Bauer,$^{8,9,10}$
P.~van~der~Werf,$^{11}$
R.~Leiton,$^{2}$}
\newauthor{M.~Aravena,$^{12}$
A.~M.~Swinbank,$^{13,14}$
M.~J.~Micha{\l}owski,$^{15}$
A.~M.~Mu{\~n}oz-Arancibia,$^{2}$}
\newauthor{G.~Orellana,$^{2}$
T.~M.~Hughes,$^{2,16,17,18}$ 
D.~Farrah,$^{19,20}$
G.~De~Zotti,$^{21}$
M.~A.~Lara-L\'opez,$^{22}$}
\newauthor{S.~Eales,$^{23}$
L.~Dunne.$^{23,24}$}\\
$^{1}$Departamento de Astronom\'ia, Universidad de Chile, Casilla 36-D, Santiago, Chile.\\
$^{2}$Instituto de F\'isica y Astronom\'ia, Universidad de Valpara\'iso, Avda. Gran Breta\~na 1111, Valpara\'iso, Chile\\
$^{3}$Department of Astronomy, University of Maryland, College Park, MD 20742, USA\\
$^{4}$CASSACA, National Astronomical Observatories, Chinese Academy of Sciences, Beijing 100012, China\\
$^{5}$Sterrenkundig Observatorium, Universiteit Gent, Krijgslaan 281 S9, B-9000 Gent, Belgium\\
$^{6}$Joint ALMA Observatory, Alonso de C\'ordova 3107, Vitacura 763-0355, Santiago, Chile\\
$^{7}$European Southern Observatory, Alonso de C\'ordova 3107, Casilla 19001, Vitacura, Santiago, Chile\\
$^{8}$Instituto de Astrof\'isica, Facultad de F\'isica, Pontificia Universidad Cat\'olica de Chile, 306, Santiago 22, Chile\\
$^{9}$Millennium Institute of Astrophysics (MAS), Nuncio Monse\~nor S\'otero Sanz 100, Providencia, Santiago, Chile\\
$^{10}$Space Science Institute, 4750 Walnut Street, Suite 205, Boulder, Colorado 80301\\
$^{11}$Leiden Observatory, Leiden University, P.O. Box 9513, NL-2300 RA Leiden, The Netherlands\\
$^{12}$N\'ucleo de Astronom\'ia, Facultad de Ingenier\'ia y Ciencias, Universidad Diego Portales, Av. Ej\'ercito 441, Santiago, Chile\\
$^{13}$Centre for Extragalactic Astronomy, Department of Physics, Durham University, South Road, Durham DH1 3LE, UK\\ 
$^{14}$Institute for Computational Cosmology, Durham University, South Road, Durham DH1 3LE, UK\\
$^{15}$Astronomical Observatory Institute, Faculty of Physics, Adam Mickiewicz University, ul.~S{\l}oneczna 36, 60-286 Pozna{\'n}, Poland\\
$^{16}$CAS Key Laboratory for Research in Galaxies and Cosmology, Department of Astronomy,\\ University of Science and Technology of China, Hefei 230026, China\\
$^{17}$School of Astronomy and Space Science, University of Science and Technology of China, Hefei 230026, China\\
$^{18}$Chinese Academy of Sciences South America Center for Astronomy, China-Chile Joint Center for Astronomy,\\ Camino El Observatorio \#1515, Las Condes, Santiago, Chile\\
$^{19}$Department of Physics and Astronomy, University of Hawaii, 2505 Correa Road, Honolulu, HI 96822, USA\\
$^{20}$Institute for Astronomy, 2680 Woodlawn Drive, University of Hawaii, Honolulu, HI 96822, USA\\
$^{21}$INAF, Osservatorio Astronomico di Padova, Vicolo dell'Osservatorio 5, I-35122 Padova, Italy\\
$^{22}$Dark Cosmology Centre, Niels Bohr Institute, University of Copenhagen, Juliane Maries Vej 30, DK-2100 Copenhagen, Denmark\\
$^{23}$School of Physics and Astronomy, Cardiff University, Queens Buildings, The Parade, Cardiff CF24 3AA, UK\\
$^{24}$Institute for Astronomy, University of Edinburgh, Royal Observatory, Blackford Hill, Edinburgh EH9 3HJ, UK\\
}

\date{Accepted XXX. Received YYY; in original form ZZZ}

\pubyear{2018}

\begin{document}
\label{firstpage}
\pagerange{\pageref{firstpage}--\pageref{lastpage}}
\maketitle

\begin{abstract} 

We present Atacama Large Millimeter/submillimeter Array (ALMA) resolved observations of molecular gas in galaxies up to $z=0.35$ to characterise the role of global 
galactic dynamics on the global interstellar medium (ISM) properties. These observations consist of a sub-sample of 39 galaxies taken from the Valpara\'iso ALMA Line 
Emission Survey (VALES). From the CO($J=1-0)$ emission line, we quantify the kinematic parameters by modelling the velocity fields. We find that the IR 
luminosity increases with the rotational to dispersion velocity ratio ($V_{\rm rot}/\sigma_v$, corrected for inclination). We find a dependence between $V_{\rm rot}/\sigma_v$ 
and the [C{\sc ii}]/IR ratio, suggesting that the so-called `[C{\sc ii}] deficit' is related to the dynamical state of the galaxies. We find that global pressure support is needed to
reconcile the dynamical mass estimates with the stellar masses in our systems with low $V_{\rm rot}/\sigma_v$ values. The star formation rate (SFR) is weakly correlated with 
the molecular gas fraction ($f_{\rm H_2}$) in our sample, suggesting that the release of gravitational energy from cold gas may not be the main energy source of the turbulent
motions seen in the VALES galaxies. By defining a proxy of the `star formation efficiency' parameter as the SFR divided by the CO luminosity 
(SFE$'\equiv$\,SFR/L$'_{\rm CO}$), we find a constant SFE$'$ per crossing time ($t_{\rm cross}$). We suggest that $t_{\rm cross}$ may be the controlling timescale 
in which the star formation occurs in dusty $z\sim0.03-0.35$ galaxies.
\end{abstract}




\section{Introduction} 

The star formation activity is one of the main processes that drives cosmic evolution of galaxies. Stars produce
heavy elements via nucleosynthesis, which are expelled into the ISM during their late stages of evolution, 
enriching the gas with metals and dust (see e.g. \citealt{Nozawa2013}). Thus, star formation is directly involved in the processes
the growth and evolution of galaxies to the formation of planets through cosmic time. Nevertheless, our knowledge about the 
physical processes that dominate the formation of stars starting from pristine gas is far from 
complete, mainly because of the wide range of physical processes are involved.

\citet{Schmidt1959} was the first to propose a power-law relationship between the star formation activity 
of galaxies and their gas content. This relationship was confirmed later by \citet{Kennicutt1998a,Kennicutt1998b},
who revealed a clear relationship between the disk-averaged total galaxy gas (atomic plus molecular) 
surface density ($\Sigma_{\rm gas}$) and the rate of star formation per surface area ($\Sigma_{\rm SFR}$), 
the Kennicutt-Schmidt relationship (hereafter, KS law). The KS law describes how efficiently galaxies turn their 
gas into stars. It has been used to constrain theoretical models and as a critical input to numerical simulations for 
galaxy evolution models (e.g. \citealt{Springel2003,Krumholz2005,Vogelsverger2014,Schaye2015}). Using this 
relationship we can compute the time at which a given galaxy would convert all of its current gas mass content $M_{\rm gas}$
if it maintains its present star formation rate (SFR), this timescale is called the depletion time: $t_{\rm dep}\equiv M_{\rm gas}/$SFR.

Since \citet{Kennicutt1998a,Kennicutt1998b}'s work, the KS law has been tested in numerous spatially-resolved
surveys on local galaxies during the last decades (e.g. \citealt{Wong2002,Kennicutt2007,Bigiel2008, Villanueva2017}). 
These surveys have allowed us to trace the SFR surface density ($\Sigma_{\rm SFR}$), atomic gas 
surface density ($\Sigma_{\rm HI}$), molecular gas surface density ($\Sigma_{\rm{H_2}}$) and study how these 
quantities relate to each other (e.g. \citealt{Leroy2008,Leroy2013}). One of the first conclusions extracted from these 
observations was that star formation in galaxies is more strongly correlated with $\Sigma_{\rm{H_2}}$ than $\Sigma_{\rm HI}$ 
(especially at $\Sigma_{\rm gas}$\,>\,10\,M$_\odot$\,pc$^{-2}$), with an observed molecular gas depletion time of $t_{\rm dep}\approx1-2$\,Gyr.

When additional data from high star-forming galaxies are included, the KS law shows an apparent bimodal behaviour where `disks' and `starburst' galaxies appear to fill the 
$\Sigma_{\rm H_2}-\Sigma_{\rm SFR}$ plane in different loci \citep{Daddi2010}. Nevertheless, by comparing $\Sigma_{\rm SFR}$ with $\Sigma_{\rm H_2}$
per galaxy free-fall time ($t_{\rm ff}$) and/or orbital time ($t_{\rm orb}$) a single power-law relationship can be recovered (e.g. \citealt{Daddi2010,Krumholz2012}).
The $\Sigma_{\rm SFR}-\Sigma_{\rm H_2}/t_{\rm ff}$ relation can be interpreted as dependence of the star formation law on the local volume density
of the gas, whilst the $\Sigma_{\rm SFR}-\Sigma_{\rm H_2}/t_{\rm orb}$ relation suggests that the star formation law is affected by the global rotation
of the galaxy. Thus, the relevant timescale gives us critical information about the physical processes that may control the formation of stars.

However, by exploiting the VALES survey in the local Universe ($z<$0.3; see \S~\ref{sec:VALES_sample}), \citet{Cheng2018} 
showed that the bimodality seen in the KS law may also be the result of the assumptions, and thus, the uncertainties behind the estimates of the 
molecular gas mass ($M_{\rm H_2}$).

The absence of an electric dipole moment in the hydrogen molecule (H$_2$) implies that direct detections of 
cold H$_2$ gas are difficult to be obtained (e.g. \citealt{PapaSea1999,Bothwell2013}) and
tracers of the molecular gas are needed. One of the methods --and perhaps the most common one-- 
to estimate the molecular gas content is through the carbon monoxide ($^{12}$C$^{16}$O, hereafter CO) 
line luminosity (e.g. \citealt{Solomon1987,Downes1998,SolomonVandenBout2005,Bolatto2013}) of rotational
low-$J$ transitions (e.g. $J=1-0$ or $J=2-1$). Because the CO emission line is generally optically thick ($\tau_{\rm CO}\approx1$), 
its brightness temperature (T$_{\rm b}$) is related to the temperature of the optically thick gas sheet,
not the column density of the gas. Thus the mass of the self gravitating entity, such as a molecular cloud,
is related to the emission line-width, which reflects the velocity dispersion of the gas \citep{Bolatto2013}.

Assuming that the CO luminosity ($L'_{\rm CO}$) of an entire galaxy comes from an ensemble of non-overlapping 
virialized emitting clouds, then if: (1) the intrinsic brightness temperature of those clouds is mostly independent of the
cloud size; (2) these clouds follow the size-line width relationship \citep{Larson1981,Heyer2009}; and (3) the clouds have a 
similar surface density. Then the molecular gas to CO luminosity relation can be expressed 
as $M_{\rm H_2}=\alpha_{\rm CO}L'_{\rm CO}$, where $M_{\rm H_2}$ is defined to include the helium mass, so that
$M_{\rm H_2}=M_{\rm gas, cloud}$, the total gas mass (hence, the virial mass) for molecular clouds \citep{SolomonVandenBout2005}
and $\alpha_{\rm CO}$ is the CO-to-H$_2$ conversion factor.
This is the so-called `mist' model \citep{Dickman1986}. Within the Milky Way, the observed relation between virial mass
and CO line luminosity for Galactic giant molecular clouds (GMCs; \citealt{Solomon1987}) yields 
$\alpha_{\rm CO}\approx4.6$\,M$_\odot$\,(K\,km\,s$^{-1}$\,pc$^2$)$^{-1}$.

Although the mist model estimates the molecular gas content successfully in the Milky Way, it overestimates 
the gas mass in more dynamically disrupted systems, such as Ultra Luminous Infrared Galaxies (ULIRGs; \citealt{Downes1998}).
Unlike Galactic clouds or gas distributed in the disk of `normal' galaxies, CO emission maps from ULIRGs show that 
the molecular gas is contained in dense rotating disks or rings. The CO emission may not come from individual 
virialized clouds, but from a filled inter-cloud medium, so the line-width is determined by the total dynamical 
mass ($M_{\rm dyn}$) in the region (gas and stars). The optically thick CO line emission may trace a medium 
bound by the gravitational potential around the galactic centre \citep{Downes1993,Solomon1997}. In order to 
estimate the $M_{\rm H_2}$ content from $L'_{\rm CO}$ in those systems a different approach is required. 
\citet{Downes1998} used kinematic and radiative transfer models to derive $M_{\rm H_2}/L'_{\rm CO}$ ratios in ULIRGs,
where most of the CO flux is assumed to come from a warm inter-cloud medium. The models yield 
$\alpha_{\rm CO}\approx0.8$\,M$_\odot$\,(K\,km\,s$^{-1}$\,pc$^2$)$^{-1}$, a ratio which is roughly six times lower 
than the standard $\alpha_{\rm CO}$ value for the Milky Way. This $\alpha_{\rm CO}$ value is usually adopted to estimate 
the molecular gas content in other non-virialized environments such as galaxy mergers.

On the other hand, from numerical simulations, galaxies that have similar physical conditions have similar CO-to-H$_2$ factors. 
This seems to be independent of galaxy morphology or evolutionary state. Thus, rather than bimodal distribution of `disk'  and 
`ULIRG' $\alpha_{\rm CO}$ values, simulations suggest that there is a continuum of conversion values that vary with galactic 
environment \citep{Narayanan2012}.

Therefore, spatially resolved studies of the molecular gas content and its kinematics in galaxies are critical to 
understand the physical processes that determine the CO-to-H$_2$ conversion factor and the star formation activity
as these two quantities seem to be dependant on the galactic dynamics.

The construction of large samples of intermediate/high-$z$ galaxies with direct molecular gas detections (via CO emission)
has remained a challenge. Beyond the local Universe, resolved CO detections are limited to the most massive/luminous yet rare 
galaxies or highly magnified objects (e.g. \citealt{Saintonge2013}). With ALMA, we are now able to study the physical conditions of the cold molecular gas in `typical'
galaxies at these redshifts and test if the actual models successfully explain the characteristics of the intermediate/high-$z$ ISM.
In this paper, we use state-of-the-art capabilities of ALMA to characterise the CO($J=1-0$) kinematics of 39 'typical' 
star-forming/mildly starburst galaxies at 0.025\,<\,$z$\,<\,0.32 drawn from the VALES survey \citep{Villanueva2017}. 
Combining these ALMA observations to auxiliary data (e.g. \citealt{Ibar2015, Hughes2017a,Hughes2017b}), we study how 
the kinematics of the cold CO(1-0) gas relate to the physical conditions of the ISM. Throughout the paper, we assume a 
$\Lambda$CDM cosmology with $\Omega_{\Lambda}$=0.73, $\Omega\rm_m$=0.27, and H$_0$=70 km\,s$^{-1}$\,Mpc$^{-1}$, 
implying a spatial resolution, determined by typical major axis of the synthesized beam in the VALES data, of $3"- 4"$ that 
corresponds to a physical scale between 2 and 17\,kpc.

\section{SAMPLE SELECTION \& OBSERVATIONS}
\subsection{VALES Survey} 
\label{sec:VALES_sample}
The VALES sample (\citealt{Villanueva2017}, hereafter V17) is taken from the $Herschel$ Astrophysical TeraHertz Large Area Survey 
($H$-ATLAS; \citealt{Eales2010,Bourne2016,Valiante2016}), which is one of the largest infra-red (IR) and submillimitre
(submm) surveys covering $\sim$600\,deg$^2$ of the sky taken by the $Herschel$ Space Observatory \citep{Pilbratt2010}.
The VALES survey covers a redshift range of 0.02$<z<$0.35, and IR-luminosity range of $L_{8-1000\mu {\rm m}} \approx 10^{10-12}$\,$L_\odot$,
thus it is an excellent galaxy sample to study the molecular gas dynamics of star-forming and `midly' starburst galaxies at low redshift.

The VALES survey is composed of ALMA observations targeting the CO(1-0) emission line in band 3 for 67 galaxies during Cycle-1 and Cycle-2,
from which 49 sources were spectroscopically detected. 

We use the V17's far-infrared (FIR; 8--1000\,${\mu}$m) luminosities, $L_{\rm{FIR}}$, which were derived
from SEDs constructed with photometry from the Infrared Astronomical Satellite (IRAS; \citealt{Neugebauer1984}), Wide-field Infrared Survey Explorer 
(WISE; \citealt{Wright2010}), and the  $Herschel$ Photoconductor Array Camera and Spectrometer (PACS; \citealt{Poglitsch2010})
and the Spectral and Photometric Imaging REceiver (SPIRE; \citealt{Griffin2010}) instruments. By assuming a \citet{Chabrier2003} initial mass function (IMF),
the SFRs are calculated following SFR$(M_\odot$\,yr$^{-1})= 10^{10}\times $\,L$_{\rm IR}$(L$_\odot$; \citealt{Kennicutt1998b}). Those values are systematically higher
than the rates estimated from fitting the SEDs with the bayesian code MAGPHYS \citep{daCunha2008} by a factor of two. However, the two estimates are well correlated despite
this systematic discrepancy (see V17 for more details).

The stellar masses ($M_*$) for our sample were calculated by modelling the SEDs from the photometry provided by the 
GAMA Panchromatic Data Release \citep{Driver2016}  --in which $all$ of our galaxies are present--
in 21 bands extending from the far-ultraviolet to far-infrared ($\sim0.1-500$\,$\mu$m). These observed SEDs 
have all been modelled with the bayesian SED fitting code MAGPHYS and presented in V17. 

\begin{figure}
\includegraphics[width=1.0\columnwidth]{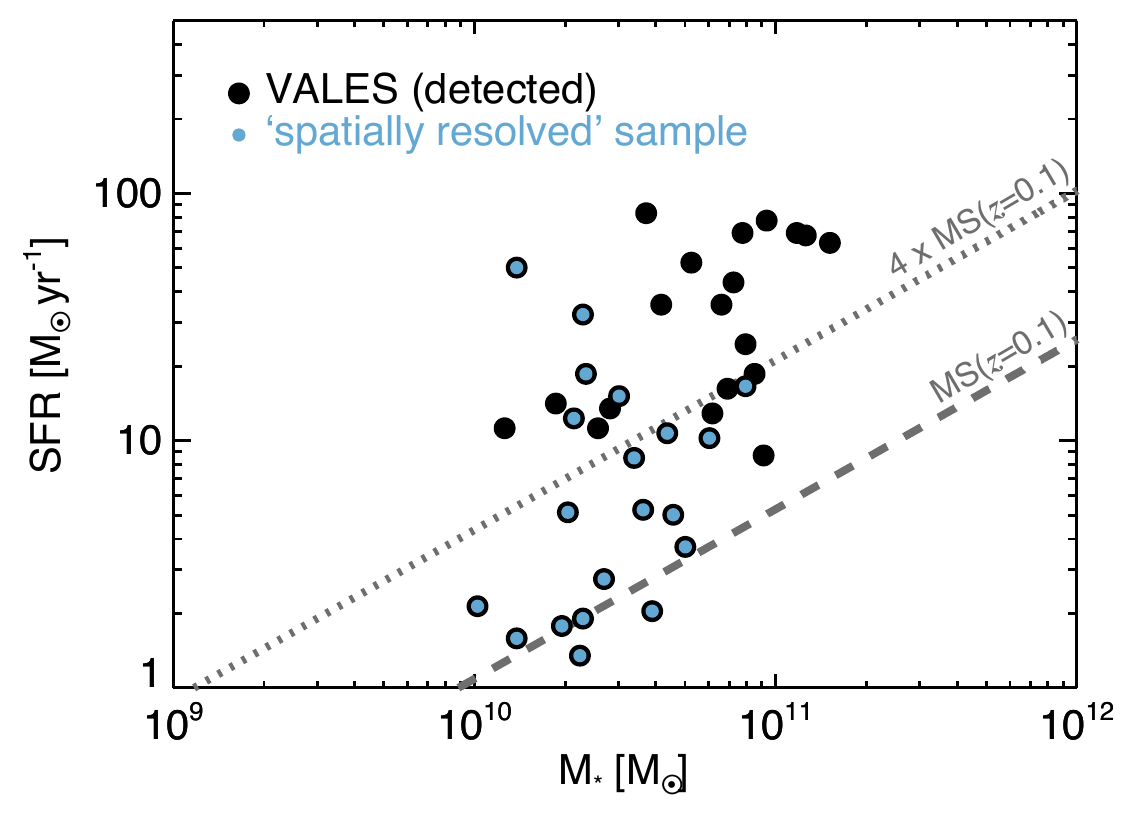}
\caption{ \label{fig:SFR_Mstar_plot}
 The SFR against the $M_*$ for the 39 galaxies which were spectroscopically detected at >\,5-sigma in datacubes with 20\,km\,s$^{-1}$ 
 fixed spectral resolution from the VALES survey \citep{Villanueva2017}.
 In blue circles we highlight the 20 sources classified as `spatially resolved' (see \S~\ref{sec:VALES_sample}, for more details).
 The dashed line represents the SFR-$M_*$ relationship for `main-sequence' star-forming galaxies at $z=0.1$ following \citet{Genzel2015}.
 The dotted line represents 4$\times$ the SFR value expected for a `main-sequence'  star-forming galaxy at a given stellar mass at $z=0.1$.}
\end{figure}

The observations, data reduction and analysis are presented in detail for the complete sample in V17,
whilst the [C\,{\sc ii}] luminosity data is presented in \citet{Ibar2015}.

The analysis presented in V17 shows ALMA cubes binned at different spectral resolutions (from 20 to 100\,km\,s$^{-1}$)
in order to boost the signal to noise (S/N) for spectral detectability. However, the use of low or variable spectral resolution observations to 
derive and/or analyse galactic kinematics may lead to erroneous conclusions (see \S~\ref{sec:rel_effects}). Thus, we kept the spectral resolution
fixed at 20\,km\,s$^{-1}$ despite of the degrade of S/N in order to minimize spectral resolution effects in our dynamical analysis. 

Out of the 49 galaxies that were spectroscopically detected in CO(1-0) by V17, we find that only 39 of them are 
spectroscopically detected at a 5\,$\sigma$ significance after fixing the spectral resolution at 20\,km\,s$^{-1}$ to all sources.
We show these 39 galaxies in the SFR$-M_*$ plane in Fig.~\ref{fig:SFR_Mstar_plot}. Our systems sample the SFRs and stellar masses in the range of 
$1-84$\,$M_\odot$\,yr$^{-1}$ and $1-15 \times 10^{10}$\,$M_\odot$, respectively. We note that the galaxies with high SFR also tend to have high 
$M_*$.

Out of these 39 galaxies, 20 are considered as `spatially resolved' (R) by following these criteria; (1) that the observed CO(1-0) emission extends for more 
than $\sqrt{2}$ times the major axis of the synthesized beam; and (2) the observations should have been taken with a projected synthesized beam smaller than 8\,kpc. 
The other 19 sources are classified as `compact' (C). We show the corresponding galaxy 
classification in the top-right of each CO(1-0) intensity map (Fig.~\ref{fig:maps}). In the forthcoming of this work, in order to guarantee 
enough independent pixels to be fitted within each galaxy map, we just analyse and model the kinematics of the galaxies considered as
`resolved'.

To classify our sources as `normal' star-forming or starburst galaxies we use the parametrization defined by \citet{Genzel2015}
for the specific star formation rate (sSFR$\equiv$SFR/$M_*$; $\log$[sSFR$(z,M_*)$]$=-1.12 + 1.14z - 0.19z^2 - (0.3 + 0.13z) \times (\log M_* -10.5)$\,Gyr$^{-1}$).
Galaxies with |\,sSFR/sSFR$(z,M_*)$\,|$\leq$4 are classified as `normal' star-forming galaxies, whilst all the galaxies with sSFR\,>\,4\,sSFR$(z,M_*)$ 
are labelled as `starburst'. We use the SFR, stellar mass and redshift of each source to perform this classification.
In Fig.~\ref{fig:SFR_Mstar_plot}, the dashed line shows the `main-sequence' of star-forming galaxies at $z=0.1$. As an example, 
the dotted line in Fig.~\ref{fig:SFR_Mstar_plot} represents our chosen sSFR criterion for galaxies at $z=0.1$.

We also use V17's morphological classification scheme to assume a bimodal CO-to-H$_2$ conversion factor 
of 0.8 or 4.6\,$M_\odot$\,(K\,km\,s$^{-1}$\,pc$^2$)$^{-1}$ depending on whether a galaxy is classified as a `merger' or `disk', respectively. 
This classification is based on visual inspection of the galaxy images extracted by using the GAMA Panchromatic Swarp Imager tool\footnote{http://gama-psi.icrar.org/psi.php}.
We note that in our `resolved' sample, just three galaxies (HATLASJ084630.7+005055, HATLASJ085748.0+004641, HATLASJ090750.0+010141) are classified as `mergers' 
by the morphological criterion.
We do not attempt to perform a kinematic classification of mergers (e.g. \citealt{Shapiro2008, Forster2009,Swinbank2012a,Molina2017}) given 
that our low spatial resolution tends to smooth the emission and kinematic deviations, making galaxy intensity and velocity fields appear more 
disky than they actually are \citep{Bellocchi2012}.

The mean molecular gas fraction [$f_{\rm H_2}\equiv M_{\rm H_2}/(M_{\rm H_2}+M_*)$] of the `resolved' sample is $0.22$ within a range of
$0.06-0.44$ with a typical relative error for each measurement of $\sim$12\%.

\subsection{Galaxy Dynamics} 
\label{sec:galaxy_dyn}
To measure the dynamics of each galaxy, we fit the CO(1-0) emission line ($\nu_{\rm{rest}}=115.271$\,GHz)
following the approach presented in \citet{Swinbank2012a}. We use a $\chi^2$ minimisation procedure, 
estimating the noise per spectral channel from a surrounding area that does not contain source emission. 
For a given pixel, we first attempt to identify a CO(1-0) emission line within a squared region that contains the synthesized
beam size around that pixel and we take the average spectrum within that region. 

Then, we fit a gaussian profile to the spectrum and we impose a S/N\,>\,5 threshold to the best-fit to detect the emission line.
If this criterion is not fulfilled, then the squared region around that pixel is increased by one pixel per side and we search for 
any emission line again. After this iteration, if the criterion is still not achieved, then we skip to the next pixel.  

red Considering that we have not applied any spectral filtering for imaging purposes, the fitted line widths correspond to the intrinsic
line widths (no deconvolution needed). Nevertheless, in order to consider if an emission line is sufficiently sampled, we only take into account 
those fits in which the fitted line width is larger than $\sqrt{2}$ times the channel width  ($\approx 28$\,km\,s$^{-1}$, e.g. Fig.~\ref{fig:maps}).
The spectral resolution is therefore impeding narrower velocity dispersion measurements.
We caution that, this masking procedure may lead an overestimated average velocity dispersion value for each galaxy. 
\\
\\

\begin{table} 
        \centering
        \setlength\tabcolsep{4pt}
     
        TABLE 1: $K-$band BROADBAND PROPERTIES\\
        \begin{tabular}{lcccccc}
                \\
                \hline
                \hline
                ID  & $\mu_{0,K}$ & $r_{1/2,K}$ & $n_S$ & PA$_K$ & $e$ & $\chi^2_\nu$ \\
                    & mag/$\prime \prime^2$ & $\prime \prime$ &  & & deg &  \\
                (1) & (2) & (3) & (4) & (5) & (6) & (7) \\
		\hline
		HATLASJ083601.5+002617 & 15.5 &  5.09 & 1.93 &   2.1 & 0.61 & 1.09 \\ 	 
		HATLASJ083745.1-005141 & 15.5 &  6.26 & 2.46 &  62.8 & 0.19 & 0.92 \\ 	 
		HATLASJ084217.7+021222 & 12.3 &  0.63 & 2.47 & 168.3 & 0.22 & 0.54 \\      
		HATLASJ084350.7+005535 & 13.5 &  1.38 & 2.61 &   0.0 & 0.57 & 1.12 \\	 
		HATLASJ084428.3+020349 &  4.2 & 23.49 & 8.92 & 101.1 & 0.38 & 1.57 \\	 
		HATLASJ084428.3+020657 & 15.6 &  2.04 & 1.28 &  58.6 & 0.77 & 1.63 \\	 
		HATLASJ084630.7+005055 &  0.6 &  0.67 & 8.44 & 141.5 & 0.19 & 1.05 \\	 
		HATLASJ084907.0-005139 &  9.7 &  1.06 & 4.95 & 136.4 & 0.34 & 1.11 \\ 	 
		HATLASJ085111.5+013006 & 11.6 &  5.20 & 3.82 & 114.8 & 0.77 & 1.42 \\	 
		HATLASJ085112.9+010342 & 13.6 &  2.68 & 2.82 & 115.6 & 0.53 & 1.16 \\	 
		HATLASJ085340.7+013348 & 16.9 &  6.68 & 2.18 &  27.4 & 0.13 & 1.17 \\	 
		HATLASJ085346.4+001252 & 14.9 &  3.31 & 1.93 &  46.0 & 0.77 & 1.07 \\	 
		HATLASJ085356.5+001256 & 17.8 &  4.56 & 1.56 &  57.4 & 0.29 & 1.08 \\	 
		HATLASJ085450.2+021207 & 14.0 &  3.62 & 2.58 & 150.3 & 0.52 & 1.48 \\	 
		HATLASJ085616.0+005237 & 13.9 &  0.97 & 2.54 &  78.1 & 0.10 & 1.05 \\	 
		HATLASJ085748.0+004641 & 10.1 &  0.72 & 3.48 & 125.3 & 0.10 & 1.28 \\	 
		HATLASJ085828.5+003815 &  8.8 &  7.51 & 5.93 & 121.0 & 0.25 & 1.19 \\	        
		HATLASJ085836.0+013149 & --   &  --   & --   & --    & --   & --   \\      
		HATLASJ090004.9+000447 & 12.5 &  1.85 & 2.84 &  47.6 & 0.22 & 1.47 \\      
		HATLASJ090750.0+010141 &  8.2 &  1.49 & 5.40 &  66.3 & 0.28 & 1.89 \\	 
		HATLASJ091205.8+002655 &  9.8 &  0.97 & 4.04 &  52.2 & 0.07 & 1.24 \\	 
                \hline
        \end{tabular}
        \caption{\label{tab:table2}
    	GAMA's morphological $K-$band photometric parameters for the `resolved' galaxy sub-sample from VALES.
	$\mu_{0,K}$ is the central surface brightness value. $r_{1/2,K}$ and $n_S$ are the half-light radius and the S\'ersic photometric index, respectively. 
	PA$_K$ is the position angle of the major axis. The ellipticity `$e$' is derived from the semi-major and minor axis ratio ($e\equiv1-b/a$).
	The chi-square of the best two-dimensional fitted photometric model is given in the last column (see \S\ref{sec:gama-galfit} for more details).}
    	
\end{table}

\begin{landscape}
\begin{table} 
	\centering
	\setlength\tabcolsep{4pt}
	TABLE 2: GALAXY PROPERTIES\\
	\begin{tabular}{lcccccccccccccc} 
		\\
		\hline
		\hline
		$H$ATLAS-DR1 ID & RA & Dec & $z_{\rm spec}$ & $\log$\,$M_\star$ & $\log$\,$L_{\rm{FIR}}$ & $L_{\rm{[C\,II]}}$ & $L'_{\rm{CO}}$ & $\theta_{\rm FWHM}$ & $r_{1/2,\rm CO}$ & inc. & $\sigma_v$ & $V_{\rm rot}$ & $\chi^2_\nu$ & Class \\
		   & J2000 & J2000 & & $M_{\odot}$ & $L_{\odot}$ & $\times$10$^{8}$\,$L_{\odot}$ & $\times$10$^{10}$\,$L_{\odot}$ & kpc & kpc & deg & km/s & km/s & & \\
		(1)& (2) & (3) & (4) & (5) & (6) & (7) & (8) & (9) & (10) & (11) & (12) & (13) & (14) & (15) \\
		\hline
		HATLASJ083601.5+002617   & 08:36:01.6 &   +00:26:18.1 & 0.03322 & 10.59$\pm$0.1 & 10.31$\pm$0.02 & 0.97$\pm$0.02 & 0.104$\pm$0.004 & 2.22 & 3.4$\pm$0.1 & 80.8$\pm$0.1 &  24$\pm$2 & 180$\pm$3  & 0.13 & R \\
           	HATLASJ083745.1$-$005141 & 08:37:45.2 & $-$00:51:40.9 & 0.03059 & 10.35$\pm$0.1 & 10.13$\pm$0.03 & 0.73$\pm$0.01 & 0.034$\pm$0.003 & 2.06 & 4.4$\pm$0.3 & 57.2$\pm$0.1 &  25$\pm$1  & 115$\pm$1  & 0.14 & R \\
             	HATLASJ083831.9+000045   & 08:38:31.9 &   +00:00:45.0 & 0.07806 & 10.27$\pm$0.1 & 11.15$\pm$0.01 & 2.43$\pm$0.10 & 0.250$\pm$0.019 & 6.05 & --          & --           &  --        &  --        & --   & C \\
             	HATLASJ084217.7+021222   & 08:42:17.9 &   +02:12:23.4 & 0.09602 & 10.53$\pm$0.1 & 10.93$\pm$0.04 & 2.28$\pm$0.11 & 0.249$\pm$0.020 & 5.58 & 12.0$\pm$3.1 & 77.5$\pm$0.2 &  35$\pm$7  &  75$\pm$3  & 0.16 & R \\
		HATLASJ084305.0+010858   & 08:43:05.1 &   +01:08:56.0 & 0.07770 & 10.41$\pm$0.2 & 11.05$\pm$0.03 & --            & 0.166$\pm$0.017 & 6.13 & --          & --           &  --        &  --        & --   & C \\     
		HATLASJ084350.7+005535   & 08:43:50.8 &   +00:55:34.8 & 0.07294 & 10.64$\pm$0.1 & 11.03$\pm$0.01 & 1.70$\pm$0.09 & 0.191$\pm$0.016 & 5.52 & 4.2$\pm$0.3 & 67.3$\pm$0.7 &  70$\pm$8 &  58$\pm$3  & 0.71 & R \\    
		HATLASJ084428.3+020349   & 08:44:28.4 &   +02:03:49.8 & 0.02538 & 10.29$\pm$0.1 & 10.25$\pm$0.01 & 0.33$\pm$0.01 & 0.041$\pm$0.003 & 1.76 & 2.4$\pm$0.4 & 80.0$\pm$0.2 &  59$\pm$14 &  69$\pm$6  & 0.86 & R \\   
		HATLASJ084428.3+020657   & 08:44:28.4 &   +02:06:57.4 & 0.07864 & 10.78$\pm$0.1 & 11.01$\pm$0.03 & 4.51$\pm$0.14 & 0.392$\pm$0.051 & 6.22 & 6.3$\pm$0.5 & 83.5$\pm$0.2 &  39$\pm$2  & 162$\pm$3  & 2.50 & R \\   
		HATLASJ084630.7+005055   & 08:46:30.9 &   +00:50:53.3 & 0.13232 & 10.36$\pm$0.1 & 11.51$\pm$0.02 & 6.22$\pm$0.63 & 0.463$\pm$0.042 & 7.51 & 4.5$\pm$0.5 & 33.3$\pm$0.1 &  37$\pm$28 & 215$\pm$8  & 20.6 & R \\   
		HATLASJ084907.0$-$005139 & 08:49:07.1 & $-$00:51:37.7 & 0.06979 & 10.48$\pm$0.1 & 11.18$\pm$0.01 & 2.28$\pm$0.09 & 0.279$\pm$0.022 & 5.27 & 4.7$\pm$1.3 & 45.9$\pm$0.3 &  54$\pm$3  & 108$\pm$4  & 0.33 & R \\
        	HATLASJ085111.5+013006   & 08:51:11.4 &   +01:30:06.9 & 0.05937 & 10.56$\pm$0.1 & 10.72$\pm$0.02 & 2.66$\pm$0.07 & 0.198$\pm$0.007 & 4.63 & 6.5$\pm$0.4 & 76.2$\pm$0.1 &  31$\pm$8  & 207$\pm$4  & 0.22 & R \\
		HATLASJ085112.9+010342   & 08:51:12.8 &   +01:03:43.7 & 0.02669 & 10.14$\pm$0.1 & 10.20$\pm$0.01 & 0.24$\pm$0.01 & 0.020$\pm$0.003 & 1.85 & 1.1$\pm$0.3 & 58.0$\pm$0.9 &  43$\pm$19 &  81$\pm$4  & 0.74 & R \\   
		HATLASJ085234.4+013419   & 08:52:33.9 &   +01:34:22.7 & 0.19500 & 10.57$\pm$0.1 & 11.92$\pm$0.01 & --            & 1.999$\pm$0.012 & 14.9 & --          & --           &  --        &  --        & --   & C \\
	     	HATLASJ085340.7+013348   & 08:53:40.7 &   +01:33:47.9 & 0.04101 & 10.36$\pm$0.1 & 10.28$\pm$0.03 & 0.95$\pm$0.02 & 0.061$\pm$0.003 & 2.95 & 3.6$\pm$0.3 & 39.0$\pm$0.2 &  24$\pm$3  & 181$\pm$14 & 0.13 & R \\
        	HATLASJ085346.4+001252   & 08:53:46.3 &   +00:12:52.4 & 0.05044 & 10.31$\pm$0.1 & 10.71$\pm$0.01 & 2.18$\pm$0.04 & 0.076$\pm$0.002 & 3.57 & 6.4$\pm$0.4 & 89.7$\pm$0.2 &  33$\pm$5  & 134$\pm$4  & 0.26 & R \\
             	HATLASJ085356.5+001256   & 08:53:56.3 &   +00:12:56.3 & 0.05084 & 10.01$\pm$0.1 & 10.33$\pm$0.03 & 1.41$\pm$0.04 & 0.068$\pm$0.002 & 3.60 & 2.6$\pm$0.2 & 52.2$\pm$0.1 &  25$\pm$4  & 109$\pm$2  & 0.14 & R \\
             	HATLASJ085450.2+021207   & 08:54:50.2 &   +02:12:08.3 & 0.05831 & 10.66$\pm$0.1 & 10.70$\pm$0.02 & 2.30$\pm$0.08 & 0.202$\pm$0.019 & 4.66 & 3.9$\pm$0.1 & 70.4$\pm$0.1 &  39$\pm$16 & 287$\pm$6  & 1.52 & R \\
             	HATLASJ085616.0+005237   & 08:56:16.0 &   +00:52:36.2 & 0.16916 & 10.96$\pm$0.1 & 10.94$\pm$0.01 & --            & 0.443$\pm$0.076 & 10.4 & --          & --           &  --        & --         & --   & C \\  
		HATLASJ085748.0+004641   & 08:57:48.0 &   +00:46:38.7 & 0.07177 & 10.37$\pm$0.1 & 11.27$\pm$0.01 & 4.69$\pm$0.09 & 0.276$\pm$0.014 & 5.57 & 4.6$\pm$0.3 & 70.2$\pm$0.1 &  51$\pm$5  &  43$\pm$3  & 0.10 & R \\
		HATLASJ085828.5+003815   & 08:58:28.6 &   +00:38:14.8 & 0.05236 & 10.43$\pm$0.1 & 10.44$\pm$0.02 & 0.94$\pm$0.03 & 0.043$\pm$0.005 & 3.72 & 2.6$\pm$0.2 & 52.3$\pm$0.1 &  22$\pm$2  & 159$\pm$2  & 0.16 & R \\
		HATLASJ085836.0+013149   & 08:58:36.0 &   +01:31:49.0 & 0.10677 & 10.90$\pm$0.1 & 11.22$\pm$0.01 & 5.30$\pm$0.21 & 0.554$\pm$0.011 & 6.17 & 11.1 $\pm$0.8 & 80.0$\pm$0.1 &  27$\pm$4  &  91$\pm$1  & 0.19 & R \\
		HATLASJ090004.9+000447   & 09:00:05.0 &   +00:04:46.8 & 0.05386 & 10.70$\pm$0.1 & 10.57$\pm$0.02 & 1.86$\pm$0.06 & 0.153$\pm$0.022 & 3.80 & 2.7$\pm$0.1 & 42.5$\pm$0.2 &  25$\pm$2  & 193$\pm$10 & 0.22 & R \\  
		HATLASJ090750.0+010141   & 09:07:50.1 &   +01:01:41.8 & 0.12808 & 10.14$\pm$0.1 & 11.70$\pm$0.01 & 9.33$\pm$0.40 & 0.535$\pm$0.045 & 7.36 & 10.3 $\pm$0.6 & 44.8$\pm$1.4 &  58$\pm$6  &  35$\pm$5  & 0.12 & R \\  
		HATLASJ090949.6+014847   & 09:09:49.6 &   +01:48:46.0 & 0.18186 & 10.89$\pm$0.1 & 11.84$\pm$0.02 & 13.8$\pm$0.68 & 1.364$\pm$0.093 & 12.7 & --          & --           &  --        &  --        & --   & C \\   
		HATLASJ091157.2+014453   & 09:11:57.2 &   +01:44:53.9 & 0.16945 & 10.90$\pm$0.2 & 11.39$\pm$0.01 & --            & 0.737$\pm$0.072 & 11.0 & --          & --           &  --        &  --        & --   & C \\   
		HATLASJ091205.8+002655   & 09:12:05.8 &   +00:26:55.6 & 0.05446 & 10.33$\pm$0.1 & 11.09$\pm$0.01 & 1.45$\pm$0.05 & 0.187$\pm$0.011 & 3.94 & 2.6$\pm$0.3 & 21.0$\pm$0.5 &  79$\pm$24 & 116$\pm$12 & 0.11 & R \\   
		HATLASJ091420.0+000509   & 09:14:20.0 &   +00:05:10.0 & 0.20216 & 10.62$\pm$0.1 & 11.55$\pm$0.01 & --            & 0.667$\pm$0.114 & 13.0 & --          & --           &  --        &  --        & --   & C \\    
		HATLASJ091956.9+013852   & 09:19:57.0 &   +01:38:51.6 & 0.17635 & 10.45$\pm$0.1 & 11.13$\pm$0.01 & --            & 0.365$\pm$0.048 & 11.7 & --          & --           &  --        &  --        & --   & C \\
              	HATLASJ113858.4$-$001629 & 11:38:58.5 & $-$00:16:30.2 & 0.16370 & 10.84$\pm$0.1 & 11.21$\pm$0.01 & --            & 0.546$\pm$0.129 & 8.94 & --          & --           &  --        &  --        & --   & C \\
              	HATLASJ114343.9+000203   & 11:43:44.1 &   +00:02:02.5 & 0.18716 & 10.10$\pm$0.1 & 11.05$\pm$0.01 & --            & 0.485$\pm$0.089 & 10.1 & --          & --           &  --        &  --        & --   & C \\
		HATLASJ114625.0$-$014511 & 11:46:25.0 & $-$01:45:13.0 & 0.16450 & 10.72$\pm$0.1 & 11.72$\pm$0.01 & --            & 0.861$\pm$0.084 & 8.91 & --          & --           &  --        &  --        & --   & C \\ 
		HATLASJ121141.8$-$015730 & 12:11:41.8 & $-$01:57:29.7 & 0.31704 & 11.18$\pm$0.1 & 11.80$\pm$0.01 & --            & 0.210           & 15.1 & --          & --           &  --        &  --        & --   & C \\ 
		HATLASJ121253.5$-$002203 & 12:12:53.5 & $-$00:22:04.4 & 0.18548 & 10.79$\pm$0.1 & 11.11$\pm$0.01 & --            & 0.447$\pm$0.065 & 9.71 & --          & --           &  --        &  --        & --   & C \\     
		HATLASJ121427.3+005819   & 12:14:27.4 &   +00:58:18.3 & 0.18045 & 10.93$\pm$0.1 & 11.27$\pm$0.01 & --            & 0.460$\pm$0.069 & 9.63 & --          & --           &  --        &  --        & --   & C \\     
		HATLASJ121446.4$-$011155 & 12:14:46.5 & $-$01:11:55.6 & 0.17971 & 10.82$\pm$0.1 & 11.55$\pm$0.01 & --            & 0.765$\pm$0.094 & 9.45 & --          & --           &  --        &  --        & --   & C \\    
		HATLASJ140912.3$-$013454 & 14:09:12.5 & $-$01:34:54.9 & 0.26492 & 10.97$\pm$0.1 & 11.89$\pm$0.01 & --            & 1.494$\pm$0.231 & 9.17 & --          & --           &  --        &  --        & --   & C \\  
		HATLASJ141008.0+005106   & 14:10:08.0 &   +00:51:06.9 & 0.25641 & 11.10$\pm$0.1 & 11.83$\pm$0.01 & --            & 1.311$\pm$0.295 & 8.80 & --          & --           &  --        &  --        & --   & C \\
             	HATLASJ142057.9+015233   & 14:20:58.0 &   +01:52:32.1 & 0.26462 & 10.86$\pm$0.1 & 11.64$\pm$0.01 & --            & 1.238$\pm$0.231 & 9.55 & --          & --           &  --        &  --        & --   & C \\
		HATLASJ142517.1+010546   & 14:25:17.1 &   +01:05:46.6 & 0.28069 & 11.07$\pm$0.1 & 11.84$\pm$0.01 & --            & 1.714$\pm$0.237 & 9.98 & --          & --           &  --        &  --        & --   & C \\
		\hline                                                                                        
	\end{tabular}
	\caption{\label{tab:table1}                                                             
	Properties of the galaxies with resolved emission from VALES. The FIR luminosities are calculated across the 8--1000\,${\mu}$m wavelength range. $\theta_{\rm FWHM}$ 
	is the synthesized beam major axis size. The CO(1-0) half-light radii ($r_{1/2, \rm CO}$) are deconvolved by the synthesized beam. The inclination angle is 
	defined as the angle between the line-of-sight (LOS) and the plane perpendicular to the galactic disk (for a face-on galaxy, inc = 0\,deg.). $\sigma_v$  is the median velocity dispersion 
	corrected for ``beam smearing'' effects; see \S\ref{sec:dispersion velocity}. $V_{\rm rot}$ is the rotational velocity at 2 times the CO(1-0) half-light radius corrected
	for inclination. $\chi^2_\nu$ is the reduced chi-square of the best two-dimensional fit. The galaxy classification in the final column denotes `Resolved' (R) or `Compact' (C)
	(see \S\ref{sec:VALES_sample} for more details).}

\end{table}
\end{landscape}

\section{METHODS}
\subsection{GAMA's morphological models} 
\label{sec:gama-galfit}
With the advent of the multiple IFS surveys at high redshift (e.g. \citealt{Forster2009,Wisnioski2015,Stott2016}), 
kinematic models have experienced a rapid development and becoming more complex
by taking into account multiple galaxy components and adding multiple degrees of freedom (e.g. 
\citealt{Swinbank2017}). The latter increases the parameter degeneracy, especially regarding inclination 
angle when low spatially resolved observations are analysed. Thus, additional information must be considered
in order to derive robust kinematic parameters from the observed velocity fields. With the aim to 
minimise parameter degeneracy, we supported our kinematic analysis by taking into account previous 
S\'ersic photometry models \citep{Sersic1963} available for the GAMA survey data (Table~\ref{tab:table2}; 
\citealt{Liske2015}). Those models are produced by using SIGMA (Structural Investigation of Galaxies via 
Model Analysis; \citealt{Kelvin2012}) on Sloan Digital Sky Survey (SDSS) and UKIRT Infrared Deep Sky Survey (UKIDSS) 
imaging data. We use the $K$-band image models to characterise stellar component of each galaxy through
the half-light radius ($r_{1/2,K}$), the orientation of major axis indicated by the position angle (PA$_K$), and the inclination angle derived
from the minor to major axis ratio (b/a). We use this inclination value to constraint the galactic inclination of the molecular gas content in the kinematic modelling.
We note, however, that the error estimates produced by SIGMA are determined from the covariance 
matrix used in the fitting procedure. As a result, the uncertainty of the inclination value tend to be underestimated
\citep{Haubler2007,Bruce2012}. Therefore, we adopt more reasonable error to the galactic inclination and discuss its choice in the following subsection. 
Out of the 20 resolved galaxies analysed in this work, 19 sources have this morphological GAMA modelling.
We do not use the inclination value derived for HATLASJ085836.0+013149 from its morphological model as it implies an unrealistic central 
surface brightness magnitude value of $-$18\,mag\,arcsec$^{-2}$. This galaxy was analysed without constraint on the kinematic parameters.

\subsection{Inclination angles} 
\label{sec:inc_angles}
The correct estimate of inclination angles is a critical issue for kinematic analyses. This parameter is used to correct the 
observed velocity field, which is the projected component of the intrinsic velocity field of the galaxy across the line-of-sight. 
With the aim to take into account the galactic `disk thickness', we model the galaxies in our sample as oblate spheroid systems.
By using the minor-major axis ratio (b/a) taken from the GAMA data, the galaxy inclination angle can be expressed as: 

\begin{equation}
    {\rm cos^2(i)}=\frac{{\rm (b/a)^2}-{\rm q_0^2}}{1-{\rm q_0^2}},
	\label{eq:inc_eqn}
\end{equation}

\noindent where `\textbf{$i$}' is the galaxy inclination angle and ${\rm q_0}$ is the axis ratio of the galaxy as if it would 
be seen as an edge-on system \citep{Holmberg1958}. In the thin-disk approximation, i.e. ${\rm q_0=0}$, the Equation~\ref{eq:inc_eqn} is
reduced to the simplistic approximation ${\rm (b/a)=cos(i)}$. Although we have no information about the `disk thickness'
for our sample, we adopt ${\rm q_0=0.14}$, which is the mean b/a ratio found in edge-on disk galaxies at 
low redshift ($z<0.05$; \citealt{Mosenkov2015}). We consider a conservative approach for the inclination angle uncertainties of 10\% in order to get 
realistic error estimates for the inclination angles and use them instead of the underestimated values derived from SIGMA 
(see~\ref{sec:gama-galfit}), as suggested by the results of the Monte Carlo methodology used by \citet{Epinat2012}. We use the 
inclination angles derived from SIGMA as initial guesses for our kinematic analysis and we 
allow them to vary within a 3\,$\sigma$ range. For the galaxies without a SIGMA fitting, we consider a range between 0 and 90 
degrees with an initial guess of $\rm(b/a)\sim$0.7 (i$\sim$55$^\circ$), that is the mean axis ratio derived by \citet{Law2012a} for a
randomly oriented spheroidal galaxy population. 

\subsection{Kinematic model} 
\label{sec:kinetic_model}
We attempt to model the two-dimensional velocity field by first identifying the dynamical
centre and the kinematic major axis. Considering the modest spatial resolution
of our observations and the smoothness of the intensity maps, we constrain the kinematic centre
to the CO(1-0) intensity peak location. We follow \citet{Swinbank2012a} to construct 
two-dimensional models with an input rotation curve following an arctan function
[$V(r)=\frac{2}{\pi}V_{\rm asym}$arctan(r/r$_{\rm t}$)], where $V_{\rm asym}$ is the 
asymptotic rotational velocity and r$_{\rm t}$ is the effective radius at which the 
rotation curve turns over \citep{Courteau1997}. This model has four free parameters
[$V_{\rm asym}$, r$_{\rm t}$, position angle (PA) and disk inclination] and 
a genetic algorithm \citep{Charbonneau1995} is used to find the best fit (see \citealt{Swinbank2012a}
for more details). 
The parameter uncertainties are calculated by considering an confidence limit of $\Delta \chi_\nu^2=1$.
An example of the best-fit kinematic maps and velocity residuals are shown in 
Fig.~\ref{fig:example_maps}, whilst the full sample maps are presented in the appendix (Fig.~\ref{fig:maps}). The best-fit
inclination values are given in Table~\ref{tab:table1}. The mean deviation from the best-fit models 
within the sample (indicated by the typical root-mean-squared; r.m.s) is $\langle$data\,$-$\,model$\rangle$\,=\,17$\pm$9\,km\,s$^{-1}$ 
with a range of $\langle$data\,$-$\,model$\rangle$\,=\,7--48\,km\,s$^{-1}$. We show this value for each galaxy in its residual map.

\begin{figure}
\flushleft
\includegraphics[width=0.452\columnwidth]{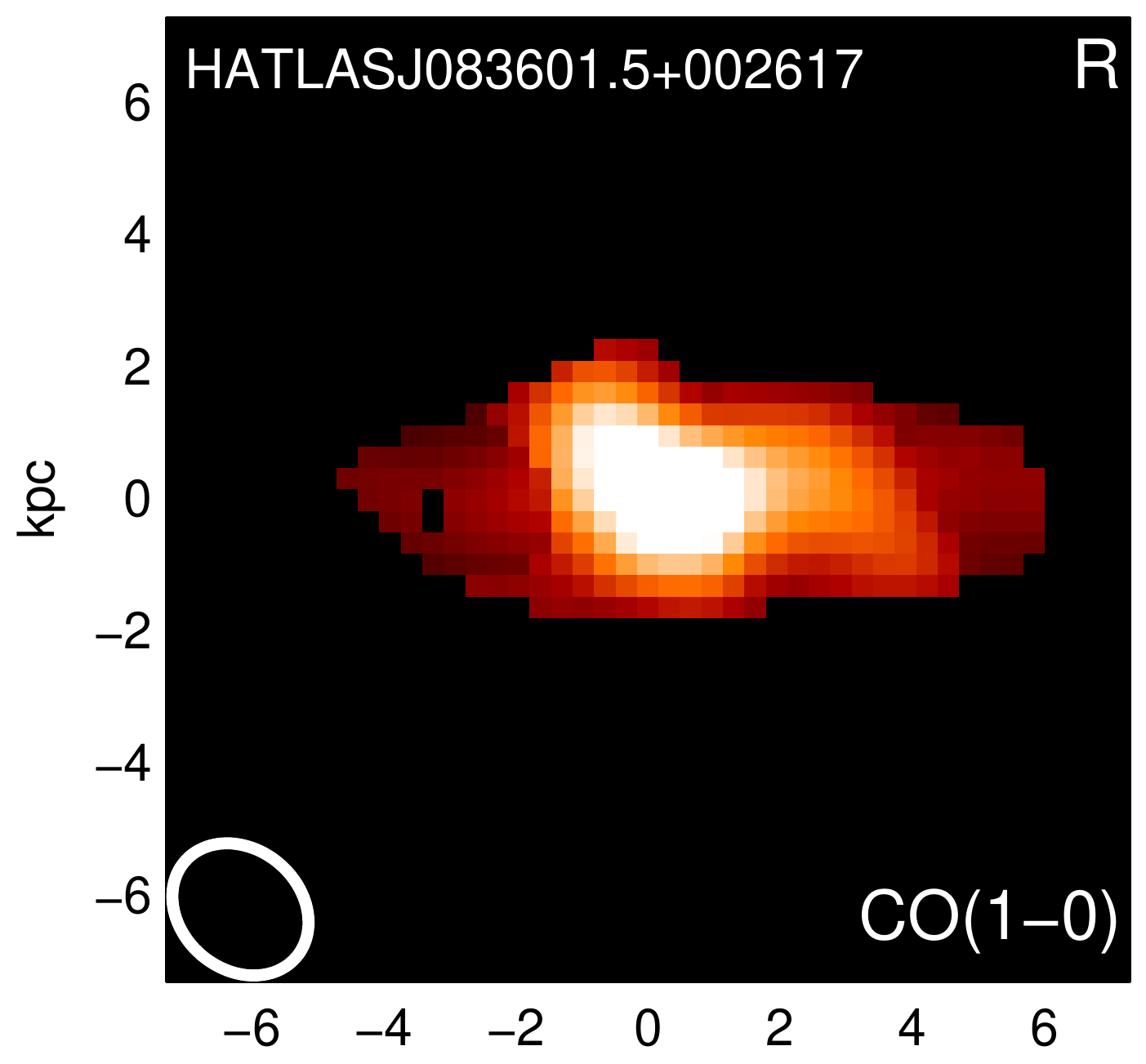}
\includegraphics[width=0.42\columnwidth]{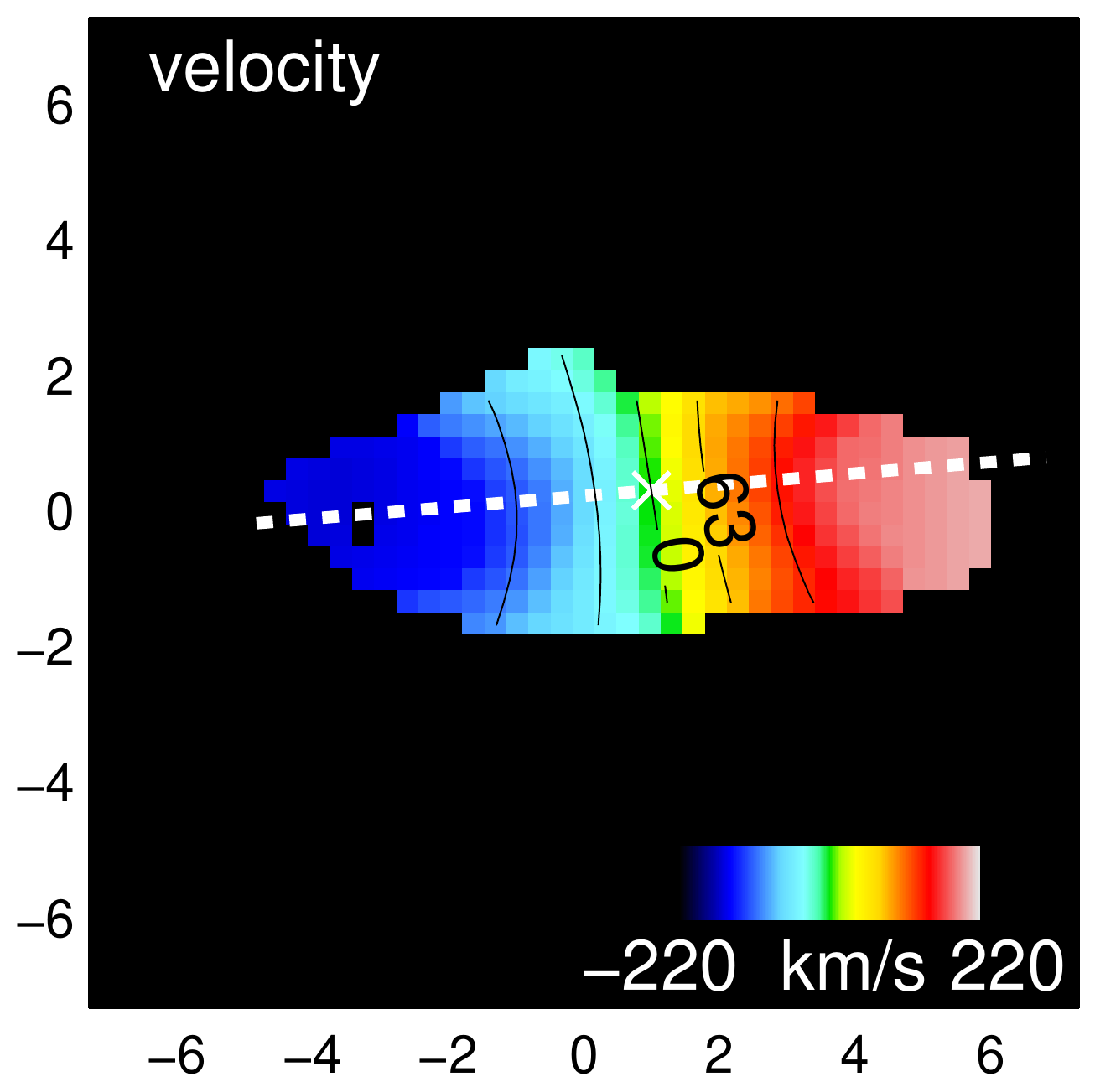}\\
\includegraphics[width=0.452\columnwidth]{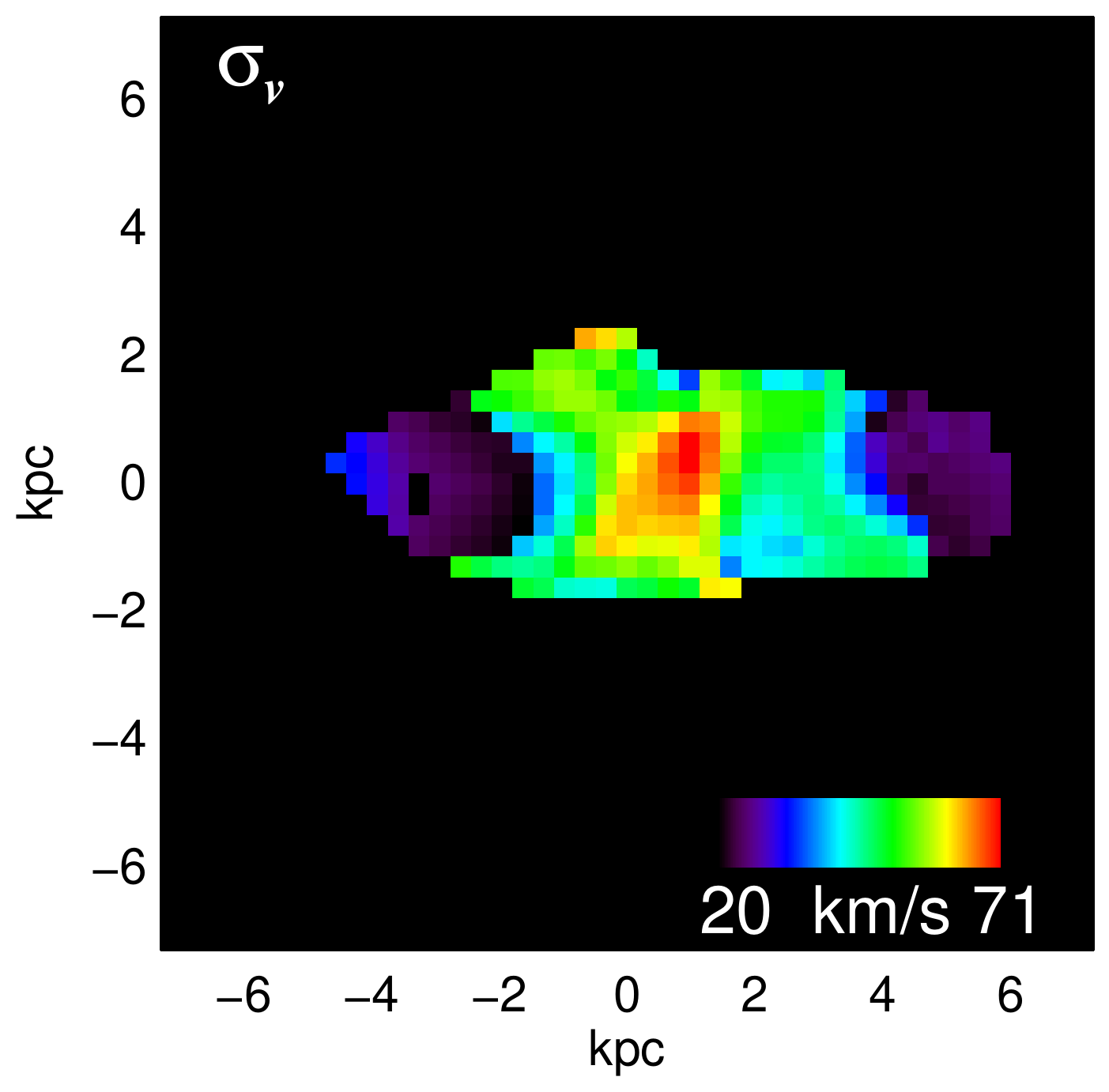}
\includegraphics[width=0.42\columnwidth]{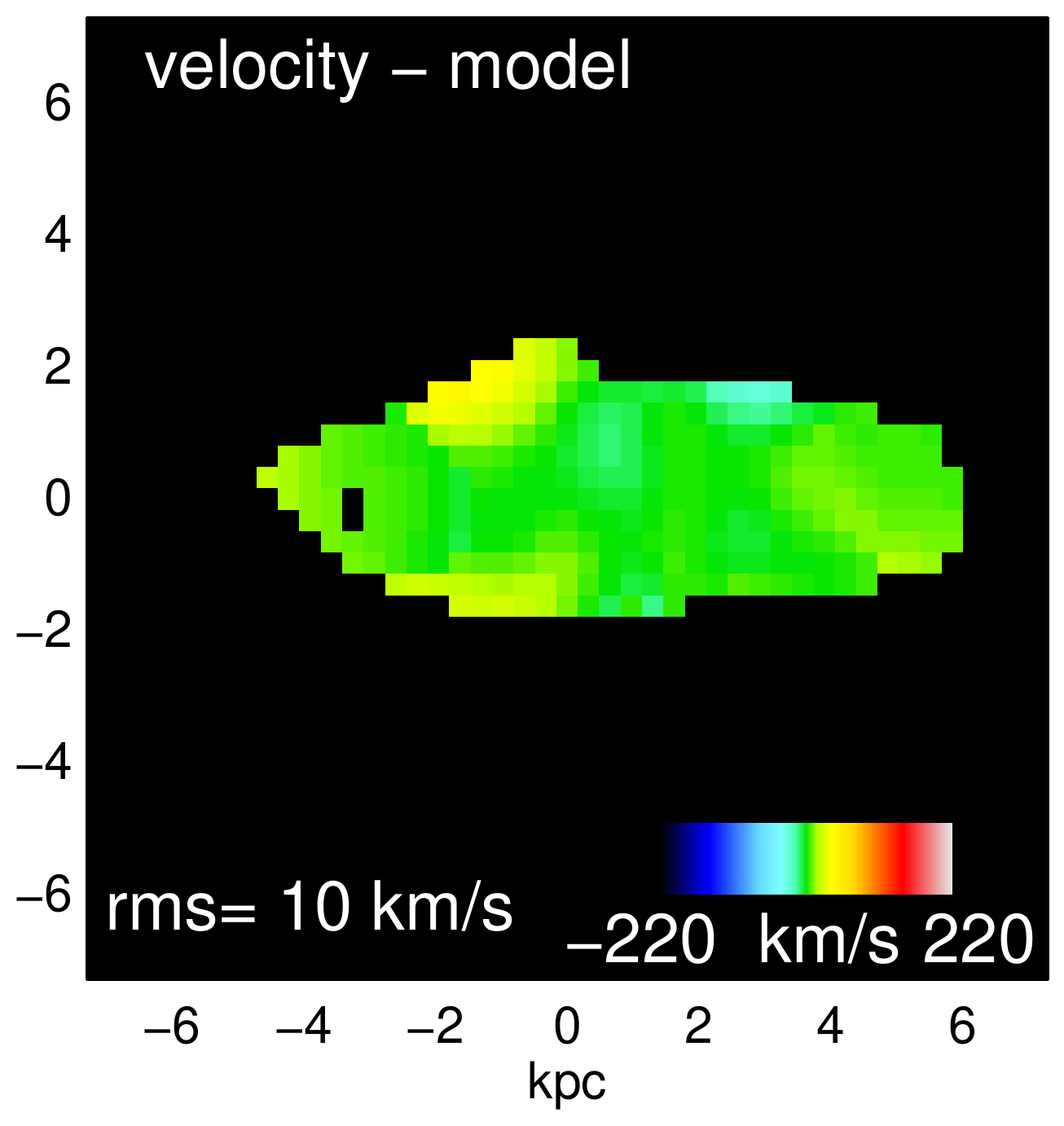}\\
\hspace{1mm}\includegraphics[width=0.459\columnwidth]{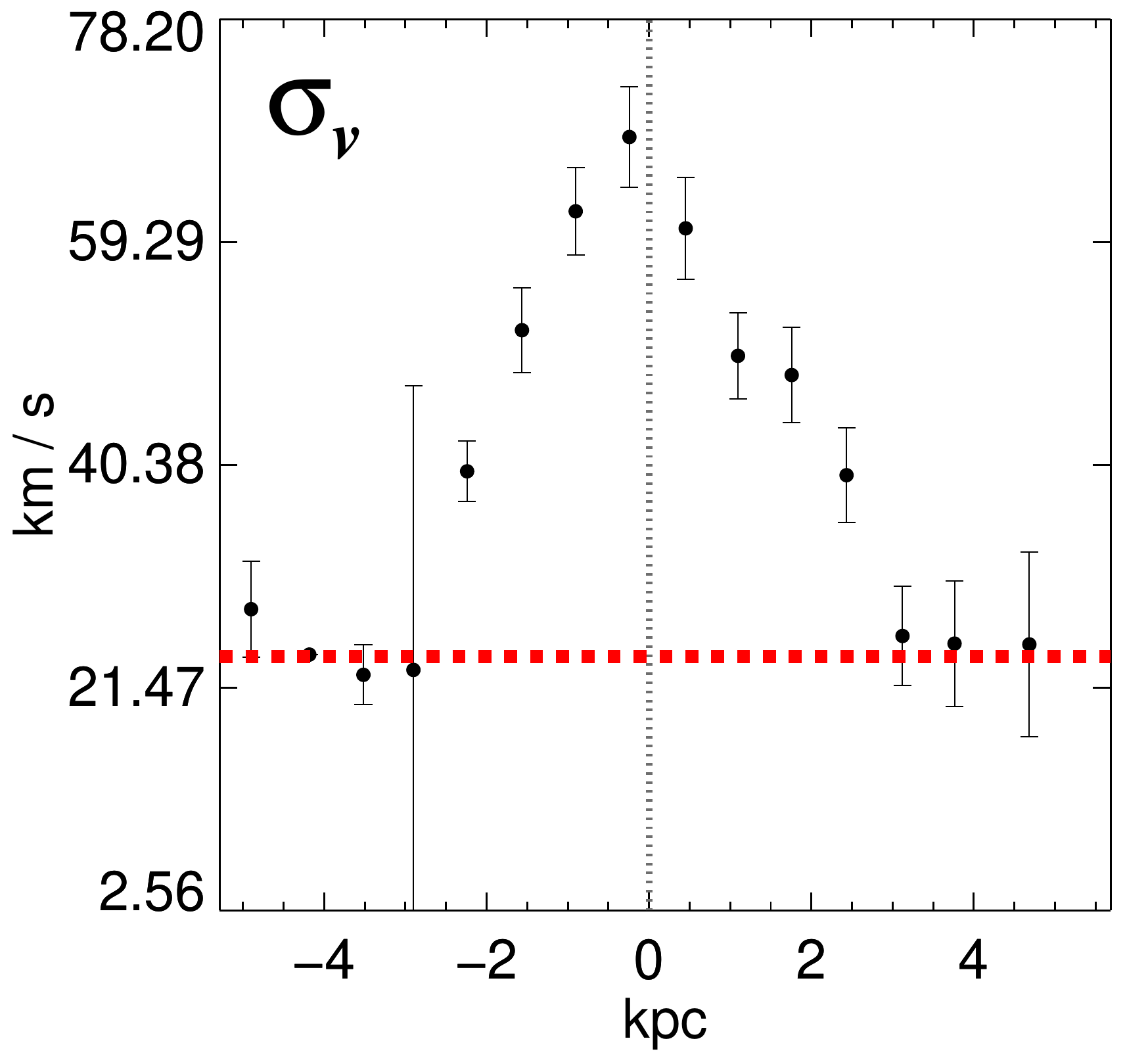}
\hspace{1.5mm} \includegraphics[width=0.462\columnwidth]{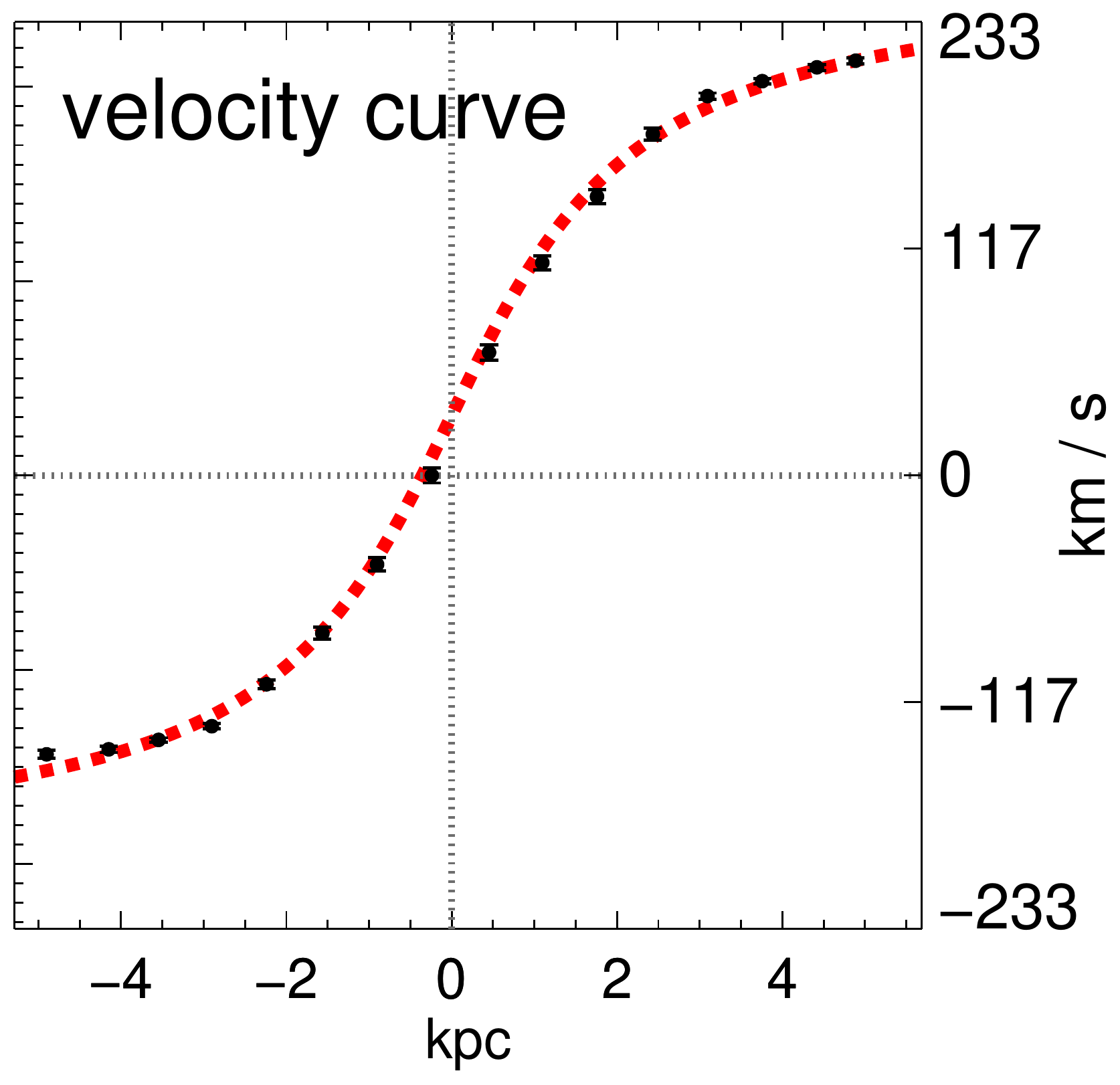}\\
\caption{ \label{fig:example_maps}
Example of the two dimensional maps and one dimensional velocity profiles for one target within our survey. 
The full sample maps, profiles figures and their explanation are shown in the appendix (Fig.~\ref{fig:maps}).
\textit{Left}: From top to bottom; CO(1-0) intensity map, LOS velocity dispersion map and one dimensional velocity dispersion profile. 
\textit{Right}: From top to bottom; rotational velocity map, residual map, and one dimensional rotational velocity profile.
}
\end{figure}

\subsection{CO(1-0) spatial extent} 
\label{sec:spatial-extent}
To measure the spatial extent of the molecular gas of each galaxy, we calculate the CO half-light radii ($r_{1/2,\rm CO}$).
These are calculated from the cubes, where the encircled CO(1-0) flux decays to half its total integrated value. 
The total integrated value is defined as the total CO(1-0) luminosity within a Petrosian radius. We adopted the 
SDSS Petrosian radius definition with R$_{\rm P,lim}=0.2$. We account for the ellipticity and position angle of the 
galaxy obtained from the best-fit disk model. The $r_{1/2,\rm CO}$ 1\,$\sigma$ errors are derived by bootstrapping
via Monte-Carlo simulations in both, measured emission line intensity and estimated dynamical parameters. The 
half-light radii are corrected for beam-smearing effects by subtracting the synthesized beam major axis width in quadrature. 
The median $r_{1/2,\rm CO}$ for our sample is $4.4 \pm 3$\,kpc (Table~\ref{tab:table1}).

\subsection{Rotation Curve \& Rotational Velocity} 
\label{sec:rotation_curve}
We use the dynamical centre and position angle derived from the best-fit dynamical model to extract the one-dimensional
rotation curve across the major kinematic axis of each galaxy. An example of the extracted rotational curves is 
presented in Fig.~\ref{fig:example_maps}, whilst the rotational curves for all the sample are  shown in the appendix 
(Fig.~\ref{fig:maps}). We define the rotational velocity corrected for inclination ($V_{\rm rot}$) as the velocity observed at two
half-light radii. We note, however, that we are observing the CO(1-0) emission line, thus the radius at which we are 
defining the representative rotational velocity of each source may not be directly related to the radius at which, for example, Integral Field Spectroscopy (IFS) surveys 
might extract rotational velocities using ionized gas dynamics (e.g. \citealt{Forster2009, Swinbank2012a, Green2014, Wisnioski2015, Stott2016}).

\subsection{Velocity Gradient Correction \& Velocity Dispersion}
\label{sec:dispersion velocity} 
As a consequence of the modest spatial resolution of our observations compared to the angular extension of the sources, 
there is a contribution to the derived line widths from the beam-smeared large-scale velocity motions across the galaxy, which
must be corrected for \citep{Davies2011}. This correction is done for each pixel where the CO(1-0) emission is detected.
We calculate the luminosity-weighted velocity gradient across the synthesized beam ($\Delta V$/$\Delta$R) in the model velocity field and we
subtract it linearly from the corresponding velocity dispersion value following Eq.\,A1 from \citet{Stott2016}. However, by using this procedure,
a $\sim20$\% residuals are expected to remain, especially on the centre of each galaxy where large velocity gradients are expected to be present \citep{Stott2016}.

In order to minimize the residual beam-smearing effects in our sample, we define the global velocity dispersion value ($\sigma_v$) for
each galaxy as the median value of the pixels at an angular distance 2 times greater than the angular extension of the synthesized beam from the
best-fitted dynamical centre. This procedure usually calculates $\sigma_v$ by considering 71\,pixels on average with a range of 6-256\,pixels. 
In the case of HATLASJ083601.5+002617 we increased the skipped area to 3 times the synthesized beam size as our method failed due to the high galaxy 
inclination angle ($\sim$80 deg.) plus a beam size not large enough to avoid the zone where velocity gradients were contributing to the 
emission line widths.

While the CO(1-0) emission line width has been traditionally used as a measure of the dynamical mass within a GMC 
(e.g. \citealt{Solomon1987}), the synthesized beam size (2--8\,kpc) within our sample is larger than the biggest
GMC size observed in galaxies ($\sim 1$\,kpc; e.g. \citealt{Swinbank2012b}); resulting in the smoothness of our 
galactic intensity maps (Fig.~\ref{fig:maps}). Thus, throughout our work, we interpret the CO(1-0) emission line width as a 
tracer of the molecular gas random motions seen over a resolution element area. This is the key property of our `resolved' 
sample as we can study the dynamics of the molecular gas directly. This opens a window of dynamical analyses which is not necessary the same as
those performed in IFS galaxy surveys which use (mainly) the ionized gas to characterise the dynamical state of galaxies.

\subsection{Spatial and spectral resolutions effects}
\label{sec:rel_effects}

In order to estimate the effect of the spatial and spectral resolution for the VALES sample
on the kinematic parameters, we use ALMA Band-3 observations with higher resolution of 
$\sim$\,0$\farcs$5 ($\sim$\,kpc scale at z$\,\sim\,0.1-0.2$) and 12\,km\,s$^{-1}$ towards three VALES galaxies (Ibar et al.\ in prep.).
The high resolution of those observations allows us to study in detail how spectral resolution and beam-smearing effects affect the derived kinematic parameters.

\begin{figure}
\centering
\includegraphics[width=1.05\columnwidth]{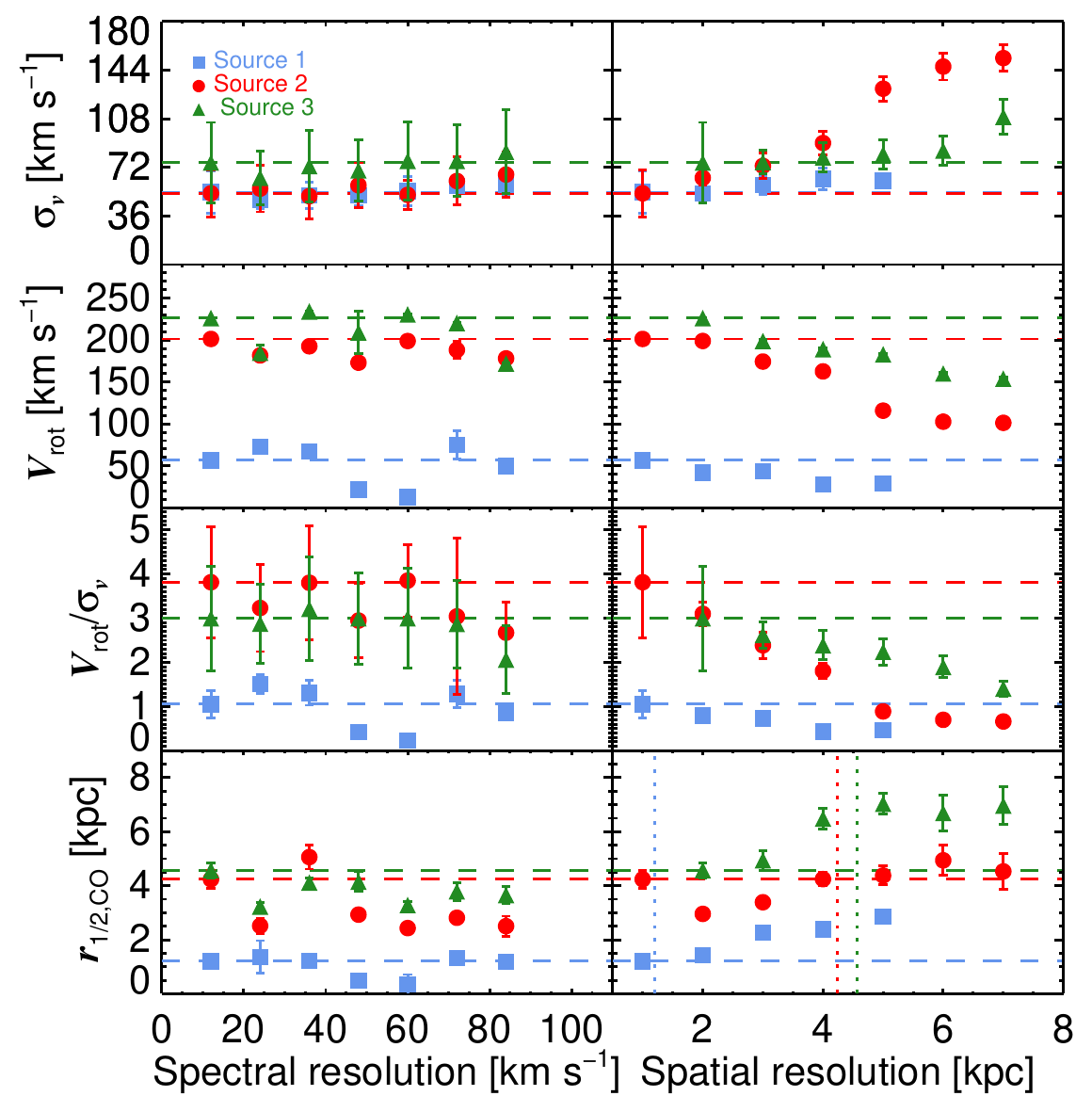}
\caption{ \label{fig:rel_effects}
	 Velocity dispersion, rotational velocity, rotational velocity to velocity dispersion ratio ($V_{\rm rot}/\sigma_v$) and CO(1-0) intensity half-light radius (rows)
	 as a function of the spectral and spatial resolution (columns). Those values were derived from mock data-cubes produced by the convolution of a three 
	 dimensional gaussian kernel with the original observations. The spatial resolution corresponds to the projected major axis (FWHM) of the synthesized beam.
	  The blue, red and green horizontal dashed lines represent the kinematic `fiducial' values for each  source. The blue, red and green vertical dotted lines represent the 
	 `fiducial' $r_{1/2,\rm CO}$ values for each galaxy (see \S~\ref{sec:rel_effects} for more details).}
\end{figure}

We create mock-observations by spatially degrading the images using two-dimensional Gaussian kernel, while also re-binning the spectral  channels to mimic 
lower spectral resolutions. The channel width is increased by 12\,km\,s$^{-1}$ per step between $\sim12-84$\,km\,s$^{-1}$, whilst the spatial resolution is degraded
by 1\,kpc per step between $\sim1-7$\,kpc (up to $\sim 3$ times the `fiducial' half-light radius). From those mock data-cubes we fit the CO(1-0) emission line,
we derive its best-fit kinematic model and calculate the $V_{\rm rot}$, $\sigma_v$ and $r_{1/2, \rm CO}$ following the procedures described in the 
previous sections, but we keep the position angle fixed to the value obtained for the data-cube with higher spatial and spectral resolutions.
In Fig.~\ref{fig:rel_effects} we show how the fitted kinematic parameters (rows) depend on spectral resolution (left column)
at fixed $\sim$1\,kpc scale and spatial resolution (right column) at fixed 12\,km\,s$^{-1}$ for the three sources. We consider the `fiducial'
value of each kinematic parameter for each source as the values derived for the data-cubes with higher spectral and spatial resolutions 
(12\,km\,s$^{-1}$ and $\sim$1\,kpc), and are represented by the horizontal dashed lines in each plot. The fiducial values for the three
galaxies are; $V_{\rm rot}=$ 56, 200 and 226\,km\,s$^{-1}$; $\sigma_v=$ 54, 53 and 76\,km\,s$^{-1}$; and $r_{1/2,\rm CO}=$ 1.2, 4.2 and 4.6\,kpc.

\begin{figure*}
\centering
\includegraphics{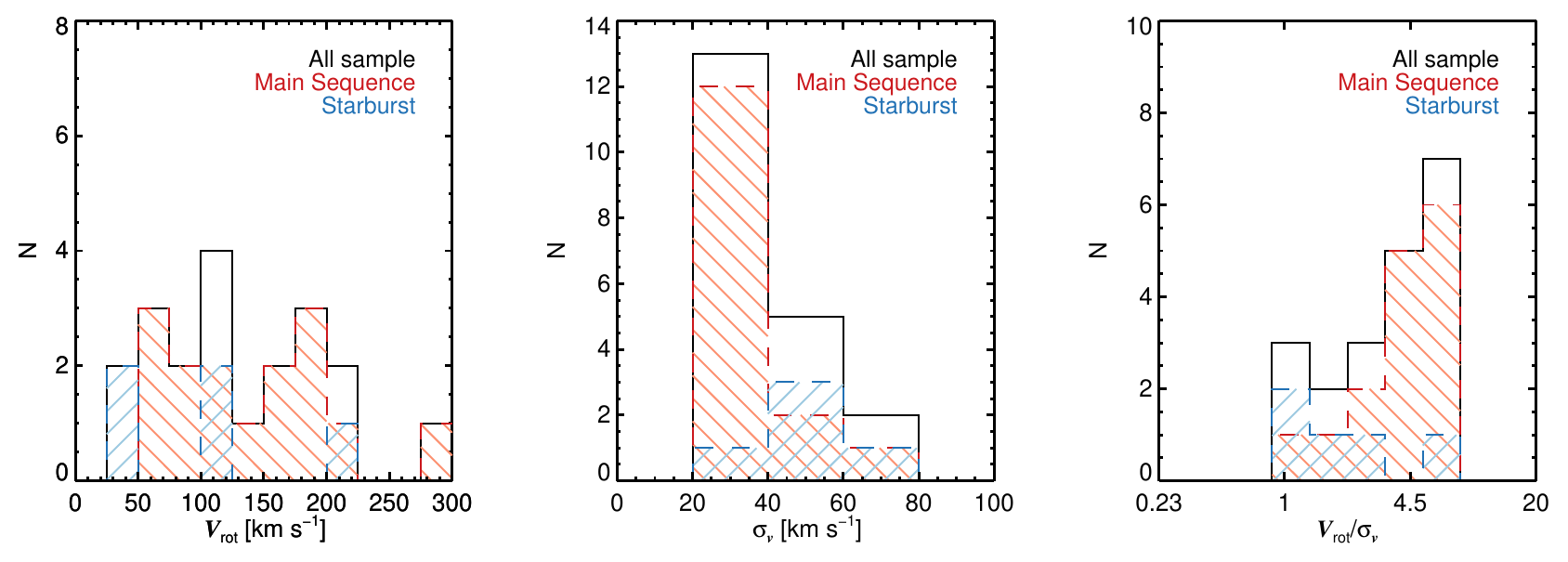}
\caption{ \label{fig:histograms}
The distribution of the rotational velocity ($V_{\rm rot}$; \textit{Left}), velocity dispersion ($\sigma_v$; \textit{middle}) and $V_{\rm rot}/\sigma_v$ (\textit{Right}) within our sample. 
In the three panels we also show the distributions for the `normal' star-forming galaxies (dashed-red) and `starburst' galaxies (dashed-blue). This classification was done by following
the same procedure done by V17 for the VALES survey (see \S~\ref{sec:VALES_sample}). Our resolved sample shows a wide ride of rotation velocities and velocity dispersions.}
\end{figure*}

In Fig.~\ref{fig:rel_effects}, we see how the measured galactic velocity dispersion remains constant when the spectral resolution is degraded.
We also see an increase of the velocity dispersion
when we spatially degrade the cubes, however, we note that the galaxy with the lowest `fiducial' rotational velocity value is also the galaxy less affected
by spatial resolution effect. This is consistent with the picture in which the velocity gradient within the beam area contributes to the emission line width 
represented by the velocity dispersion. We note also that galaxy mass and inclination may also affect the $\sigma_v$ estimation (e.g. \citealt{Burkert2016}).

In the second row of Fig.~\ref{fig:rel_effects} we measure $V_{\rm rot}$ for each data-cube. Although we can recover nearly the same $V_{\rm rot}$
value regardless of the spectral resolution, we can see how it varies when we spatially degrade the cubes. At poor spatial resolution, lower rotational velocity
values are recovered. This effect is expected as the observed emission line is the result of the convolution of the emission lines produced within the beam area. 
This convolution flavour brighter emission lines which are mainly produced in the central part of the galaxy where $V_{\rm rot}$ is lower.

In the third row of Fig.~\ref{fig:rel_effects} we show the variation of the $V_{\rm rot}/\sigma_v$ ratio as a function of spectral and spatial resolutions. 
We see how this ratio is not affected by the increase of the channel width. However, we
observe a decrease of the $V_{\rm rot}/\sigma_v$ ratio with lower spatial resolution. This is produced by a combination of both effects, the underestimation
and overestimation of  the $V_{\rm rot}$ and $\sigma_v$ values, respectively. However the way in which the $V_{\rm rot}/\sigma_v$ ratio decreases seems to be 
different for each target, suggesting that the internal kinematics of each galaxy may affect the derived $V_{\rm rot}/\sigma_v$ ratio through the convolution with 
the synthesized beam.

In the four row of Fig.~\ref{fig:rel_effects} we see how $r_{1/2,\rm CO}$ does not vary significantly with spectral resolution in any source. The gain of flux from the outskirts
of each target seems to be marginal compared to the total flux of the source.
On the other hand, we see a clear increase of $r_{1/2,\rm CO}$ when we lower the spatial resolution. We note that the derived half-light radii tend to suffer an appreciable
increase of their value when the synthesized beam size becomes comparable to the `fiducial' $r_{1/2,\rm CO}$ value for each galaxy (dotted vertical lines).

As a summary, the velocity dispersion and half-light radius parameters seem to be saturated to a minimum value limited by the spatial 
resolution. The $V_{\rm rot}/\sigma_v$ ratio tend to decrease towards low spatial resolution. However, dispersion dominated sources 
seem to be less affected by this effect. Thus, high spatial resolution data is required to obtain reliable estimates of those parameters.
We find no trend between the spectral resolution and the kinematic estimates from our observations.

Taking into account the resolution effects discussed above, we set the spectral resolution
to 20\,km\,s$^{-1}$, the maximum spectral resolution possible for our observations. We expect that spectral resolution effects do not strongly influence the conclusions of our work.
We set this spectral resolution regardless of the spatial resolution effects inherent in our observations 
which may imply an overestimation of the observed $\sigma_v$ and $r_{1/2,\rm CO}$ values and an underestimation of the $V_{\rm rot}$ value for our sources.

\section{RESULTS \& DISCUSSION}

\subsection{Morphological and kinematic properties}
\label{sec:morph_kin_prop} 

We show the CO(1-0) intensity, velocity and line of sight velocity dispersion maps for our sample in the appendix
(Fig.~\ref{fig:maps}). The intensity maps show smooth distributions of emission with no level of
clumpiness except for HATLASJ085340.7+013348 source. Despite the low resolution
data, most of our sources show a rotational pattern in their velocity maps (Fig.~\ref{fig:maps}), 
with the larger rotational velocity values being preferentially measured in galaxies at lower $z$. 
We note that this bias effect may be mainly produced by the IR flux selection criteria used within the VALES sample 
(see \S~\ref{sec:VALES_sample}). In particular, for our resolved sample, the flux criterion selects $0.02<z<0.2$ `normal' 
star-forming rotating disk-like galaxies, whilst it also selects $0.1<z<0.35$ starburst galaxies with high velocity dispersion (Table~\ref{tab:table1}). 

We note that we find a median $r_{1/2, K} / r_{1/2,\rm CO}$ ratio of $\sim$1, i.e., the molecular gas component shows an spatial extension
comparable to stellar component in our galaxies. This is consistent with molecular gas observations of galaxies in the local universe (e.g. \citealt{Bolatto2017}).
We note that the $r_{1/2, K} / r_{1/2,\rm CO}$ median ratio is lower than the median value ($\sim1.6$) reported by V17 for the VALES sample. We note that this 
difference could be explained by considering that our emission line fitting routine is able to find CO emission at larger radius than the V17's procedure. Nevertheless, we 
calculate the CO and $K-$band half-light radius by taking into account the projection effects (i.e. galactic PA and inclination angles), whilst V17 do not consider for such effects.

In Fig.~\ref{fig:histograms} we show the distribution of $V_{\rm rot}$, $\sigma_v$, and the $V_{\rm rot}/\sigma_v$ ratio for our resolved sample.
The $V_{\rm rot}$ values ranges from 35-287\,km\,s$^{-1}$. The starburst and `normal' star-forming galaxies show
rotational velocities across the full range of the $V_{\rm rot}$ distribution. The velocity dispersion values ranges from 22-79\,km\,s$^{-1}$.
We find median velocity dispersion values of 31 and 53\,km\,s$^{-1}$ for the `normal' star-forming and starburst
galaxies, respectively. However, the $\sigma_v$ values are susceptible to the procedure used to estimate them. Different methods can lead inconsistent results even when
the same sample is analysed (e.g. \citealt{Stott2016}). Thus, we perform the method developed by \citet{Wisnioski2015} to calculate the velocity dispersion values 
($\sigma_{v,{\rm W}}$) in our sample and to compare with our $\sigma_v$ values. This method calculates the velocity dispersion values across the major axis
of the galaxy, but far from the galactic centre where velocity gradients contribute to the observed line widths (see \citealt{Wisnioski2015}, for more details).

We found a median $\sigma_{v,{\rm W}}$ value of 36\,km\,s$^{-1}$, and $\sigma_{v,{\rm W}}$ ranges between $19-70$\,km\,s$^{-1}$. This median value is in 
agreement with the median $\sigma_v$ value (37\,km\,s$^{-1}$) derived by our procedure. The derived velocity dispersion ranges are also consistent 
for both methods. Thus, the slightly overestimation of the $\sigma_v$ values produced by our procedure should not change the results presented in our work. 
We caution that we can not neglect overestimation of the velocity dispersion values produced by spatial resolution effects from this analysis.

The $V_{\rm rot}/\sigma_v$ ratio range between $0.6-7.5$, with the starburst galaxies preferentially to showing the lower values. The median $V_{\rm rot}/\sigma_v$ ratio for our 
sample is 4.1, and the median $V_{\rm rot}/\sigma_v$ values for the `normal' star-forming and starburst sub-samples are 4.3 and 1.6, respectively. Our sample shows a large 
variety of $V_{\rm rot}/\sigma_v$ ratios, from high values comparable to local thin disk galaxies ($V/\sigma_v \sim10-20$  \citealt{Epinat2010, Bershady2010}), to low values 
comparable to the $V_{\rm rot}/\sigma_v$ ratios observed in $z\sim 1$ systems (e.g. $V/\sigma_v \sim 2-5$ \citealt{Forster2009, Wisnioski2015, Stott2016}).

\begin{figure}
 \centering
 \includegraphics{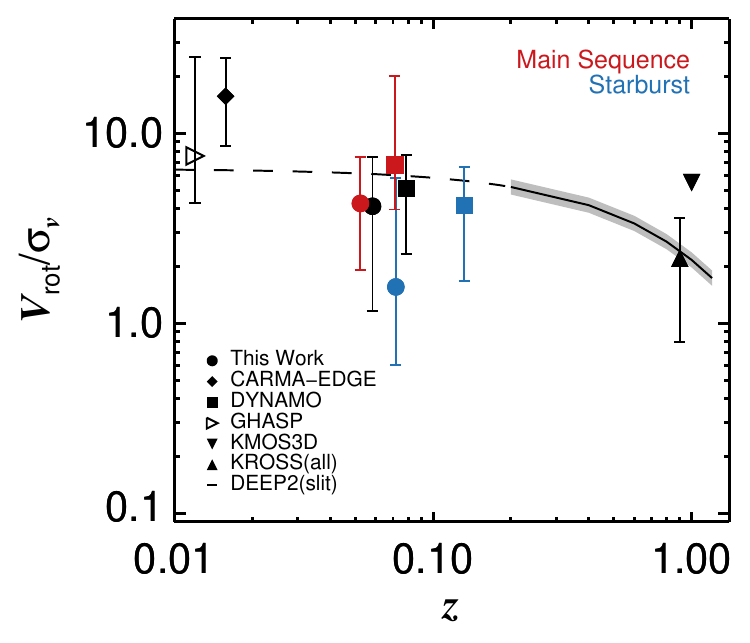}
 \caption{ \label{fig:kinetic_evolution}
 	Evolution of the $V_{\rm rot}/\sigma_v$ ratio at $z\approx0.01-1.0$. The symbols represent the median values for each survey and the error bars correspond to 
 	the 1$\sigma$ region calculated from the 16th and 84th percentiles for each population. The CARMA-EDGE kinetic data are extracted by using the same procedure 
 	explained in previous sections but assuming thin disk geometry (see Appendix~\ref{appendix2}, for more details). We classify our sources and the DYNAMO galaxies
 	as `starburst' or `normal' star-forming galaxy following the same procedure done by V17 for the VALES survey (see \S~\ref{sec:VALES_sample}). The KMOS3D data 
 	correspond to the median value for `main sequence' rotationally supported star-forming disk galaxies at $z \sim 1$, whilst the KROSS data correspond to the median
 	value for all sample, i.e., including `main sequence' dispersion dominated galaxies. The black line and the shaded area represent the best-fit and 1$\sigma$ region 
 	measured for the single slit DEEP2 survey. The dashed line represents the extrapolation of the best-fit to the DEEP2 survey data to lower redshifts}
\end{figure}

In Fig.~\ref{fig:kinetic_evolution} we study the evolution of the $V_{\rm rot}/\sigma_v$ ratio at  $z=0.10-1.0$. We compare with the median $V_{\rm rot}/\sigma_v$ values estimated
for the GHASP \citep{Epinat2010}, CARMA-EDGE \citep{Bolatto2017}, DYNAMO \citep{Green2014}, KMOS3D \citep{Wisnioski2015}, and KROSS \citep{Stott2016} surveys. The
continuous line and the grey-shaded area represent the best-fit relation and the 1$\sigma$ region estimated from the DEEP2 survey \citep{Kassin2012} at $z=0.2-1.0$, respectively. 
The dashed line represent an extrapolation of this relation at low-$z$. DEEP2 is the only long-slit survey considered in Fig.~\ref{fig:kinetic_evolution}. We just consider the galaxies 
with stellar masses between $M_*=10^{10-11}$\,$M_\odot$, approximately the same stellar mass range covered by our sample (see Fig.~\ref{fig:SFR_Mstar_plot}). We also plot the 
median $V_{\rm rot}/\sigma_v$ values for the galaxies classified as `starburst' and `normal' galaxies within our sample and the DYNAMO sample as both surveys study 
star-forming galaxies the same epoch. However, the DYNAMO SFRs are based on dust-corrected $H\alpha$ emission line measurements, whilst the SFR estimates for our sample
are estimated by applying SED fitting. We also note that our sample and the CARMA-EDGE survey observe molecular gas kinematics, whilst GHASP, DYNAMO, KMOS3D and 
KROSS surveys study ionized gas kinematics.

The median $V_{\rm rot}/\sigma_v$ value for our sample is slightly lower but still consistent with the expected value at $z\sim0.06$. This value
is also comparable with the median value found for the KMOS3D sample of `main-sequence' rotating disks star-forming galaxies at $z\sim1$. However,
the median $V_{\rm rot}/\sigma_v$ value of our survey is highly influenced by the low $V_{\rm rot}/\sigma_v$ ratios measured for our starburst galaxies
(Fig~\ref{fig:histograms}). If we do not consider those starburst systems, we find that the median $V_{\rm rot}/\sigma_v$ value for the `normal' star-forming 
galaxies in our sample is consistent with the expected value for local galaxies. It is also consistent with the median $V_{\rm rot}/\sigma_v$ value measured 
for `normal' star-forming galaxies within the DYNAMO survey at nearly the same epoch.

Nevertheless, the median $V_{\rm rot}/\sigma_v$ for our starburst galaxies is $\sim 2.7\times$ lower than the median value observed for the DYNAMO starburst galaxies at the same 
redshift. Although both values are consistent within 1$\sigma$ error. A difference between the spatial extension of the ionized gas compared to the molecular gas across the galaxy 
may explain this discrepancy. An extended ionized gas component would allow to measure $V_{\rm rot}$ in the flat part of the rotation curve whilst the molecular gas observations 
would not allow to do it (e.g. HATLASJ084217.7+021222). On the other hand, different procedures used to calculate $\sigma_v$ may also explain this discrepancy. However, the different
spatial resolution at which both surveys were made is likely to be producing the discrepancy between both $V_{\rm rot}/\sigma_v$ ratios. The DYNAMO galaxies were observed in
natural seeing conditions ($\theta_{\rm FWHM}=0\farcs9-4''$), whilst our sample was observed at $\theta_{\rm FWHM}=3"-4"$.

Regardless of the discrepancy of the median $V_{\rm rot}/\sigma_v$ measured for our sample and the DYNAMO survey, Fig~\ref{fig:kinetic_evolution} show that
starburst galaxies at $z\sim0.1-0.2$ present typical $V_{\rm rot}/\sigma_v$ which are consistent with median the $V_{\rm rot}/\sigma_v$ values presented for the 
KMOS3D and KROSS surveys at $z\sim1$ \citep{Wisnioski2015,Stott2016}. However, high spatial resolution observations of a large sample of the low-$z$ starburst
galaxies is needed to test this result.

\subsection{Luminosity dependence on galactic kinematics}
\subsubsection{CO(1-0) luminosity}
\label{sec:lum_co_kin} 

The CO(1-0) luminosity has been widely used as an estimator of the H$_2$ mass \citep{Bolatto2013}. Through a dynamically calibrated 
CO-to-H$_2$ conversion factor, reliable molecular mass estimates can be achieved (e.g. \citealt{Solomon1987,Downes1998}).
Thus, depending on the dynamical model, we may expect some dependence of the CO luminosity on the galactic dynamics. 

\begin{figure}
 \centering
 \includegraphics[width=0.8\columnwidth]{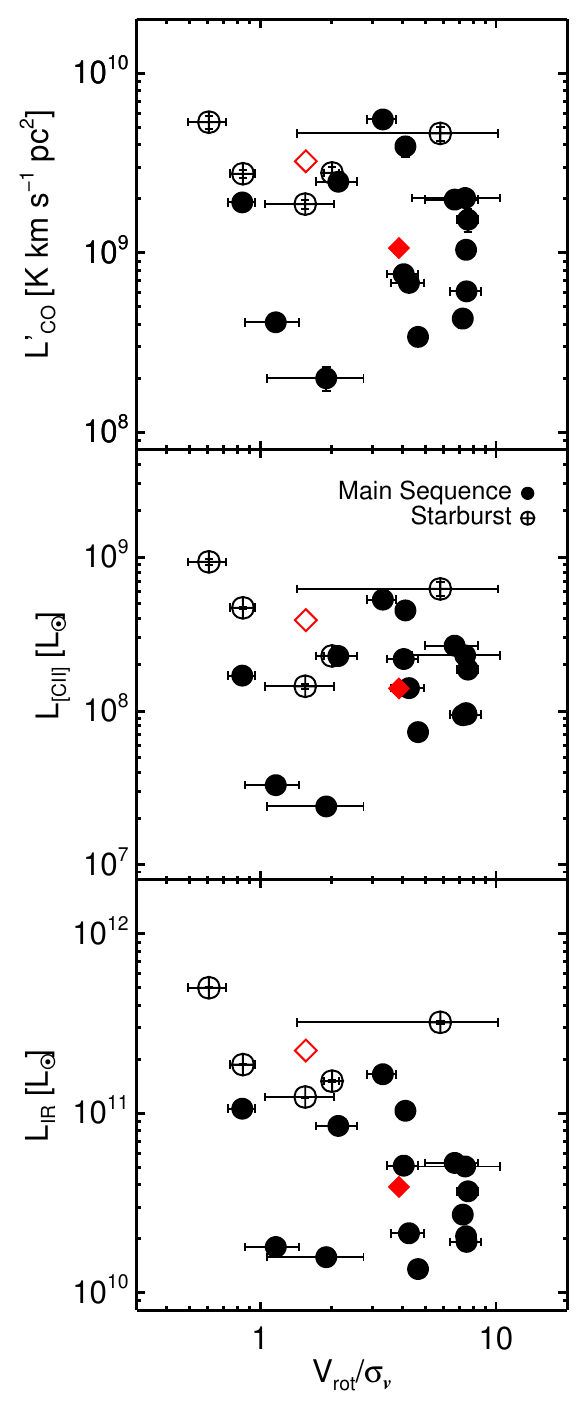}
 \caption{ \label{fig:Lall_kin}
 	From top to bottom: CO(1-0), [C\,{\sc ii}] and IR luminosities as a function of the $V_{\rm rot}/\sigma_v$ ratio for our sample. 
 	We also classify the sources as  `starburst' or `normal' star-forming galaxy. The red open and filled diamonds represent the
 	mean value (in log-space) for the `starburst' and `normal' star-forming galaxies respectively in each panel. We a tentative anti-correlation between
 	the $L_{\rm IR}$ luminosity and the $V_{\rm rot}/\sigma_v$. Galaxies with greater pressure support, reflected by the low $V_{\rm rot}/\sigma_v$ ratio, 
 	tend to show higher CO(1-0), [C\,{\sc ii}] and $L_{\rm IR}$ luminosities. The tentative anti-correlations may suggest a smooth transition between `starburst'
 	 and `normal' star-forming galaxies within our VALES sample.
 	}
\end{figure}

In the top panel of Fig.~\ref{fig:Lall_kin} we show the galactic $L'_{\rm CO}$ as a function of the rotational velocity to dispersion
velocity ratio ($V_{\rm rot}/\sigma_v$). The Spearman's rank correlation coefficient ($\rho_{\rm Spearman}$) is $-0.23$ with a 
probability of 32\% that the correlation is produced by chance. Thus, we find a tentative weak correlation between $L'_{\rm CO}$ and $V_{\rm rot}/\sigma_v$.
in our data, suggesting that the CO luminosity might tend to decrease at higher $V_{\rm rot}/\sigma_v$.
Considering that $V_{\rm rot}/\sigma_v$ measures the level of support given by ordered versus disordered motion support within a 
galaxy, then we suggest that turbulence supported galaxies tend to have greater $L'_{\rm CO}$. 

The high $L'_{\rm CO}$ values may reflect high molecular gas masses in systems with low $V_{\rm rot}/\sigma_v$ values. On the other hand,
we may also expect that systems with high SFRs produce more UV photons which heats the gas through the photoelectric effect on dust grains.
This change of gas temperature may also affect the CO-to-H$_2$ conversion factor \citep{Bolatto2013}. However, we lack of the adequate observations 
to test this.

We also note that low $V_{\rm rot}/\sigma_v$ ratios can be present in both, disk-like galaxies and major-merger systems \citep{Molina2017}, thus, the weak 
correlation found in Fig.~\ref{fig:Lall_kin} suggests that the increase of the CO(1-0) luminosity may not be associated only to major merger events in agreement with 
previous results from numerical simulations \citep{Shetty2011b, Narayanan2012, Papa2012}. This weak correlation also suggests that turbulence may play
a role in the enhancement of  $L'_{\rm CO}$ in galaxies. Nevertheless, higher spatial resolution CO(1-0) observations are required to properly discard
or validate the possible trend between CO(1-0) luminosity and $V_{\rm rot}/\sigma_v$. 

\subsubsection{{\rm [C\,{\sc ii}]} luminosity} 
\label{sec:lum_cii_kin}

The [C\,{\sc ii}] $\lambda$157.74\,$\mu$m emission line ($\nu_{\rm rest}=1900.54$\,GHz) is a far-infrared fine-structure line with a low ionization 
potential (11.26\,eV) that makes it a key participant in the cooling of the warm and diffuse ISM to the cold and dense clouds \citep{Dalgarno1972}.
This emission line is a tracer of all the different stages of evolution of the ISM and detailed characterisation of its emergence
has been made for the Milky Way and local galaxies (e.g. \citealt{Kramer2013,Pineda2013,Pineda2014}) suggesting that different ISM phases produce 
roughly comparable contributions to the [C\,{\sc ii}] luminosity \citep{Madden1993}. However, such detailed characterisations are impeded by observational
limitations in distant galaxies which are typically detected in a single telescope beam. Thus, the [C\,{\sc ii}] line intensity is related to an average quantity 
that arises from a mix of the ISM phases (e.g. \citealt{Gullberg2015}, and references therein). Nevertheless, physical properties of the gaseous 
components of the ISM may be characterised by studying correlations between the [C\,{\sc ii}] emission with various galaxy properties [e.g. 
CO(1-0), $L_{\rm IR}$; \citealt{Ibar2015,Hughes2017a}].

In the middle panel of Fig~\ref{fig:Lall_kin} we show the [C\,{\sc ii}] luminosity as a function of the $V_{\rm rot}/\sigma_v$ ratio for our galaxies.
We find a weak correlation between these two quantities. We measure a  $\rho_{\rm Spearman}=-0.16$ with a
probability of 50\% that the correlation is produced by chance. This may indicates that galaxies with lower $V_{\rm rot}/\sigma_v$ values have higher [C\,{\sc ii}] luminosity.
However, we do not attempt to fit the data as we just have two galaxies measured [C\,{\sc ii}] luminosity at $L_{\rm [C\,II]}< 7 \times  10^7 L_\odot$. 
We need more [C{\sc ii}] luminosity measurements, especially at $L_{\rm [C\,II]}<7 \times 10^7 L_\odot$, in order to discard or validate the possible trend
between [C{\sc ii}] luminosity and $V_{\rm rot}/\sigma_v$. [C{\sc ii}] spatially resolved observations would be also useful in order to account for 
extended and/or nuclear emission effects  (e.g. \citealt{DiazSantos2014}).

\subsubsection{IR luminosity \& the $L_{[\rm C\, II]}/L_{\rm IR}$ deficit} 
\label{sec:lum_ir_kin}

Infrared luminosities are commonly used as a tracer of the star formation activity in galaxies. It can be understood as the emitted UV radiation from young stars
which is re-processed by dust. In the limit of complete obscuration the re-emitted L$_{\rm IR}$ will effectively provide a bolometric measure of the SFR
\citep{Kennicutt1998a}. However, if the attenuation of the stellar light is not completely re-processed, then the IR emission may underestimate the SFR.
Applying SED fitting methods, the IR emission can be also used as a tracer of dust temperature (T$_{\rm dust}$) and mass ($M_{\rm dust}$; e.g. \citealt{Draine2007, Ibar2015}).
 
In the bottom panel of Fig~\ref{fig:Lall_kin} we show the $L_{\rm IR}$ compared to the $V_{\rm rot}/\sigma_v$ ratio for our sources. The data present an anti-correlation
with $\rho_{\rm Spearman}=-0.44$ with a probability of 5\% that the correlation is produced by chance. 
Sources with greater L$_{\rm IR}$ have lower $V_{\rm rot}/\sigma_v$ values, indicating that high IR-luminosities are likely to be present in systems where pressure support becomes
comparable and even greater than rotational support. We note that $L_{\rm IR}$ show strong anti-correlation with the $V_{\rm rot}/\sigma_v$ ratio than the CO luminosity.
This suggests that the L$_{\rm IR}$/$L'_{\rm CO}$ ratio correlates with the $V_{\rm rot}/\sigma_v$ values. We will discuss this further in \S~\ref{sec:kennicutt}.

The IR luminosity has also been traditionally compared to the [C{\sc ii}] luminosity (e.g \citealt{Stacey1991}). The [C{\sc ii}] luminosity to IR luminosity ratio 
($L_{\rm [C\,II]}/L_{\rm IR}$) is found to be roughly constant for local star-forming galaxies with $L_{\rm IR}<10^{11}$\,L$_\odot$, but decreases at higher luminosities
(e.g. \citealt{Stacey1991, Malhotra1997}). This is the so-called `[C{\sc ii}] deficit'. However, the intricate decomposition of the [C{\sc ii}] emission into the different ISM
phases complicates the interpretation of this correlation (e.g. \citealt{Ibar2015}). Therefore, additional comparisons with other galactic properties are needed. Considering that 
our `resolved' VALES sample covers the  $10^{10-12}$\,L$_\odot$ IR luminosity range, it is an ideal sample to study the `[C{\sc ii}] deficit' from a kinematic point of view.

\begin{figure}
 \centering
 \includegraphics[width=1.0\columnwidth]{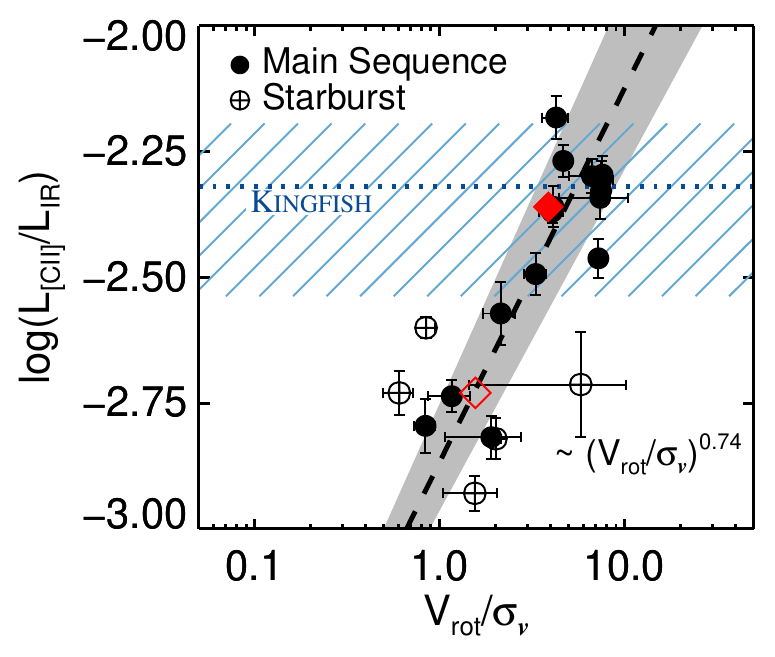}
 \caption{ \label{fig:CII-IR_ratio}
 	$L_{\rm [C\,II]}/L_{\rm IR}$ ratio as a function of the $V_{\rm rot}/\sigma_v$ ratio
 	for our sample. We also classify the sources as `starburst' or `normal' star-forming galaxy.
 	The dashed line represents the best power-law fit to the data and the grey-shaded area represents its 1\,$\sigma$ error. The best-fit slope is presented 
 	in the plot. The red open and filled diamond represent the mean value (in log-space) for the `starburst' and `normal' star-forming galaxies,
 	respectively. The horizontal dotted blue line and the blue-dashed area represent the median $L_{\rm [C\,II]}/L_{\rm IR}$ ratio and its 1\,$\sigma$ region 
 	for the \texttt{KINGFISH} survey data regardless of the kinematics, respectively \citep{Smith2017}. We find an increase of the [C\,{\sc ii}]/IR ratio when the 
 	$V_{\rm rot}/\sigma_v$ ratio increases.}
\end{figure}

In Fig.~\ref{fig:CII-IR_ratio} we show the $L_{\rm [C\,II]}/L_{\rm IR}$  as a function of the $V_{\rm rot}/\sigma_v$ ratio. We find that 
$L_{\rm [C\,II]}/L_{\rm IR}$ increases at high $V_{\rm rot}/\sigma_v$ ratios, but shows a significant scatter at low $V_{\rm rot}/\sigma_v$ values.
This correlation has $\rho_{\rm Spearman}=0.76$ with a probability of $0.0001$\% that the correlation is produced by chance.
We note that this probability is significantly lower than the $L_{\rm [C\,II]}-V_{\rm rot}/\sigma_v$ and $L_{\rm IR}-V_{\rm rot}/\sigma_v$ Spearman correlation's probabilities.
The data are well-represented by a power-law with best-fit slope of 0.74$\pm$0.14. Considering that a high $V_{\rm rot}/\sigma_v$ value suggests a
host galaxy with a dominant disk geometry, then our finding is consistent with \citet{Ibar2015}, who found that galaxies presenting a prominent disk
show higher $L_{\rm [C\,II]}/L_{\rm IR}$ ratios than those that do not present disky morphologies. 

In Fig.~\ref{fig:CII-IR_ratio} we also compare our measured $L_{\rm [C\,II]}/L_{\rm IR}$ ratios with the values derived for the Key Insights on Nearby Galaxies--a Far-Infrared
Survey with $Herschel$ (\texttt{KINGFISH}; \citealt{Kennicutt2011}). These $L_{\rm [C\,II]}/L_{\rm IR}$ ratios are measured from over $\sim$15000 resolved regions within 54 
nearby ($d\leq30$\,Mpc) galaxies \citep{Smith2017} and we represent the median $L_{\rm [C\,II]}/L_{\rm IR}$ ratio of the sample and its 1\,$\sigma$ region with the dotted 
blue line and the blue dashed-area, respectively. A sub-sample of eight galaxies from the \texttt{KINGFISH} survey have measured molecular gas dynamics 
from the HERA CO Line Emission Survey (HERACLES; \citealt{Leroy2009, Mogotsi2016}), and accurate rotation curves derived through H\,{\sc i} observations from
The H\,{\sc i} Nearby Galaxy Survey (THINGS; \citealt{Walter2008,deblok2008}). Those observations suggest $V_{\rm rot}/\sigma_v \gtrsim 10$ for this sub-sample. However, as we 
can not assume that this sub-sample is representative from the complete survey, we do not assume any constraint in the $V_{\rm rot}/\sigma_v$ ratio for the \texttt{KINGFISH} data.

We find that the VALES galaxies with $V_{\rm rot}/\sigma_v \gtrsim 3$ present similar $L_{\rm [C\,II]}/L_{\rm IR}$ ratios compared to the \texttt{KINGFISH} data. However,
the VALES galaxies with $V_{\rm rot}/\sigma_v \lesssim 3$ tend to show even lower $L_{\rm [C\,II]}/L_{\rm IR}$ values. This is independent whether the galaxy was classified as
`normal' star-forming galaxy or `starburst'. 
 
We note that our sample is not significantly contaminated by AGNs (V17) and the [C{\sc ii}] emission is likely to be
optically thin within the galaxies of our sample as based on photodissociation region (PDR) modelling \citep{Hughes2017a}, suggesting
that these two possible effects are not substantially affecting the trend observed in Fig.~\ref{fig:CII-IR_ratio}. An increase of the
star formation efficiency seems not to produce the trend seen between the $L_{\rm [C\,II]}/L_{\rm IR}$ ratio with the $V_{\rm rot}/\sigma_v$
ratio, as most of the galaxies shown in the bottom panel of Fig.~\ref{fig:Lall_kin} form stars at apparently the same efficiency (V17).

\subsection{PDR modelling \& molecular gas kinematics}
\label{sec:PDR_models} 

PDR modelling has been traditionally used to derive the physical properties of the gaseous components of the ISM
(e.g. \citealt{Tielens1985}). Although each PDR code has its own unique model setup and output, it usually determines
the physical parameters by solving chemical and energy balance while also solving the respective radiative transfer equations \citep{Rollig2007}. 

For the VALES survey, \citet{Hughes2017a} applied the PDR model of \citet{Kaufman1999,Kaufman2006}, which is an updated version of the PDR model of 
\citet{Tielens1985}. The model treats PDR regions as homogeneous infinite plane slabs of hydrogen with physical conditions characterised by the hydrogen 
nuclei density ($n_H$) and the strength of the incident $FUV$ radiation field, $G_0$, which is normalised to the Habing Field \citep{Habing1968}. The model covers
a density range of 10$<n_H<10^7$\,cm$^{-3}$ and FUV radiation field strength range of $10^{0.5}<G_0<10^{6.5}$.
In this model, the gas is assumed to be collisionally heated via the ejection of photoelectrons from dust grains and polycyclic aromatic hydrocarbon (PAH) 
molecules by $FUV$ photons, and gas cooling from line emission is predicted by simultaneously solving the chemical and energy equilibrium in the slab.

\citet{Hughes2017a} assumed that the galactic emission comes from a single PDR. They compare the predicted $L_{\rm [C\,II]}/L_{\rm IR}$ and $L'_{\rm CO}/L_{\rm IR}$ 
luminosity ratios with the observed quantities. However, since the fragment of the [C\,{\sc ii}] emission produced in PDRs with respect to the total galactic
emission is observed to vary between 0.5--0.7 (e.g. \citealt{Stacey1991,Malhotra2001,Oberst2006, Stacey2010a}). They also consider two additional models in which 
they adjust the parameters to match to the 50\% and 70\% values of the total [C\,{\sc ii}] luminosity for each galaxy. In these two models, they also consider the missing
CO(1-0) flux emitted along different line-of-sight by multiplying their observed CO(1-0) emission by a factor of two (see \citealt{Hughes2017a} for more details). 
Although these assumptions can modify the values of the derived PDR parameters, in the remaining analysis we will only consider the possible trends seen between
$n_H$ and $G_0$ with respect to the molecular gas kinematics regardless the absolute values for each quantities in each model.

\begin{figure}
 \centering
 \includegraphics[width=1.0\columnwidth]{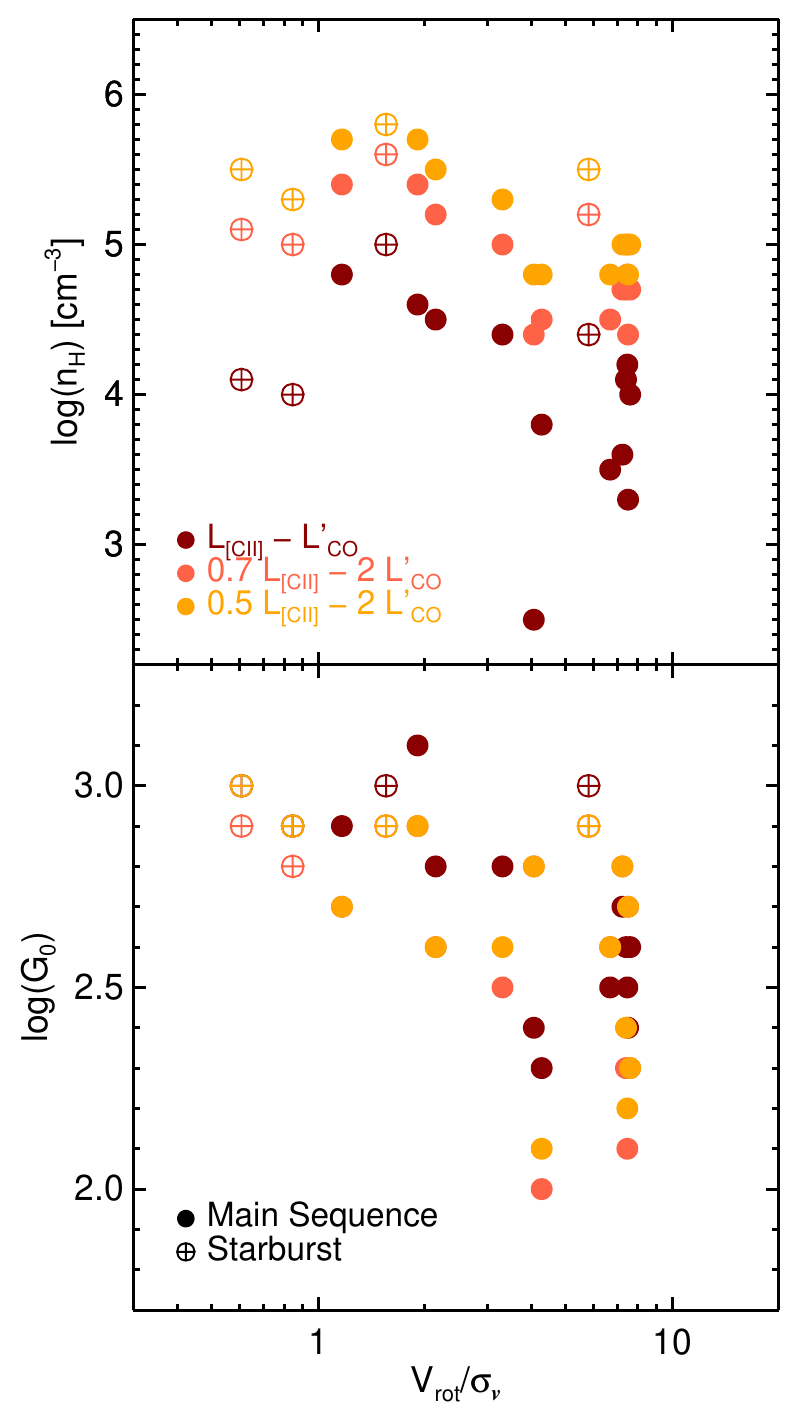}
 \caption{ \label{fig:PDR_plot}
 	Hydrogen nuclei density ($n_H$, \textit{top}) and incident FUV radiation strength ($G_0$, \textit{bottom}) as a function of $V_{\rm rot}/\sigma_v$ ratio for our sample. 
 	We colour-code the data depending on the [C\,{\sc ii}] and CO(1-0) luminosities used to constrain the physical parameters through the PDR modelling 
 	(see \S~\ref{sec:PDR_models} for more details). We also classify the sources as `starburst' or `normal' star-forming galaxy. 
 	Regardless of the PDR model, galaxies with low $V_{\rm rot}/\sigma_v$ ratio tend to show higher hydrogen nuclei density and $G_0$ values. We note that $G_0$ 
 	seems to be insensitive to the [C\,{\sc ii}] and CO(1-0) luminosities assumed to constrain the PDR model within our sample.
 	}
\end{figure}

In Fig.~\ref{fig:PDR_plot} we compare $n_H$ and $G_0$ as a function of the $V_{\rm rot}/\sigma_v$ ratio for each model within our sample. We do not consider the galaxies with
$n_H\lesssim10^2$\,cm$^{-3}$ due to the high degeneracy of the model in the parameter space \citep{Hughes2017a}. In the top panel we see how the hydrogen nuclei density 
may increase at low $V_{\rm rot}/\sigma_v$ for each PDR modelling. This is consistent with the picture in which higher density environments usually show higher velocity 
dispersions. In the bottom panel, the incident $FUV$ radiation strength also increases at low $V_{\rm rot}/\sigma_v$ for each PDR modelling. 
$G_0$ seems to be nearly independent of the assumed [C\,{\sc ii}] and CO(1-0) luminosities to constrain the PDR model, the 
variation of $G_0$ across our sample may reflect the variation of SFR through the IR-luminosity. Therefore, the trend between $G_0$ and $V_{\rm rot}/\sigma_v$
seen in Fig.~\ref{fig:PDR_plot} may reflect the $L_{\rm IR}-V_{\rm rot}/\sigma_v$ correlation observed in \S~\ref{sec:lum_ir_kin}. 

If systems with high $G_0$ -- i.e. high SFR (or $L_{\rm IR}$) and low $V_{\rm rot}/\sigma_v$ -- have ionized most of their atomic carbon content within the PDRs,
then this should result in an inefficient gas cooling through the [C\,{\sc ii}] emission line and a lack of the observed [C\,{\sc ii}] luminosity compared to the IR luminosity.
This may explain the `[C\,{\sc ii}] deficit' correlation with galactic dynamics found in \S~\ref{sec:lum_ir_kin}.

\subsection{Dynamical Masses of Turbulent Thick Galactic Disks}
\label{sec:mdyn_gasfrac} 

The dynamical mass estimate ($M_{\rm dyn}$) is a major tool that allows to measure the mass of galaxies, and a simple way to probe the existence 
of dark matter haloes (e.g. \citealt{Gnerucci2011}). By considering galaxies as thin disks, in which all the material is supported by rotation
[$M_{\rm dyn,thin}(r) = \frac{V_{\rm rot}^2(r)r}{G}$], the dynamical mass can be easily derived from the two-dimensional kinematic modelling (e.g. \citealt{Genzel2011}).
However, galaxies with low $V_{\rm rot}/\sigma_v$ ratio are believed to be well-represented by galactic thick disks (e.g. \citealt{Glazebrook2013}). In those galaxies,
a considerable pressure support is needed to be taken into account in order to calculate reliable dynamical mass estimates \citet{Burkert2010}.

In order to test whether the galaxies in our sample are better represented by galactic thick disks rather than thin disks we calculate their dynamical masses and
compare with their stellar masses (Table~\ref{tab:table1}). Following \citet{Burkert2010}, we model our galaxies as turbulent galactic gas disks in which
pressure support cannot be neglected. In this model, the $observable$ rotational velocity is given by:

\begin{equation}
    V^2_{\rm rot}=V^2_0+2\sigma_v^2\frac{d \ln \Sigma}{d \ln r},	
    \label{eq:vrot_eqn}
\end{equation}

\noindent where $V_0$ is the zero-pressure velocity curve ($V_0^2\equiv r\times d\Phi/dr$), which traces the gravitational potential of the galaxy; $\sigma_v$
is the one-dimensional velocity dispersion of the gas, and $\Sigma$ is the total mass surface density profile of the galaxy. 
In order to derive an explicit model from of Eq.~\ref{eq:vrot_eqn} we need to make some assumptions about the total mass surface density distribution $\Sigma(r)$.
Assuming both, that $\Sigma$ follows the stellar mass surface density profile ($\Sigma_*$) and constant $K-$band mass-to-light ratio  
across the galactic disk ($\Upsilon_K$), then $\Sigma$ can be approximated by the $K$-band surface brightness distribution ($\mu_K$), i.e. 
$\Sigma(r) \approx \Sigma_*(r) \approx \Upsilon_K \mu_K(r)$. Considering that $\mu_K$ is well-described by a S\'ersic profile \citep{Sersic1963}, Eq.~\ref{eq:vrot_eqn}
can be written as:

\begin{equation}
    V^2_{\rm rot}=V^2_0-\frac{2\sigma_v^2b_{n_S}}{n_S}(\frac{r}{r_e})^{1/n_S},
	\label{eq:vrot_eqn_final
	}
\end{equation}

\noindent where $n_S$ is the S\'ersic index and $b_{n_S}$ is the S\'ersic coefficient, which sets $r_e$ as the half-light radius. 	
We note that the case $n_S=1$ is equivalent to an exponential profile. In this model, the dynamical mass is traced by $V^2_0$ rather than $V^2_{\rm rot}$:

\begin{equation}
    M_{\rm dyn,thick}(r) = \frac{V_0^2(r)r}{G}.
	\label{eq:Mdyn_eqn}
\end{equation}

We note that the $K-$band S\'ersic model parameters for our resolved sample are listed in 
Table~\ref{tab:table2} for each galaxy. For the galaxy without $K-$band modelling, we assume an exponential $\Sigma_\star$ profile.

\begin{figure}
 \centering
 \includegraphics[width=1.0\columnwidth]{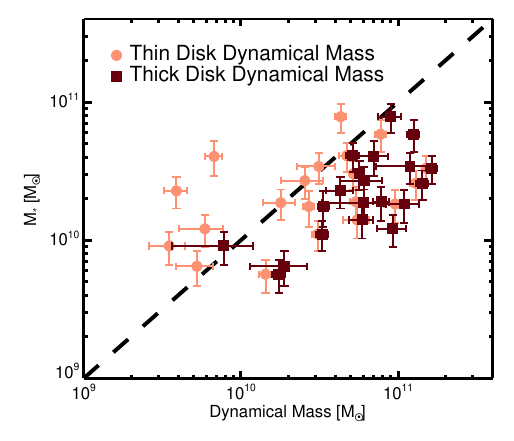}
 \caption{ \label{fig:mdyn_plot}
 	Encircled stellar masses at the radii at which $V_{\rm rot}$ is extracted (2\,$r_{\rm 1/2,CO}$) as a function of the dynamical masses encircled
 	at the same radius. The light-red circles show the dynamical masses assuming a thin disk model, i.e. the total mass is traced just by the 
 	observed rotational velocity. The dark red squares show the dynamical masses assuming a thick disk model in which the surface density 
 	profile of each galaxy is traced by the $K-$band surface brightness also assuming a constant mass-to-light ratio. The encircled stellar mass
 	content is calculated by considering the $\mu_K$ profile (see Eq.~\ref{eq:Mstar_eqn}). The dashed line corresponds to the 1:1 ratio. We clearly
 	see how the thin disk approximation fails to estimate the total mass in five systems as their truncated stellar masses have greater values.  }
\end{figure}

In Fig.~\ref{fig:mdyn_plot} we show the dynamical masses calculated by assuming a thin disk model  and a thick disk model
(Eq.~\ref{eq:Mdyn_eqn}) within $r\leq2r_{\rm 1/2,CO}$. We note that we have considered the radius at which we extracted $V_{\rm rot}$. We compare these dynamical masses with the 
stellar masses truncated at the same radius and normalized to the stellar mass values derived in V17 (Table~\ref{tab:table1}):

\begin{equation}
    M_*(r) \equiv  M_* \frac{\int_{S(<r)} \Sigma_*(r) dS}{\int_{S} \Sigma_*(r) dS} \approx M_* \frac{\int^r_0\mu_K(r)rdr}{\int^\infty_0\mu_K(r)rdr}. 
	\label{eq:Mstar_eqn}
\end{equation}

The dashed line in Fig.~\ref{fig:mdyn_plot} represents the 1:1 ratio between both quantities. Clearly, the thin disk dynamical mass model underestimates the total 
mass for five of our systems, as it predicts masses lower than the stellar masses. On the other hand, the thick disk dynamical mass model estimates
masses greater than the stellar masses, with just one target showing stellar mass nearly equal to their estimated dynamical mass within 1\,$\sigma$ error.
This suggests that these five VALES galaxies with lower $V_{\rm rot}/\sigma_v$ ratio may be better represented by thick galactic disk, while the rest of the sample is best 
described by thin galactic disk. We note that the existence of gaseous thick disks at the observed redshift range may indicate a late assembly of the thick disk stellar 
component in those systems \citep{Bournaud2009}.

The validity of this result depends on the assumption that the $K-$band surface brightness traces $\Sigma_*$  by considering a mass-to-light ratio which does not
vary as a function of galactocentric radius. We explore the effect produced by a different mass distribution in Appendix~\ref{appendix1}, where we show that it has 
a negligible effect when considering, for example, an exponential disk mass profile. We conclude that the considerable pressure support predicted by the high velocity
dispersion values is a key ingredient to obtain reliable conclusions from the modelling. This may be especially important in the systems which present the highest gas fractions
(see \S~\ref{sec:presion_support}).

We caution, however, that spatial resolutions effects may produce an overestimation or underestimation of the dynamical mass values derived from the thin 
and thick disk models. The $V_{\rm rot}$ values can be underestimated by beam smearing, especially in cases when the rotation curve beyond the turn-over 
radius is not observed (e.g. HATLASJ084217.7+021222). On the other hand, overestimated $r_{\rm 1/2,CO}$ and $\sigma_v$ values are expected to be 
calculated due to the same effect. The result of the competition between both effects is uncertain. Thus, high spatial resolution observations are required to 
obtain more accurate dynamical mass estimates.

\subsection{Gravitationally stable disks}
\label{sec:disk_support} 
Gravitational stability analysis is usually used to explain the formation and growth of internal galactic sub-structures at low (e.g. \citealt{Lowe1994}) and high redshifts
\citep{Swinbank2012b,Wisnioski2012}, between other major topics (e.g. \citealt{Kennicutt1998b}).  In thin galactic disks, gravitational stability was first 
studied by \citet{Toomre1964}, who derived a simple criterion that can be quantified through the stability parameter: 

\begin{equation}
    Q_{\rm Toomre}\equiv\frac{\kappa\sigma_v}{\pi G \Sigma_{\rm gas}},
	\label{eq:Qtoomre}
\end{equation}

\noindent where, $\kappa \equiv (2\Omega/r)$\,$ d(r^2\Omega)/dr=av_c/r$ is the epicyclic frequency, usually expressed as a function of orbital frequency ($\Omega$)
or the circular velocity $v_c$ at some radius $r$ with $a=\sqrt{2}$ for a flat rotational curve; $\sigma_v$ is the measure of the random motions of the gas; $\Sigma_{\rm gas}$ is 
the gas surface density; and $G$ is the gravitational constant. If $Q_{\rm Toomre}<1$, then the system is prone to develop local gravitational instabilities. 
Otherwise ($Q_{\rm Toomre}>1$), the system is not susceptible to local gravitational collapse.

Since the \citet{Toomre1964}'s earlier work, the $Q_{\rm Toomre}$ parameter has been generalized to include different physical effects such as galactic disk thickness 
($Q_{\rm thick}$, e.g. \citealt{Goldreich1965, Romeo1992}) and/or multiple galactic components ($Q_M$, e.g. \citealt{Jog1984,Jog1996,Rafikov2001, RW2011}):

\begin{equation}
    Q_{\rm thick} = T Q_{\rm Toomre},
	\label{eq:Qthick}
\end{equation}

\begin{equation}
    \frac{1}{Q_M} = \sum^M_{k=1} \frac{W_k}{Q_{{\rm Toomre},k}},
	\label{eq:Qn}
\end{equation}

\noindent where, $T$ represents the stabilizing effect of the disk thickness, and ranges between 1--1.5 depending on the velocity dispersion anisotropy 
($\sigma_{v,z}/\sigma_{v,R}$; \citealt{RW2011}). $Q_{{\rm Toomre},k}$ is the Toomre parameter of component $k$, $M$ is the total number of different galactic 
components considered in the analysis, and $W_k$ is a weighting factor that is higher for the component with smallest $Q_{\rm Toomre}$ value (see \citealt{Romeo2013}
 for more details). Other physical effects such as gas dissipation (e.g. \citealt{Elmegreen2011}), and supersonically turbulence (e.g. \citealt{Romeo2010}) can be included 
 to derive other generalized $Q$ parameters, however they required assumptions on how the gas dissipates energy across different scales and its beyond the scope of this
 work test those assumptions. Thus, in order to maintain simplicity, we just test the $Q_{\rm Toomre}$, $Q_{\rm thick}$ and the $Q_M$ stability parameters and for the $Q_M$
parameter we just consider the stellar and molecular gas galactic components ($M=2$).

In order to proceed further, by assuming that; (1) the system is supported by rotation; (2) the galactic mass budget is dominated by the gas and stars at the radii in which $V_{\rm rot}$ 
is derived; and (3) the gas within that radii is principally in the form of molecular gas; then the $Q_{\rm Toomre}$ (hereafter, $Q_{\rm gas}$) can be rewritten as a function of the 
molecular gas kinematics and the molecular gas fraction \citep{Genzel2011}: 

\begin{equation}
    Q_{\rm gas} \approx \sqrt{2}\frac{\sigma_v}{v_c} f_{\rm H_2}^{-1}.
	\label{eq:Qtoomre_kin}
\end{equation}

By following an analogous procedure, we find similar formulas for the $Q_{\rm thick}$ and $Q_M$ (hereafter, $Q_2$) parameters:

\begin{equation}
    Q_{\rm thick} \approx T \sqrt{2}\frac{\sigma_v}{v_c} f_{\rm H_2}^{-1},
	\label{eq:Qthick_kin}
\end{equation}

\begin{equation}
    Q_{2} \approx \left\{ \begin{array}{ll} \renewcommand{\arraystretch}{2.5}
    \sqrt{2}\frac{\sigma_v}{v_c} [f_{\rm H_2} + \frac{2}{1+ s^2} (1-f_{\rm H_2})]^{-1}  &\mbox{ if  $s > \frac{1}{f_{\rm H_2}} - 1$;} \\
    \sqrt{2}\frac{\sigma_v}{v_c} [\frac{2s}{1+ s^2} f_{\rm H_2} + \frac{1}{s} (1-f_{\rm H_2})]^{-1} &\mbox{ otherwise,}
    \end{array} \right.
    \label{eq:Qn_kin}
\end{equation}

\noindent where `$s$' is the stellar to molecular gas velocity dispersion ratio ($s \equiv \sigma_*/\sigma_v \geq 1$) and the conditioning 
represents the $Q_{\rm stars} > Q_{\rm gas}$ requirement (see \citealt{Romeo2013}, for more details).

\begin{figure}
 \centering
 \includegraphics[width=1.0\columnwidth]{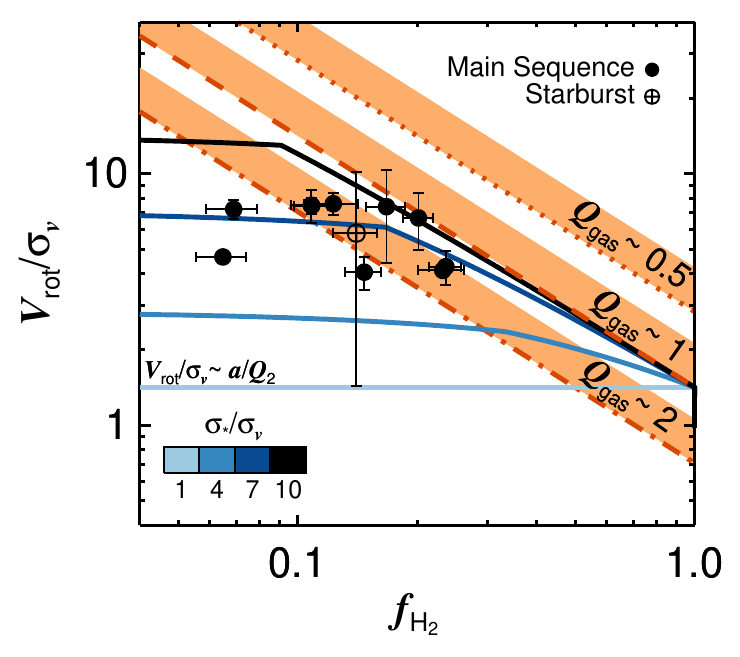}
 \caption{ \label{fig:ratio_gasfrac}
 $V_{\rm rot}/\sigma_v$ as a function of the molecular gas fraction. The orange dotted-dashed, dashed and dotted lines represent the 
 \citet{Toomre1964}'s $Q_{\rm gas}$ values of 2, 1 and 0.5 for thin disk galaxies, respectively. The orange-shaded areas represent the possible $Q_{\rm thick}$ values given
 the mentioned $Q_{\rm gas}$ values. The black, dark-blue, blue and light-blue lines shows the $Q_2 = 1$ values for different $\sigma_*/\sigma_v$ ratios listed in the colorbar.
 We also classify the sources as `starburst' or `normal' star-forming galaxy.}
\end{figure}

In order to fulfil assumption (1), we choose $M_{\rm dyn,thin}/M_{\rm dyn,thick} > 0.5$ as a somewhat crude criterion to select galaxies that are mainly
supported by rotation. Regardless of the density profile of the galaxies, this criterion can be traduced into a threshold to the measured $V_{\rm rot}/\sigma_v$
ratio ($V_{\rm rot}/\sigma_v \gtrsim 2$ in our case). Thus, within our resolved sample, we just find 11 galaxies consistent with being rotationally supported.

In Fig.~\ref{fig:ratio_gasfrac} we show $f_{\rm H_2}$ as a function of $V_{\rm rot}/\sigma_v$ for the rotationally supported galaxies in our sample. 
The orange dot-dashed, dashed and dotted lines represent the $Q_{\rm gas} =2$, 1, 0.5 values, respectively. Three galaxies are consistent with $Q_{\rm gas}\sim1$ 
(within 1$\sigma$ range), whilst eight galaxies have $Q_{\rm gas}\gtrsim2$. The majority of our rotationally supported systems seems to be gravitationally unstable 
within the thin disk single component approximation. Although the poor spatial resolution of our observations smooth the CO intensity maps, we note that the unique source that
shows some degree of clumpiness in its CO intensity map (HATLASJ085340.7+013348) is consistent with being susceptible to gravitational instabilities.

The next step is to include the disk thickness effect in our analysis. In order to do that, we use Eq.~\ref{eq:Qthick} to compute $Q_{\rm thick}$ from the 
$Q_{\rm gas}=2$, 1, 0.5 values. As we do not have velocity dispersion anisotropy estimates for our sample to determine the $T$ factor, we assume $T$ values
between the limit ranges (1$\leq T \leq$1.5; \citealt{RW2011}) and we present the possible $Q_{\rm thick}$ values as the orange-shaded areas in Fig.~\ref{fig:ratio_gasfrac}.
From our sample, is clearly that we can not differentiate the disk thickness effect through the gravitational stability analysis as our kinematic estimates are not enough accurate.

As a final step, we consider a two-component gravitational stability analysis in which the main components are the molecular gas and the stars.
We note that the two-component system is more unstable than either component in the system by itself \citep{Jog1996}. However, in order to use the two-component 
gravitational stability criterion ($Q_2$), we must measure $\sigma_*$, the velocity dispersion of the stars, or in equivalence the $\sigma_*/\sigma_v$ ratio. As we lack of 
that information for our rotationally supported galaxies, we just assume four different values of $\sigma_*/\sigma_v$ between the range indicated in the colorbar in 
Fig.~\ref{fig:ratio_gasfrac}. We note that $\sigma_*/\sigma_v = 1$ is the minimum value that can be assumed within this model (see \citealt{Romeo2013}
for more details). On the other hand, a maximum range value of $\sigma_*/\sigma_v = 10$ may be appropriate for local spiral galaxies. For example, the expected value for the 
Milky Way is in the range of $4<\sigma^{\rm MW}_*/\sigma^{\rm MW}_v<8$, whether we consider the stellar velocity dispersion of the thin or thick disk as the representative 
$\sigma^{\rm MW}_*$ value \citep{Glazebrook2013}. In the remaining of our analysis we kept fixed $Q_2$ to the unity value.

At high $\sigma_*/\sigma_v$ ratio, the molecular gas component is more susceptible to gravitational instabilities than stars, and therefore, the gravitational stability of the system
is dictated by the gaseous component alone ($Q_2 \approx Q_{\rm gas}$) at least if $f_{\rm H_2}$ is not low enough. The latter case implies $Q_2 \approx Q_{\rm stars}$.
In Fig.~\ref{fig:ratio_gasfrac}, the $Q_2 \approx Q_{\rm gas}$ case is better represented by the black line, which approach to the orange-dashed line ($Q_{\rm gas} = 1$) at 
$f_{\rm H_2} \gtrsim 0.1$.  For molecular gas fractions below that value, the gravitational stability of the system is dictated mainly by the stellar component from which we do
not have any information. We also note that when the $\sigma_*/\sigma_v$ ratio decreases, $Q_2$ approximates to $Q_{\rm gas}$ at higher $f_{\rm H_2}$ values 
as the gravitational effect of the stellar component becomes more significant.

At $\sigma_*/\sigma_v \sim 1$, from Eq.~\ref{eq:Qn_kin} we can see that the $Q_2$ parameter does not depends on $f_{\rm H_2}$. In this limit, the
two-component system behave as a single component fluid in which the gravitational criterion is dictated by the total surface density of the system. 
In Fig.~\ref{fig:ratio_gasfrac} it is better represented by the light-blue line. In this limit, the $Q_2$ value can be recovered by measuring the kinematics 
of the galaxy and the shape of the rotation curve (accounted by the factor $a$) at a given radii. Nevertheless, if the dark matter 
component is not negligible, i.e. assumption (2) is incorrect, then an additional baryonic mass fraction needs to be accounted. 

Within two-fluid component framework, two of the rotationally supported galaxies are consistent with being gravitational unstable systems (within 1$\sigma$ range). 
The seven galaxies with the lower $f_{\rm H_2}$ values are likely to be in the $Q_2 \approx Q_{\rm stars}$ regime. Thus, we can not determine if these systems 
are gravitationally stable or not. We also note that the remaining galaxies can be consistent with being gravitationally unstable or not depending on its $\sigma_*$ value.

Therefore, from the gravitational stability analysis we conclude that two galaxies are consistent with being marginally gravitationally unstable disk, but observations 
of the stellar dynamics are required to determine the gravitational stability for the remaining nine cases.

This result disagrees with \citet{White2017} who found a $Q_{\rm gas}\sim1$ trend in their sample of local star-forming galaxies taken from the DYNAMO 
survey ($z\sim 0.06-0.08$ \& $z\sim 0.12-0.16$). They found this trend by fitting the  Eq.~\ref{eq:Qtoomre_kin} (their Eq. 10) to their sample, but by considering 
ionized gas kinematics instead molecular gas kinematics. 

We note that our conservative choice of $\alpha_{\rm CO}$ values (see \S~\ref{sec:galaxy_dyn}) tends to overestimate the molecular gas 
reservoir for most of the galaxies within our sample. Using a lower CO-to-H$_2$ conversion factor would imply greater $Q_{\rm gas}$, $Q_{\rm thick}$ and $Q_2$
values in these galaxies.

Nevertheless, we have analysed the galaxies in which the assumption (1) is likely to be correct, however we can not determine if assumptions (2) and (3)
are correct. We stress that H{\sc i} observations are required in order to test assumptions (2) and (3).

\subsection{Energy sources of turbulent motions}
\label{sec:presion_support} 

The origin of the energy sources of the random motions in galactic disks are unclear, low and high-$z$ galaxy observations show 
a positive correlation between the ionized gas turbulence with the measured $\Sigma_{\rm SFR}$, with larger scatter at high-$z$ 
(e.g. \citealt{Lehnert2009, Johnson2017, Zhou2017}). These observations favour a models in which stellar feedback is driving those random motions. 
However, observations also suggest that other energy sources may contribute to produce turbulence in the ISM \citep{Zhou2017}.

From a theoretical perspective, two possible scenarios have been proposed. In the first scenario, the star-formation is determined by the requirement to maintain
hydrostatic balance through the input of energy from supernovae feedback. In this model, stars are produced efficiently by the gravitational collapse of 
gas within GMCs and the GMCs are treated as bound entities that are hydrodynamically decoupled from the galactic disk \citep{FQH2013}. Therefore, 
the production of stars is limited by the formation of GMCs and this process is driven by the self-gravity of the gas, and not by a combination of the 
gravitational potential of gas and stars from the galactic disk. Thus, in the feedback-driven model, it is expected that SFR$\propto\sigma_v^2$. In the 
second scenario, the turbulence is expected to be driven by the release of gravitational energy of the gas which is accreted through the disk 
\citep{Krumholz2016}. The accretion of the gas is ultimately powered by gravitational instabilities through the galactic disk that are regulated by the 
gravitational potential of stars and gas. This model also assumes a star formation law in which the star formation rate per molecular gas mass is 
represented by $\epsilon_{\rm ff}/t_{\rm ff}$, the efficiency per free-fall time ($\epsilon_{\rm ff}\approx0.01$, e.g. \citealt{KrumholzTan2007,Krumholz2012}). 
The $t_{\rm ff}$ is estimated by assuming that the star-forming gas density is set by the total gravitational potential of the ISM, rather than by the properties
of hydrodynamically decoupled GMCs \citep{Krumholz2012}. In this gravity-driven model, the SFR vary as SFR$\propto\sigma_v f_{\rm gas}^2$, 
where $f_{\rm gas}$ is the mid-plane galactic gas fraction. 
 
 \begin{figure}
 \centering
 \includegraphics[width=0.9\columnwidth]{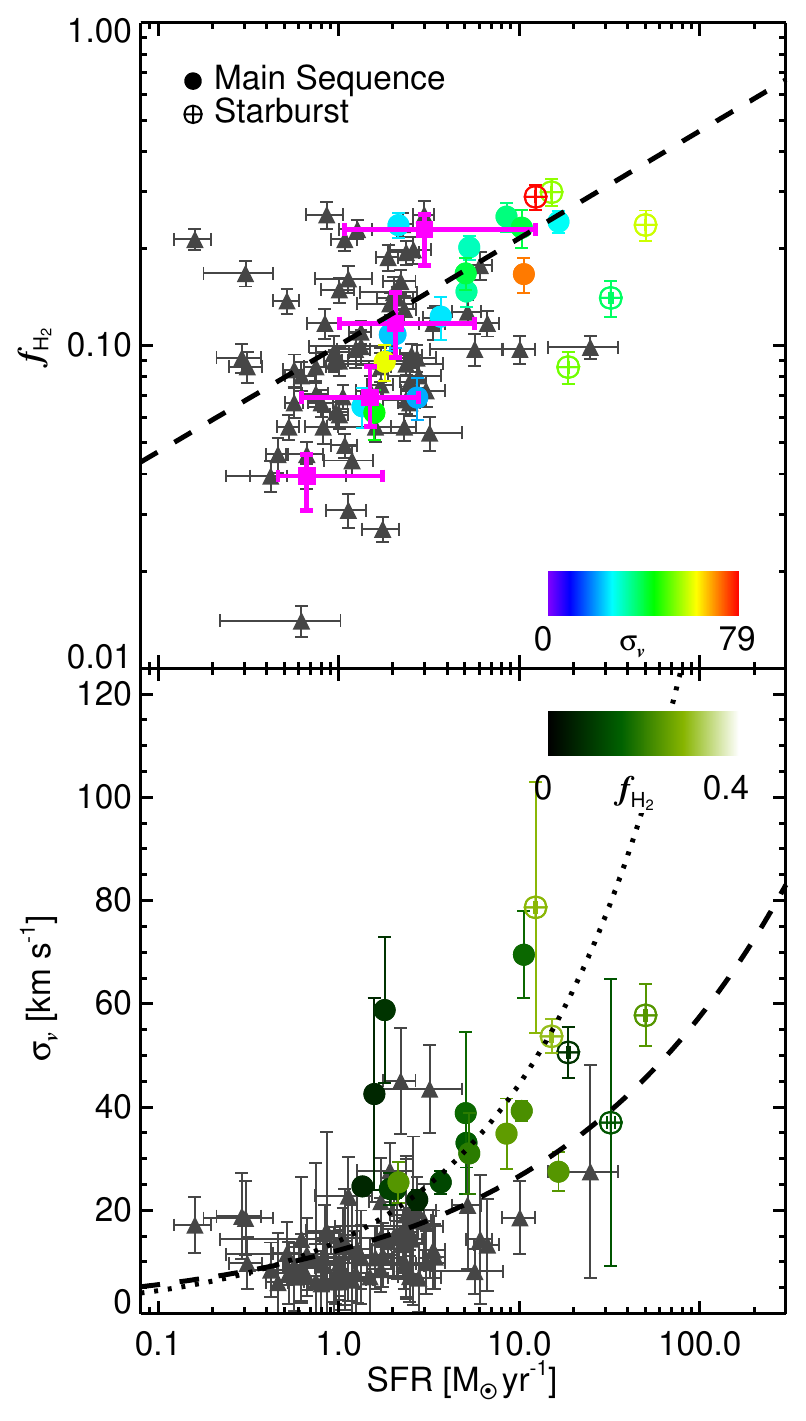}
 \caption{ \label{fig:krumholz}\textit{Top:} Molecular gas fraction as a function of SFR, colour-coded by velocity dispersion.
 \textit{Bottom:} Velocity dispersion as a function of SFR. We colour-coded the VALES galaxies by molecular gas fraction. In both panels the VALES galaxies
 are classified as `normal' (filled circles) and `starburst' (circles with plus sign) star-forming galaxies. The dark grey
 triangles represent the galaxies from the EDGE-CALIFA survey \citep{Bolatto2017}. In the top panel, the magenta squares represent the median $f_{\rm H_2}$ 
 and SFR values per $\log_{10}(f_{\rm H_2})$= 0.4 bin.
The dashed line represents the best-fitted gravity-driven model (SFR$\propto \sigma_v f_{\rm gas}^2$) for the VALES galaxies; \citealt{Krumholz2016}), 
assuming $f_{\rm gas}\approx f_{\rm H_2}$. The dotted line in the bottom panel represents the best-fitted feedback-driven model for the VALES galaxies 
(SFR$\propto \sigma_v^2$; \citealt{FQH2013}).}
\end{figure} 

With the aim to test both models, in the top and bottom panels of Fig~\ref{fig:krumholz} we show $f_{\rm H_2}$ and $\sigma_v$ as a function of the SFR, for the VALES  
survey and the Extragalactic Database for Galaxy Evolution Survey selected from the Calar Alto Legacy Integral Field Area sample (EDGE-CALIFA) \citep{Bolatto2017}. 
The VALES data are colour-coded in each panel by the velocity dispersion and the molecular gas fraction, respectively. The EDGE-CALIFA data are represented by the 
dark-grey triangles. The EDGE-CALIFA data were modelled using the same procedure described for the VALES data (see \S~\ref{appendix2} for more details). In top 
panel, we also show the median $f_{\rm H_2}$ and SFR values per $\log_{10}(f_{\rm H_2})$= 0.4 bin combining the data from both surveys. The median values
suggest that the SFR is weakly correlated with $f_{\rm H_2}$. Systems with high molecular gas fraction may tend to have high SFR, although 
the scatter is considerable. On the other hand, in the bottom panel, we observe that systems with high SFRs also tend to present high $\sigma_v$ values. We note that
we do not find any correlation between inclination angles and velocity dispersions within Fig~\ref{fig:krumholz}.

We represent the best-fitted gravity-driven and feedback-driven models by the dashed and dotted lines in Fig.~\ref{fig:krumholz}, respectively. 
The gravity-driven model gives a poor description to the data. At high molecular gas fractions ($f_{\rm H_2}\gtrsim0.1$), galaxies tend to present a large variety 
of SFRs than the predicted by this model. This suggests that the release of the gravitational energy from the molecular gas may not be the main source of energy that
support the $\sigma_v$ values observed in the VALES and EDGE-CALIFA surveys. The feedback-driven model may also not explain the loci of the galaxies in the 
SFR$-f_{\rm H_2}-\sigma_v$ phase space as it tends to overestimate the $\sigma_v$ values for most of the systems with SFR$\gtrsim2$\,$M\odot$\,yr$^{-1}$. The 
distribution of the VALES and EDGE-CALIFA galaxies in Fig.~\ref{fig:krumholz} suggest that different energy sources may sustain the observed supersonic turbulence.

We stress that the scatter behind Fig.~\ref{fig:krumholz} might be induced by a handful of effects, including the bimodal CO-to-H$_2$ conversion
factor (see \S~\ref{sec:VALES_sample}) used to calculate the molecular gas masses and, therefore, the molecular gas fractions. On the other hand, poor spatial resolution
could potentially bias $\sigma_v$ towards higher values. It may contribute to the high velocity dispersion values seen in the VALES galaxies with higher SFRs. 
Spatial resolution effects may favour models which accounts for a higher dependence of the SFR with $\sigma_v$.

Despite of the weakness of the median trend observed between SFR and $f_{\rm H_2}$, we note that this result is in contradiction with \citet{Green2010} results. They found that 
velocity dispersion values measured from a sample of 65 star-forming galaxies at $z\sim0.1$ seems to be correlated with their SFRs but not with the gas fraction. However, 
as a difference with our work, they estimated the velocity dispersion values from the ionized gas kinematics traced by the H$\alpha$ emission line, whist we are observing 
the molecular gas kinematics. Moreover, \citet{Green2010} calculated the gas mass content for their sample by converting the measured $\Sigma_{\rm SFR}$ through the 
application of the KS law, this assumes that the molecular and ionized gas are spatially correlated, but also that the galaxies in their sample follow the KS law, which is not 
straightforward to assume (see \S~\ref{sec:kennicutt}). On the other hand, we measure the molecular gas content from the CO line emission by applying the CO-to-H$_2$ 
conversion factor. Nevertheless, we stress that spatially resolved observations of gas-rich systems ($f_{\rm H_2} \gtrsim 0.3$) are needed to refute or probe possible trends 
between SFR and $f_{\rm H_2}$.

It should be mentioned that the model developed by \citet{Krumholz2016} relates the SFR with the mid-plane galactic gas fraction and its velocity 
dispersion rather than the molecular gas fraction and the molecular gas velocity dispersion which are shown in Fig~\ref{fig:krumholz}. We note that it is not straightforward to 
expect that those quantities are related between each other. The model also assumes that the stellar velocity dispersion should be comparable with the velocity dispersion
of the gas. Taking into account these caveats, in order to produce a more complete observational test, atomic gas (H{\sc i}) and stellar kinematic observations are needed.

\begin{figure*}
 \centering
 \includegraphics[width=2.0\columnwidth]{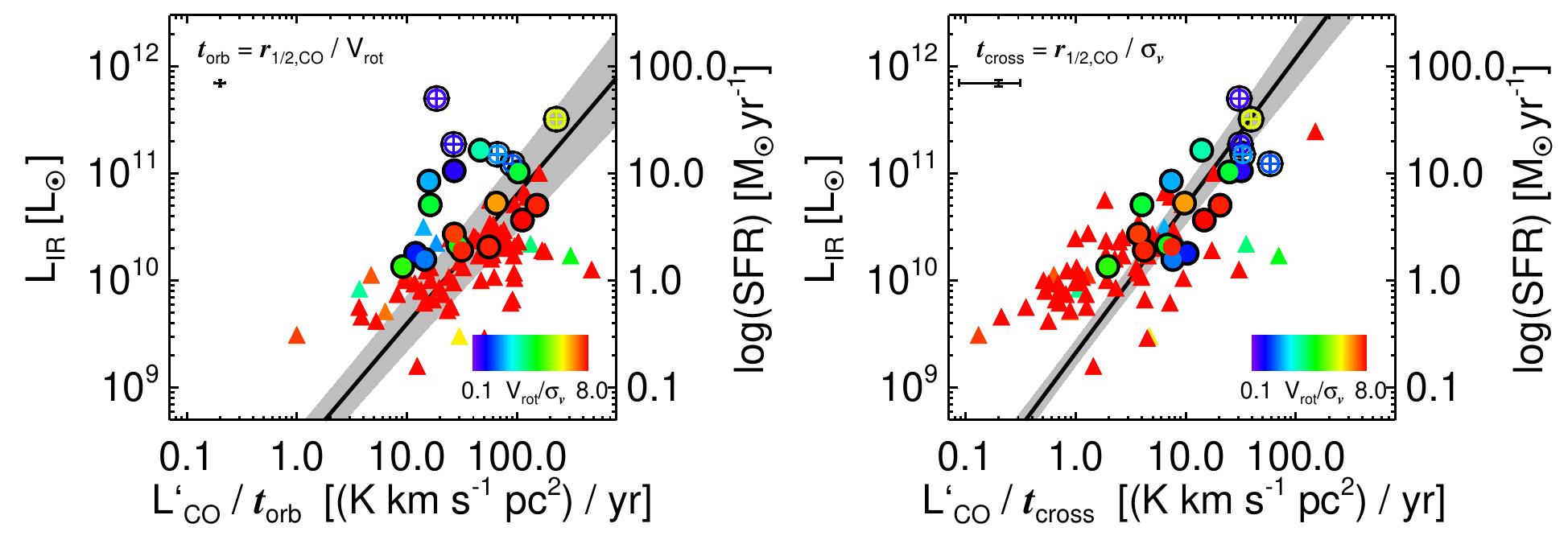}
 \caption{ \label{fig:ks_law}
 The IR luminosity as a function of the CO luminosity divided by the orbital timescale (left panel) and the crossing time (right panel) 
  In both panels the VALES galaxies are classified as `normal' (filled circles) and `starburst' (circles with plus sign) galaxies as in Fig.~\ref{fig:Lall_kin}.
  The triangles represent the EDGE-CALIFA galaxies \citep{Bolatto2017}. The VALES and CARMA-EDGE data presented in both panels are colour-coded 
  by the observed $V_{\rm rot}/\sigma_v$ ratio. The line and the grey shaded area represent the best-fitted power-law function and its 1\,$\sigma$ error 
  respectively in each panel. For the SFR-$L'_{\rm CO}/\tau_{\rm orb}$ plot (left panel) we find a best-fit slope of 1.21$\pm$0.14, whilst
 for the SFR-$L'_{\rm CO}/\tau_{\rm cross}$ plot (right panel) we find a best-fit slope of 1.38$\pm$0.13.
 }
\end{figure*}  

\subsection{Kennicutt-Schmidt Law Efficiency \& Depletion Times}
\label{sec:kennicutt} 

The Kennicutt-Schmidt law \citep{Kennicutt1998a,Kennicutt1998b} describes the power-law relationship between
the galaxy star formation rate surface density and the disk-averaged total gas surface density. It describes how 
efficiently galaxies turn their gas into stars. For local galaxies, this correlation is well-fitted by $\Sigma_{\rm SFR}\propto \Sigma_{\rm gas}^{1.4}$.
\citep{Kennicutt1998b}. Although a tight relation can be found when $\Sigma_{\rm SFR}$ is compared with the molecular gas surface density $\Sigma_{\rm H_2}$
rather than $\Sigma_{\rm gas}$ (e.g. \citealt{Bigiel2008,Leroy2008,Leroy2013}), also the slope is changed ($\Sigma_{\rm SFR}\propto \Sigma_{\rm H_2}$).

However, the KS law shows an apparent bimodal behaviour where `disks' and `starburst' galaxies fill the $\Sigma_{\rm H_2}-\Sigma_{\rm SFR}$
plane in two parallel sequences \citep{Daddi2010}. Nevertheless, by computing $\Sigma_{\rm H_2}/t_{\rm ff}$ and/or $\Sigma_{\rm H_2}/t_{\rm orb}$ relationships, a 
single power-law relation can be obtained (e.g. \citealt{Daddi2010,Krumholz2012}). The $\Sigma_{\rm SFR}-\Sigma_{\rm H_2}/t_{\rm ff}$ relation can be 
interpreted as dependence of the star formation law on the local volume density of the gas, whilst the $\Sigma_{\rm SFR}-\Sigma_{\rm H_2}/t_{\rm orb}$ 
relation suggests that the star formation law is affected by the global rotation of the galaxy. Thus, the relevant timescale gives us critical information about 
the physical processes that may control the formation of stars.

Although we are considering just galaxies where their CO luminosity is spatially resolved (see ~\ref{sec:VALES_sample}), we do not have any information 
of the spatial extent of the IR luminosity, i.e. the SFR. Therefore, to study the KS law and its dependence on different timescales we need to assume 
the spatial extent of the SFR within each galaxy. However, in order to avoid the need of this assumption, instead of using the surface density quantities
$\Sigma_{\rm H_2}$ and $\Sigma_{\rm SFR}$, we use the spatially integrated variables. We also try not to assume a specific CO-to-H$_2$ conversion factor,
thus, we finally use the SFR (from $L_{\rm IR}$) and $L'_{\rm CO}$  galactic quantities (Table.~\ref{tab:table1}).

In Fig.~\ref{fig:ks_law} we investigate whether the star formation activity occurs on a timescale set by the orbital time ($t_{\rm orb}\equiv r_{\rm 1/2,CO}/V_{\rm rot}$; left panel)
or the crossing time ($t_{\rm cross}\equiv r_{\rm 1/2,CO}/\sigma_v$; right panel) by studying the SFR$-L'_{\rm CO}/t_{\rm orb}$ and SFR$-L'_{\rm CO}/t_{\rm cross}$
correlations, respectively. We just consider these two timescales as they can be calculated directly from our molecular gas ALMA observations. We also include the data
presented in the EDGE-CALIFA survey \citep{Bolatto2017}. These data were modelled using the same procedure described for the VALES data (see \S~\ref{appendix2} for more details).

In the left panel of Fig.~\ref{fig:ks_law} we find that galaxies with high SFRs also tend to present high $L'_{\rm CO}/t_{\rm orb}$ ratios. The Spearman's rank correlation
coefficient is 0.52 with a significance of its deviation from zero of 1.3$\times 10^{-7}$. Thus, we find a correlation between SFR and $L'_{\rm CO}/t_{\rm orb}$ in our data. 
The data are fitted by a power-law function with best-fit slope of 1.21$\pm$0.14. However, we note that the VALES galaxies with low $V_{\rm rot}/\sigma_v$ ratio
tend to lie above this fit, suggesting an enhanced star-formation efficiency per orbital time in these systems, in contradiction with \citet{Daddi2010}.
Although this correlation is often used to suggest that global galactic rotation may affect the star formation process (e.g. \citealt{Silk1997}),
we note that, perhaps, this explanation may not be the unique.

Galaxies with higher SFRs are expected to be more massive and have large gas content. Also, a massive galaxy is expected to rotate faster in order to balance its self-gravity. 
Therefore, the SFR-$L'_{\rm CO}/t_{\rm orb}$ correlation may reflect the mass-velocity trend in galaxies. We note that the location of the galaxies with low 
$V_{\rm rot}/\sigma_v$ ratio in the SFR-$L'_{\rm CO}/t_{\rm orb}$ plane can also be explained by the same kind of argument, but by considering a more sophisticated 
hydrodynamical balance against the gravitational force. In \S~\ref{sec:mdyn_gasfrac} we found that five systems are better represented by thick galactic disks than thin galactic disks. 
In thick disks, the gravitational force is balanced by both, negative radial pressure gradients and rotational support. Thus, thick galactic disks present lower $V_{\rm rot}$ values 
compared to thin galactic disks with equivalent mass. Therefore, the apparently enhanced star formation efficiency per orbital time observed in the SFR-$L'_{\rm CO}/t_{\rm orb}$ 
plane for galaxies with low $V_{\rm rot}/\sigma_v$ ratio in our sample may be produced by a relative reduction of the rotational velocity as pressure support is not 
negligible across their galactic disk.

Within the thick disk model \citep{Burkert2010}, the relevance of the pressure support is reflected by the velocity dispersion value (Eq.~\ref{eq:vrot_eqn}). Thus, one 
possible way to test the pressure support influence on the star formation is to test dependence of the SFR on the timescale given by $\sigma_v$, the crossing time.
We remind that the velocity dispersion range observed in our sample is $\sigma_v\sim 22-79$\,km\,s$^{-1}$ (Table~\ref{tab:table1}).
In the right panel of Fig.~\ref{fig:ks_law} we plot the SFR as a function of $L'_{\rm CO}/t_{\rm cross}$. We also find that galaxies with high SFR tend to have greater 
$L'_{\rm CO}/t_{\rm orb}$ ratio. The Spearman's rank correlation coefficient is 0.69 with a significance of  5.6$\times 10^{-13}$, also suggesting a correlation between both 
quantities. The best-fit power law function is 1.38$\pm$0.13. The SFR-$L'_{\rm CO}/t_{\rm cross}$ represents a reasonably better fit to data. However, we caution
that the fitting procedures are highly sensitive on whether we include the starburst galaxies or not in our data, i.e., the parameter errors may be underestimated.

\begin{figure*}
 \centering
 \includegraphics[width=2.0\columnwidth]{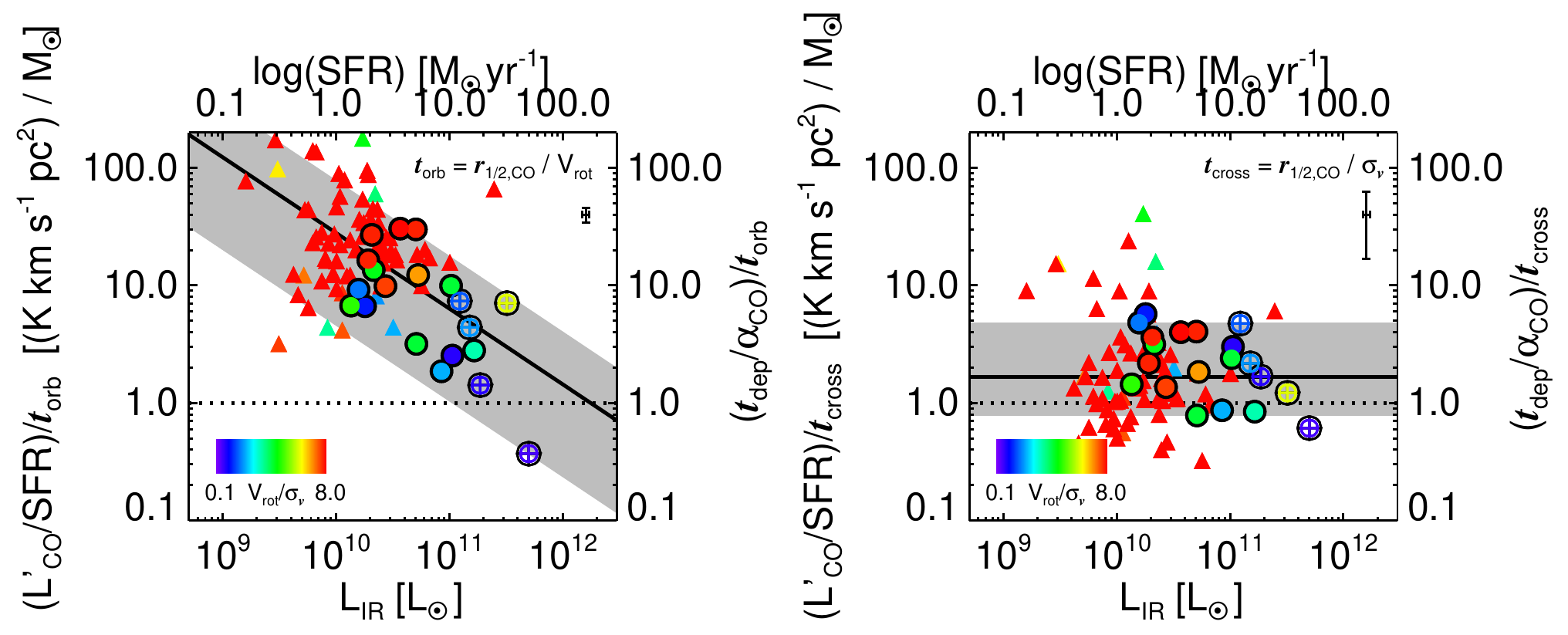}
 \caption{ \label{fig:t_dep}
 The $L'_{\rm CO}/L'_{\rm IR}$ ratio (SFE$'^{-1}$) divided by the orbital time (left) and crossing time (right) as a function of the IR-luminosity (SFR). 
 This can be interpreted as the molecular gas depletion time divided by the respective timescale (orbital or crossing time) as a function of the SFR,
 without any assumption of the CO-to-H$_2$ conversion value. The data presented in both panels are coded in the same way as Fig.~\ref{fig:ks_law}).
 The dotted lines represent a 1:1 ratio. In the left panel, the black line shows the best-fitted power-law function and the 1\,$\sigma$ error is represented by the
 grey-shaded region. In the right panel, the black line shows the median value and the grey shaded region also represents the 1\,$\sigma$ region. Interestingly,
  the data is consistent with a star formation law in which our proxy of depletion time is fixed per crossing time.}
\end{figure*}

Another way to study the star-formation law is by defining the `star formation efficiency' parameter as the star formation rate divided by the CO(1-0) luminosity 
(SFE$'\equiv$\,SFR/L$'_{\rm CO}$),  i.e. a proxy of the molecular depletion time without assuming any $\alpha_{\rm CO}$ factor can be obtained by calculating
 L$'_{\rm CO}$/SFR \citep{Cheng2018}. This depletion time proxy can be compared with other timescales `$t$'  by calculating the (L$'_{\rm CO}$/SFR)/$t$ ratio.
If the star formation efficiency can be expressed as SFE$'=$SFE$'_t /t$, with the star-formation efficiency per timescale (SFE$'_t$) being constant, then the
quantity (L$'_{\rm CO}$/SFR)$/t$ should be also constant for the timescale `$t$' regardless of the SFR of the system. This can be understood as a constant depletion time 
per timescale unit. We test this in Fig~\ref{fig:t_dep} by showing our proxy of the depletion time divided by the orbital time [(L$'_{\rm CO}$/SFR)$/t_{\rm orb}$] and the crossing 
time [(L$'_{\rm CO}$/SFR)$/t_{\rm cross}$]. Both panels are plotted against the SFR. 

We can see in the left panel of Fig~\ref{fig:t_dep} how our proxy of depletion time per orbital time varies with SFR. The Spearman's rank correlation coefficient is
$-0.30$ with a significance of 0.004. We also get a best-fit power-law slope of $-0.64\pm$0.09. As mentioned before, this trend may be enhanced
 by pressure support effects which increase the observed orbital time (by decreasing $V_{\rm rot}$) as $\sigma_v$ becomes comparable to $V_{\rm rot}$.
 On the other hand, in the right panel of Fig~\ref{fig:t_dep} we can see that our proxy of depletion time per crossing time appears to be independent of the SFR. 
The Spearman's rank correlation coefficient is $-0.003$ with a significance of 0.98, suggesting not correlation between both quantities. This plot also suggests that
L$'_{\rm CO}$/SFR$\approx t_{\rm cross}$, with a median value of $\sim$1.7 and the 1\,$\sigma$ region between $\sim$0.8 and 4.8. This suggests  SFE$'_{t_{\rm cross}}\sim1$.

We note that a constant star formation efficiency per crossing time found in our work departs from the fixed efficiency per free-fall time suggested by \citet{Krumholz2012},
as their estimation of the free-fall time for their extragalactic data set varies with the rotational velocity ($t_{\rm ff,T}\propto \Omega^{-1} \propto V_{\rm rot}^{-1}$).
Nevertheless, in the $V_{\rm rot}/\sigma_v\approx1$ limit, the free-fall time calculated by \citet{Krumholz2012} becomes comparable with the crossing time  
($t_{\rm cross}\approx\Omega^{-1}$). Thus, both laws fit the extragalactic data as a simple linear function in the $V_{\rm rot}/\sigma_v \gtrsim 1$ range. 
A possible way to differentiate both laws would be by performing spatially resolved molecular gas observations in star-forming galaxies with 
$V_{\rm rot}/\sigma_v < 1$, where $t_{\rm ff,T}>t_{\rm cross}$.

Our finding also does not contradict numerical simulations in which it is found that the star formation efficiency is `well-represented' by an exponentially decreasing 
function of the angular velocity of the disk \citep{Utreras2016}.

Before finalising this study, we would like to caution that spatial resolution effects may affect our analysis of  
Fig.~\ref{fig:ks_law} and Fig~\ref{fig:t_dep}. Indeed, the low spatial resolution of our observations may lead to
an overestimation of $t_{\rm orb}$ through possible overestimated CO half-light radii and underestimated 
$V_{\rm rot}$ (\S~\ref{sec:rel_effects}). This effect might decrease the estimated gas consumption rate per  
orbital time, especially on the sources observed at lower spatial resolution, which also are the galaxies with greater pressure support
and higher $r_{\rm 1/2, CO}$ values within our sample. We note that the slope of the best-fit for the SFR-$L'_{\rm CO}/\tau_{\rm orb}$
correlation gets lower ($N=0.87\pm0.09$) if we just consider galaxies observed with a projected beam size lower than
5\,kpc within our sample. This spatial resolution limit is, for example, the value obtained for the current seeing limited
($\sim$0$\farcs6$) IFS observations at $z\sim1$. We also note that this spatial resolution threshold selects 
`normal' star-forming galaxies, but just one starburst galaxy with $V_{\rm rot}/\sigma_v \approx 2$  at $z<0.06$ within
our resolved sample. In this case, all of the selected galaxies have SFR$\lesssim$12\,$M_\odot$\,yr$^{-1}$. Thus, our conclusion remains unchanged.  

On the other hand, $t_{\rm cross}$ may be less affected by spatial resolution effects as both, $\sigma_v$ and   
$r_{\rm 1/2, CO}$ values tend to be overestimated. If we consider the galaxies observed with a projected beam size lower  
than 5\,kpc within our sample, we found a best-fit slope for the SFR-$L'_{\rm CO}/\tau_{\rm cross}$ correlation  
of $1.13\pm0.17$. However, as we mentioned earlier, this threshold just include one starburst galaxy  
with low $V_{\rm rot}/\sigma_v$ ratio and SFR. When we include galaxies observed with a spatial resolution up to
7\,kpc (six more galaxies; $V_{\rm rot}/\sigma_v \sim 1$; SFR$\lesssim$20\,$M_\odot$\,yr$^{-1}$), we obtain a slope of $1.23\pm0.12$.
In summary, our results are dependent whether we consider the systems with high SFRs and lower $V_{\rm rot}/\sigma_v$ 
ratios. Regardless of the spatial resolution effects discussed recently, the variable $\alpha_{\rm CO}$ factor should also have an affect on our analysis.

\section{Conclusions} 
We present ALMA observations of 39 flux-selected ($S_{160\mu m} \geq 100$\,mJy; $L_{\rm IR} \approx 10^{10-12}$\,$L_\odot$) galaxies with
detected CO($J=1-0$) emission, comprising `starburst' and `normal' star-forming galaxies drawn from the VALES survey (V17), at the redshift range 
of 0.02\,<\,$z$\,<\,0.35. We incorporate the exquisite multi-wavelength coverage from the GAMA survey. We found 20 galaxies with extended (`resolved') 
emission whilst 19 have `compact' (or `unresolved') emission. The spatial resolution of the sample ranges from 2 to 8\,kpc, with a fixed spectral 
resolution of 20\,km\,s$^{-1}$. We model the CO(1-0) kinematics by using a two-dimensional disk model with an arctan velocity profile and consider
disk thickness effects on the projection of the galactic disk in the observed plane. These new observations represent one of the largest samples of molecular 
gas kinematics traced by the CO of `typical' and `starburst' star-forming galaxies at intermediate redshifts.

The median $V_{\rm rot}/\sigma_v$ ratio for our sample is 4.1 and the $V_{\rm rot}/\sigma_v$ values range between $0.6-7.5$. We found median 
$V_{\rm rot}/\sigma_v$ ratios of 4.3 and 1.6 for the `normal' star-forming and starburst sub-samples, respectively. The median $V_{\rm rot}/\sigma_v$ 
value for the `normal' galaxies in our sample is consistent with the expected evolution with redshift for this ratio.

We find a tentative correlation between the L$_{\rm IR}$ luminosity with the rotation-to-pressure support ratio ($V_{\rm rot}/\sigma_v$).
That anti-correlation suggest a smooth transition of the star formation efficiency on terms of the kinematic state for `starburst' and `normal' star-forming galaxies.

We find that the [C{\sc ii}]/IR ratio decreases at low $V_{\rm rot}/\sigma_v$ ratio. The data are well-represented by a power-law with best-fit slope of 
0.74$\pm$0.14. Our finding is consistent with \citet{Ibar2015} who found that galaxies presenting a prominent disk show higher $L_{\rm [C_{II}]}/L_{\rm IR}$ 
ratios than those which do not present disky morphologies. The VALES galaxies with $V_{\rm rot}/\sigma_v \gtrsim 3$ tend to show $L_{\rm [C_{II}]}/L_{\rm IR}$ 
comparable with the values measured in the \texttt{KINGFISH} survey for nearby galaxies \citep{Smith2017}, whilst galaxies with $V_{\rm rot}/\sigma_v \lesssim 3$
tend to show lower $L_{\rm [C_{II}]}/L_{\rm IR}$ values.

We compare the physical parameters derived by PDR modeling for our sample \citep{Hughes2017a} with the $V_{\rm rot}/\sigma_v$ ratio.
We find that high hydrogen nuclei densities and high strength of the FUV radiation field are likely to be found in systems with low $V_{\rm rot}/\sigma_v$ ratio,
with the latter quantity being almost independent of the CO, [C{\sc ii}] luminosities used to constrain the PDR model.

By calculating dynamical masses following both, thin and thick turbulent disk models, we find that the thin disk model tends to underestimate the galactic total 
mass as its values are lower than the estimated stellar masses for five of our galaxies. On the other hand, the thick turbulent disk model tends to 
alleviate this conflict, suggesting that these sources with low $V_{\rm rot}/\sigma_v$ values are better represented by thick galactic disks. This also suggests 
that pressure support effects should not be neglected in high velocity dispersion galactic disks. We caution that this conclusion is strongly dependent on the 
spatial resolution of our observations.

We test if our rotationally supported galaxies are prone to develop gravitational instabilities. This is done by analysing our sources in the $f_{\rm H_2}-V_{\rm rot}/\sigma_v$
plane and comparing with expected values for a marginally stable gaseous thin disk ($Q_{\rm gas}=1$), a gaseous thick disk ($Q_{\rm thick}=1$) and a two component 
disk (stars plus gas; $Q_2=1$). From 11 galaxies classified as rotationally supported systems, we find that three galaxies are consistent with $Q_{\rm gas}\approx1$,
i.e., are prone to develop gravitational instabilities. The other eight systems have measured $Q_{\rm gas}\gtrsim1$. This conclusion is not changed 
if we apply the thick disk gravitational stability analysis as the kinematics estimates are not enough accurate. The gravitational analysis considering a galactic disk with both, 
a gaseous and stellar component, may change this result by increasing the number of galaxies consistent with being susceptible to develop gravitational instabilities, however 
stellar dynamics measurements are needed to corroborate this result.

We explore the possible origin of the energy sources of those high turbulent motions seen in our galaxies by comparing the SFRs with $f_{\rm H_2}$ and $\sigma_v$.
We find that the SFR is weakly correlated with $f_{\rm H_2}$. By comparing the data with two theoretical models in the literature, the feedback-driven and gravity-driven
models, we find that both models give a poor description of the data. This suggests that the main energy source of the supersonic turbulence observed in the VALES galaxies
seem to be neither the gravitational energy released by cold gas accreted through the galactic disk nor the energy injected into the ISM by supernovae feedback.

We study the spatially integrated star formation law dependence on galactic dynamics, avoiding assumptions about the the CO-to-H$_2$ conversion factor by studying the 
SFR$-L'_{\rm CO}/t_{\rm orb}$ and SFR$-L'_{\rm CO}/t_{\rm cross}$ relations. We find a correlation between SFR and $L'_{\rm CO}/t_{\rm orb}$, with a best-fit power-law 
slope of 1.21$\pm$0.13. We suggest that the SFR$-L'_{\rm CO}/t_{\rm orb}$ correlation is affected by the decrease of $V_{\rm rot}$ (thus, an increase of $t_{\rm orb}$) by 
pressure support which dilutes the gravitational potential in systems with low $V_{\rm rot}/\sigma_v$ ratio.

We find that our proxy of the `star formation efficiency' (SFE$' \propto$\,SFR/$L'_{\rm CO}$) is correlated with the crossing time, suggesting an 
efficiency per crossing time of $\sim$0.6. Therefore, by knowing the size, SFR, and mean velocity dispersion of a galaxy, we can estimate its molecular gas mass. 
By considering the better correlation between SFE$'$ and $t_{\rm cross}$, we propose that the crossing time may be the timescale in which the star formation occurs
in our systems. We caution, however, that the assumption of a CO-to-H$_2$ conversion factor and/or spatial resolution effects may change this result. 

Obtain deeper and higher resolution observations of the molecular gas in a large sample of highly turbulent systems is critical to confirm or refute the findings reported in our work.
It will allow to overcome spatial resolution effects which bias the velocity dispersion to higher values and to characterise the rotational velocity of the systems by observing
the flat part of the velocity curve. This will be done in future work.

\section*{Acknowledgements} 
We thank the referee for her/his careful reading and useful comments.
We acknowledge to L. Cortese and P. Papadopoulos for their generous discussions and useful comments.
This research was supported by CONICYT Chile (CONICYT-PCHA/Doctorado-Nacional/2014- 21140483)
E.I. acknowledges partial support from FONDECYT through grant N$^\circ$\,1171710. This paper makes use
of the following ALMA data: ADS/JAO.ALMA\#2012.1.01080.S, ADS/JAO.ALMA\#2013.1.00530.S and 
ADS/JAO.ALMA\#2015.1.01012.S. ALMA is a partnership of ESO (representing its member states), 
NSF (USA) and NINS (Japan), together with NRC (Canada), MOST and ASIAA (Taiwan), and KASI 
(Republic of Korea), in cooperation with the Republic of Chile. The Joint ALMA Observatory is operated 
by ESO, AUI/NRAO and NAOJ. CONICYT grants Basal-CATA PFB-06/2007 (FEB) and FONDECYT Regular
1141218 (FEB); the Ministry of Economy, Development, and Tourism's Millennium Science Initiative through
grant IC120009, awarded to The Millennium Institute of Astrophysics, MAS (FEB).
A.M.M.A.~acknowledges support from CONICYT through FONDECYT grant 3160776.
M.J.M.~acknowledges the support of the National Science Centre, Poland
through the POLONEZ grant 2015/19/P/ST9/04010; this project has
received funding from the European Union's Horizon 2020 research and
innovation programme under the Marie Sk{\l}odowska-Curie grant
agreement No. 665778.
GDZ acknowledges support  by the ASI/Physics Department of the university of Roma--Tor Vergata 
agreement n. 2016-24-H.0 for study activities of the Italian cosmology community.
This publication has received funding from the European Union's Horizon 2020 research and innovation 
programme under grant agreement No 730562 [RadioNet].
C.C. was supported from the Chinese Academy of Sciences (CAS) through the CASSACA Postdoc Grant, 
by the Visiting Scholarship Grant administered by the CAS South America Center for Astronomy (CASSACA), NAOC and
by the Young Researcher Grant of National Astronomical Observatories, Chinese Academy of Sciences.
G.O. acknowledges the support provided by CONICYT(Chile) through FONDECYT postdoctoral research grant no 3170942.
C.Y. was supported by an ESO Fellowship.
T.M.H.~acknowledges the support from the Chinese Academy of Sciences (CAS) and the National Commission for Scientific and Technological
Research of Chile (CONICYT) through a CAS-CONICYT Joint Postdoctoral Fellowship administered by the CAS South America Center for Astronomy
(CASSACA) in Santiago, Chile.





\appendix

\section{DYNAMICAL MASS WITH DIFFERENT DENSITY PROFILES}
\label{appendix1} 

In \S~\ref{sec:mdyn_gasfrac} we show that roughly half of the galaxies within our sample are best described by a thick disk model rather than a thin disk model. 
The thick disk dynamical model predicts total masses greater than the stellar masses at the same radii unlike the thin disk dynamical model. However,
in order to implement the thick disk model, the surface density of the source needs to be known. For our galaxies we assumed a surface 
density profile given by the observed surface density brightness in the $K-$band, but we also assume a constant mass-to-light ratio and light extinction 
over the galactic disk. This is likely not to be true since we expect a major concentration of dust (thus extinction) in the central part of galaxies compared to
their outskirts. With the aim to show that our choice of surface density distribution should not affect the conclusions of  \S~\ref{sec:mdyn_gasfrac}, 
in Fig.~\ref{fig:mdyn_apendix} we plot the ratio between the dynamical masses assuming an exponential surface density profile and our observational S\'ersic-like 
surface density profile as a function of the $V_{\rm rot}/\sigma_v$ ratio. When assuming an exponential surface density profile,
we obtain greater dynamical mass values on average compared to our observational S\'ersic-like surface density profiles. However, the 
median ratio of $\sim1.5$ indicates that our conclusions should not be sensitive to the chosen surface density profiles. 
We note that the possible trend between the dynamical mass ratio and the $V_{\rm rot}/\sigma_v$ ratio observed in Fig.~\ref{fig:mdyn_apendix}  is consistent
with the finding of increasing S\'ersic indices above unity at galaxies with $\log(M_*/M_\odot) > 10.5$ \citep{Wuyts2011b, Bell2012, Lang2014}. We also note that
disk truncation is expected to be enhanced in galaxies with considerable turbulent pressure support \citep{Burkert2016}.

\section{EDGE-CALIFA SURVEY}
\label{appendix2} 
In \S~\ref{sec:presion_support}, \S~\ref{sec:presion_support} and \S~\ref{sec:kennicutt} we complement our analysis by adding the EDGE-CALIFA survey 
data \citep{Bolatto2017} to our VALES data. The EDGE-CALIFA is survey based on interferometric CO($1-0$) observations made with the Combined Array 
for Millimeter-wave Astromy (CARMA) of 126 nearby ($d=23-130$\,Mpc) galaxies from the EDGE survey. This sample is selected from the CALIFA survey 
and it has on average spectral and spatial resolution of $\sim10$\,km\,s$^{-1}$ and $\sim$1.4\,kpc, respectively. Those are higher spectral and spatial resolution
observations than the ones presented in our survey (Table~\ref{tab:table1}).

From the EDGE-CALIFA survey, we analyse the galactic kinematics of the galaxies which have available their CO intensity, velocity and dispersion velocity maps
with the additional requirement that the velocity map must sample the galactic centre given from the SDSS `$igu$' multi-color image. Thus, we `just' analyse 70
galaxies from the EDGE-CALIFA survey. The kinematic analysis is done in the same manner than we did for the VALES survey, but with two differences; 
(1) we constrain the inclination angles by using the values presented in \citet{Bolatto2017}; and (2) we model these galaxies as thin galactic disks, i.e., q$_0=0.0$.
Finally, we correct the gas mass content by using our chosen CO-to-H$_2$ conversion factor, and we correct the stellar masses and  SFRs for a Chabrier IMF.

\section{KINEMATIC MAPS AND VELOCITY PROFILES}
\label{appendix3} 

In Fig.~\ref{fig:maps} we plot the kinematic maps (1st to 3rd columns), residual maps (4th column) and
one-dimensional velocity profiles (5th and 6th columns) for our sample taken from the VALES survey. Full 
information for the panels in the figure is given in its caption.

\renewcommand{\thefigure}{A1}
\begin{figure}
 \centering
 \includegraphics[width=1.0\columnwidth]{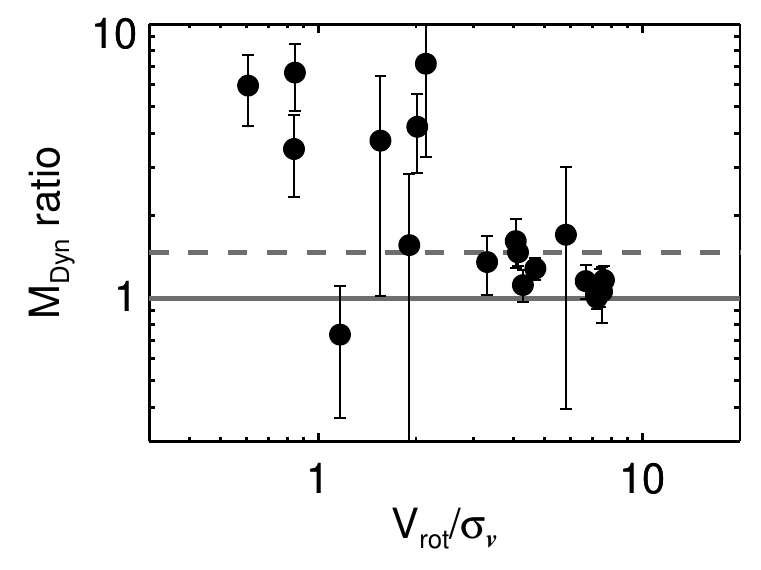}
 \caption{ \label{fig:mdyn_apendix}
 Ratio between the thick disk dynamical masses assuming an exponential surface density profile and our $K-$band converted surface density profile as a function
 of the $V_{\rm rot}/\sigma_v$ ratio for our sample. The grey line represents equality between both quantities. The dashed line represents the median value of $\sim1.5$
 for our sample. By assuming an exponential surface density profile we obtain greater dynamical mass estimates. This seems to be dependent of the
 observed $V_{\rm rot}/\sigma_v$ ratio.
 }
\end{figure}

\renewcommand{\thefigure}{C1}
\begin{figure*}
\flushleft
\includegraphics[width=0.343\columnwidth]{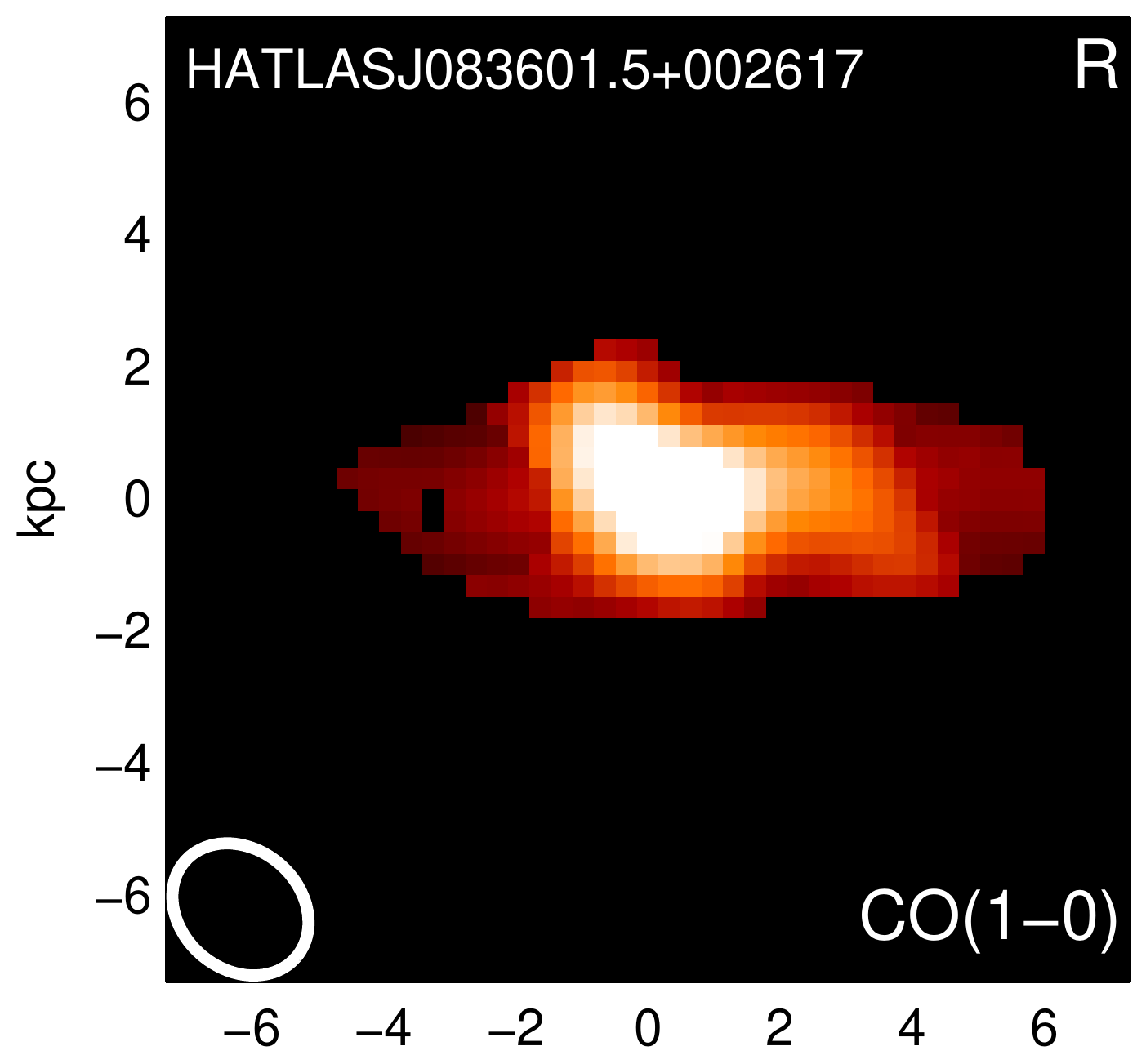}
\includegraphics[width=0.32\columnwidth]{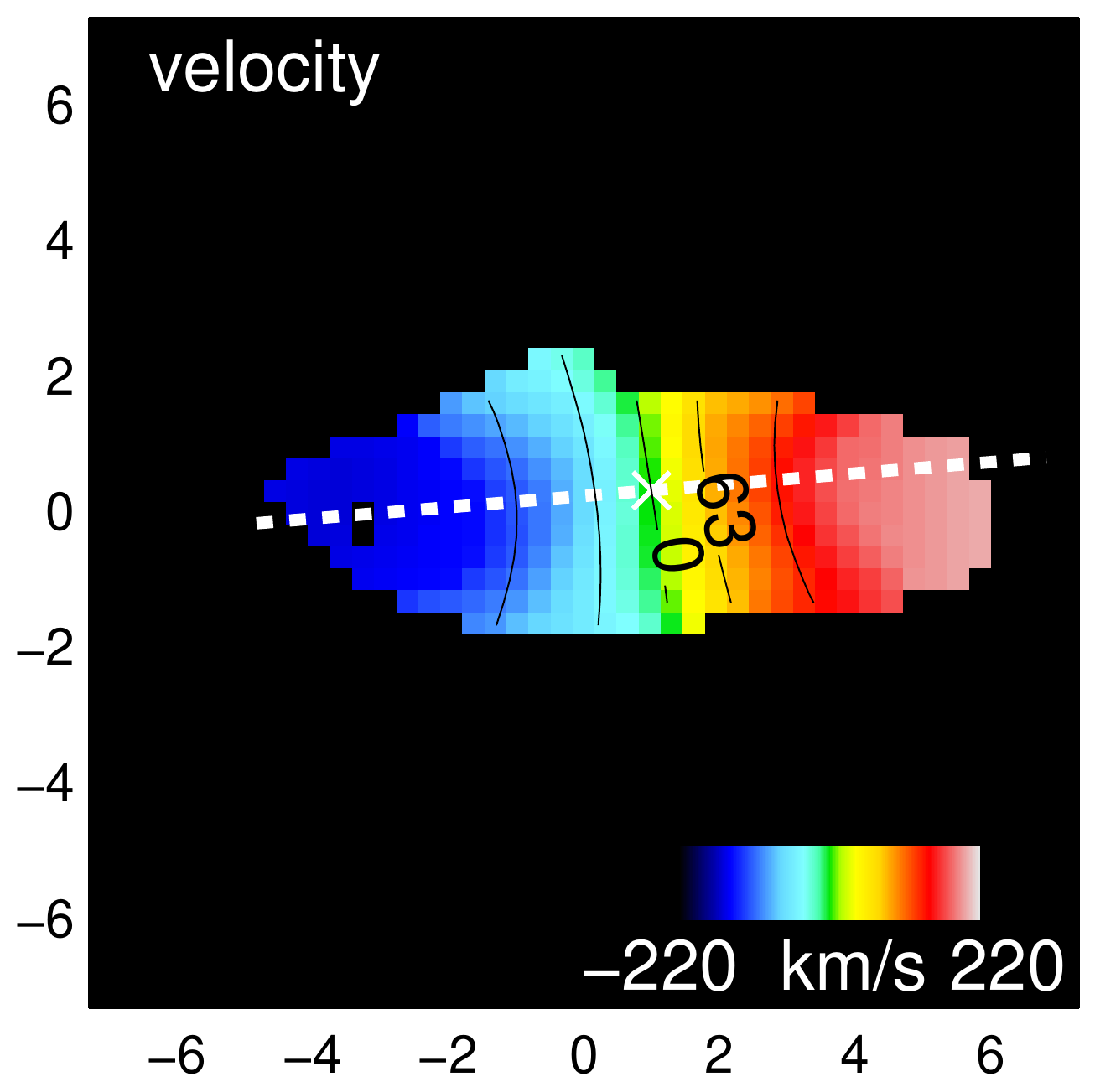}
\includegraphics[width=0.32\columnwidth]{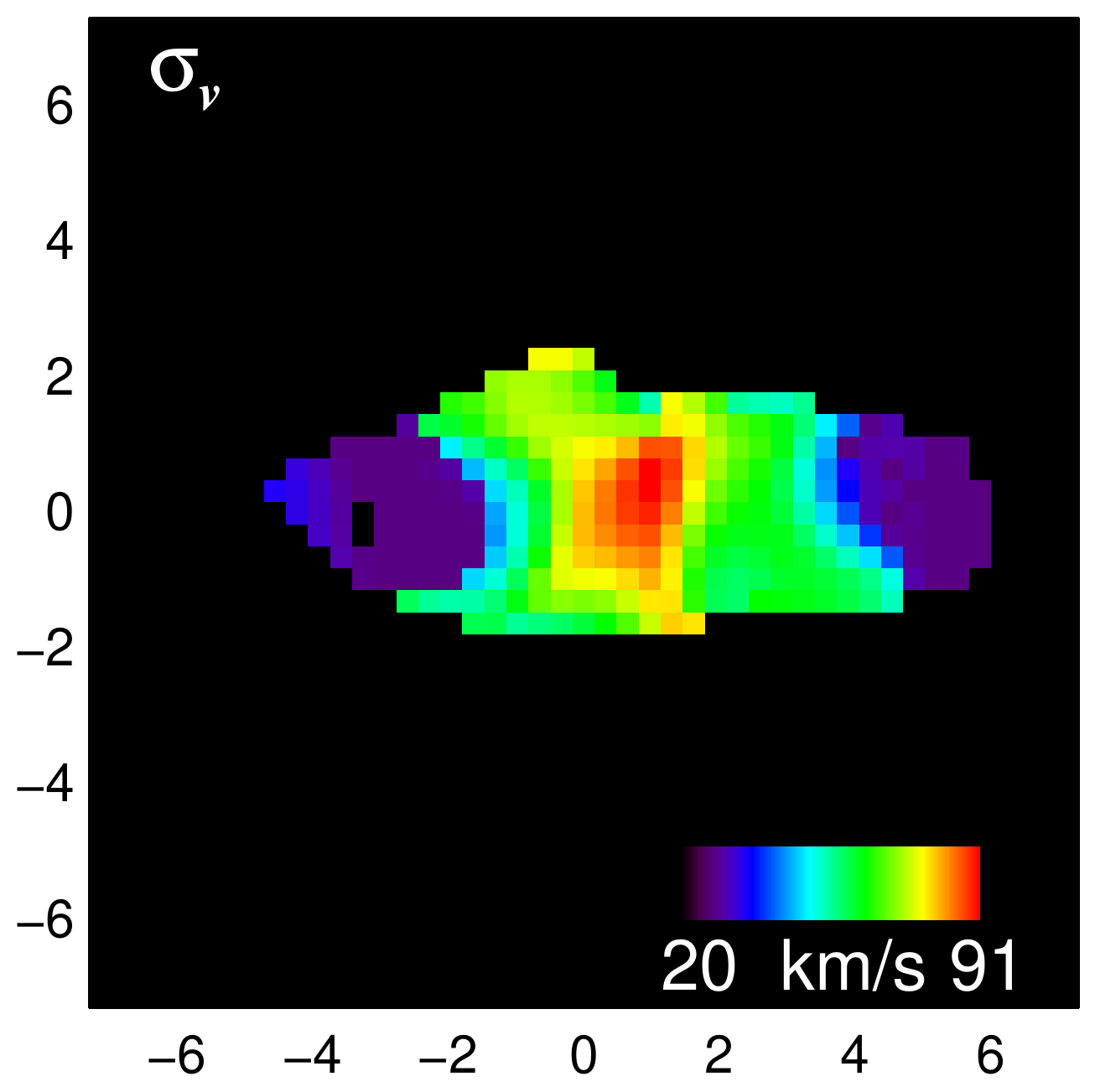}
\includegraphics[width=0.32\columnwidth]{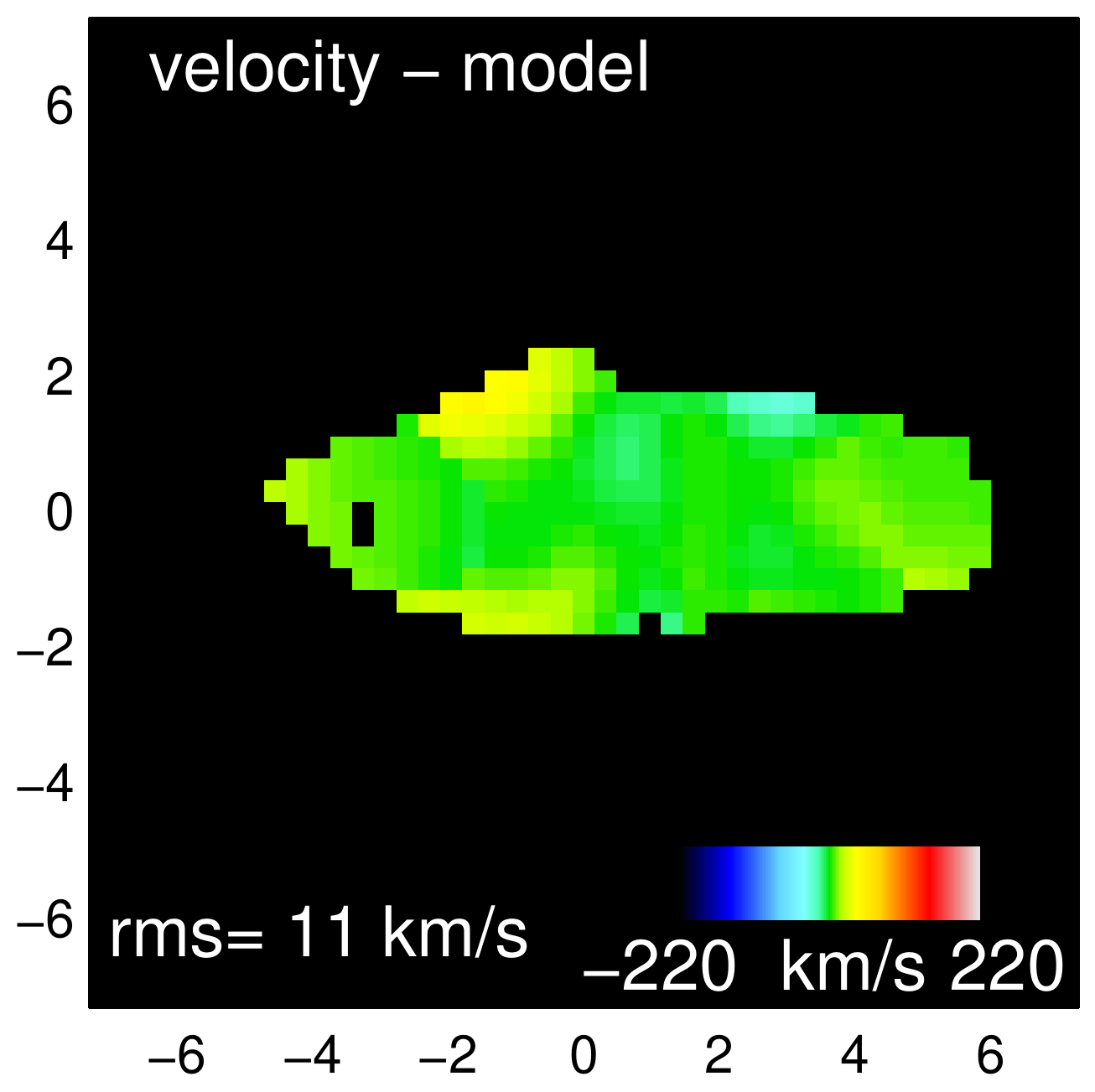}
\includegraphics[width=0.345\columnwidth]{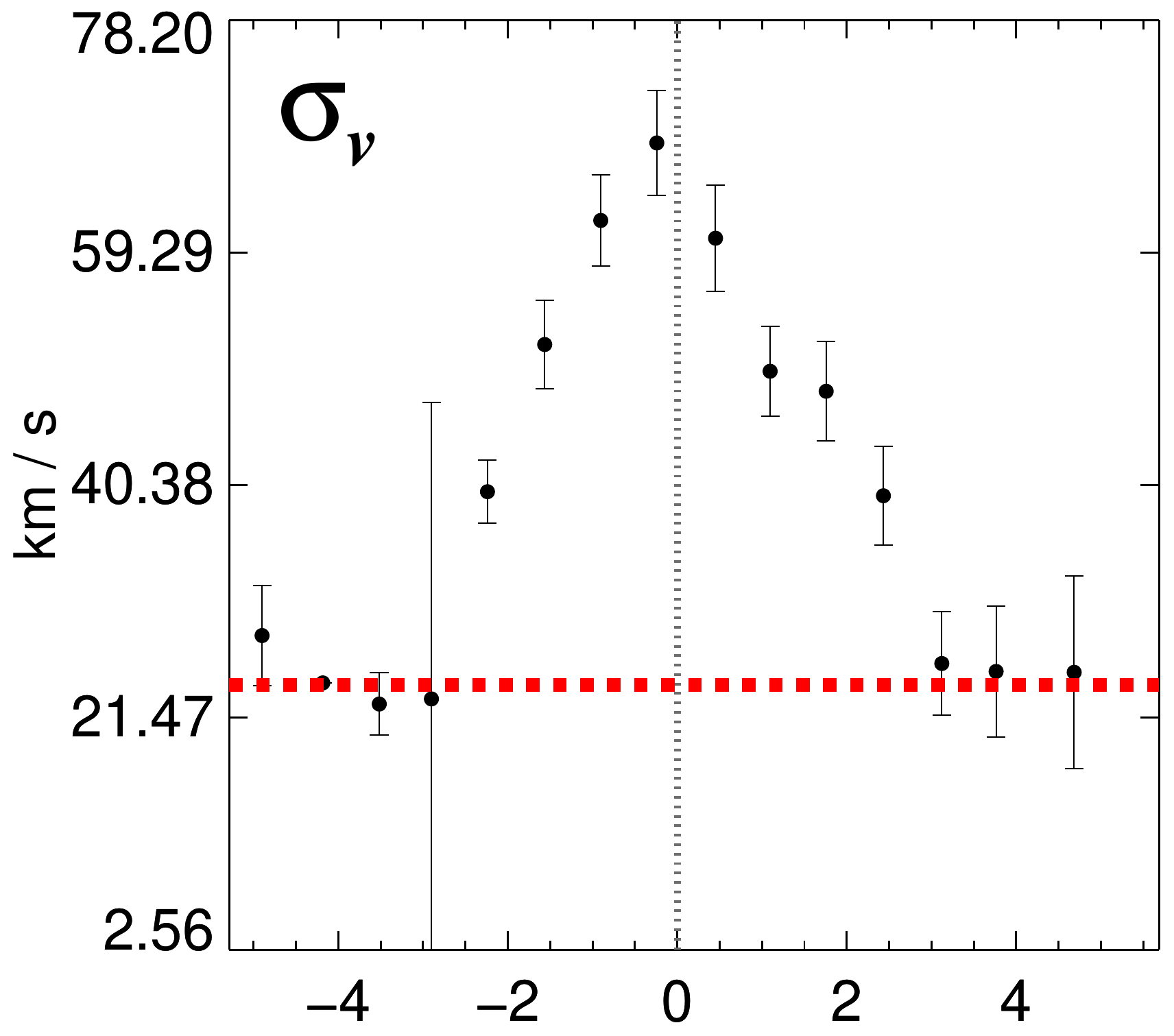}
\includegraphics[width=0.363\columnwidth]{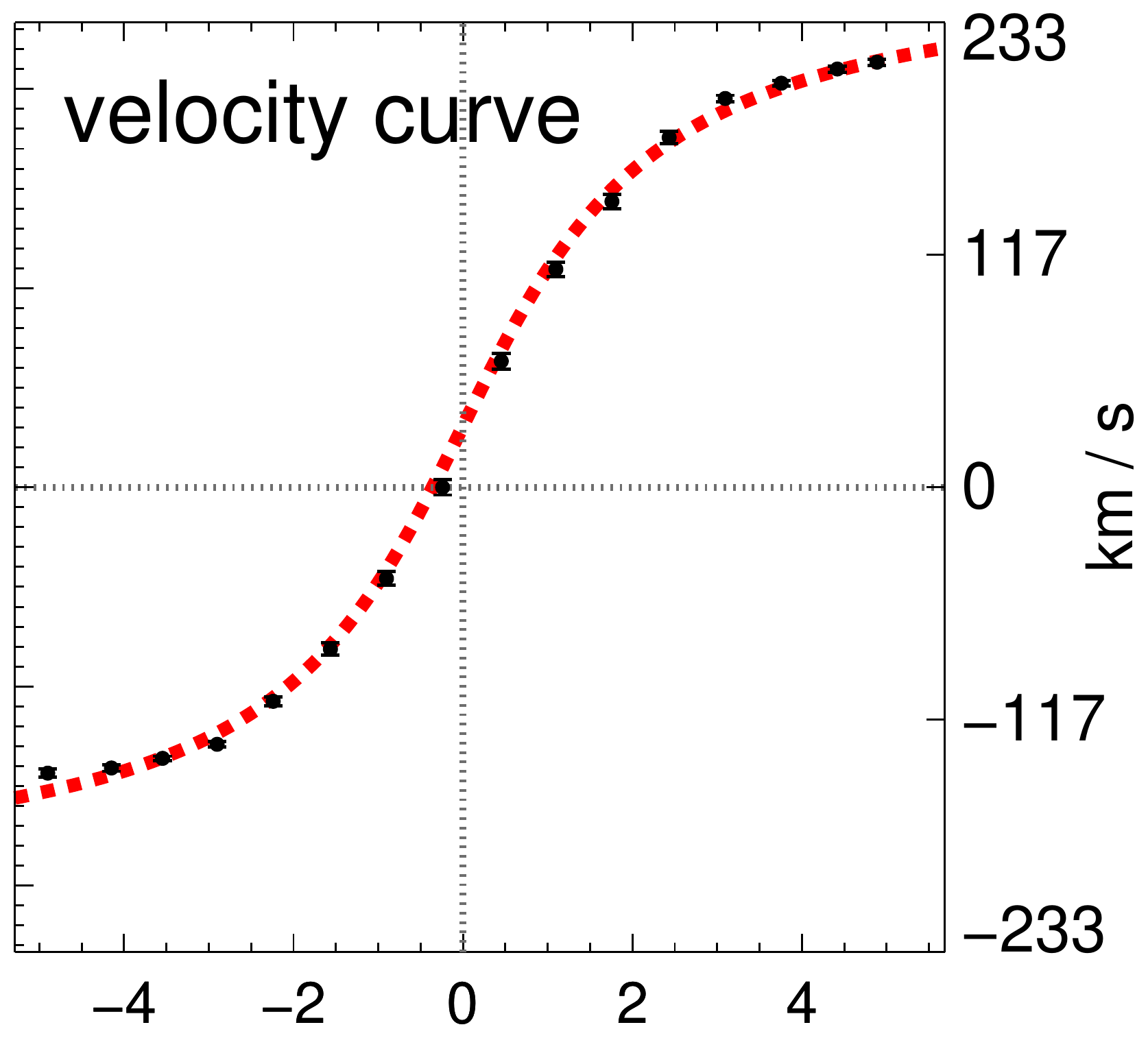}\\
\vspace{0.25mm}
\includegraphics[width=0.343\columnwidth]{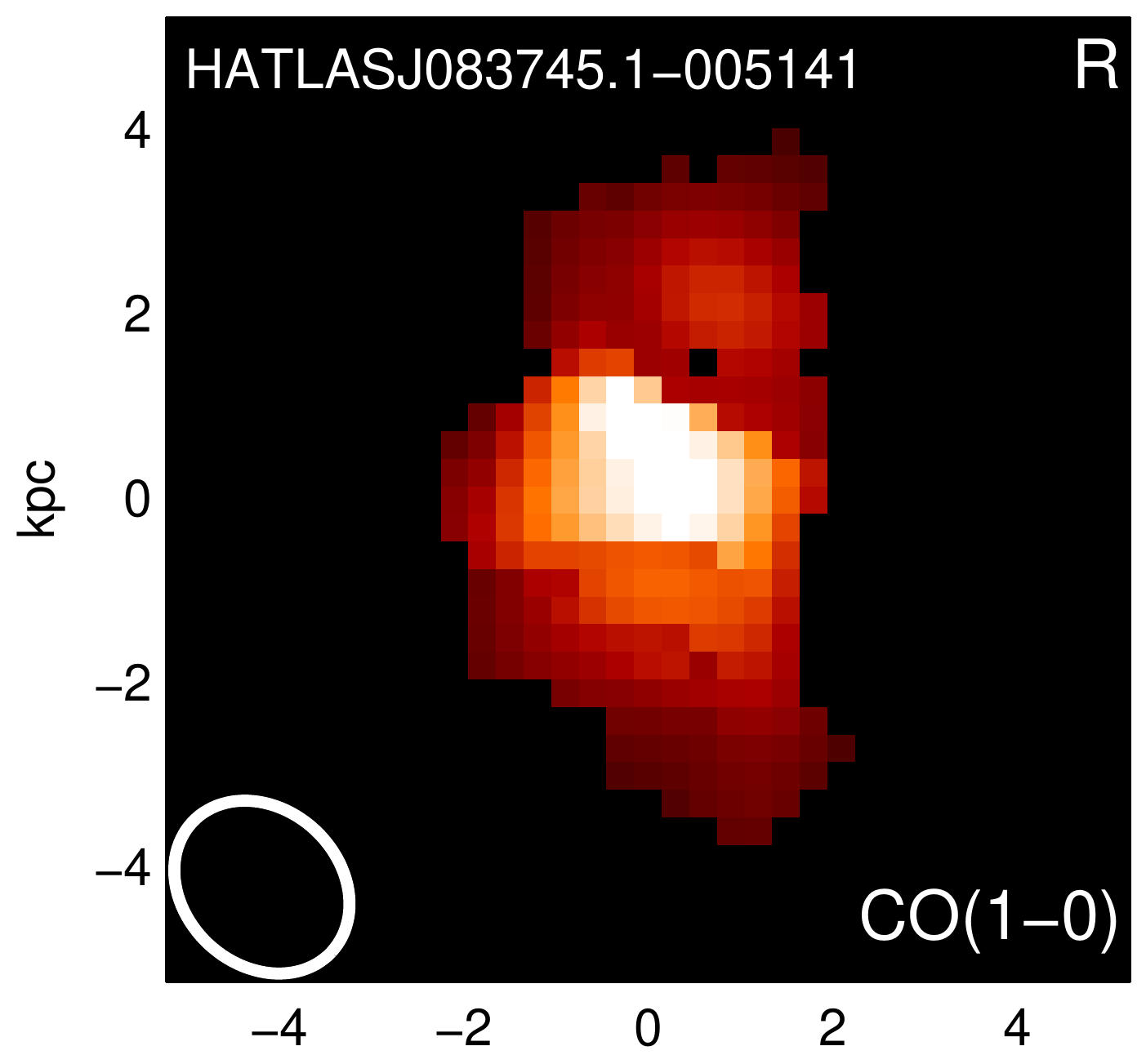}
\includegraphics[width=0.32\columnwidth]{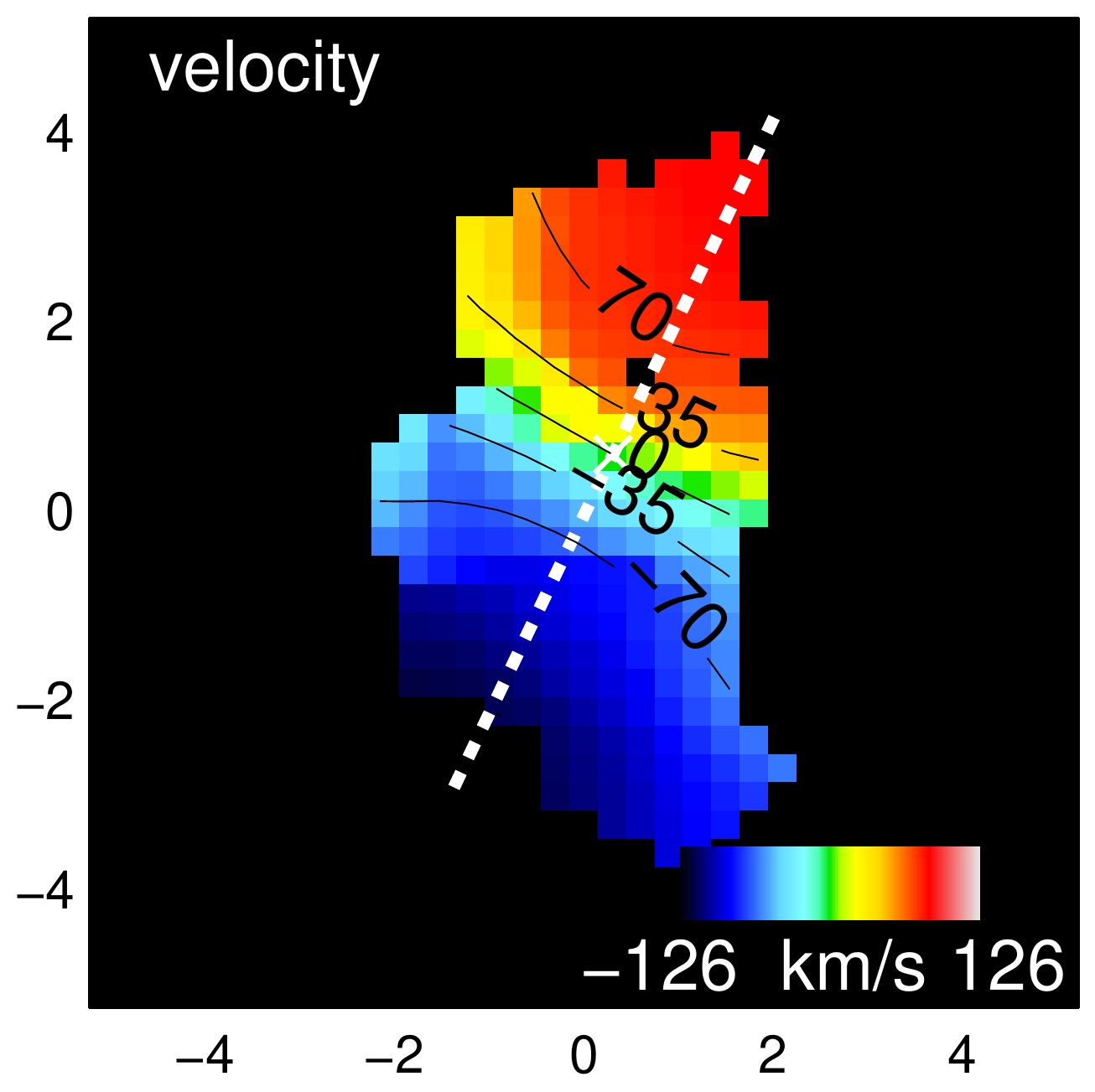}
\includegraphics[width=0.32\columnwidth]{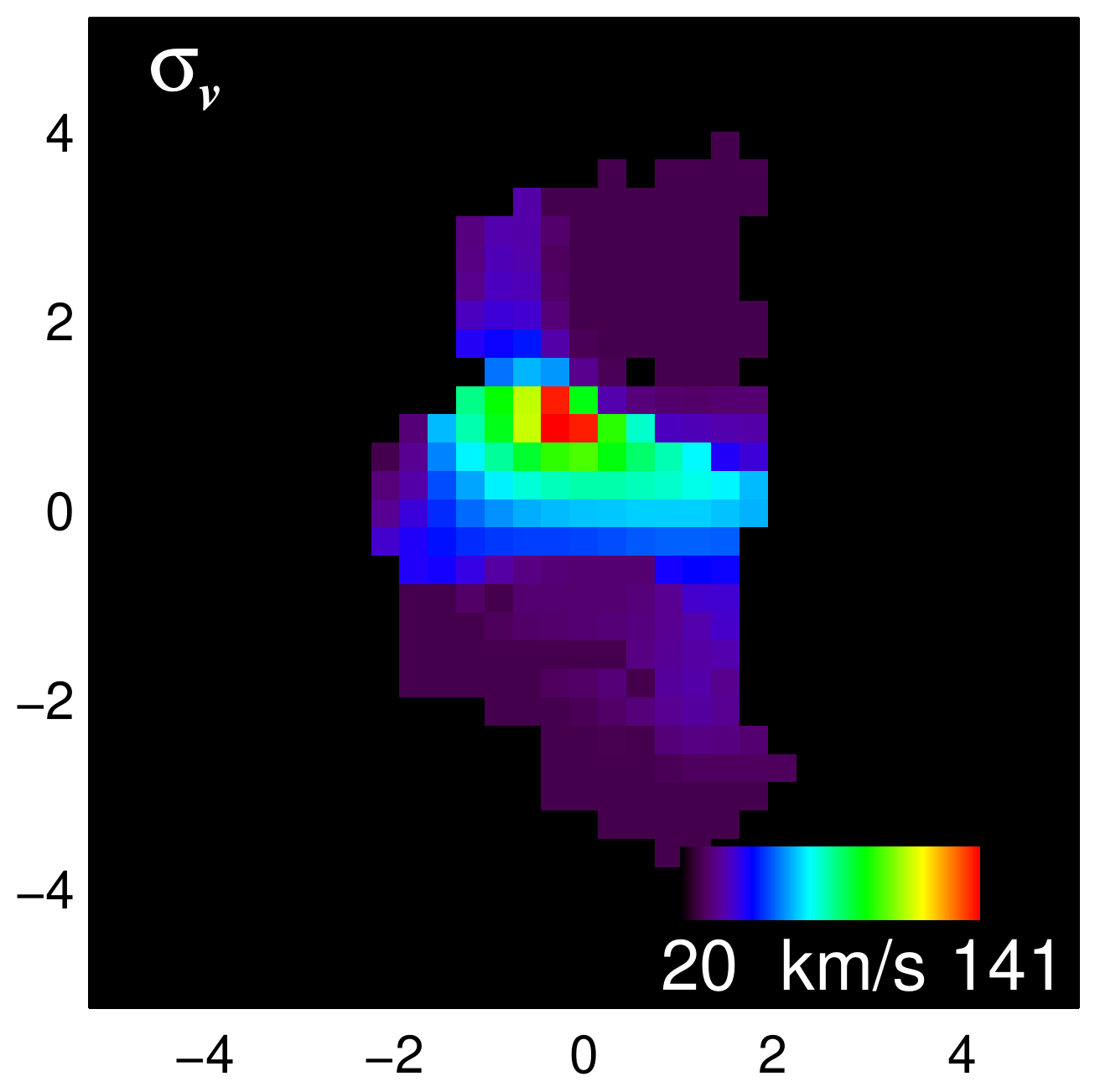}
\includegraphics[width=0.32\columnwidth]{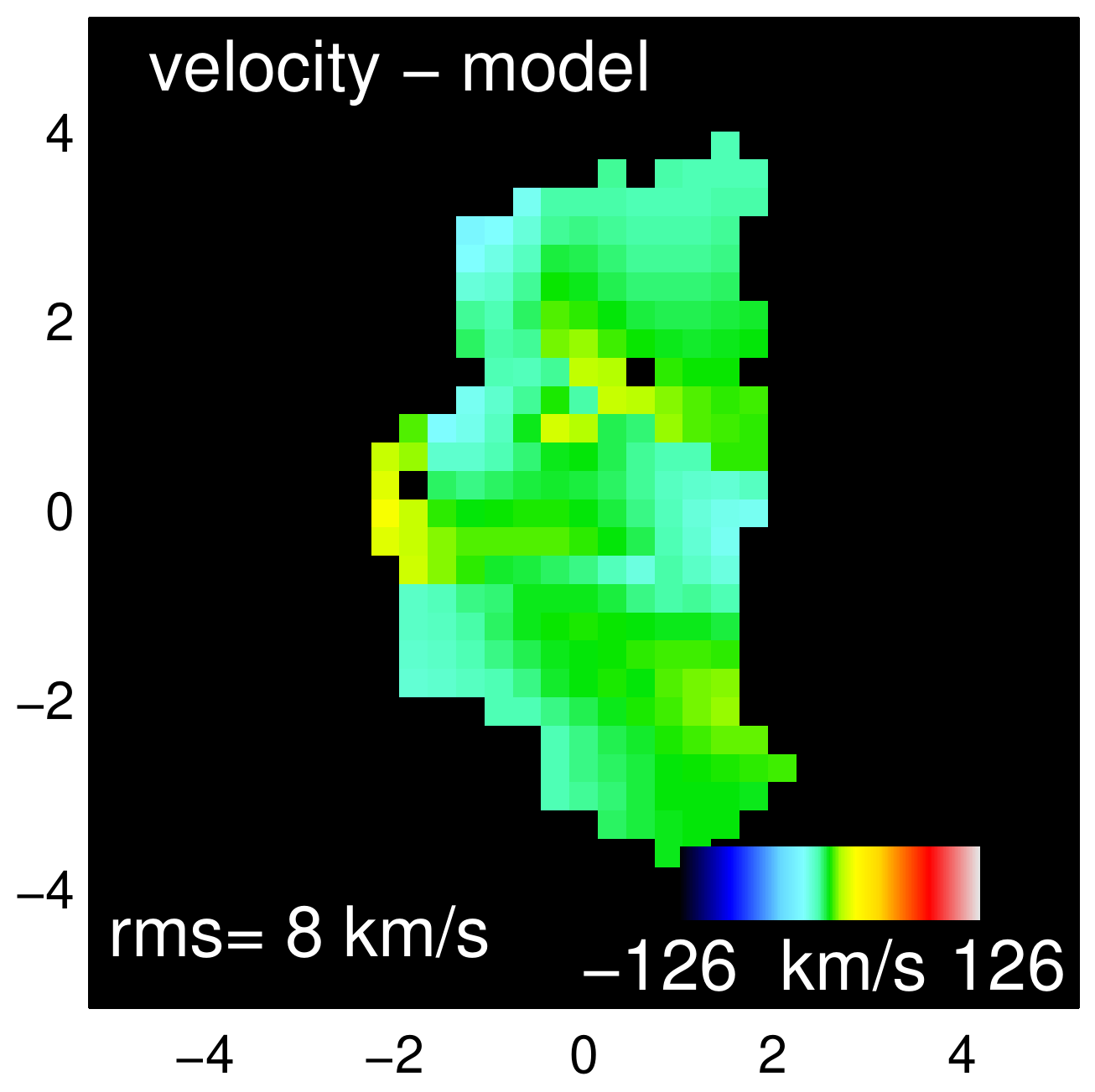}
\includegraphics[width=0.345\columnwidth]{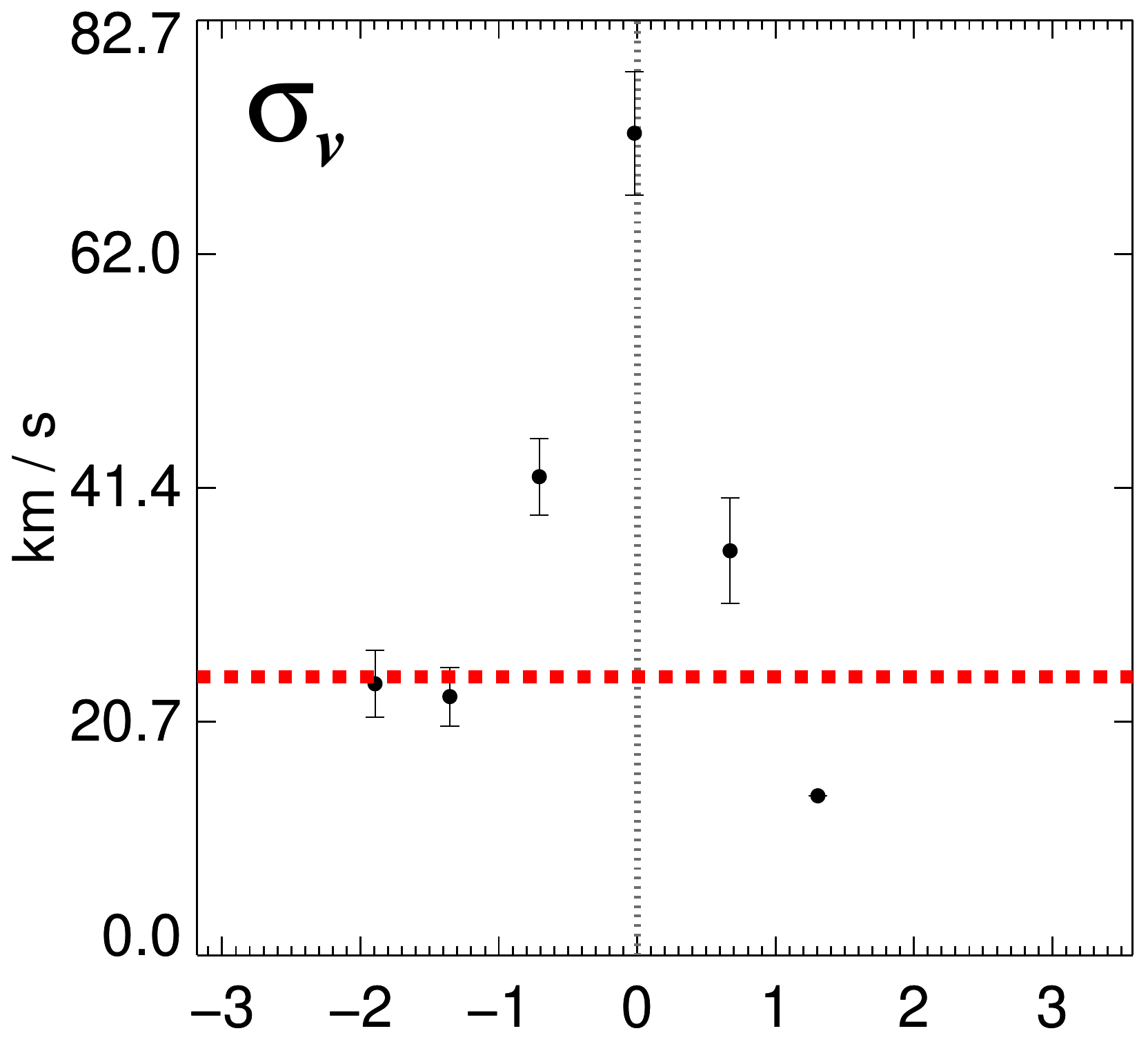}
\includegraphics[width=0.351\columnwidth]{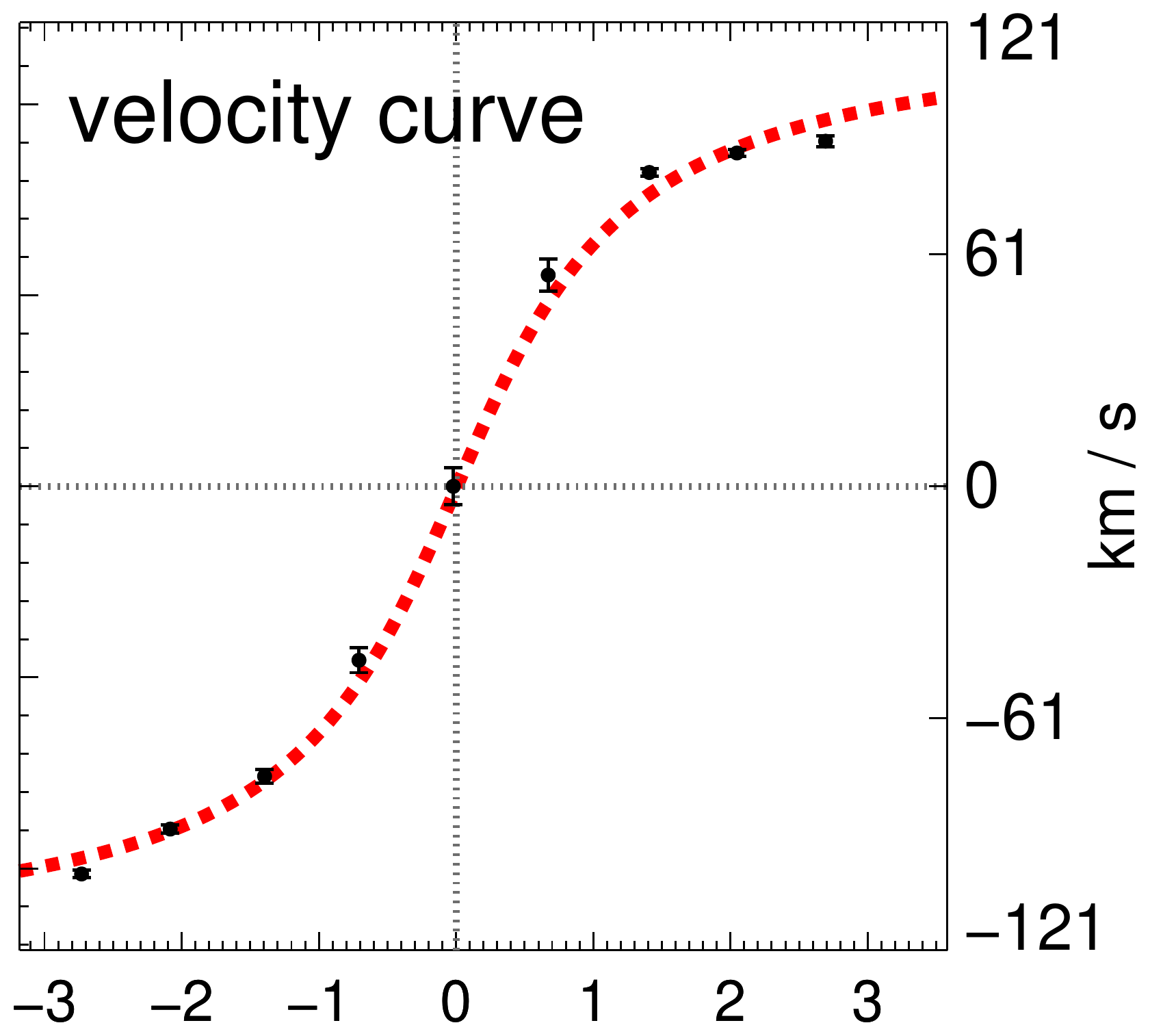}\\
\vspace{0.25mm}
\includegraphics[width=0.343\columnwidth]{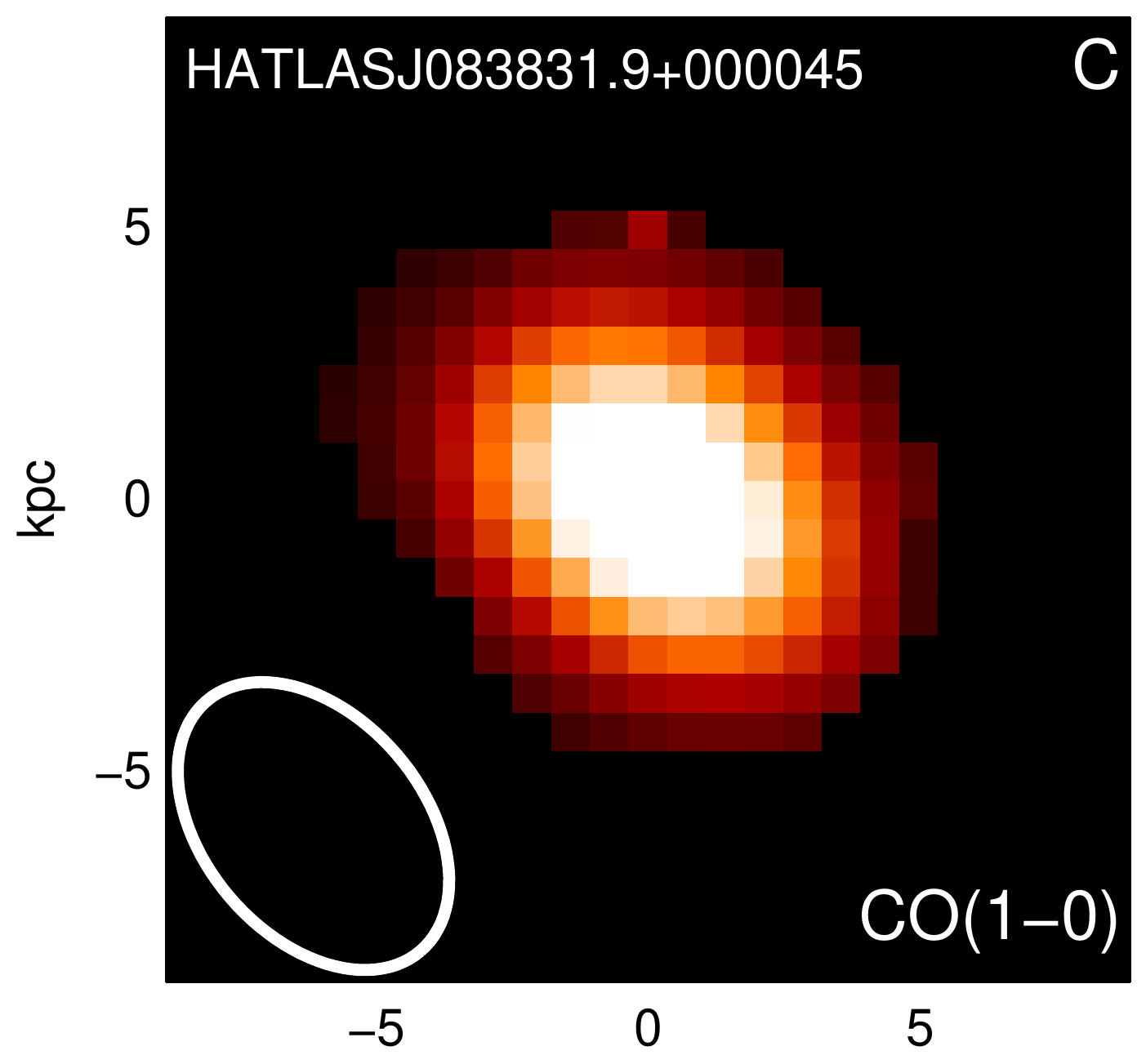}
\includegraphics[width=0.32\columnwidth]{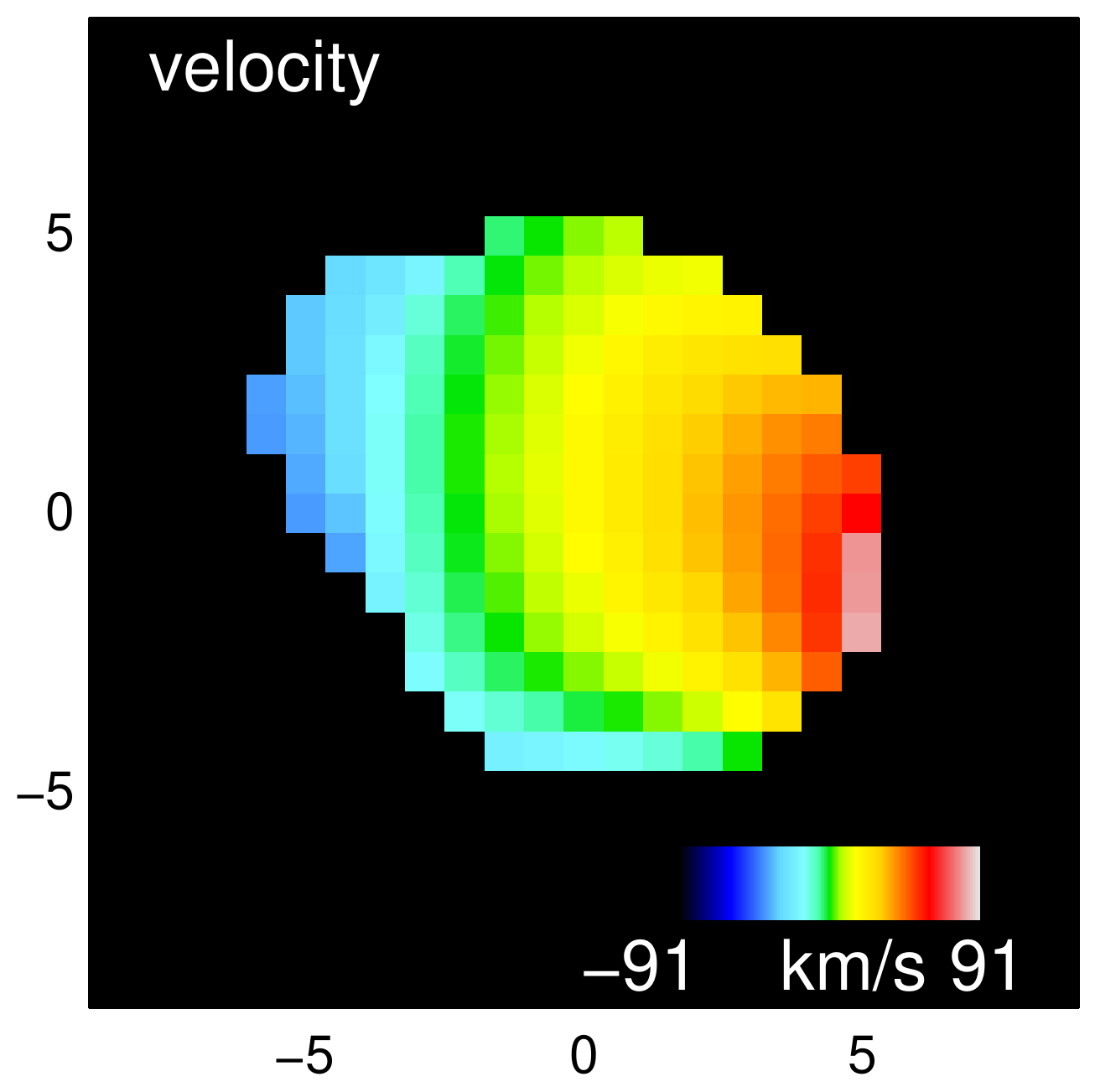}
\includegraphics[width=0.32\columnwidth]{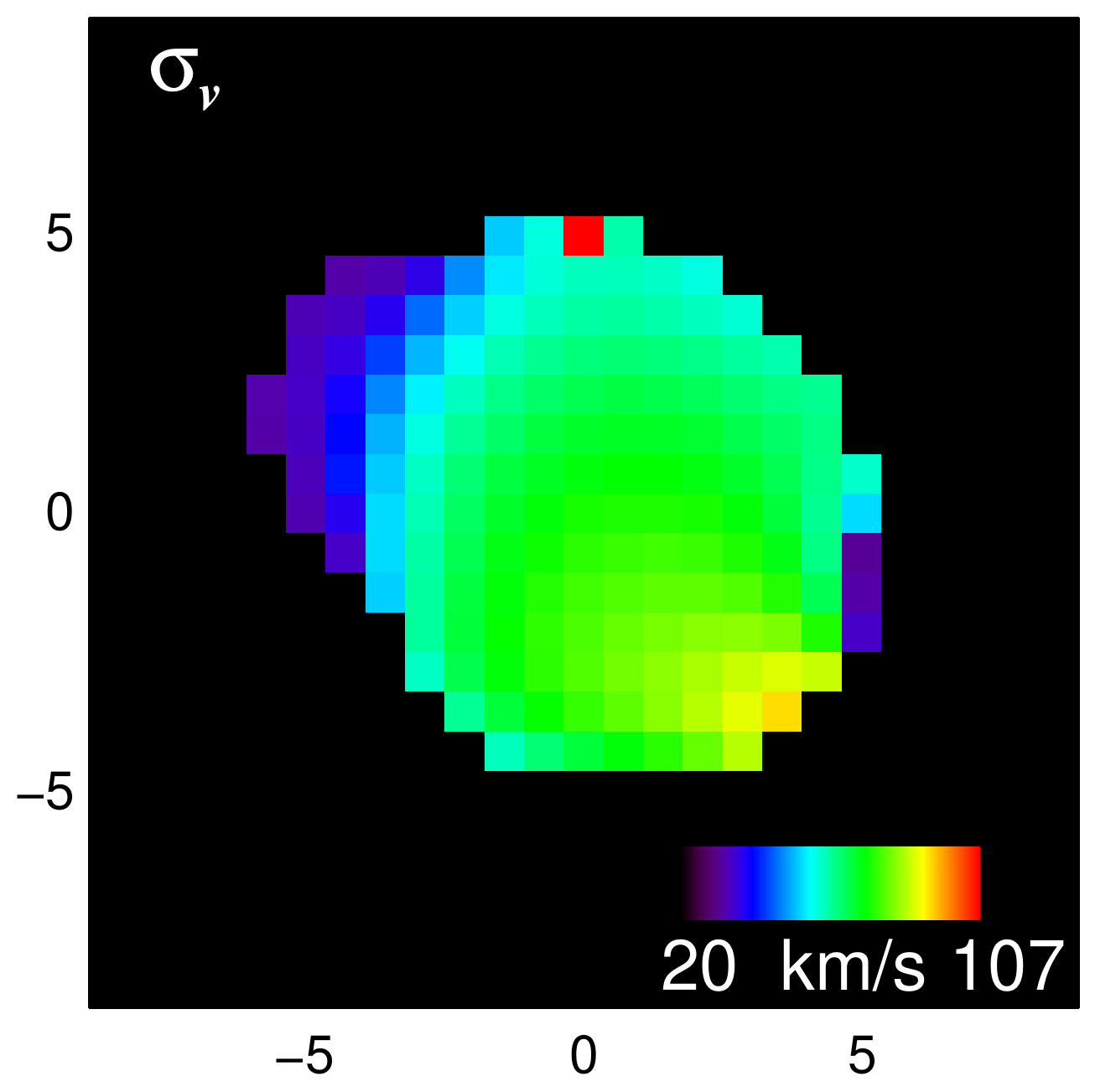}\\
\vspace{0.25mm}
\includegraphics[width=0.343\columnwidth]{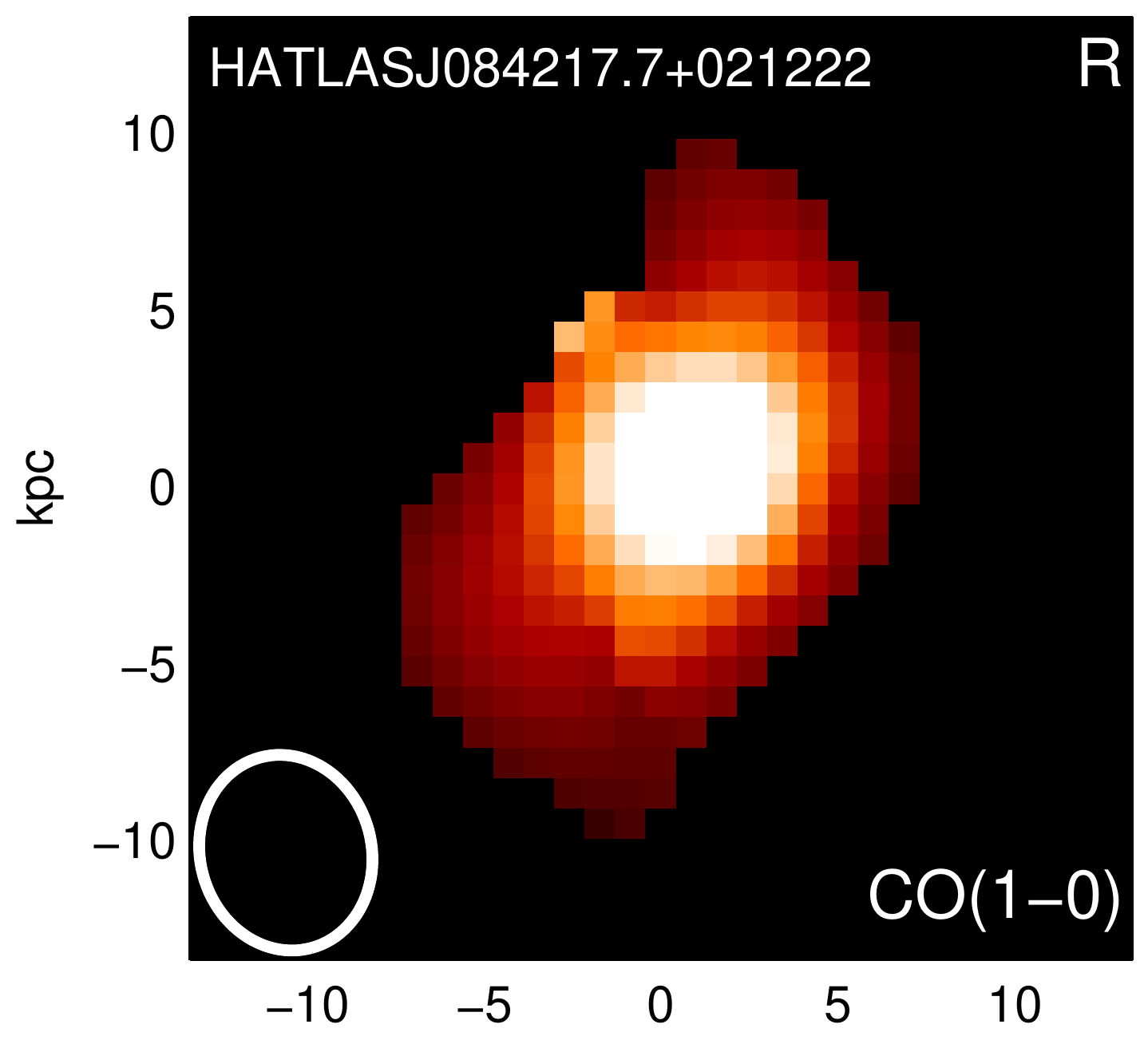}
\includegraphics[width=0.32\columnwidth]{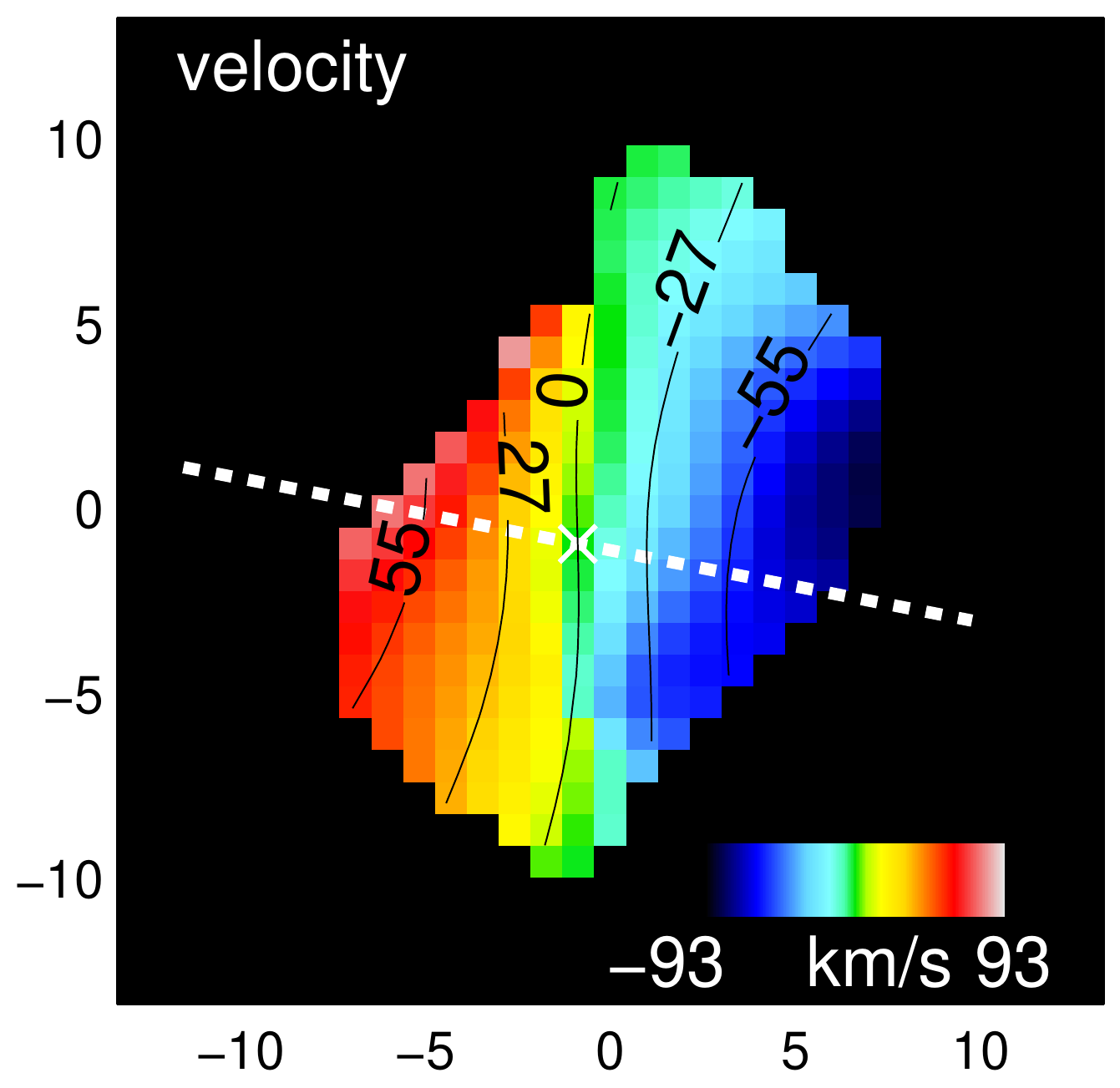}
\includegraphics[width=0.32\columnwidth]{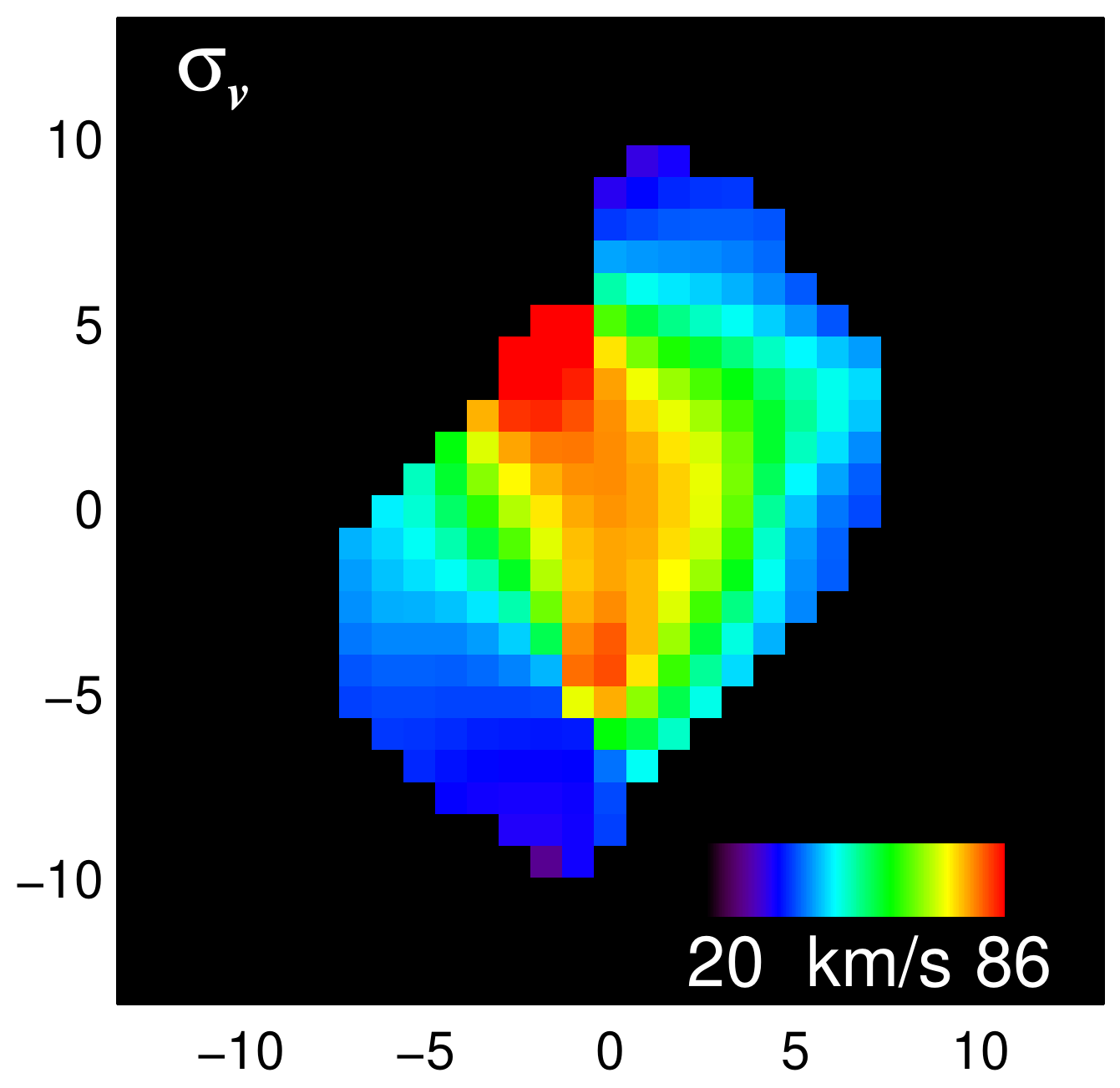}
\includegraphics[width=0.32\columnwidth]{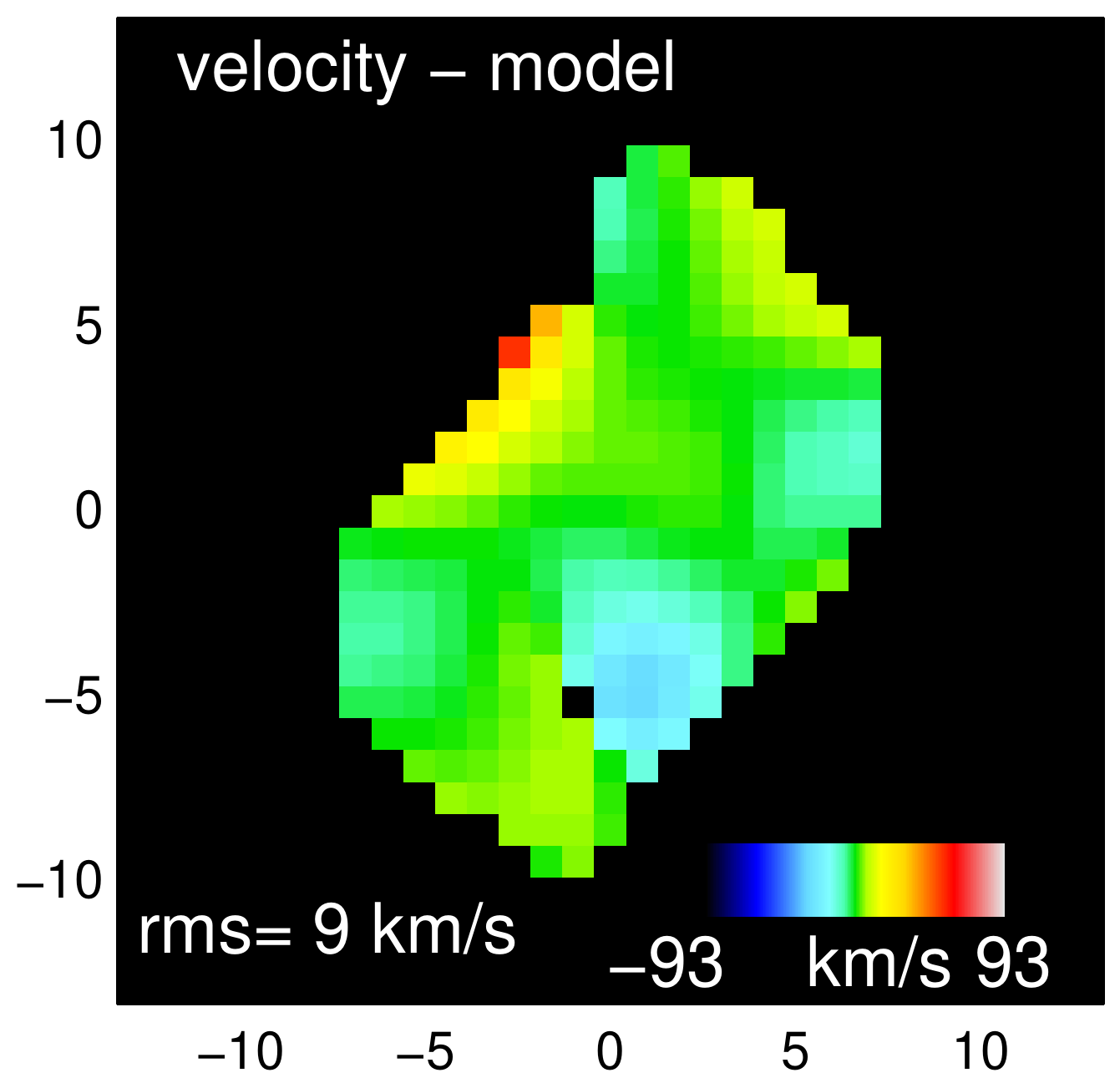}
\includegraphics[width=0.345\columnwidth]{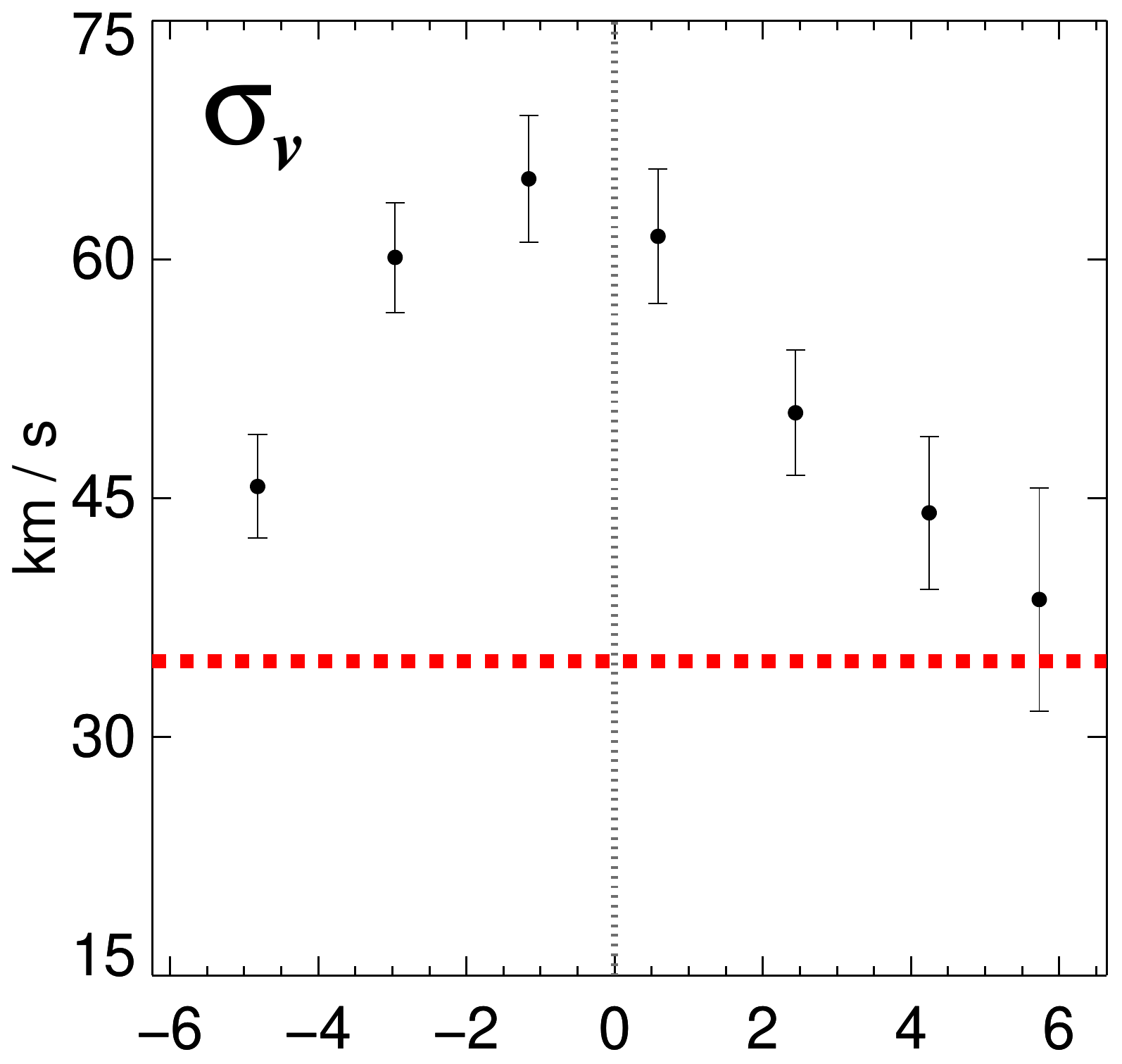}
\includegraphics[width=0.351\columnwidth]{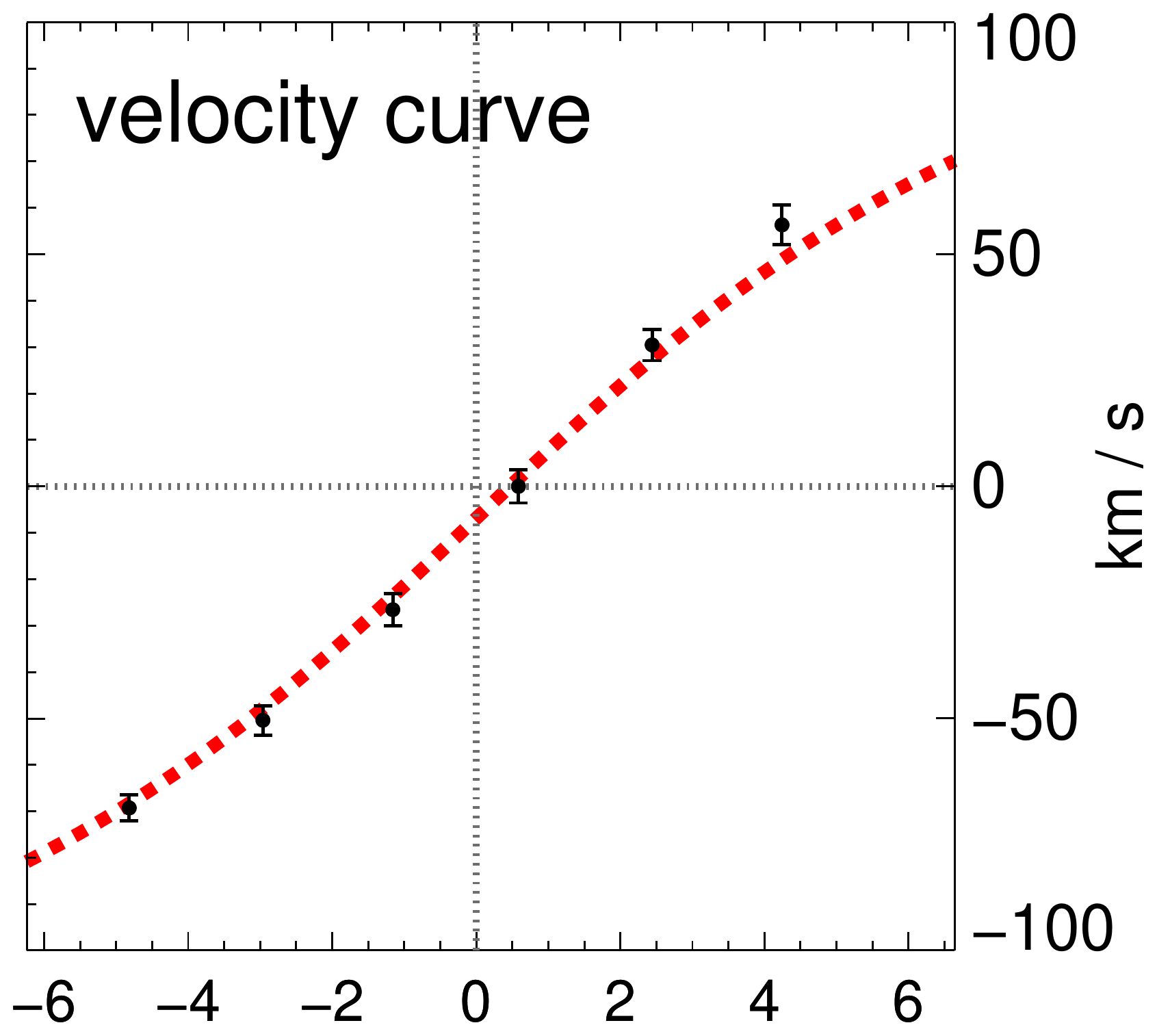}\\
\vspace{0.25mm}
\includegraphics[width=0.343\columnwidth]{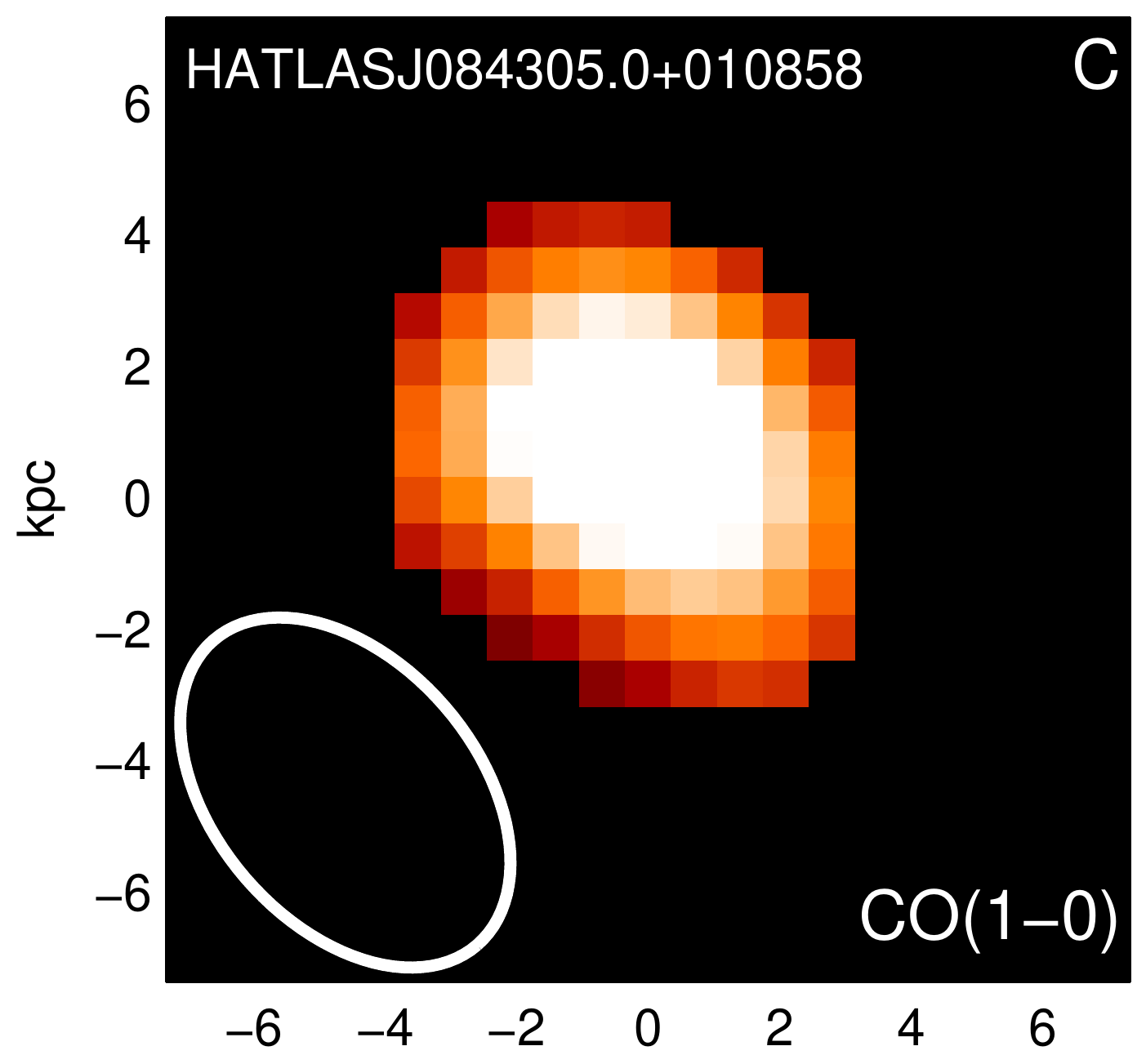}
\includegraphics[width=0.32\columnwidth]{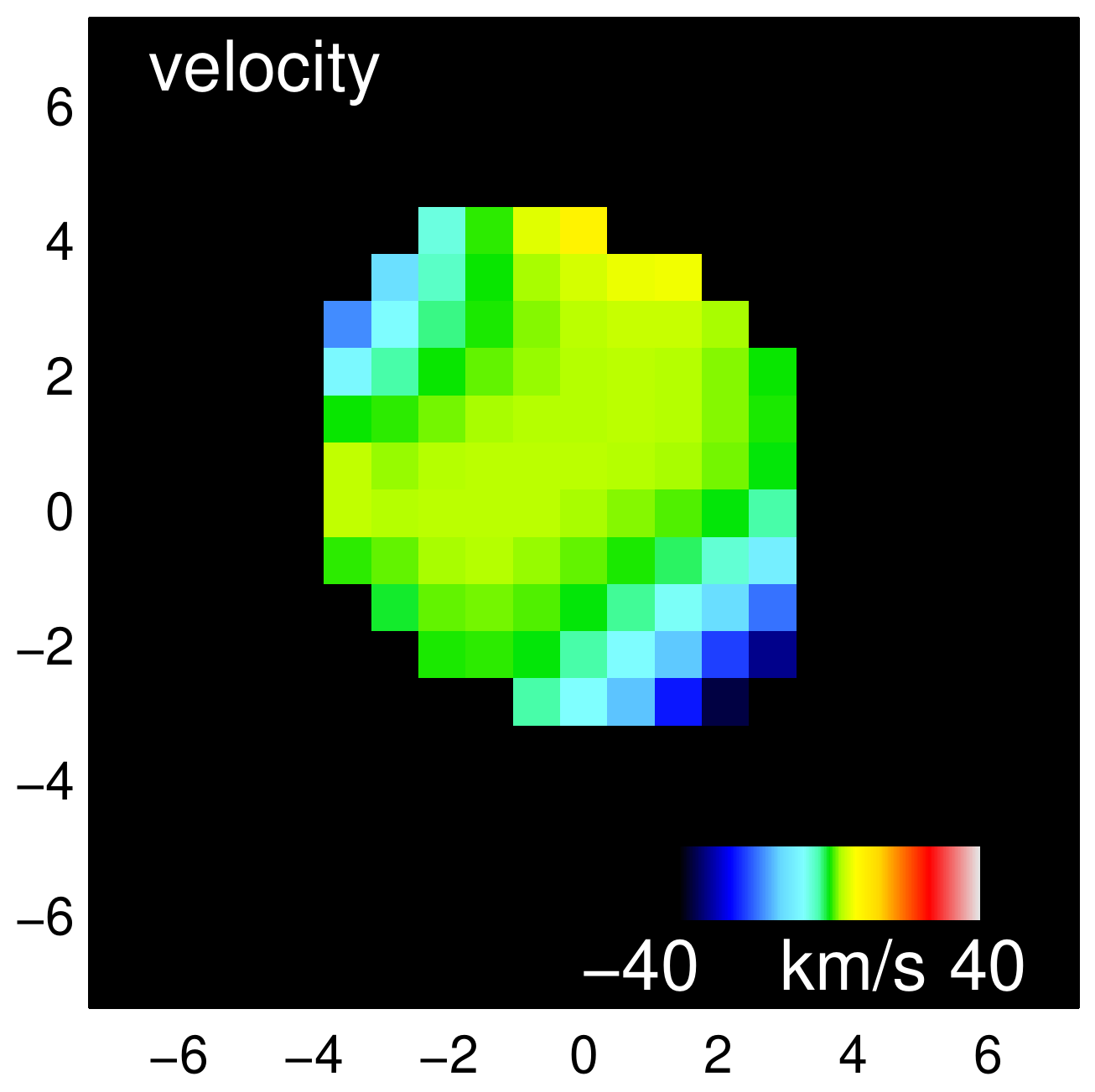}
\includegraphics[width=0.32\columnwidth]{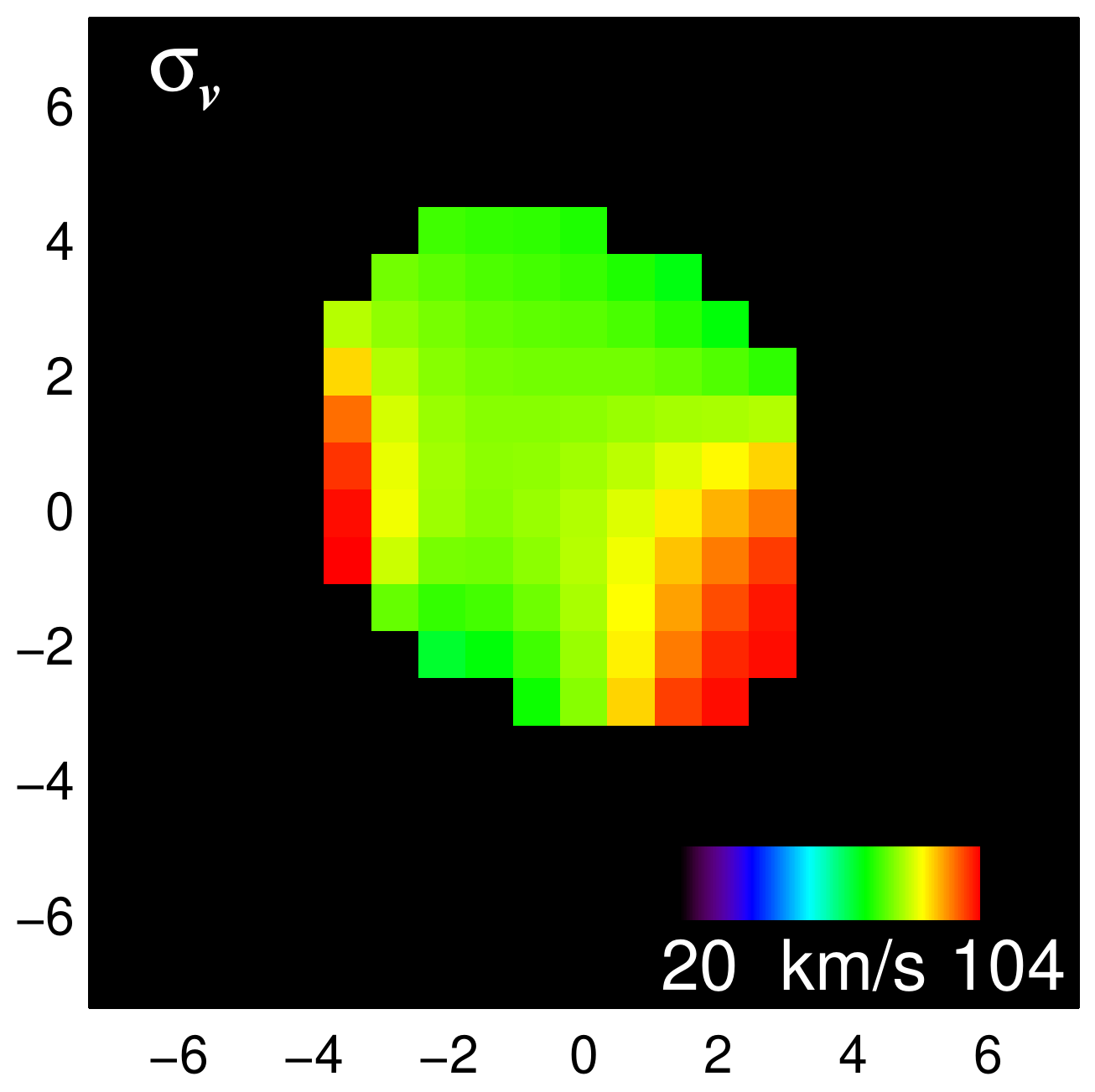}\\
\vspace{0.25mm}
\includegraphics[width=0.343\columnwidth]{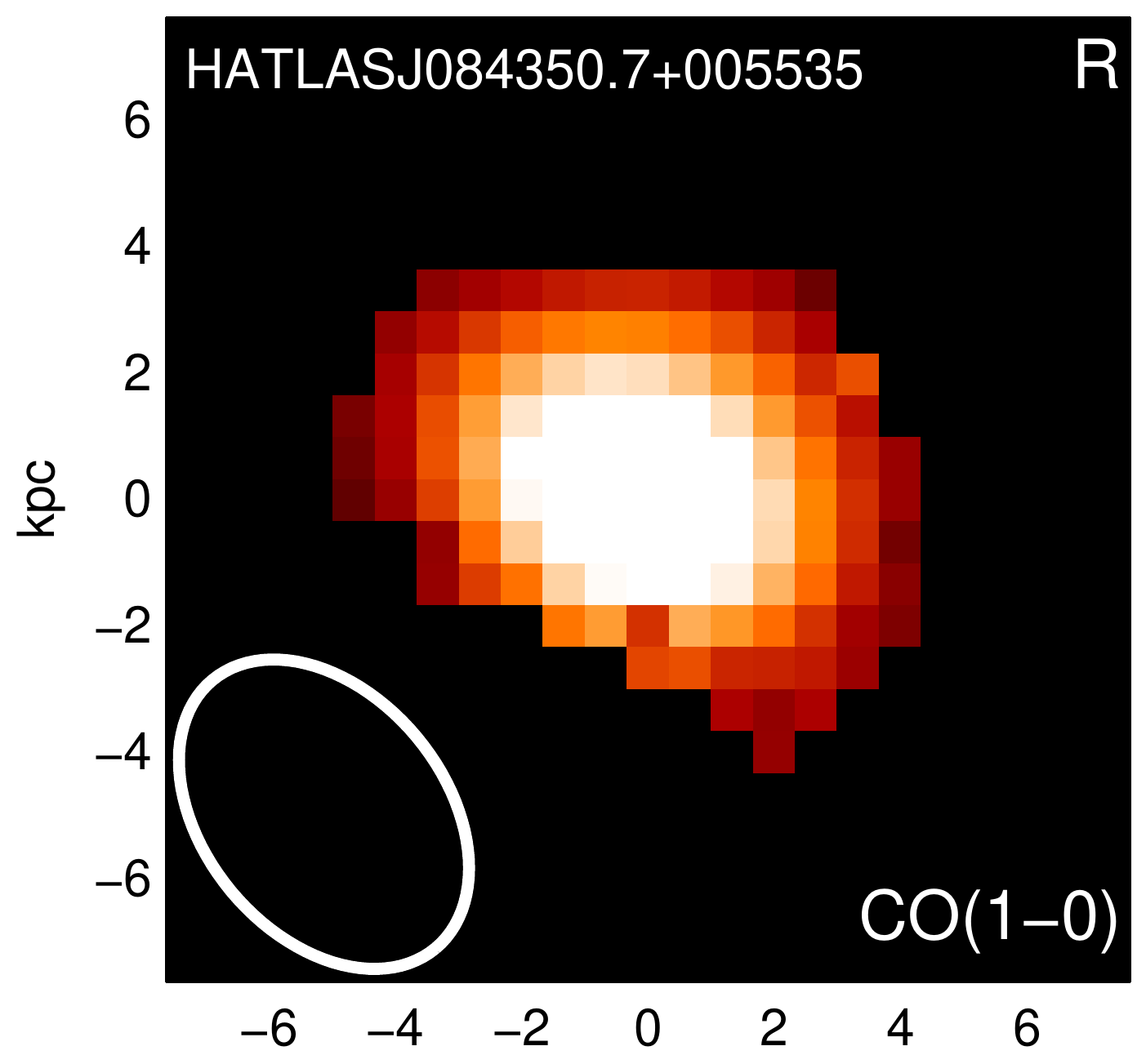}
\includegraphics[width=0.32\columnwidth]{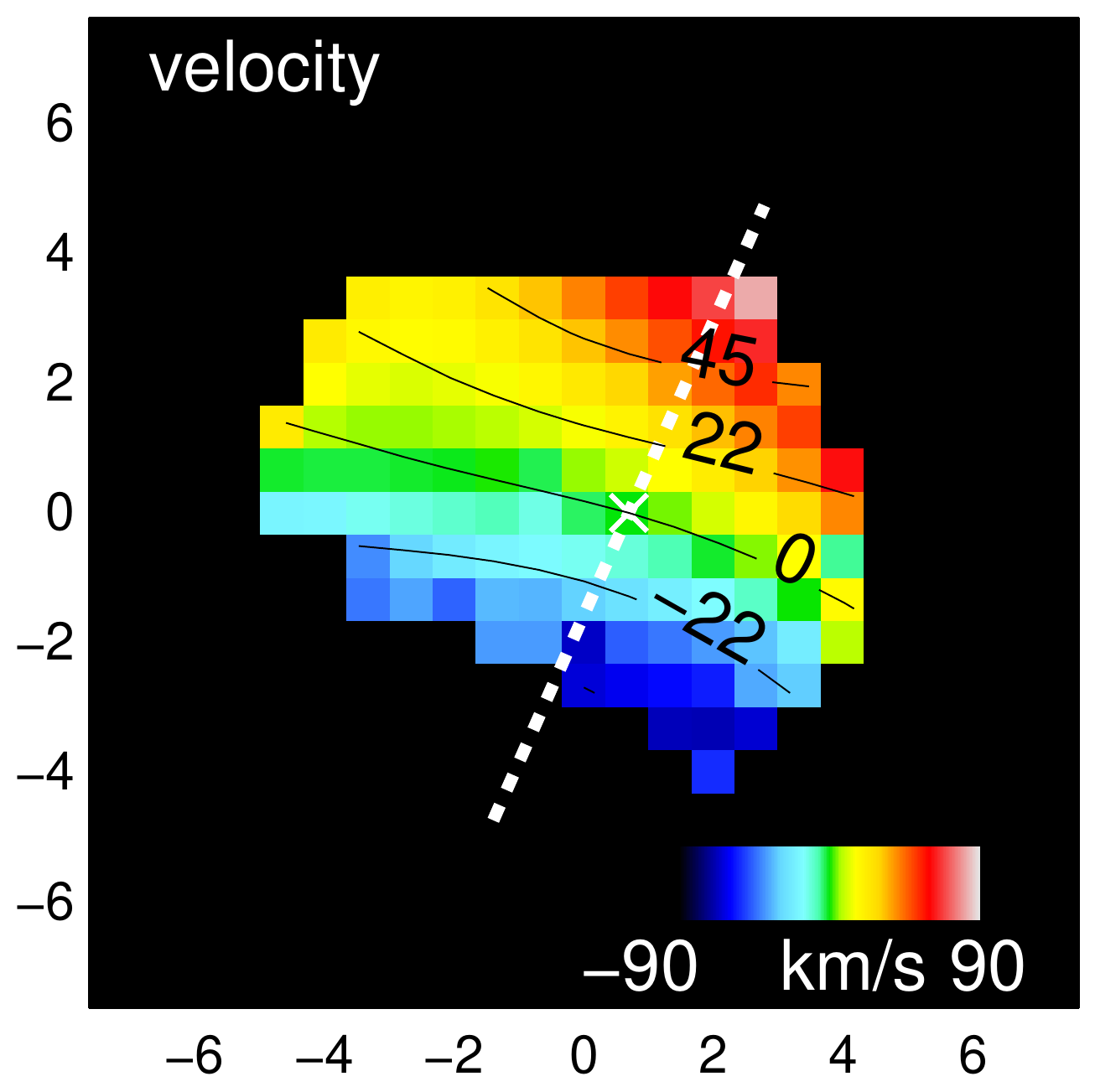}
\includegraphics[width=0.32\columnwidth]{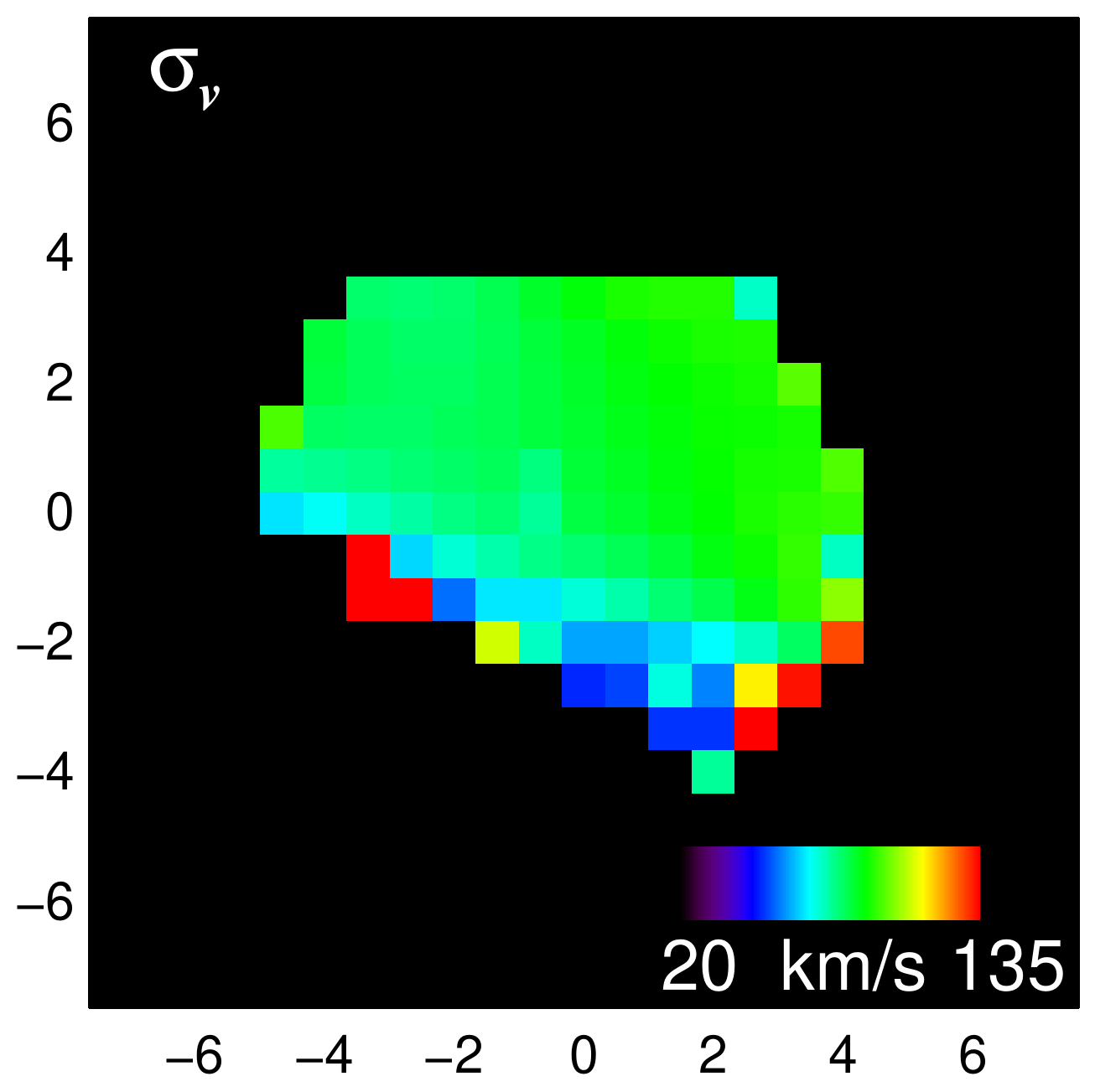}
\includegraphics[width=0.32\columnwidth]{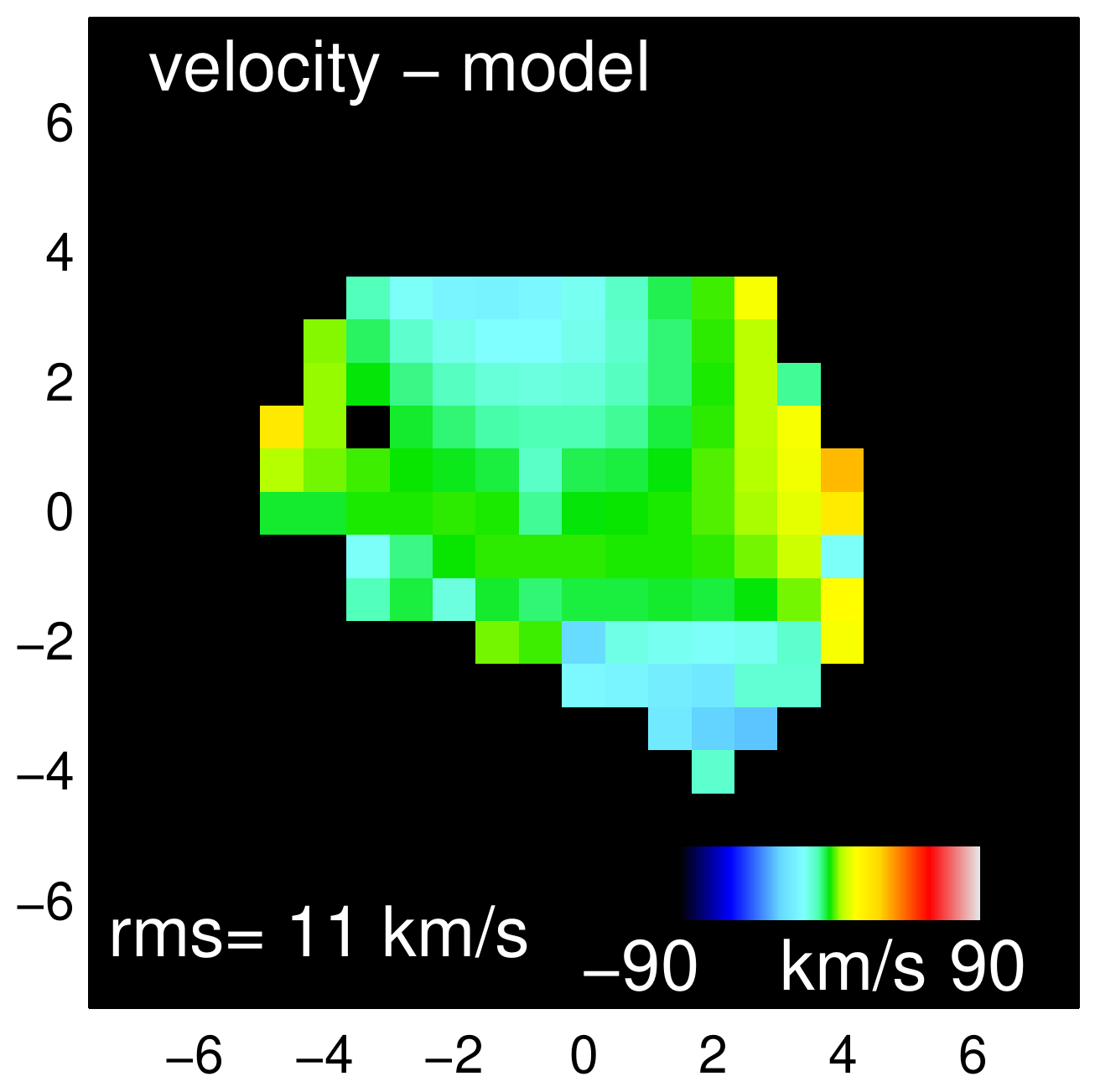}
\includegraphics[width=0.345\columnwidth]{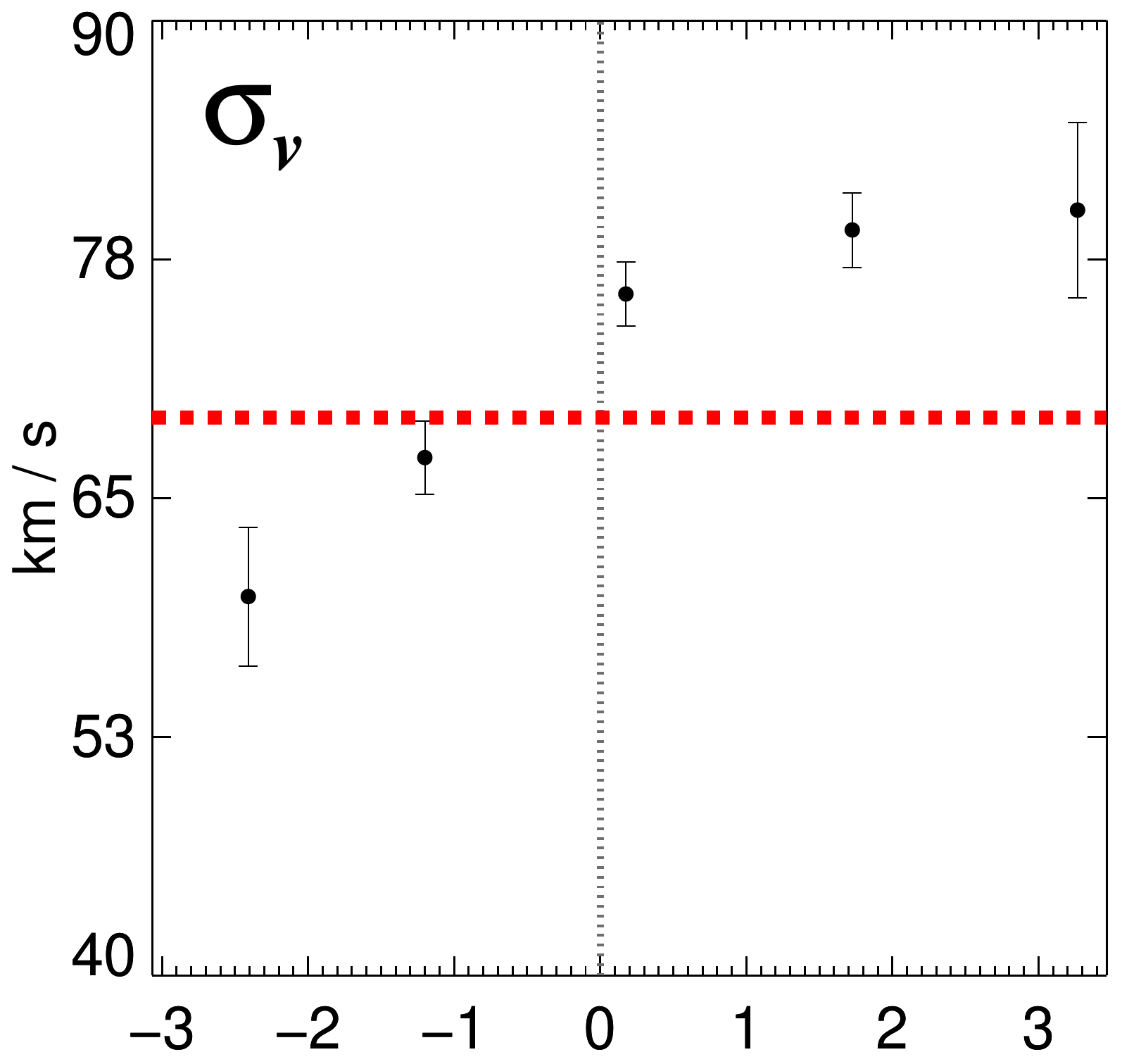}
\includegraphics[width=0.361\columnwidth]{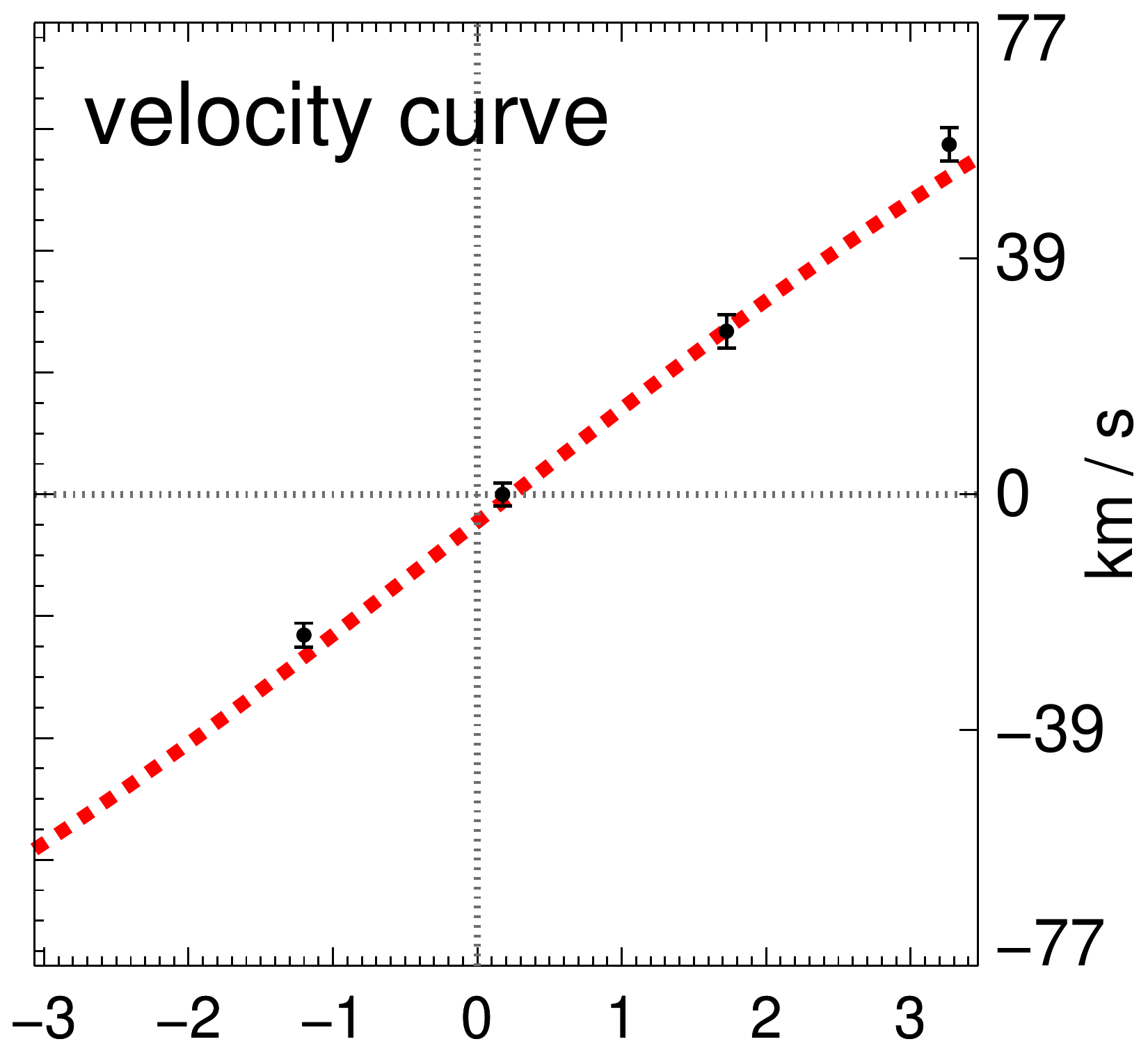}\\
\vspace{0.25mm}
\includegraphics[width=0.343\columnwidth]{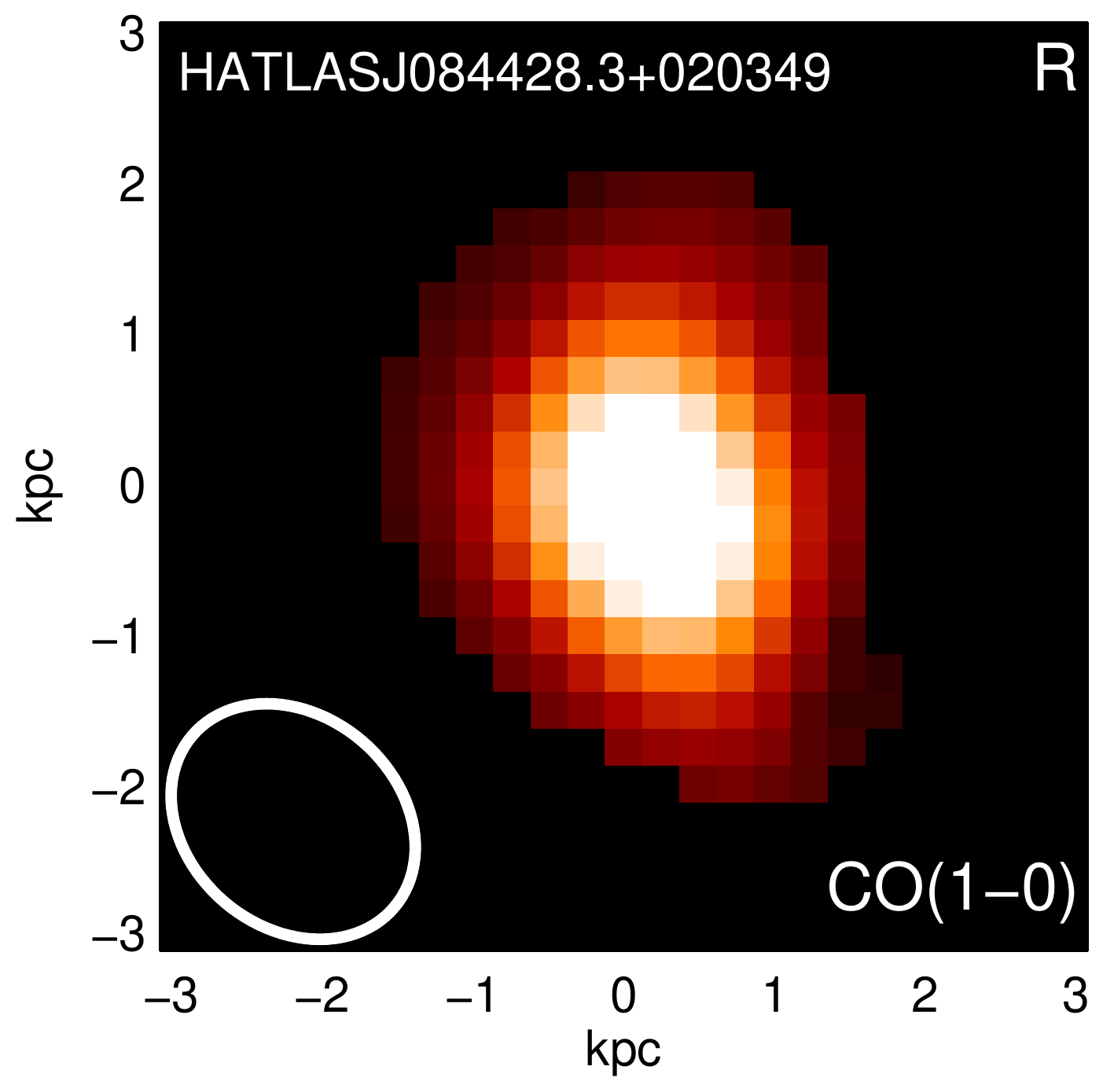}
\includegraphics[width=0.32\columnwidth]{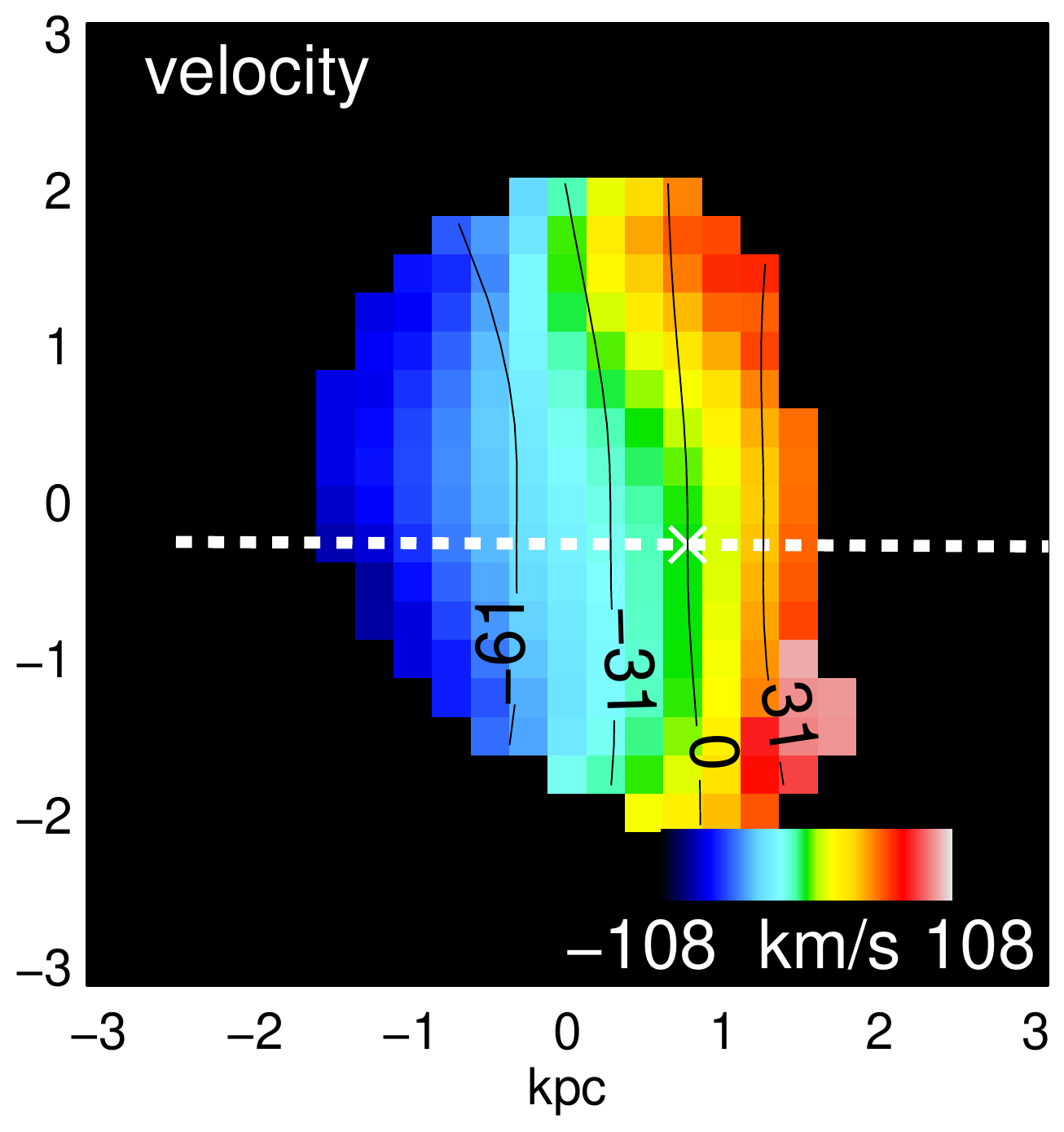}
\includegraphics[width=0.32\columnwidth]{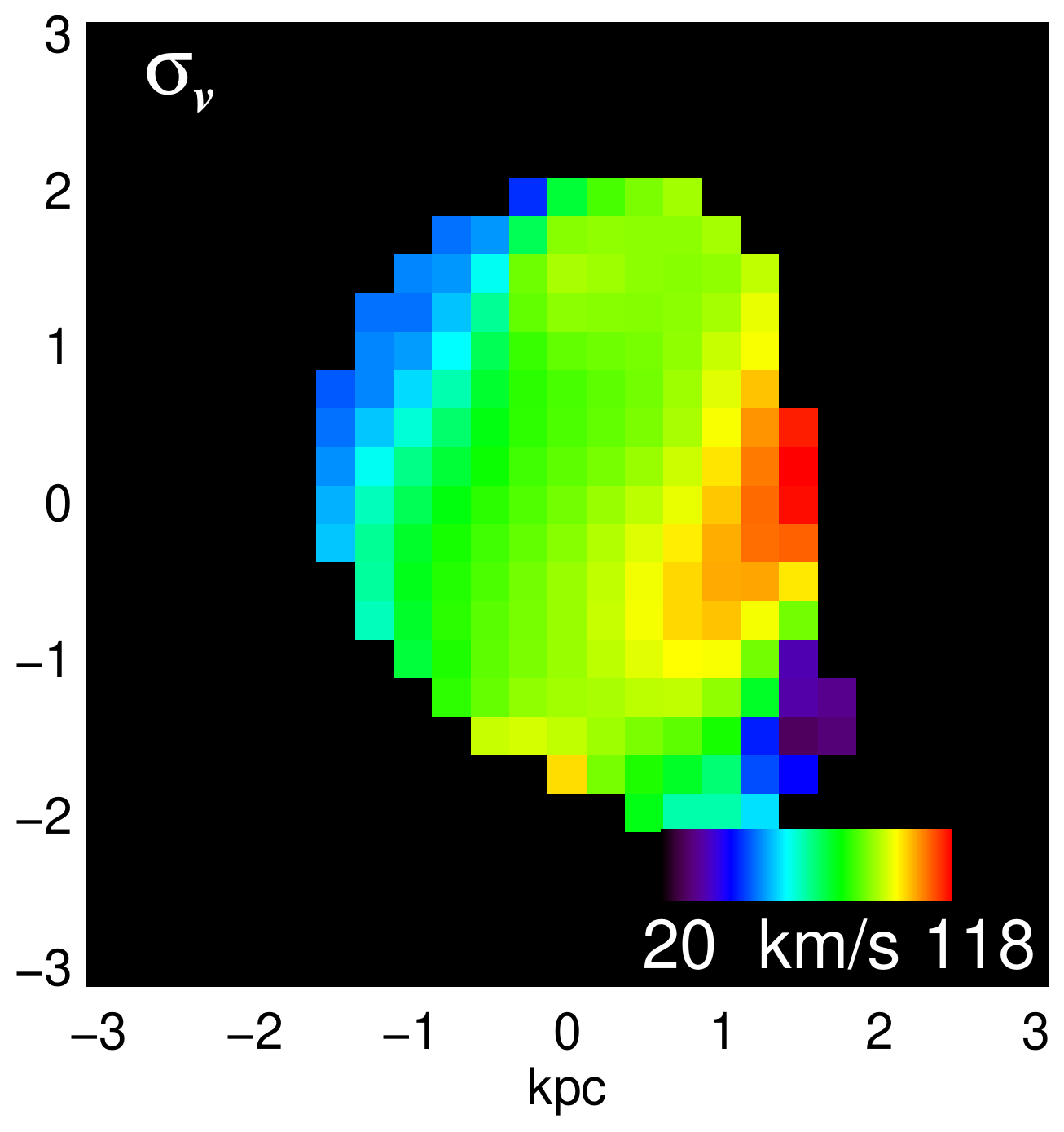}
\includegraphics[width=0.32\columnwidth]{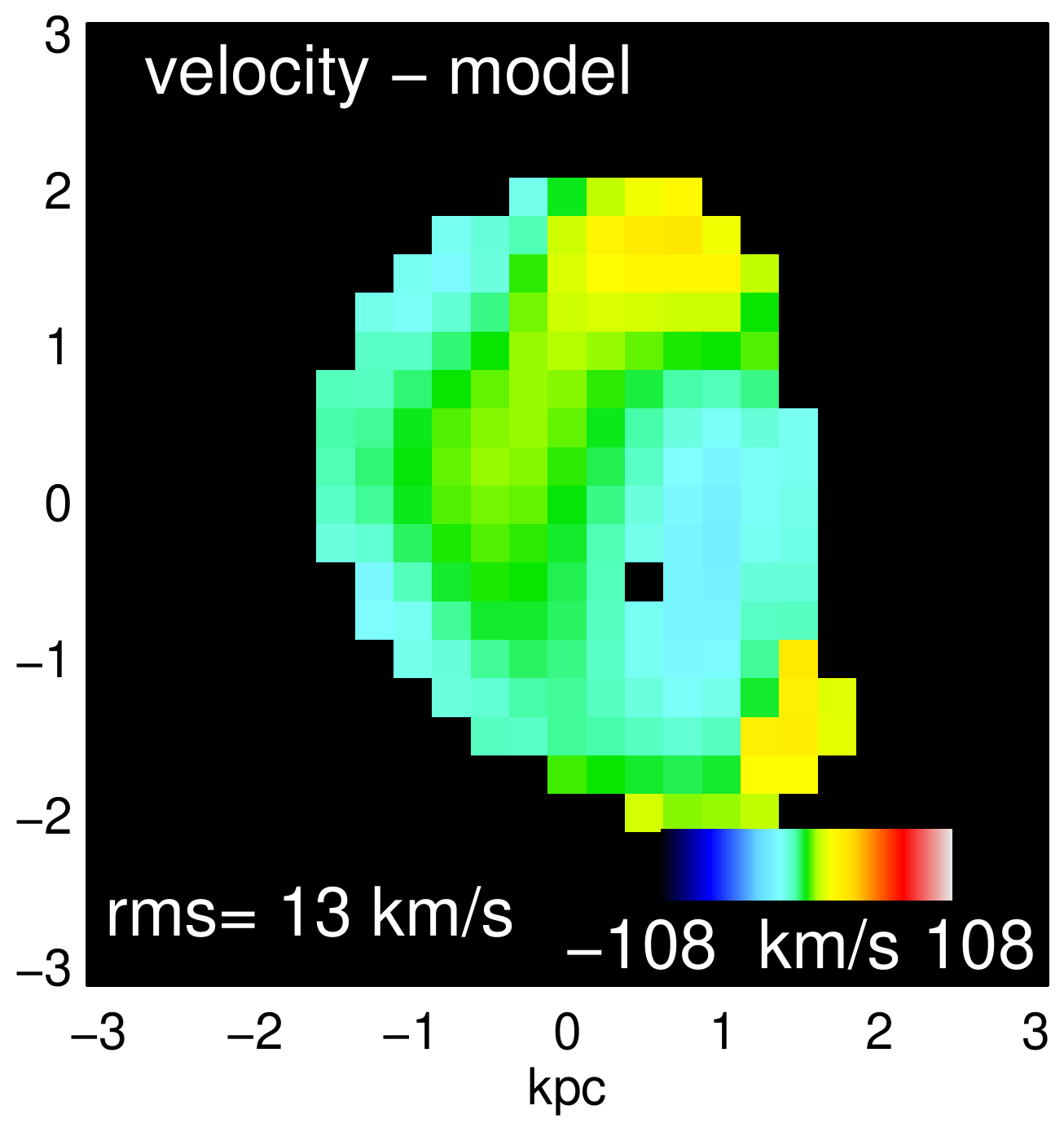}
\includegraphics[width=0.345\columnwidth]{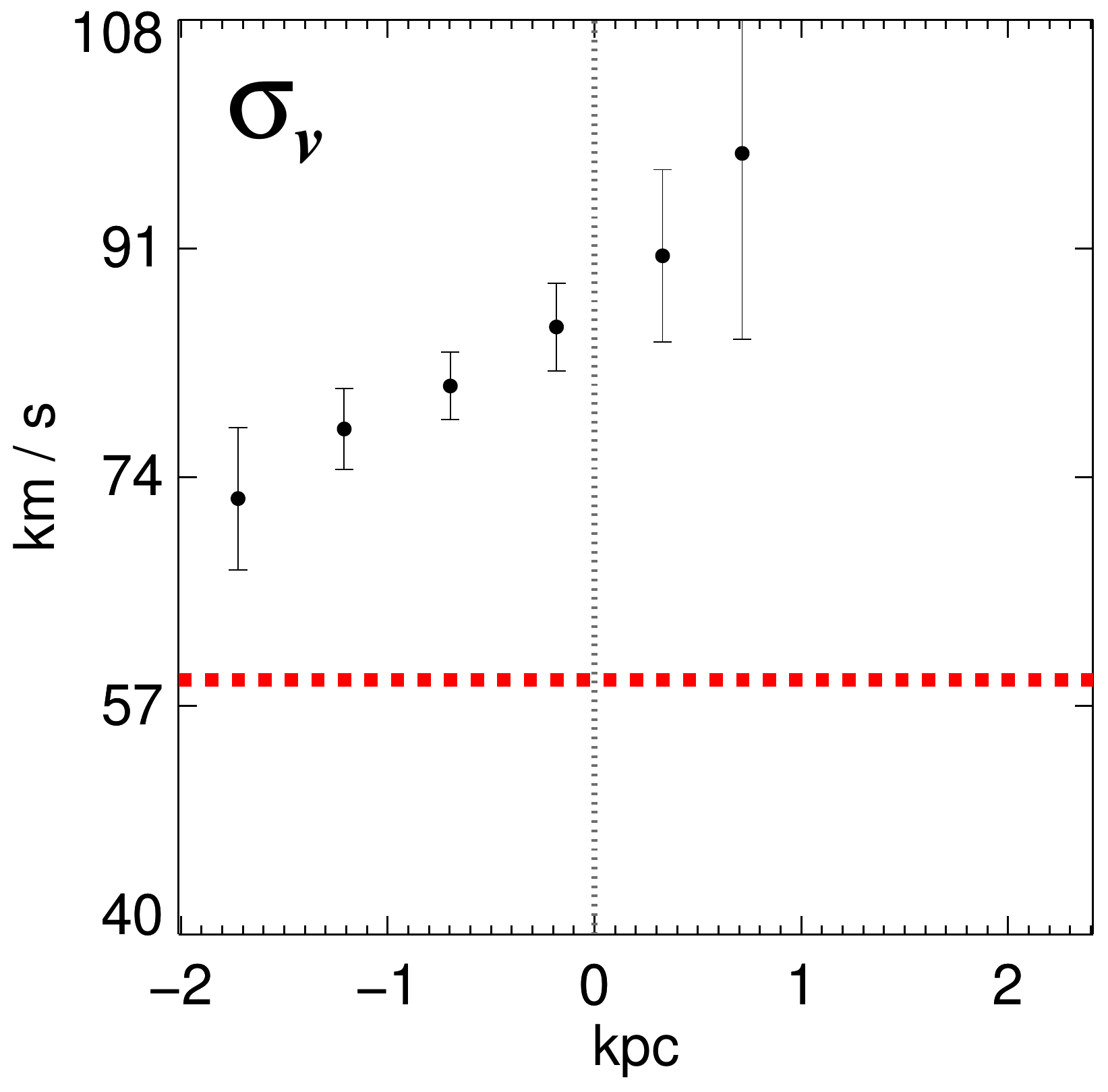}
\includegraphics[width=0.351\columnwidth]{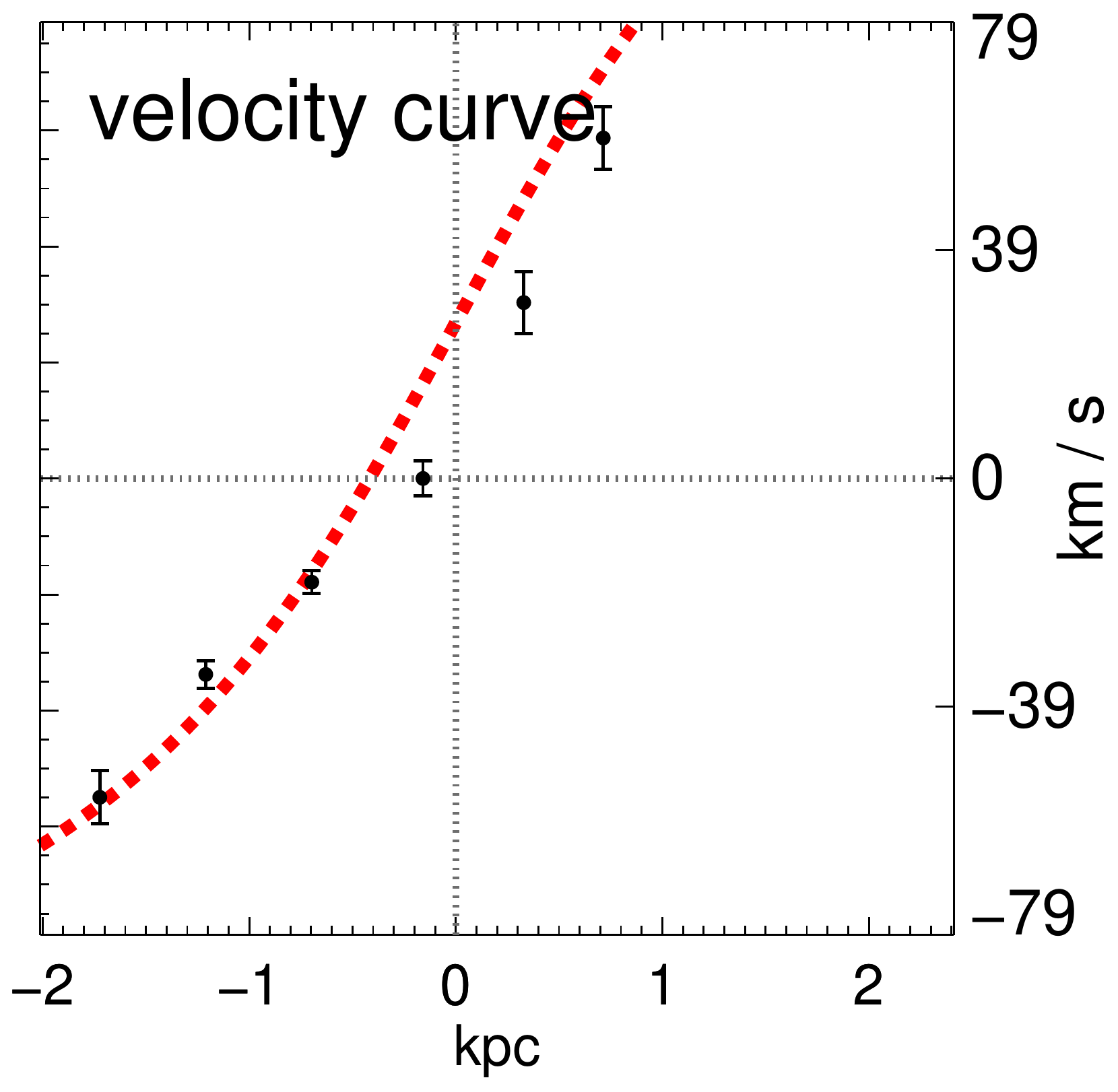}\\
\caption{\label{fig:maps}
CO(1-0) intensity, velocity, LOS velocity dispersion ($\sigma_v$), residual field, major axis 
velocity dispersion and velocity profiles (columns) for each target from our sample (rows).  
Each target is labelled as `resolved' (R) or `compact' (C) in their intensity map, `compact' galaxies
were not modelled (see \S\ref{sec:galaxy_dyn} for more details). The map also shows the 
synthesized beam. The velocity field has overplotted
 the kinematic centre, the major kinematic axis and velocity contours from their best-fit disk model.
The LOS velocity dispersion ($\sigma_v$) field is corrected for the local velocity gradient 
($\Delta$V/$\Delta$R) across the synthesized beam. The residual map is constructed by subtracting
the velocity disk model from the velocity map: the r.m.s. of these residuals are given in each panel. 
The one-dimensional profiles are derived from the two dimensional velocity fields using the best-fit kinematic
parameters and a slit width with size equal to half of the beam FWHM across the major kinematic axis. In each one-dimensional profile, 
the error bars show the 1\,$\sigma$ uncertainty and the vertical dashed grey line represents the best-fit dynamical centre.
In the velocity dispersion profile panels, the red-dashed line shows the mean galactic value (Table~\ref{tab:table1}), whilst
in the last column, the red-dashed curve shows also the best-fit for each source.
}
\end{figure*}

\begin{figure*}
\flushleft
\includegraphics[width=0.343\columnwidth]{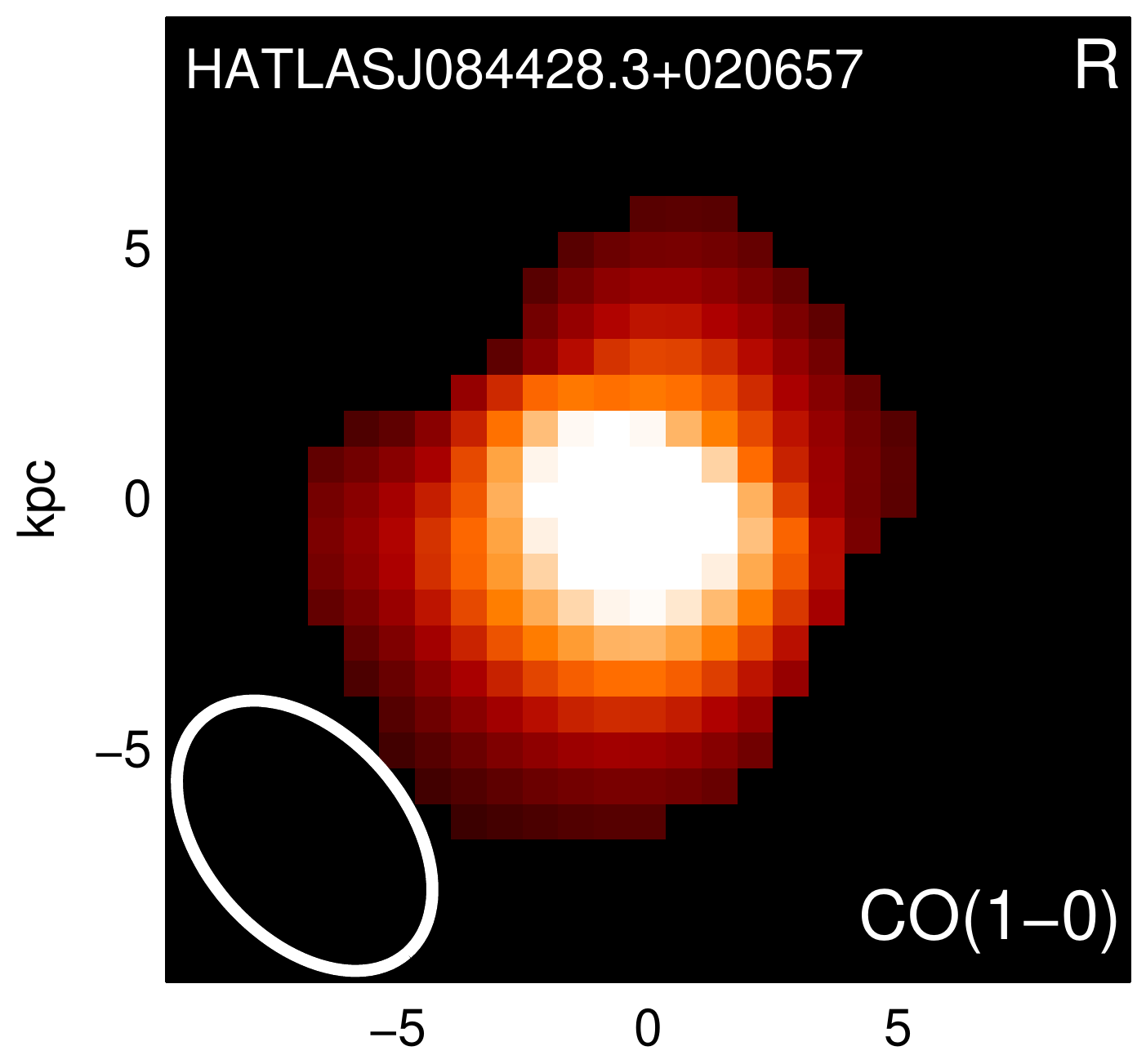}
\includegraphics[width=0.32\columnwidth]{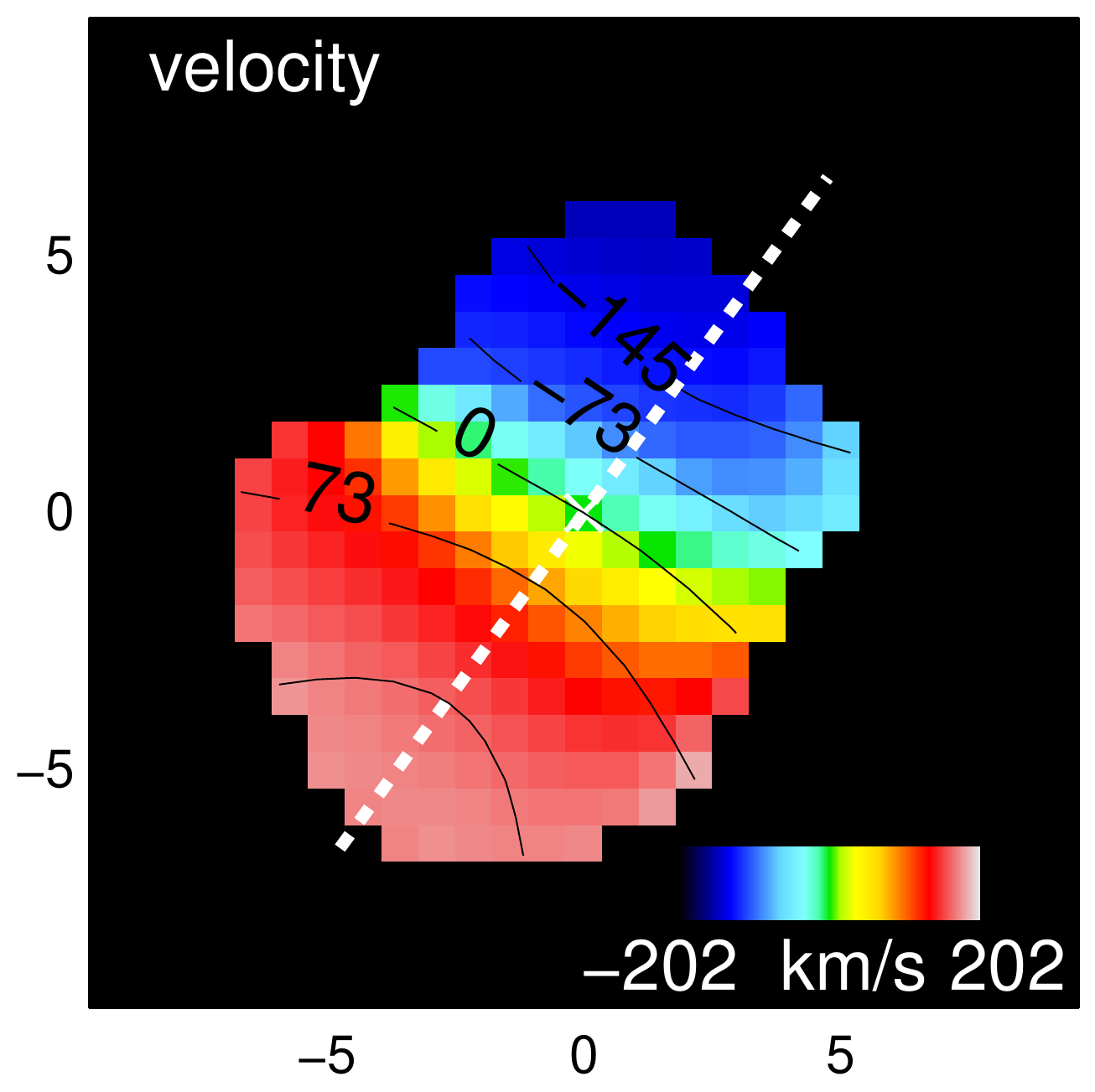}
\includegraphics[width=0.32\columnwidth]{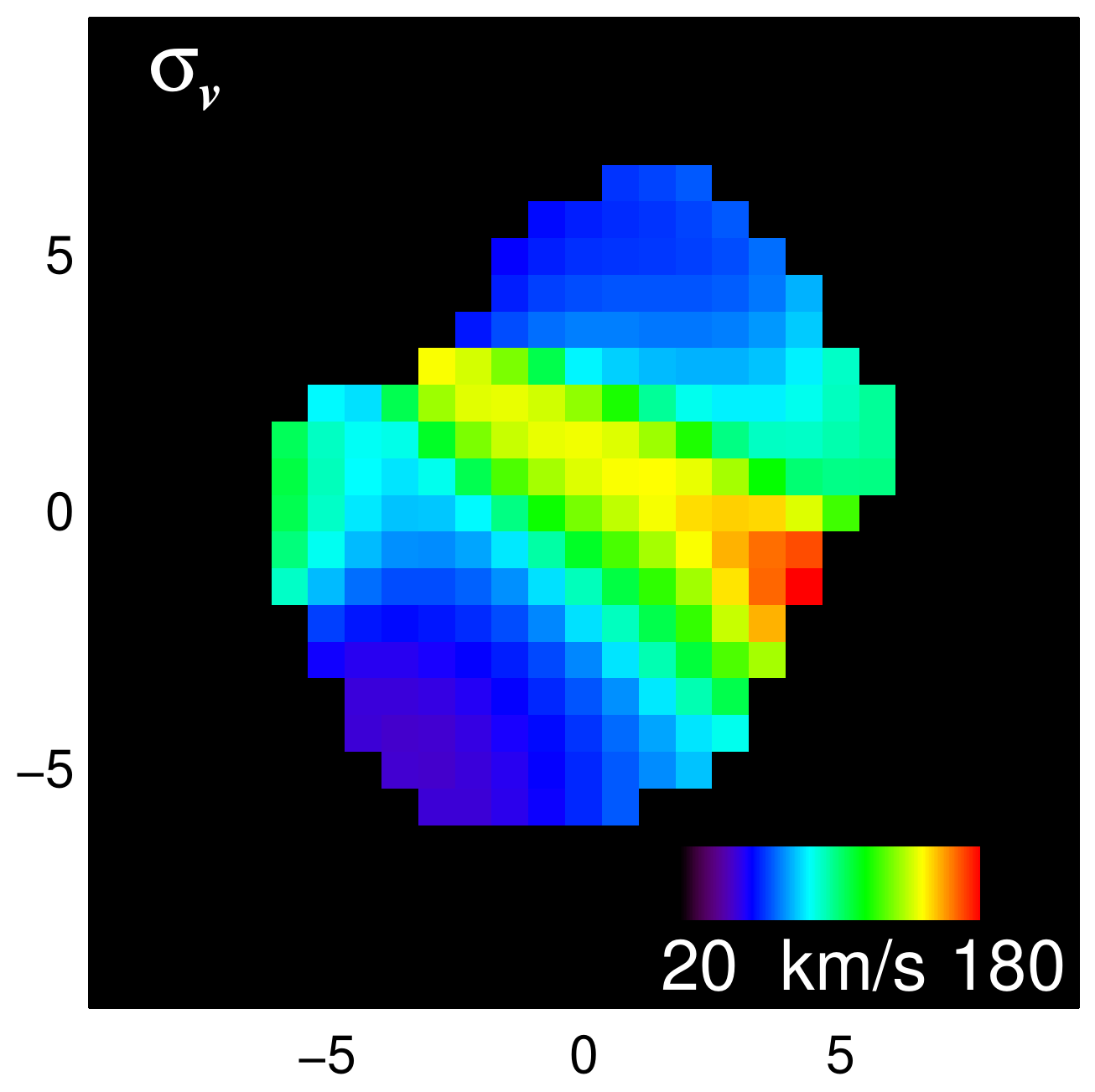}
\includegraphics[width=0.32\columnwidth]{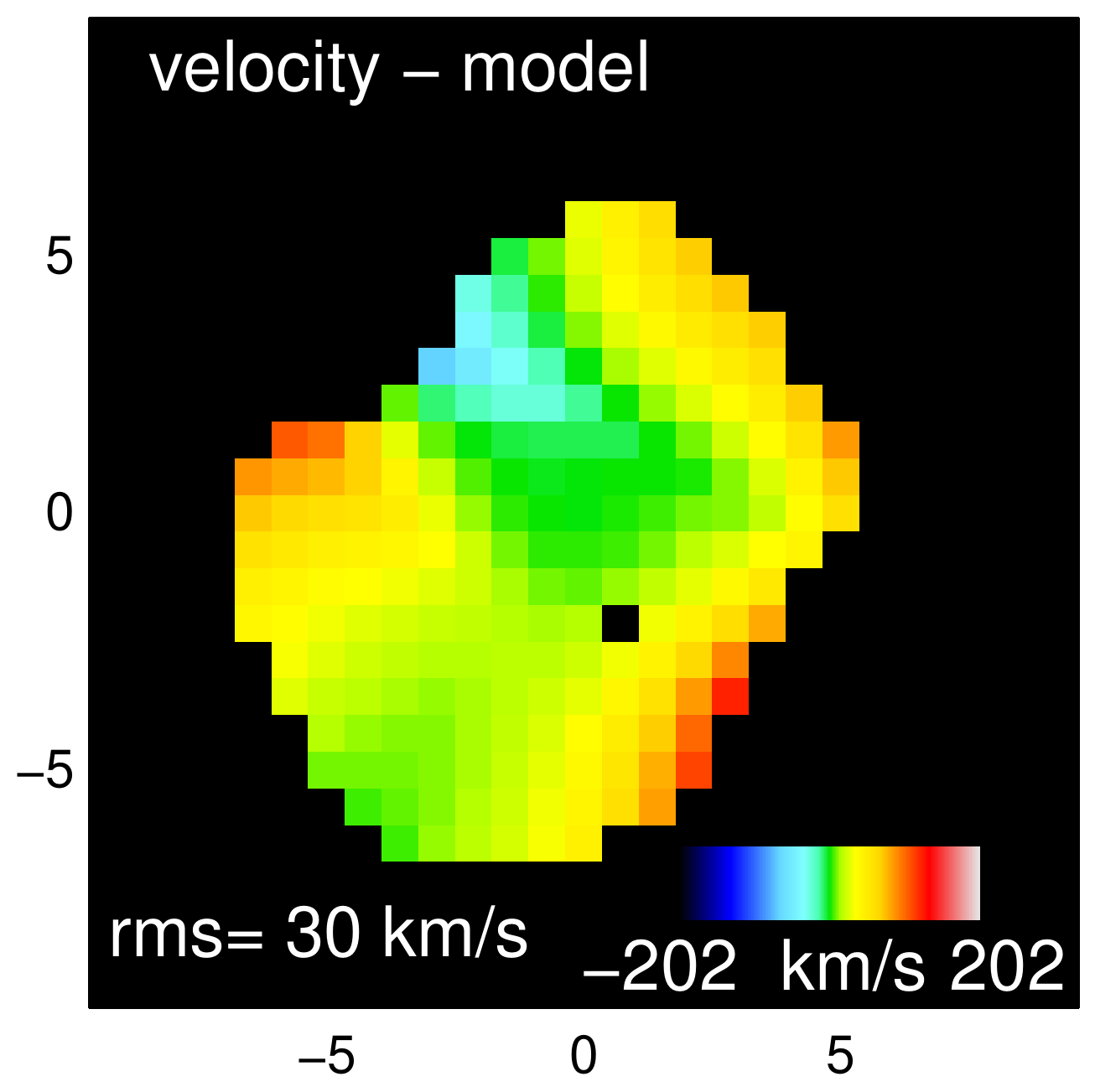}
\includegraphics[width=0.345\columnwidth]{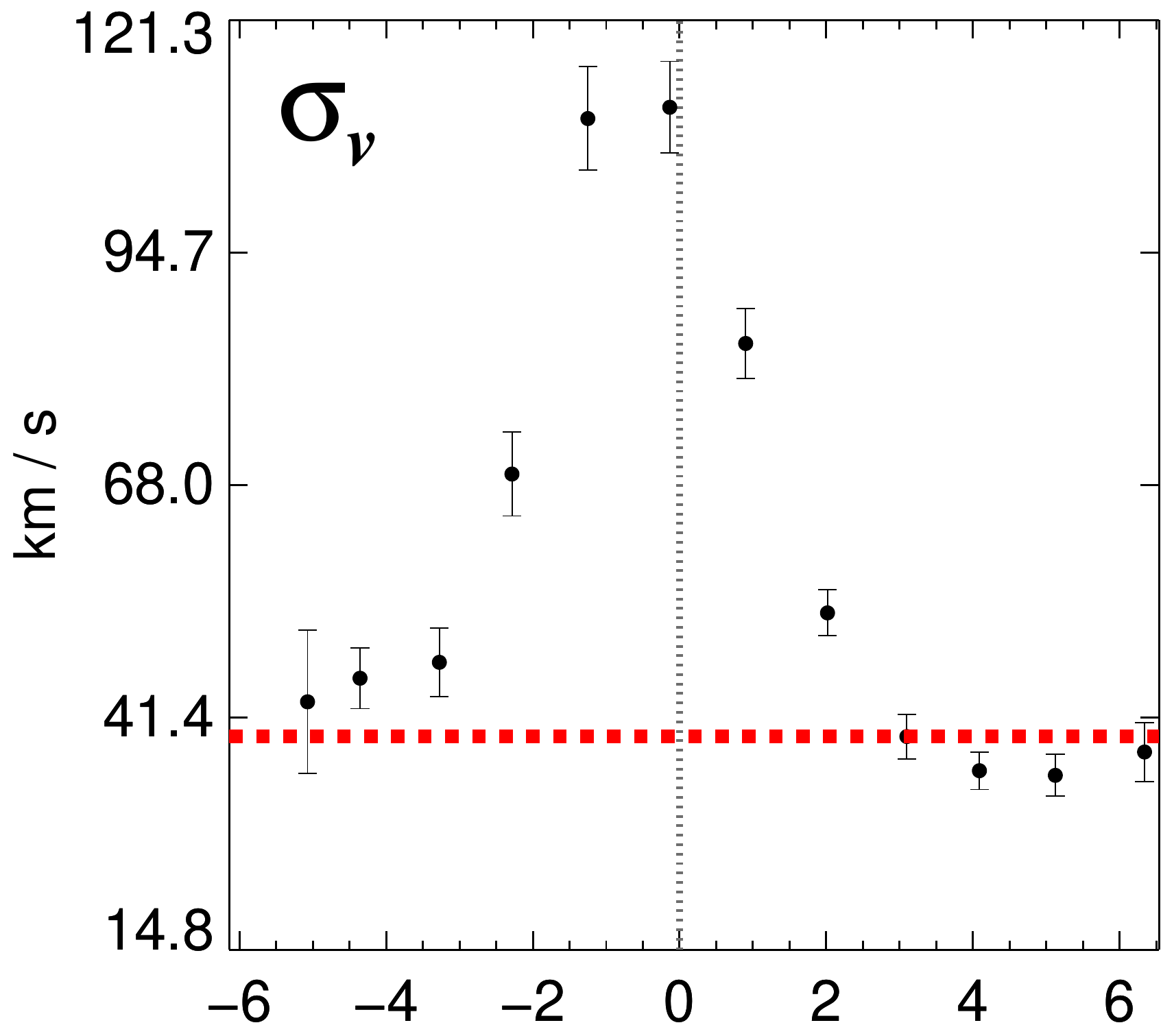}
\includegraphics[width=0.351\columnwidth]{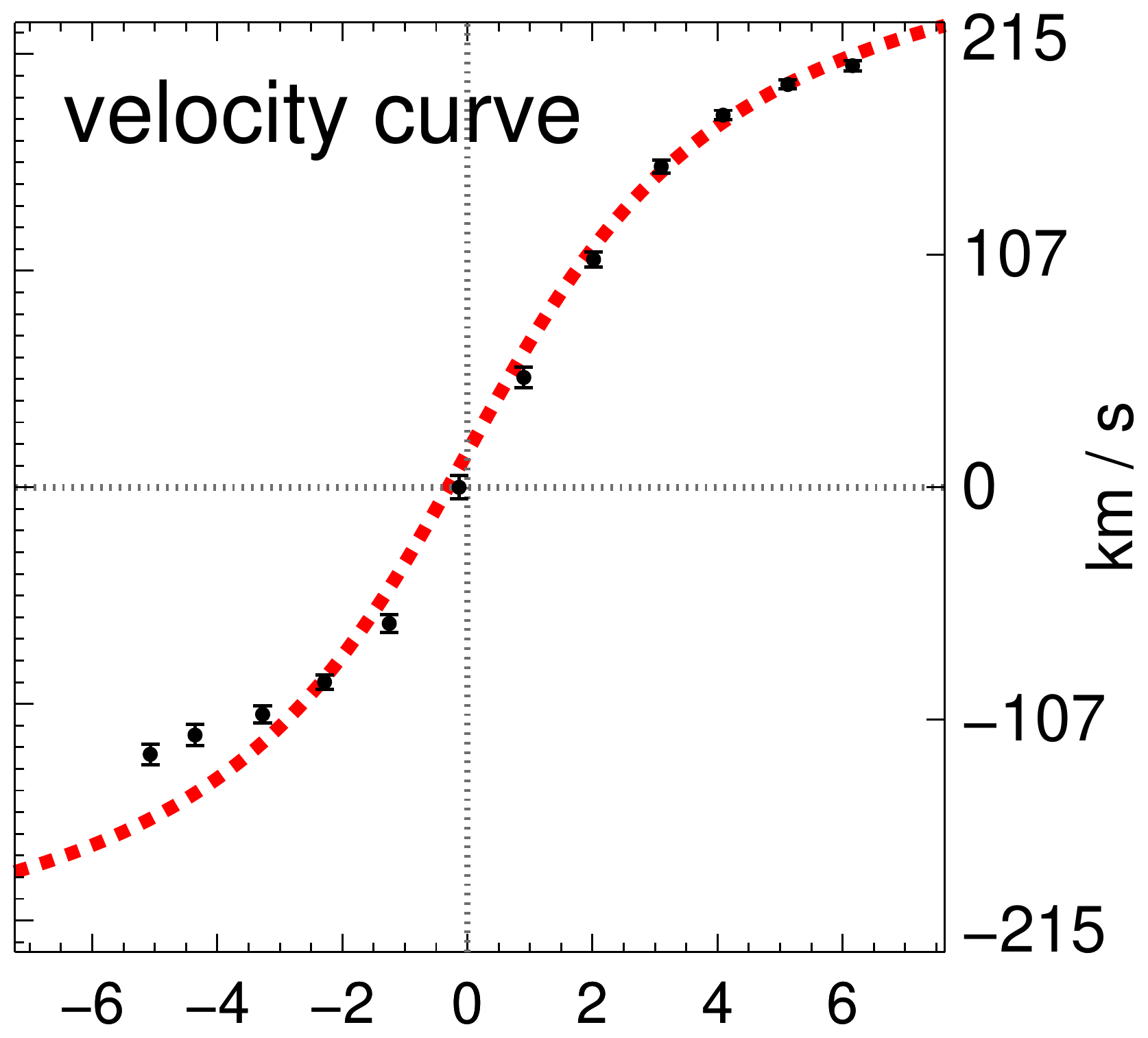}\\
\vspace{1mm}
\includegraphics[width=0.343\columnwidth]{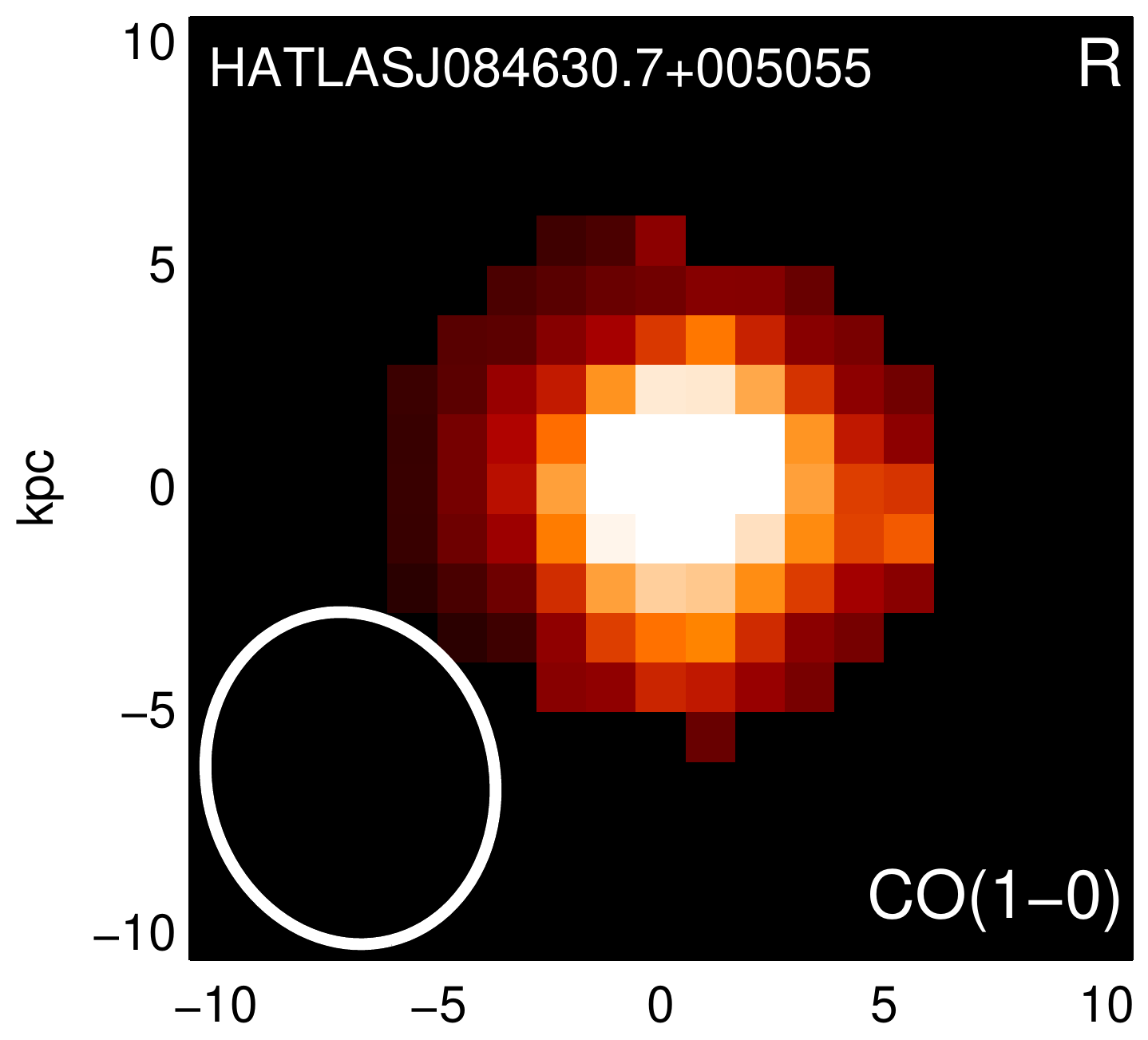}
\includegraphics[width=0.32\columnwidth]{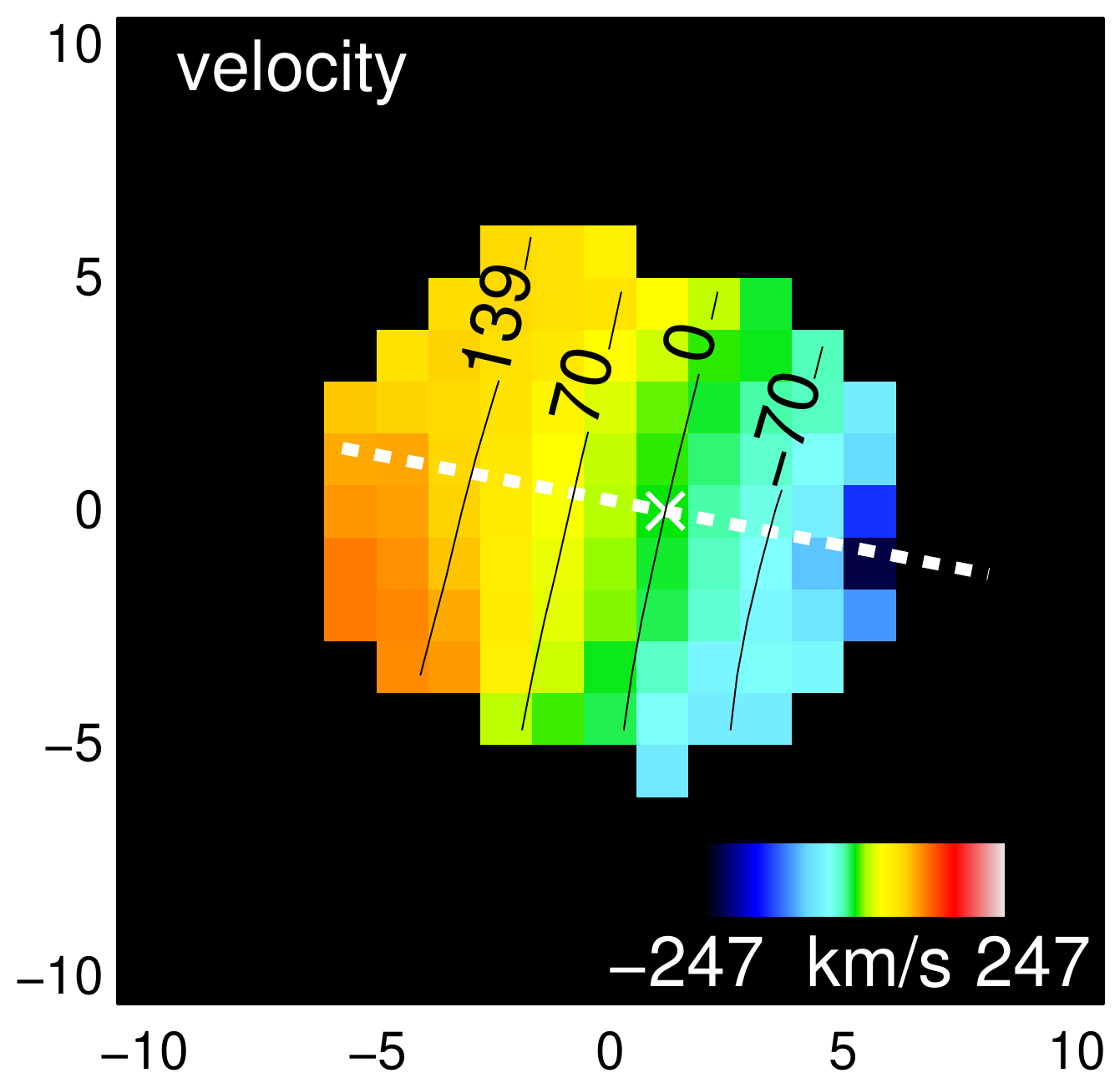}
\includegraphics[width=0.32\columnwidth]{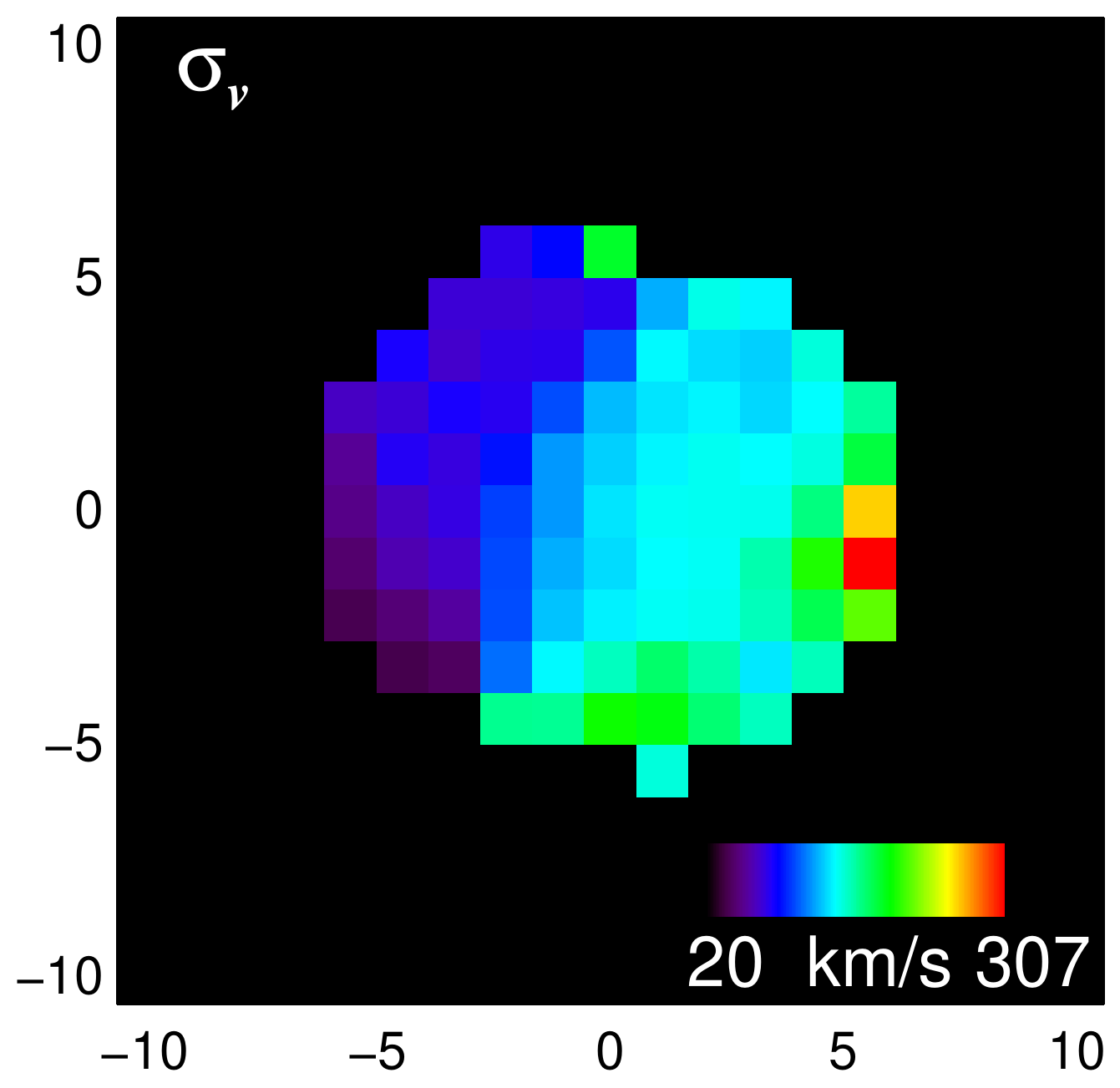}
\includegraphics[width=0.32\columnwidth]{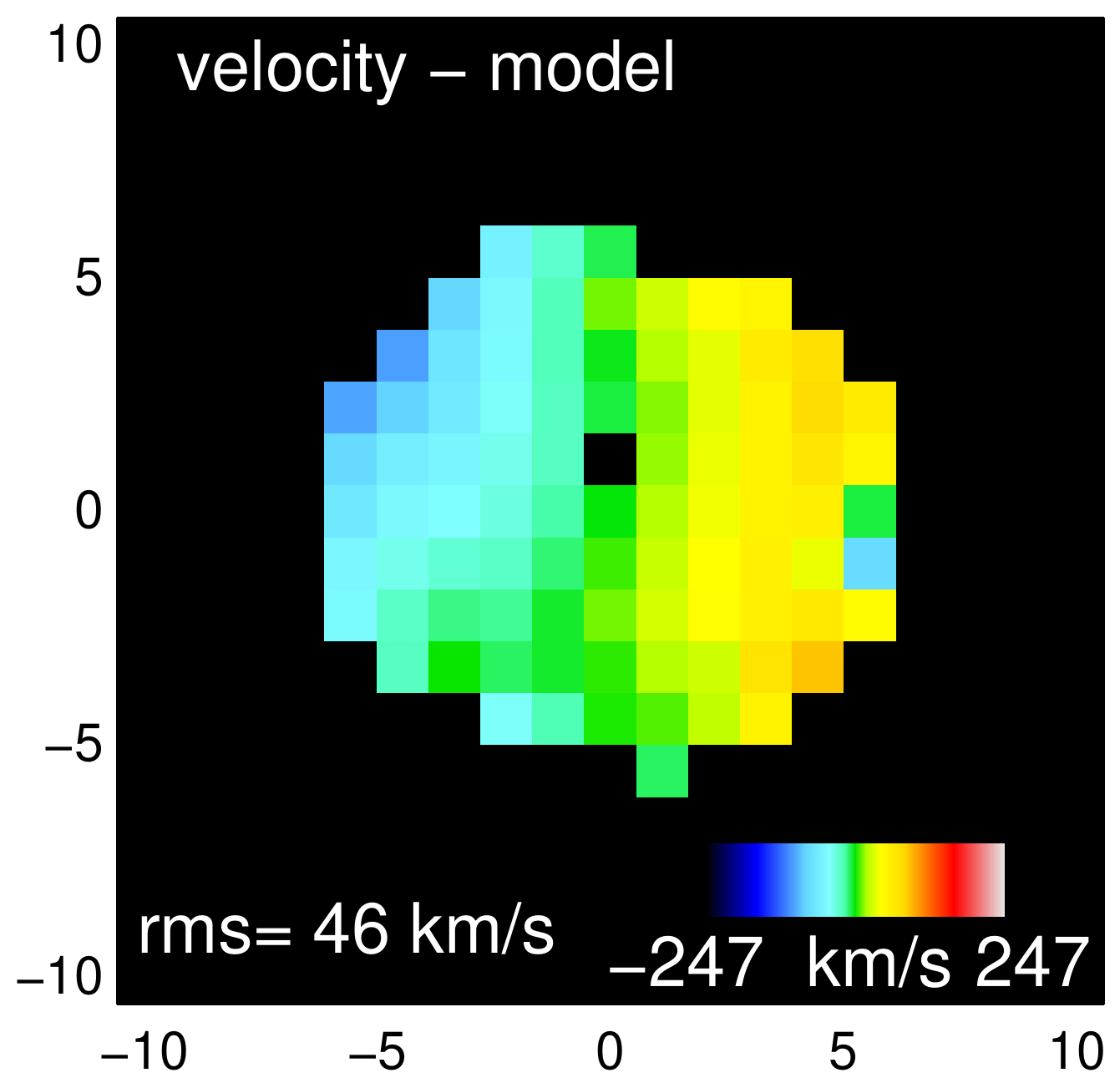}
\includegraphics[width=0.345\columnwidth]{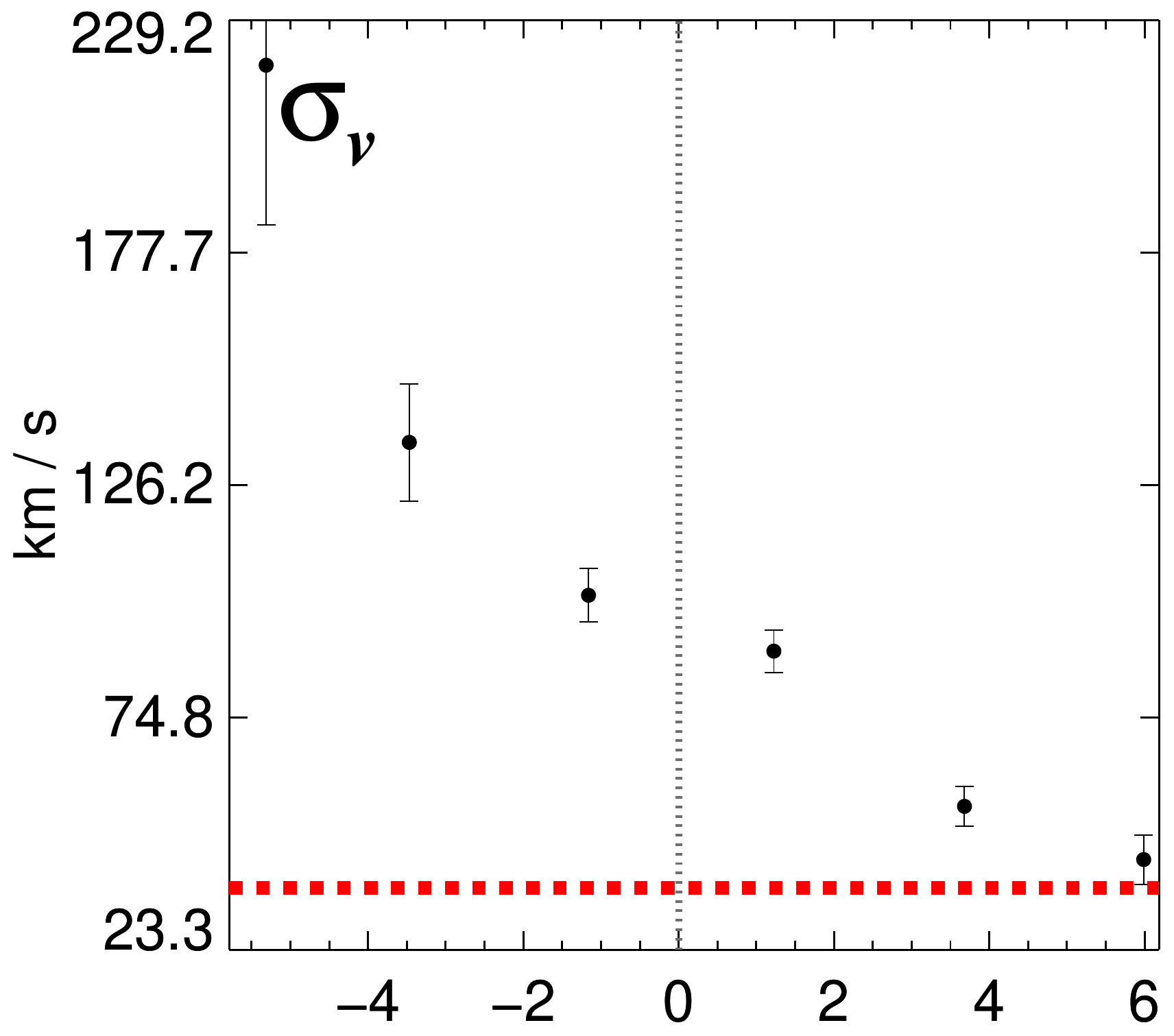}
\includegraphics[width=0.351\columnwidth]{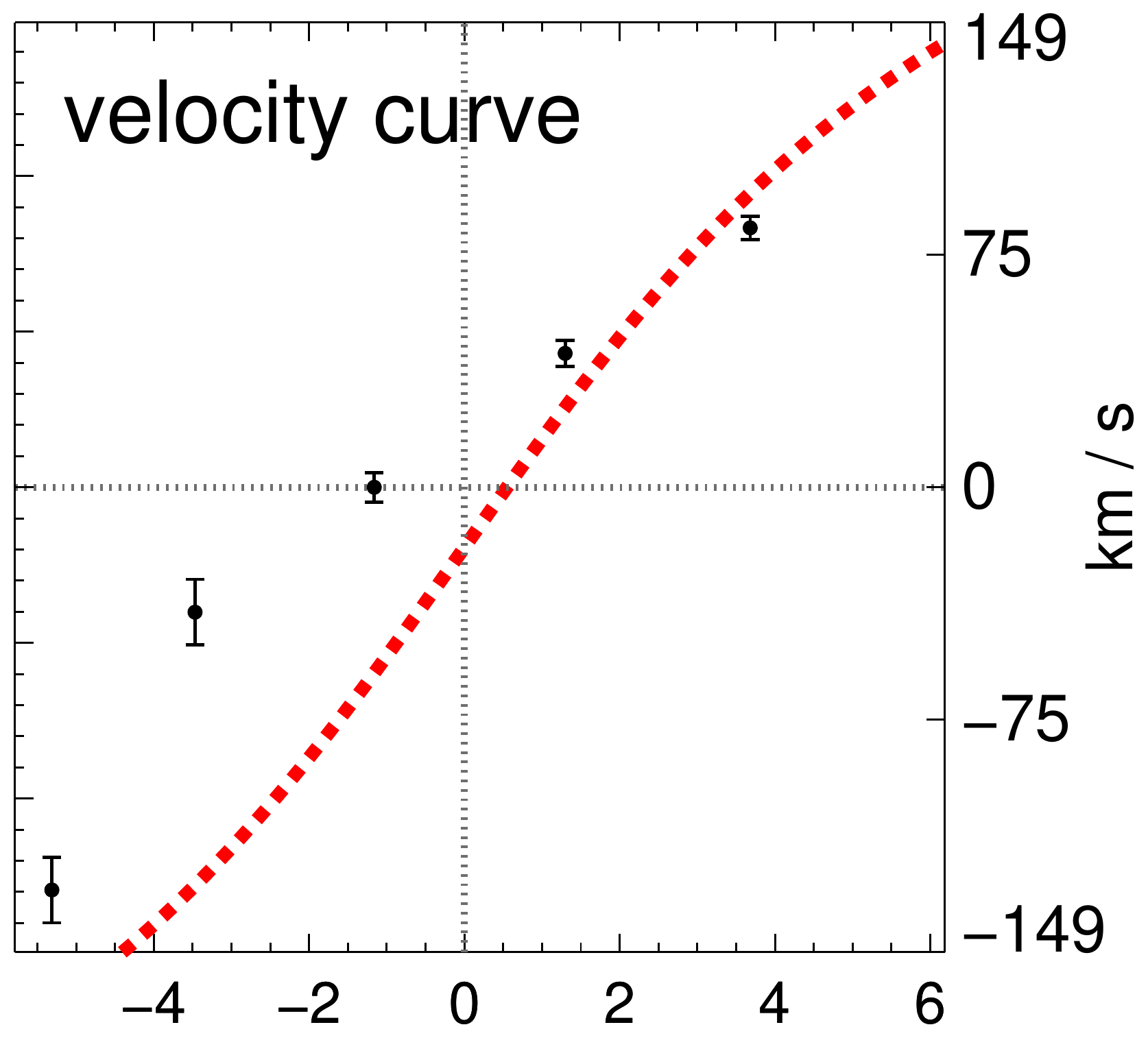}\\
\vspace{1mm}
\includegraphics[width=0.343\columnwidth]{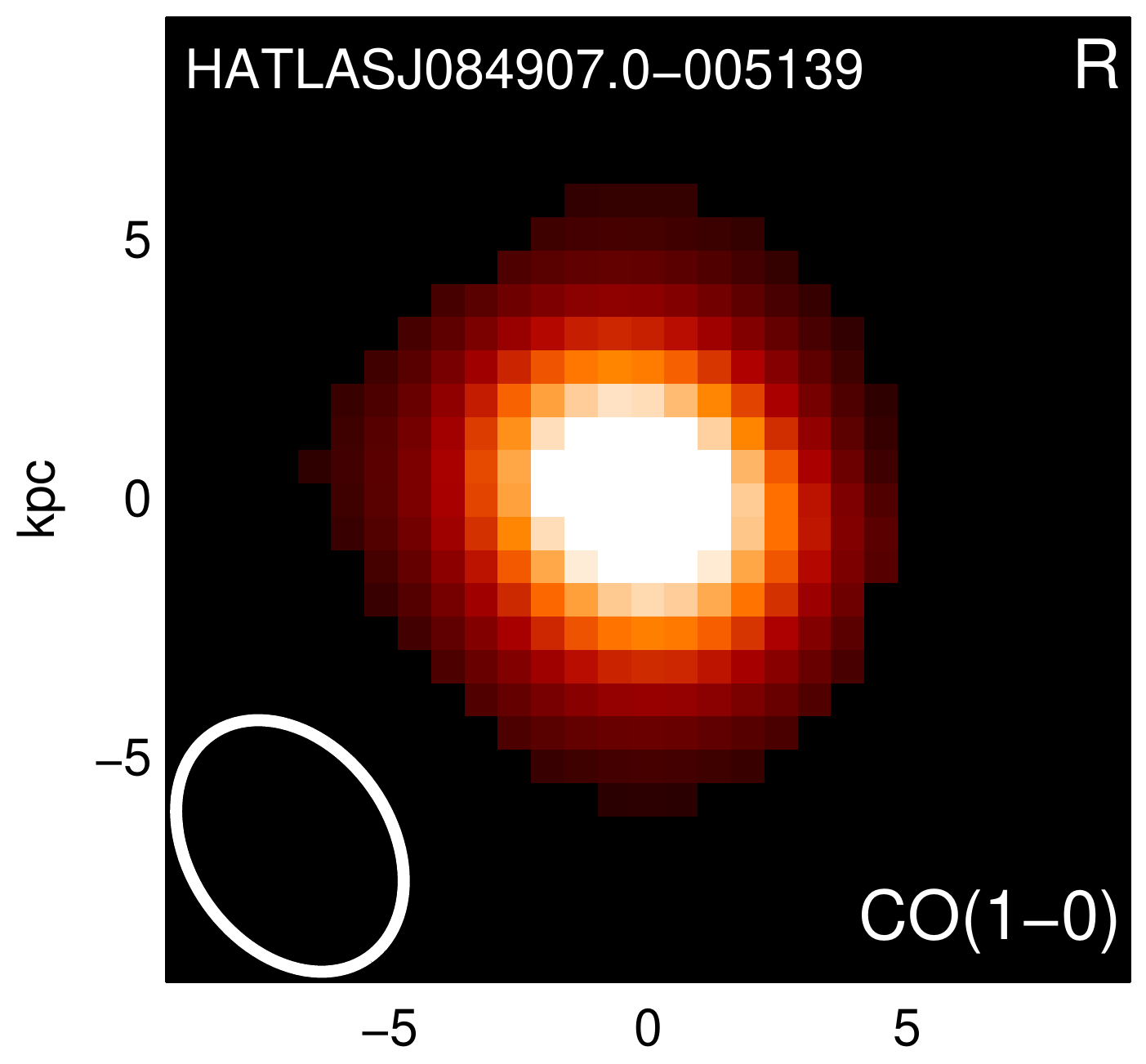}
\includegraphics[width=0.32\columnwidth]{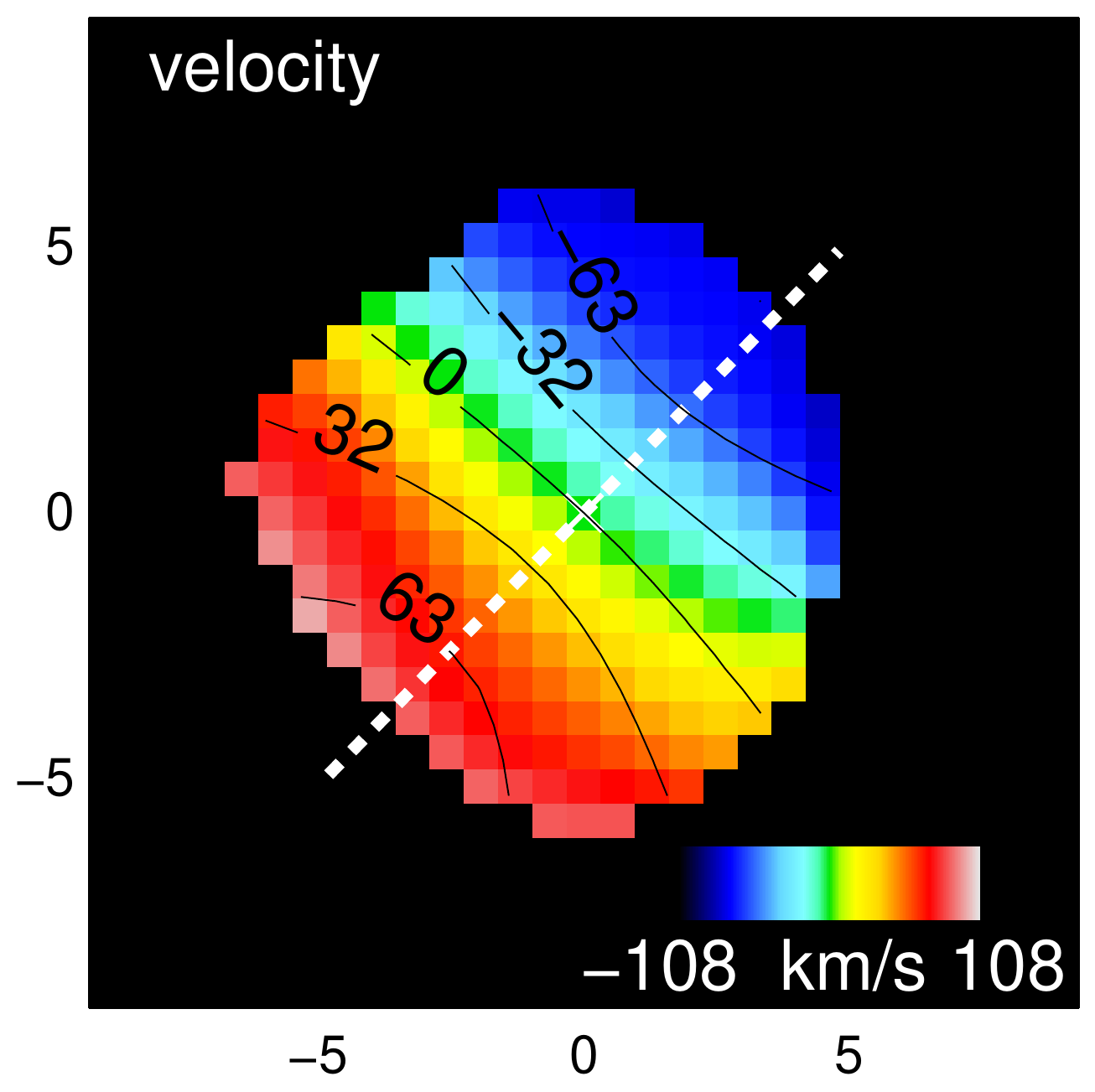}
\includegraphics[width=0.32\columnwidth]{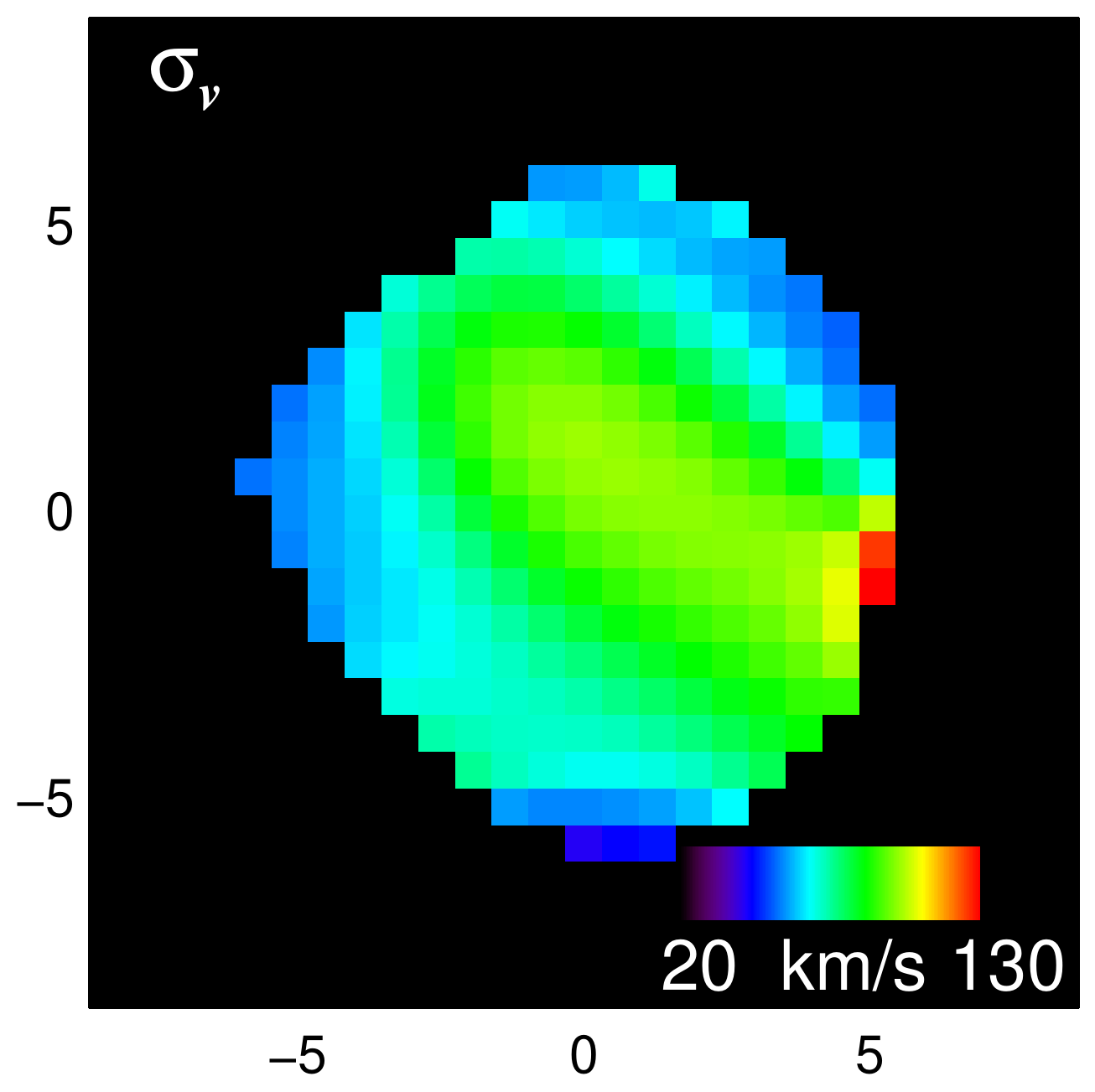}
\includegraphics[width=0.32\columnwidth]{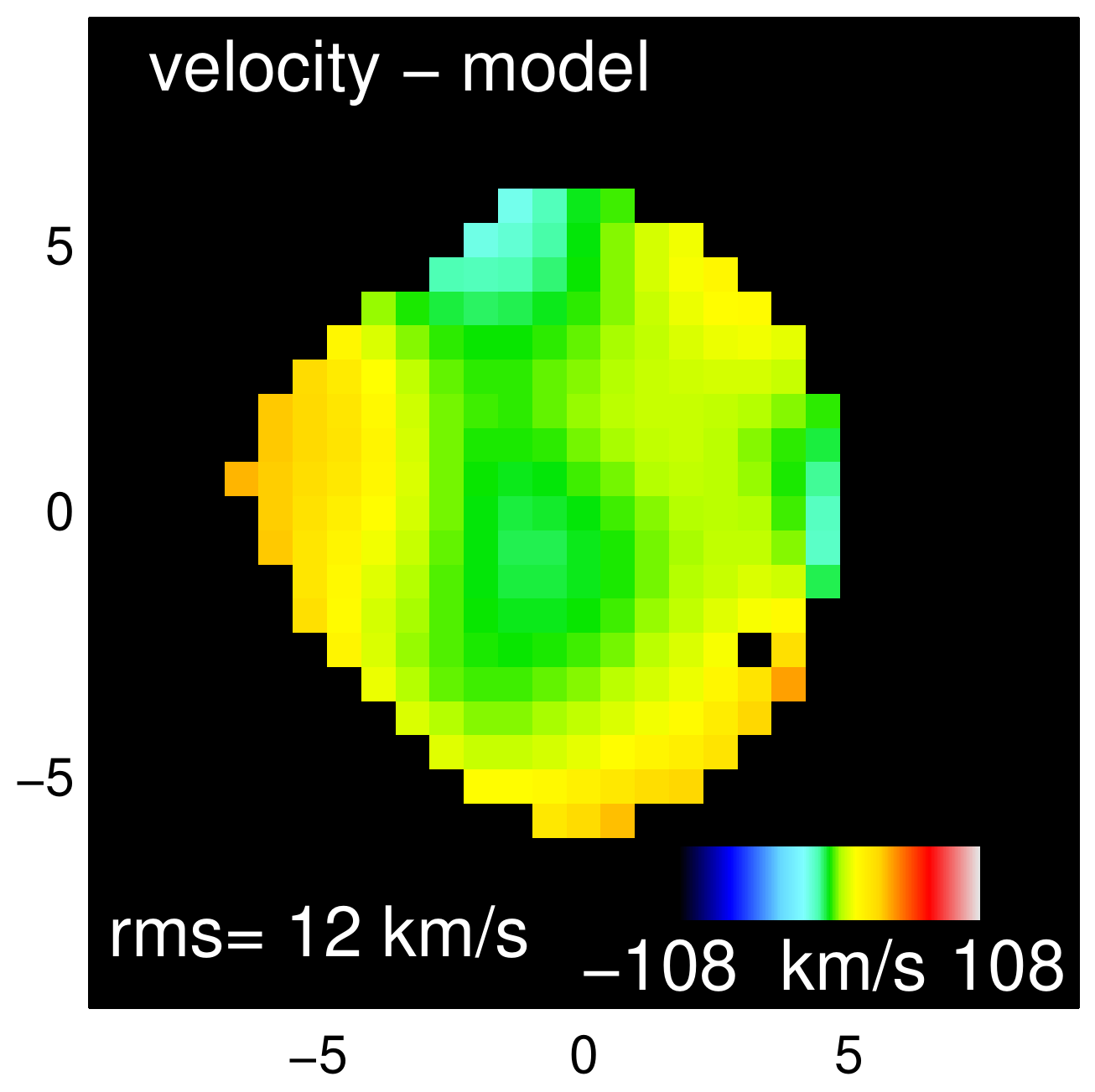}
\includegraphics[width=0.345\columnwidth]{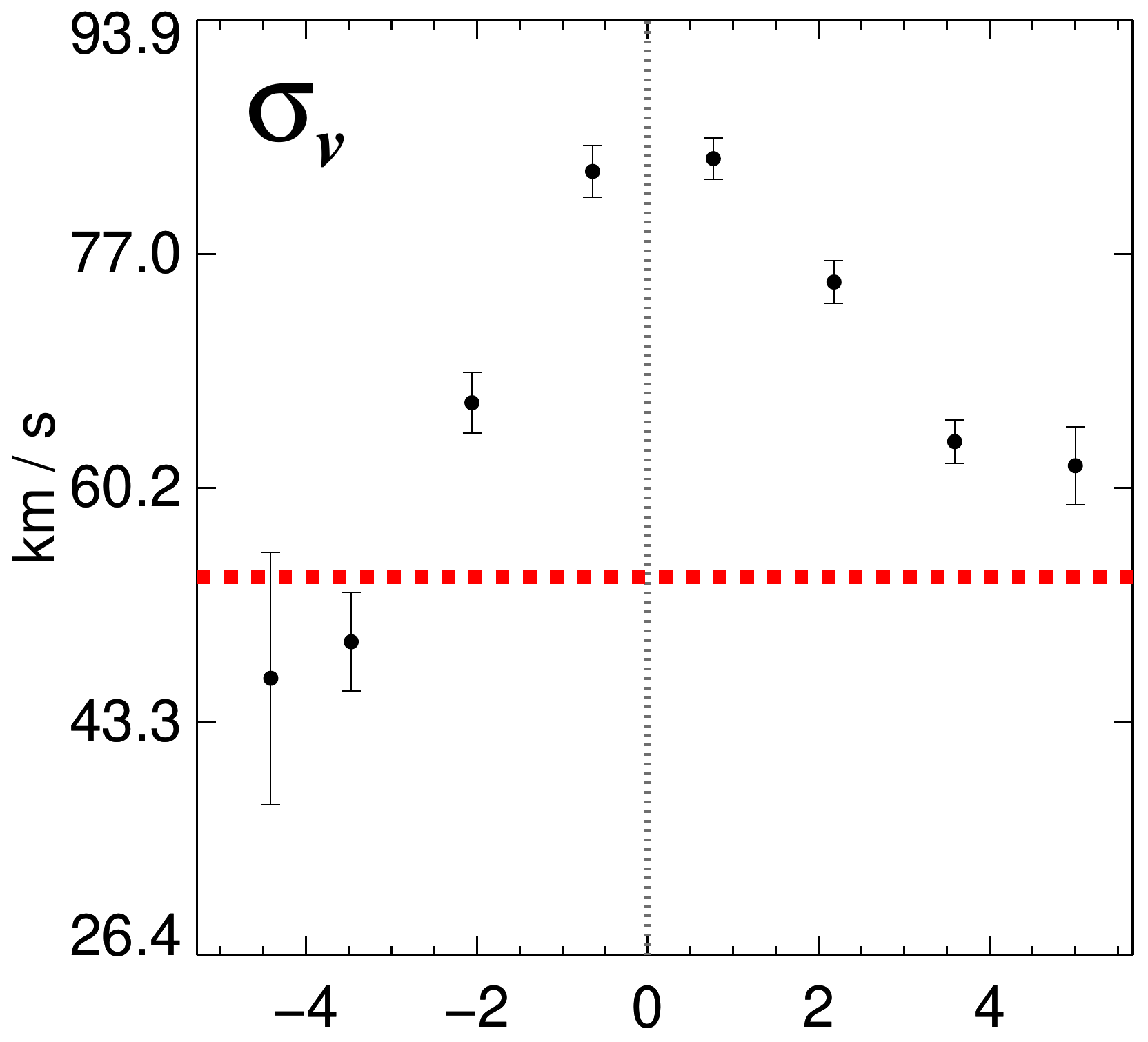}
\includegraphics[width=0.351\columnwidth]{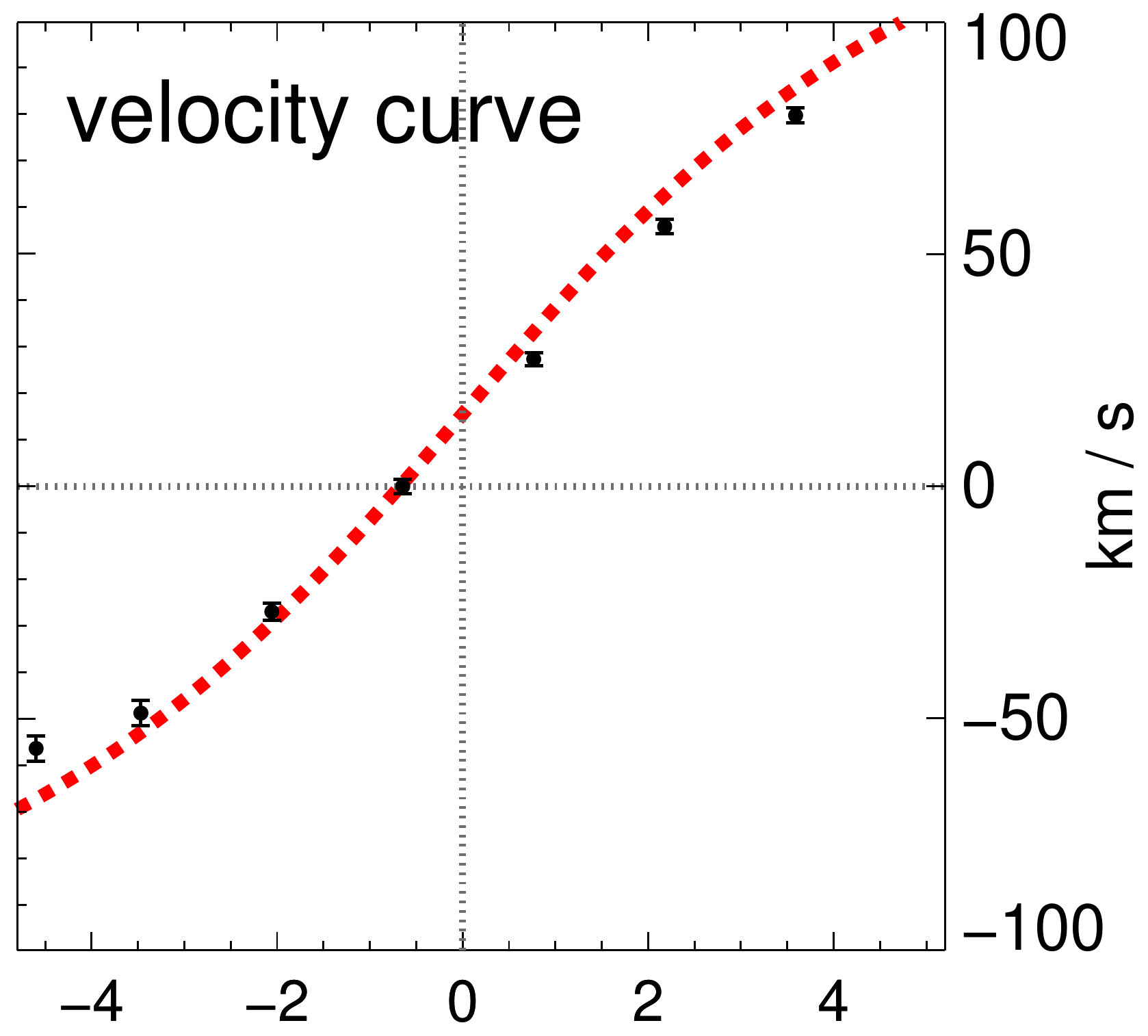}\\
\vspace{1mm}
\includegraphics[width=0.343\columnwidth]{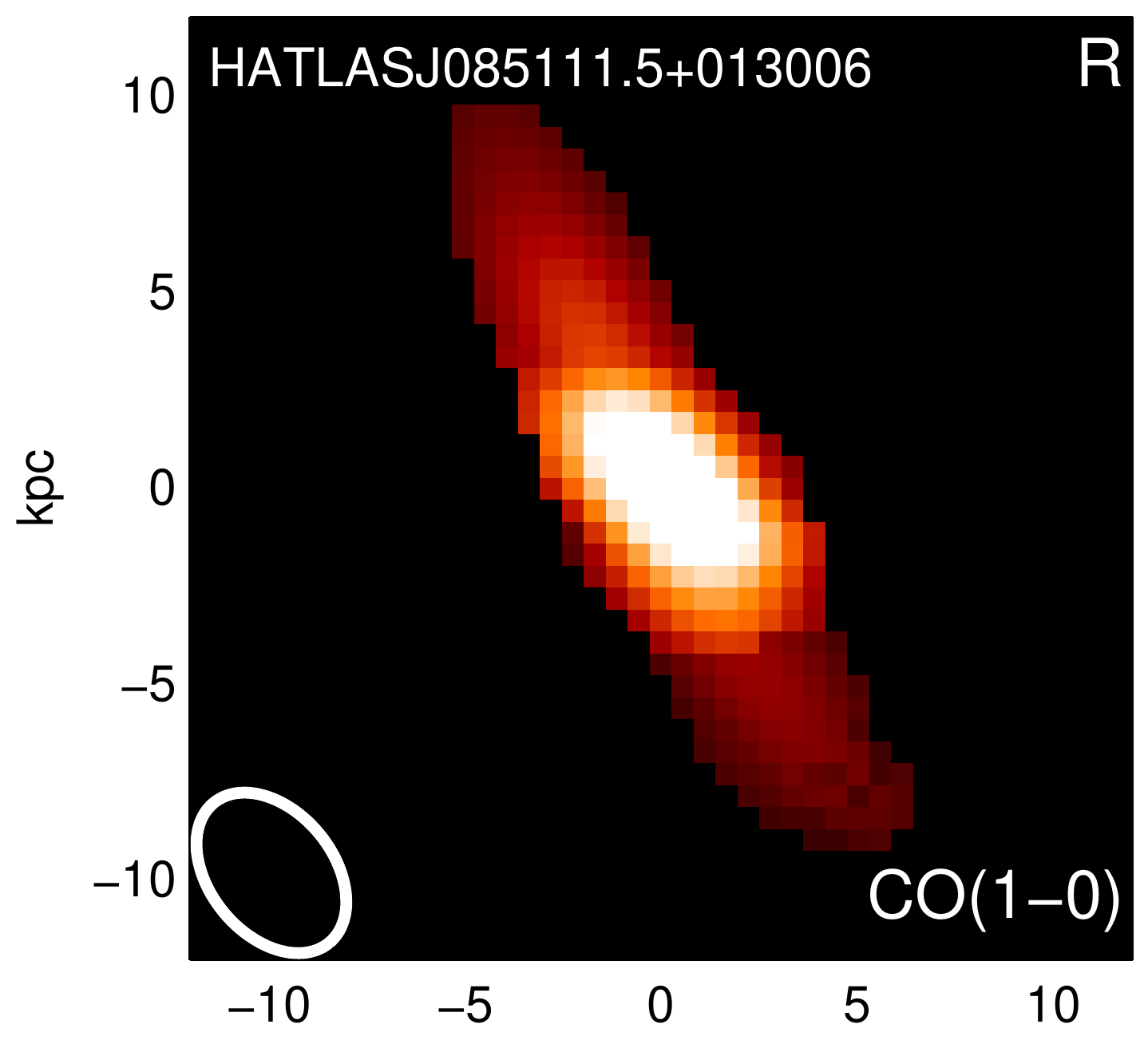}
\includegraphics[width=0.32\columnwidth]{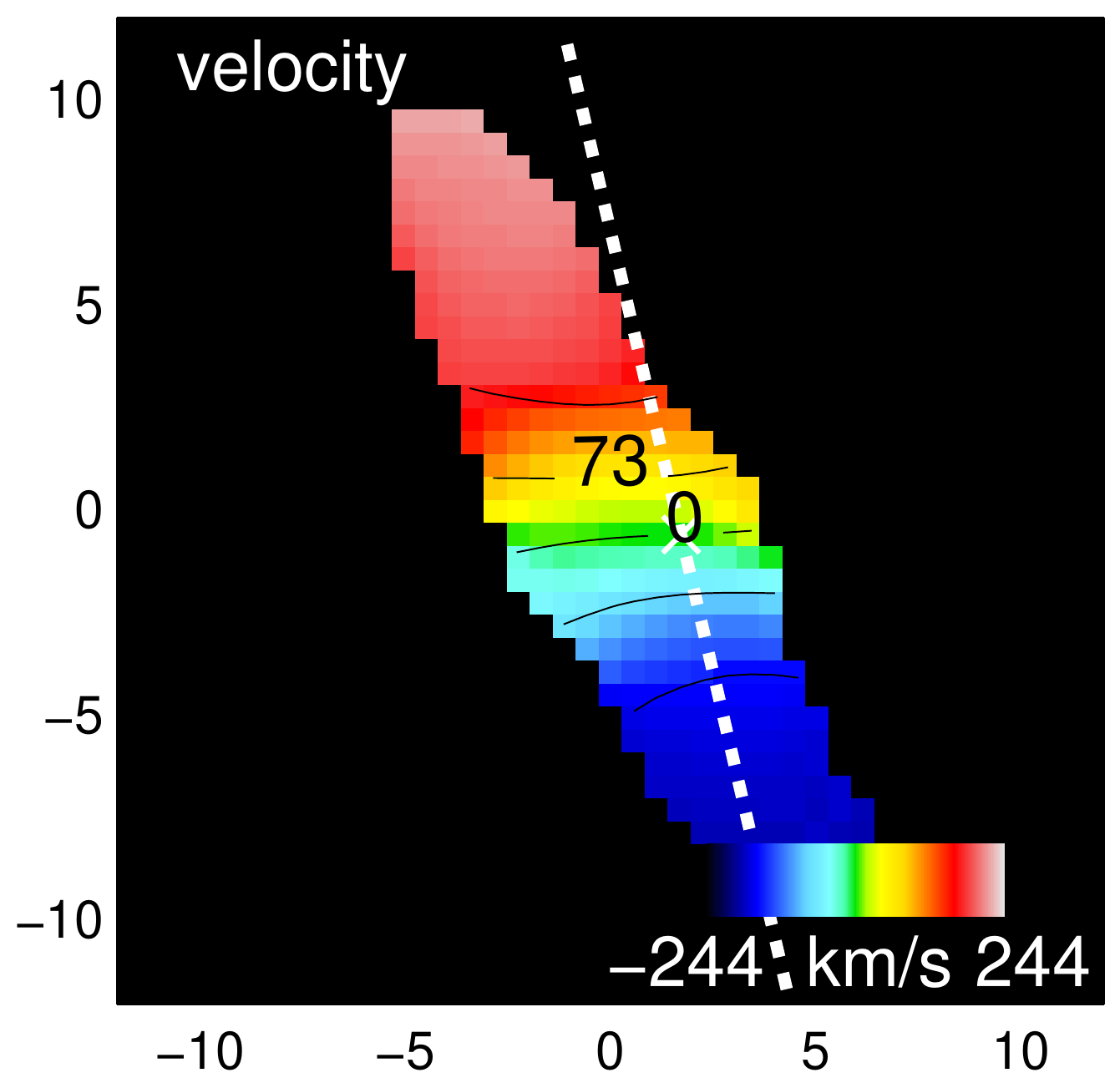}
\includegraphics[width=0.32\columnwidth]{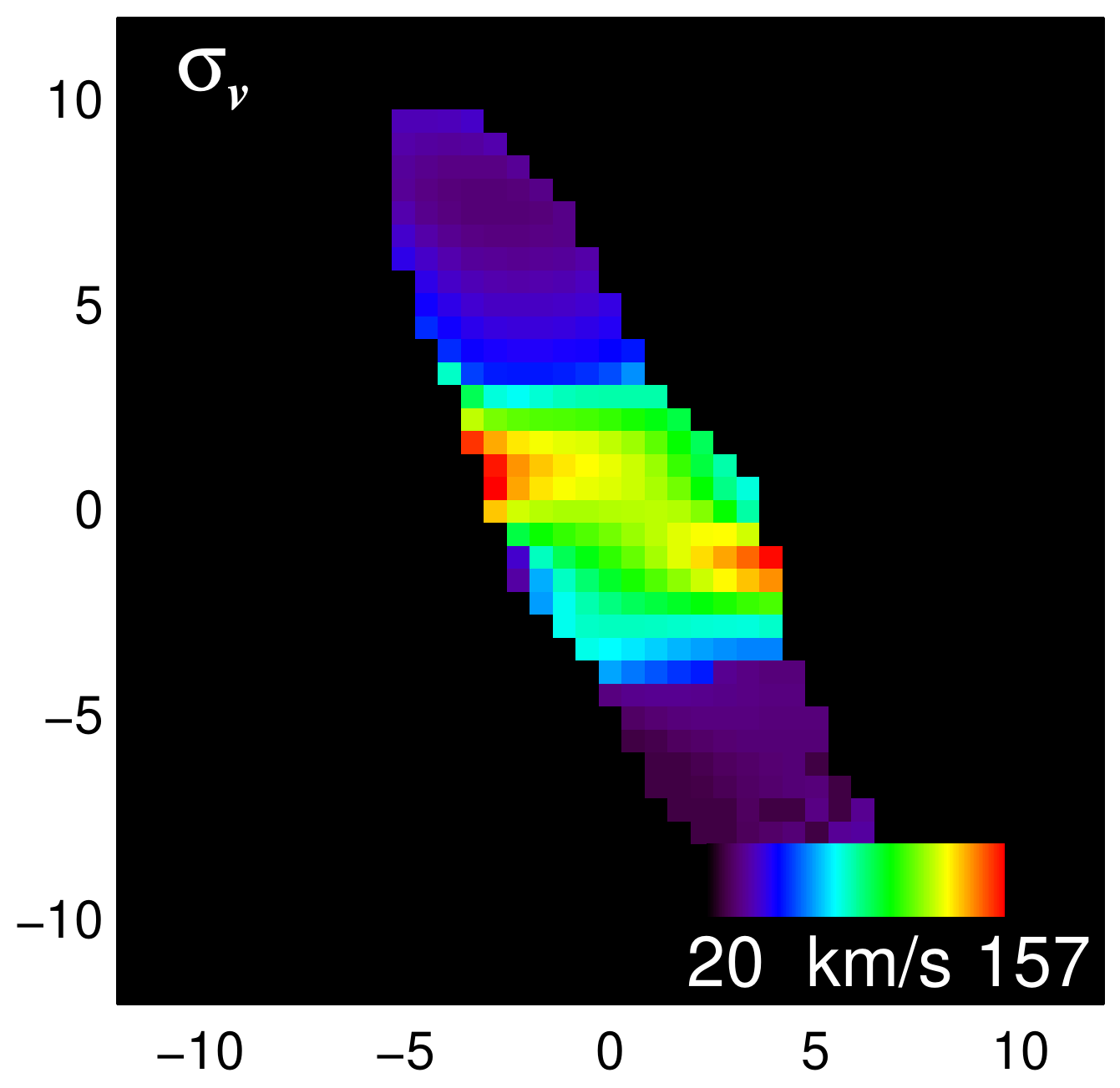}
\includegraphics[width=0.32\columnwidth]{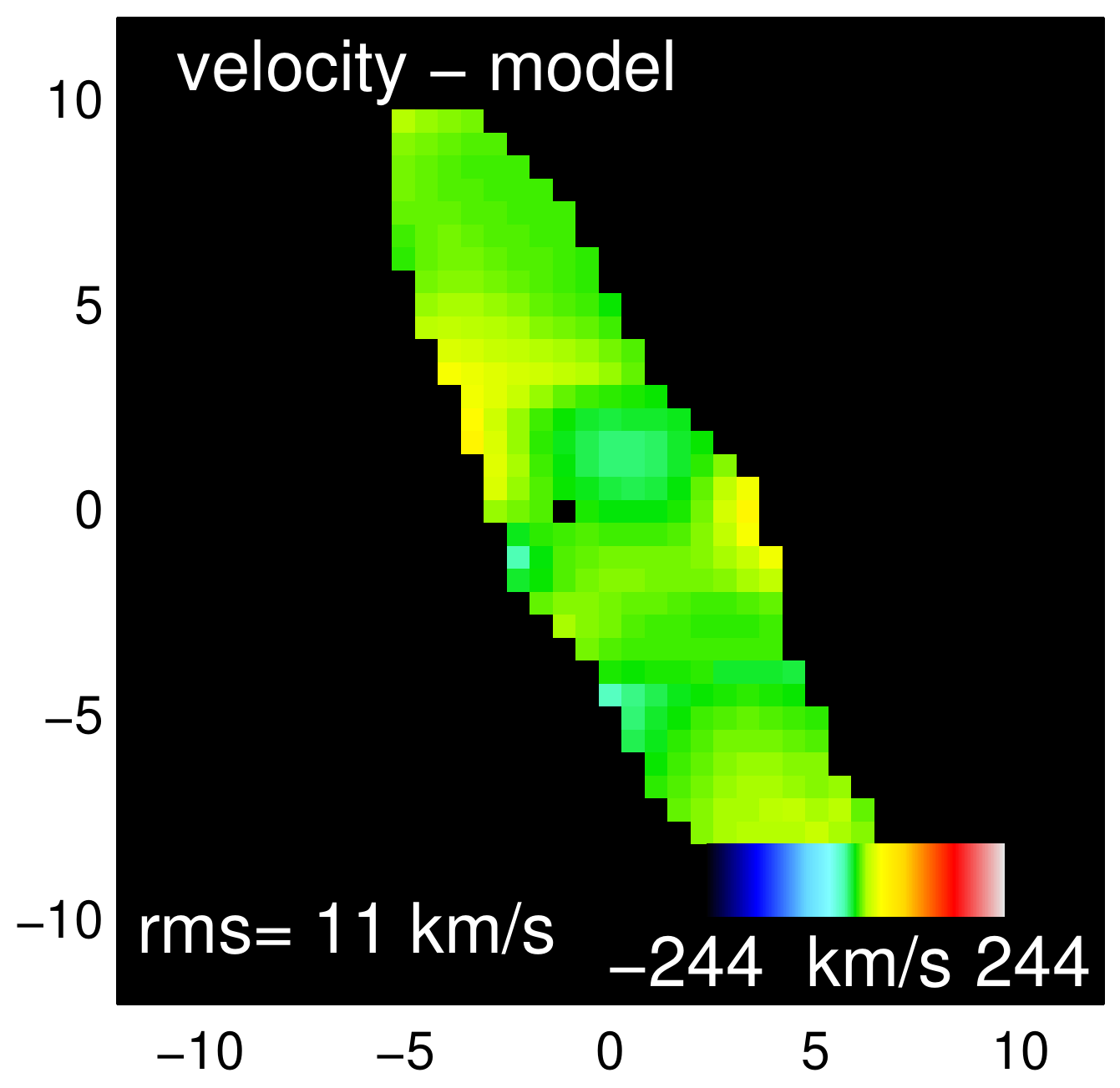}
\includegraphics[width=0.345\columnwidth]{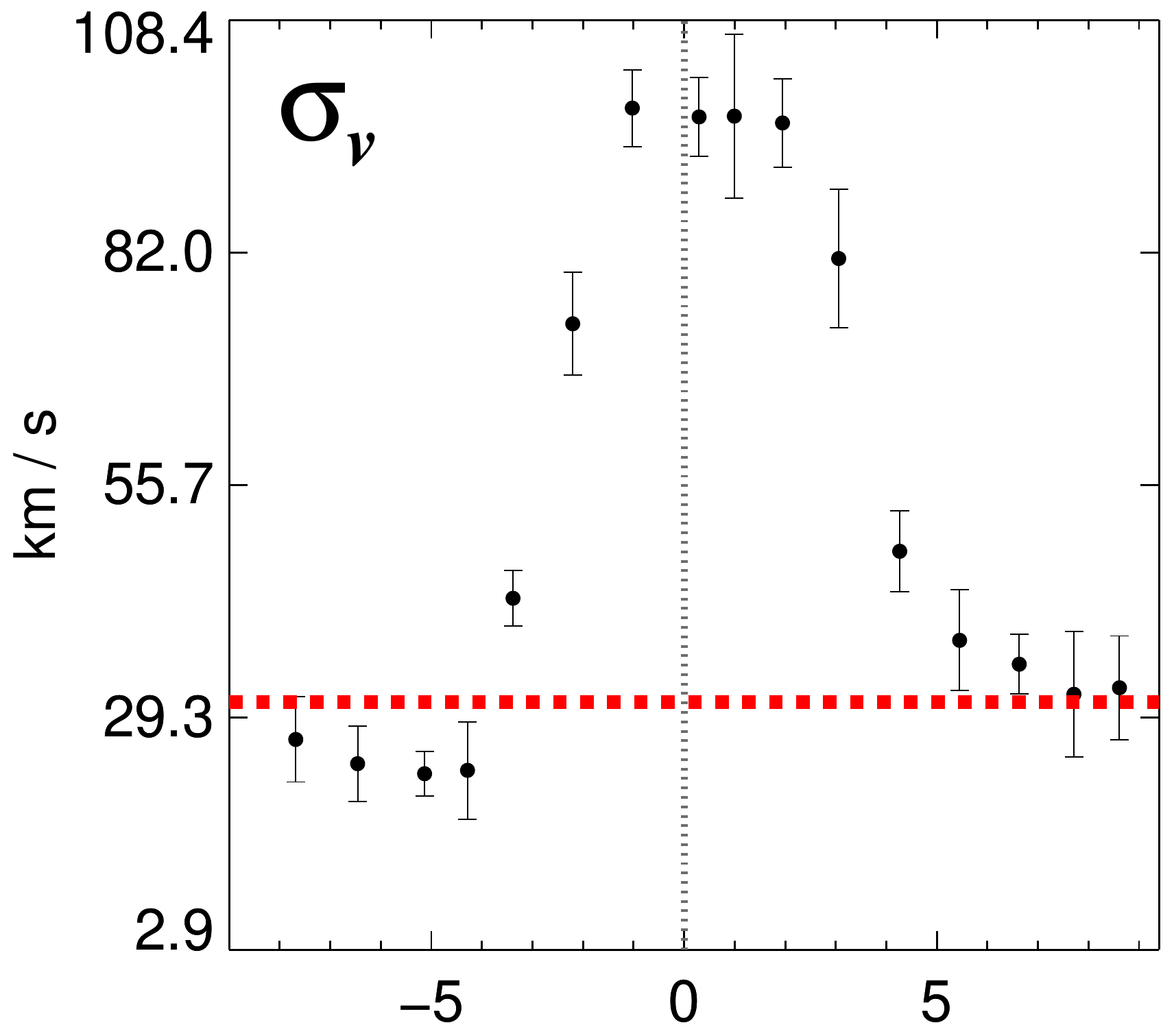}
\includegraphics[width=0.361\columnwidth]{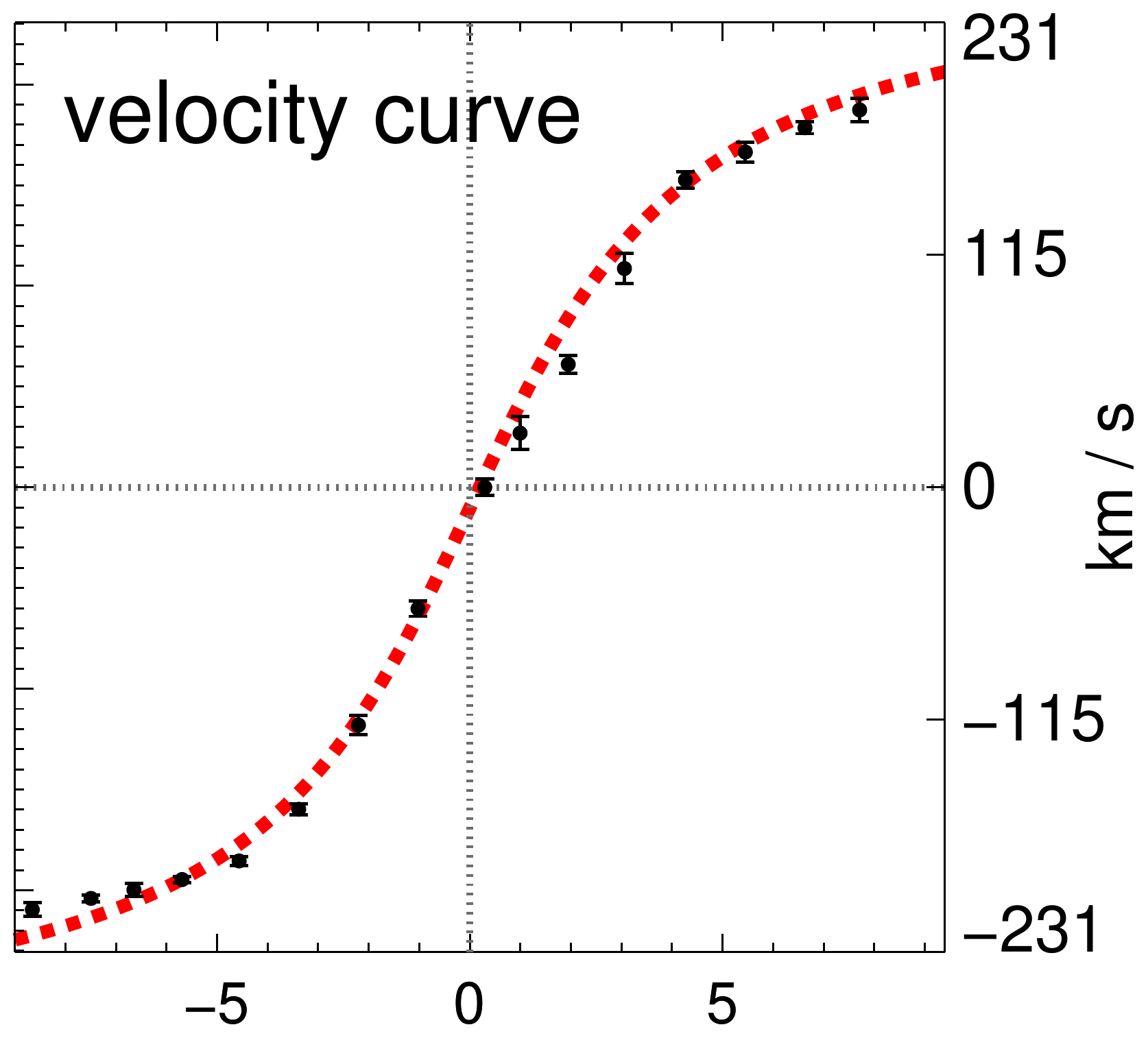}\\
\vspace{1mm}
\includegraphics[width=0.343\columnwidth]{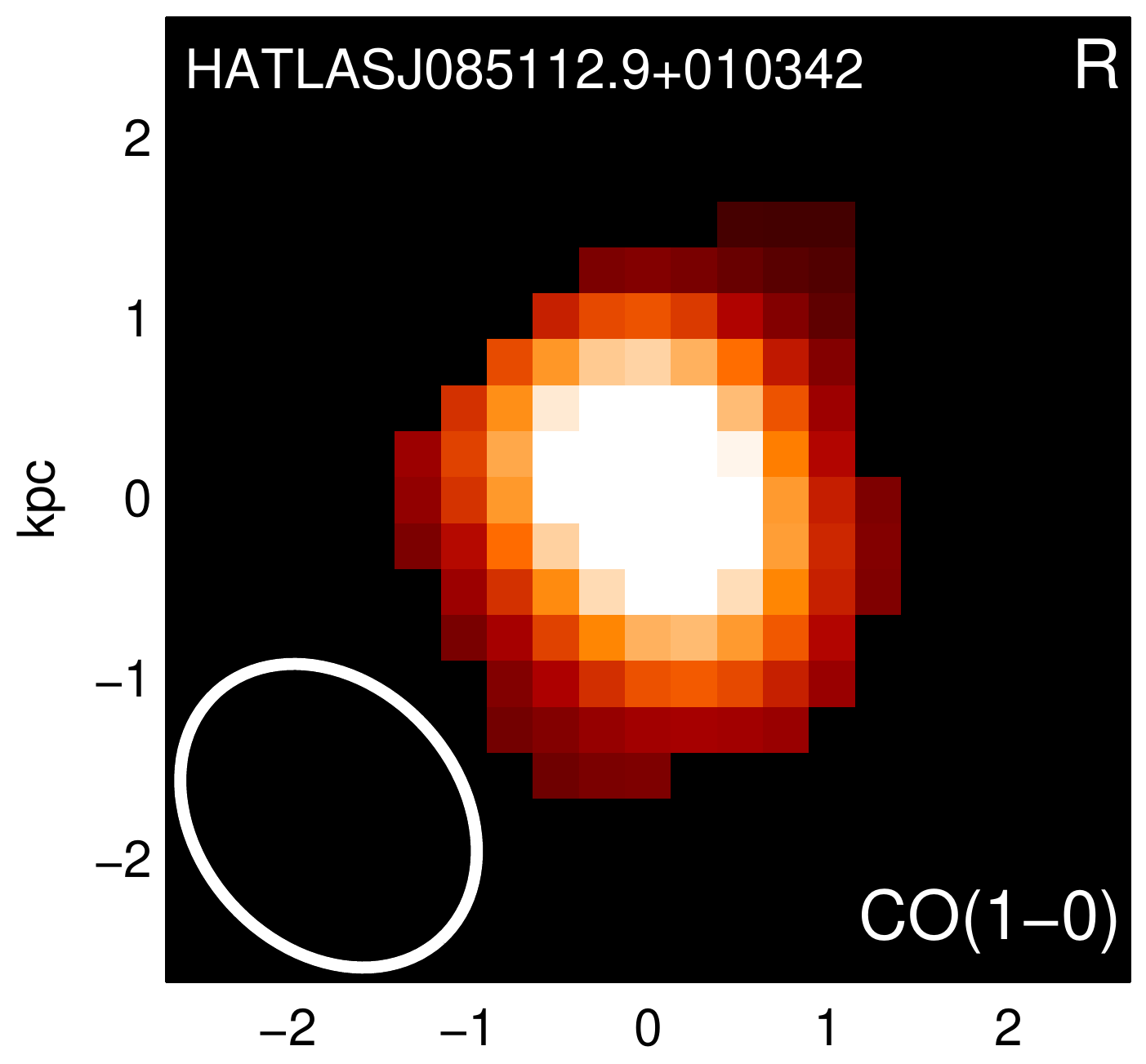}
\includegraphics[width=0.32\columnwidth]{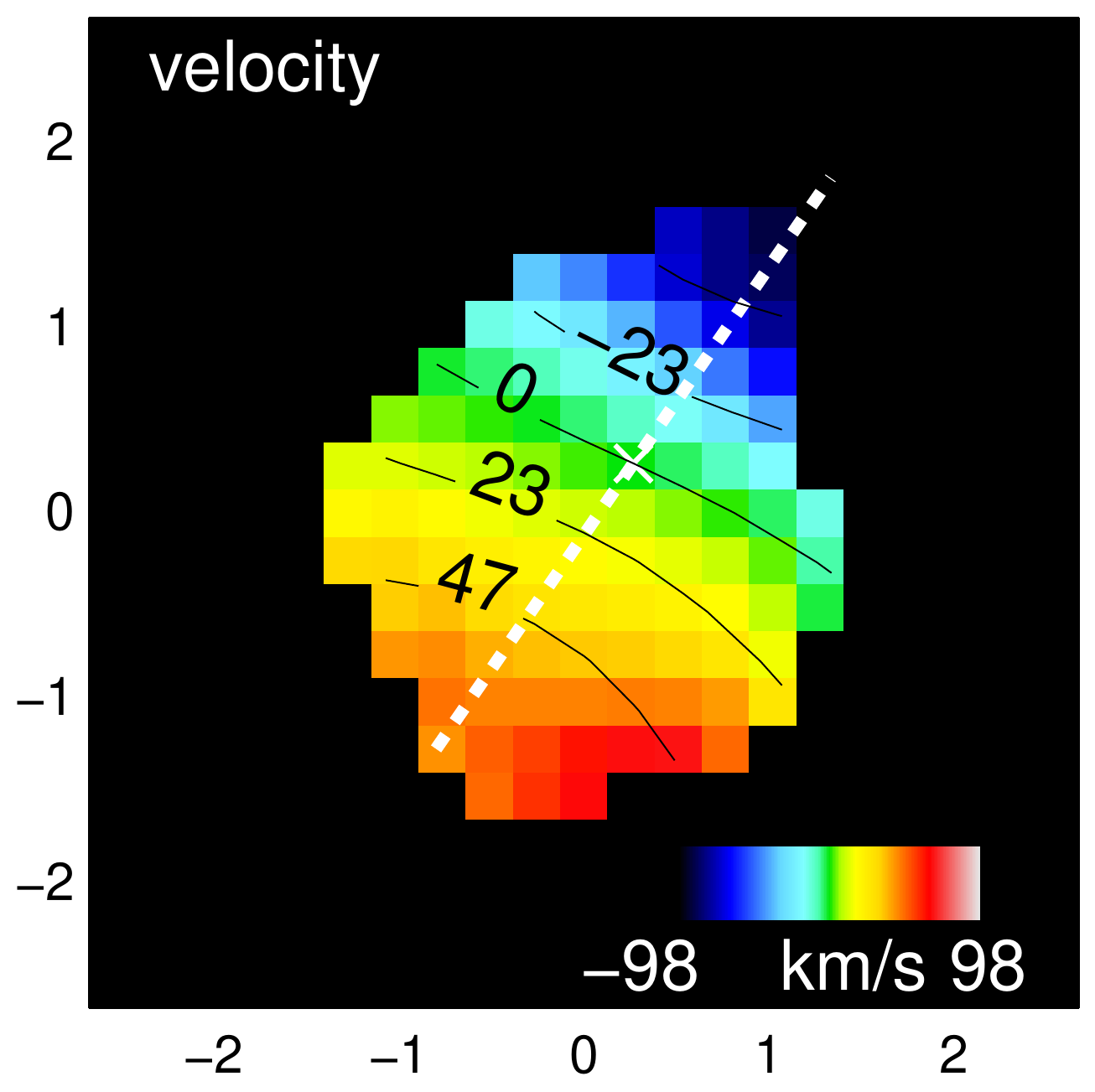}
\includegraphics[width=0.32\columnwidth]{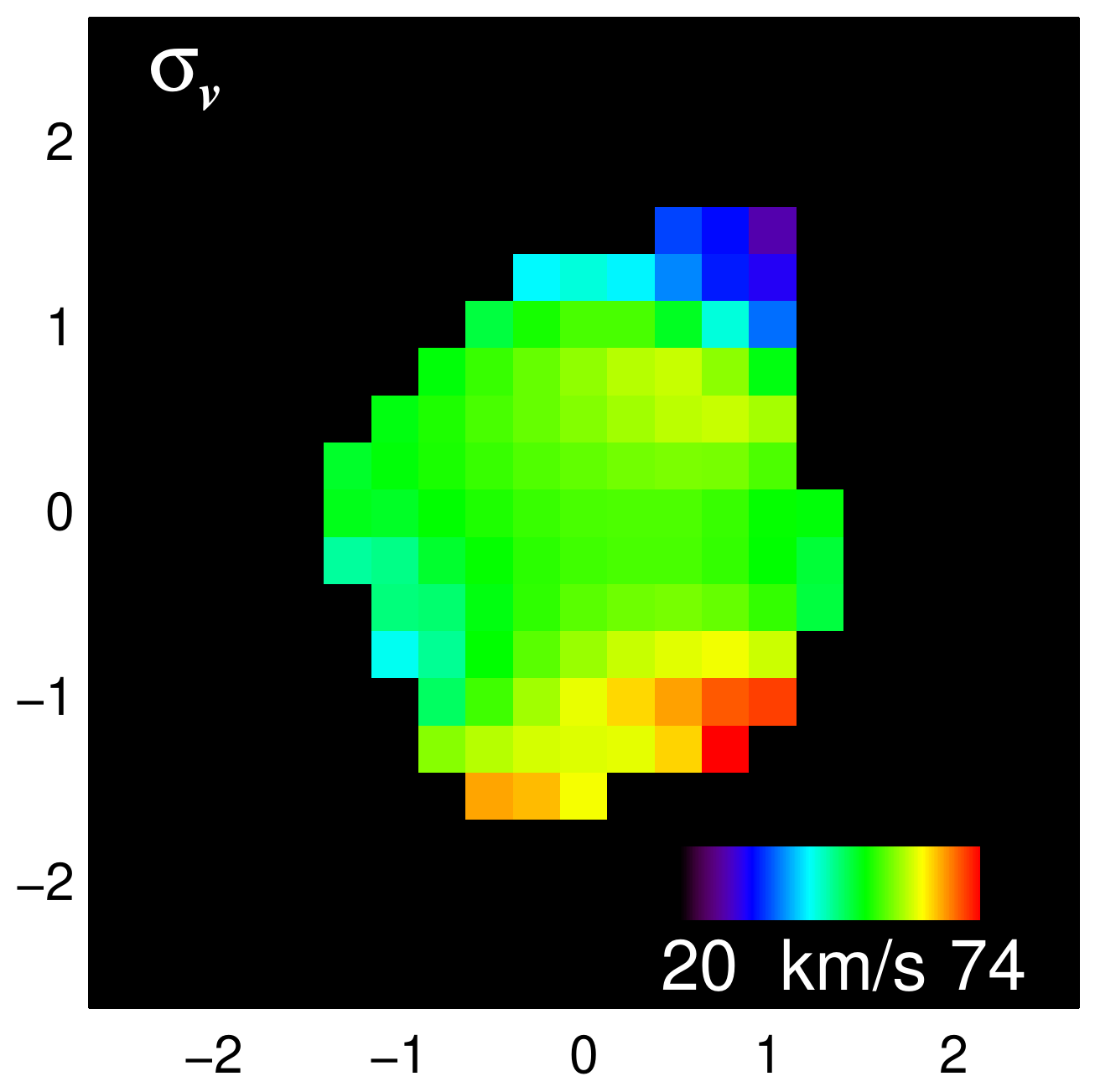}
\includegraphics[width=0.32\columnwidth]{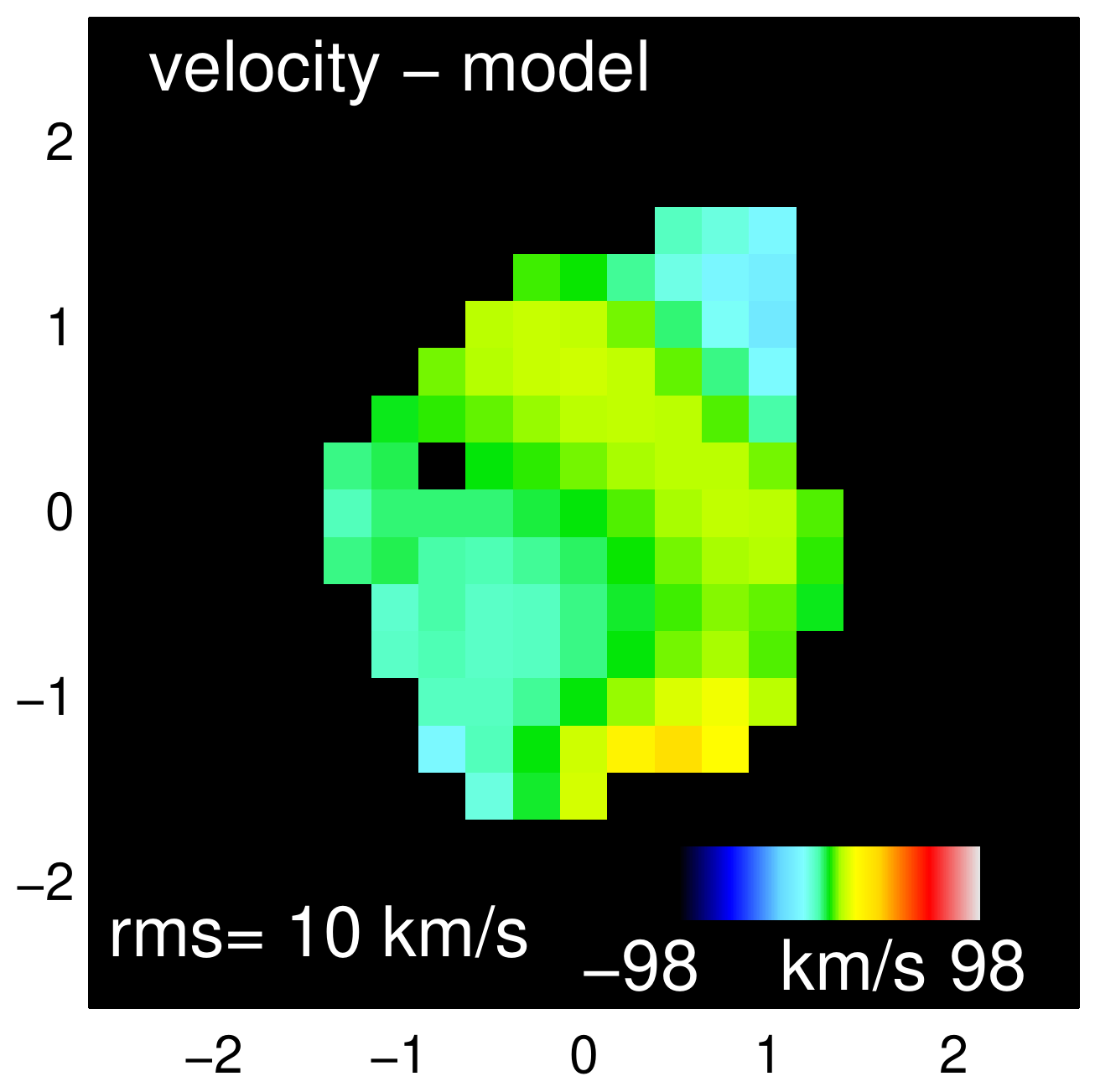}
\includegraphics[width=0.345\columnwidth]{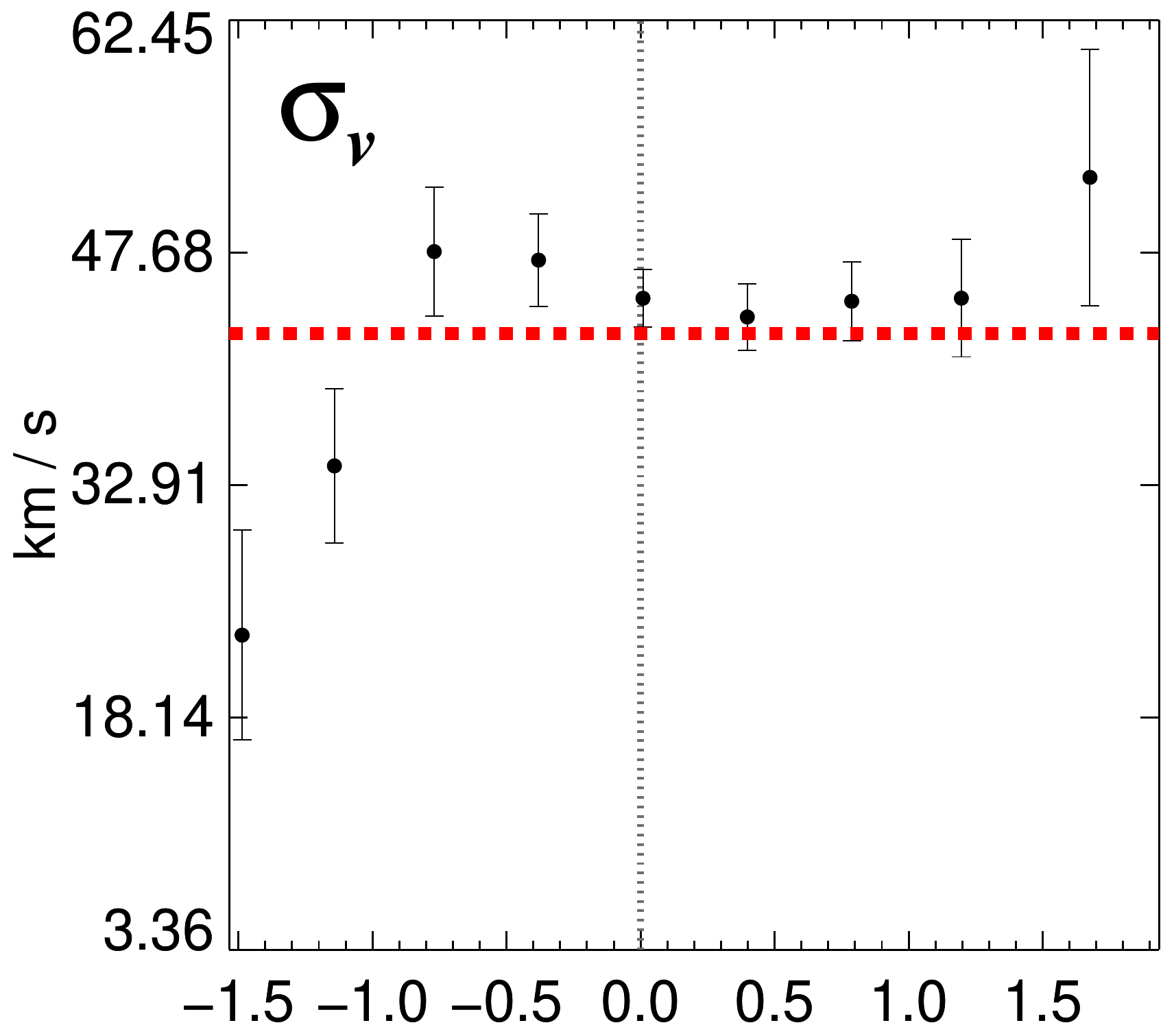}
\includegraphics[width=0.351\columnwidth]{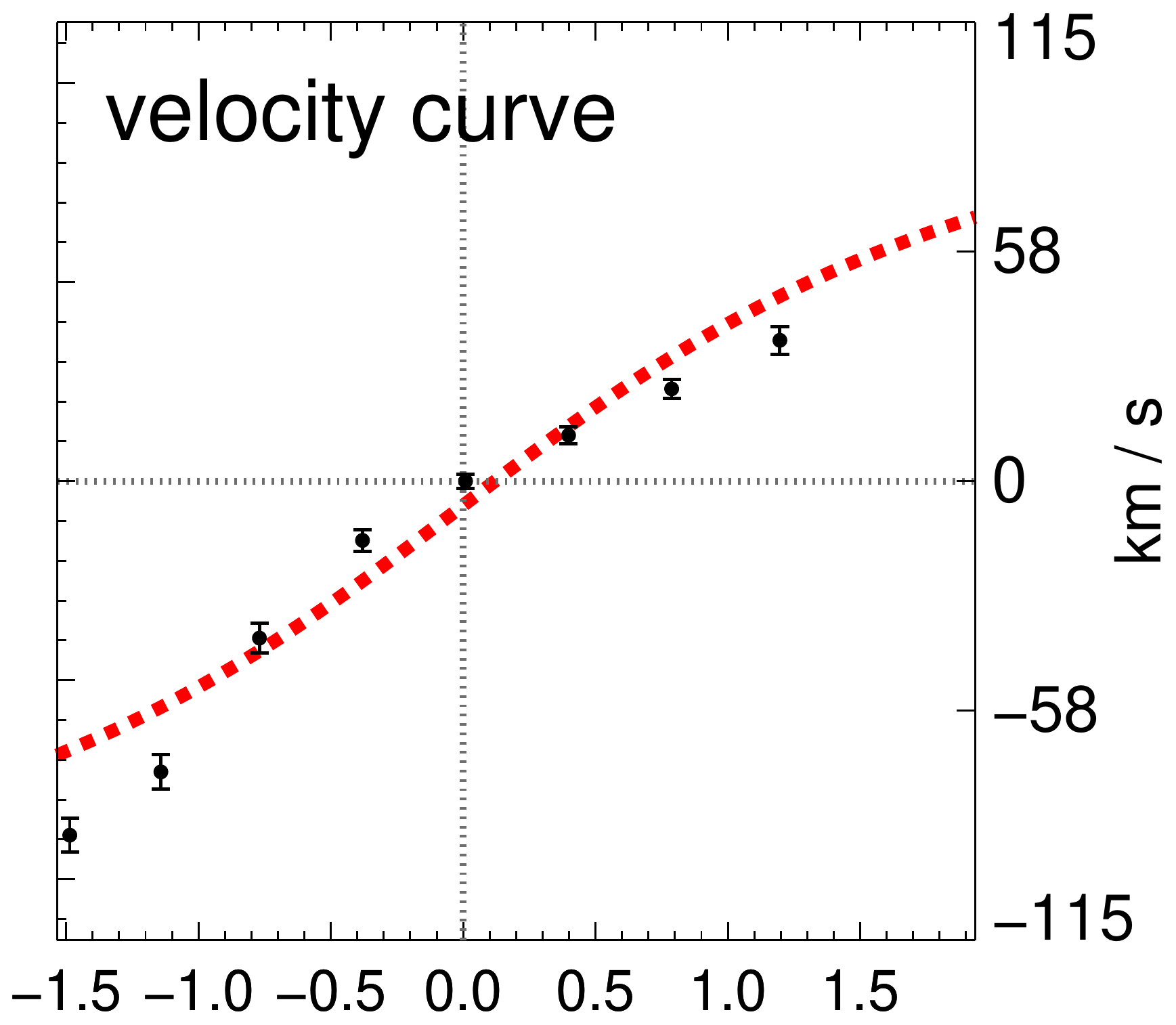}\\
\vspace{1mm}
\includegraphics[width=0.343\columnwidth]{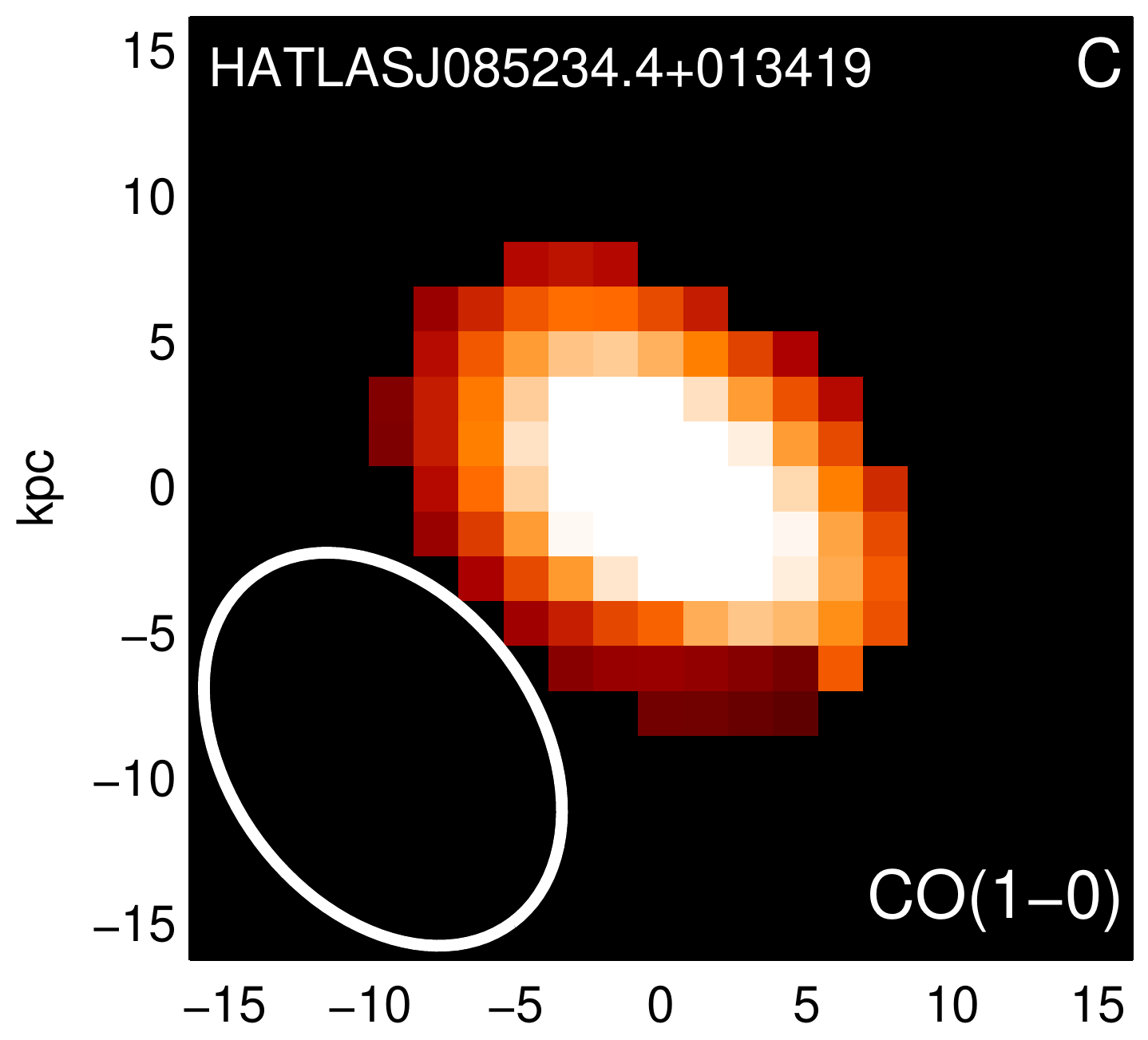}
\includegraphics[width=0.32\columnwidth]{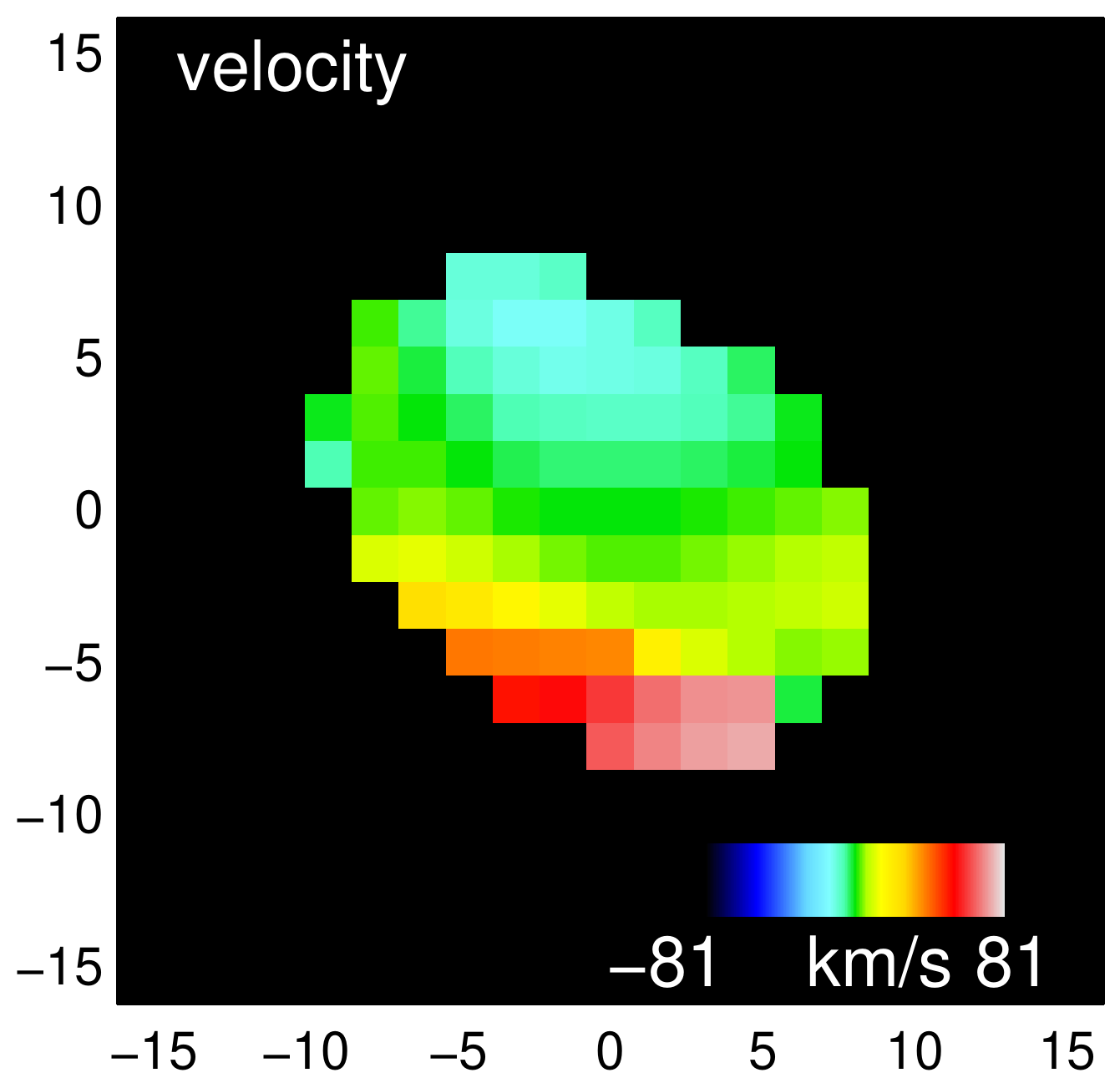}
\includegraphics[width=0.32\columnwidth]{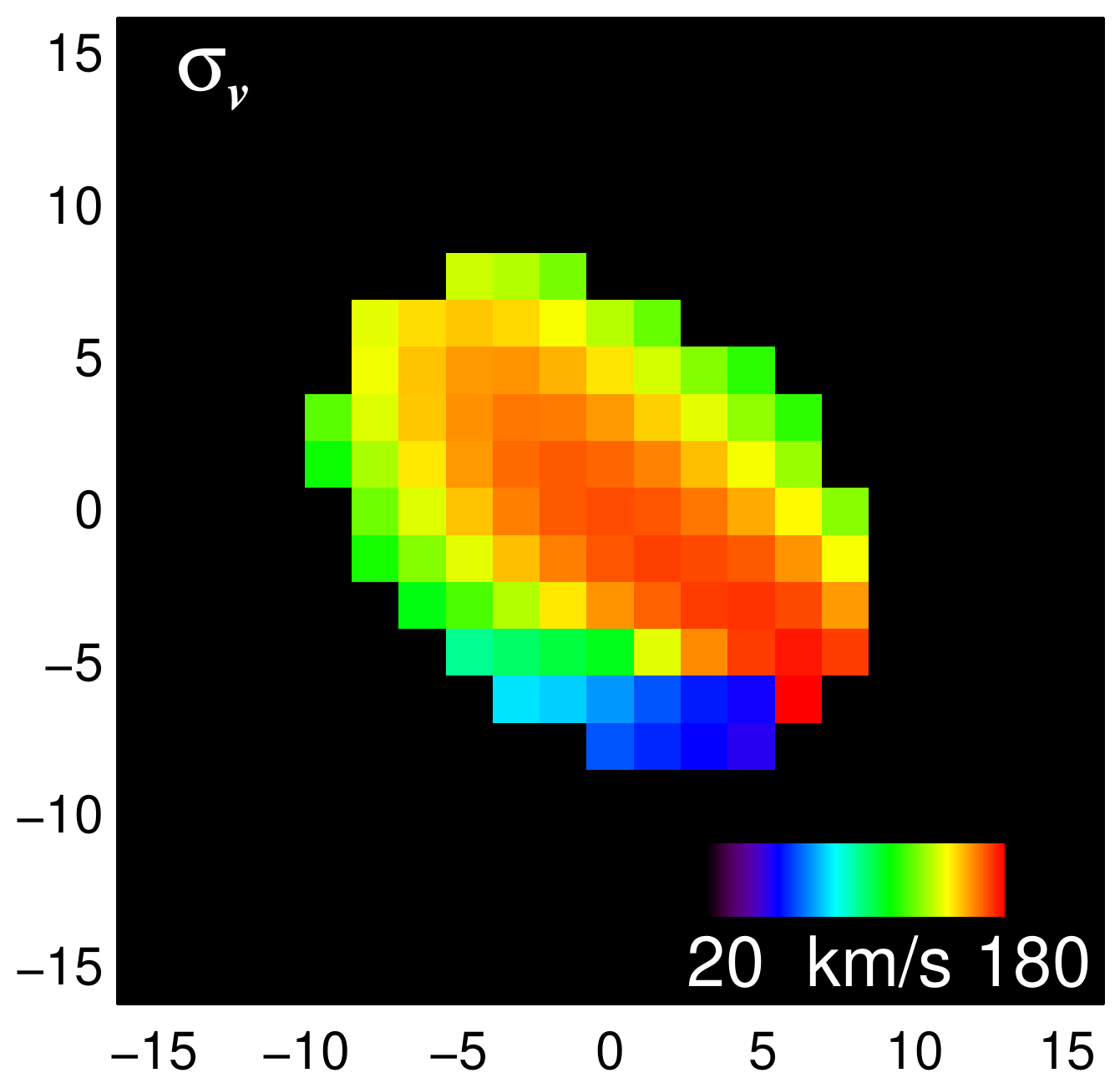}\\
\vspace{1mm}
\includegraphics[width=0.343\columnwidth]{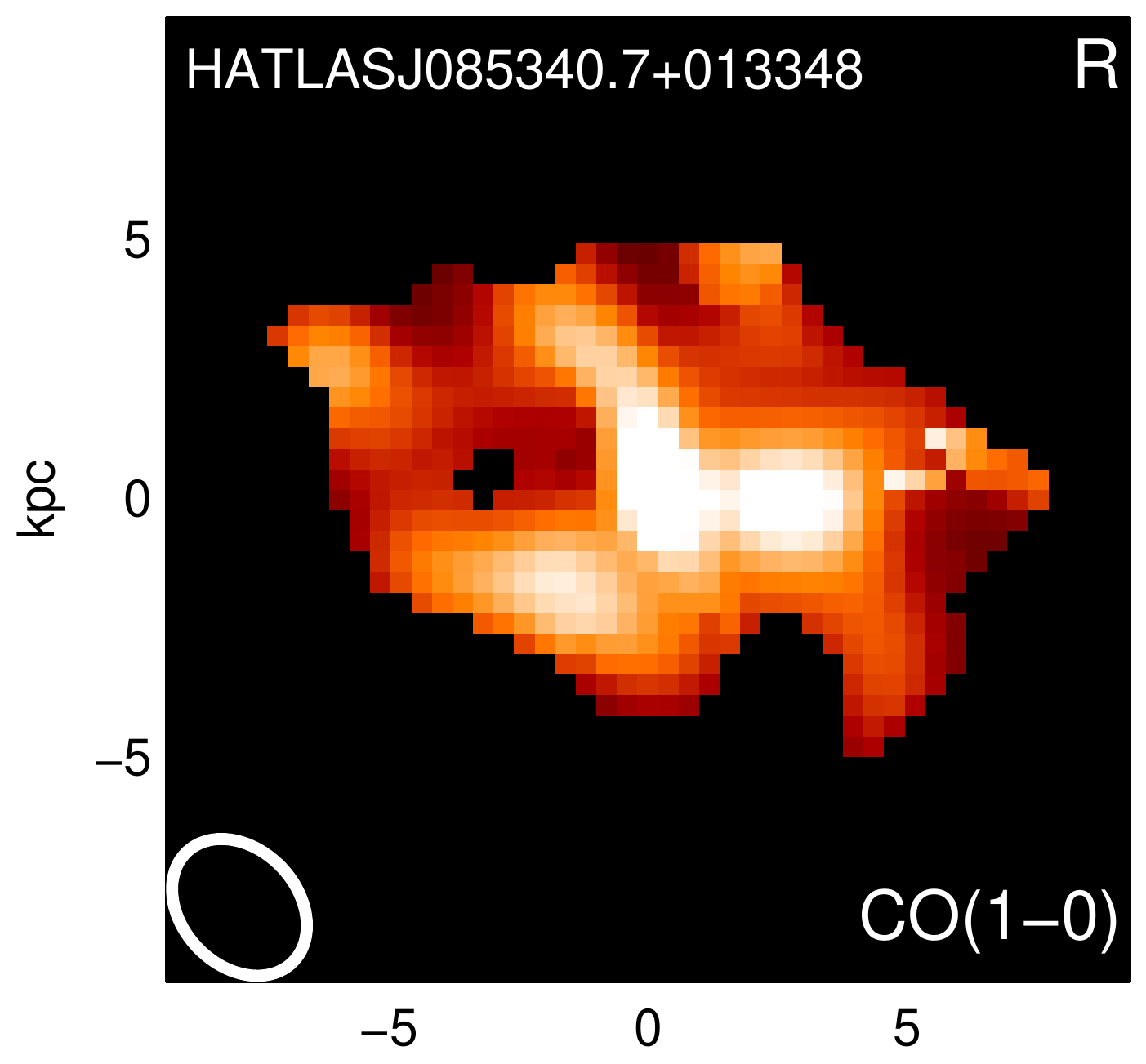}
\includegraphics[width=0.32\columnwidth]{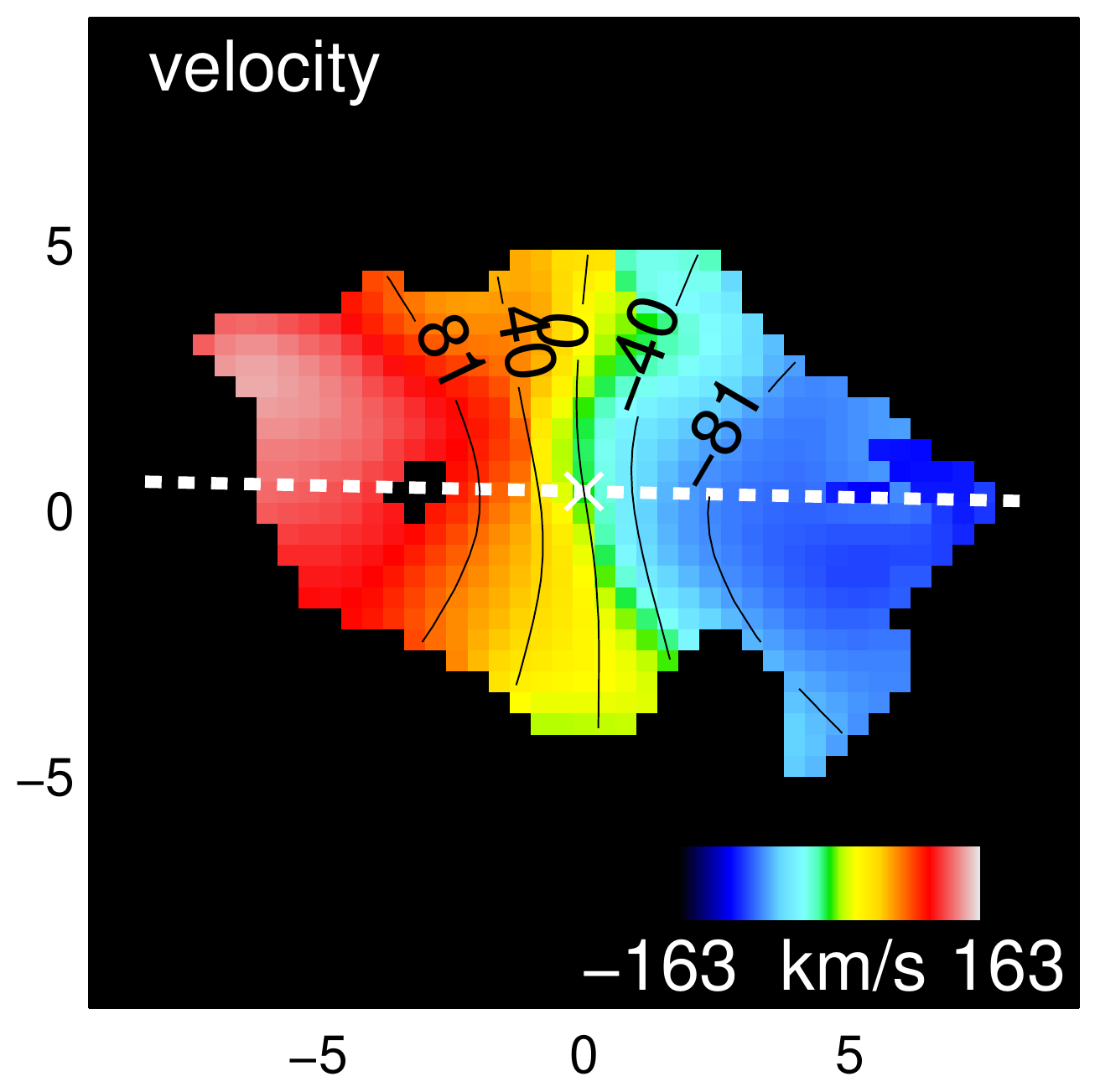}
\includegraphics[width=0.32\columnwidth]{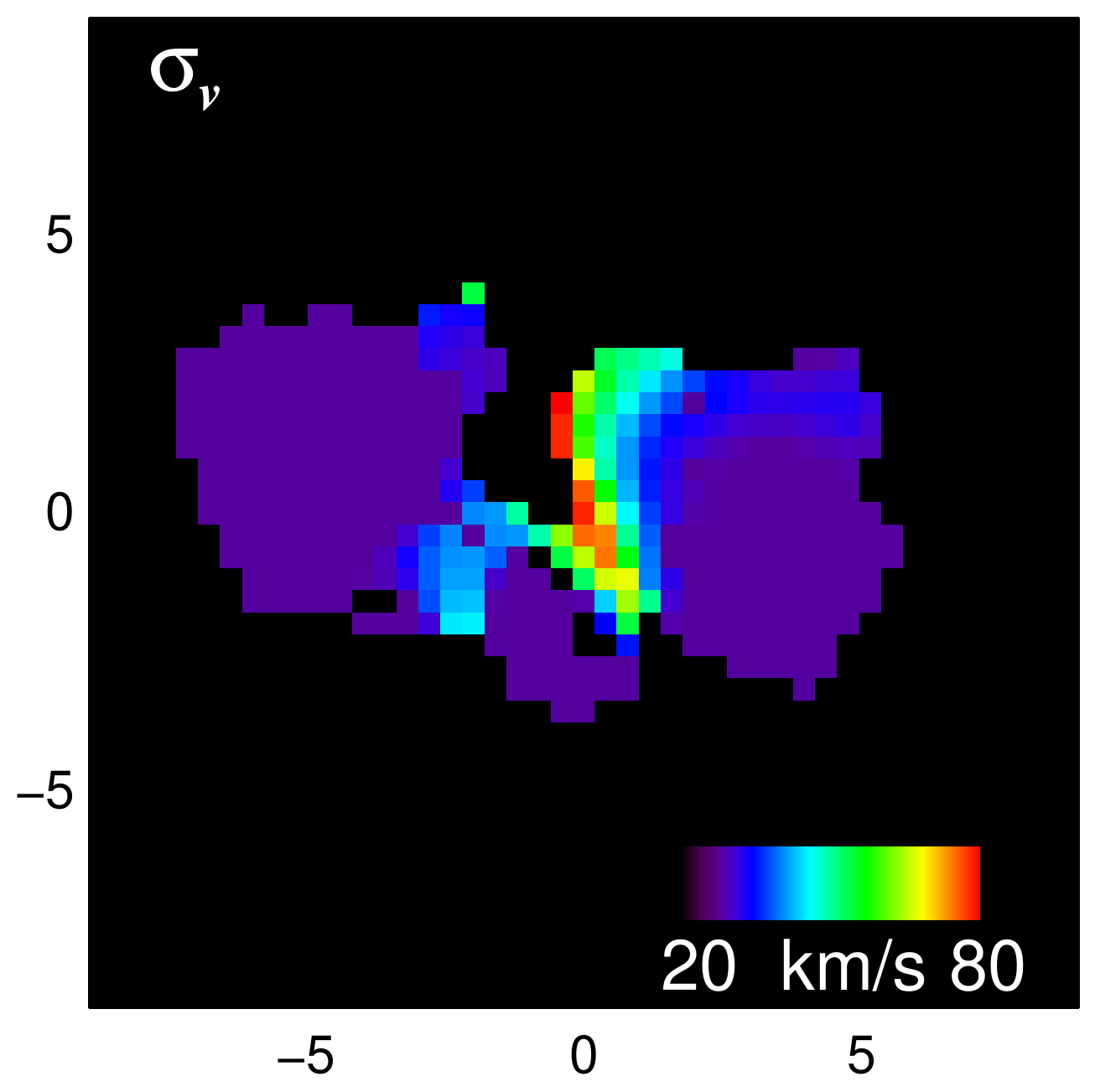}
\includegraphics[width=0.32\columnwidth]{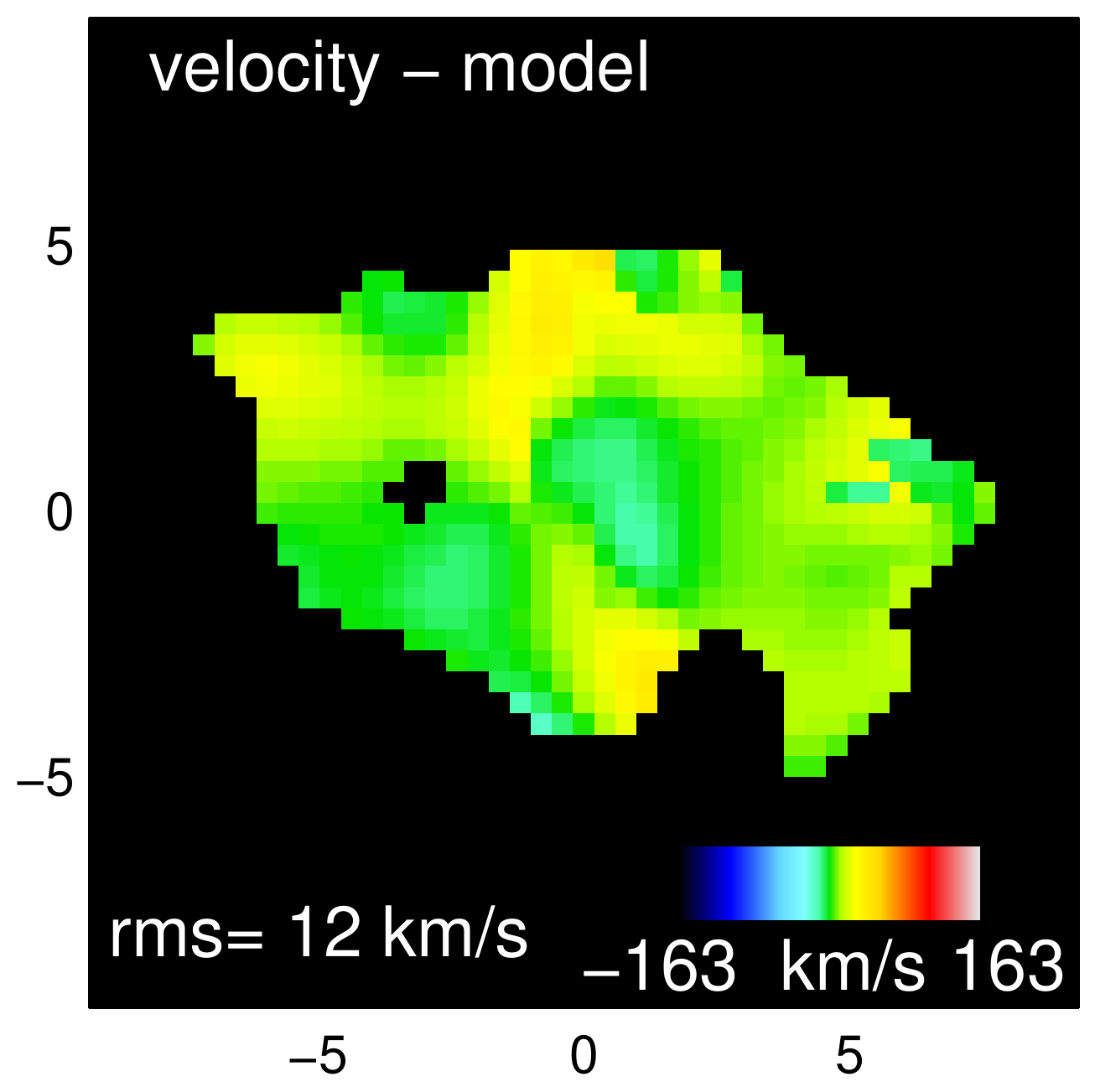}
\includegraphics[width=0.345\columnwidth]{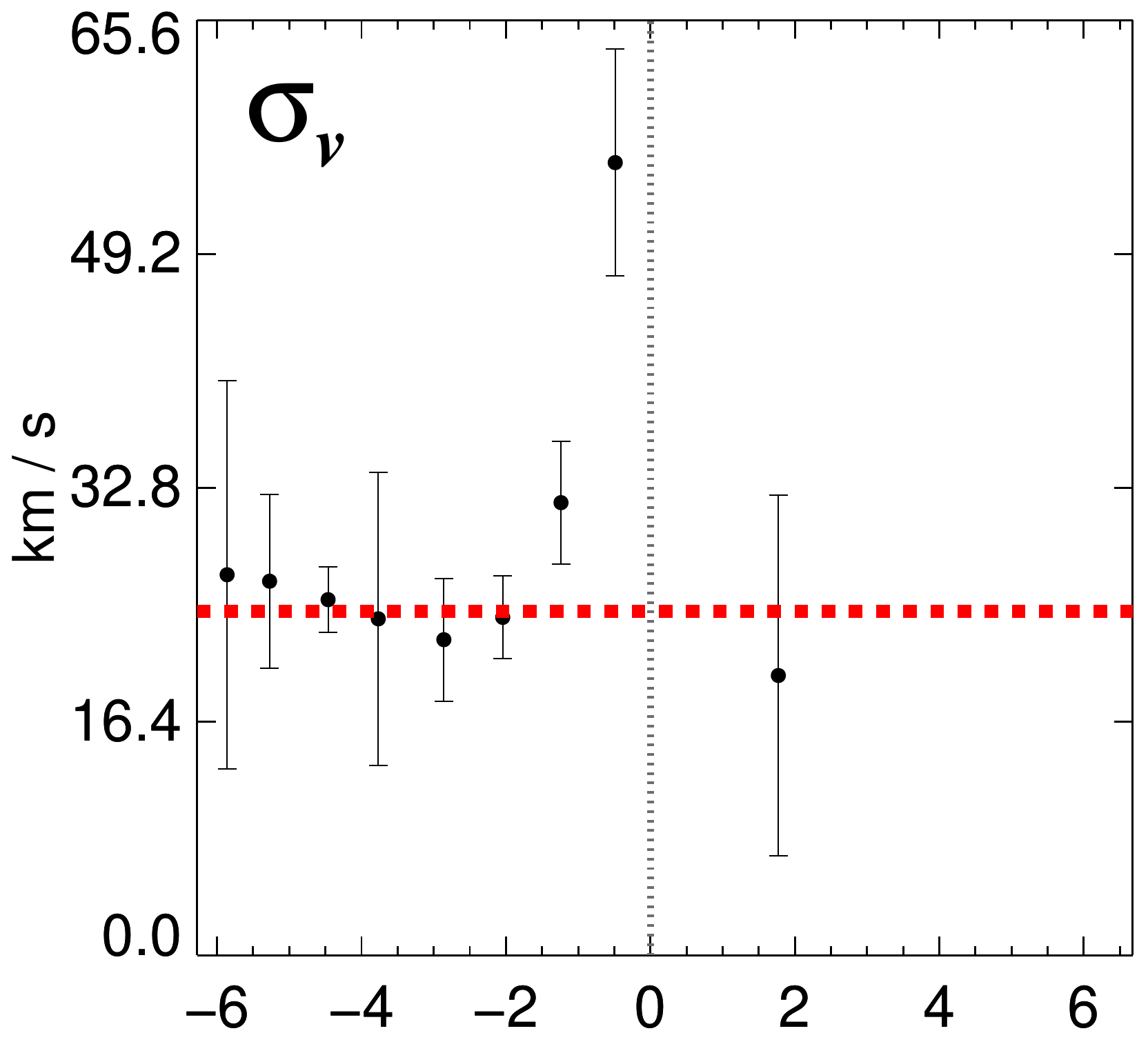}
\includegraphics[width=0.361\columnwidth]{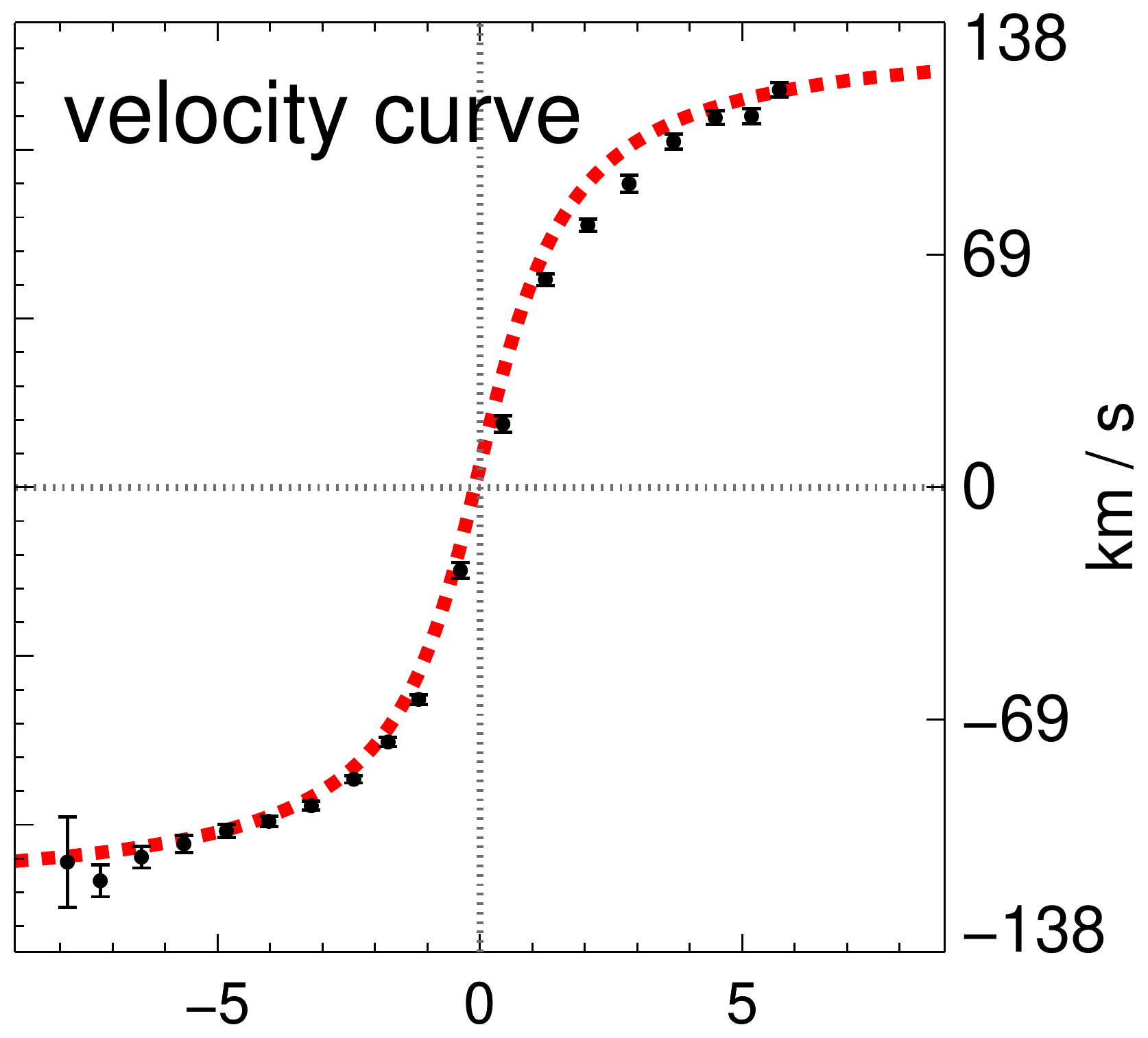}\\
\vspace{1mm}
\includegraphics[width=0.343\columnwidth]{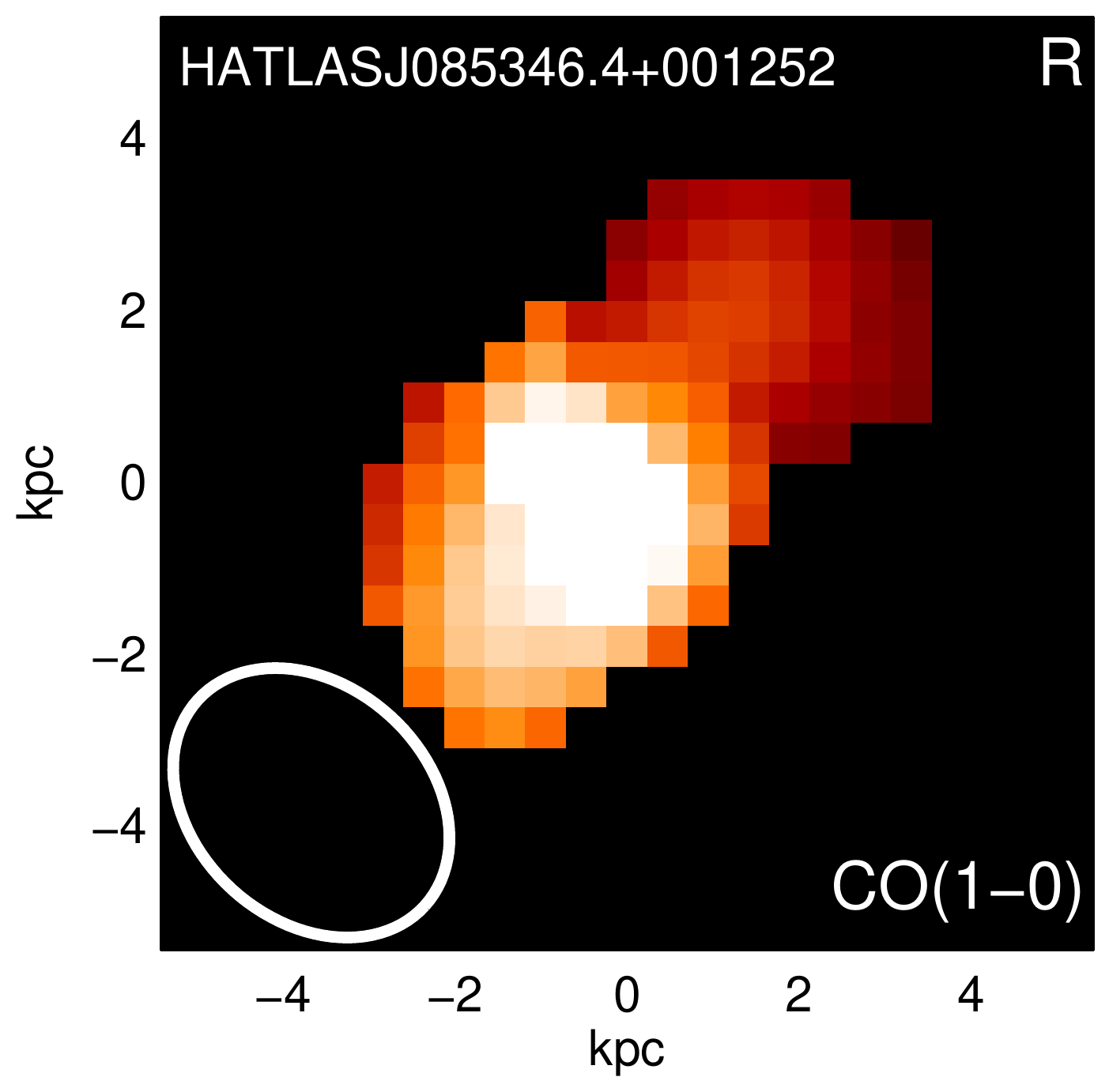}
\includegraphics[width=0.32\columnwidth]{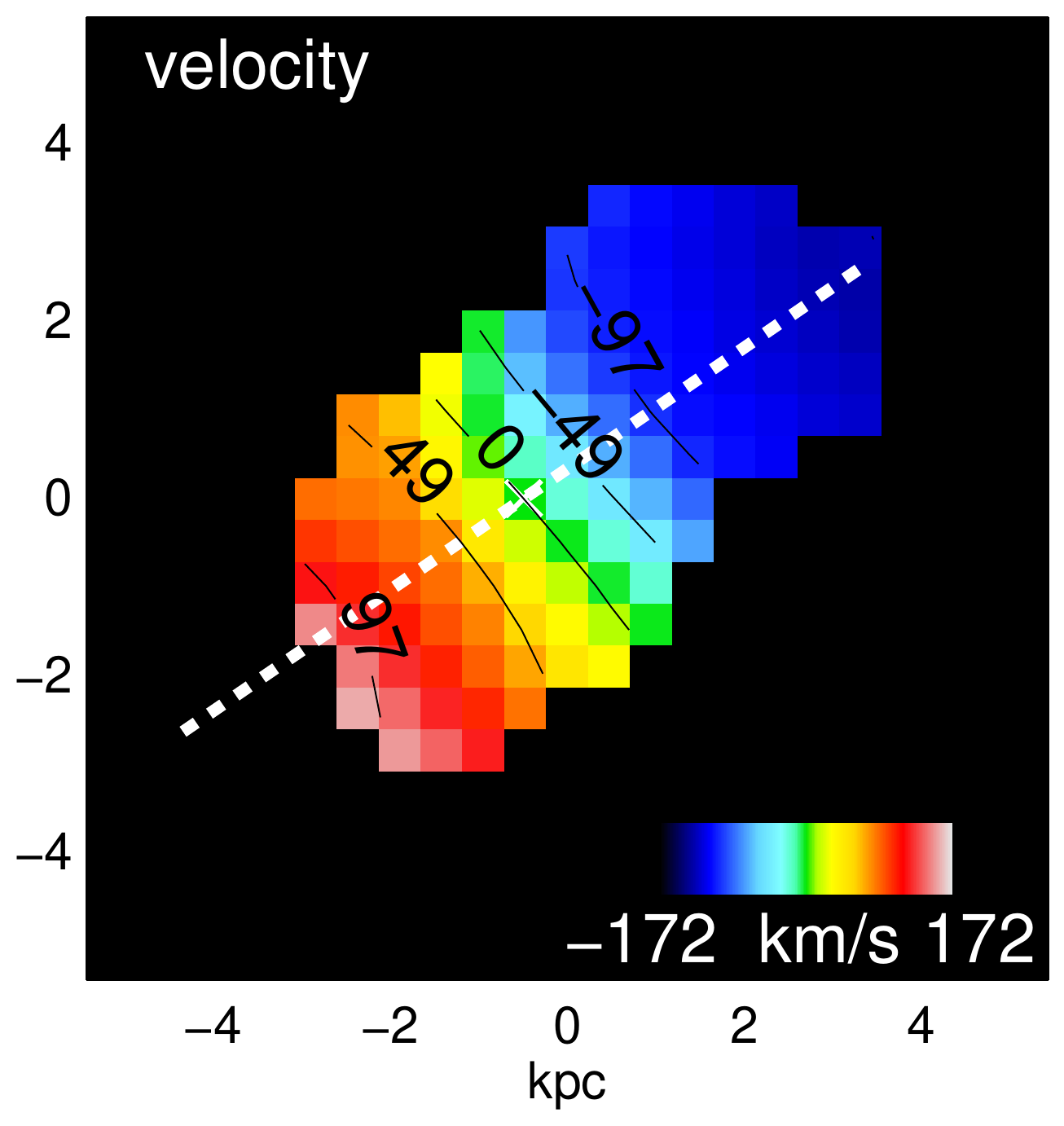}
\includegraphics[width=0.32\columnwidth]{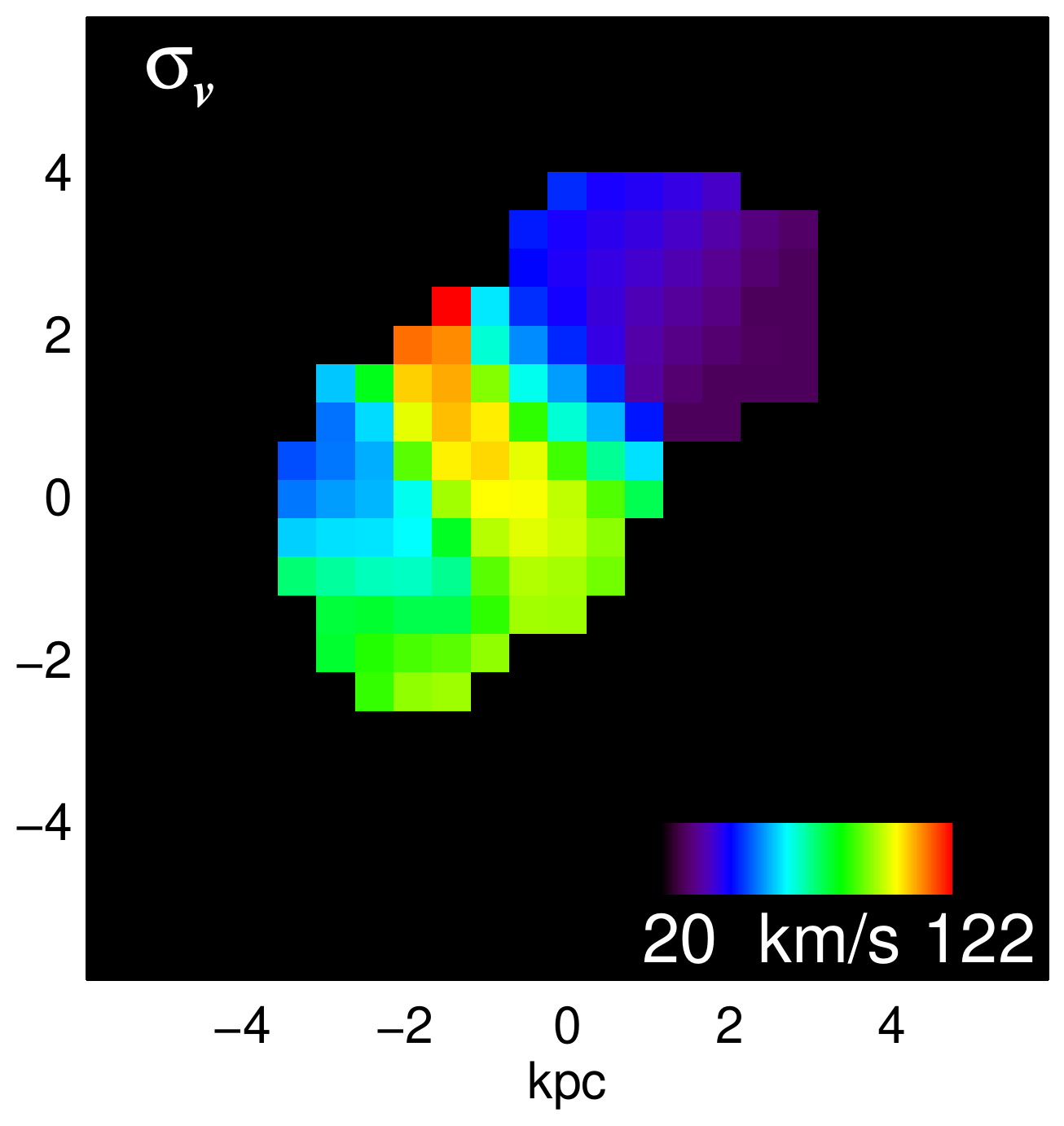}
\includegraphics[width=0.32\columnwidth]{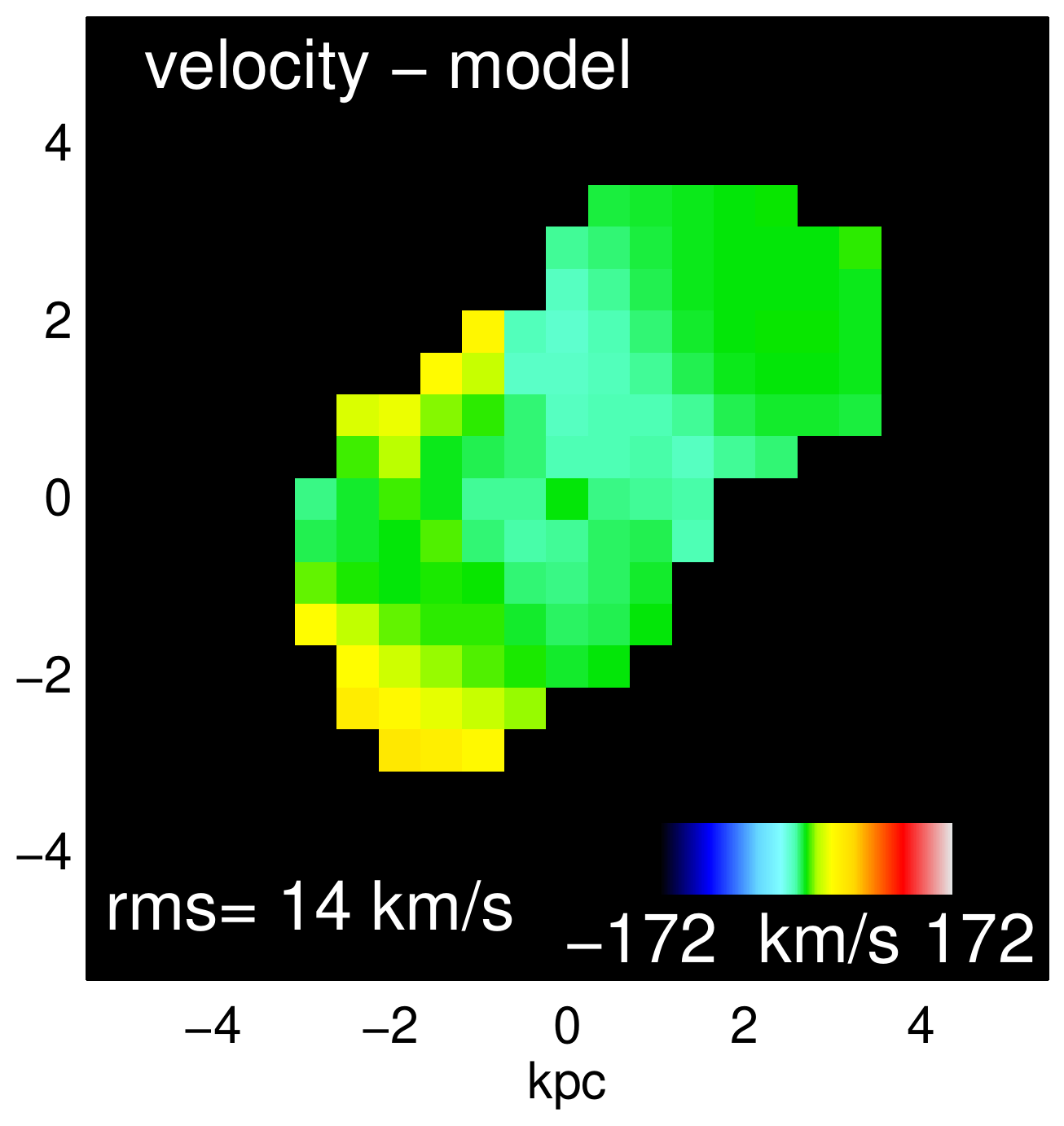}
\includegraphics[width=0.345\columnwidth]{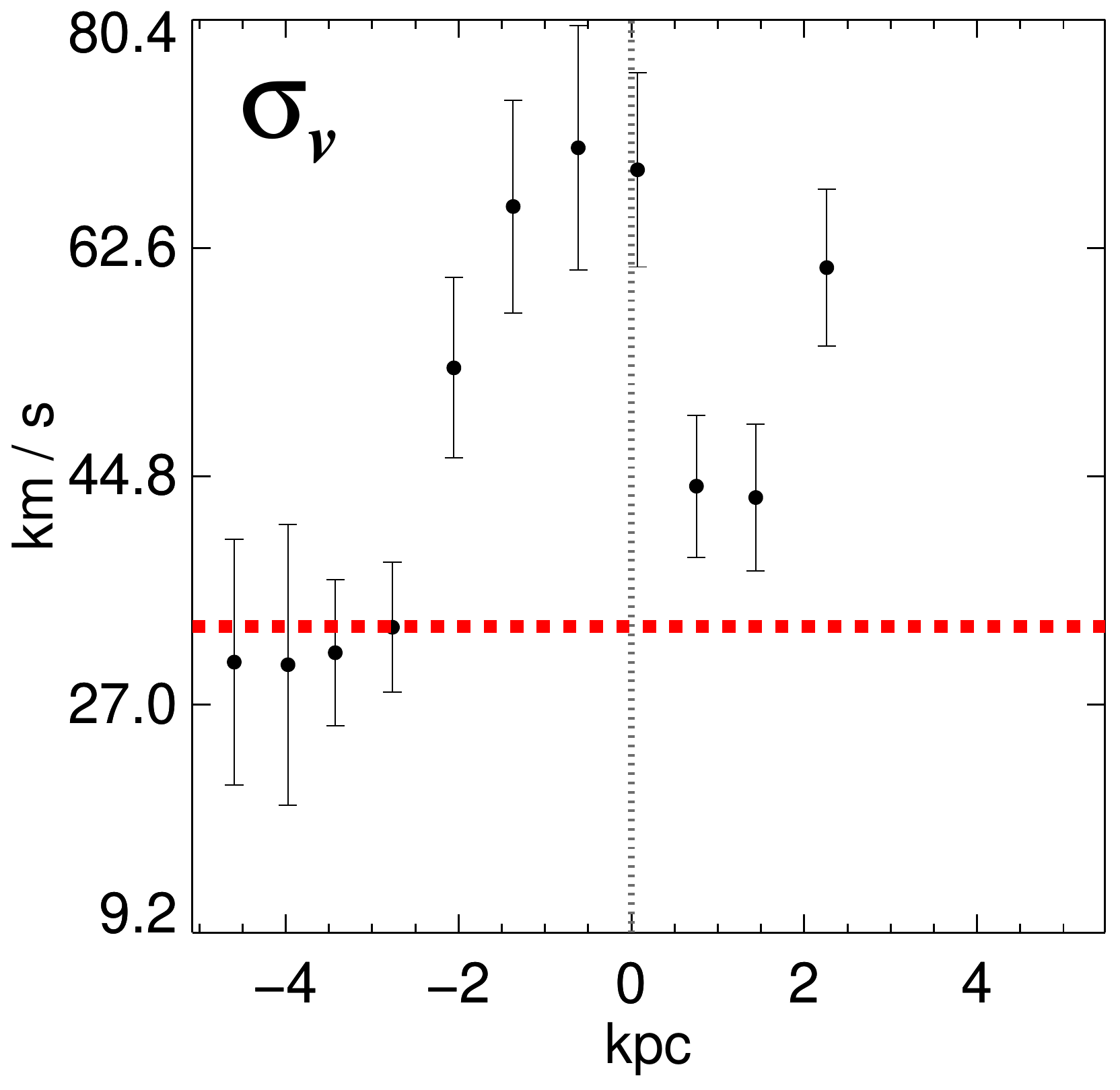}
\includegraphics[width=0.351\columnwidth]{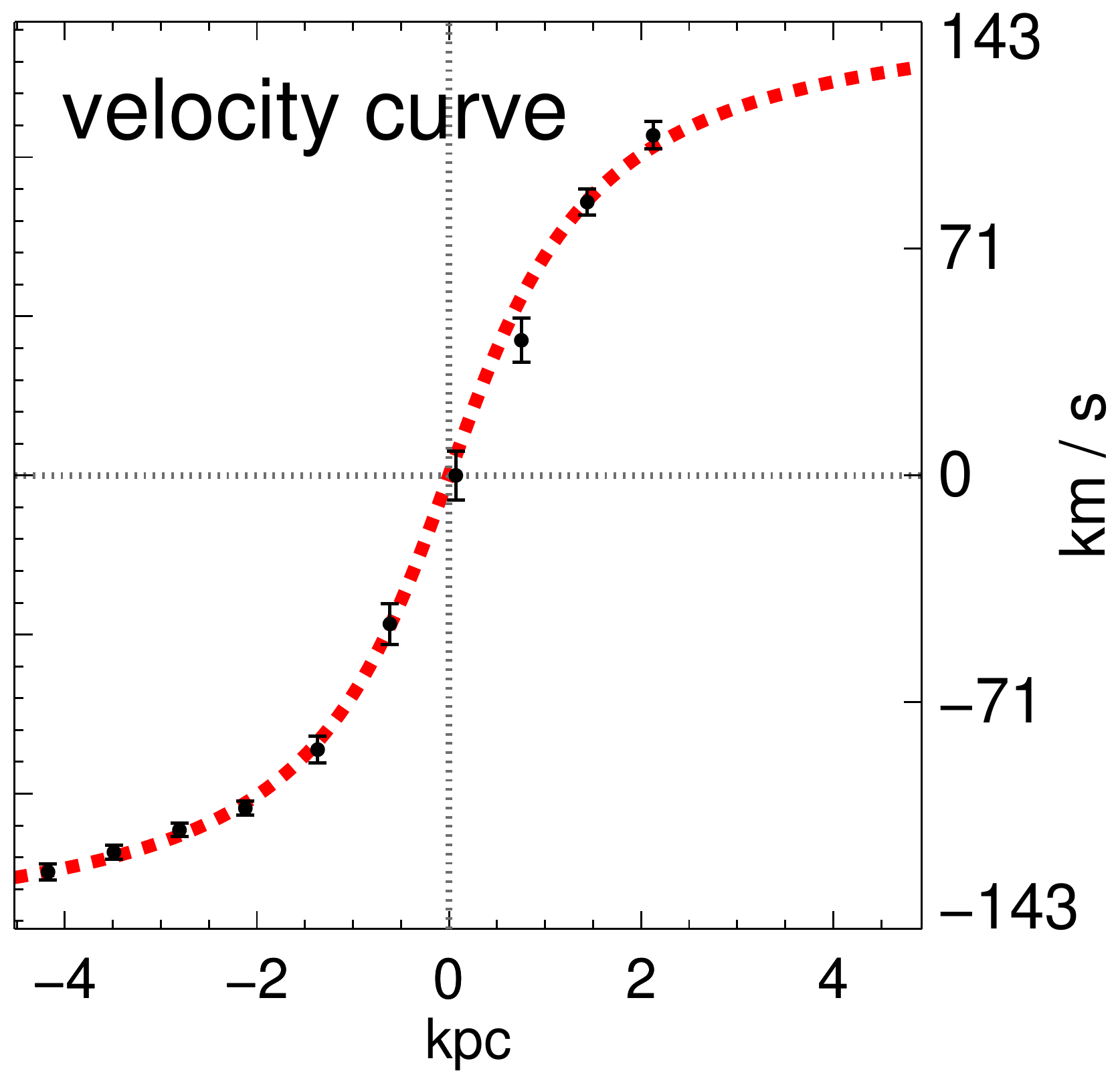}\\
\centering{\textbf{Figure C1. Continued.}}
\end{figure*}

\begin{figure*}
\flushleft
\includegraphics[width=0.343\columnwidth]{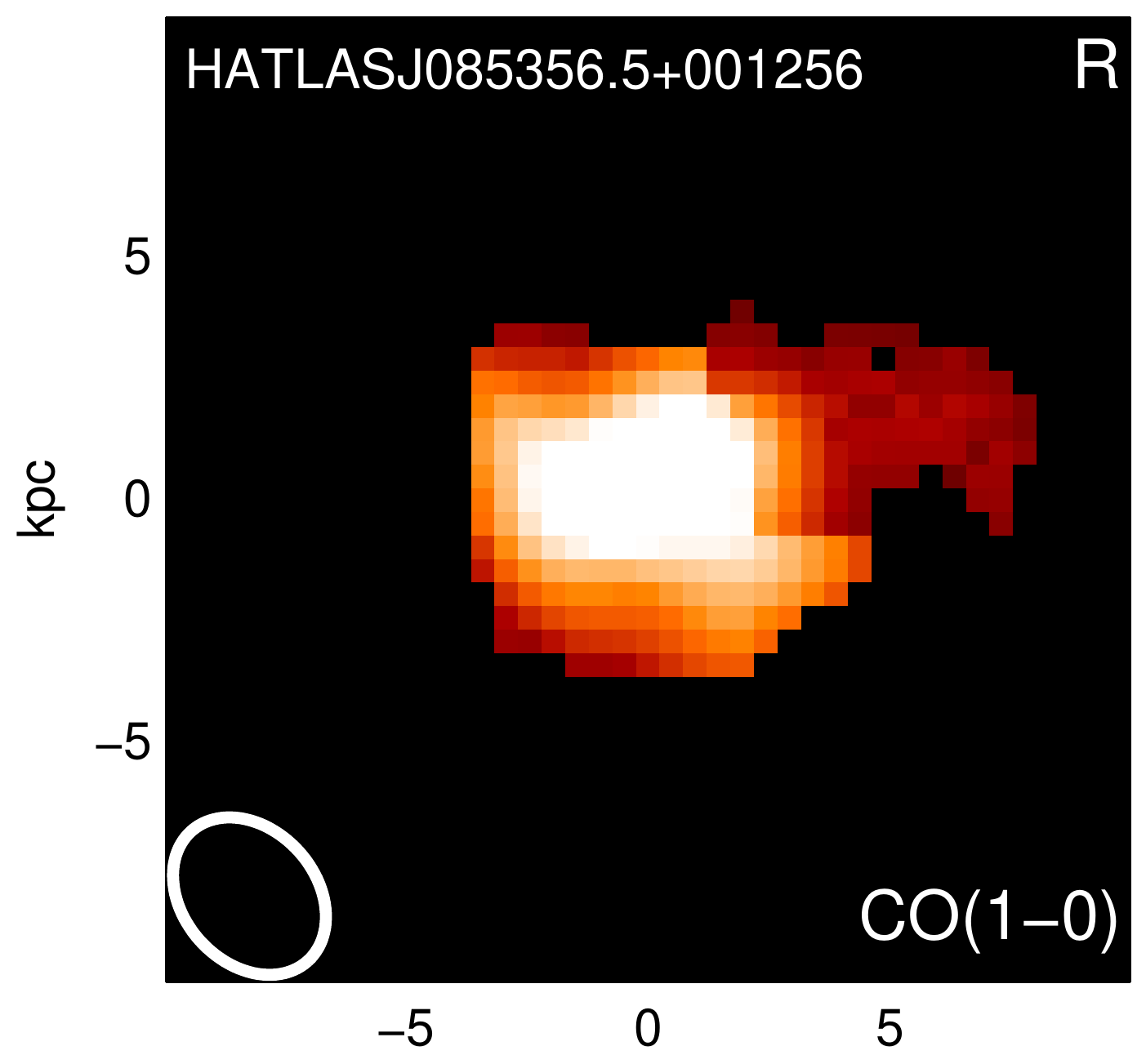}
\includegraphics[width=0.32\columnwidth]{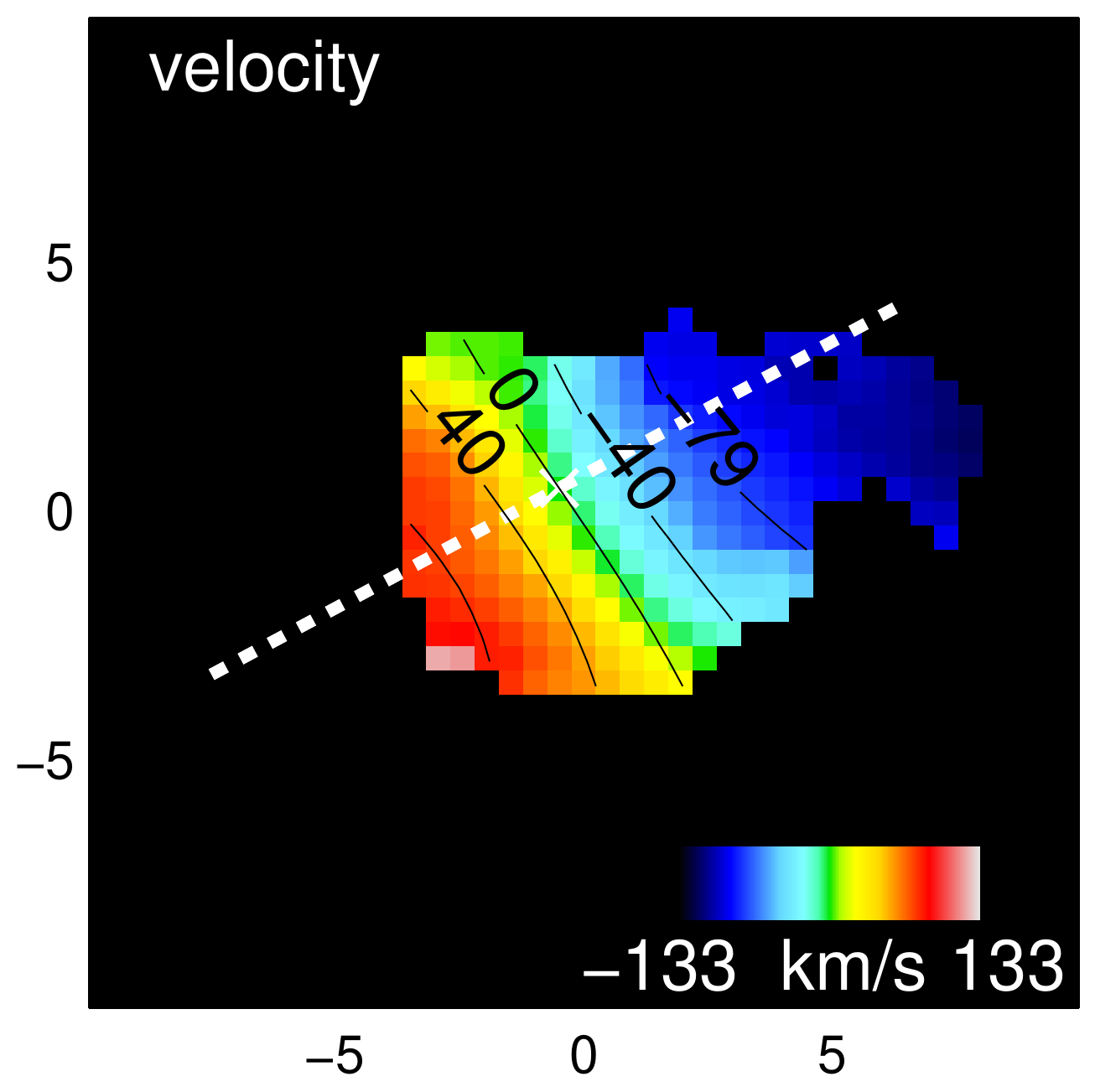}
\includegraphics[width=0.32\columnwidth]{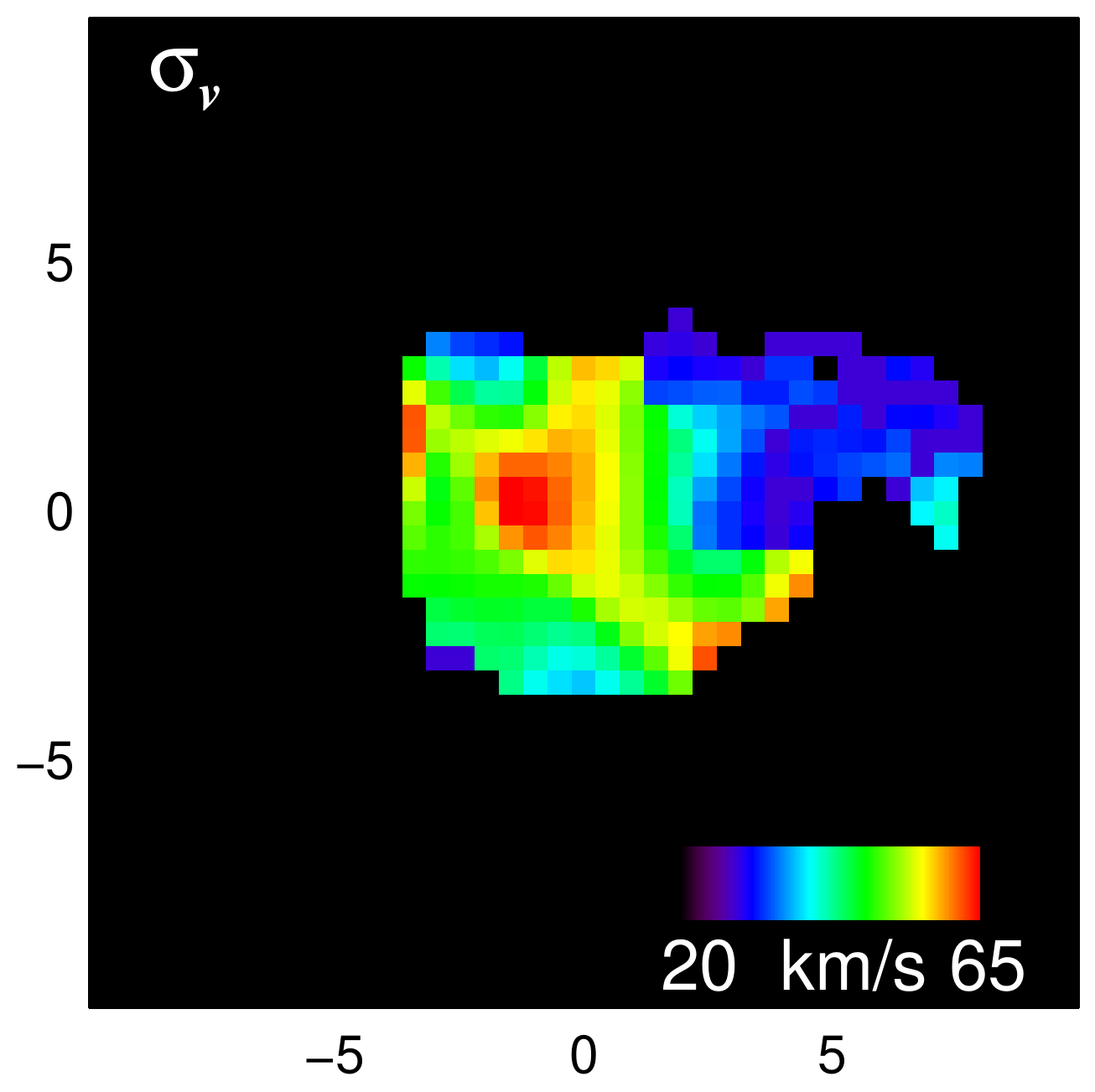}
\includegraphics[width=0.32\columnwidth]{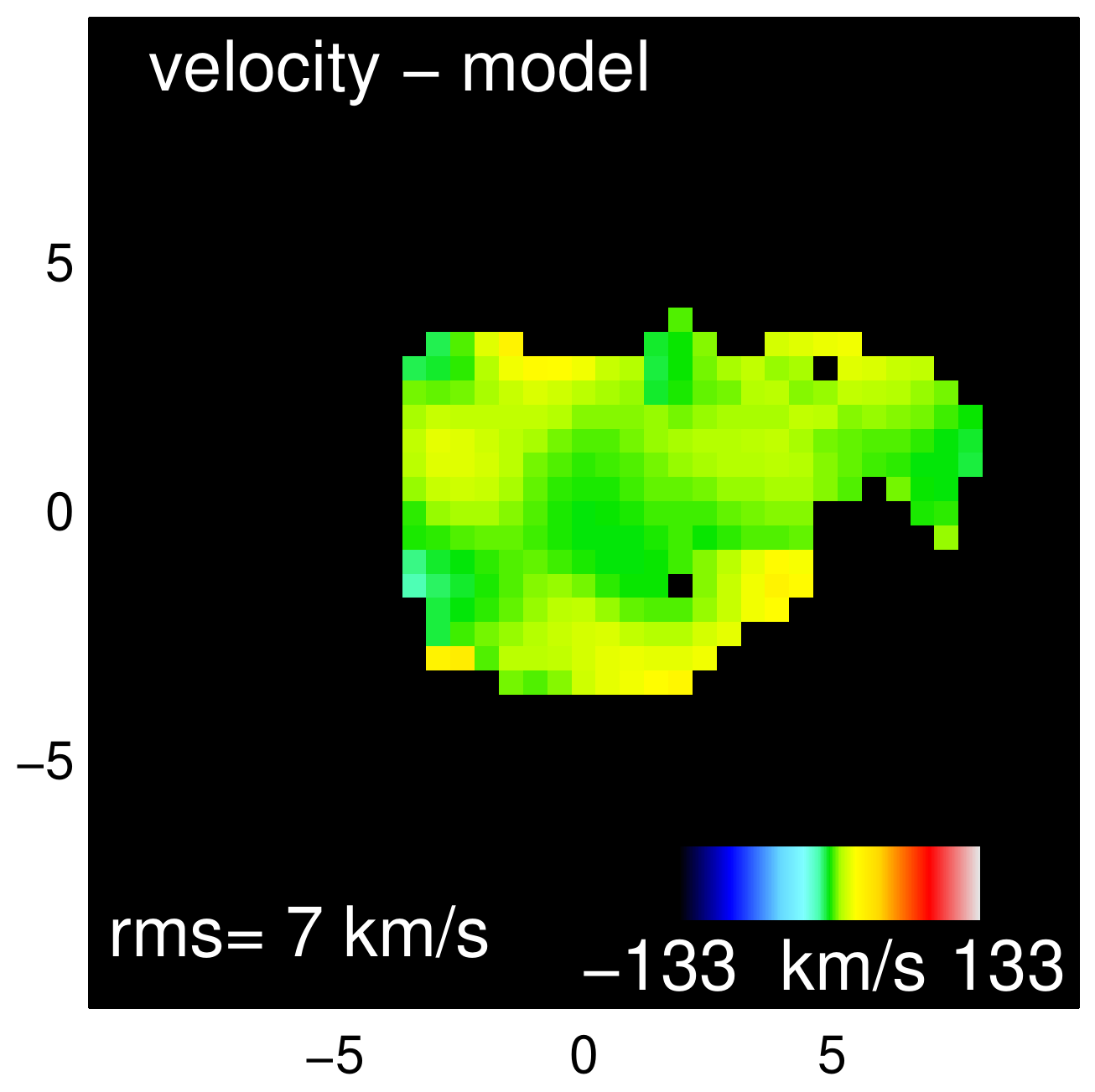}
\includegraphics[width=0.345\columnwidth]{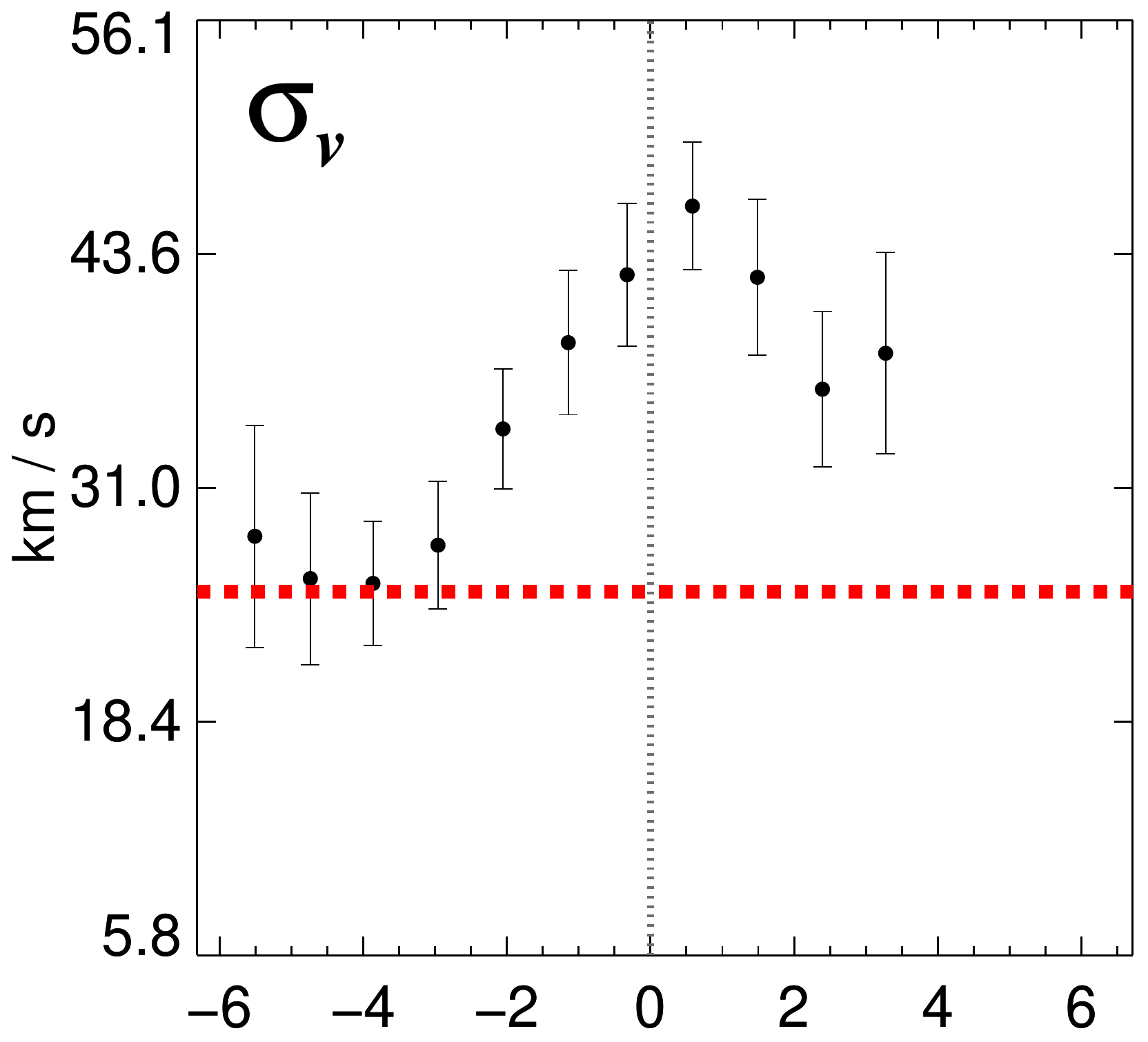}
\includegraphics[width=0.351\columnwidth]{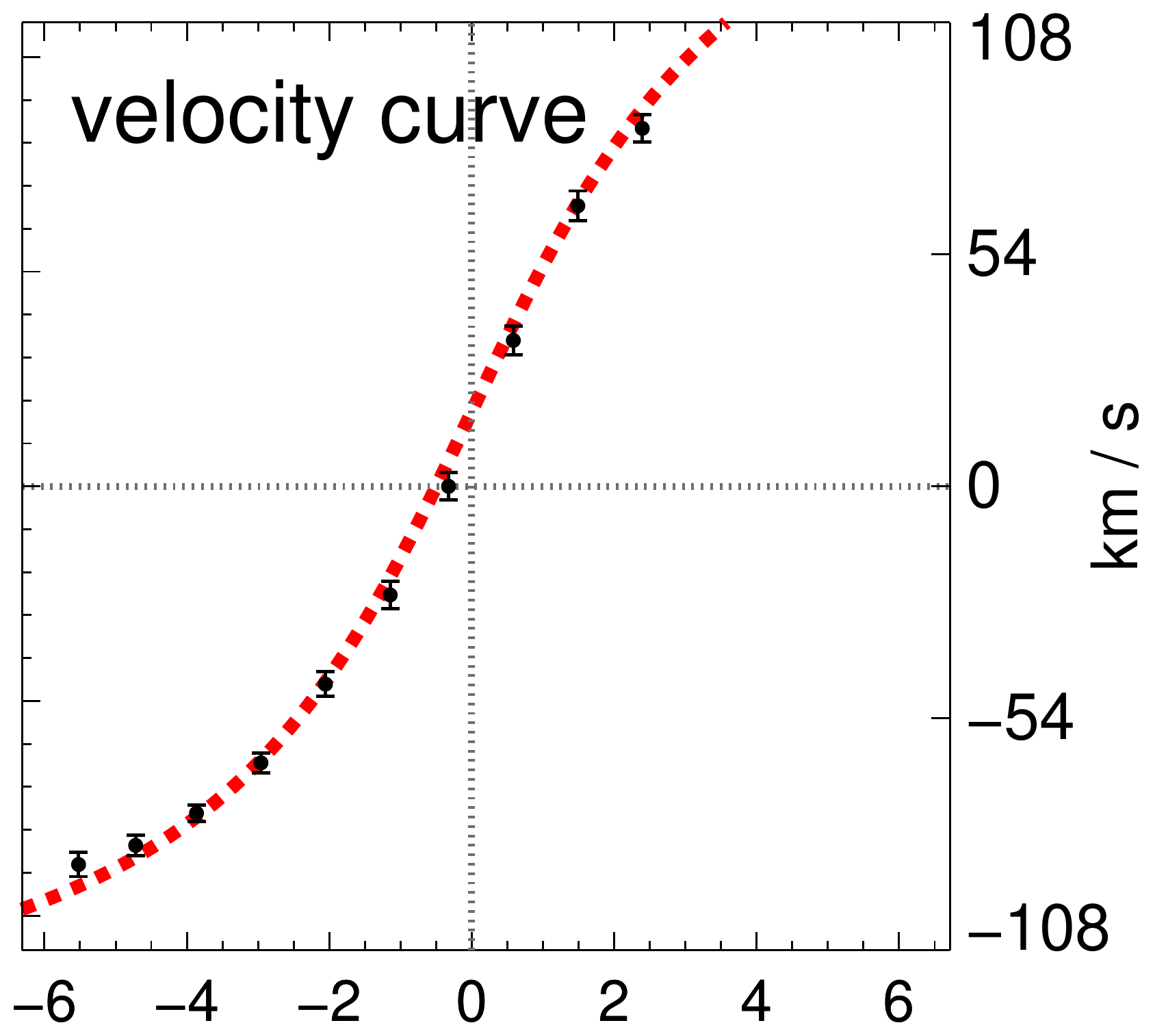}\\
\vspace{1mm}
\includegraphics[width=0.343\columnwidth]{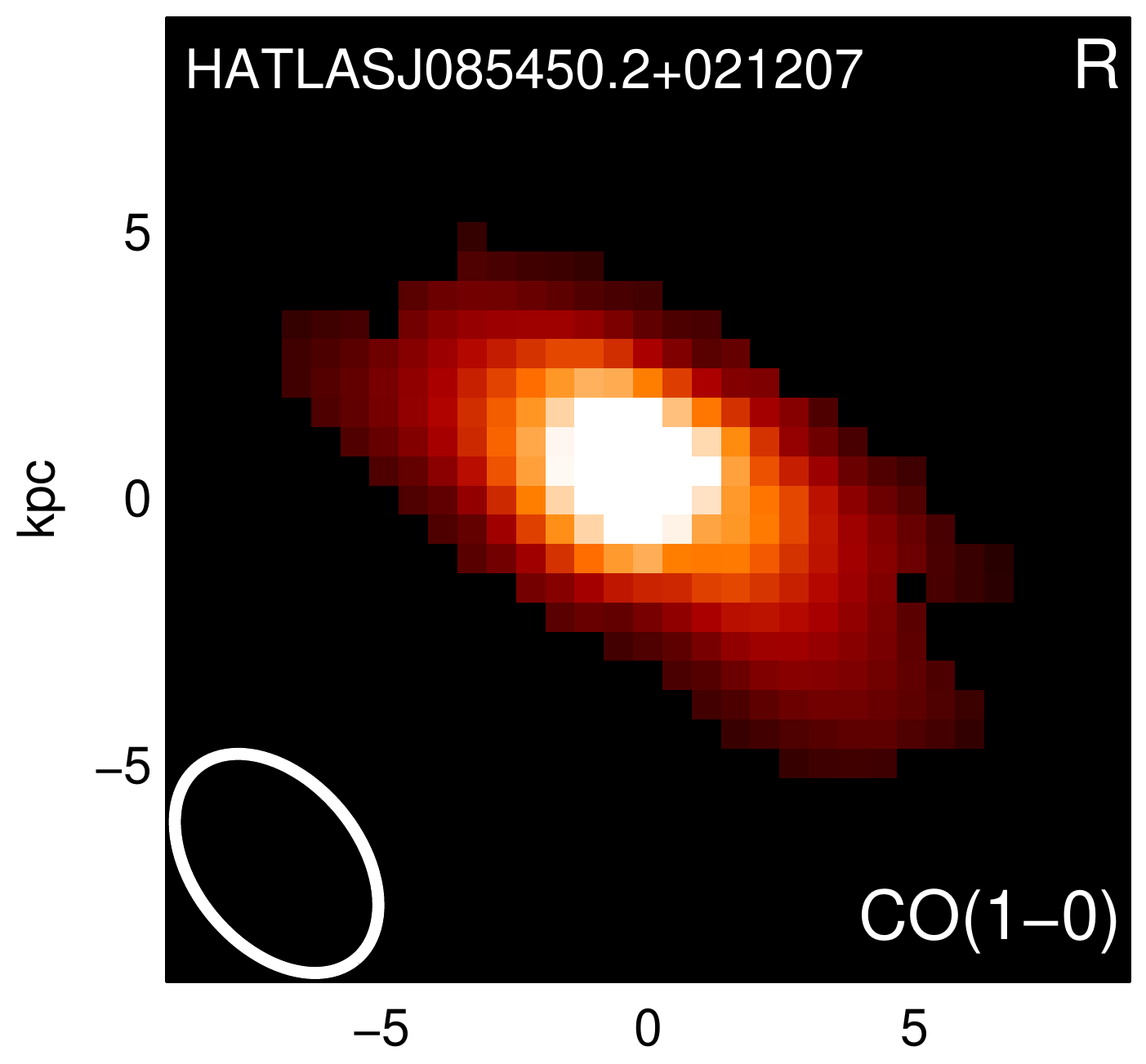}
\includegraphics[width=0.32\columnwidth]{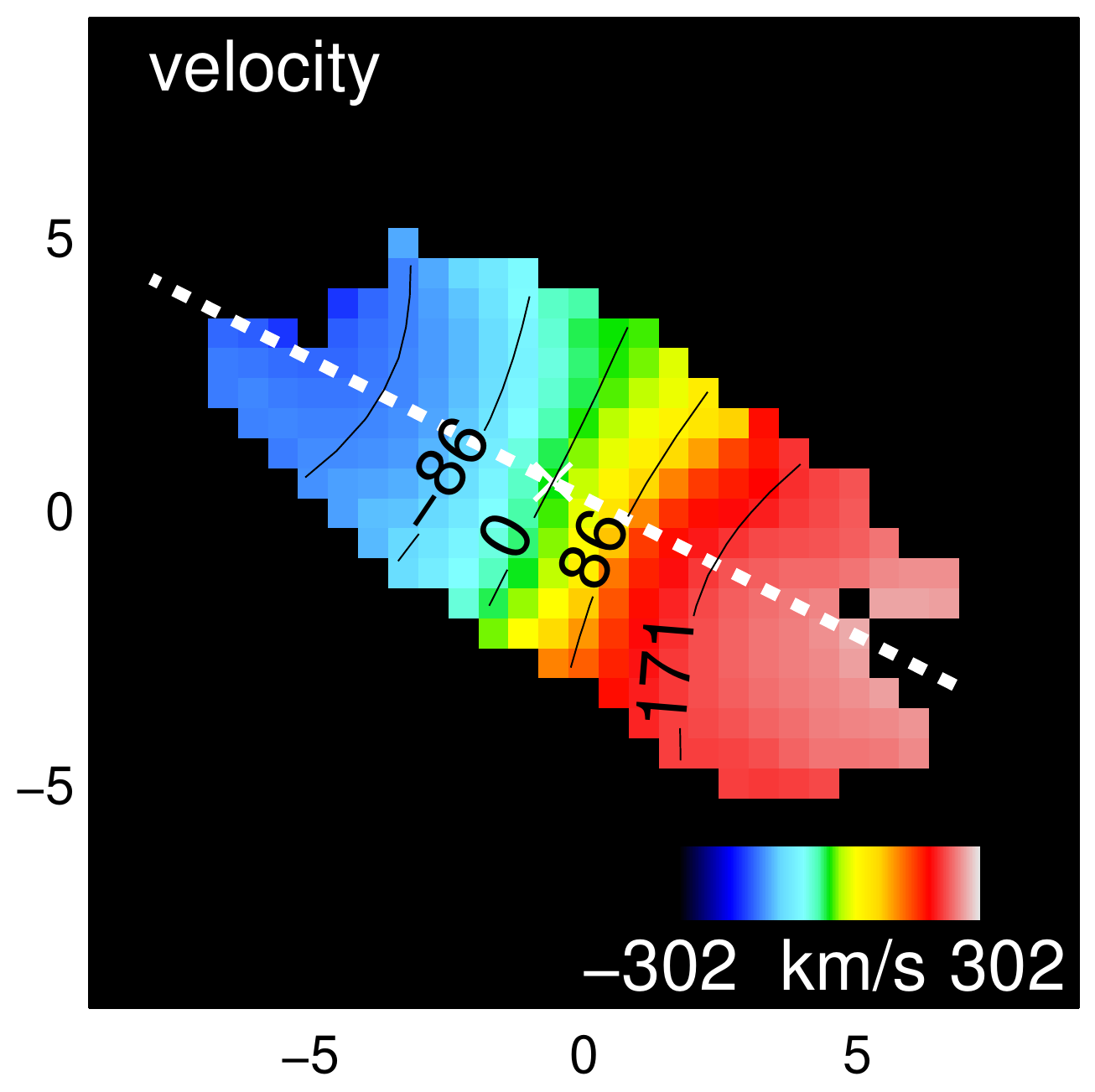}
\includegraphics[width=0.32\columnwidth]{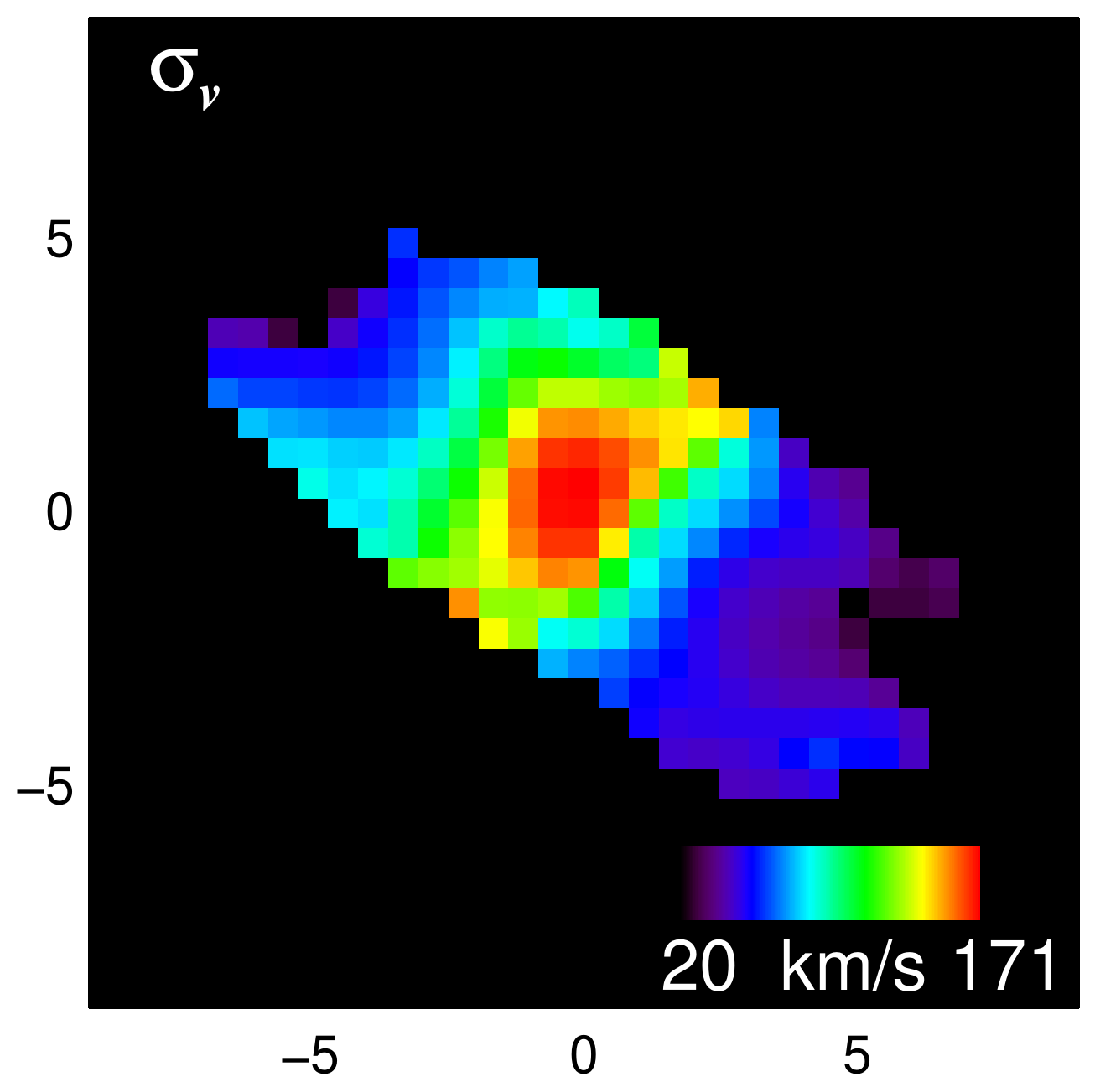}
\includegraphics[width=0.32\columnwidth]{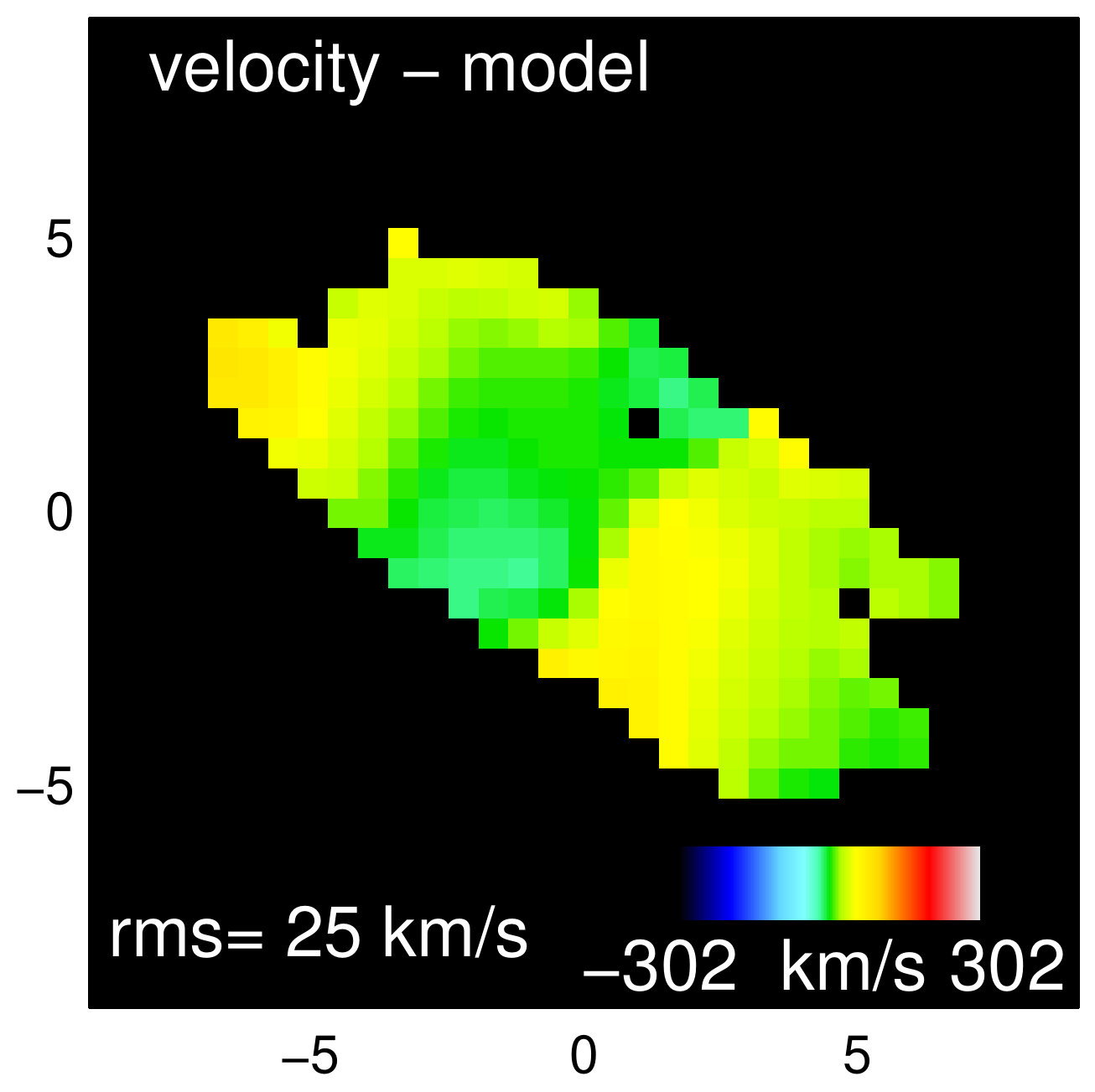}
\includegraphics[width=0.355\columnwidth]{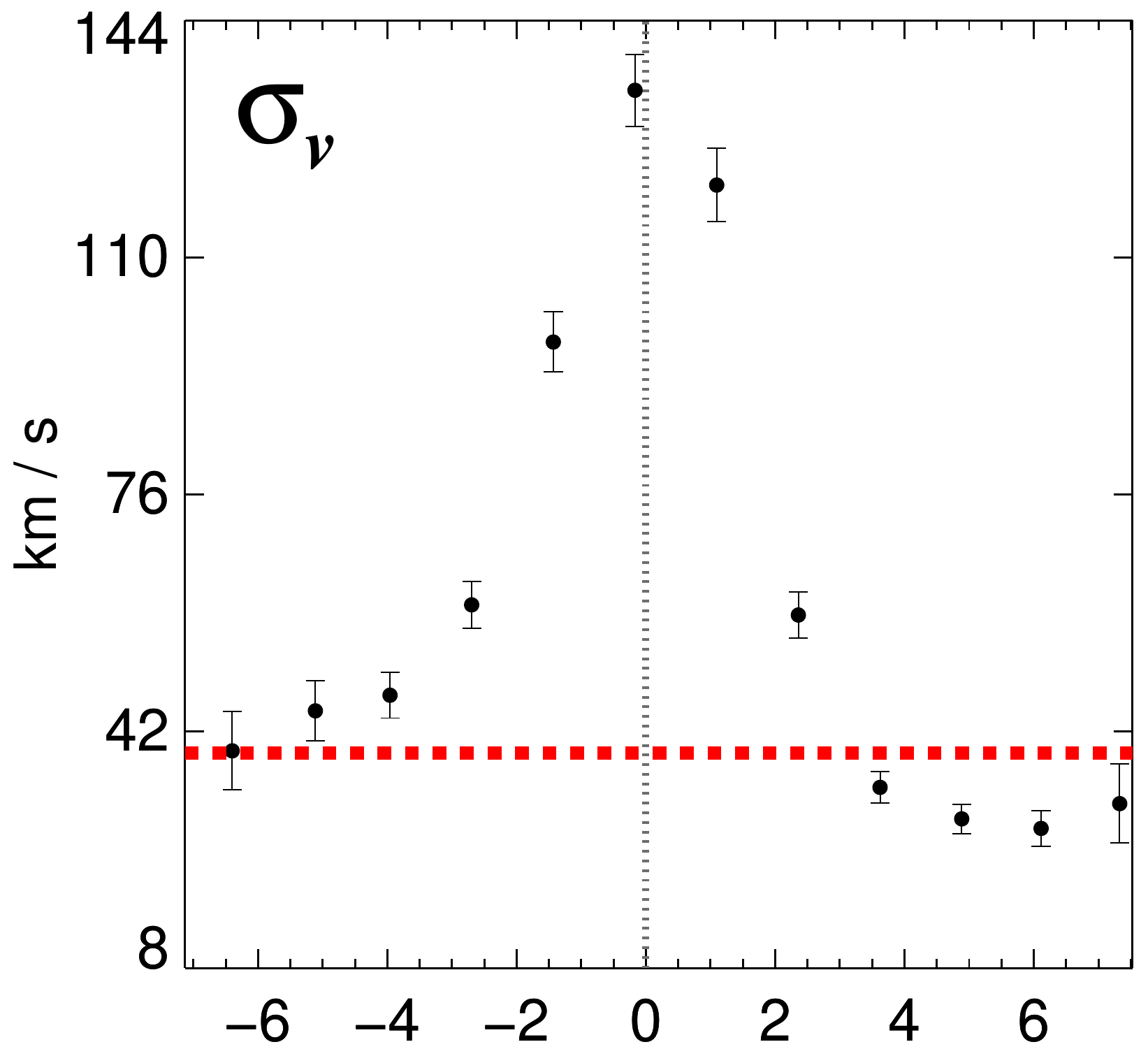}
\includegraphics[width=0.351\columnwidth]{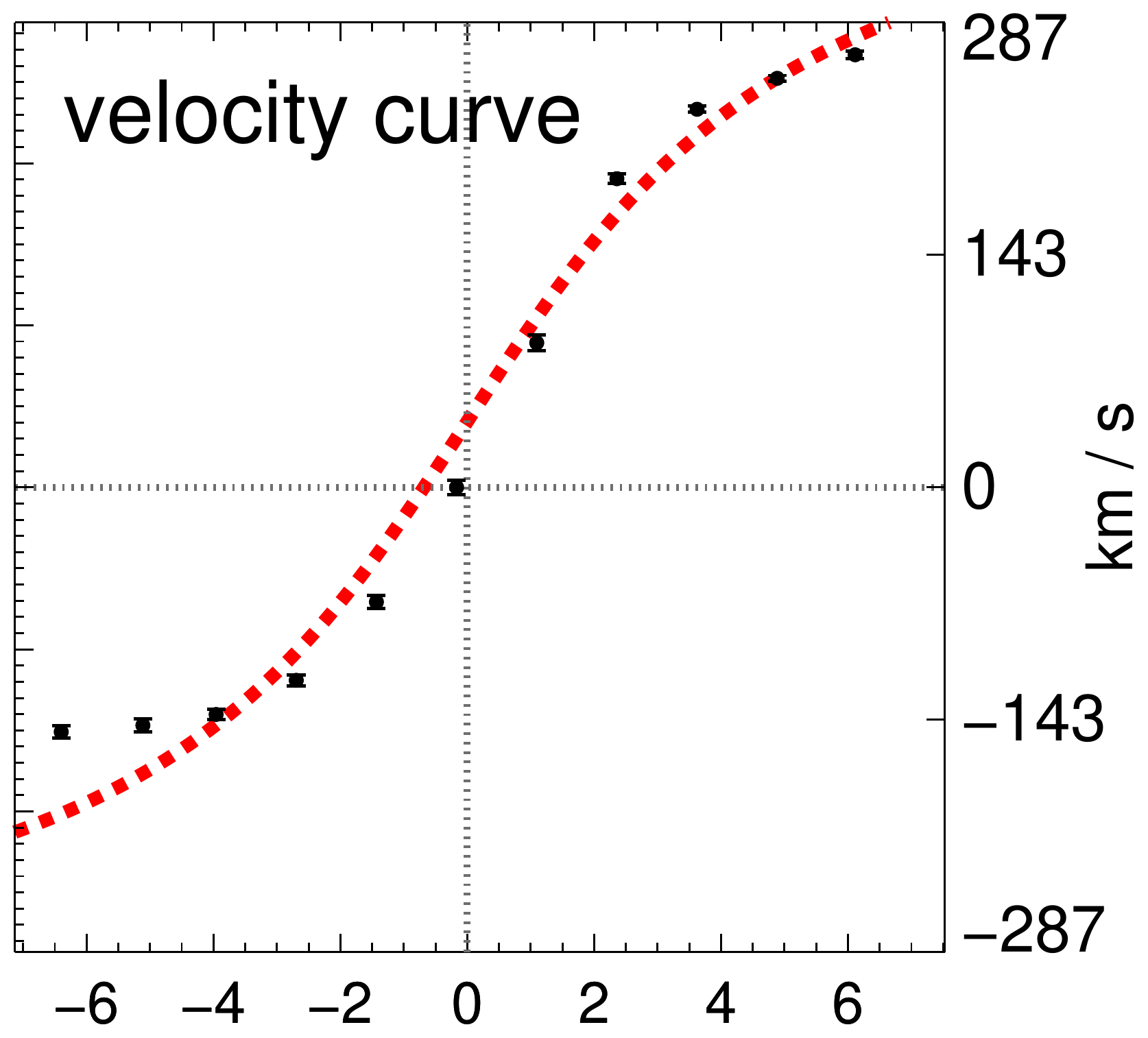}\\
\vspace{1mm}
\includegraphics[width=0.343\columnwidth]{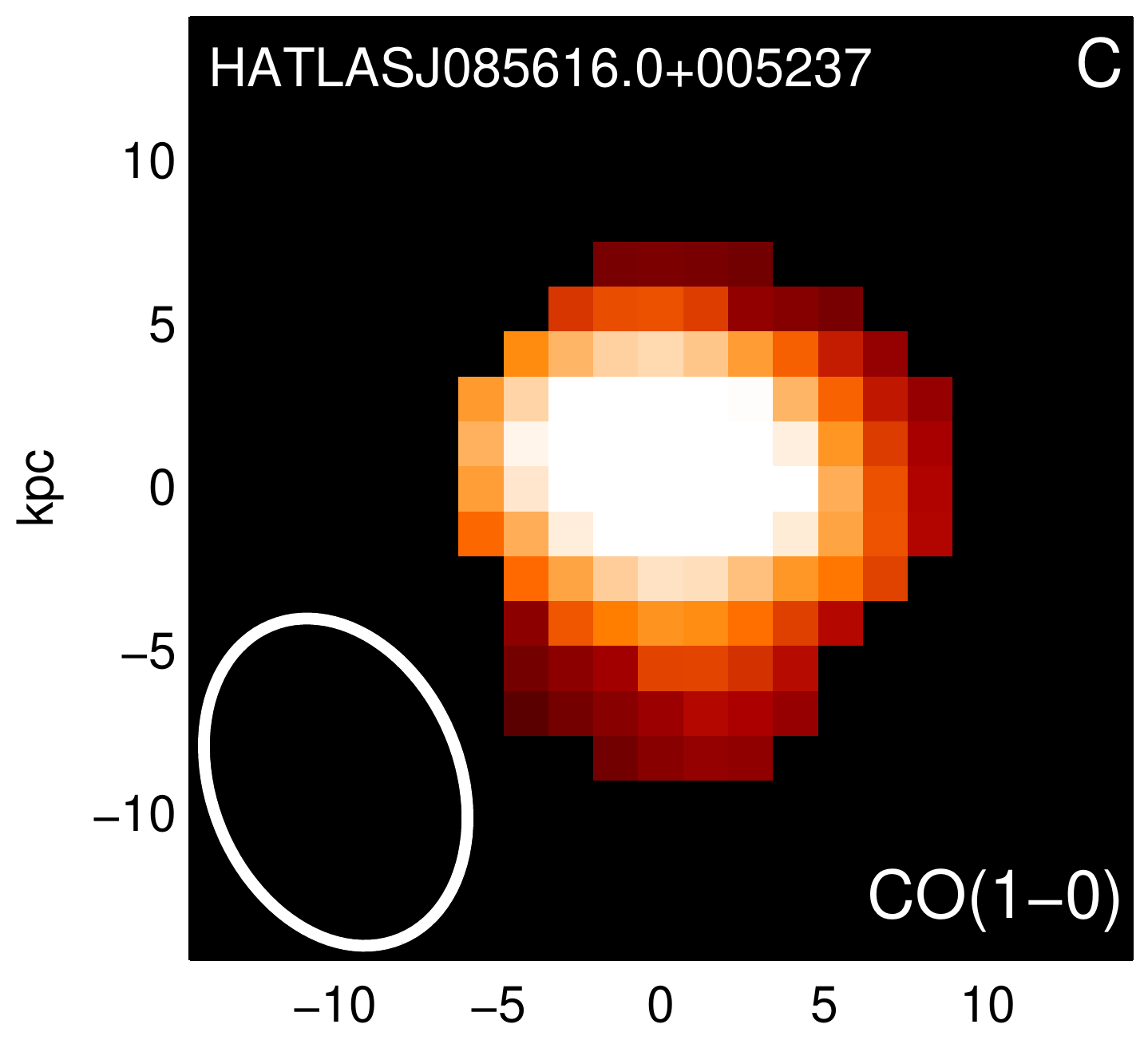}
\includegraphics[width=0.32\columnwidth]{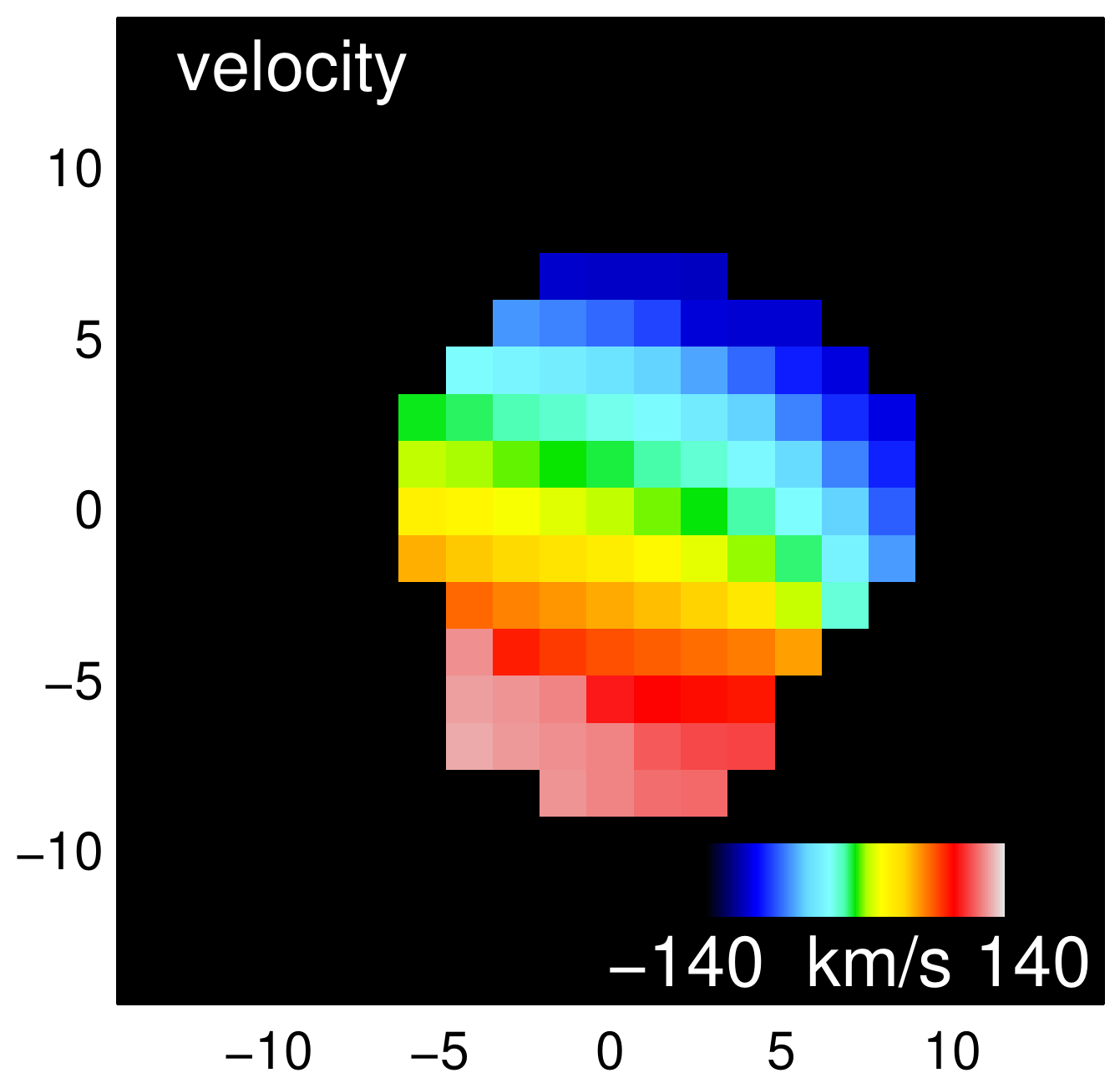}
\includegraphics[width=0.32\columnwidth]{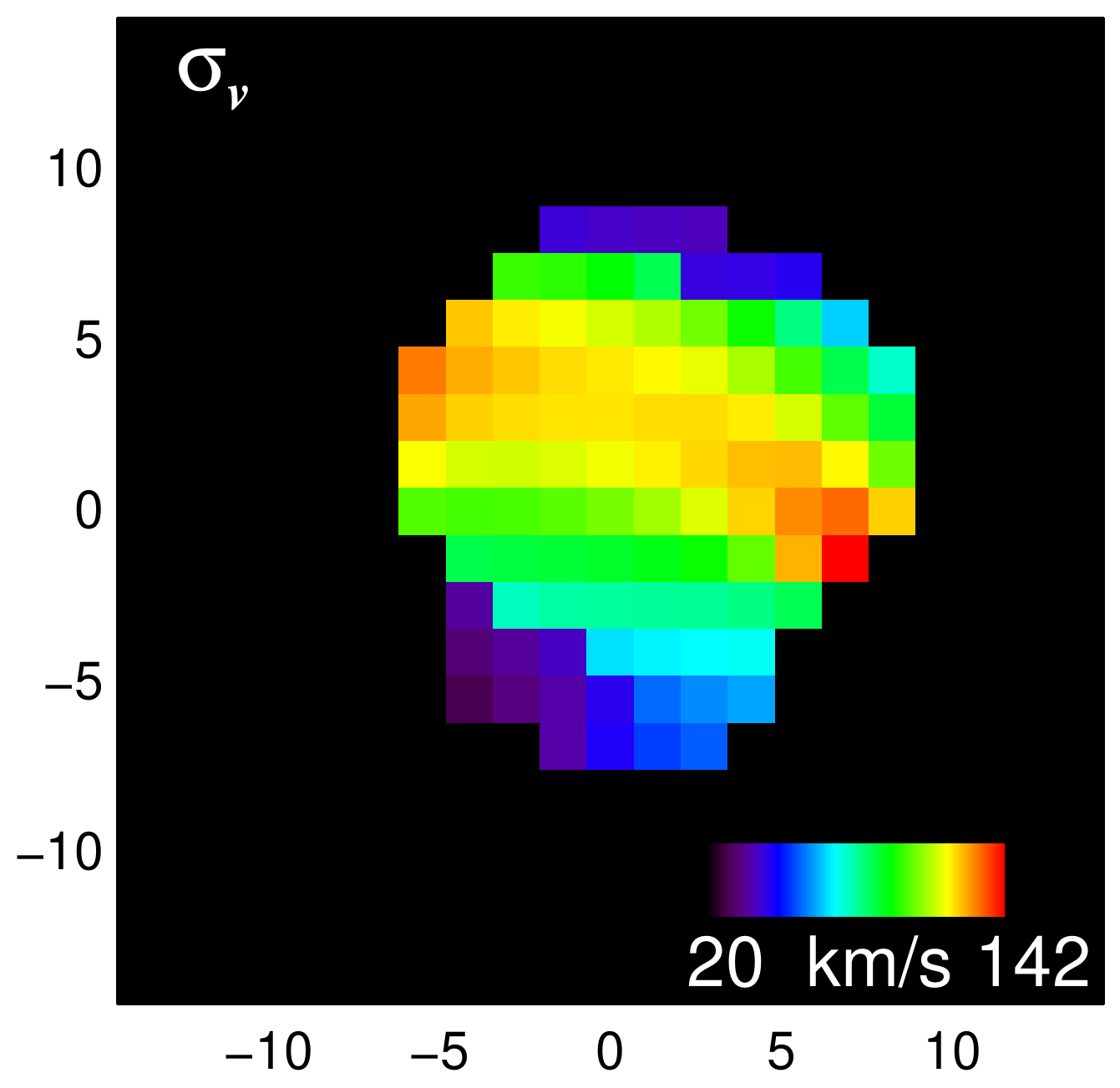}\\
\vspace{1mm}
\includegraphics[width=0.343\columnwidth]{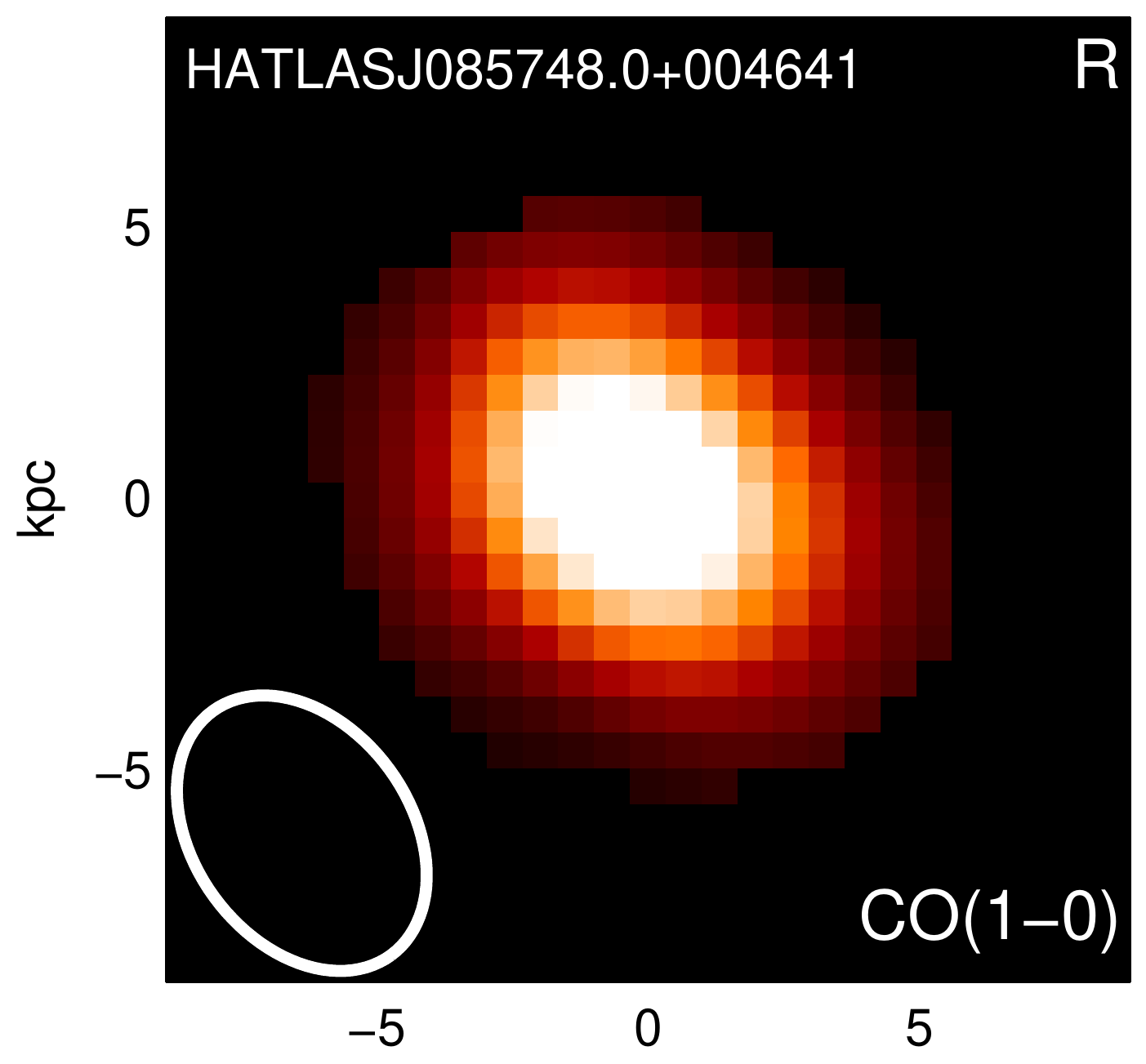}
\includegraphics[width=0.32\columnwidth]{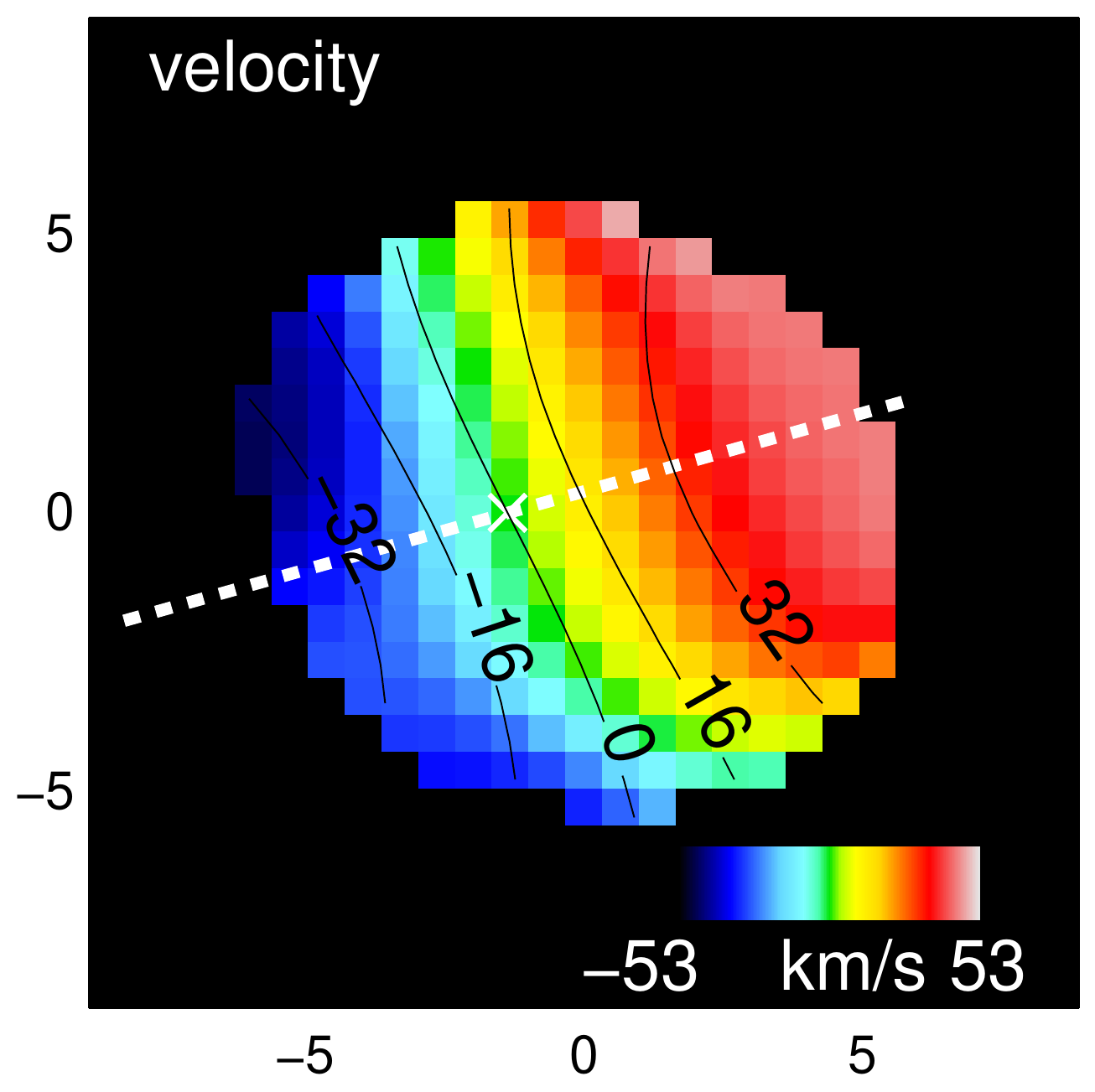}
\includegraphics[width=0.32\columnwidth]{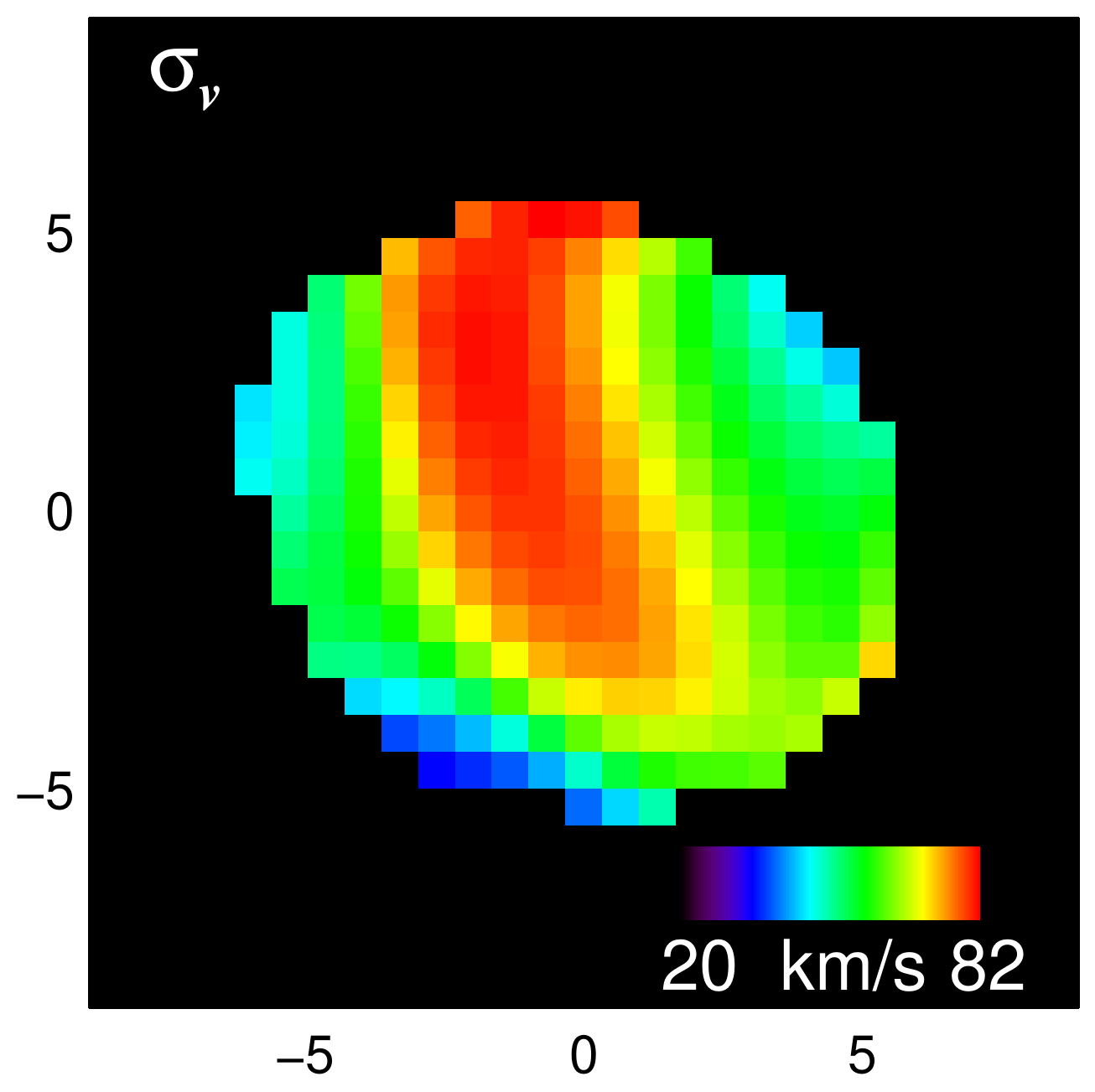}
\includegraphics[width=0.32\columnwidth]{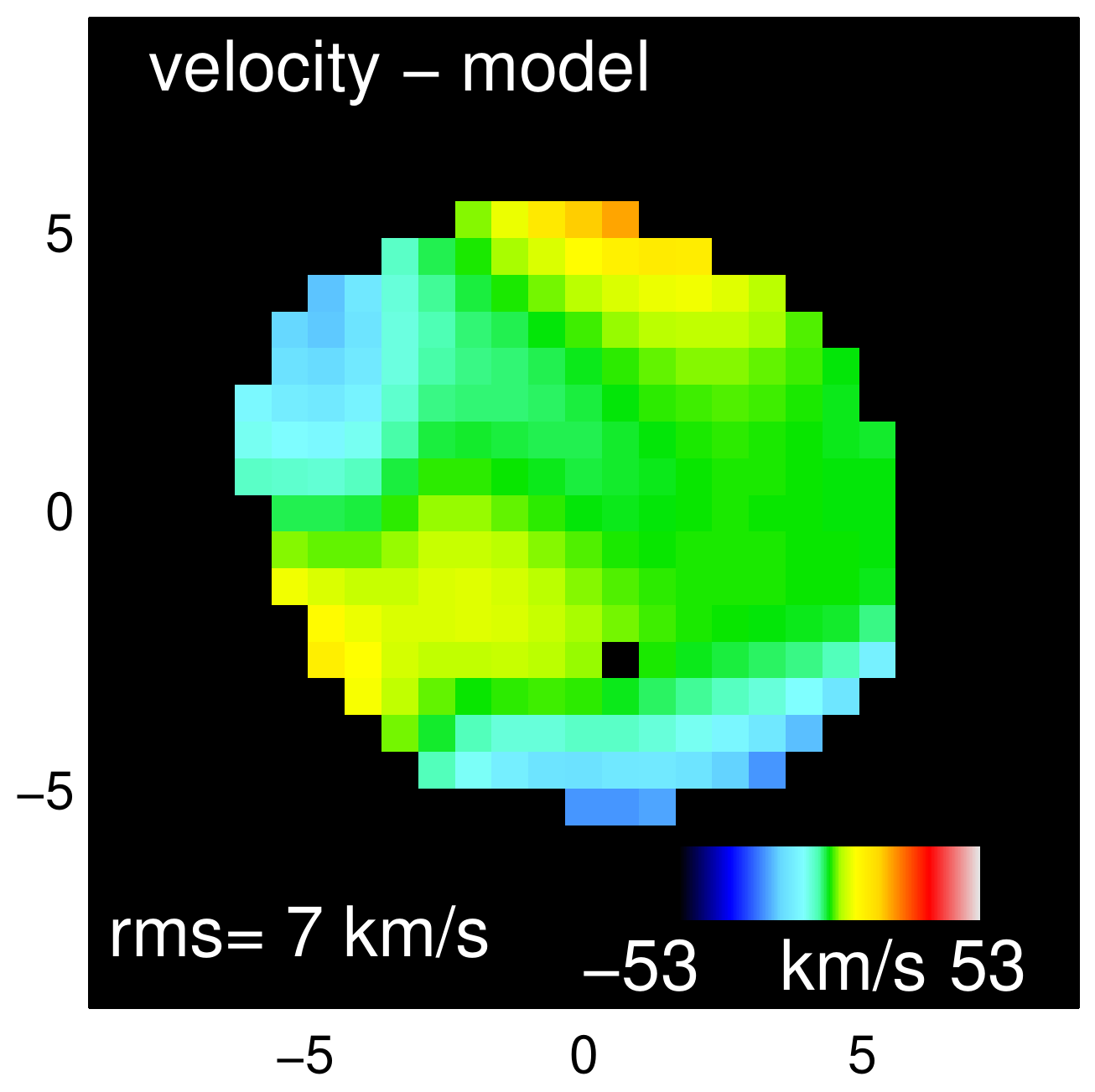}
\includegraphics[width=0.345\columnwidth]{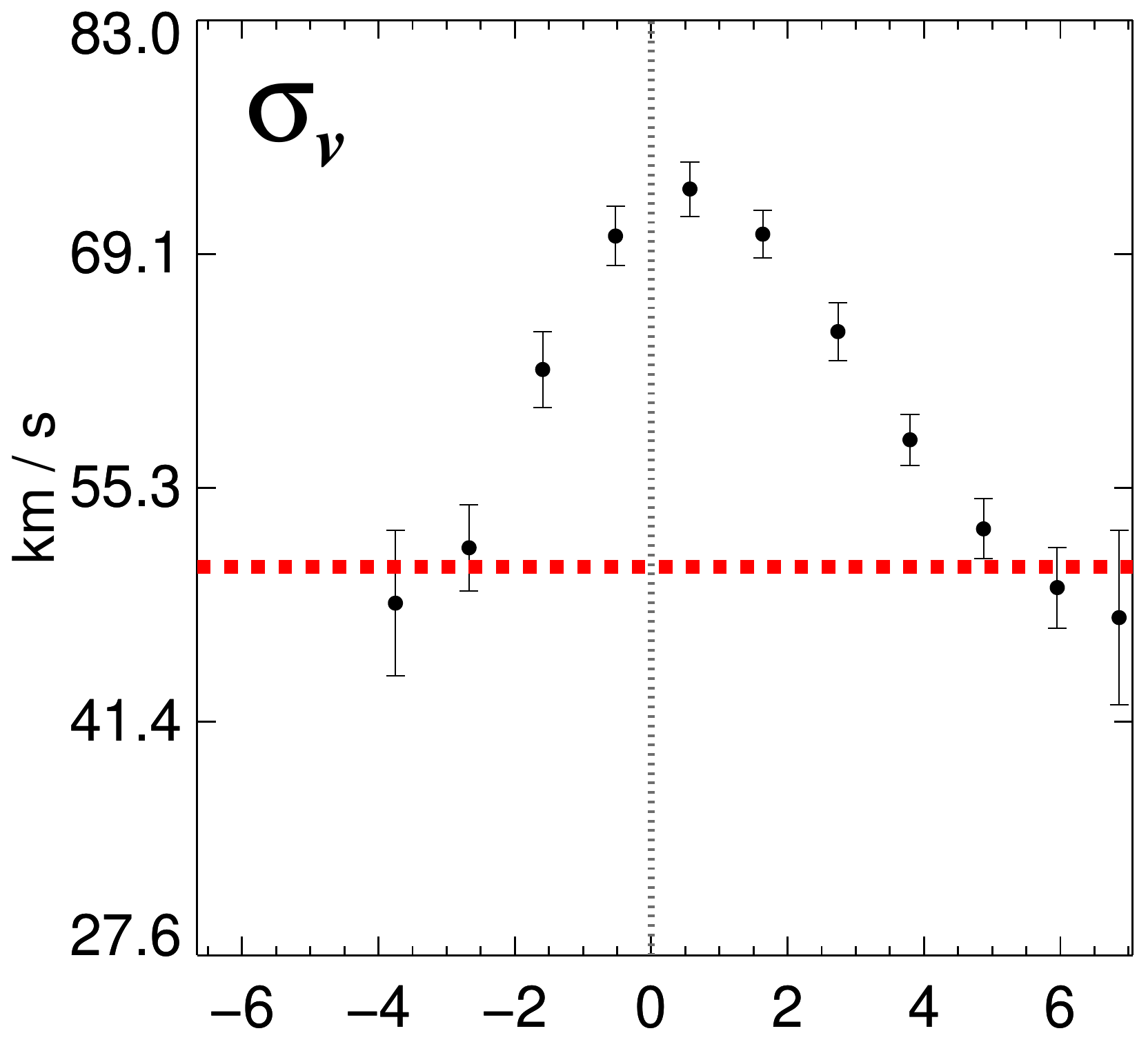}
\includegraphics[width=0.343\columnwidth]{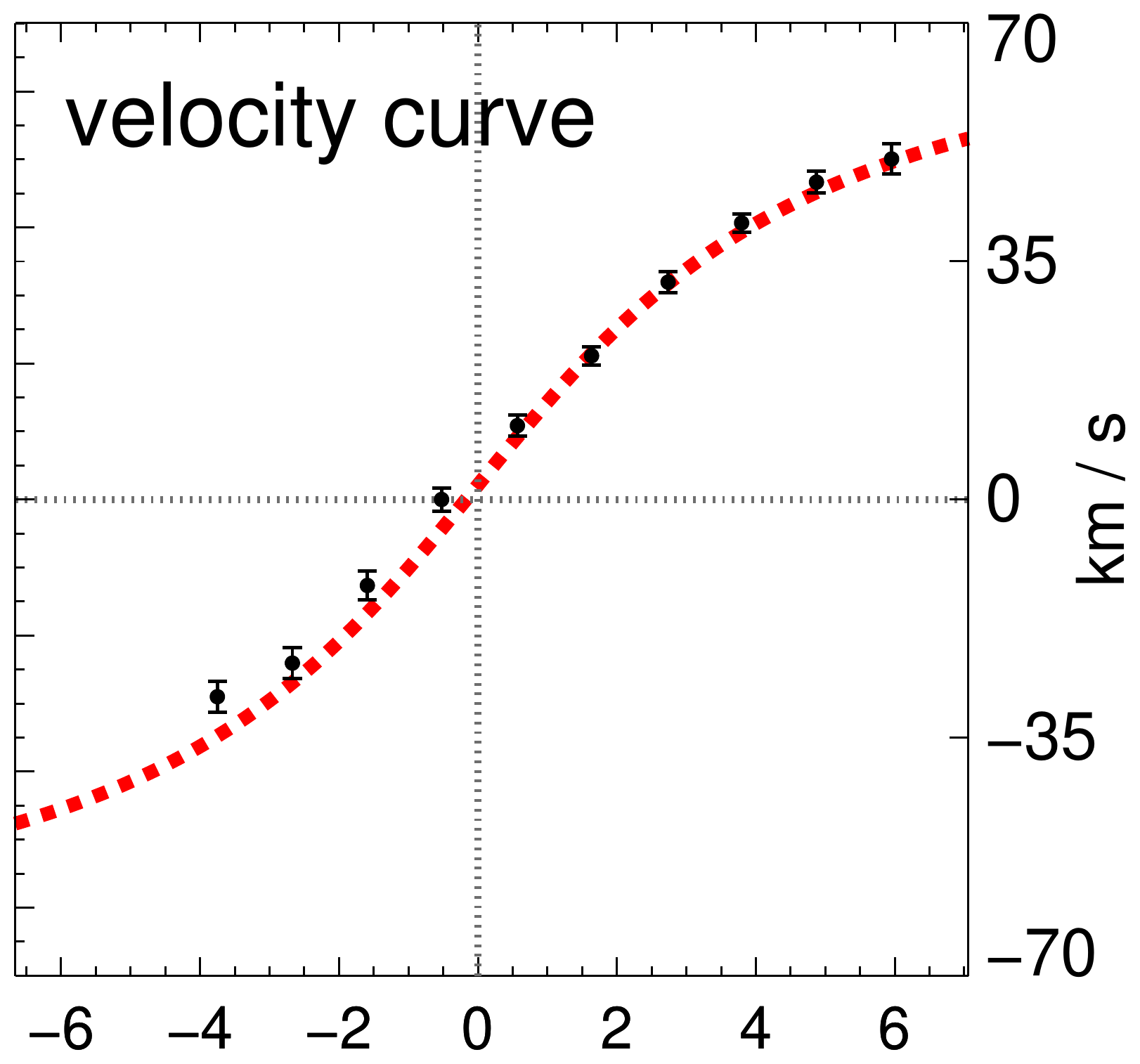}\\
\vspace{1mm}
\includegraphics[width=0.343\columnwidth]{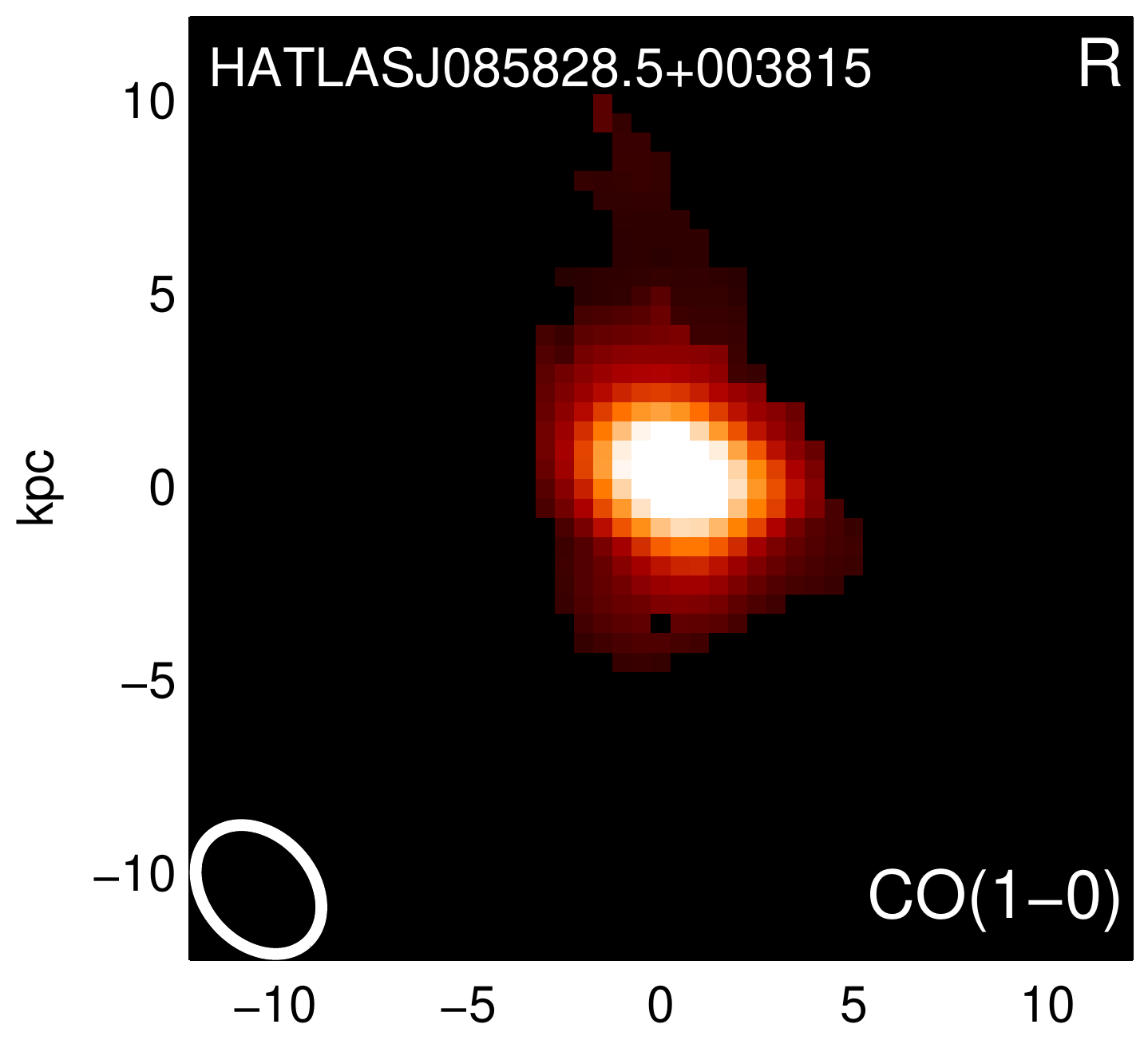}
\includegraphics[width=0.32\columnwidth]{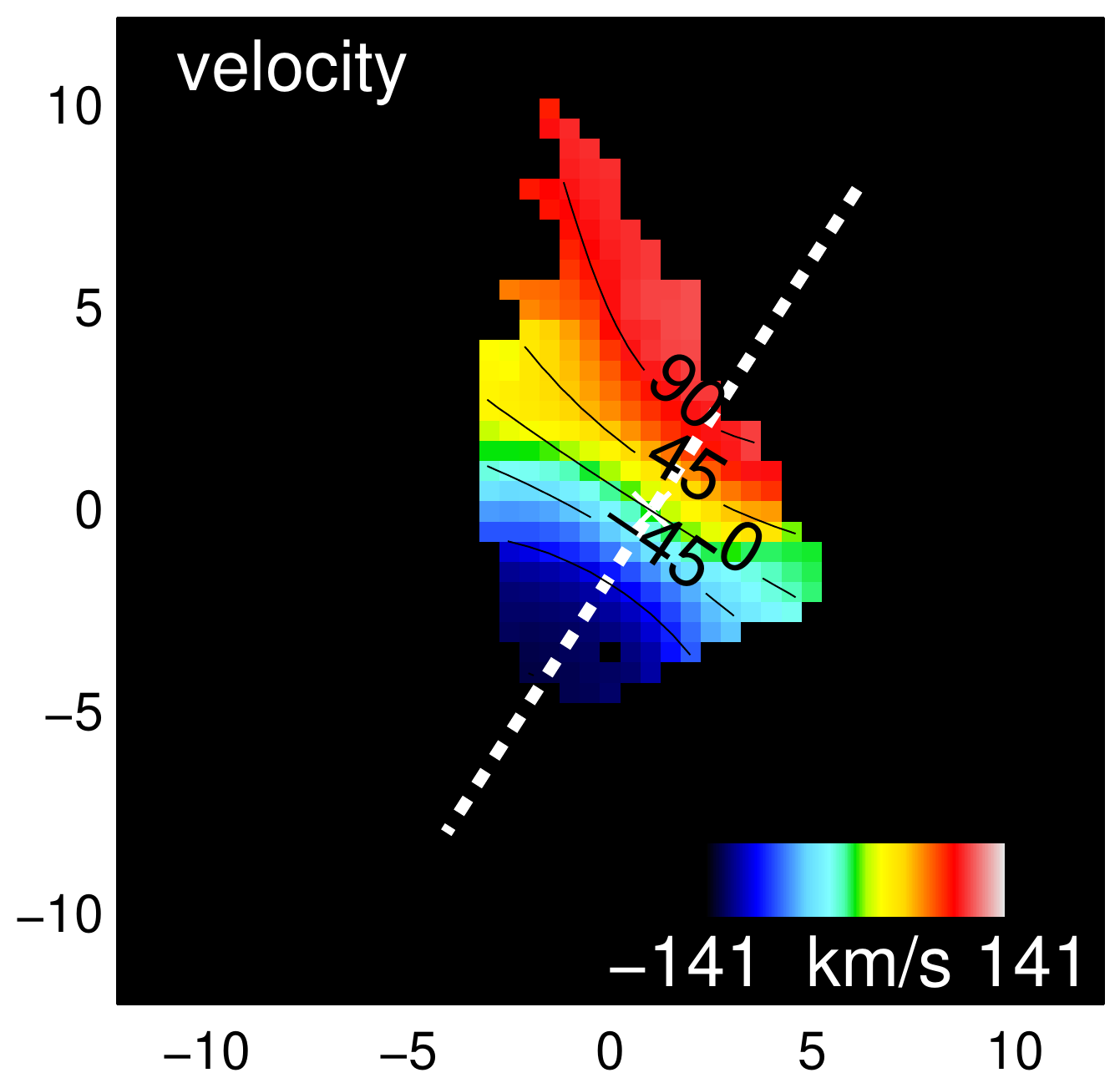}
\includegraphics[width=0.32\columnwidth]{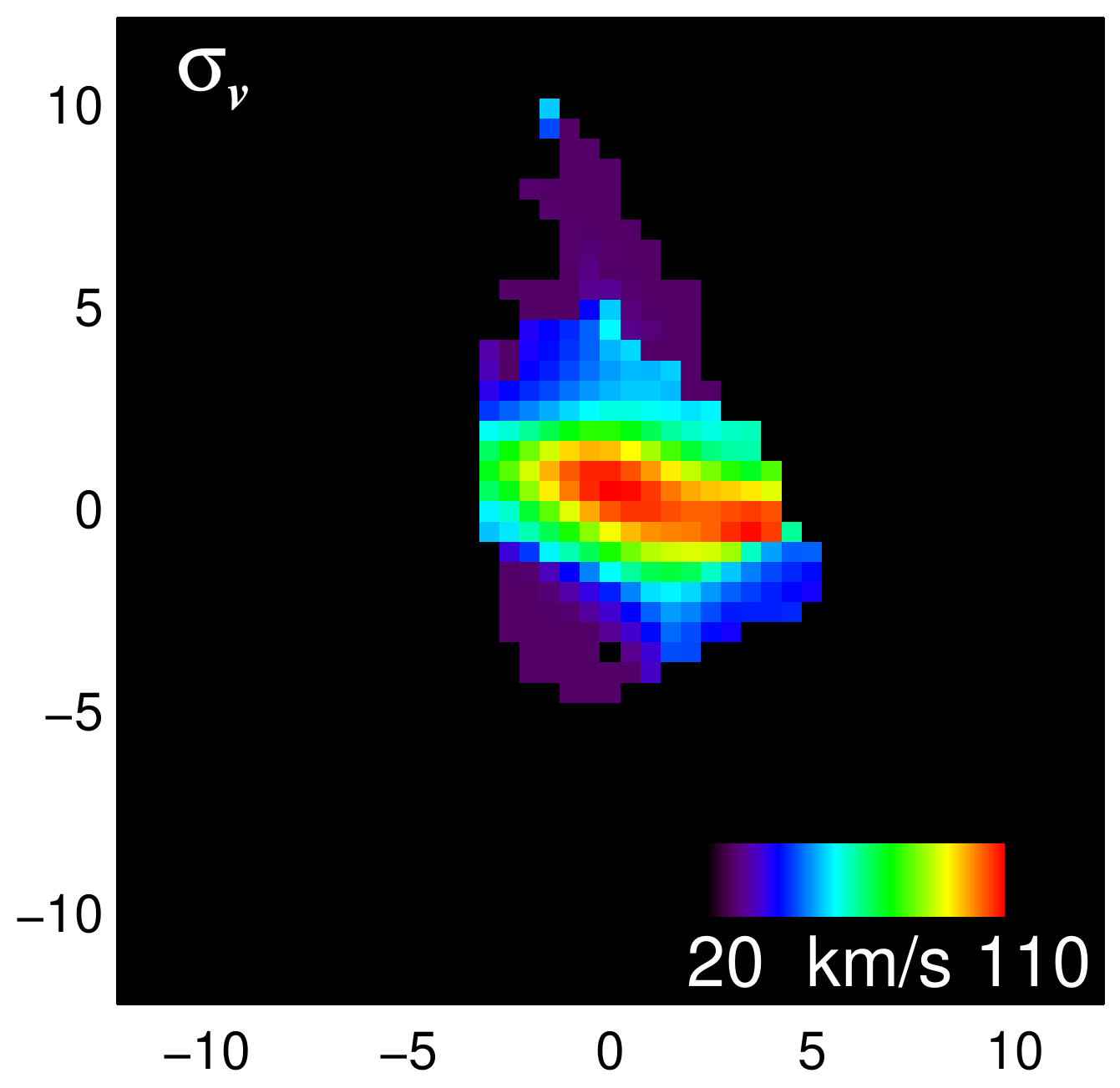}
\includegraphics[width=0.32\columnwidth]{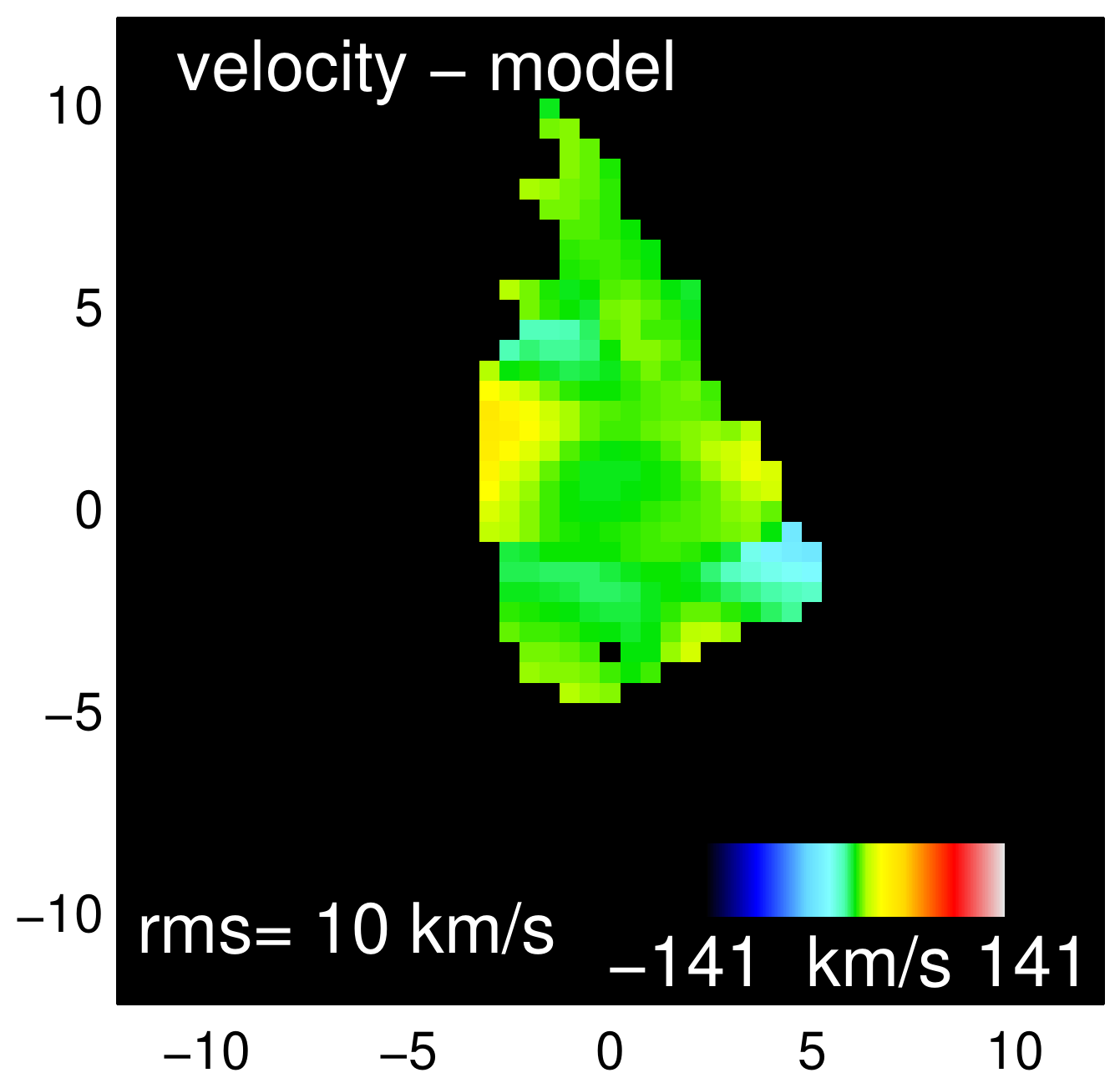}
\includegraphics[width=0.345\columnwidth]{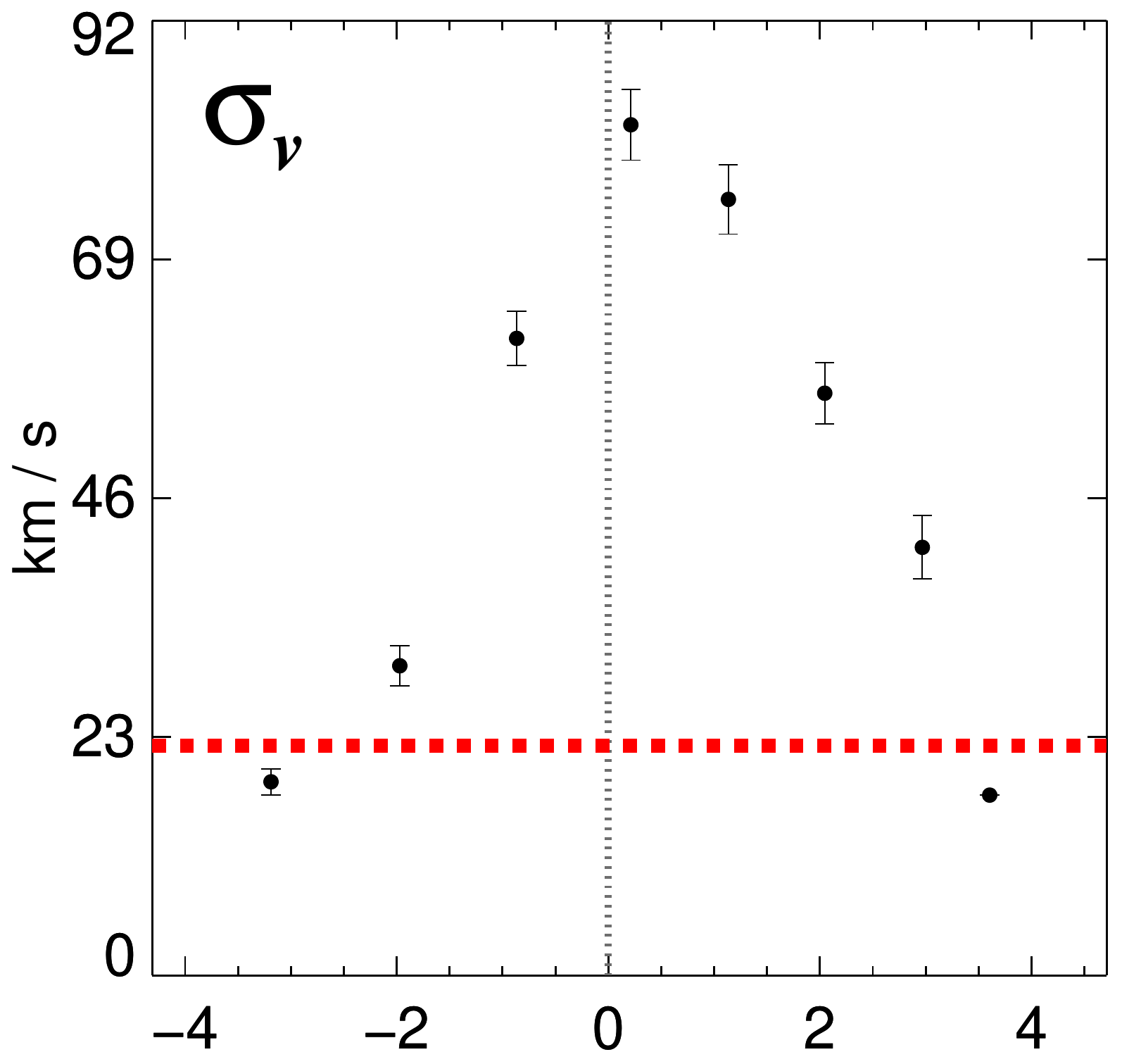}
\includegraphics[width=0.351\columnwidth]{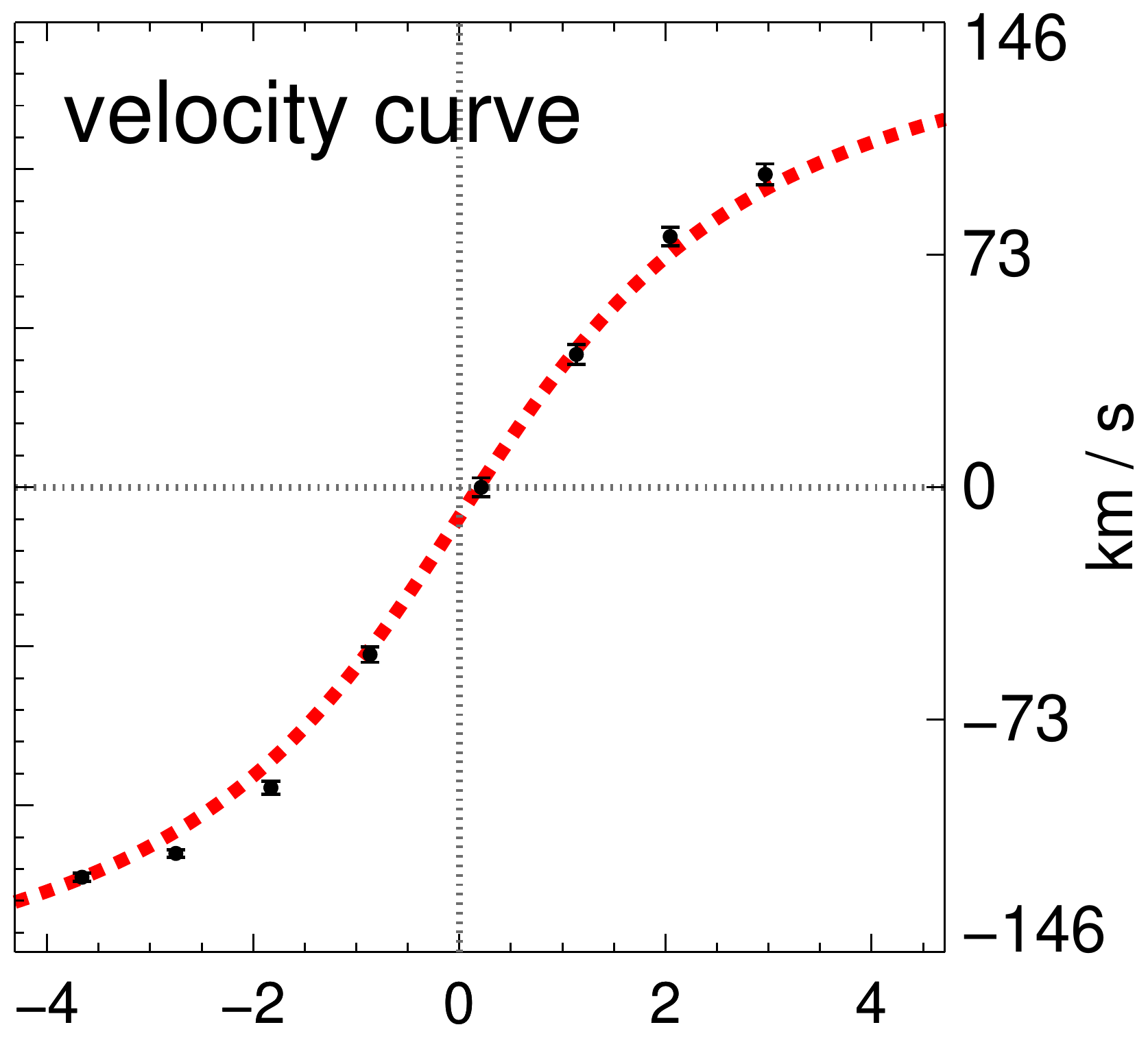}\\
\vspace{1mm}
\includegraphics[width=0.343\columnwidth]{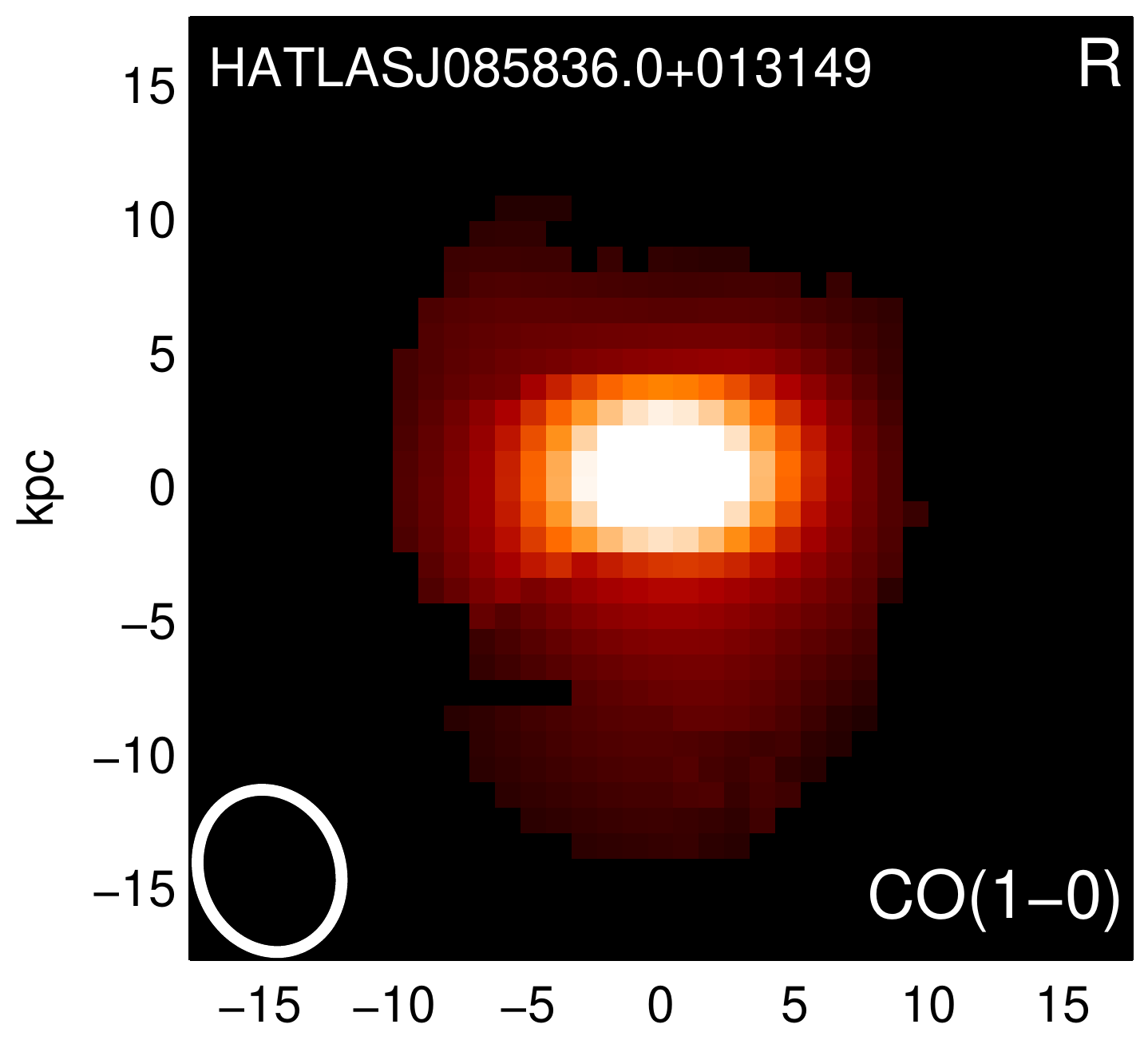}
\includegraphics[width=0.32\columnwidth]{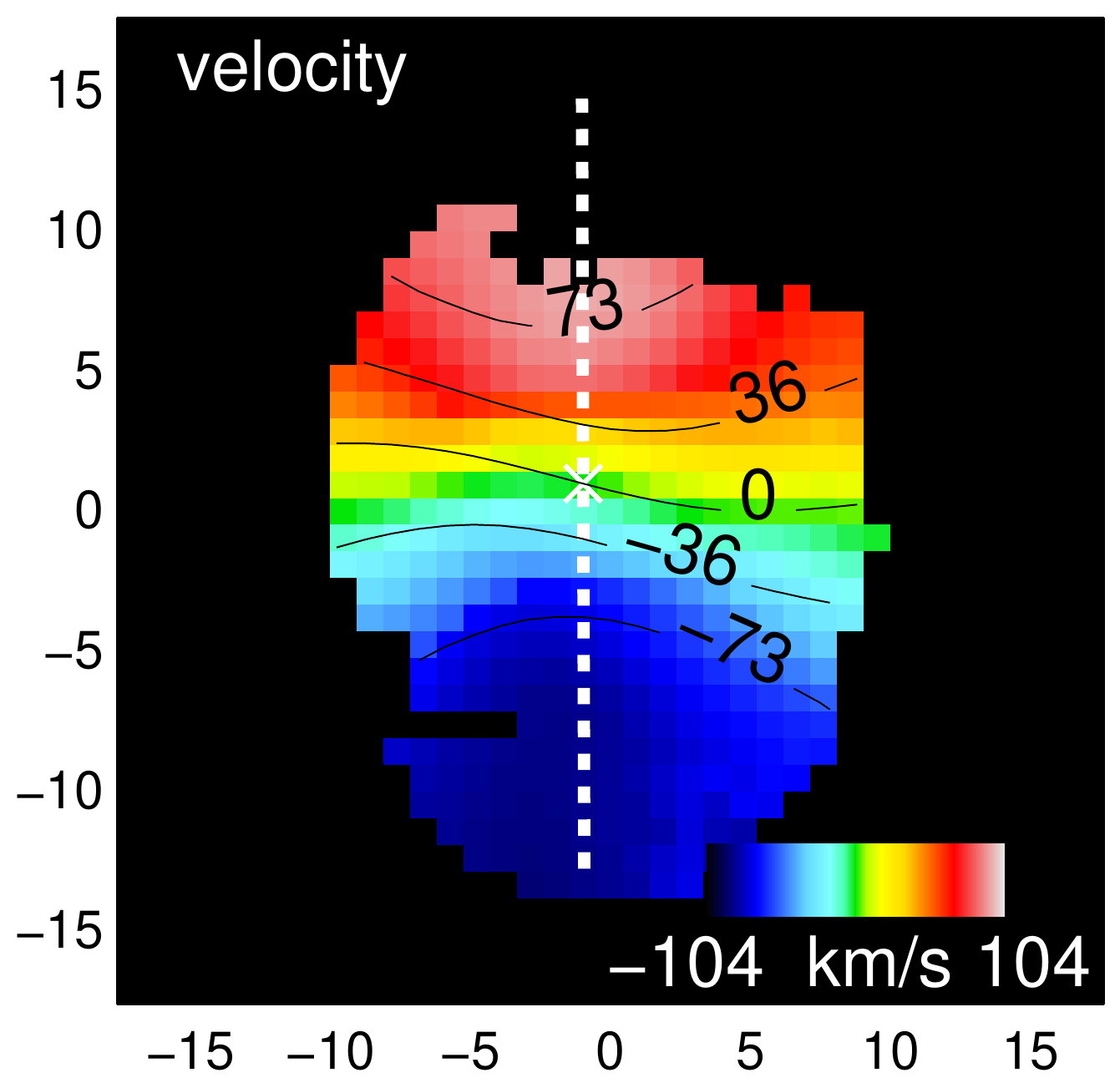}
\includegraphics[width=0.32\columnwidth]{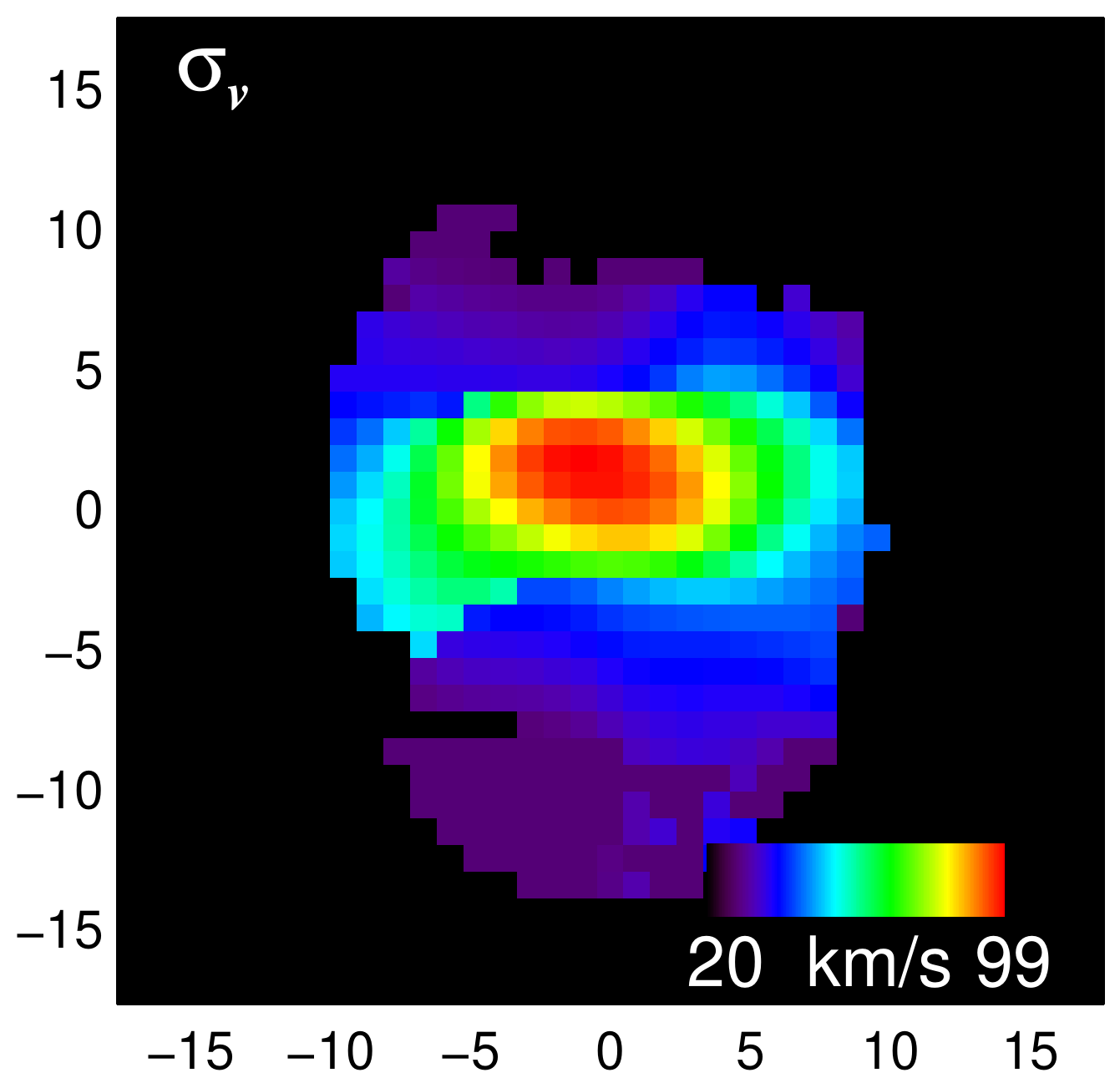}
\includegraphics[width=0.32\columnwidth]{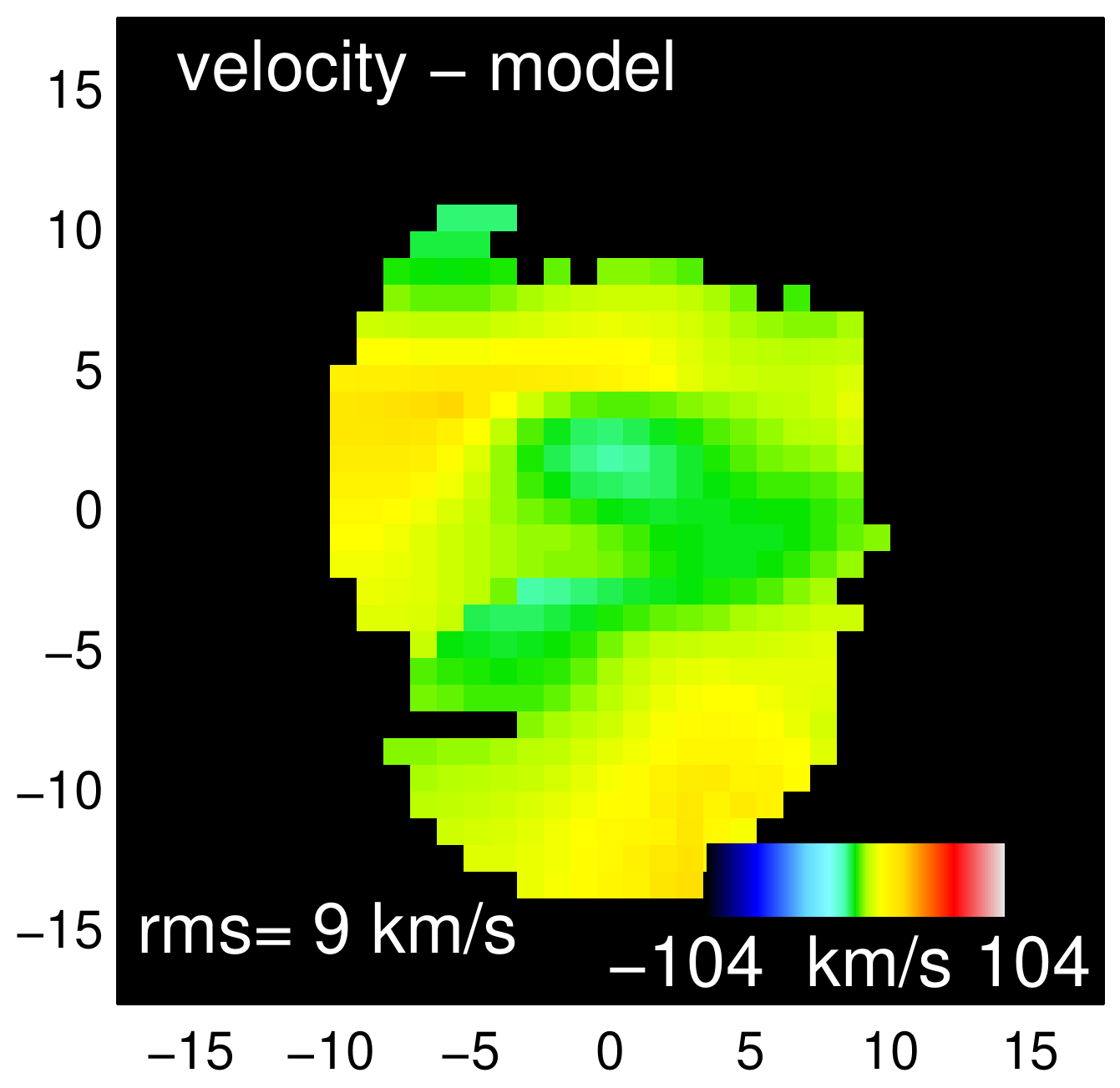}
\includegraphics[width=0.345\columnwidth]{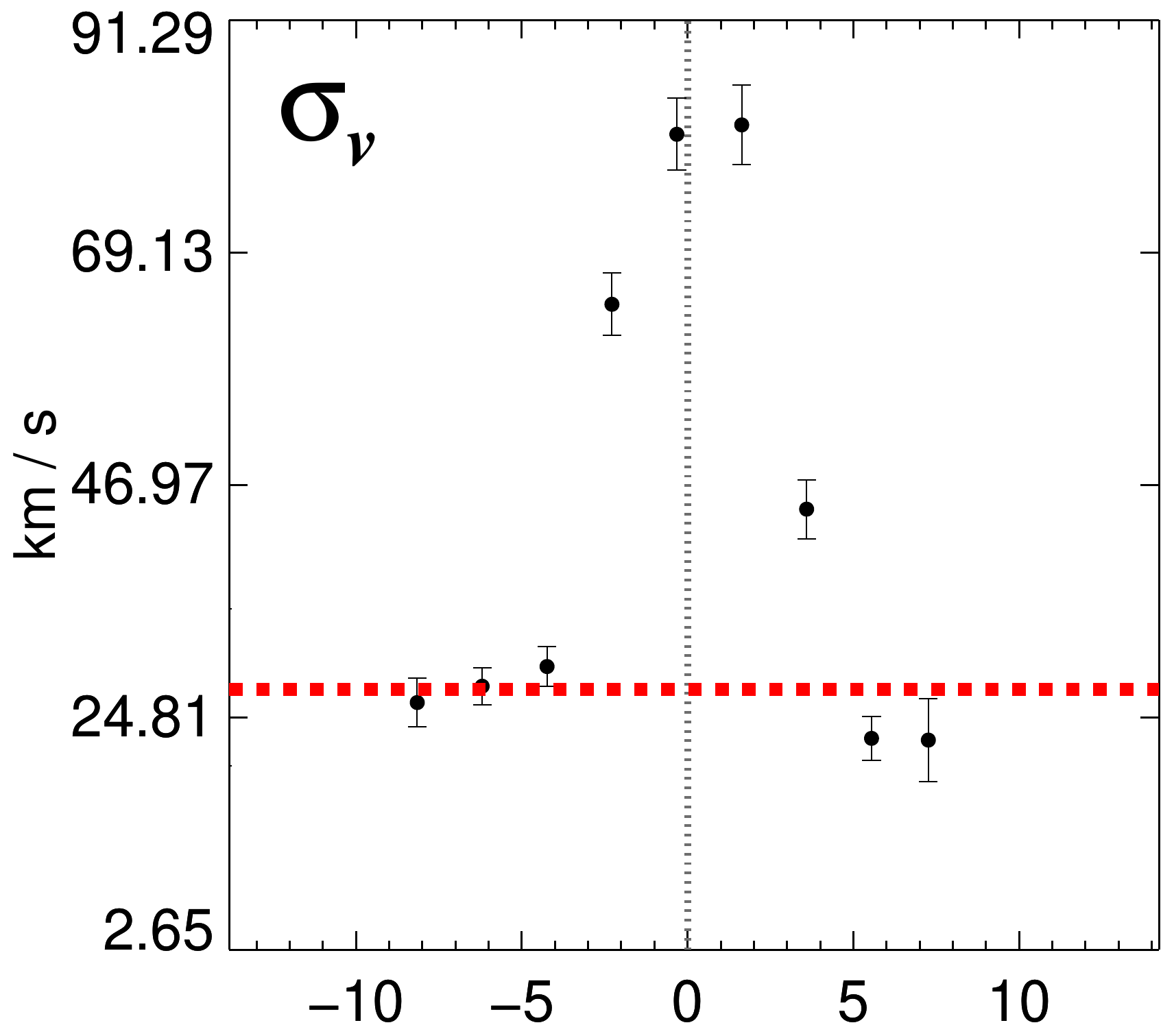}
\includegraphics[width=0.351\columnwidth]{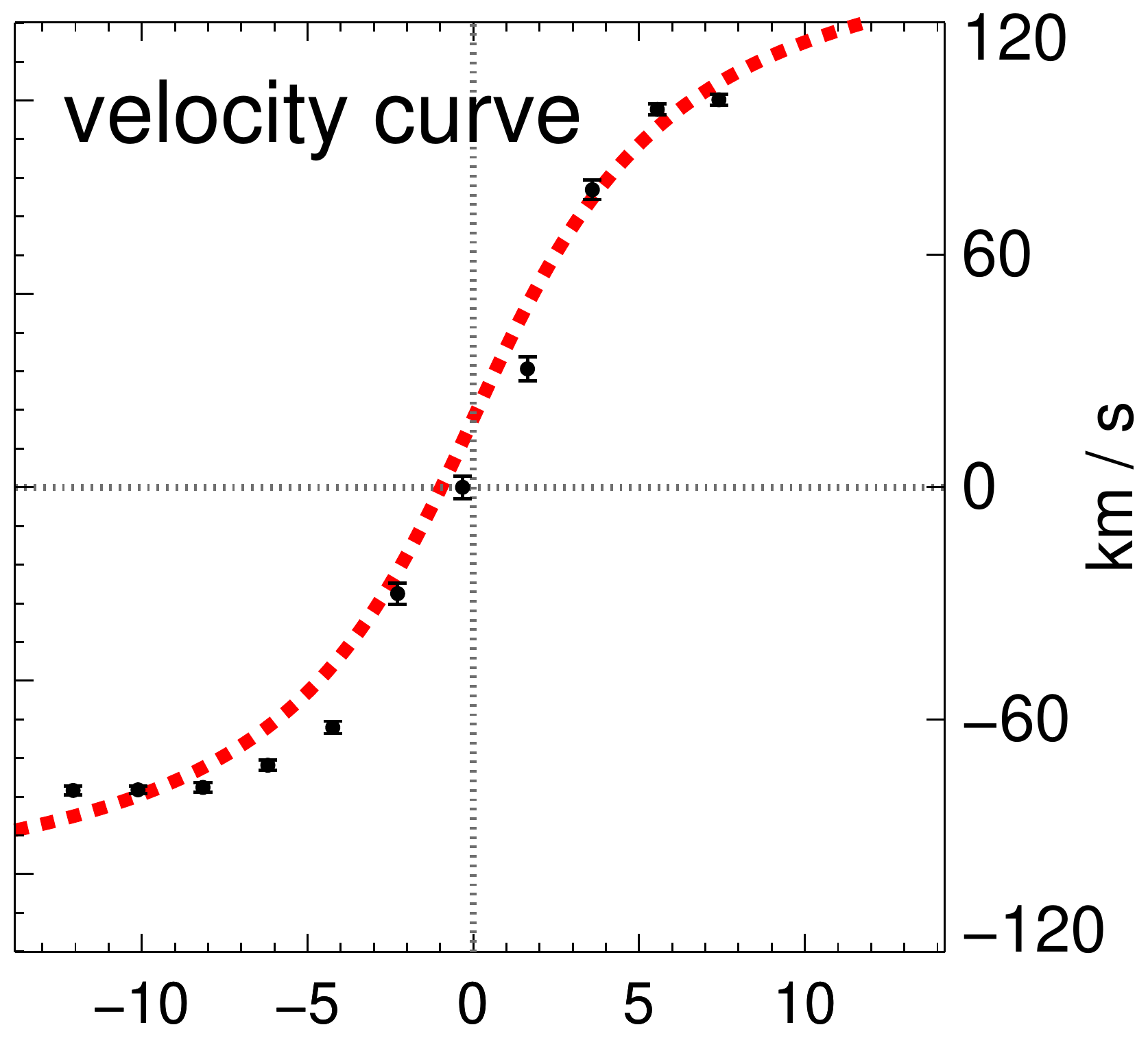}\\
\vspace{1mm}
\includegraphics[width=0.343\columnwidth]{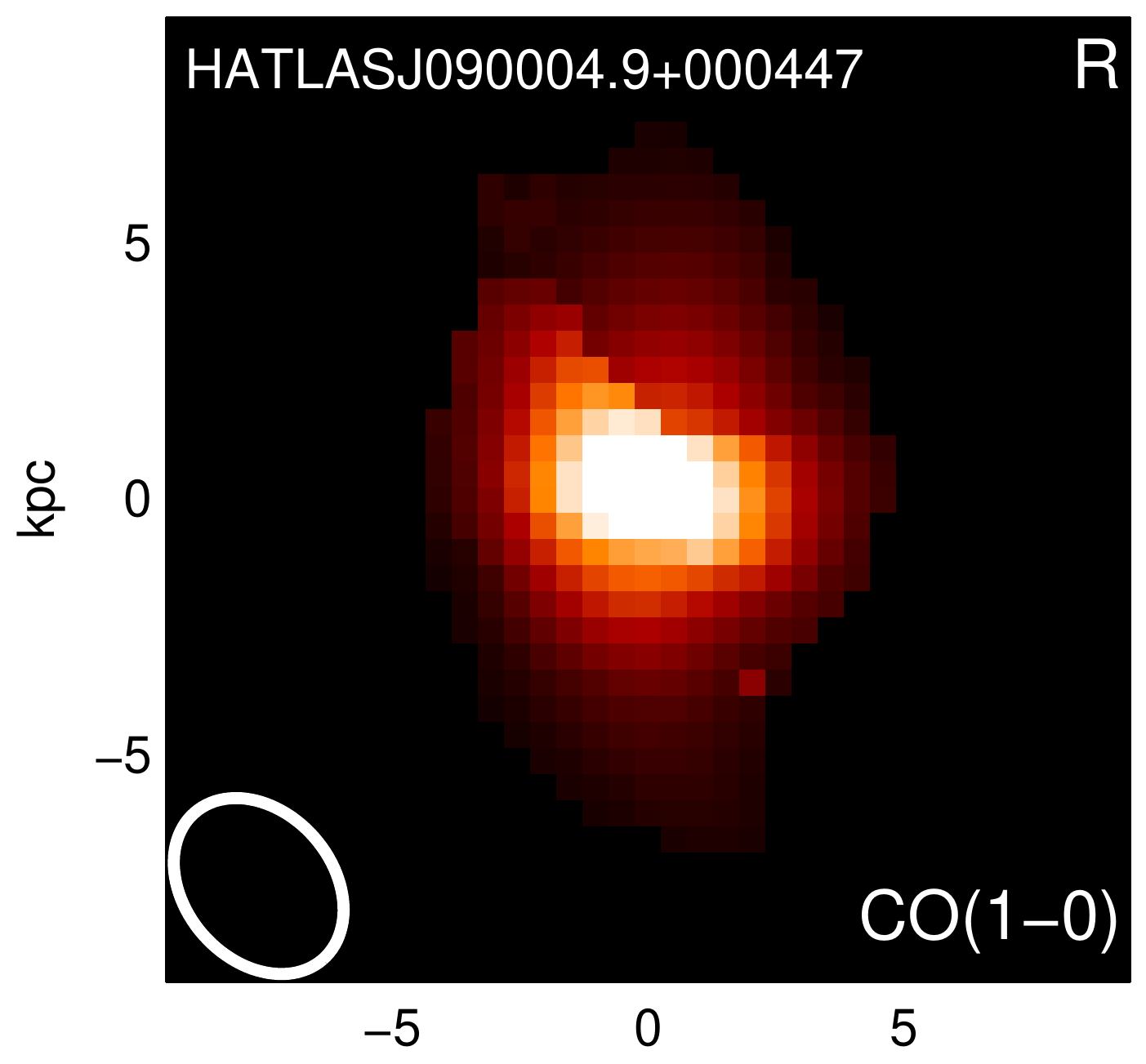}
\includegraphics[width=0.32\columnwidth]{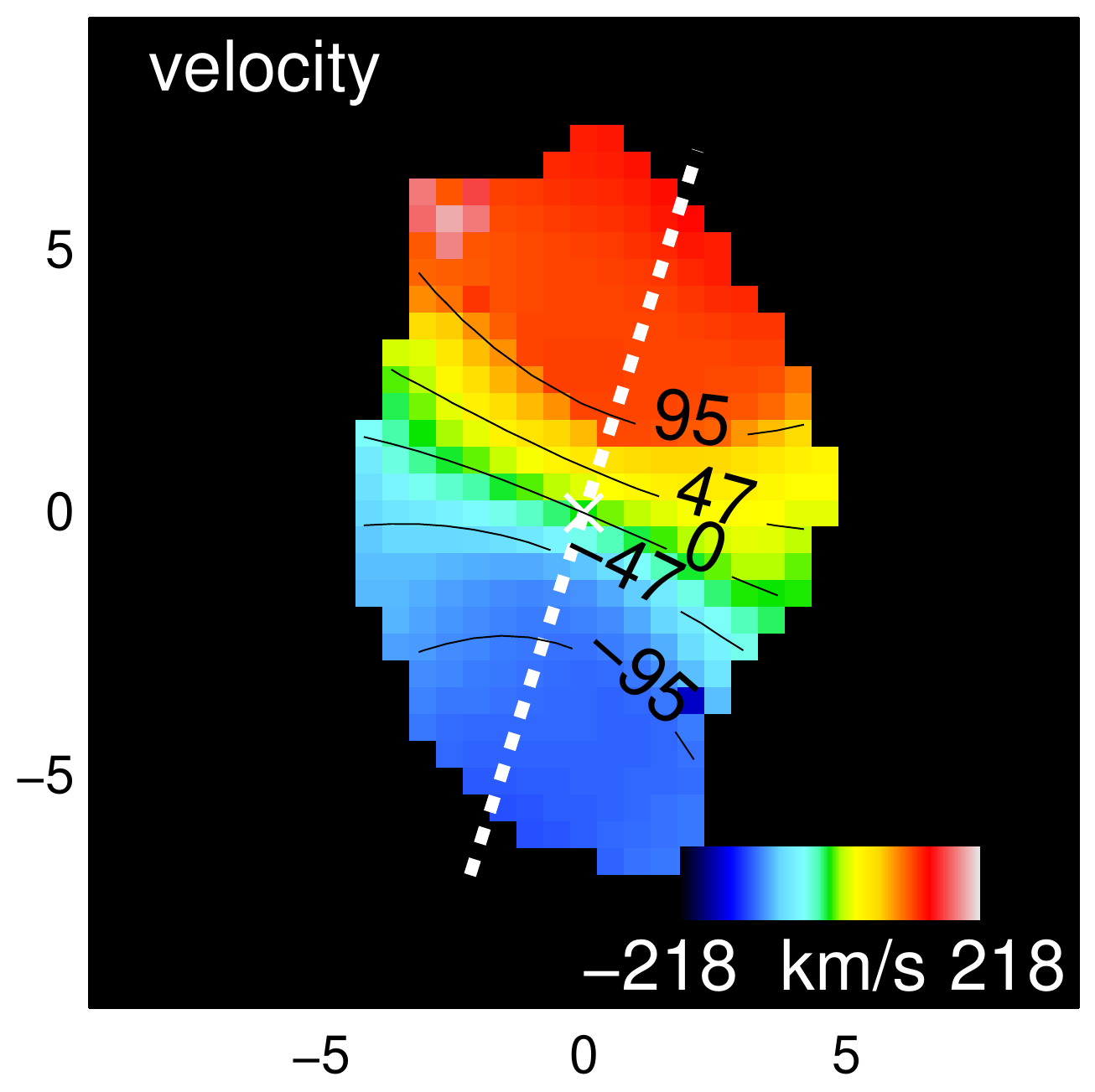}
\includegraphics[width=0.32\columnwidth]{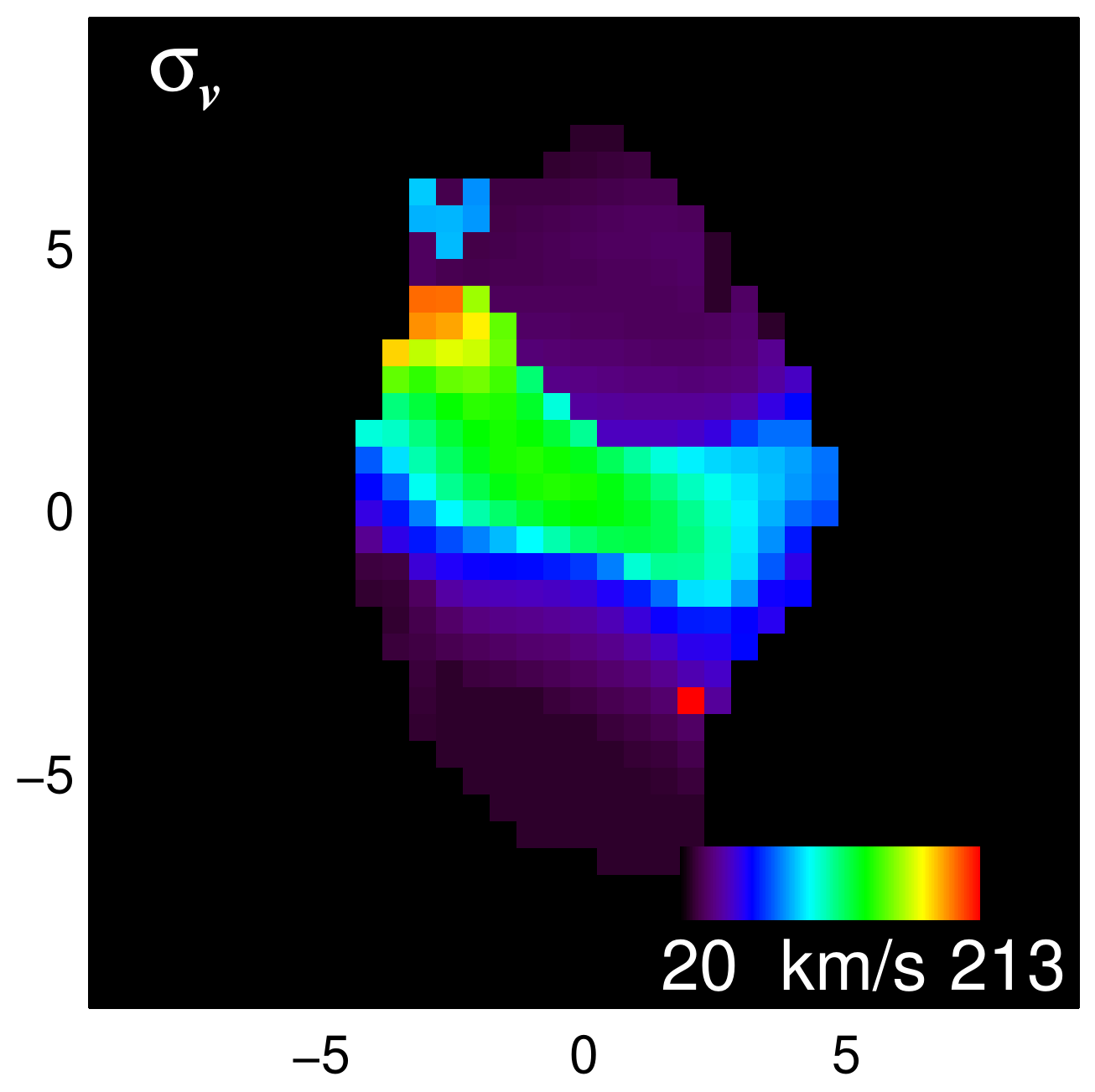}
\includegraphics[width=0.32\columnwidth]{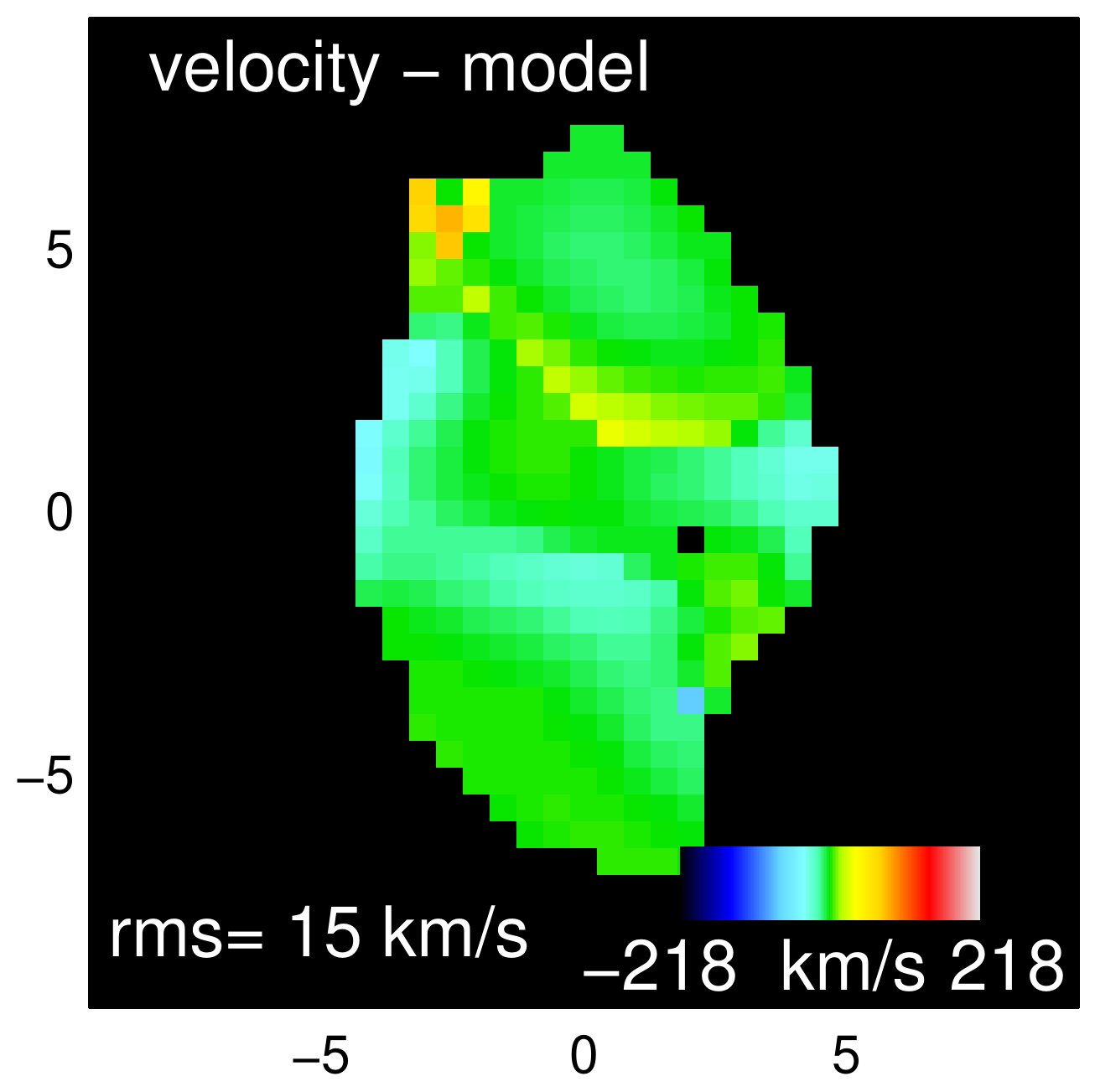}
\includegraphics[width=0.345\columnwidth]{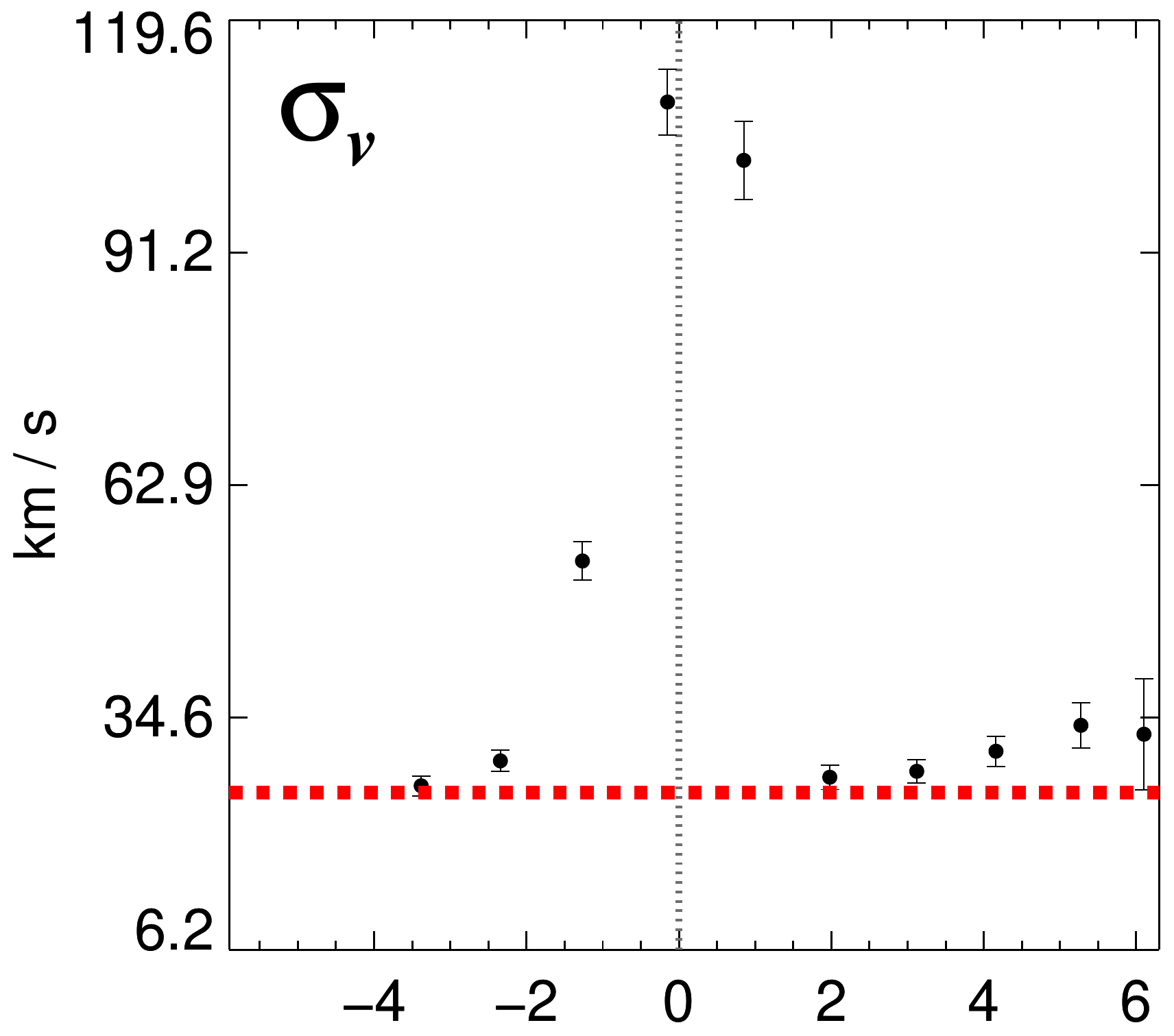}
\includegraphics[width=0.351\columnwidth]{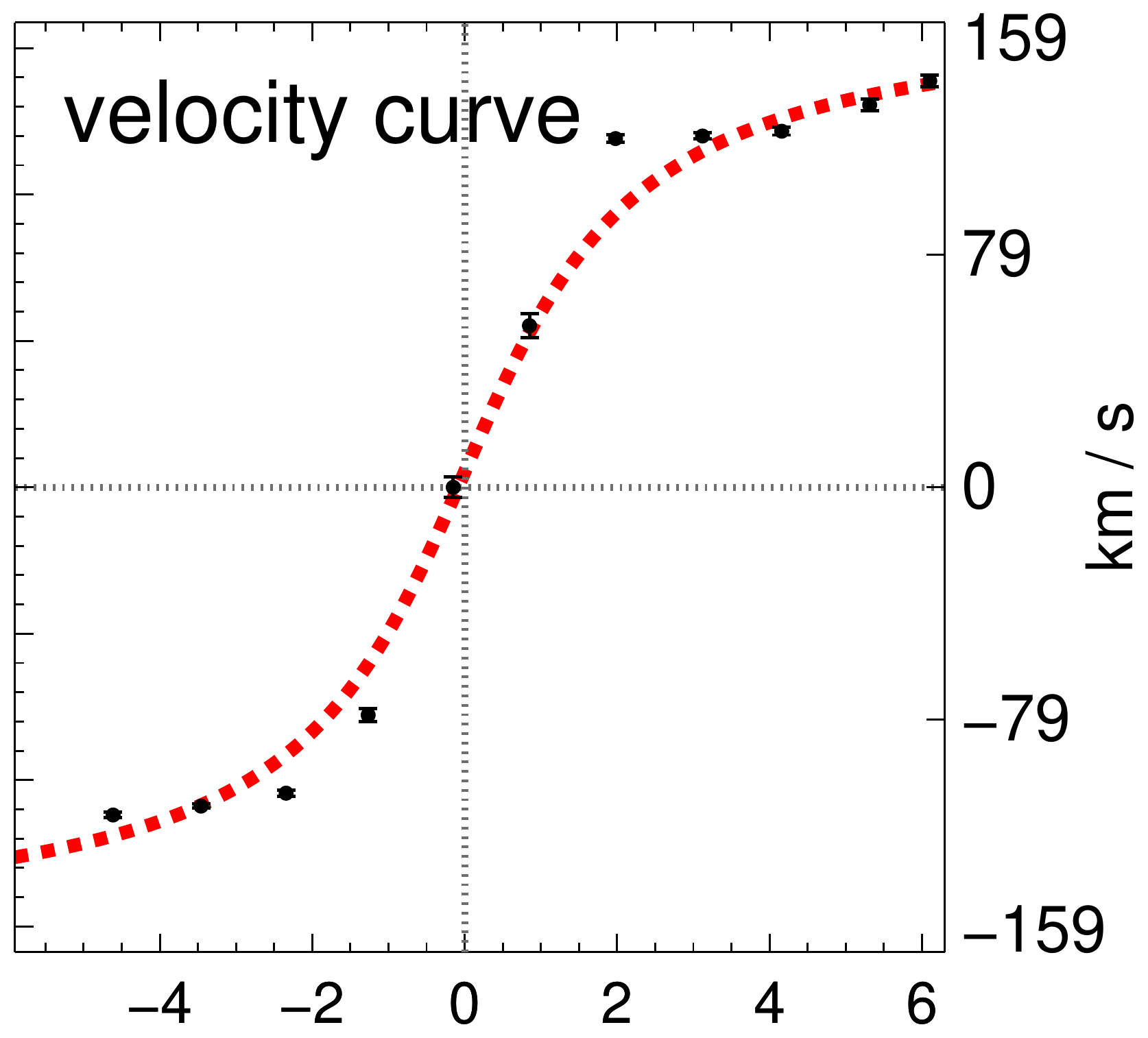}\\
\vspace{1mm}
\includegraphics[width=0.343\columnwidth]{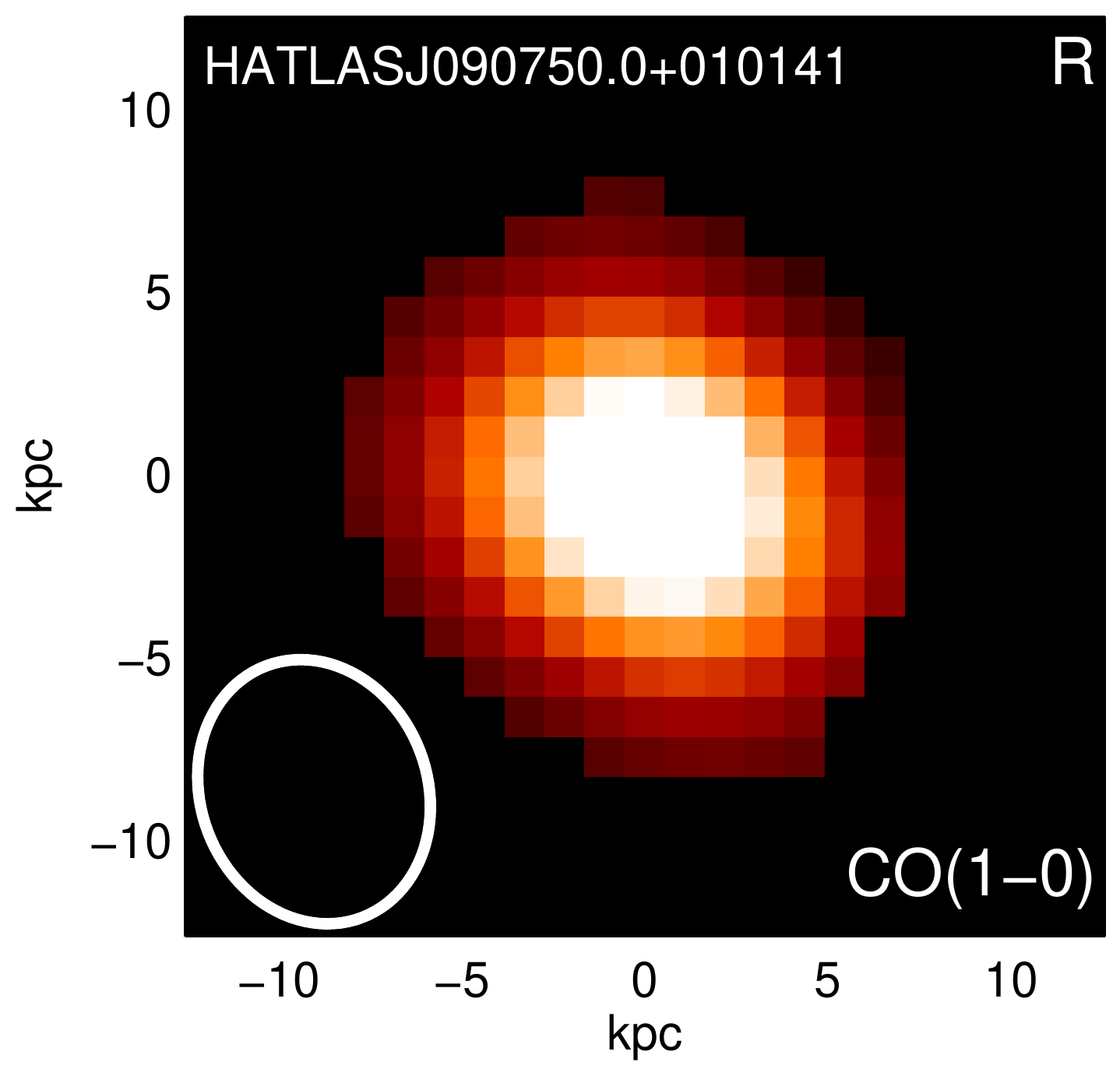}
\includegraphics[width=0.32\columnwidth]{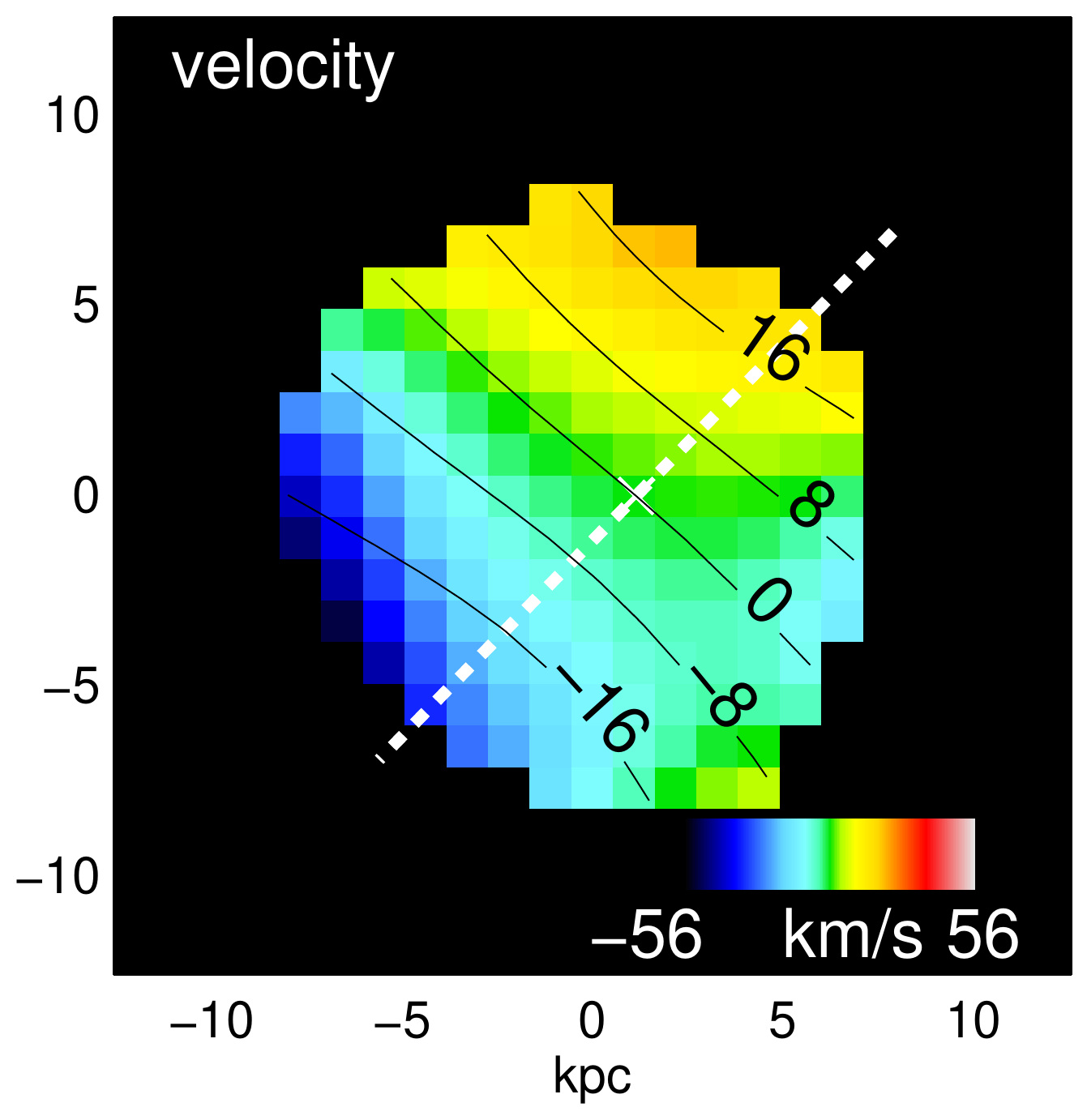}
\includegraphics[width=0.32\columnwidth]{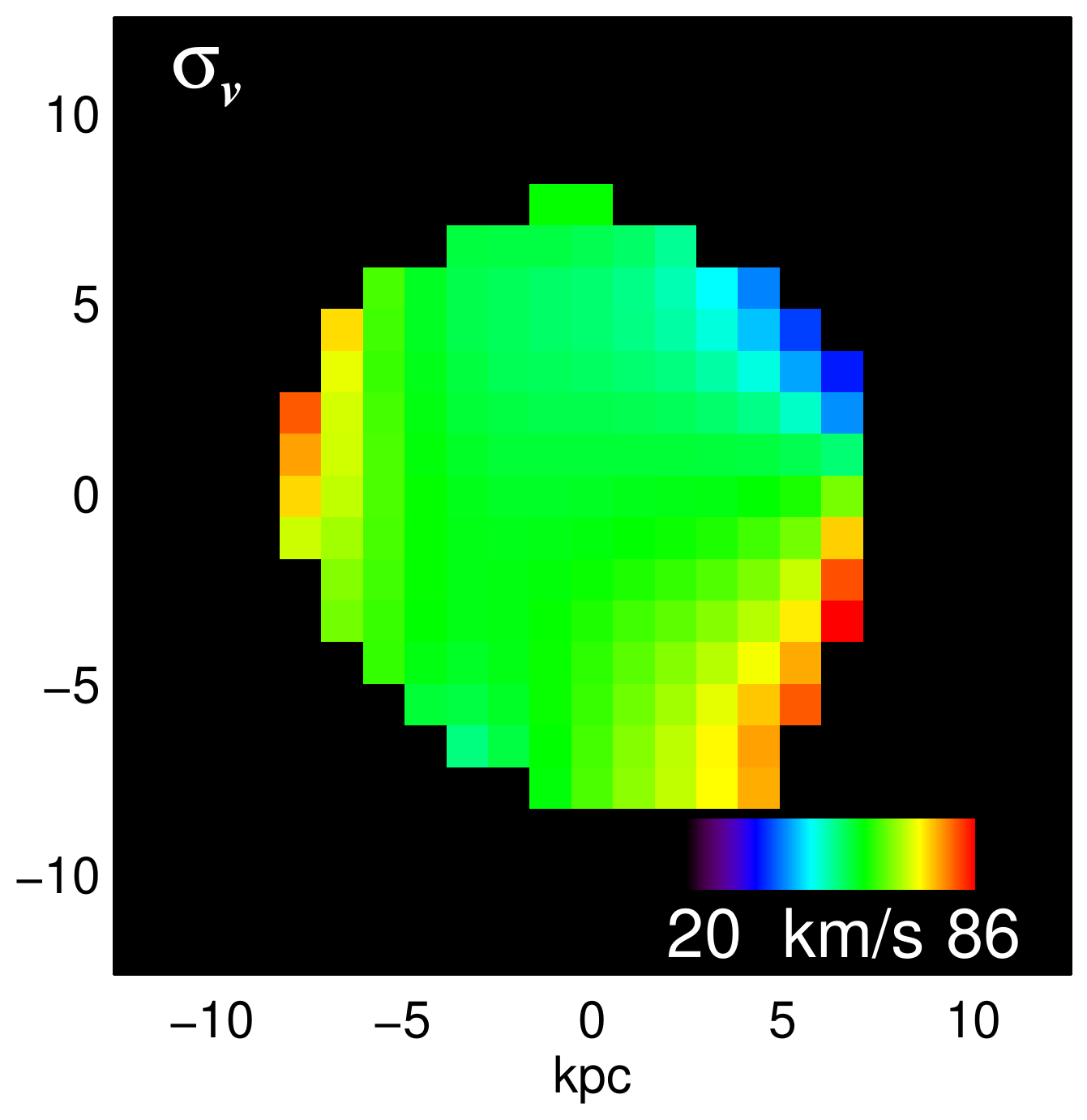}
\includegraphics[width=0.32\columnwidth]{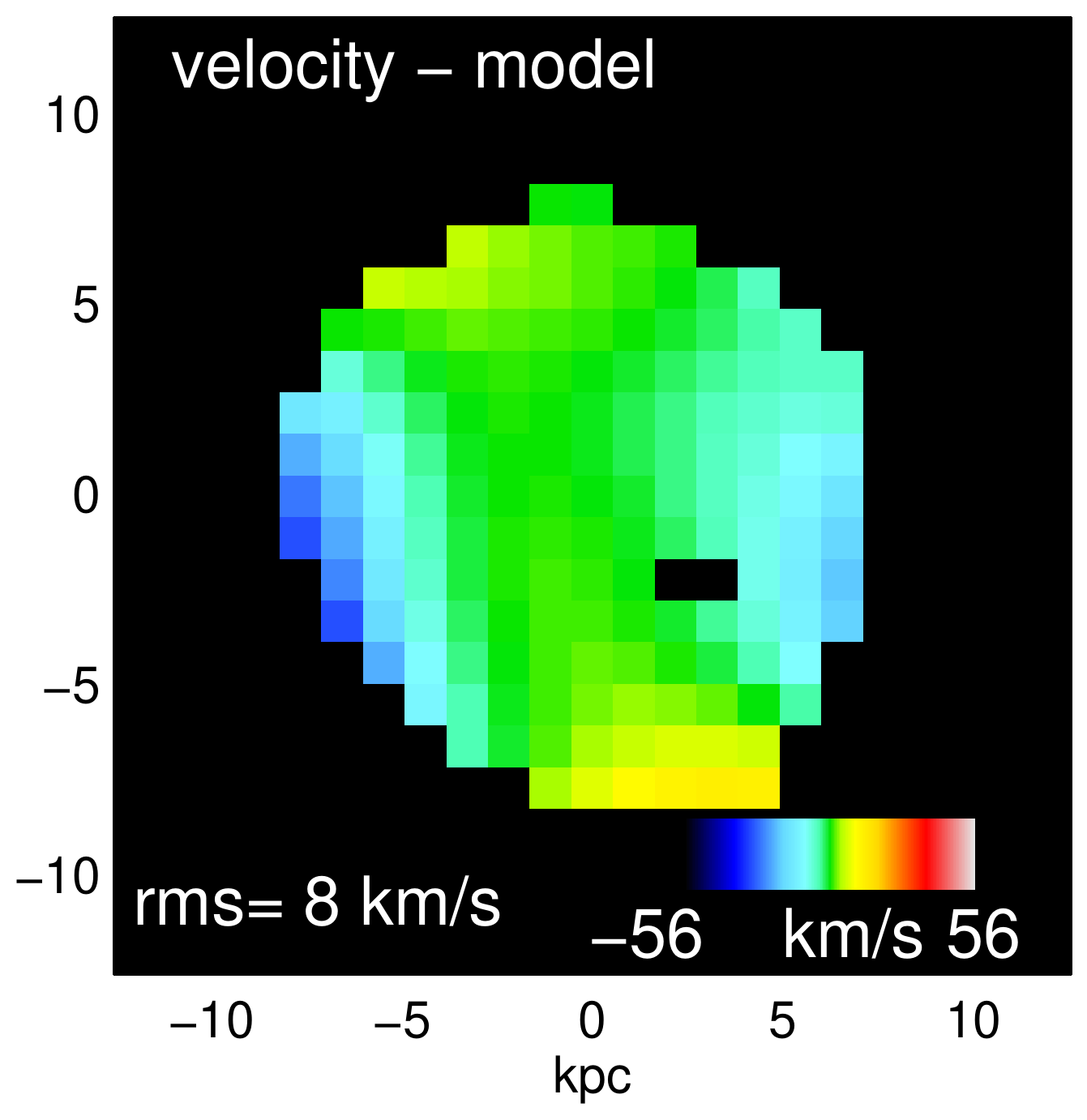}
\includegraphics[width=0.345\columnwidth]{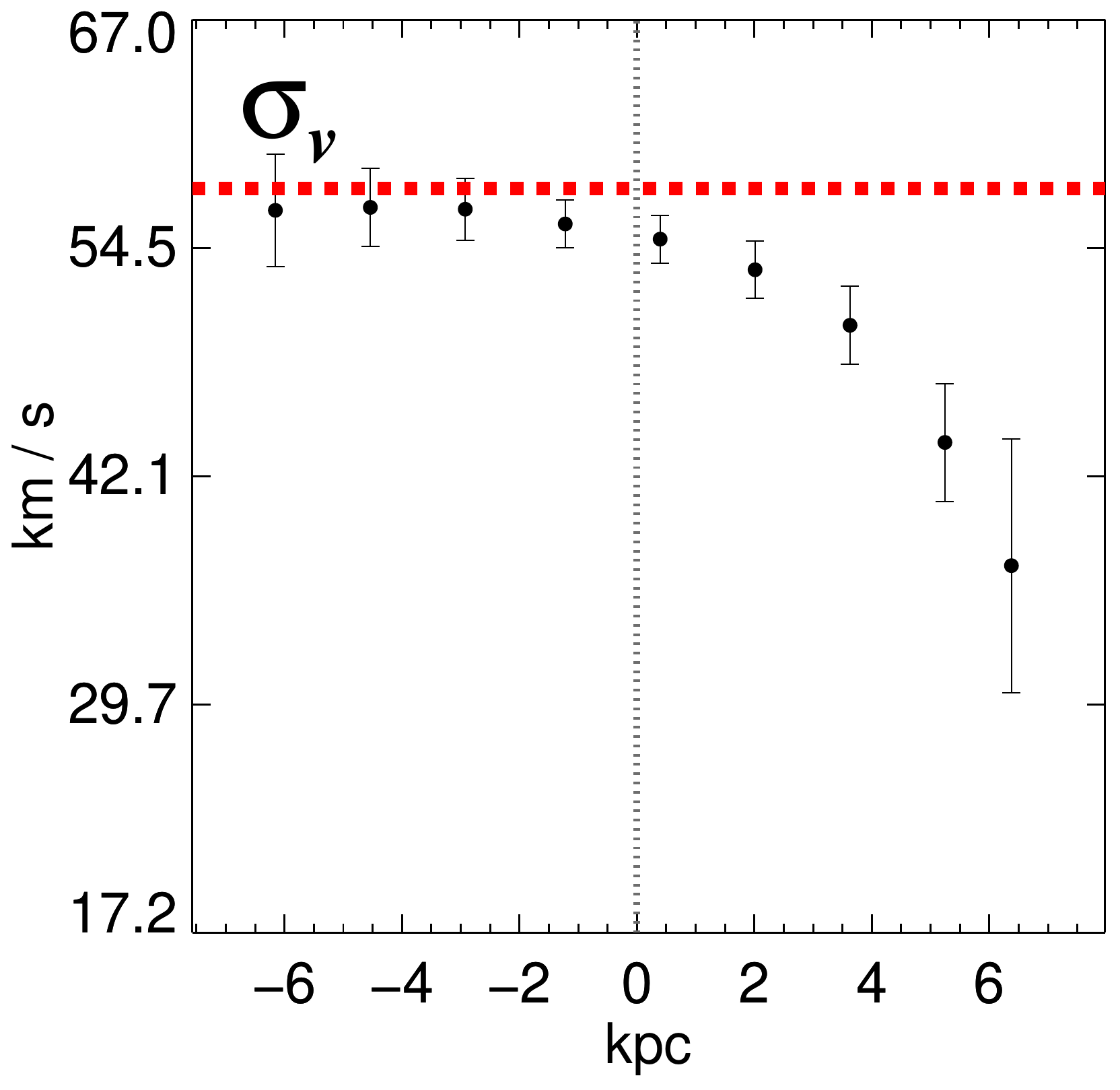}
\includegraphics[width=0.341\columnwidth]{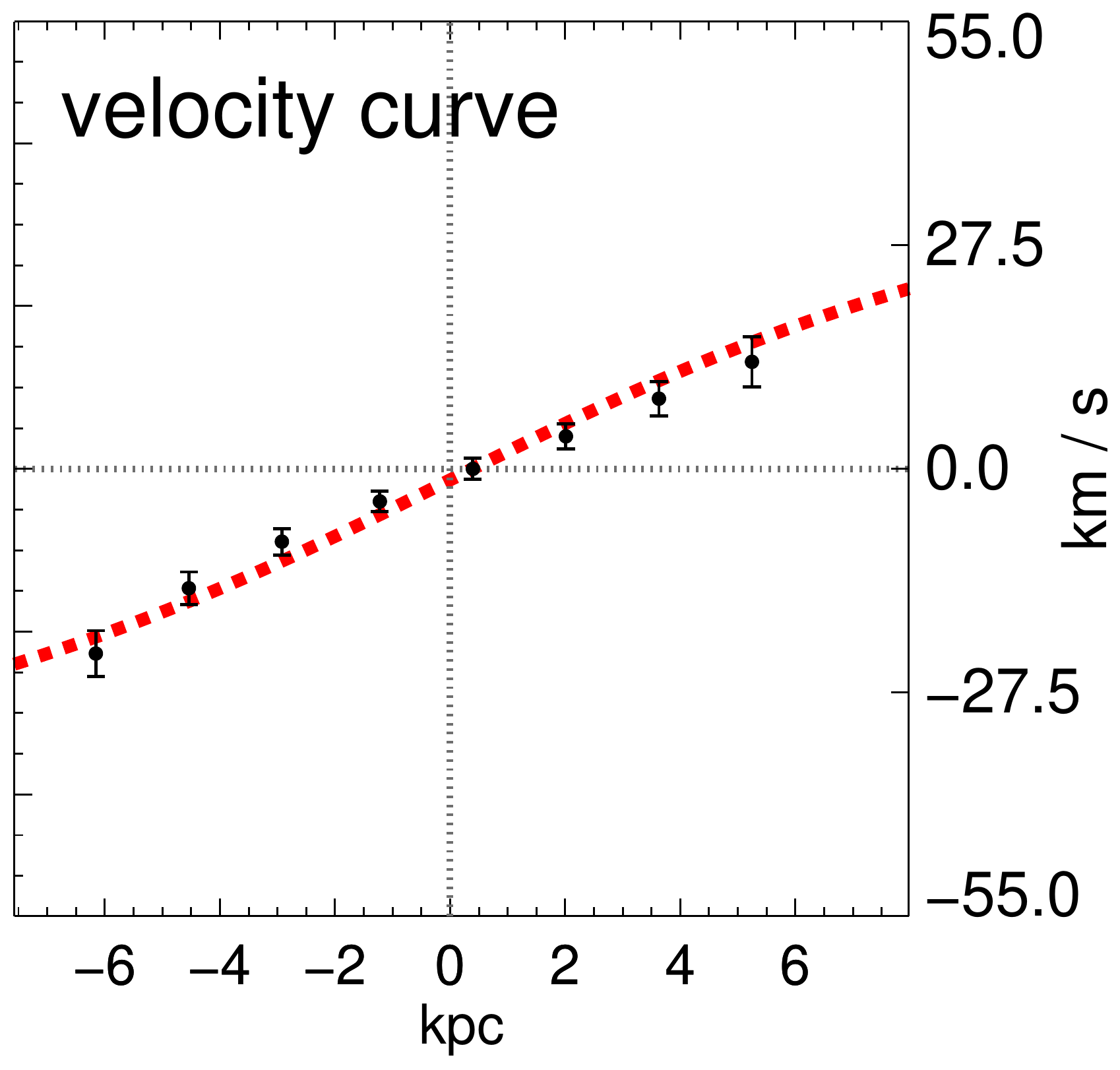}\\
\centering{\textbf{Figure C1. Continued.}}
\end{figure*}

\begin{figure*}
\flushleft
\includegraphics[width=0.343\columnwidth]{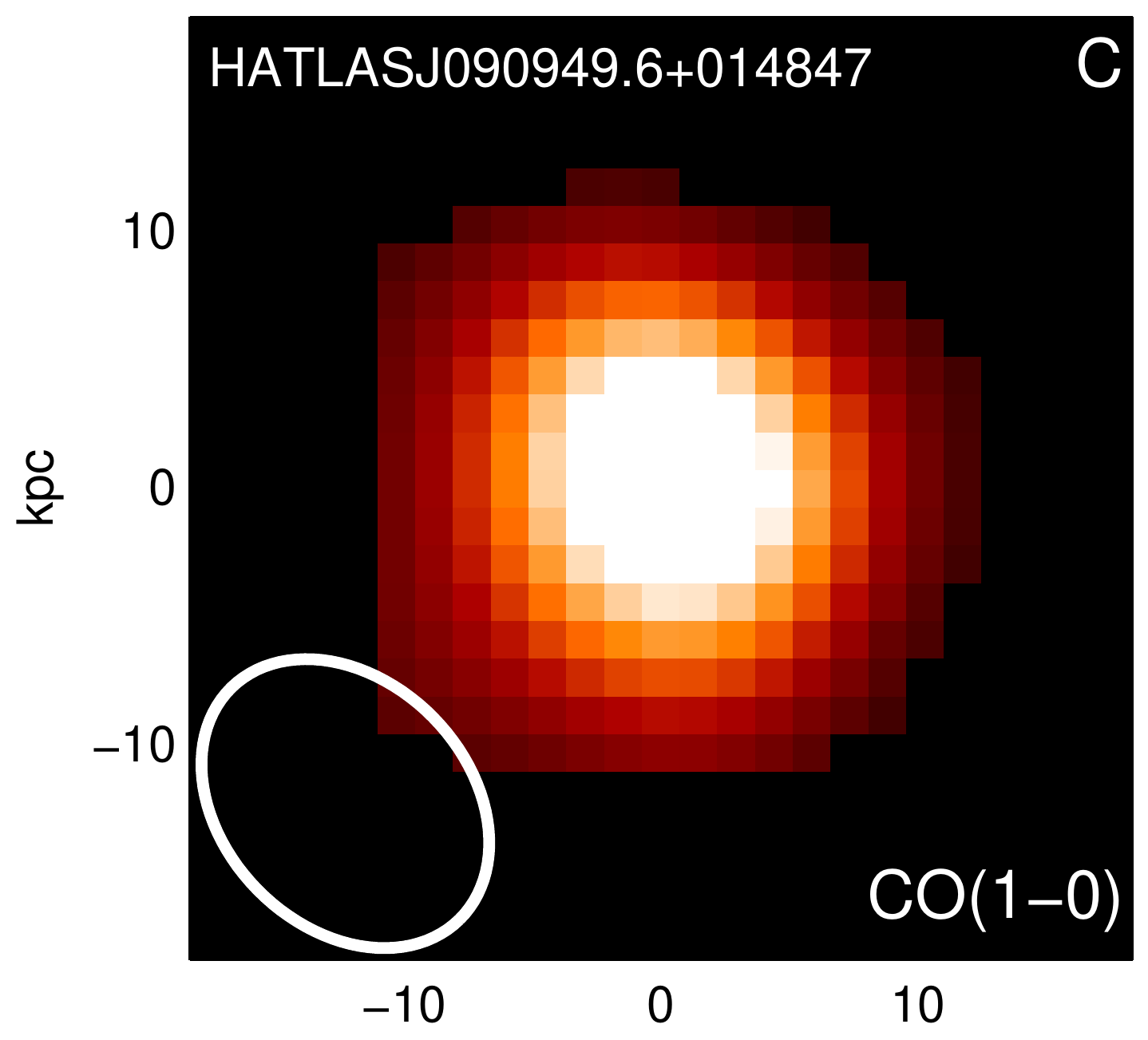}
\includegraphics[width=0.32\columnwidth]{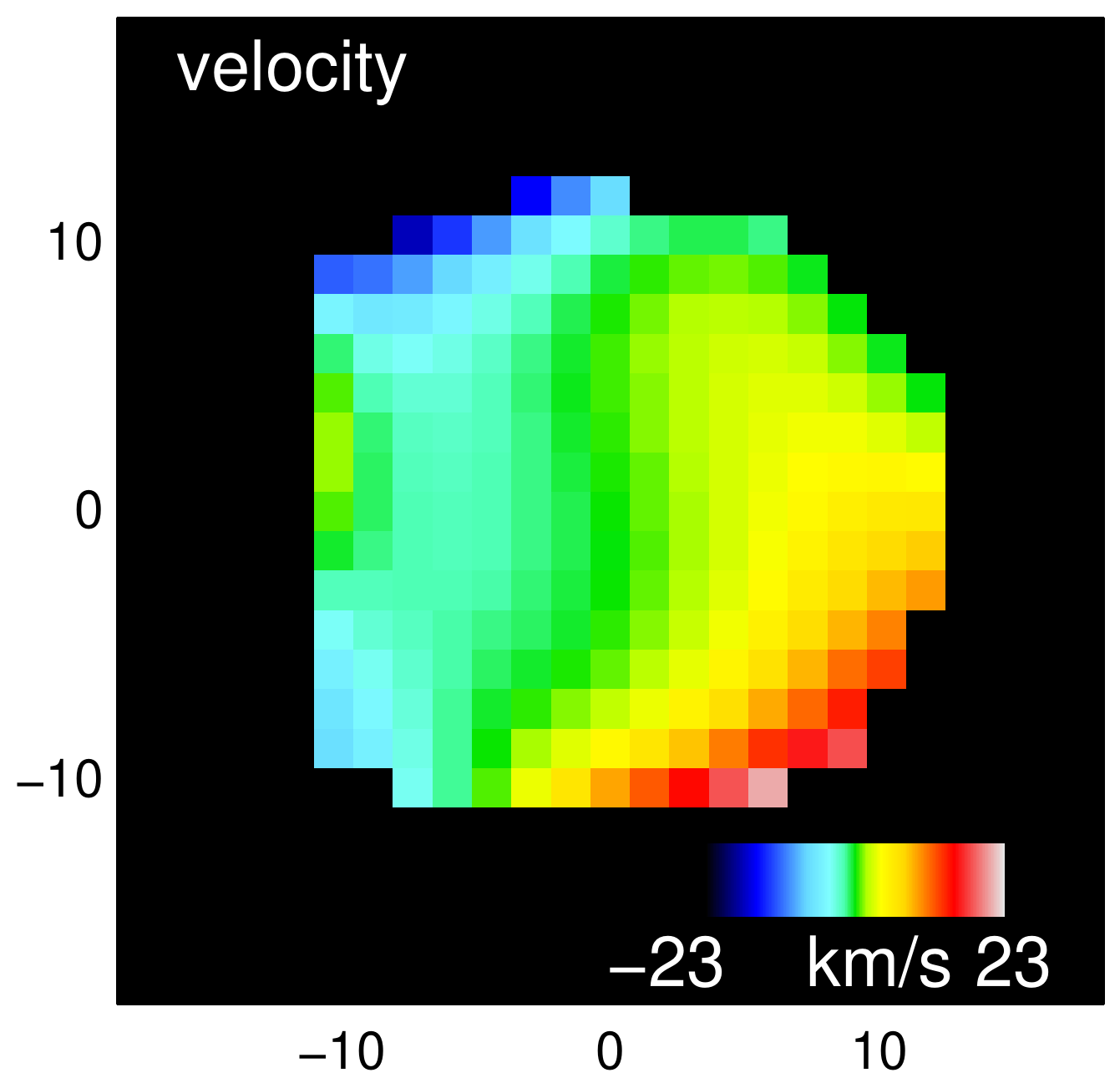}
\includegraphics[width=0.32\columnwidth]{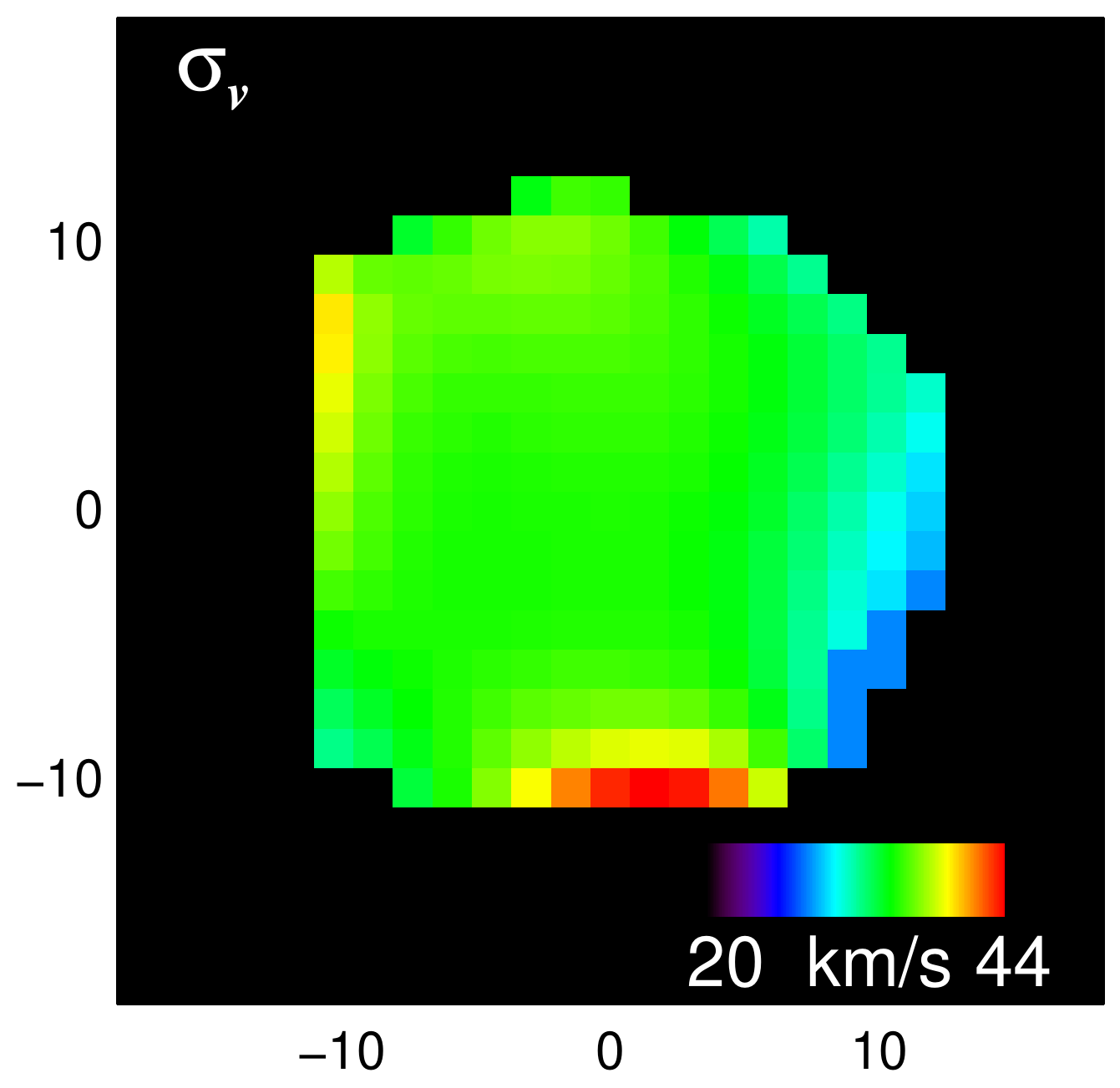}\\
\vspace{1mm}
\includegraphics[width=0.343\columnwidth]{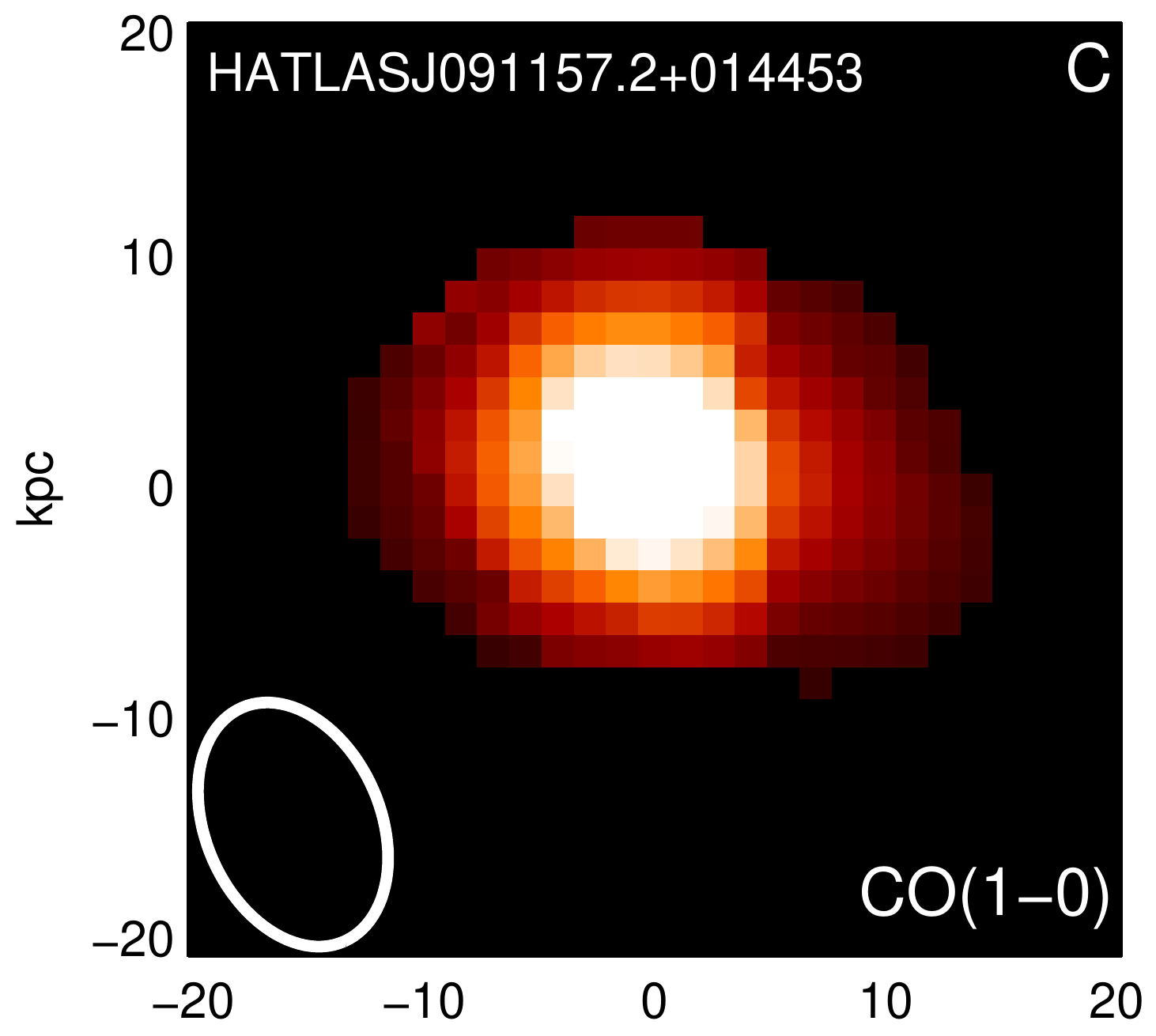}
\includegraphics[width=0.32\columnwidth]{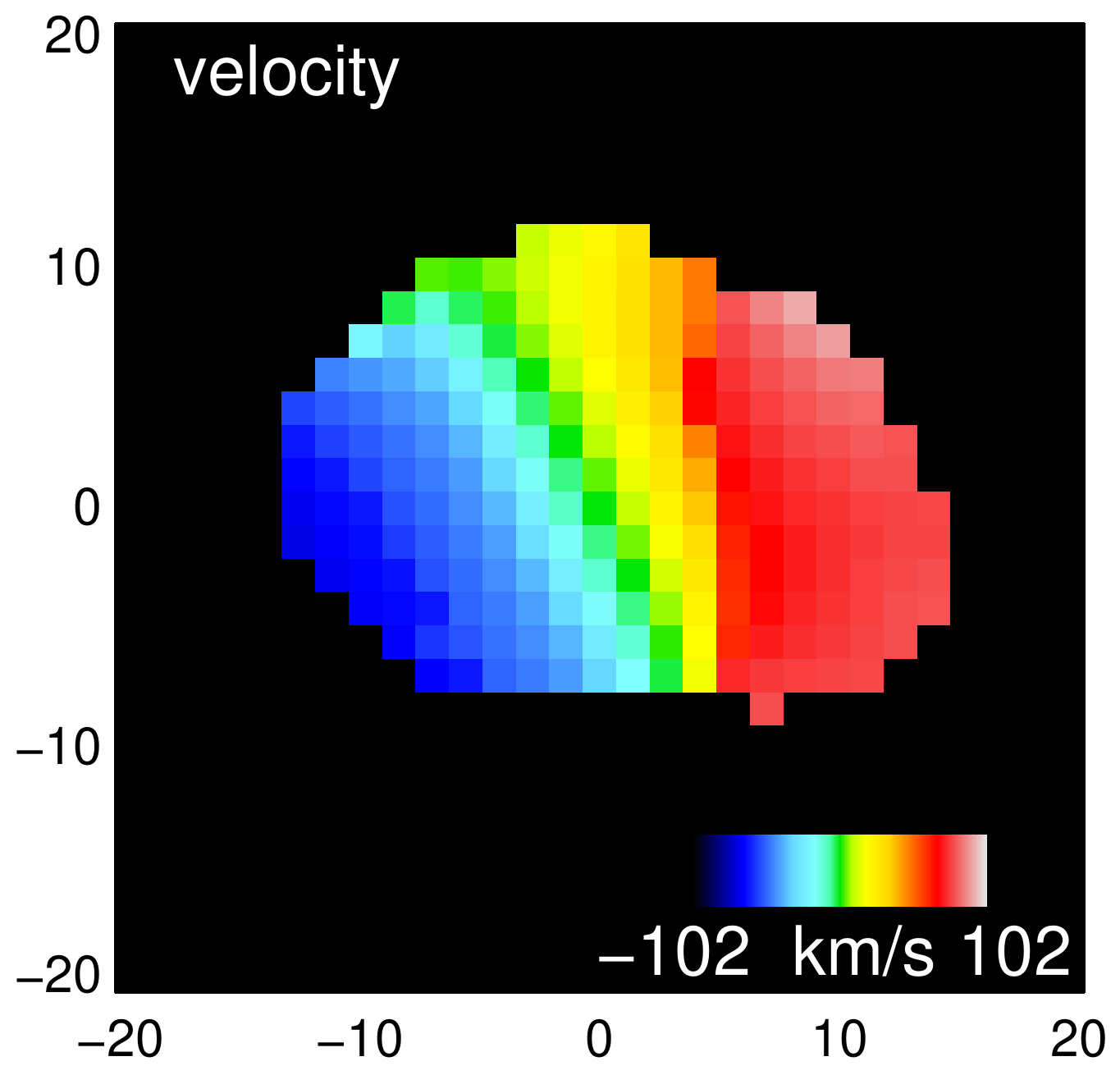}
\includegraphics[width=0.32\columnwidth]{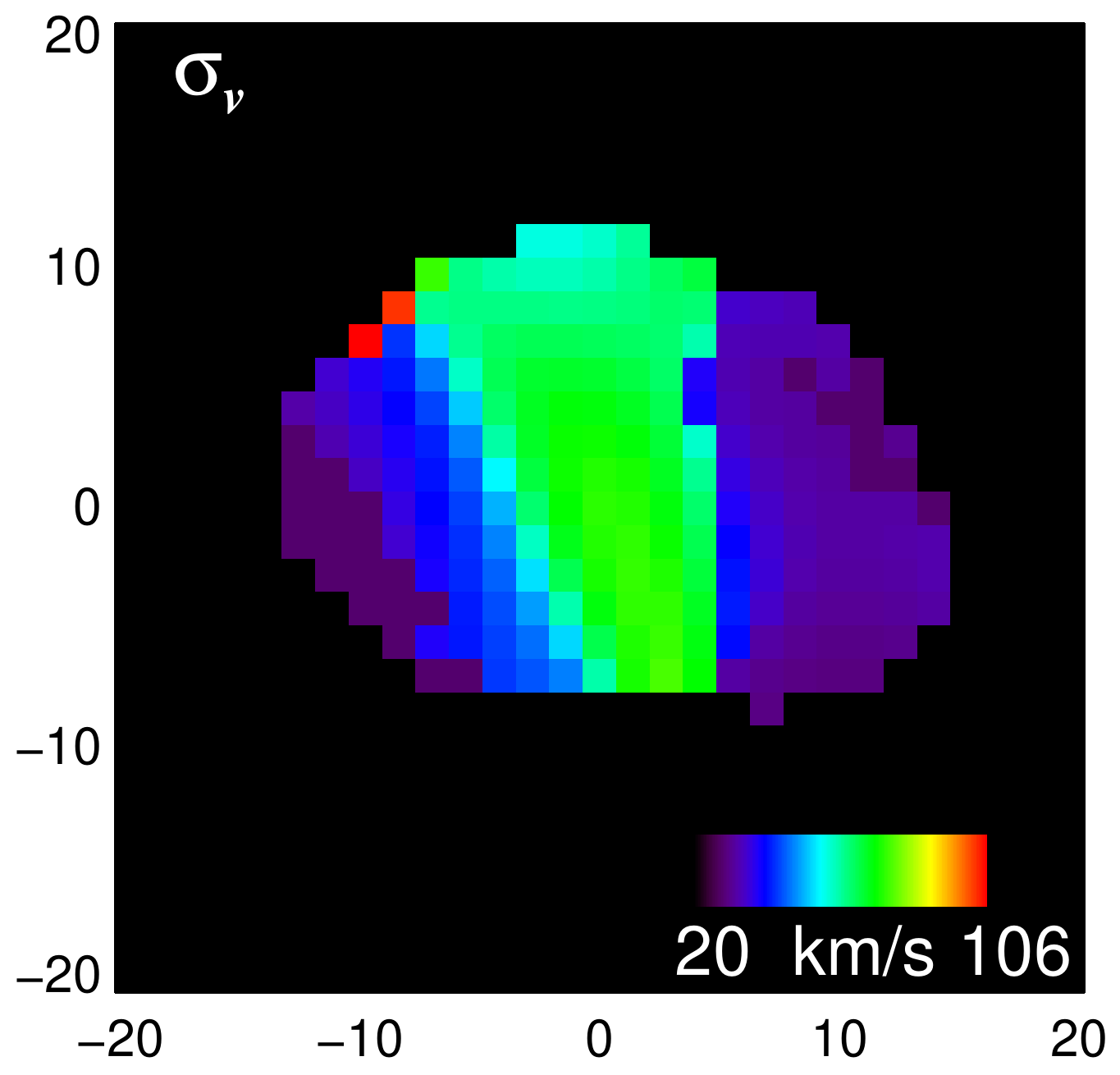}\\
\vspace{1mm}
\includegraphics[width=0.343\columnwidth]{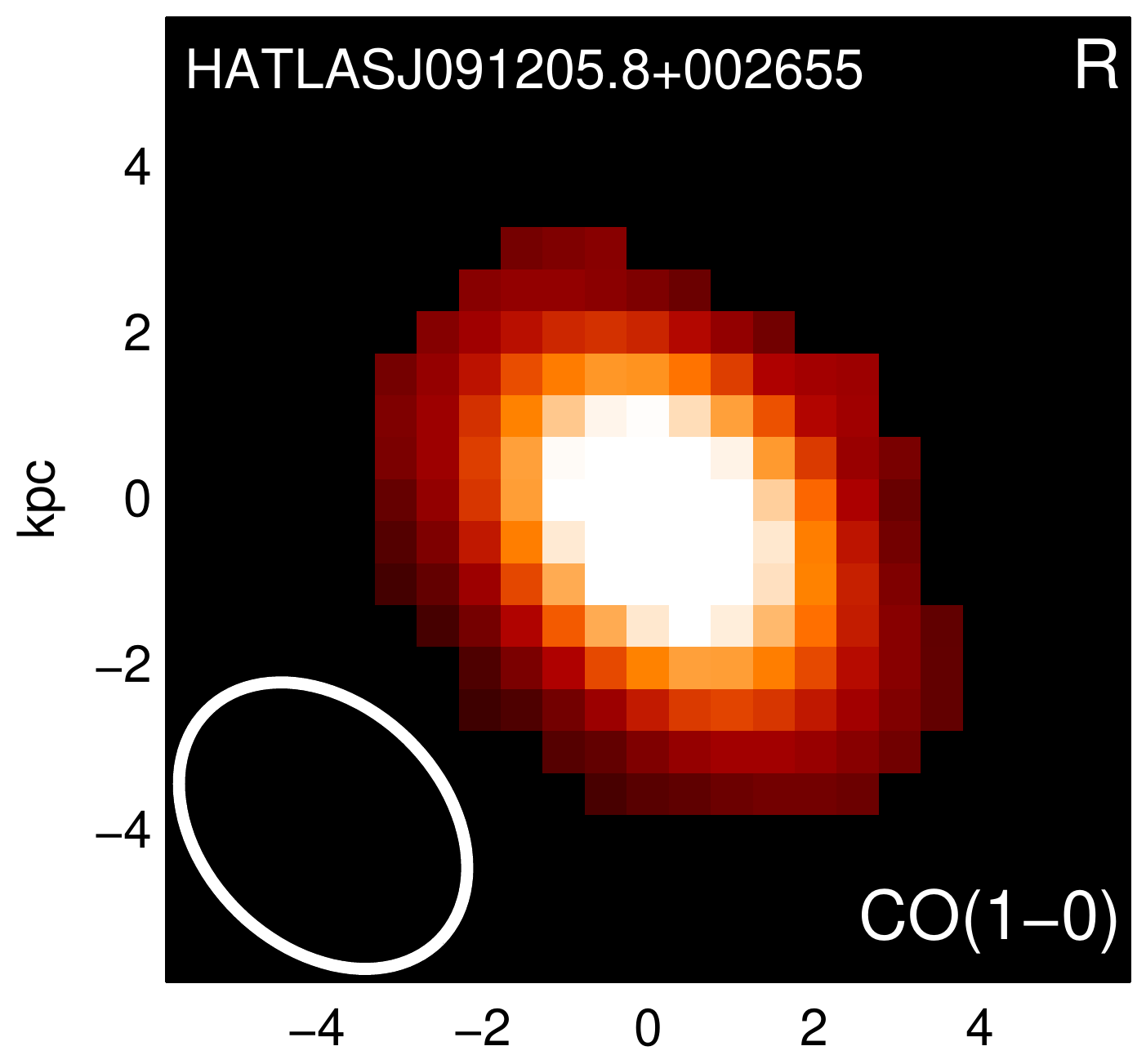}
\includegraphics[width=0.32\columnwidth]{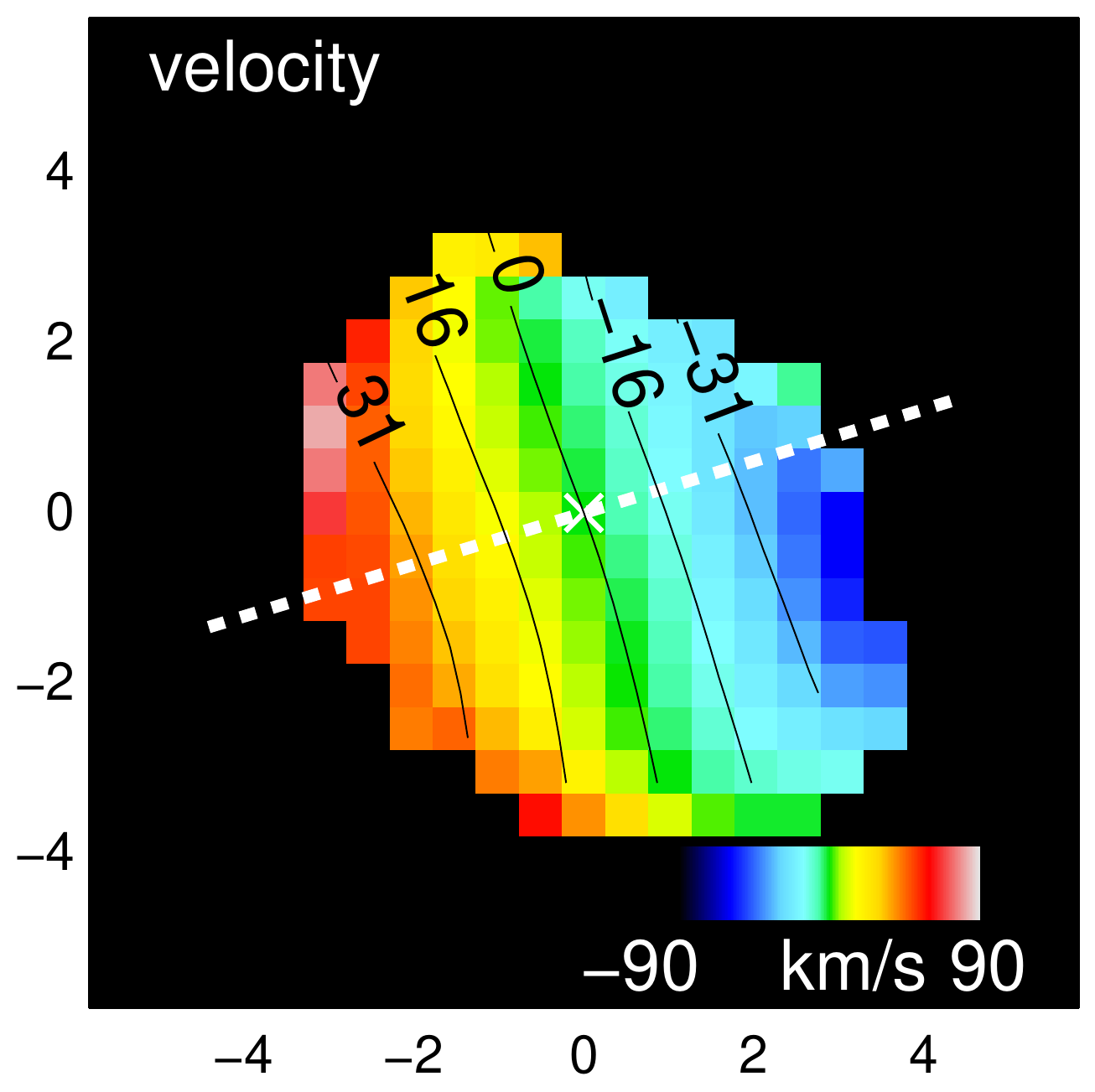}
\includegraphics[width=0.32\columnwidth]{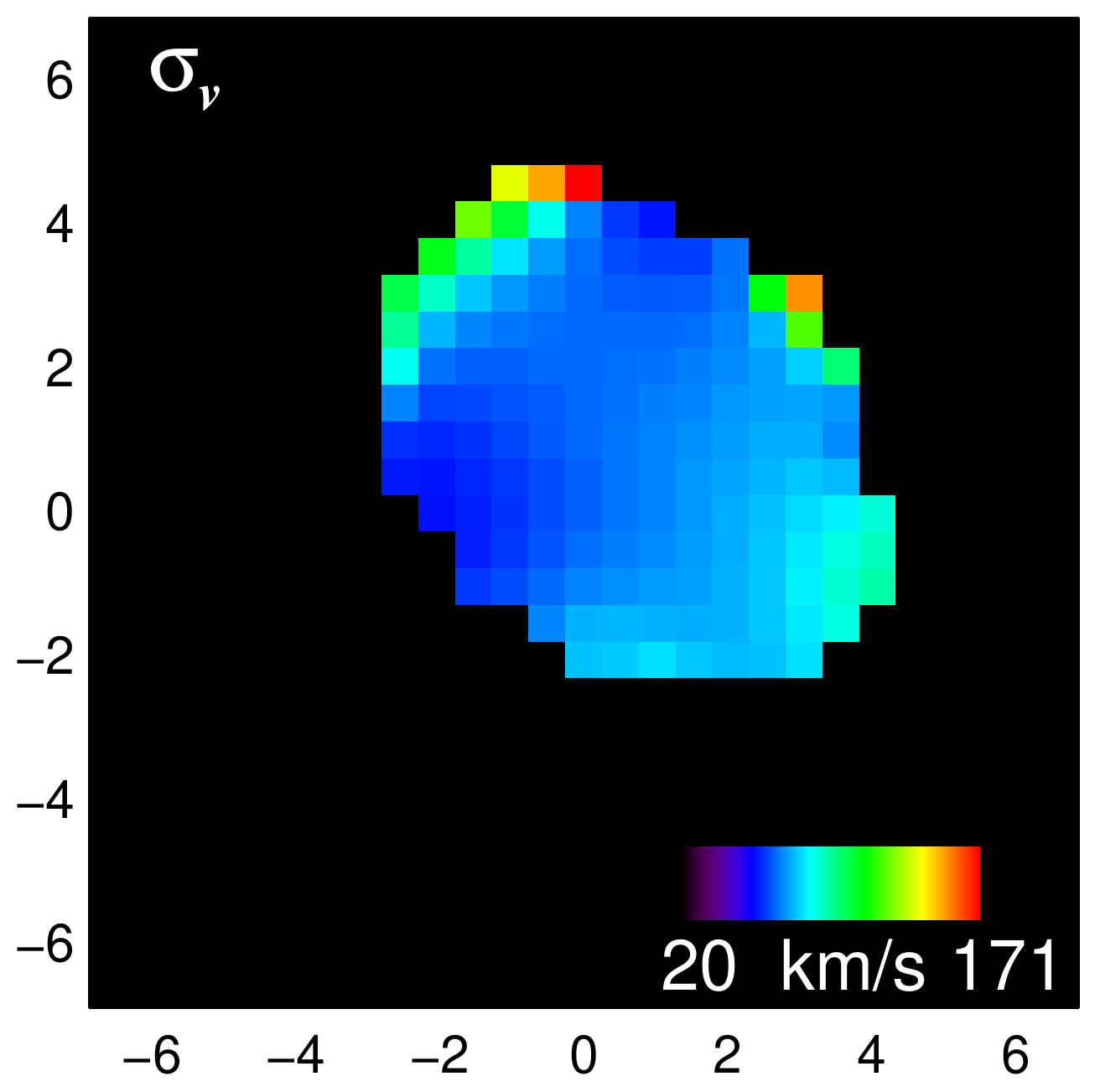}
\includegraphics[width=0.32\columnwidth]{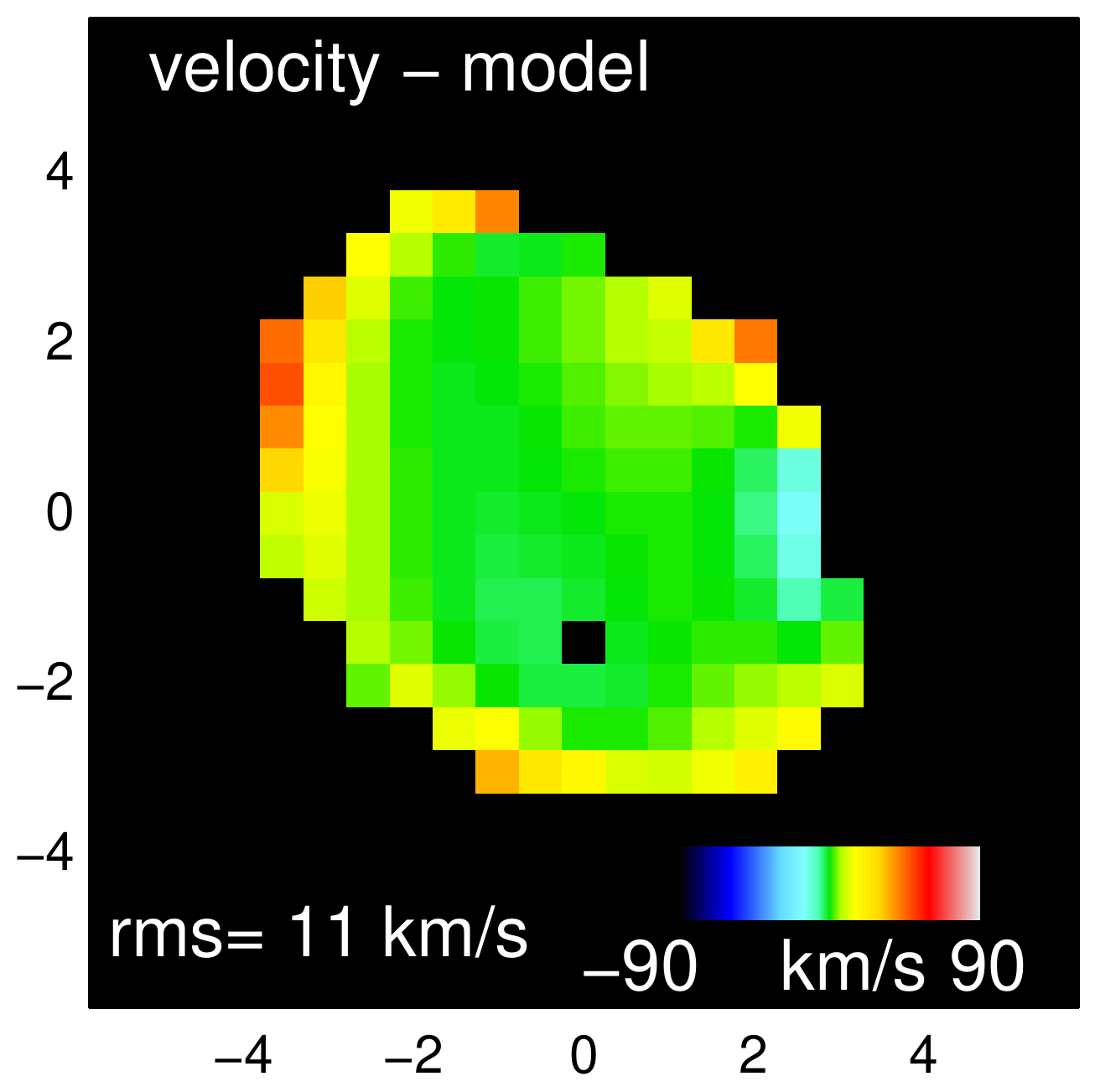}
\includegraphics[width=0.352\columnwidth]{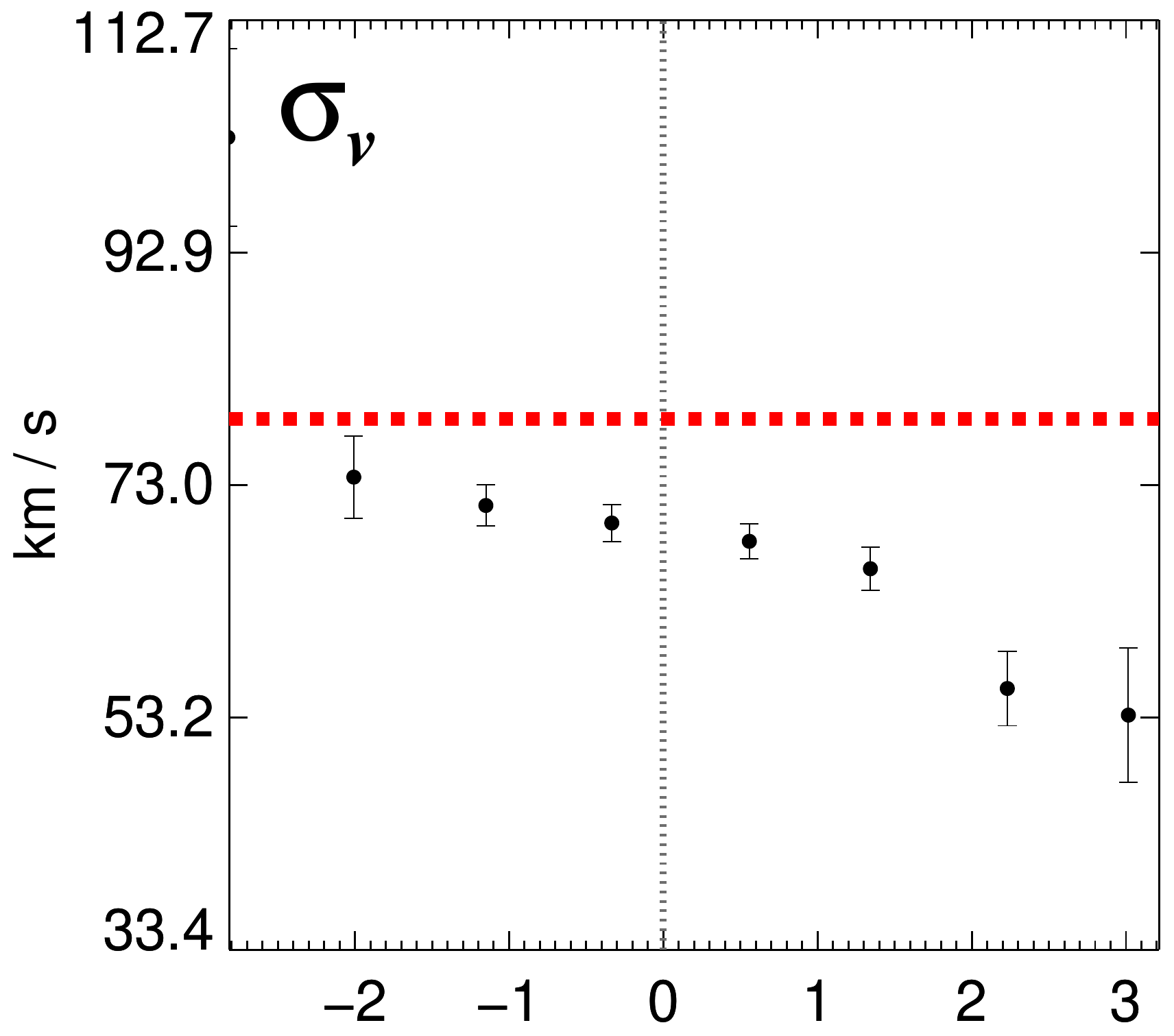}
\includegraphics[width=0.351\columnwidth]{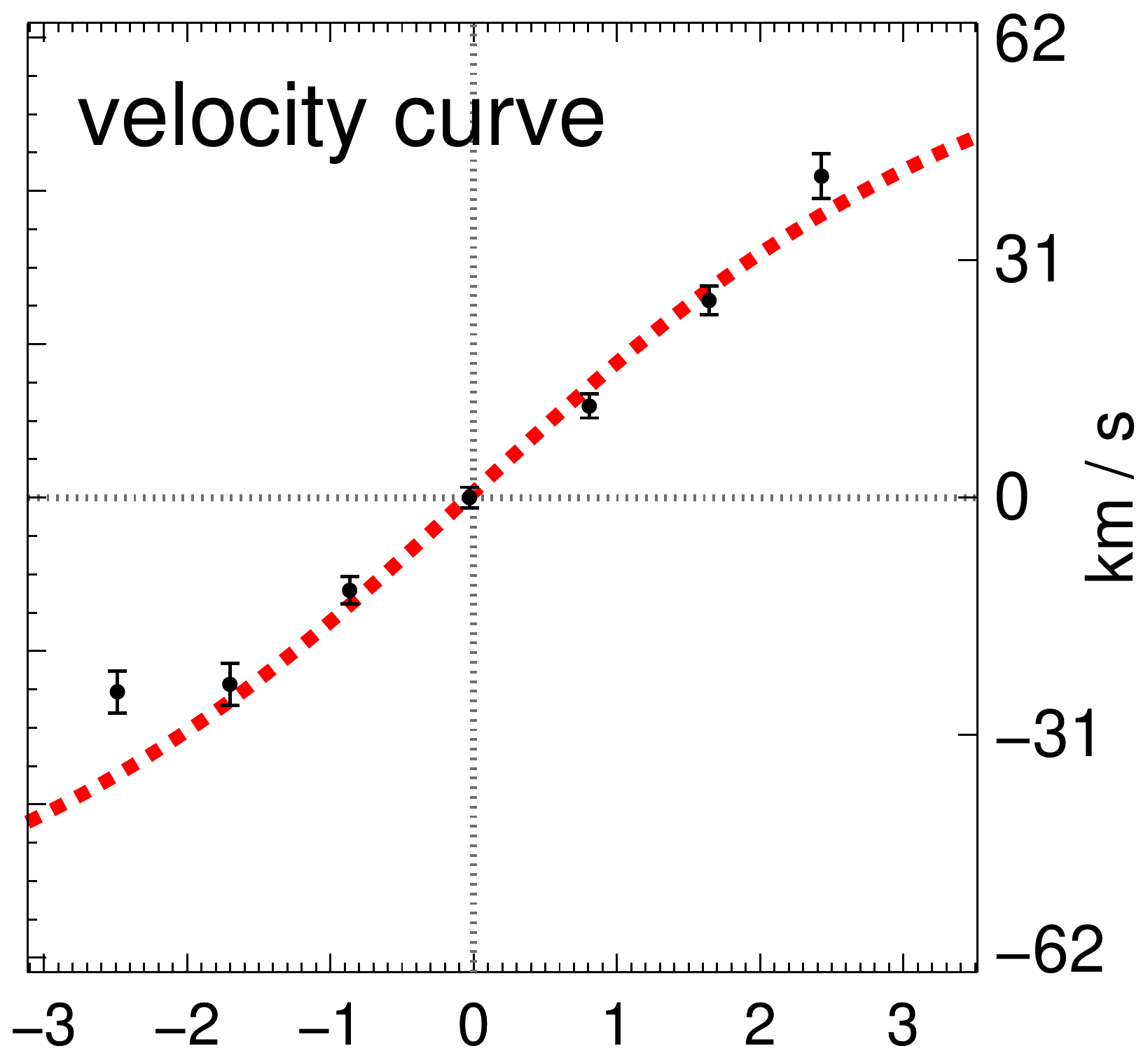}\\
\vspace{1mm}
\includegraphics[width=0.343\columnwidth]{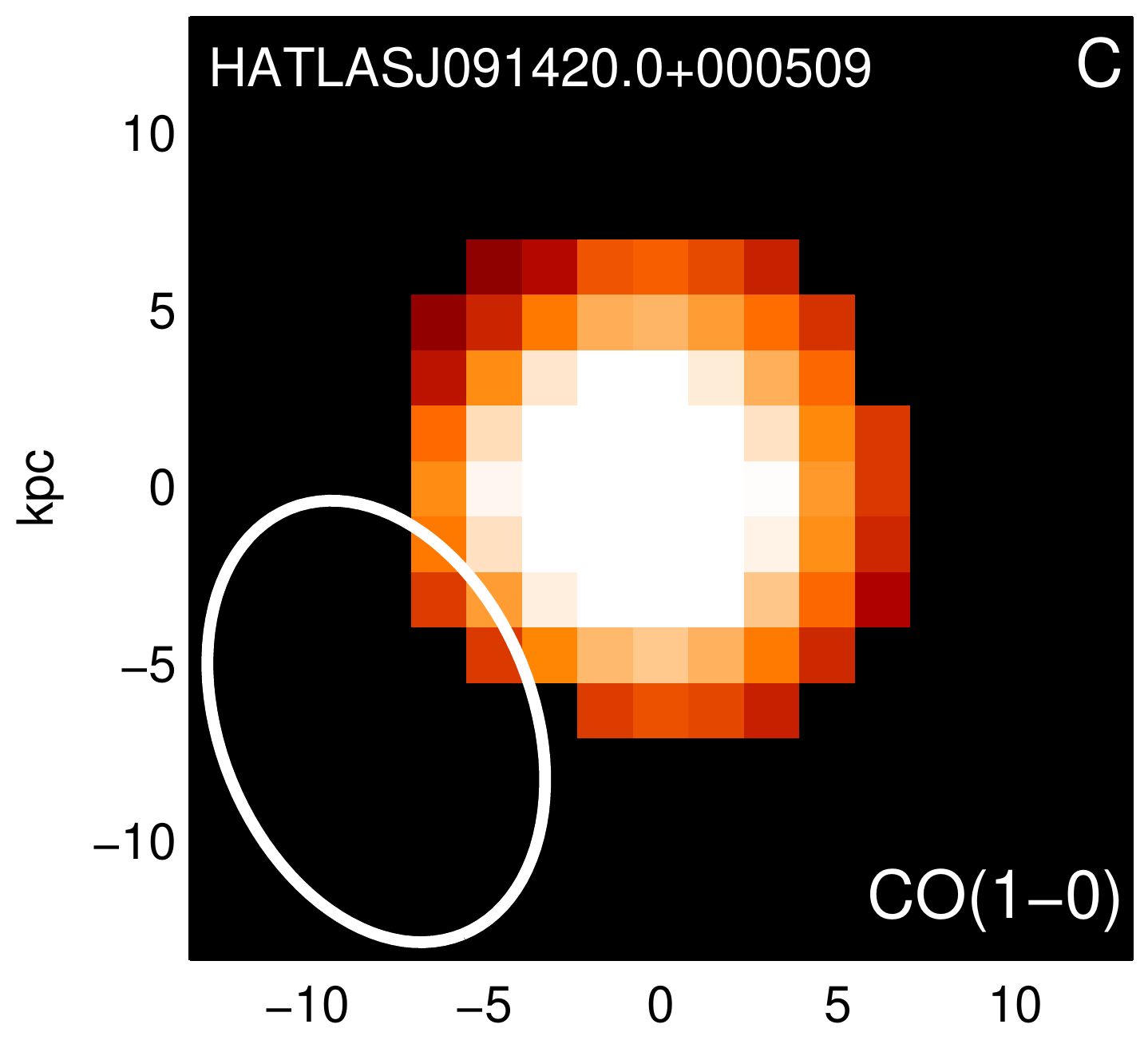}
\includegraphics[width=0.32\columnwidth]{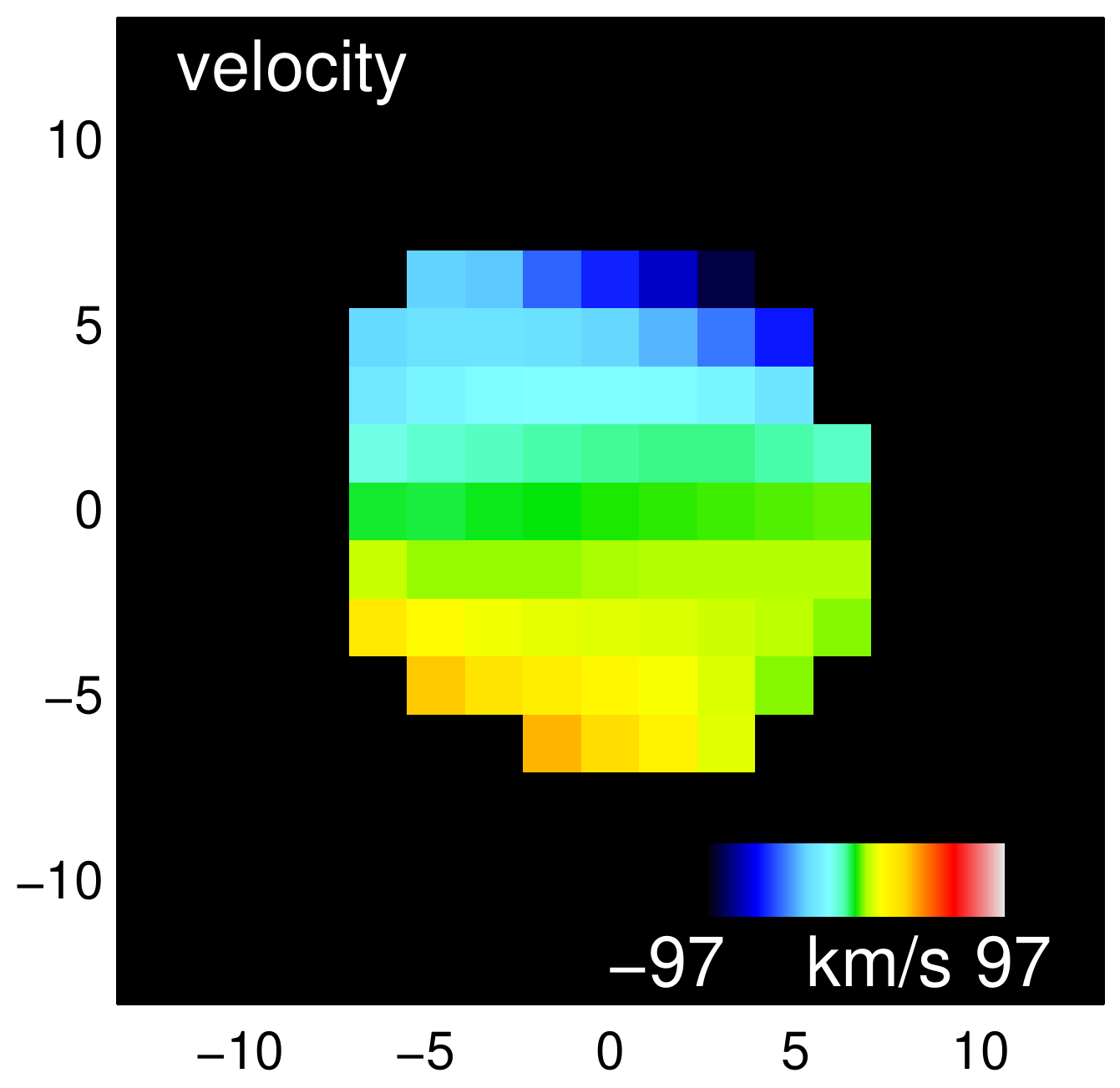}
\includegraphics[width=0.32\columnwidth]{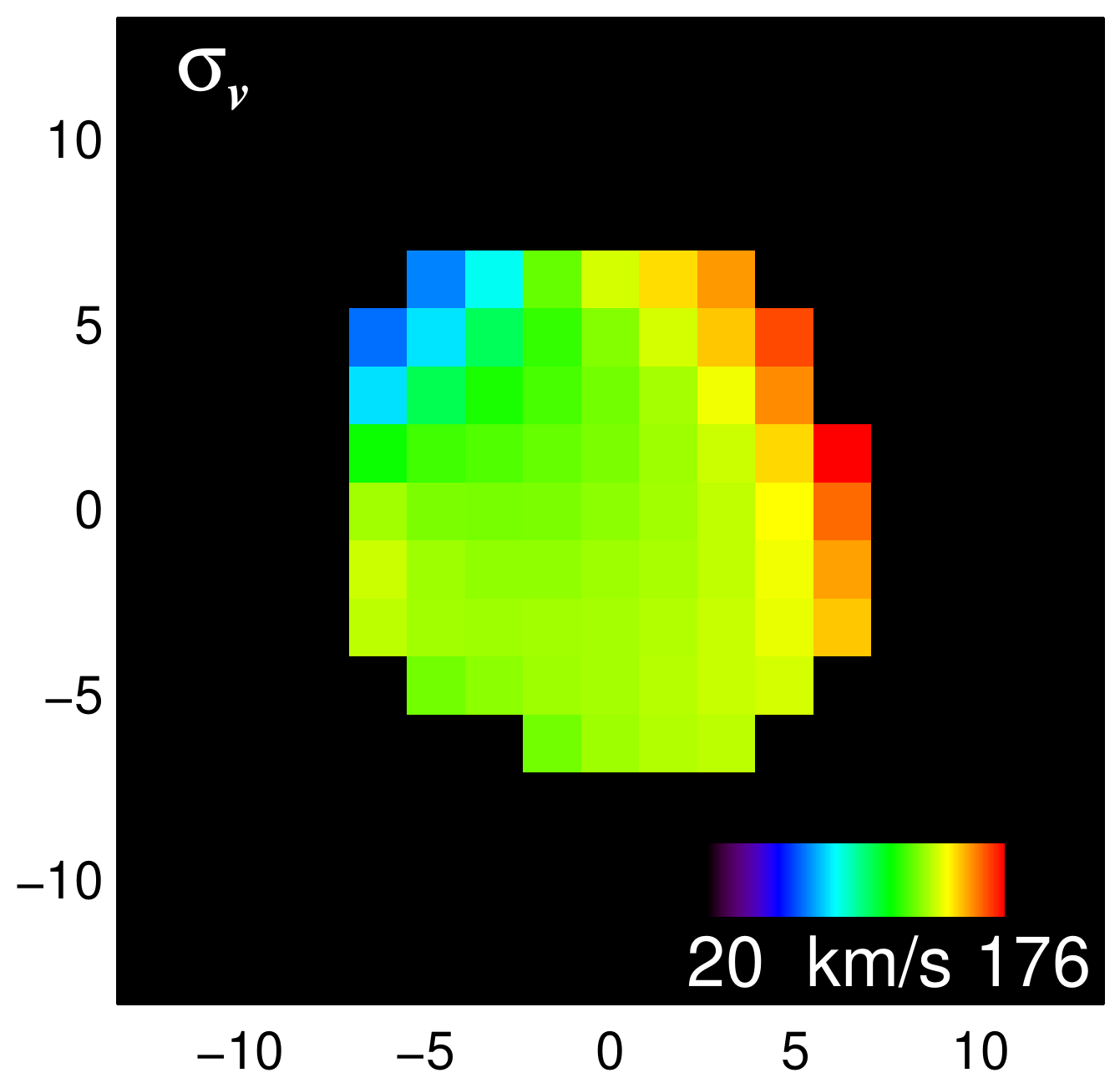}\\
\vspace{1mm}
\includegraphics[width=0.343\columnwidth]{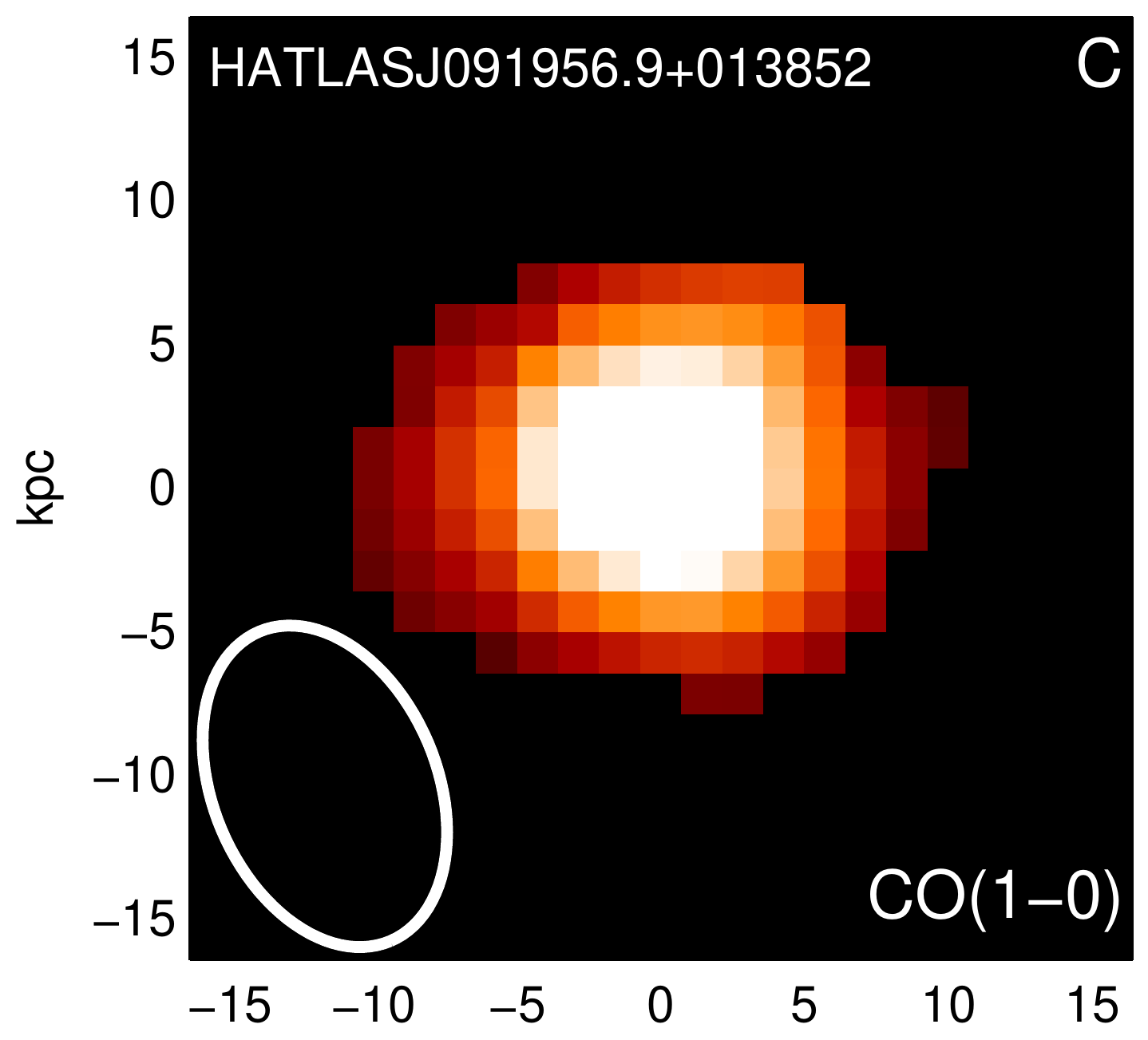}
\includegraphics[width=0.32\columnwidth]{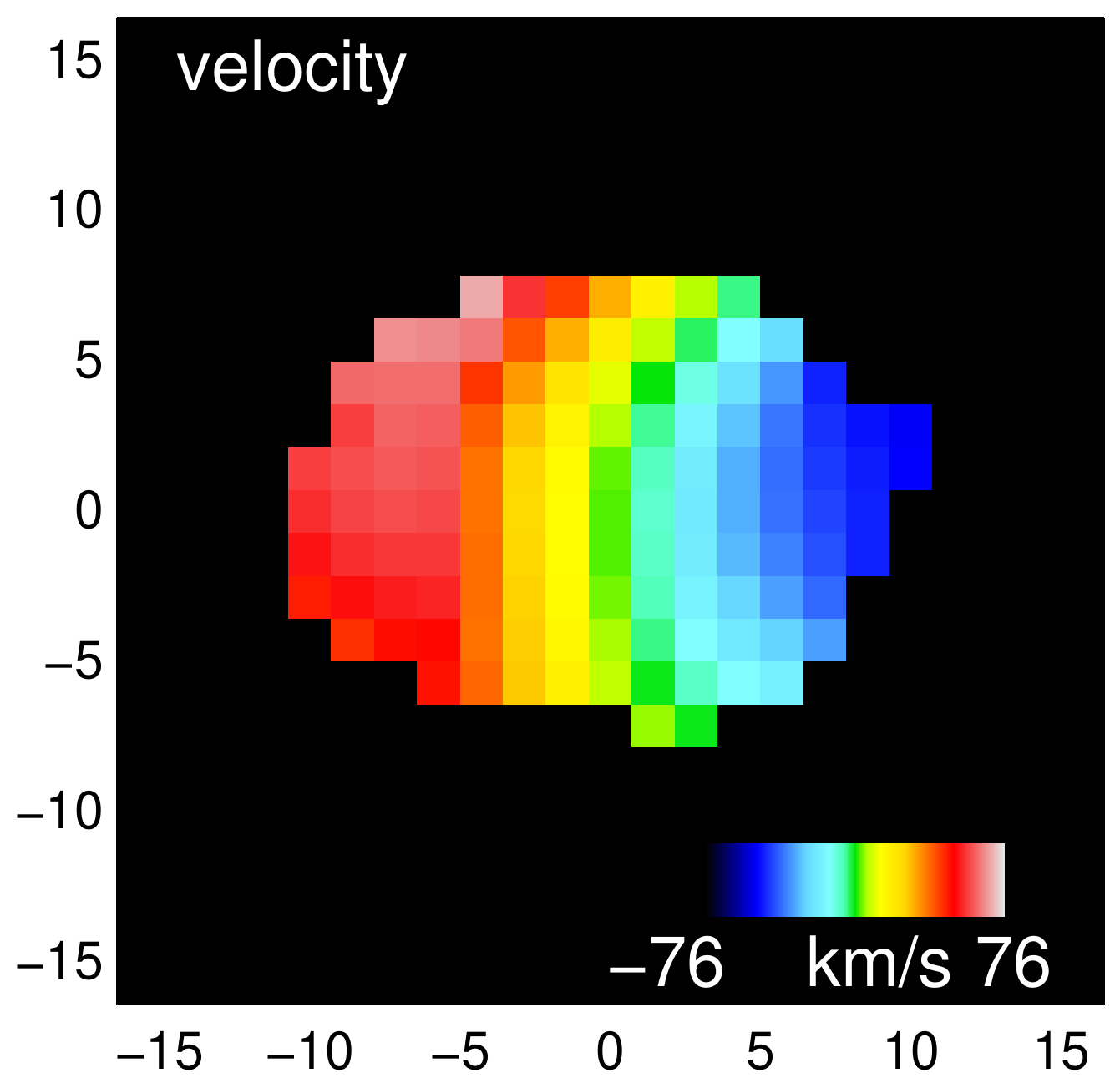}
\includegraphics[width=0.32\columnwidth]{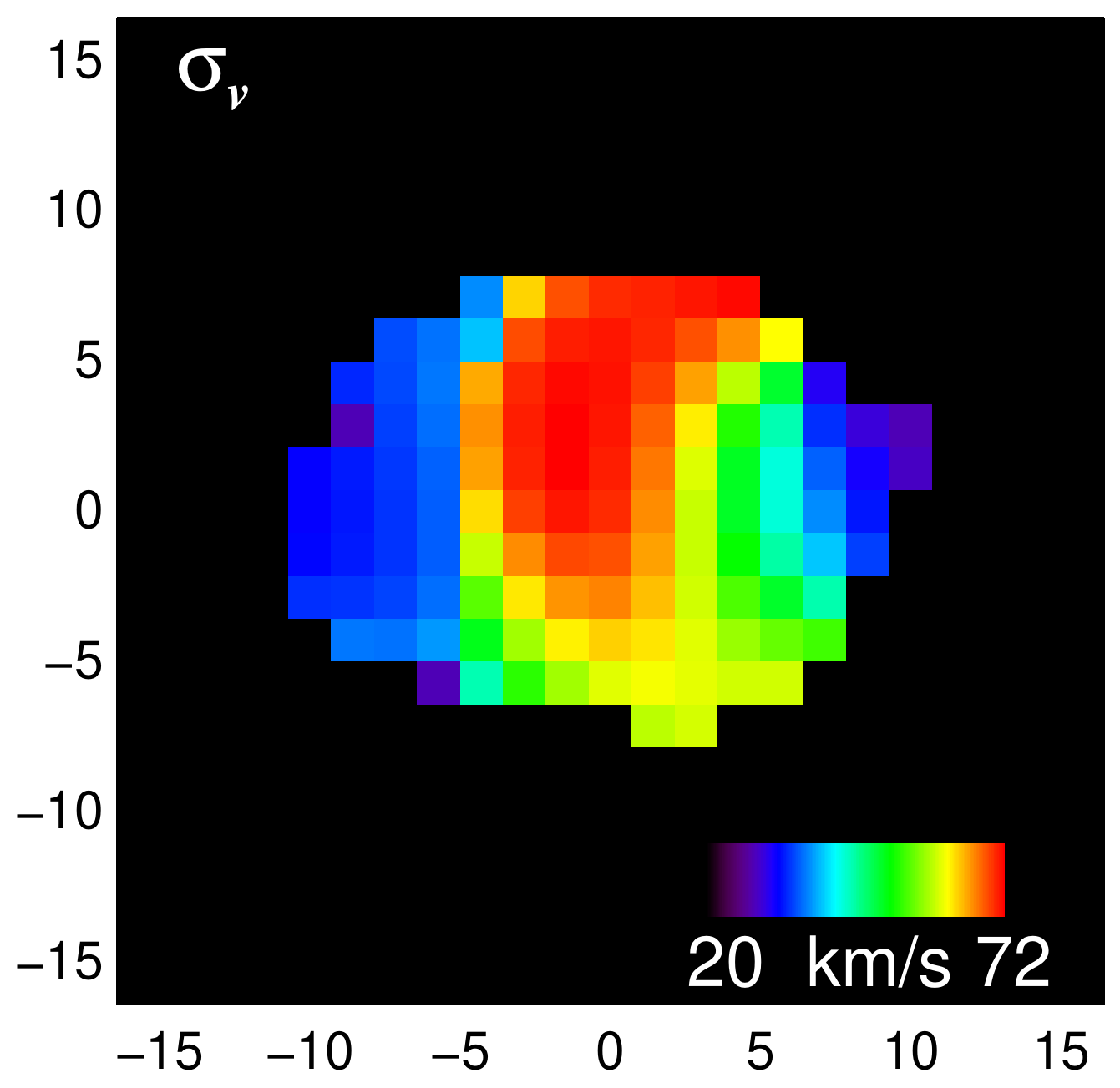}\\
\vspace{1mm}
\includegraphics[width=0.343\columnwidth]{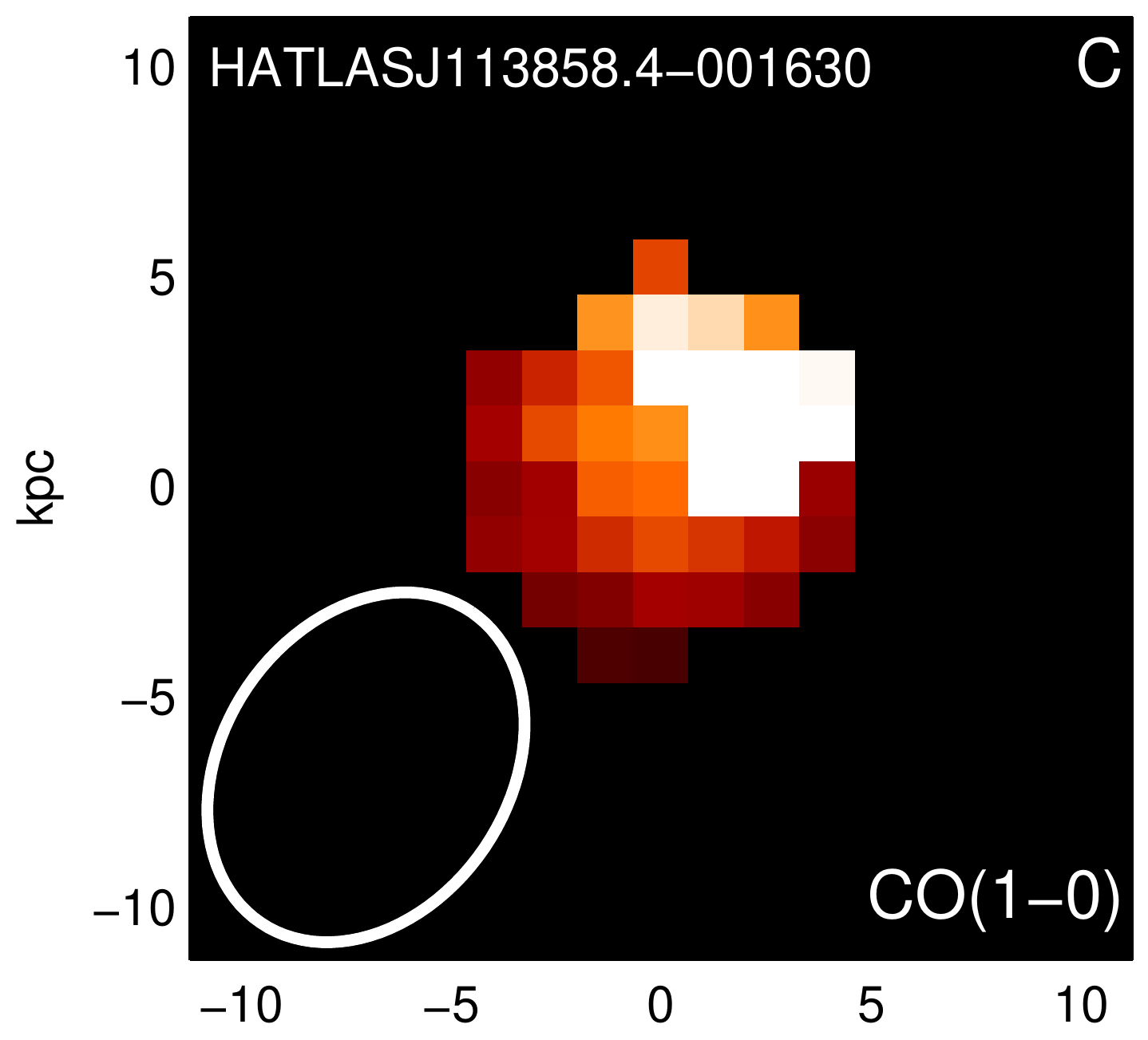}
\includegraphics[width=0.32\columnwidth]{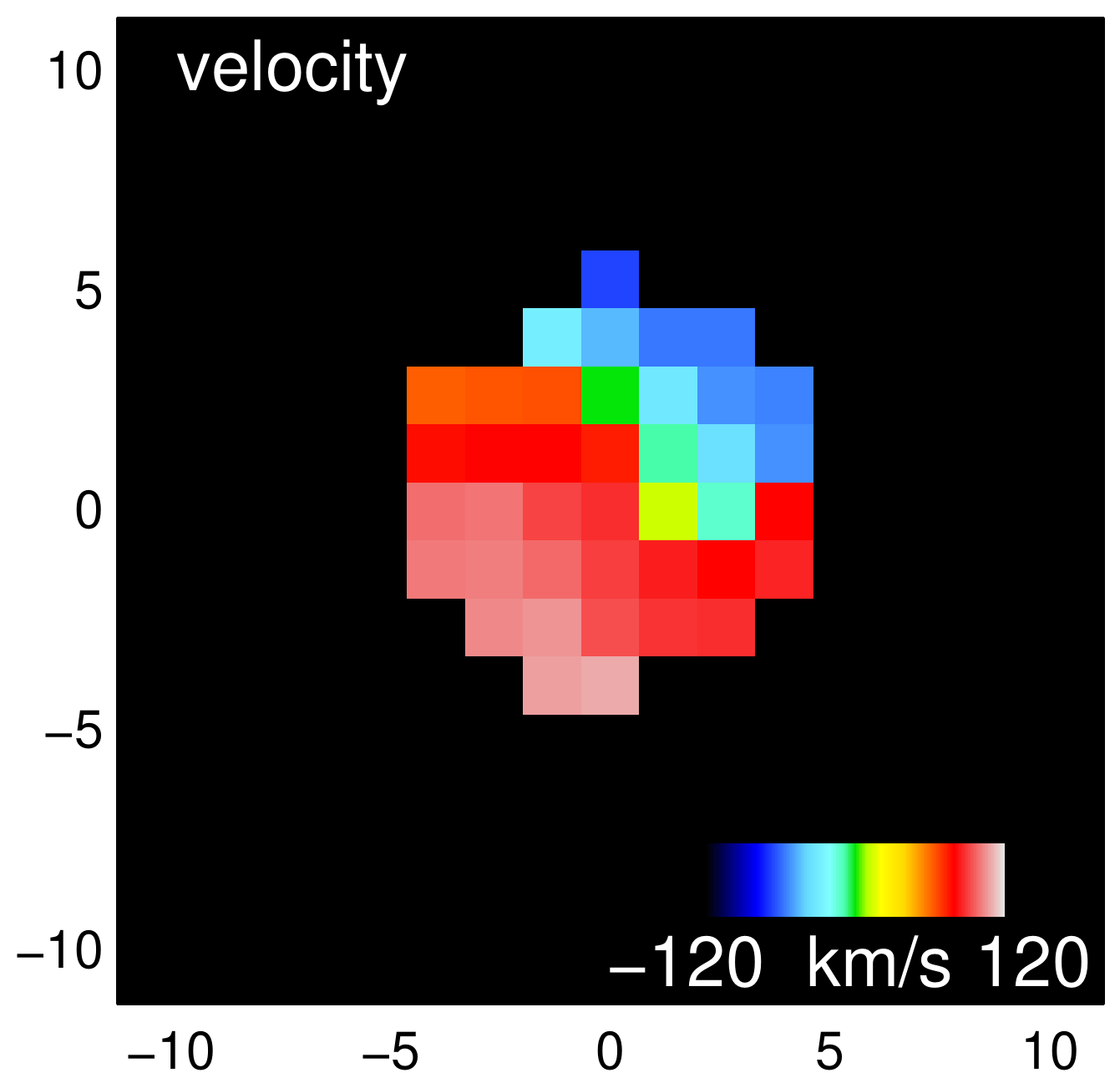}
\includegraphics[width=0.32\columnwidth]{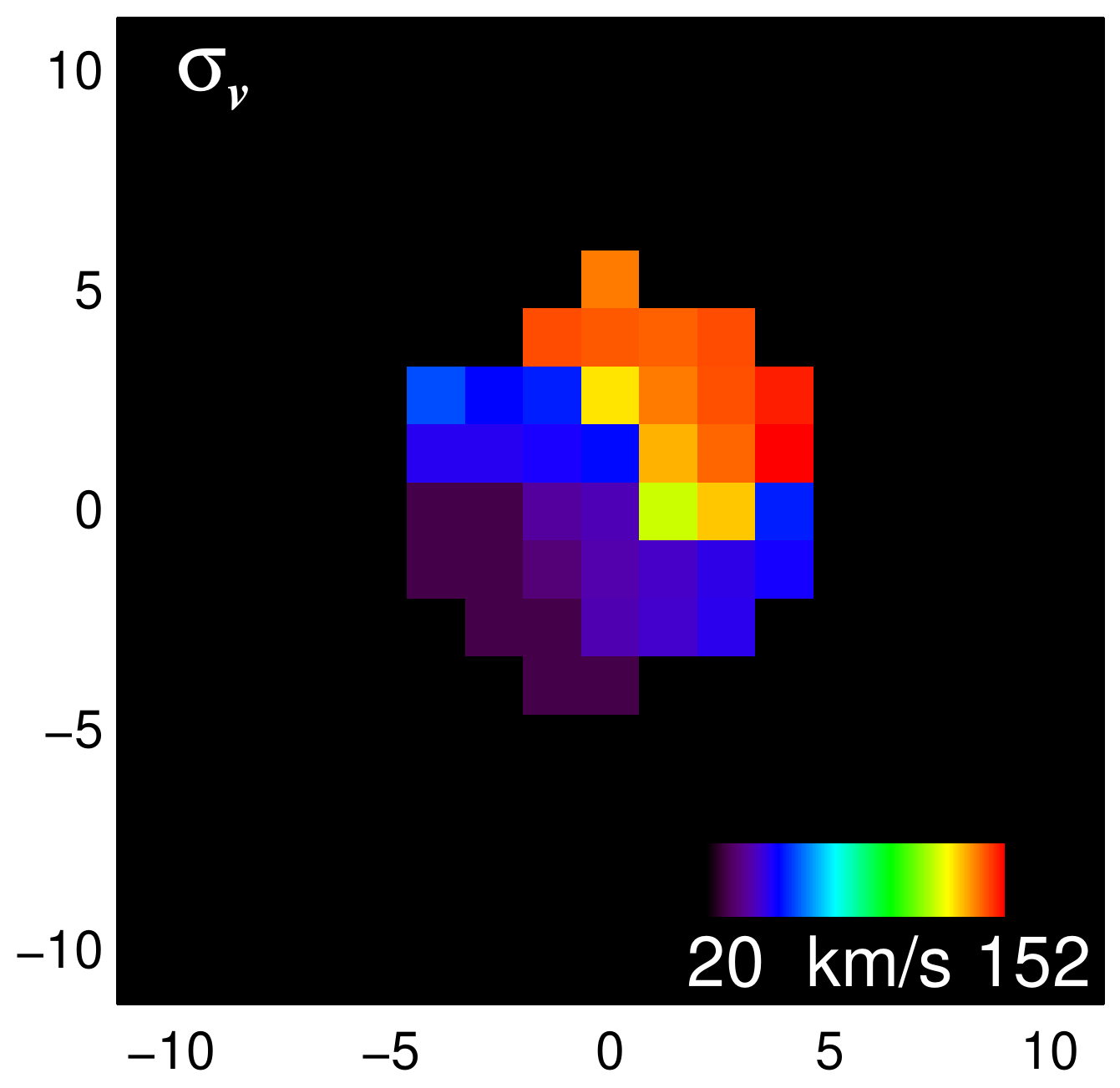}\\
\vspace{1mm}
\includegraphics[width=0.343\columnwidth]{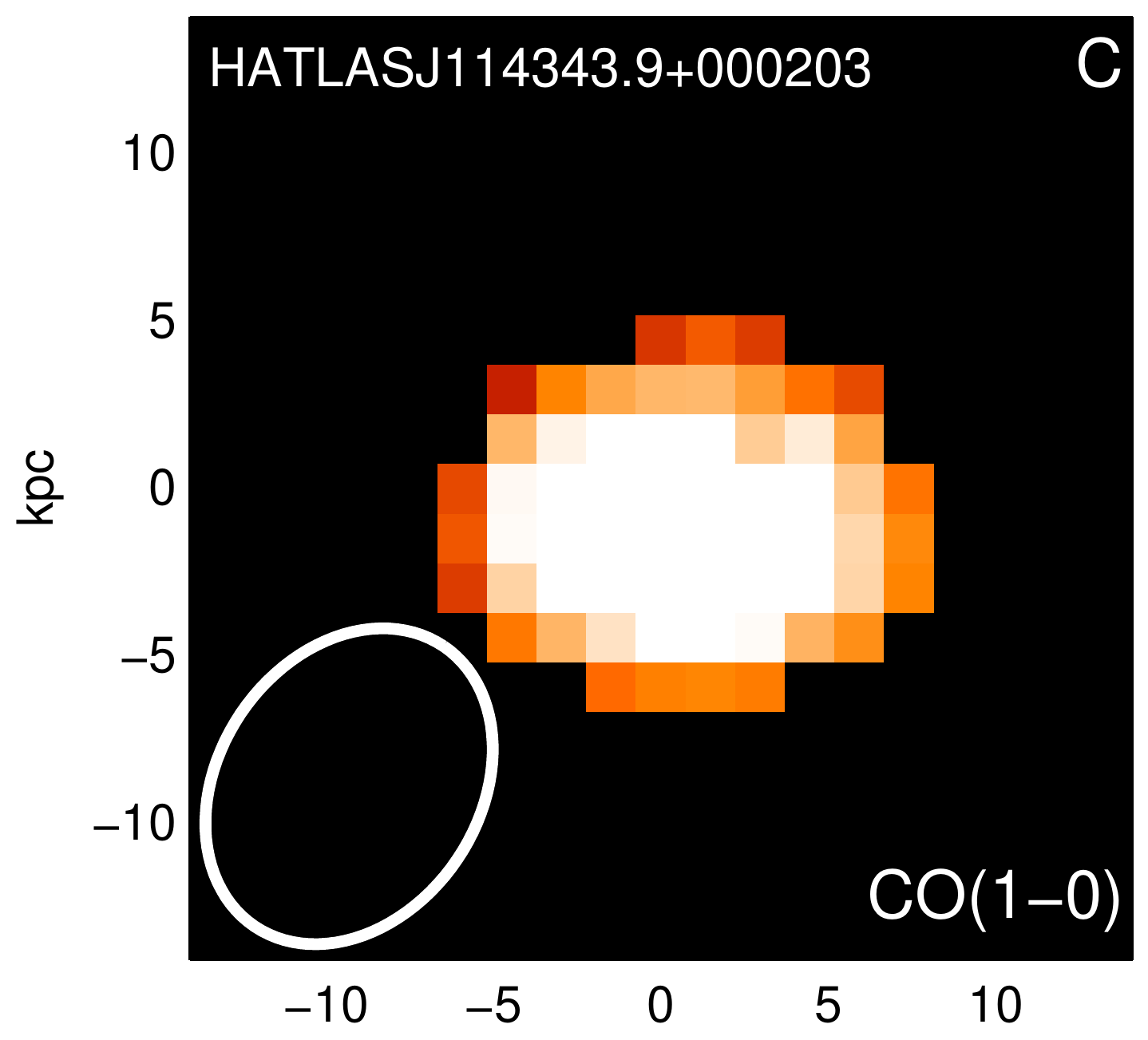}
\includegraphics[width=0.32\columnwidth]{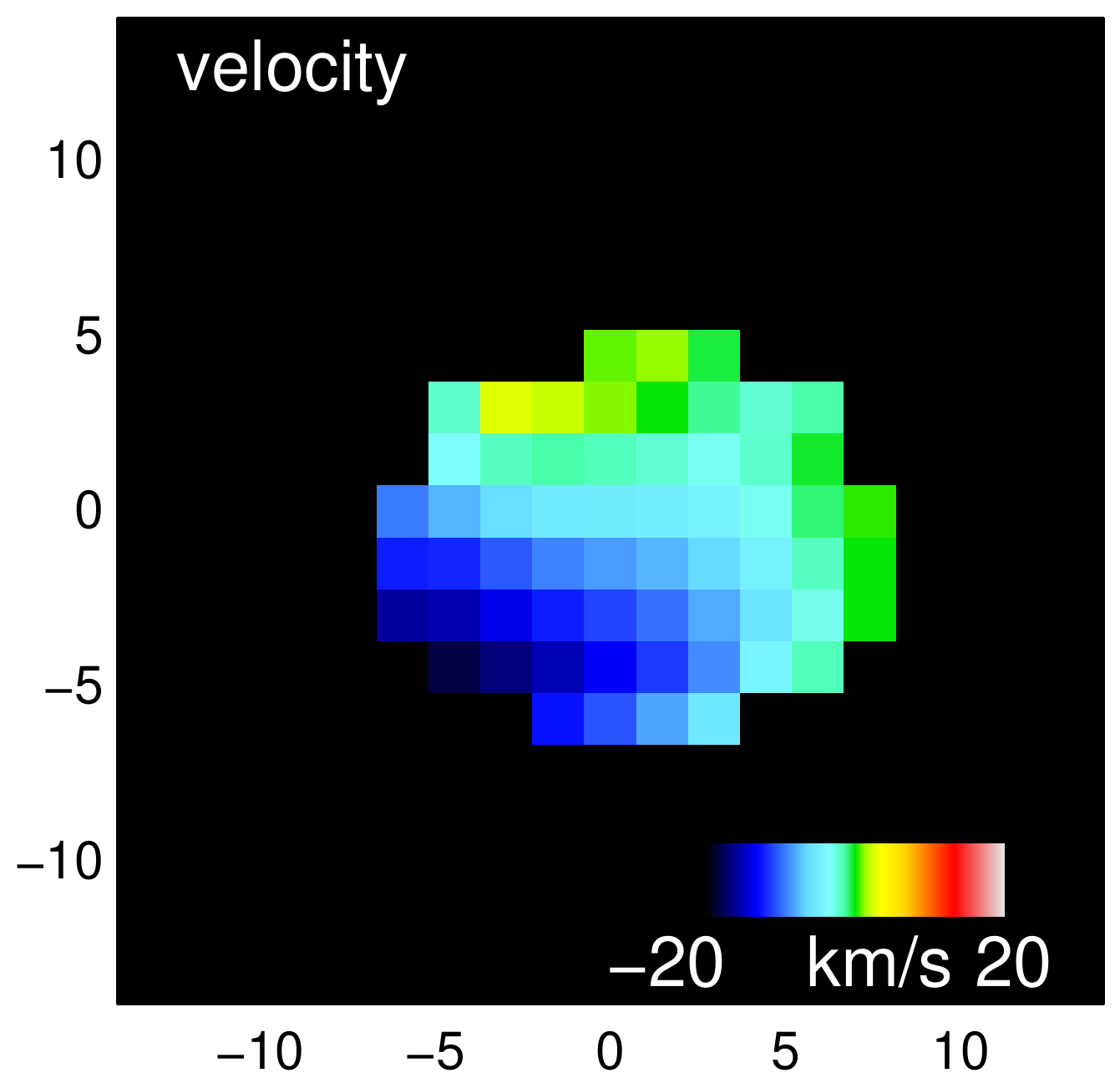}
\includegraphics[width=0.32\columnwidth]{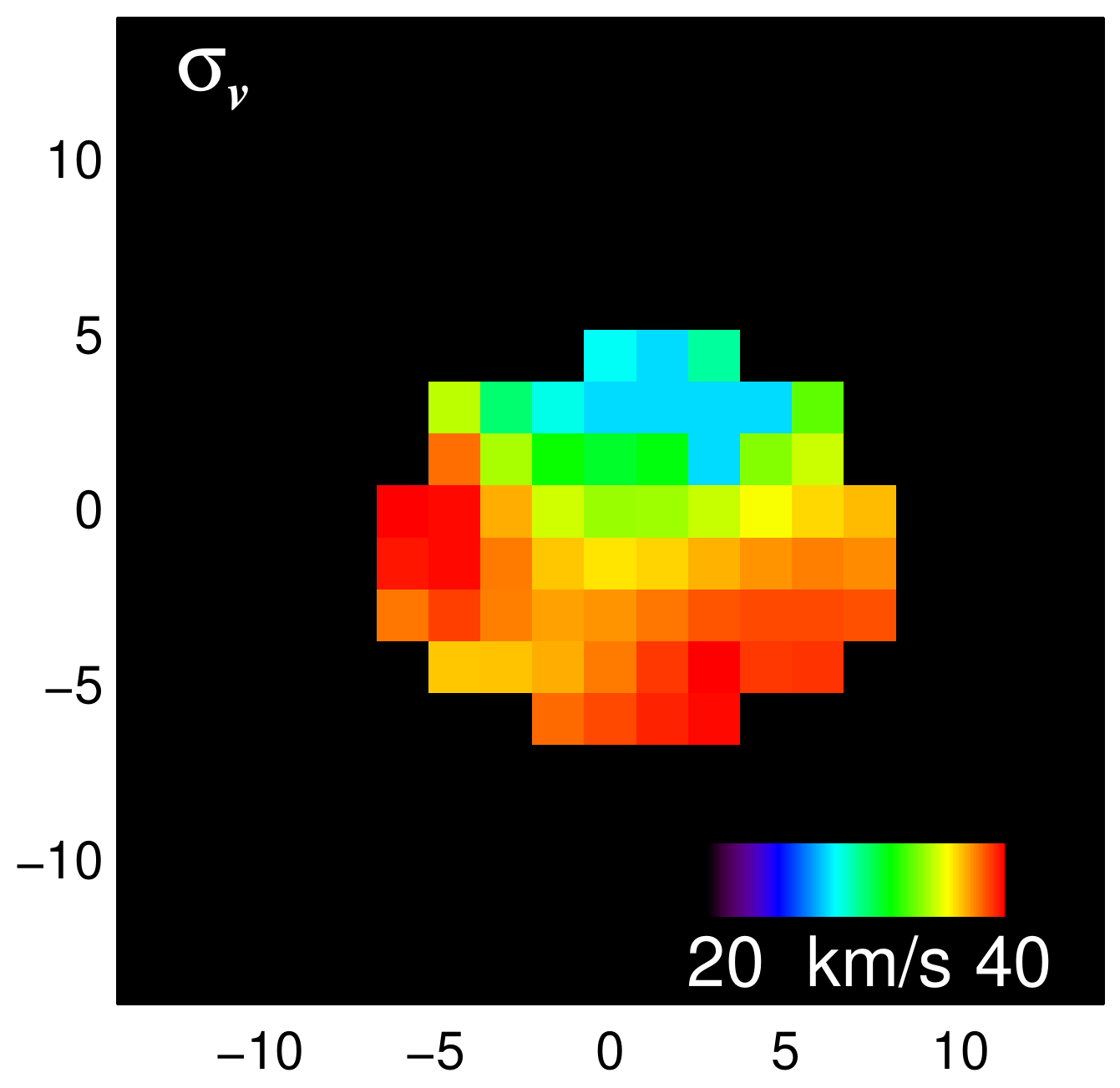}\\
\vspace{1mm}
\includegraphics[width=0.343\columnwidth]{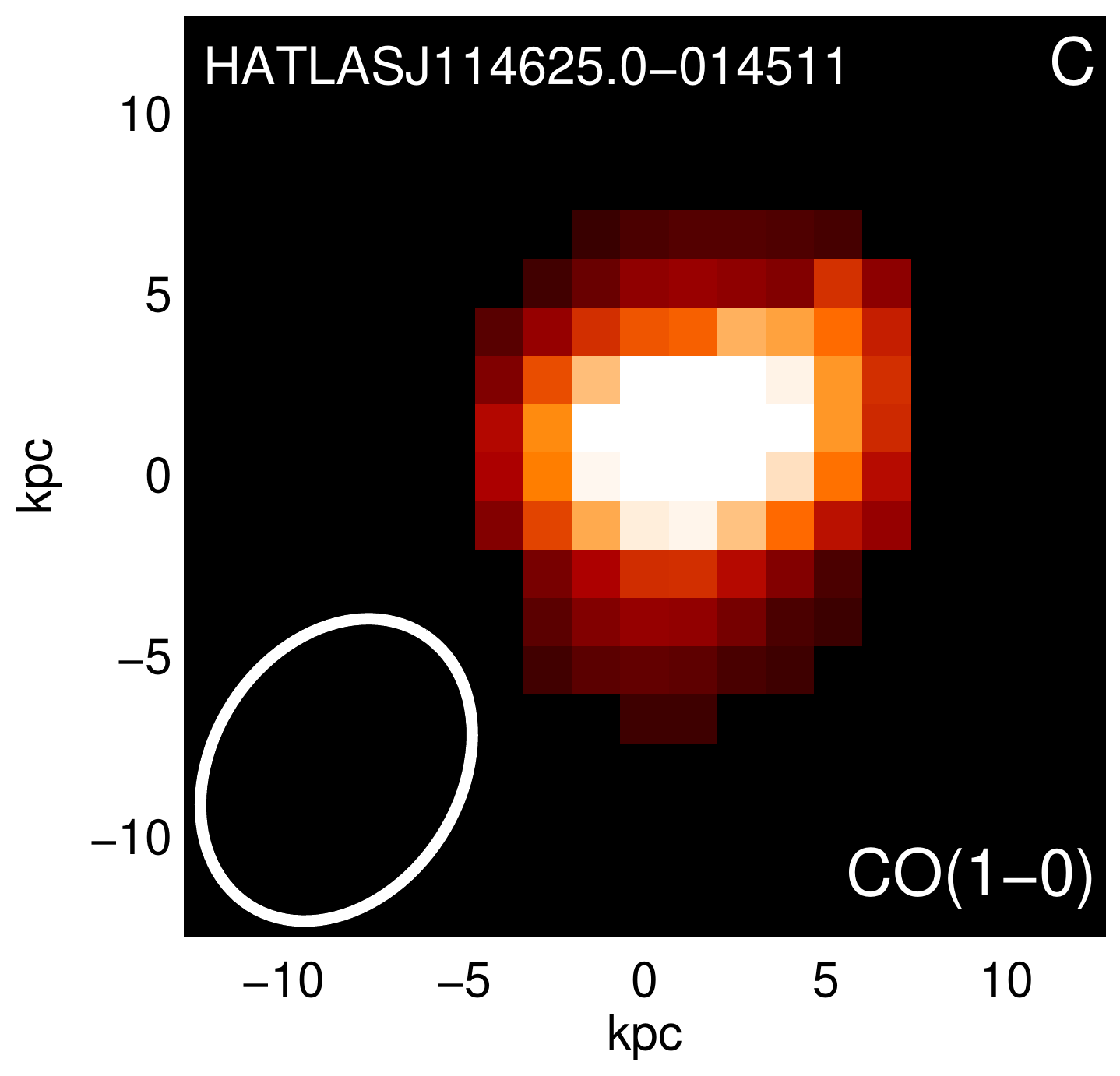}
\includegraphics[width=0.32\columnwidth]{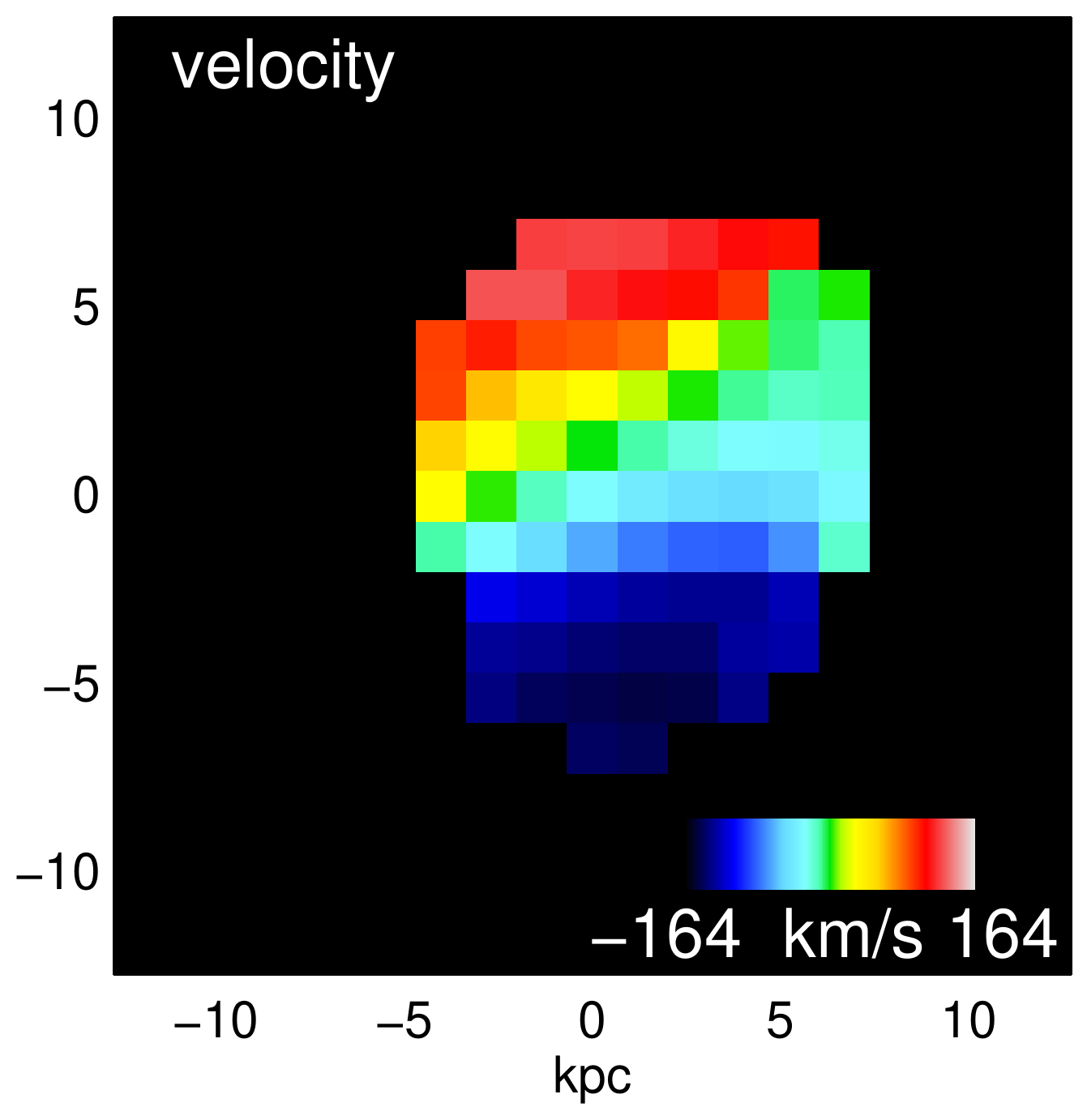}
\includegraphics[width=0.32\columnwidth]{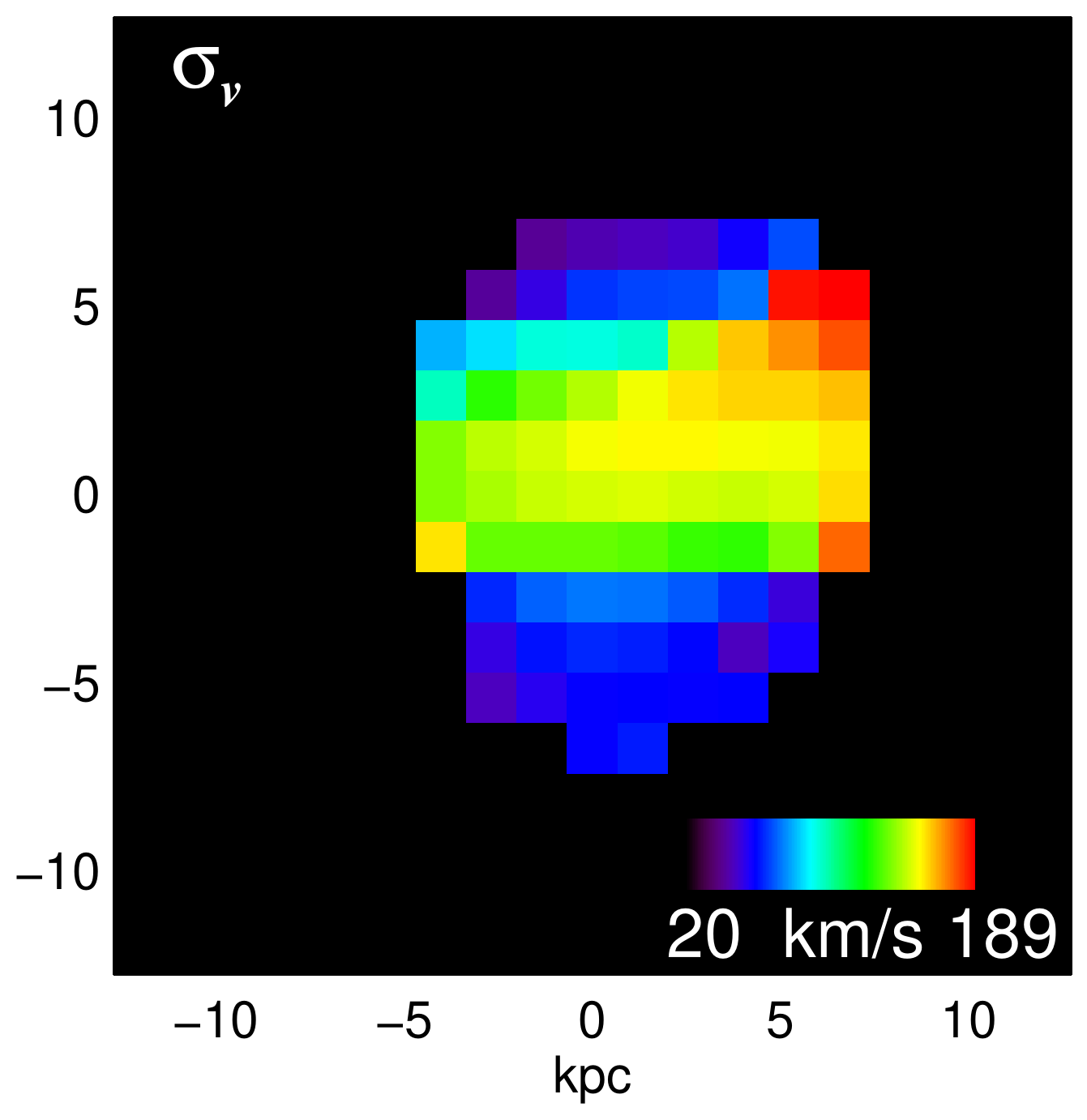}\\
\centering{\textbf{Figure C1. Continued.}}
\end{figure*}

\begin{figure*}
\flushleft
\includegraphics[width=0.343\columnwidth]{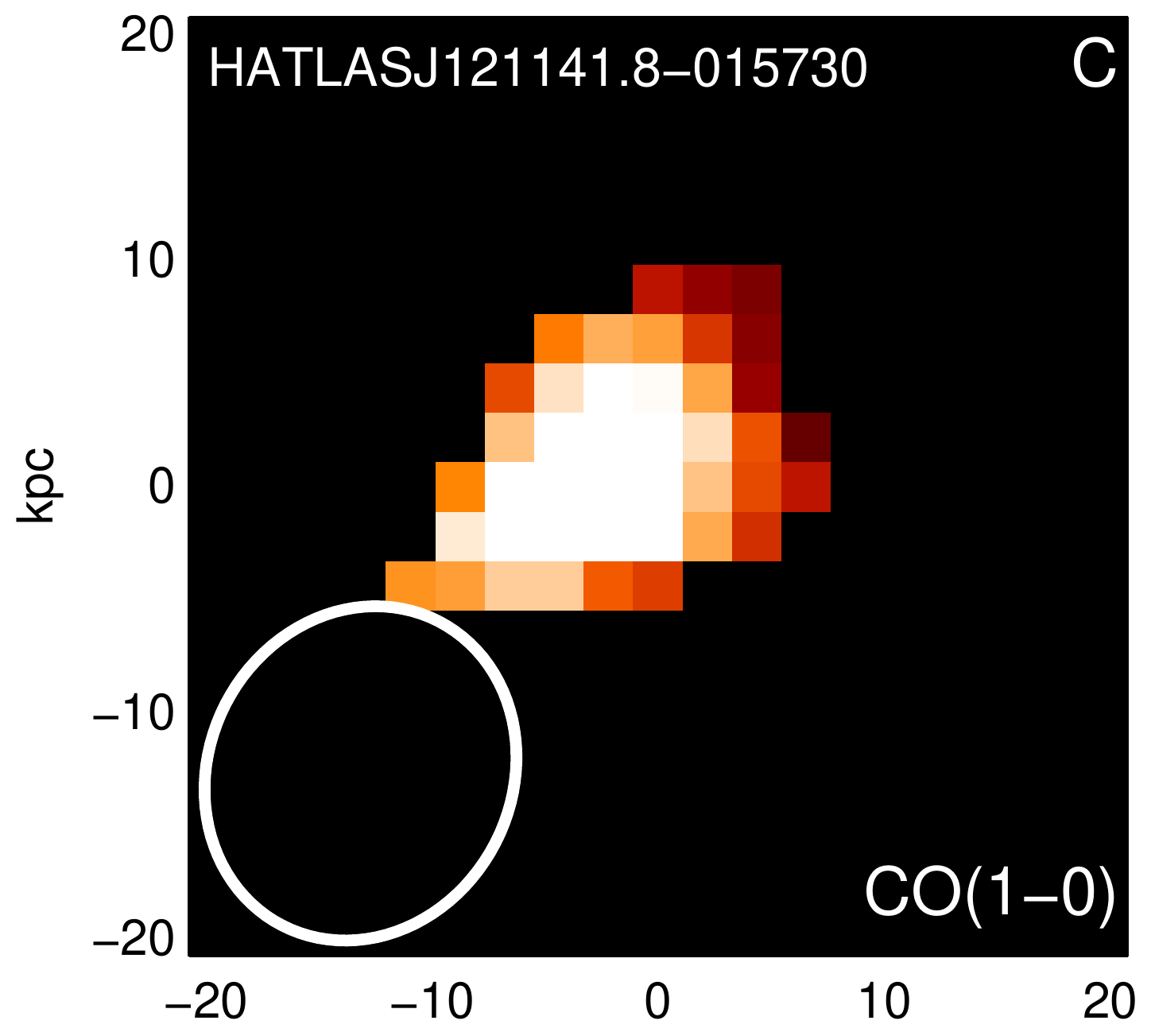}
\includegraphics[width=0.32\columnwidth]{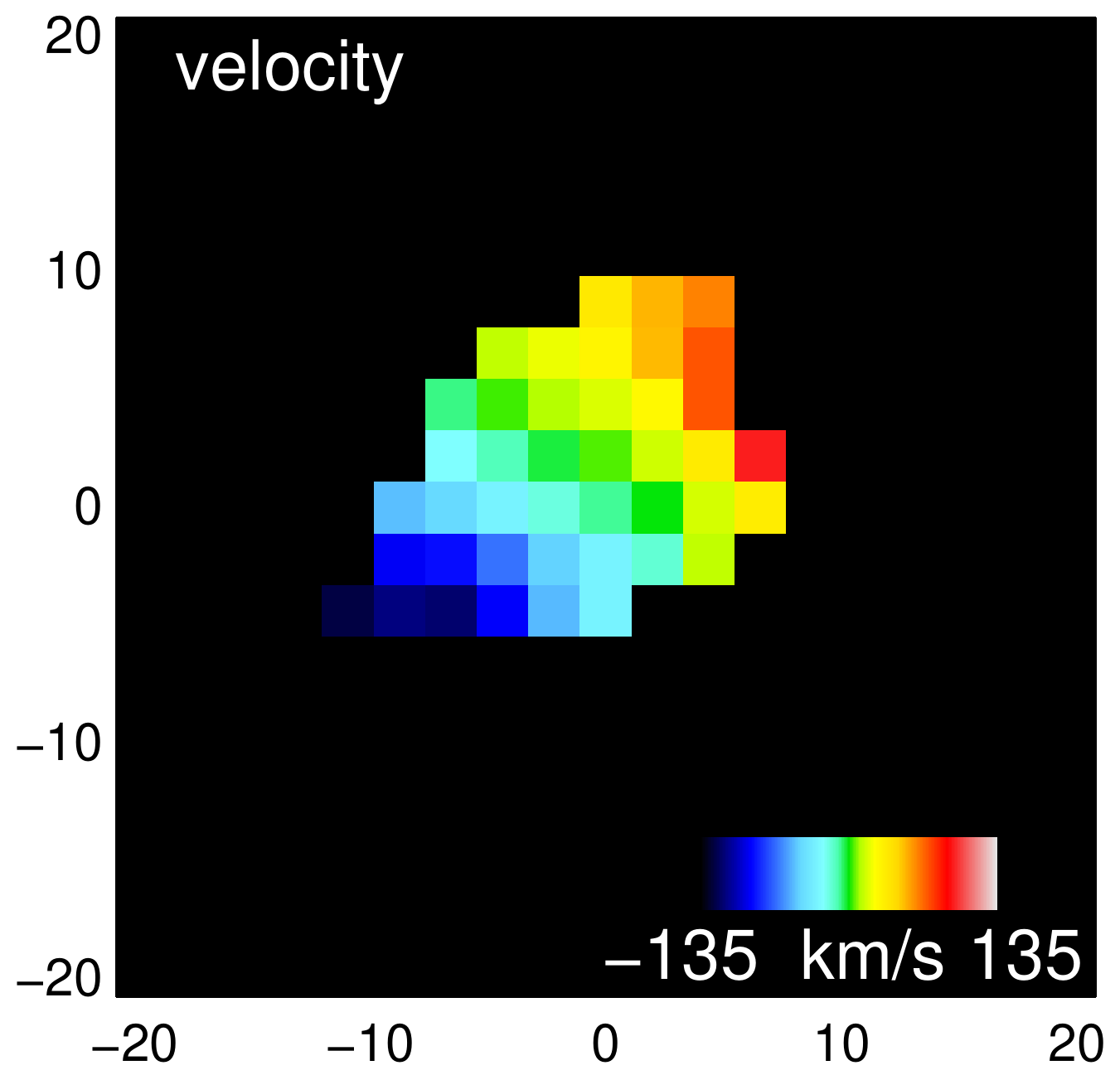}
\includegraphics[width=0.32\columnwidth]{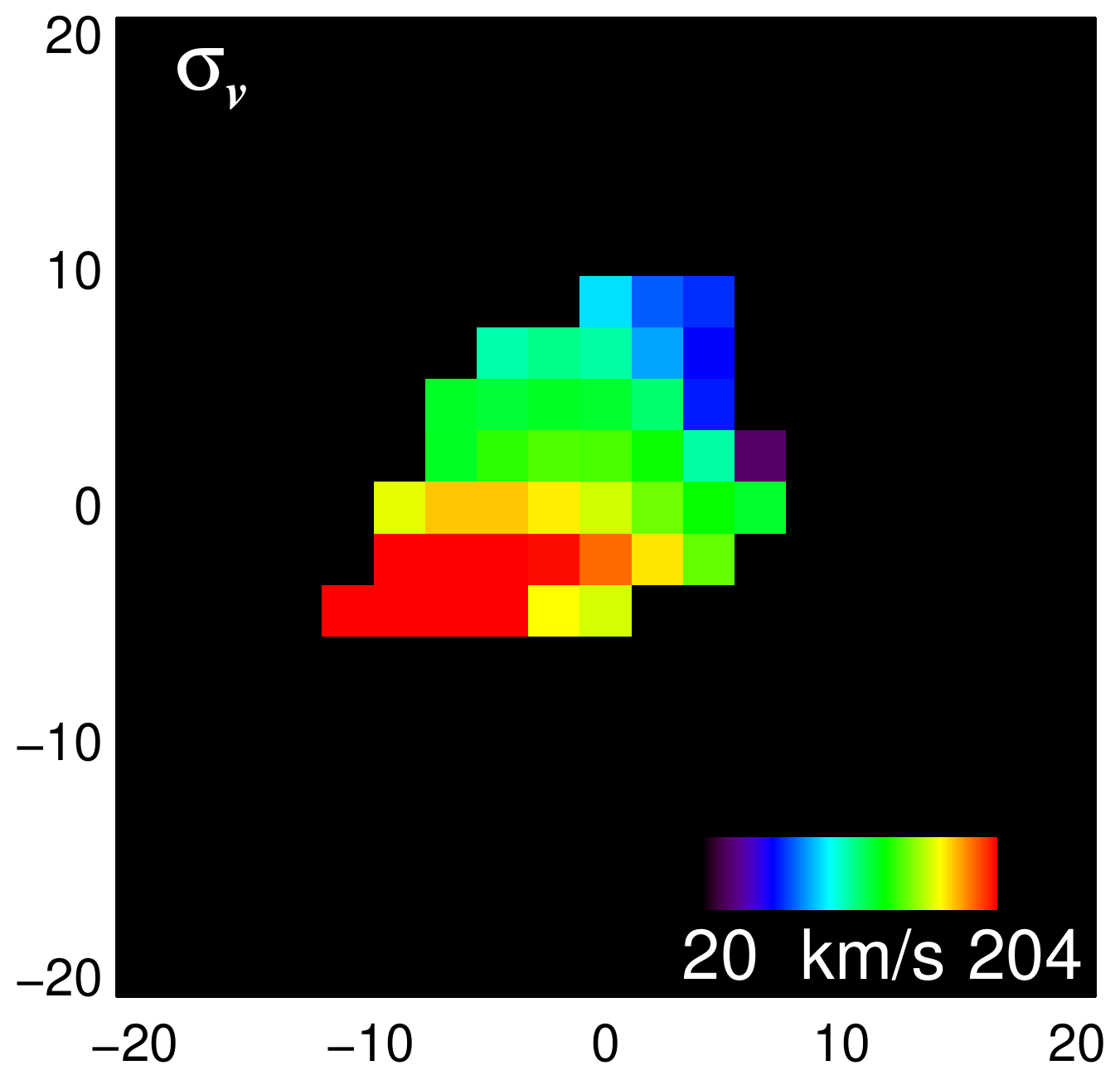}\\
\vspace{1mm}
\includegraphics[width=0.343\columnwidth]{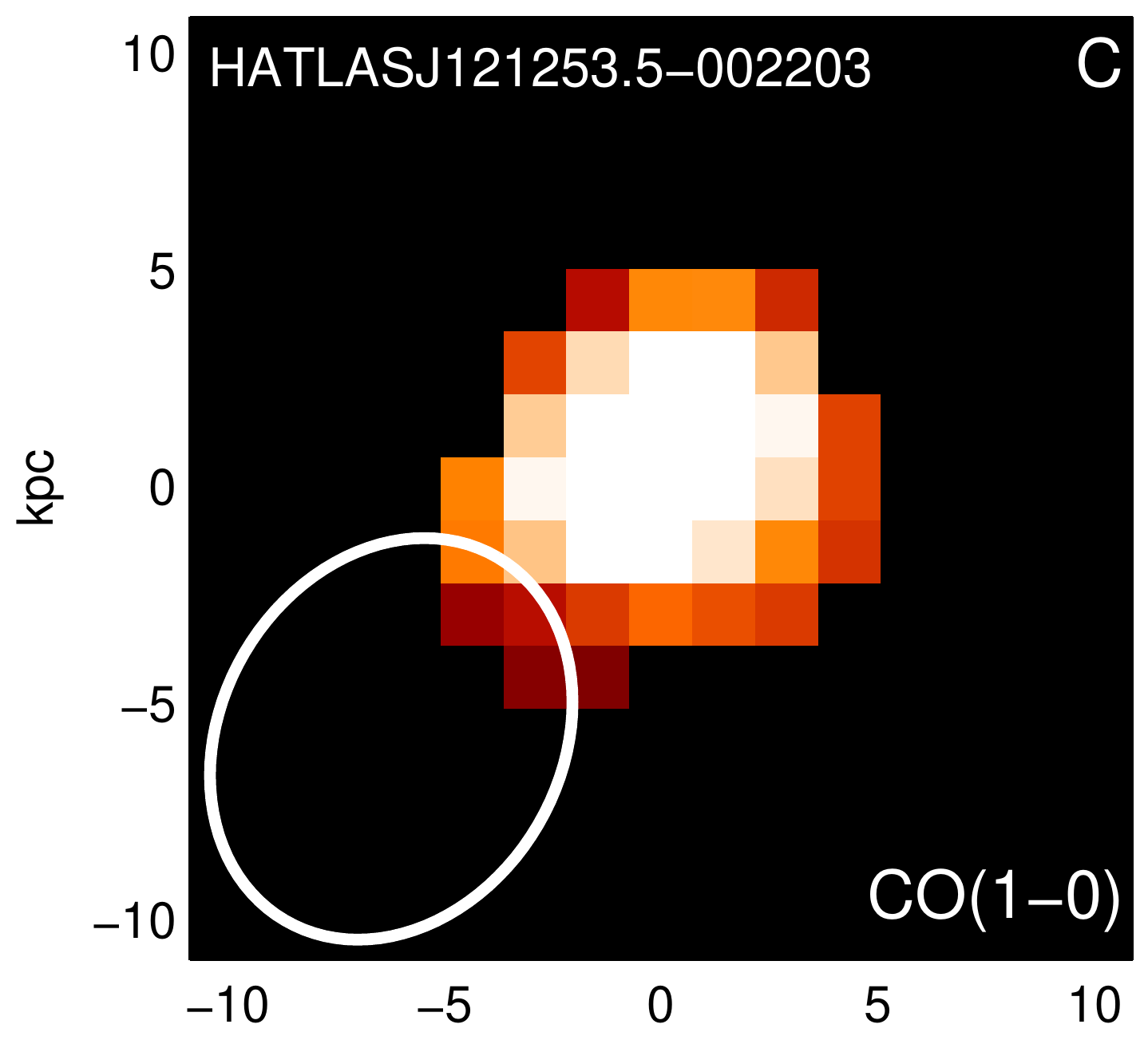}
\includegraphics[width=0.32\columnwidth]{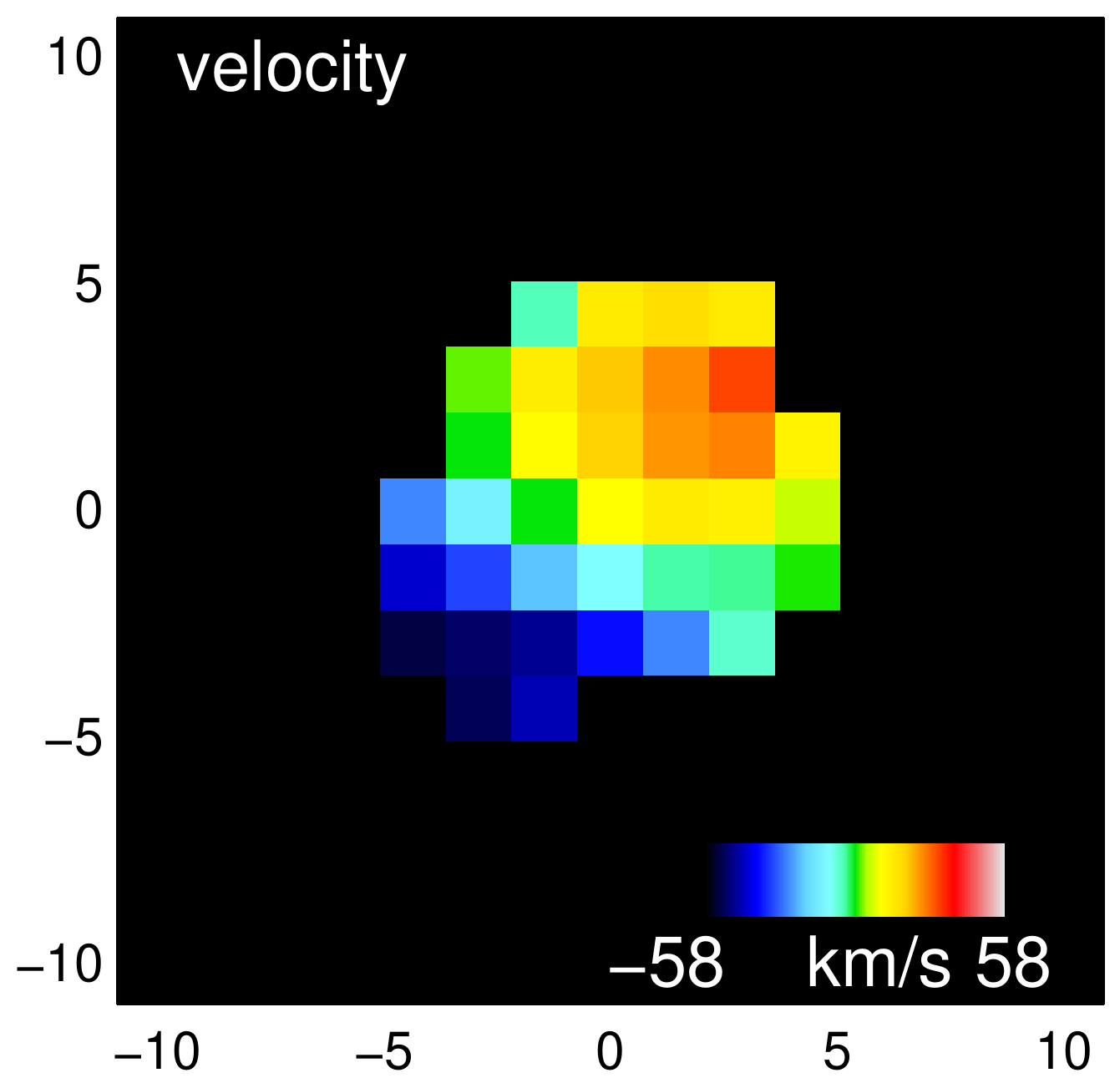}
\includegraphics[width=0.32\columnwidth]{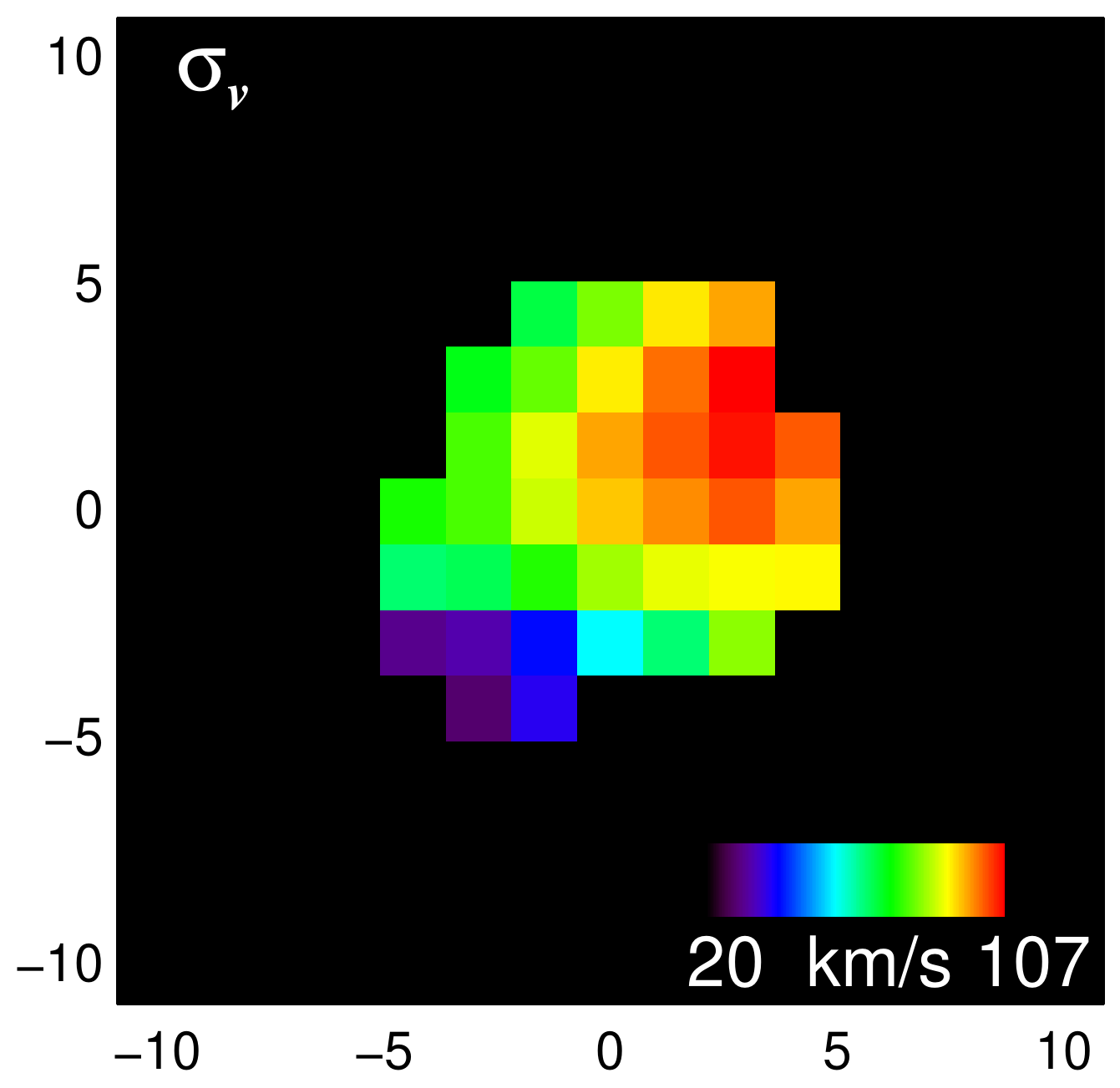}\\
\vspace{1mm}
\includegraphics[width=0.343\columnwidth]{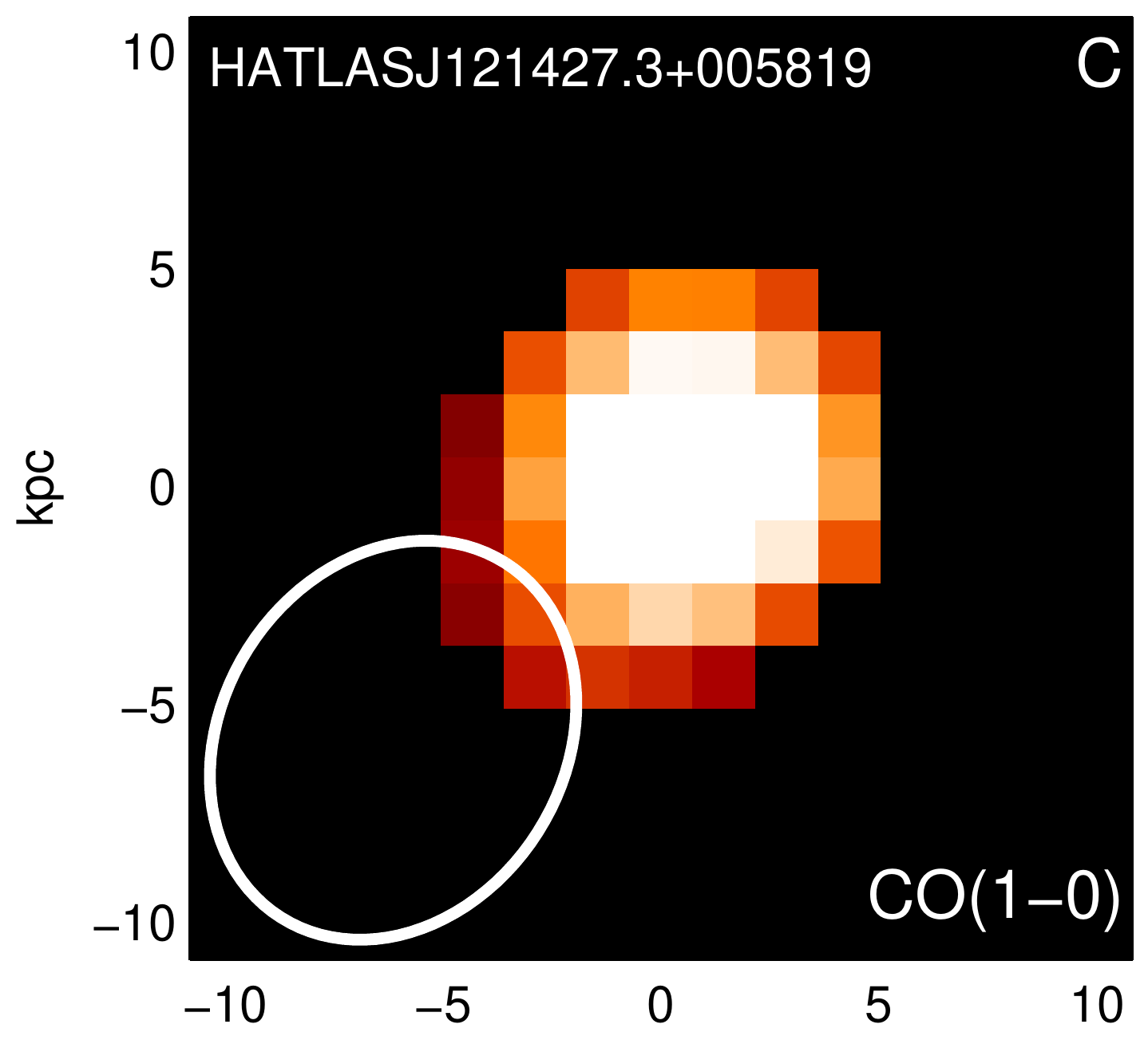}
\includegraphics[width=0.32\columnwidth]{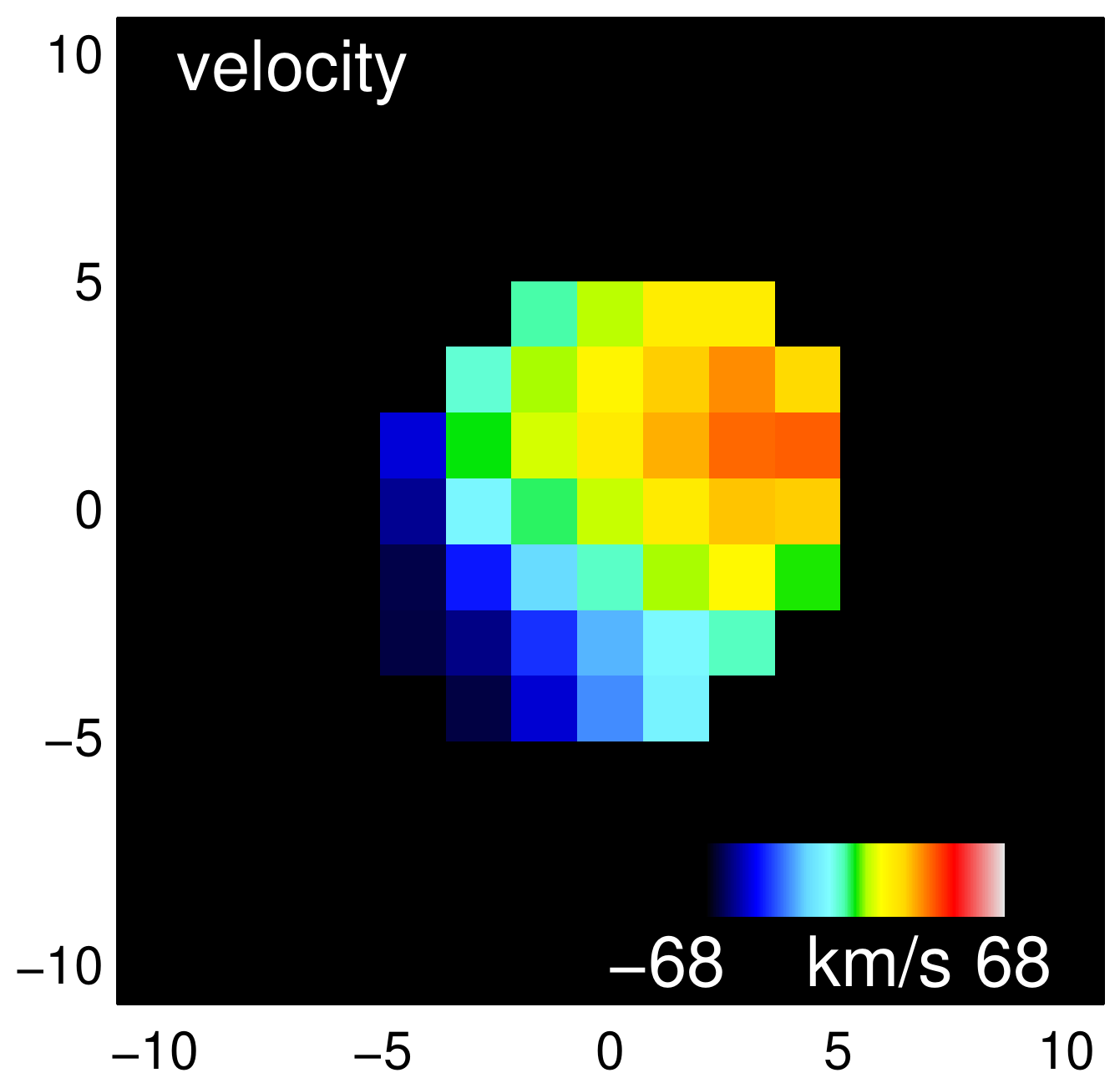}
\includegraphics[width=0.32\columnwidth]{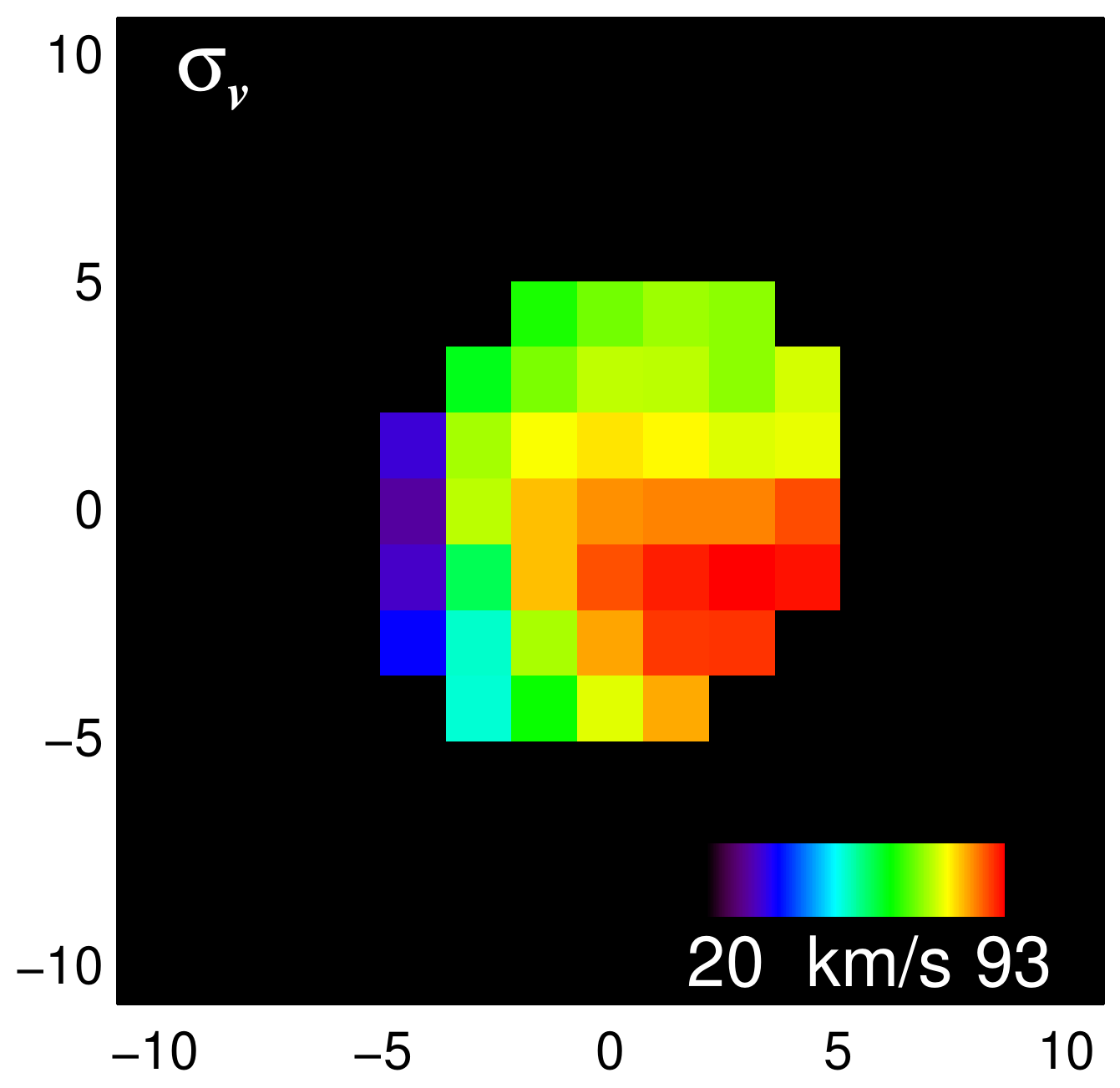}\\
\vspace{1mm}
\includegraphics[width=0.343\columnwidth]{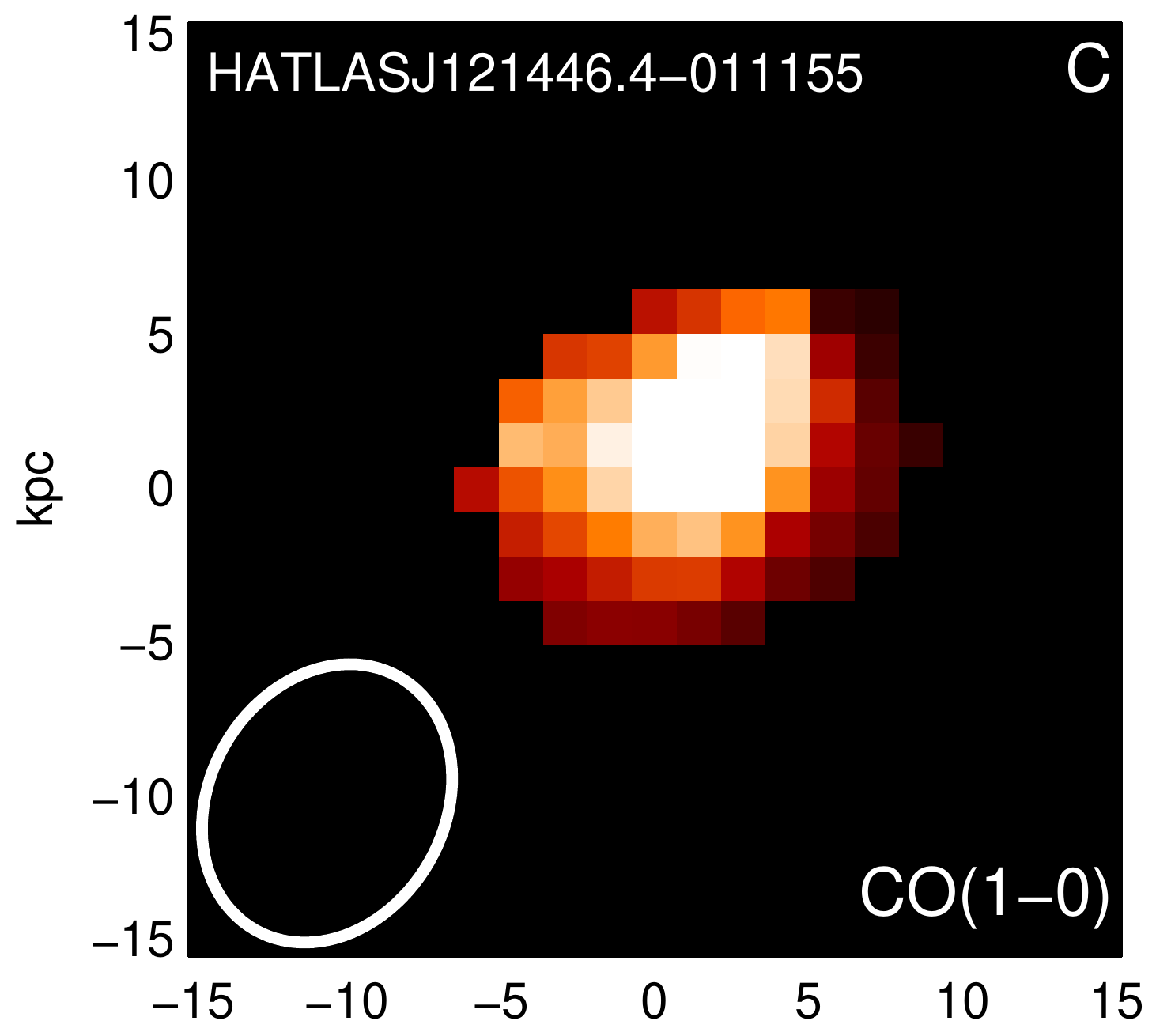}
\includegraphics[width=0.32\columnwidth]{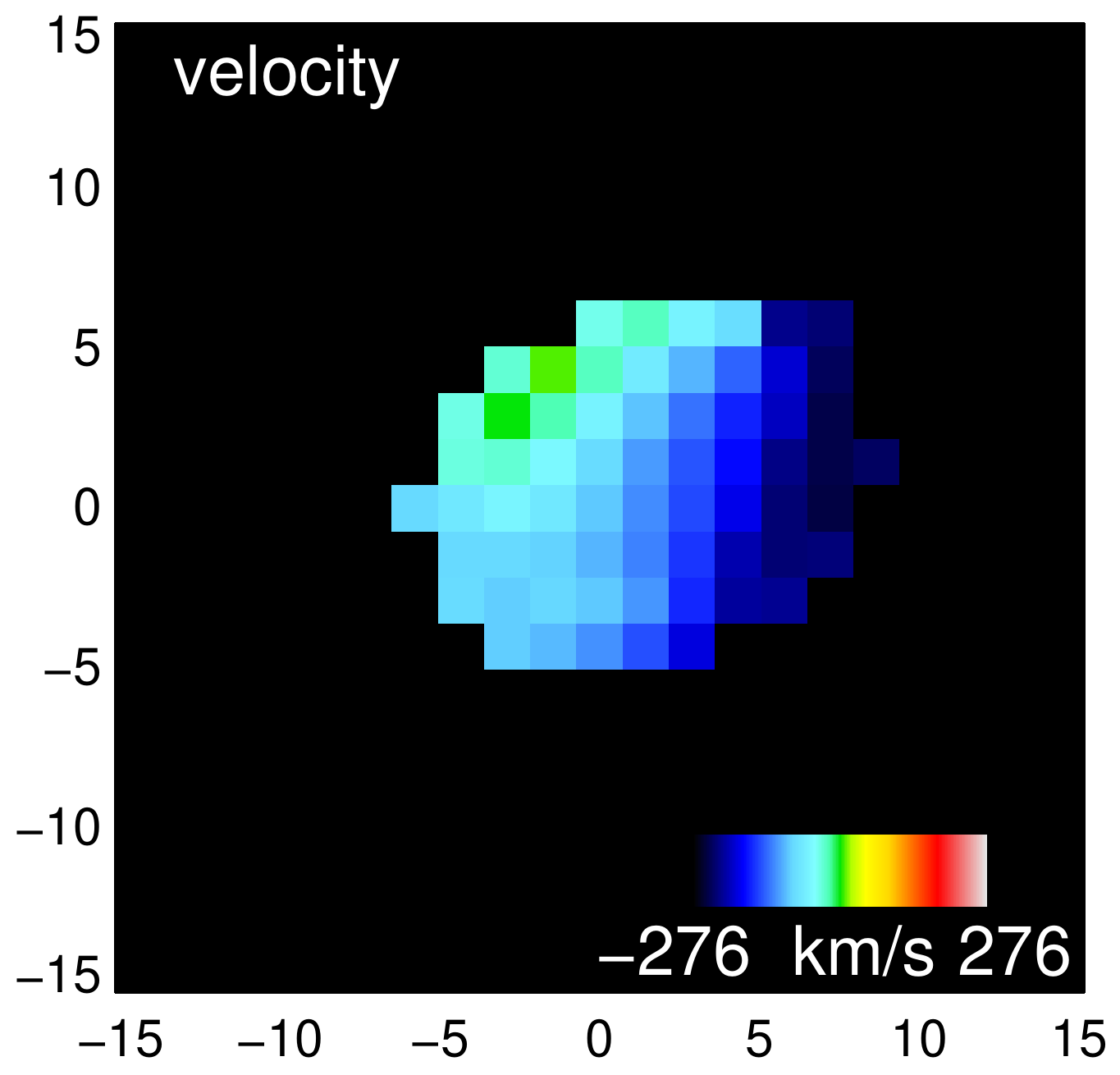}
\includegraphics[width=0.32\columnwidth]{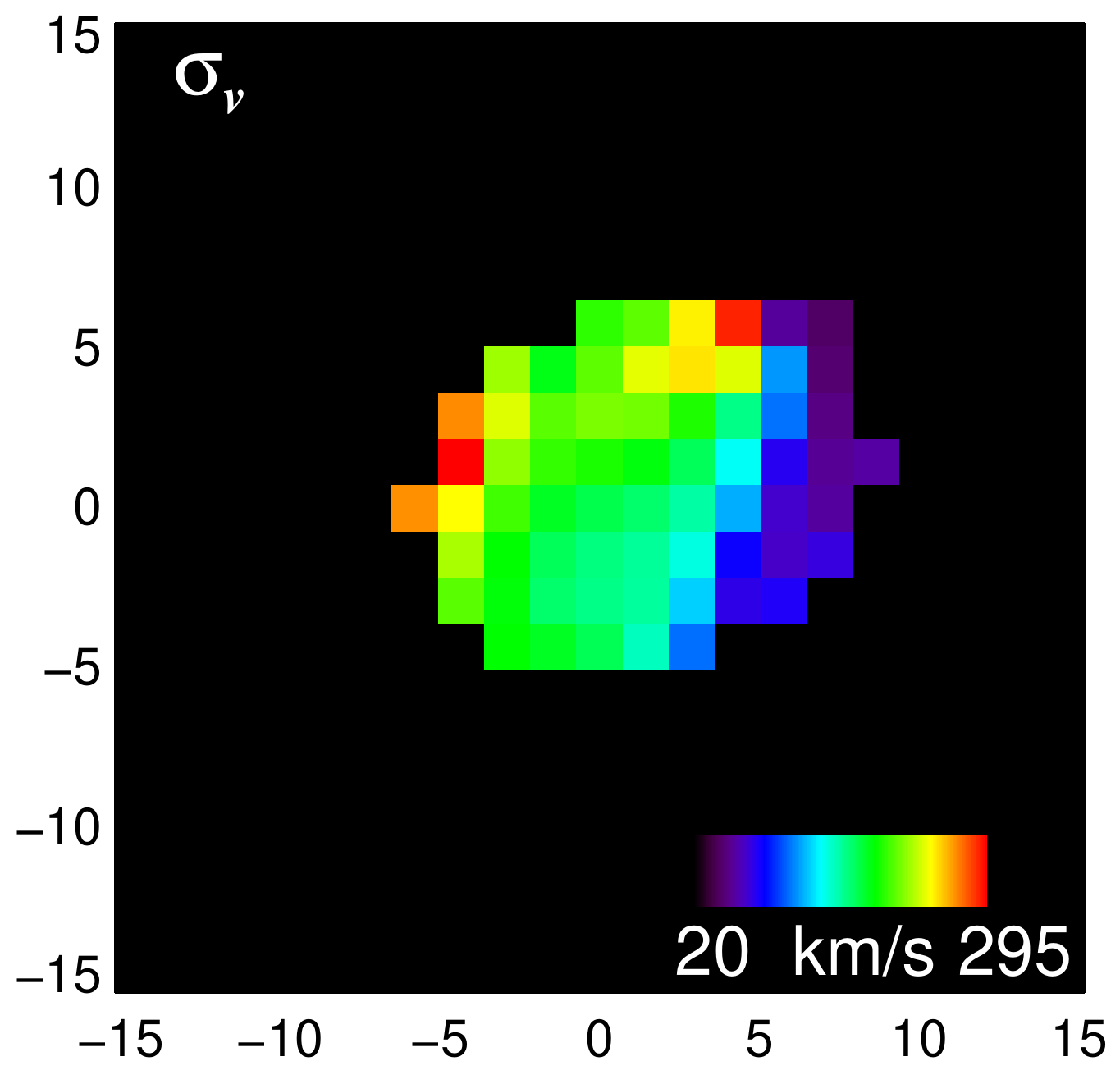}\\
\vspace{1mm}
\includegraphics[width=0.343\columnwidth]{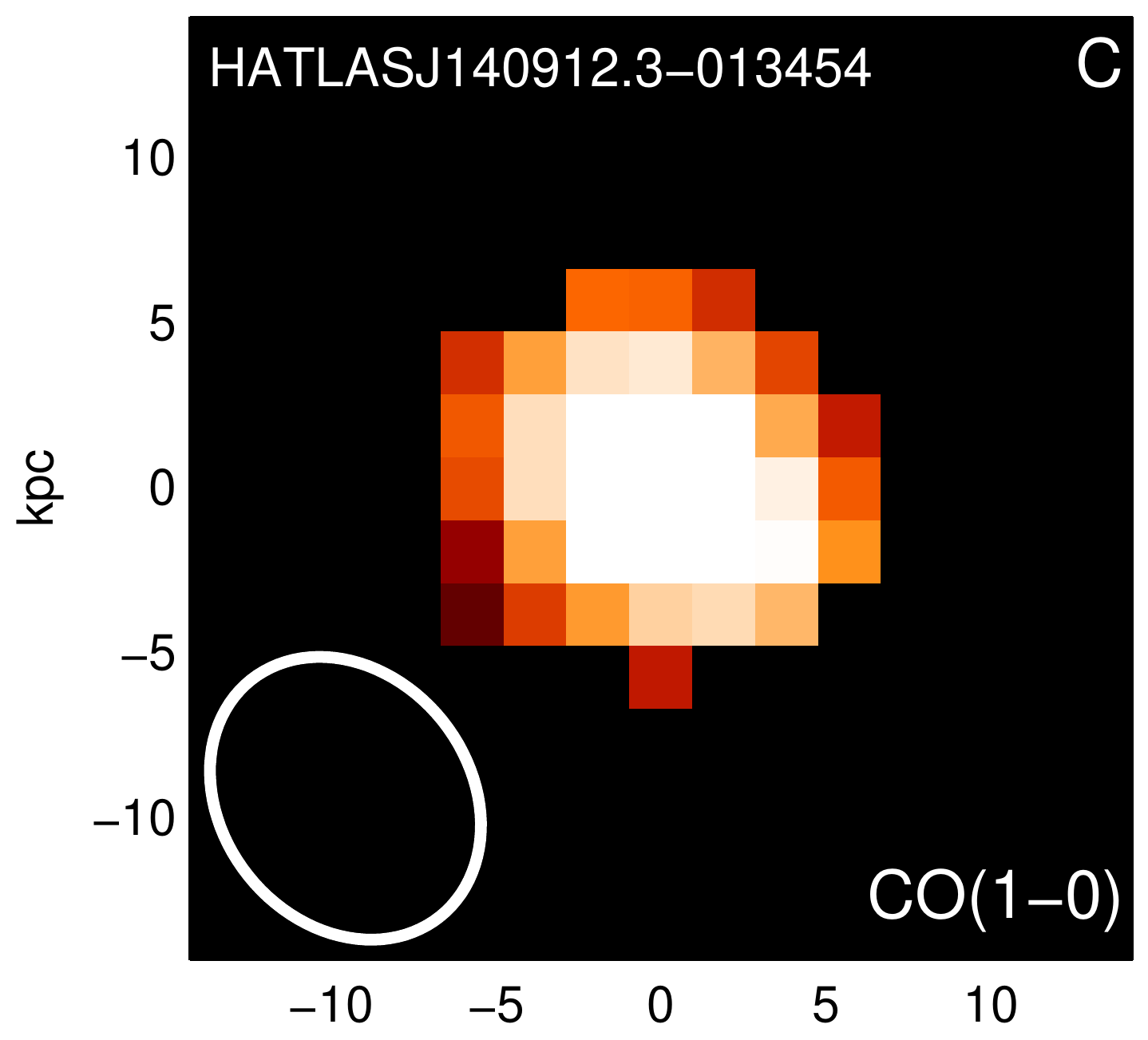}
\includegraphics[width=0.32\columnwidth]{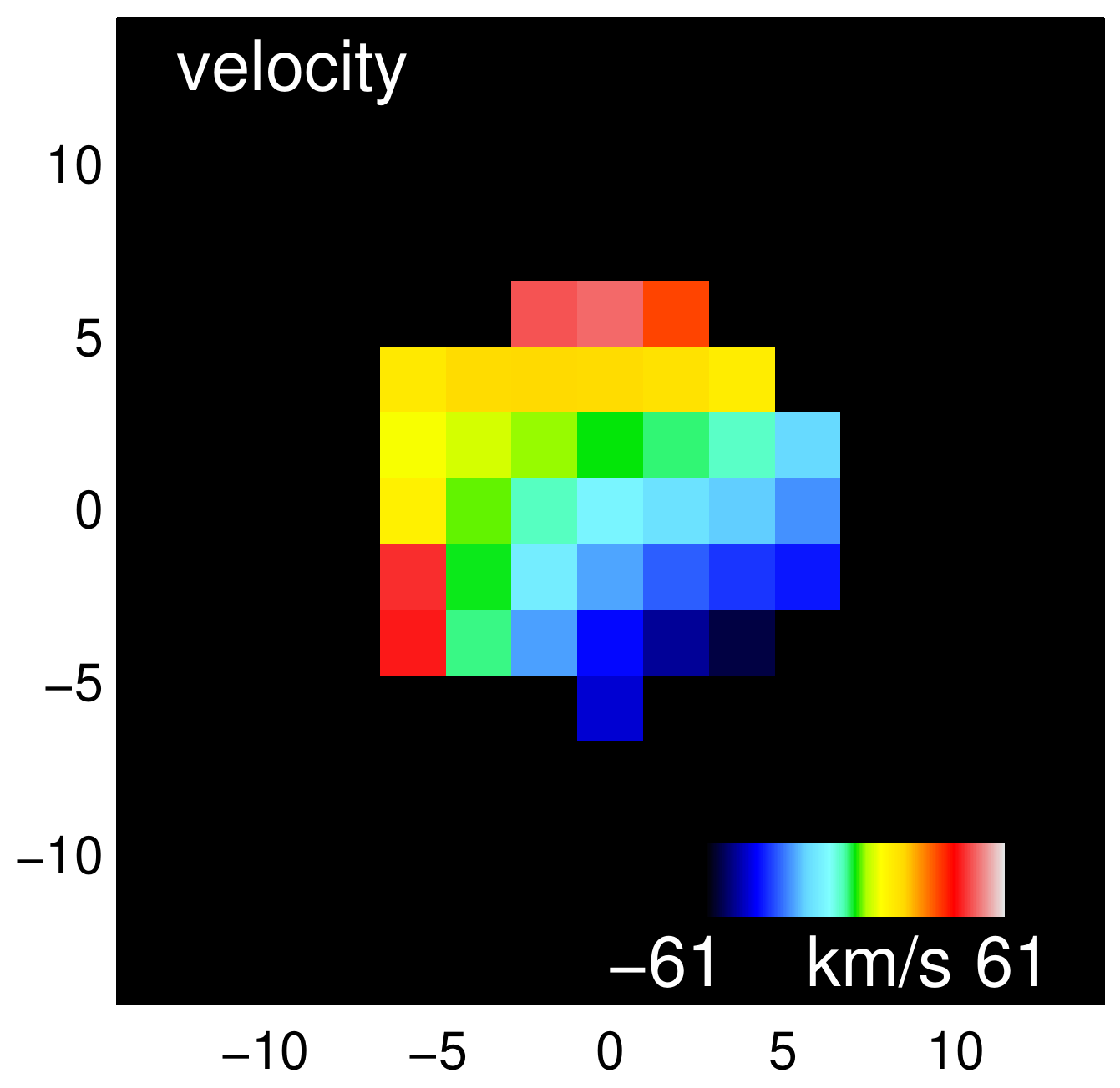}
\includegraphics[width=0.32\columnwidth]{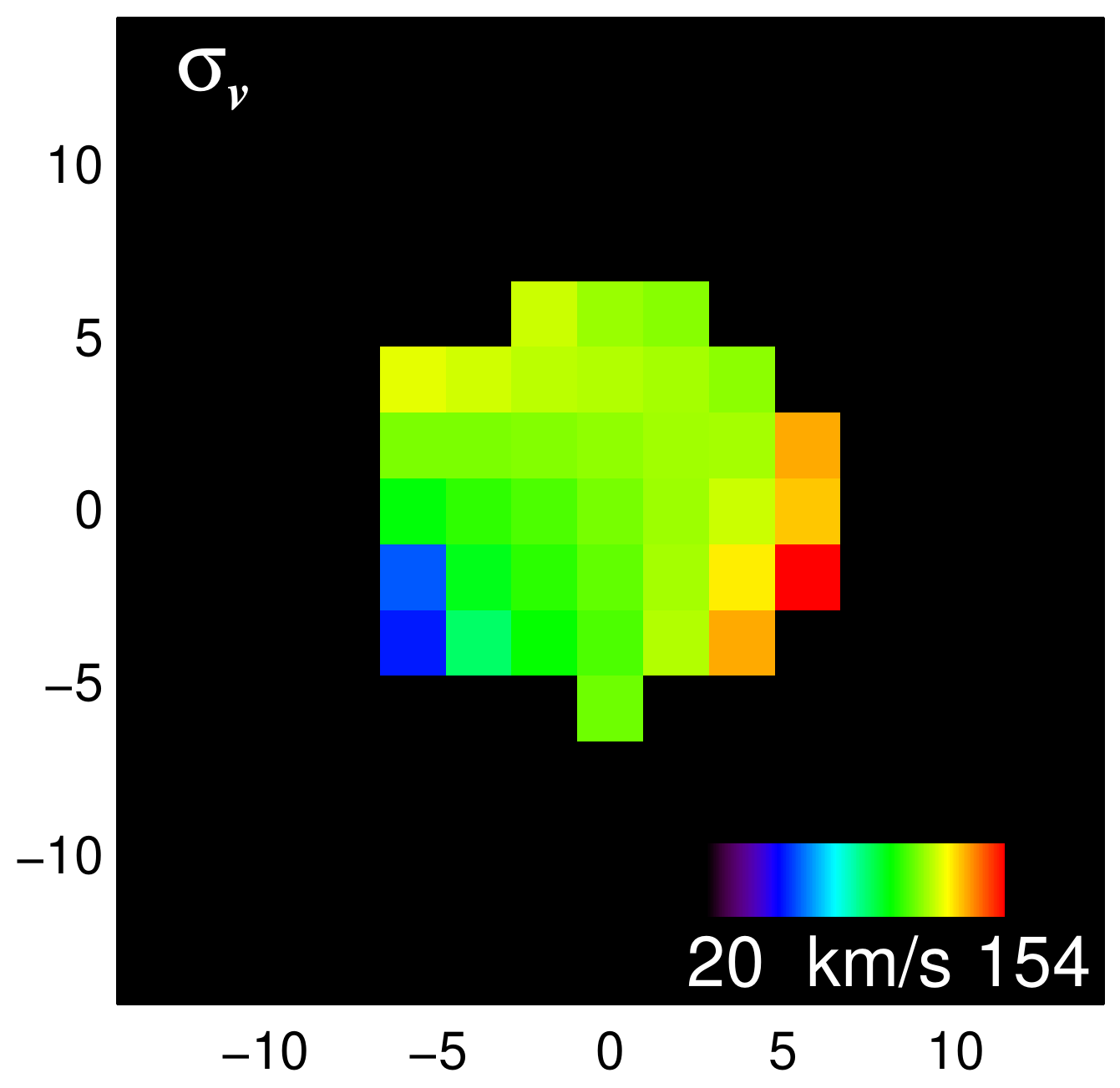}\\
\vspace{1mm}
\includegraphics[width=0.343\columnwidth]{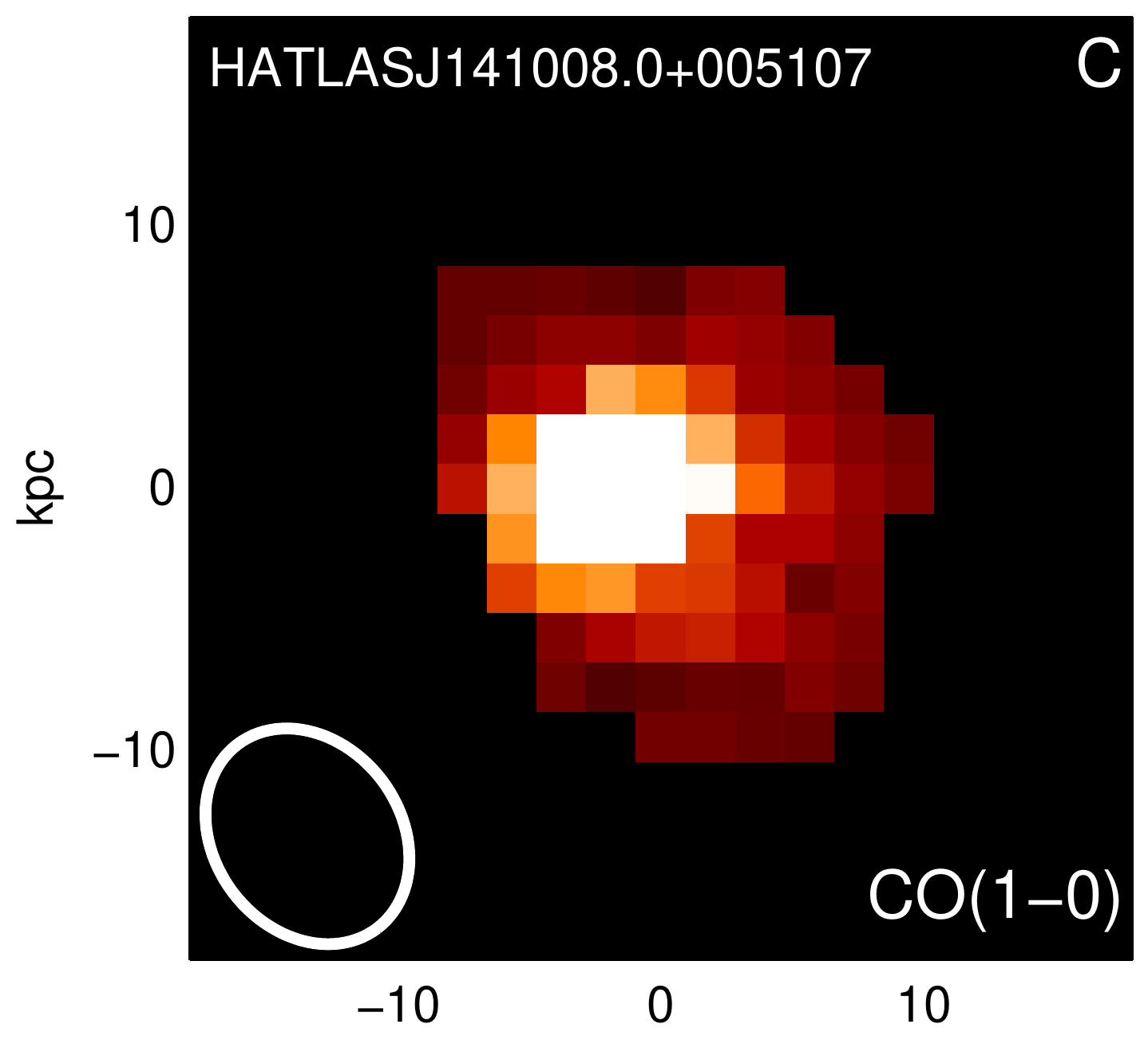}
\includegraphics[width=0.32\columnwidth]{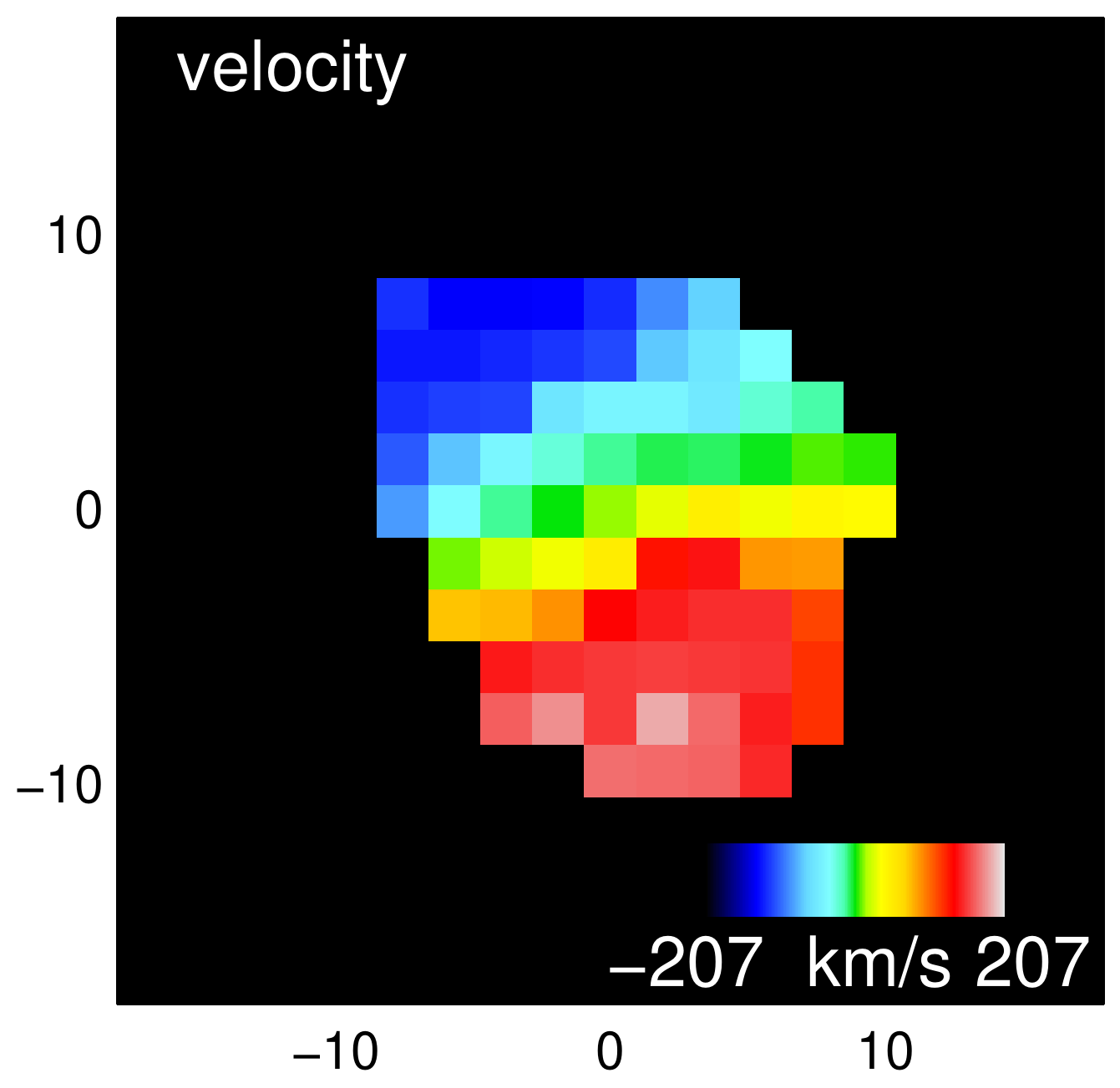}
\includegraphics[width=0.32\columnwidth]{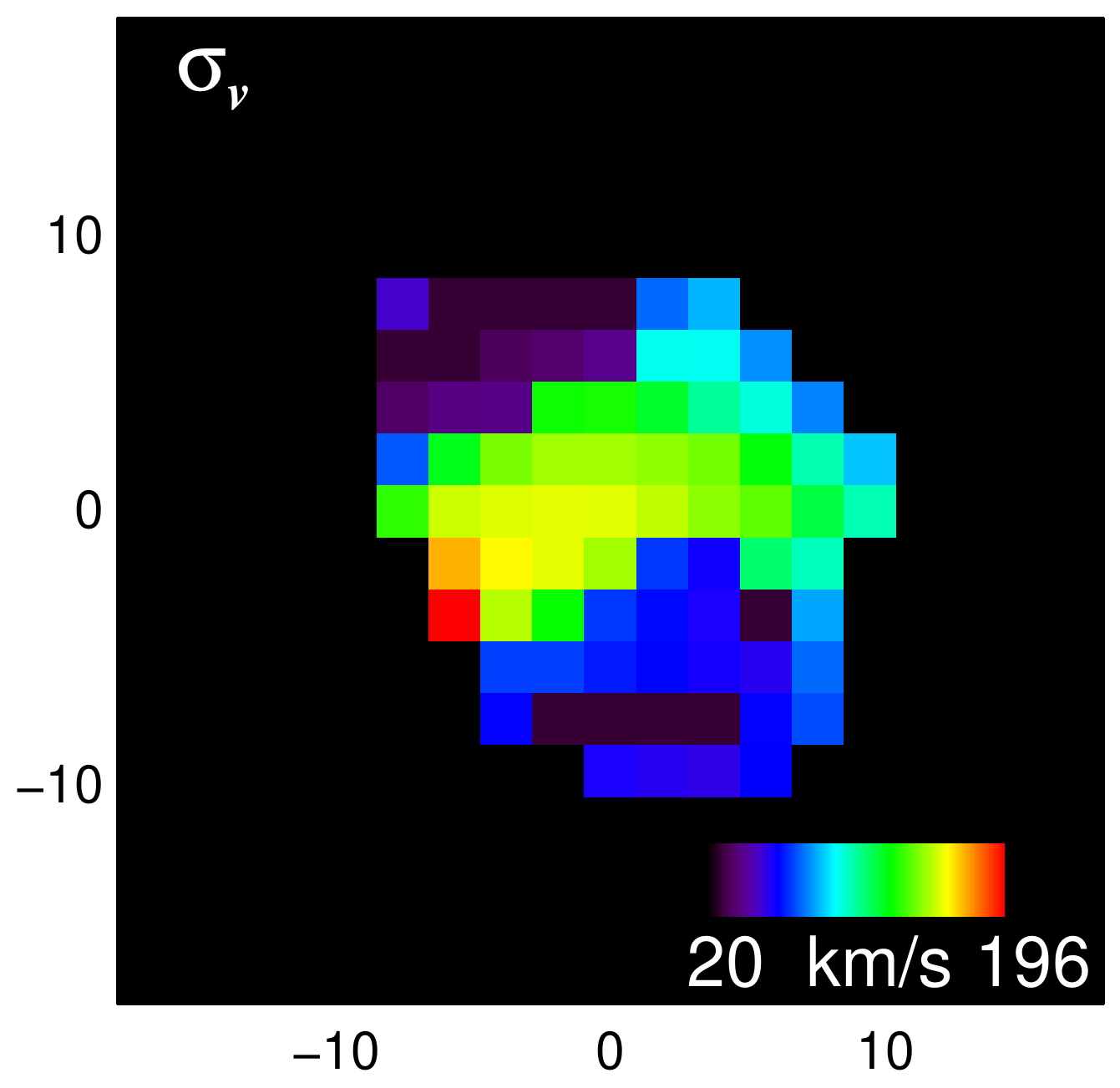}\\
\vspace{1mm}
\includegraphics[width=0.343\columnwidth]{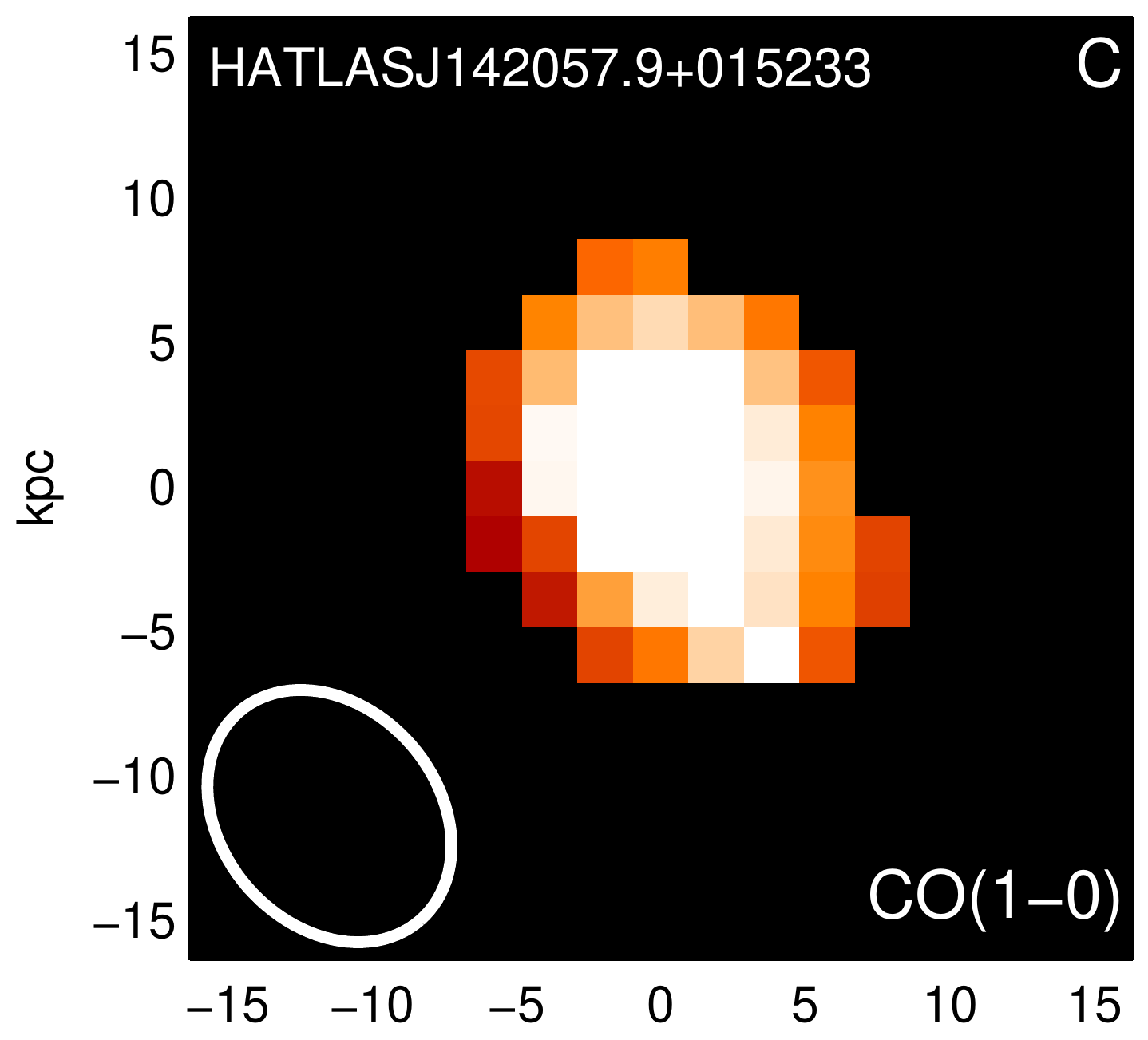}
\includegraphics[width=0.32\columnwidth]{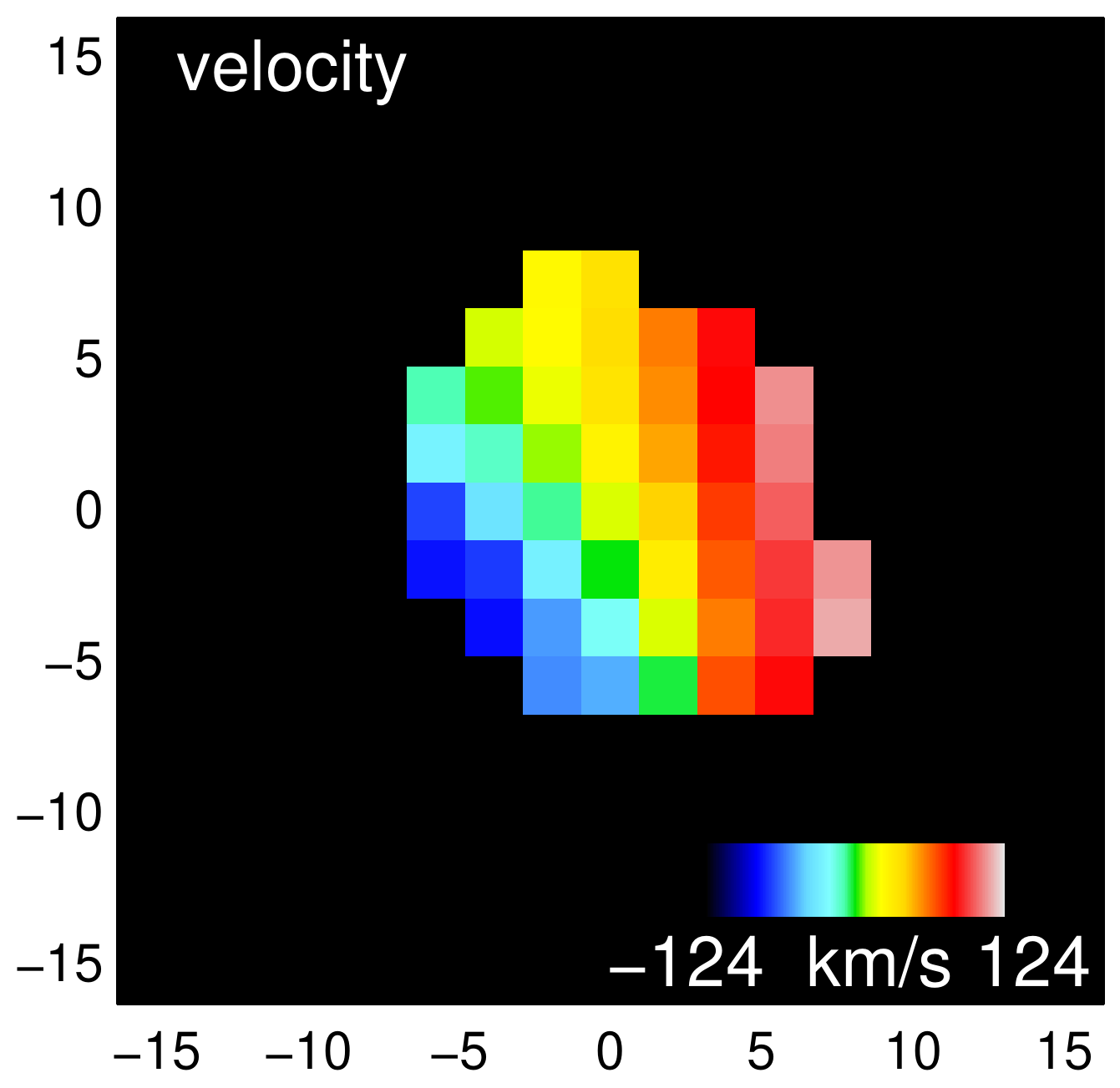}
\includegraphics[width=0.32\columnwidth]{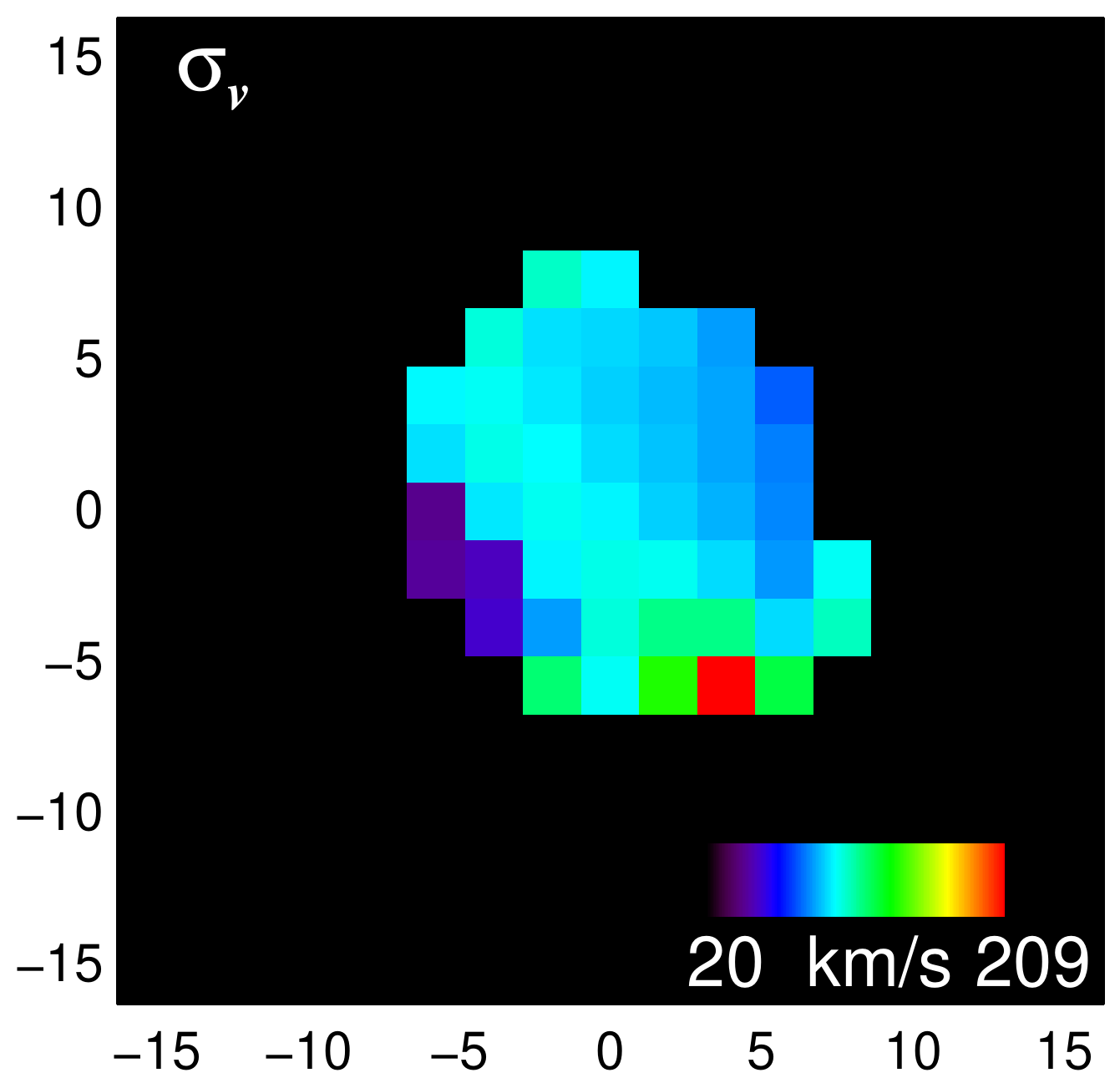}\\
\vspace{1mm}
\includegraphics[width=0.343\columnwidth]{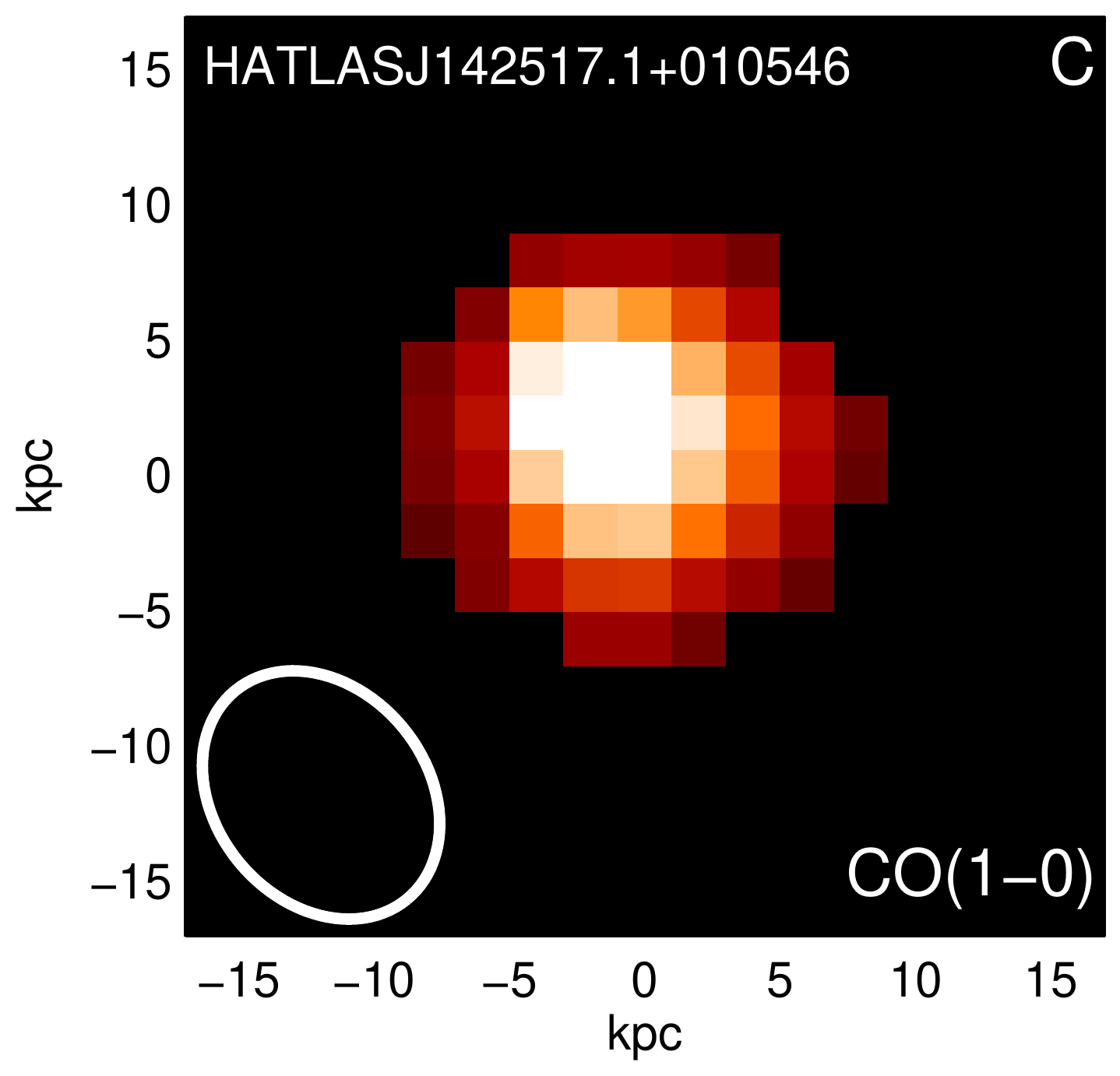}
\includegraphics[width=0.32\columnwidth]{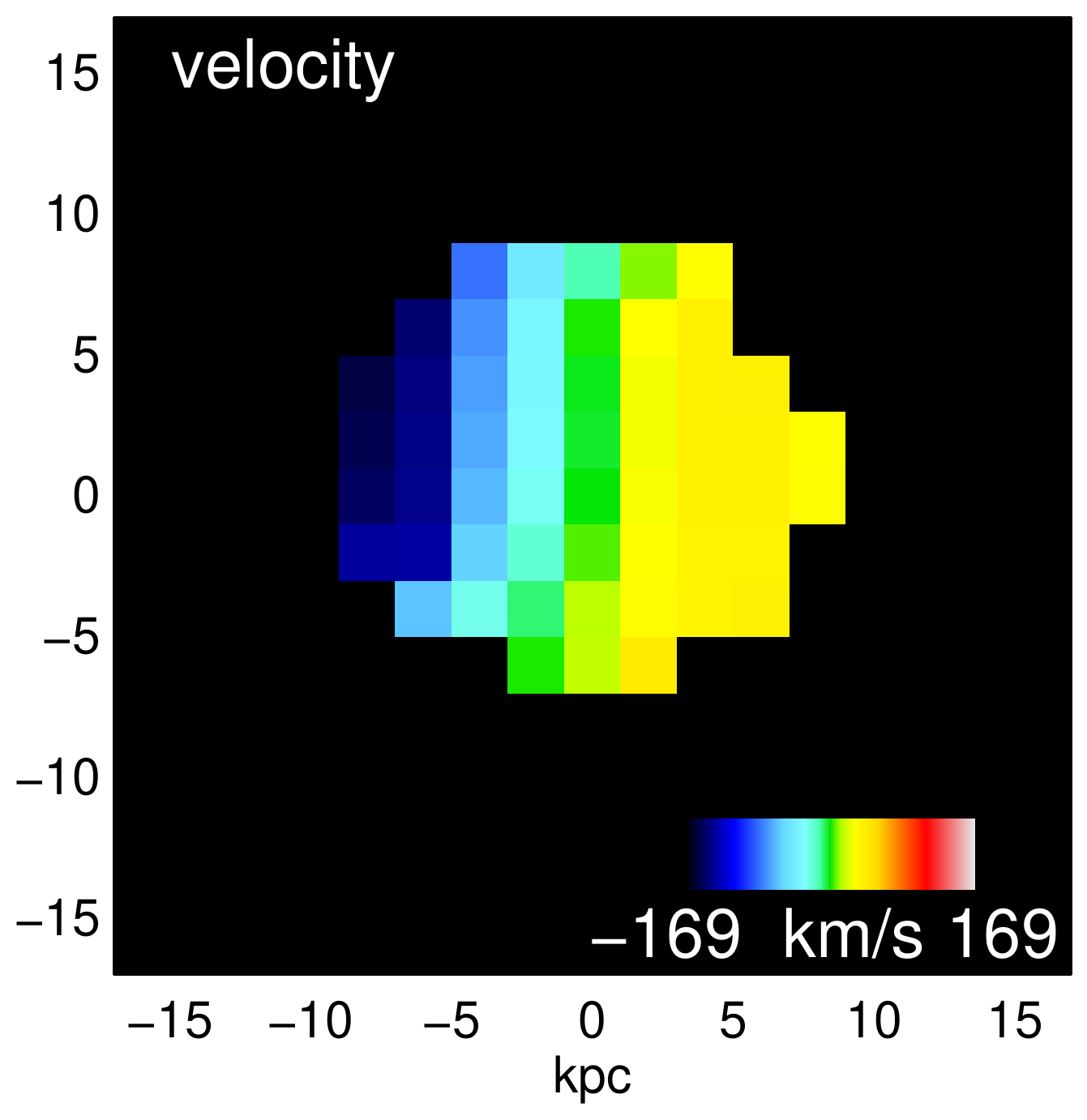}
\includegraphics[width=0.32\columnwidth]{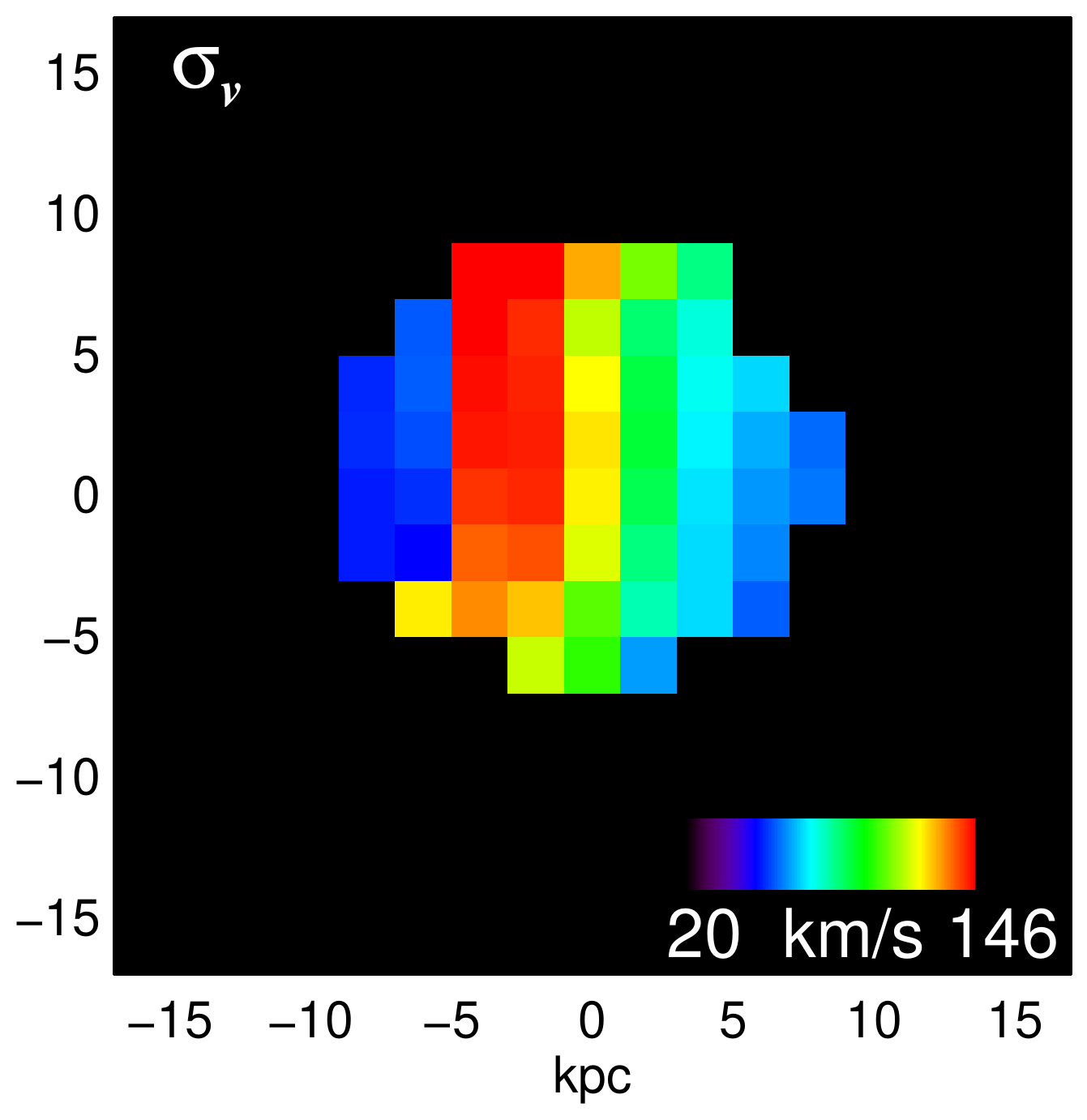}\\
\centering{\textbf{Figure C1. Continued.}}
\end{figure*}


\bsp	
\label{lastpage}

\begin{thebibliography}{99}
\bibitem[\protect\citeauthoryear{Bell et al.}{2012}]{Bell2012}
Bell, E. F., van der Wel, A., Papovich, C., et al. 2012, ApJ, 753, 167

\bibitem[\protect\citeauthoryear{Bellocchi et al.}{2012}]{Bellocchi2012}
Bellocchi, E., Arribas, S., \& Colina, L. 2012, A\&A, 542, 54

\bibitem[\protect\citeauthoryear{Bershady}{2010}]{Bershady2010}
Bershady, M., Verheijen, M., Swaters, R., et al. 2010, ApJ, 716, 198

\bibitem[\protect\citeauthoryear{Bertin \& Arnouts}{1996}]{Bertin1996}
Bertin, E. \& Arnouts, S. 1996, A\&AS, 117, 393 

\bibitem[\protect\citeauthoryear{Bigiel et al.}{2008}]{Bigiel2008}
Bigiel, F., Leroy, A., Walter, F., et al. 2008, AJ, 136, 2846

\bibitem[\protect\citeauthoryear{Bolatto et al.}{2013}]{Bolatto2013}
Bolatto, A. D., Wolfire, M. \& Leroy, A. K. 2013, ARA\&A, 51, 207

\bibitem[\protect\citeauthoryear{Bolatto et al.}{2017}]{Bolatto2017}
Bolatto, A. D., Wong, T., Utomo, D., et al. 2017, ApJ, 846, 159

\bibitem[\protect\citeauthoryear{Bothwell et al.}{2013}]{Bothwell2013}
Bothwell, M. S., Smail, I., Chapman, S. C., et al. 2013, MNRAS, 429, 3047

\bibitem[\protect\citeauthoryear{Bourne et al.}{2016}]{Bourne2016}
Bourne, N., Dunne, L., Maddox, S. J., et al. 2016, MNRAS, 462, 1714

\bibitem[\protect\citeauthoryear{Bournaud, Elmegreen \& Martig}{2009}]{Bournaud2009}
Bournaud, F., Elmegreen, B. G. \& Martig, M. 2009, ApJ, 707, 1

\bibitem[\protect\citeauthoryear{Bruce et al.}{2012}]{Bruce2012}
Bruce, V. A., Dunlop, J. S., Cirasuolo, M., et al. 2012, MNRAS, 427, 1666

\bibitem[\protect\citeauthoryear{Burkert et al.}{2010}]{Burkert2010}
Burkert, A., Genzel, R., Bouch\'e, N., et al. 2010, ApJ, 725, 2324

\bibitem[\protect\citeauthoryear{Burkert et al.}{2016}]{Burkert2016}
Burkert, A., F\"{o}rster Schreiber, N. M., Genzel, R., et al. 2016, ApJ, 826, 214

\bibitem[\protect\citeauthoryear{Chabrier}{2003}]{Chabrier2003}
Chabrier, G. 2003, PASP, 115, 763

\bibitem[\protect\citeauthoryear{Charbonneau}{1995}]{Charbonneau1995}
Charbonneau, P. 1995, ApJS, 101, 309

\bibitem[\protect\citeauthoryear{Cheng et al.}{2018}]{Cheng2018}
Cheng, C., Ibar, E., Hughes, T., et al. 2017, MNRAS, 475, 248

\bibitem[\protect\citeauthoryear{Courteau}{1997}]{Courteau1997}
Courteau, S. 1997, AJ, 114, 2402

\bibitem[\protect\citeauthoryear{da Cunha et al.}{2008}]{daCunha2008}
da Cunha, E., Charlot, S., Elbaz, D., et al. 2008, MNRAS, 388, 1595

\bibitem[\protect\citeauthoryear{Daddi et al.}{2010}]{Daddi2010}
Daddi, E., Elbaz, D., Walter, F., et al. 2010, ApJL, 714, 118

\bibitem[\protect\citeauthoryear{Davies et al.}{2011}]{Davies2011}
Davies, R., F\"{o}rster Schreiber, N. M., Cresci, G., et al. 2011, ApJ, 741, 69 

\bibitem[\protect\citeauthoryear{Dalgarno \& McCray}{1972}]{Dalgarno1972}
Dalgarno, A. \& McCray, R. A., 1972, ARA\&A, 51, 105

\bibitem[\protect\citeauthoryear{de Blok et al.}{2008}]{deblok2008}
de Blok, W., Walter, F., Brinks, E., et al. 2008, 136, 2648

\bibitem[\protect\citeauthoryear{D\'iaz-Santos et al.}{2014}]{DiazSantos2014}
D\'iaz-Santos, T., Armus, L., Charmandaris, V., et al. 2014, ApJ, 774, 68

\bibitem[\protect\citeauthoryear{Dickman}{1978}]{Dickman1978}
Dickman, R. L. 1978, ApJS, 37, 407 

\bibitem[\protect\citeauthoryear{Dickman, Snell \& Schloerb}{1986}]{Dickman1986}
Dickman, R. L., Snell, R. L. \& Schloerb, F. P. 1986, ApJ, 309, 326

\bibitem[\protect\citeauthoryear{Di Teodoro, Fraternali \& Miller}{2016}]{DiTeodoro2016}
Di Teodoro, E. M., Fraternali, F. \& Miller, S. H. 2016, A\&A, 594, 77

\bibitem[\protect\citeauthoryear{Downes, Solomon \& Radford}{1993}]{Downes1993}
Downes, D., Solomon, P. M. \& Radford, S. J. E. 1993, ApJ, 414, 13

\bibitem[\protect\citeauthoryear{Downes \& Solomon}{1998}]{Downes1998}
Downes, D. \& Solomon, P. M. 1998, ApJ, 507, 615

\bibitem[\protect\citeauthoryear{Draine \& Li}{2007}]{Draine2007}
Draine,  B. \& Li, A. 2007, ApJ, 657, 810

\bibitem[\protect\citeauthoryear{Driver et al.}{2016}]{Driver2016}
Driver, S. P., Wright, A. H., Andrews, S. K., et al. 2016, MNRAS, 455, 3911

\bibitem[\protect\citeauthoryear{Eales et al.}{2010}]{Eales2010}
Eales, S., Dunne, L., Clements, D., et al. 2010, PASP, 122, 499

\bibitem[\protect\citeauthoryear{Elmegreen}{2011}]{Elmegreen2011}
Elmegreen, B. G. 2011, ApJ, 737, 10

\bibitem[\protect\citeauthoryear{Epinat et al.}{2010}]{Epinat2010}
Epinat, B., Amram, P., Balkowski, C., \& Marcelin, M. 2010, MNRAS, 401, 2113

\bibitem[\protect\citeauthoryear{Epinat et al.}{2012}]{Epinat2012}
Epinat, B., Tasca, L., Amram, P., et al. 2012, A\&A, 539, 92 

\bibitem[\protect\citeauthoryear{Escala \& Larson}{2008}]{Escala2008}
Escala, A. \& Larson, R. B. 2008, ApJ, 685, 31 

\bibitem[\protect\citeauthoryear{Faucher-Gigu\`ere, Quataert \& Hopkins}{2013}]{FQH2013}
Faucher-Gigu\`ere, C., Quataert, E. \& Hopkins, P. 2013, MNRAS, 433, 1970

\bibitem[\protect\citeauthoryear{F\"{o}rster Schreiber et al.}{2009}]{Forster2009}
F\"{o}rster Schreiber, N. M., Genzel, R., Bouch\'e, N., et al. 2009, ApJ, 706, 1364

\bibitem[\protect\citeauthoryear{Gnerucci et al.}{2011}]{Gnerucci2011}
Gnerucci, A., Marconi, A., Cresci, G., et al. 2011, ApJ, 533, 124

\bibitem[\protect\citeauthoryear{Genzel et al.}{2008}]{Genzel2008}
Genzel, R., Burkert, A., Bouché, N., et al. 2008, ApJ, 687, 59

\bibitem[\protect\citeauthoryear{Genzel et al.}{2011}]{Genzel2011}
Genzel, R., Newman, S., Jones, T., et al. 2011, ApJ, 733, 101

\bibitem[\protect\citeauthoryear{Genzel et al.}{2015}]{Genzel2015}
Genzel, R., Tacconi, L., Lutz, D., et al., 2015, ApJ, 800, 20

\bibitem[\protect\citeauthoryear{Glazebrook}{2013}]{Glazebrook2013}
Glazebrook, K. 2013, PASA, 30, 56

\bibitem[\protect\citeauthoryear{Green et al.}{2010}]{Green2010}
Green, A., Glazebrook, K., McGregor, P., et al. 2010,  Natur, 467, 684

\bibitem[\protect\citeauthoryear{Green et al.}{2014}]{Green2014}
Green, A., Glazebrook, K., McGregor, P., et al. 2014, MNRAS, 437, 1070

\bibitem[\protect\citeauthoryear{Griffin et al.}{2010}]{Griffin2010}
Griffin, M., Abergel, A., Abreu, A., et al. 2010,  A\&A, 518, 3

\bibitem[\protect\citeauthoryear{Goldreich \& Lynden-Bell}{1965}]{Goldreich1965}
Goldreich, P. \& Lynden-Bell, D. 1965, MNRAS, 130, 97

\bibitem[\protect\citeauthoryear{Gullberg et al.}{2015}]{Gullberg2015}
Gullberg, B., De Breuck, C., Vieira, J. D., et al. 2015, MNRAS, 449, 2883

\bibitem[\protect\citeauthoryear{Habing}{1968}]{Habing1968}
Habing, H. J. 1968, Bull. Astron. Inst. Netherlands, 19, 421

\bibitem[\protect\citeauthoryear{Harrison et al.}{2017}]{Harrison2017}
Harrison, C., Johnson, H., Swinbank, A. M., et al. 2017, 467, 1965

\bibitem[\protect\citeauthoryear{H\"au\ss{}ler et al.}{2007}]{Haubler2007}
H\"au\ss{}ler, B., McIntosh, D. H., Barden, M., et al. 2007, ApJS, 172, 615

\bibitem[\protect\citeauthoryear{Heyer et al.}{2009}]{Heyer2009}
Heyer, M., Krawczyk, C., Duval, J.\& Jackson, J. M. 2009, ApJ, 699, 1092

\bibitem[\protect\citeauthoryear{Holmberg}{1958}]{Holmberg1958}
Holmberg, E. 1958, MeLuS, 136, 1

\bibitem[\protect\citeauthoryear{Hopkins}{2012}]{Hopkins2012}
Hopkins, P. F. 2012, MNRAS, 423, 2016

\bibitem[\protect\citeauthoryear{Hughes et al.}{2017a}]{Hughes2017a}
Hughes, T., Ibar, E., Villanueva, V., et al. 2017a, A\&A, 602, 49

\bibitem[\protect\citeauthoryear{Hughes et al.}{2017b}]{Hughes2017b}
Hughes, T., Ibar, E., Villanueva, V., et al. 2017b, MNRAS, 468, 103

\bibitem[\protect\citeauthoryear{Ibar et al.}{2015}]{Ibar2015}
Ibar, E., Lara-L\'opez, M. A., Herrera-Camus, R., et al. 2015, MNRAS, 449, 2498

\bibitem[\protect\citeauthoryear{Jog \& Solomon}{1984}]{Jog1984}
Jog, C. J. \& Solomon, P. M. 1984, ApJ, 276, 114 

\bibitem[\protect\citeauthoryear{Jog}{1996}]{Jog1996}
Jog, C. J. 1996, MNRAS, 278, 209

\bibitem[\protect\citeauthoryear{Johnson et al.}{2018}]{Johnson2017}
Johnson, H., Harrison, C., Swinbank, A., et al. 2018, MNRAS, 474, 5076

\bibitem[\protect\citeauthoryear{Kassin et al.}{2012}]{Kassin2012}
Kassin, S. A., Weiner, B. J., Faber, S. M., et al. 2012, ApJ, 758, 106

\bibitem[\protect\citeauthoryear{Kaufman et al.}{1999}]{Kaufman1999}
Kaufman, M. J., Wolfire, M. G., Hollenbach, D. J., \& Luhman, M. L. 1999, ApJ, 527, 795

\bibitem[\protect\citeauthoryear{Kaufman et al.}{1999}]{Kaufman1999}
Kaufman, M. J., Wolfire, M. G., Hollenbach, D. J., \& Luhman, M. L. 1999, ApJ, 527, 795

\bibitem[\protect\citeauthoryear{Kaufman et al.}{2006}]{Kaufman2006}
Kaufman, M. J., Wolfire, M. G., \& Hollenbach, D. J. 2006, ApJ, 644, 283

\bibitem[\protect\citeauthoryear{Kramer et al.}{2013}]{Kramer2013}
Kramer, C., Abreu-Vicente, J., Garc\'ia-Burillo, S., et al. 2013, A\&A, 553, 114

\bibitem[\protect\citeauthoryear{Kelvin et al.}{2012}]{Kelvin2012}
Kelvin, L. S., Driver, S. P., Robotham, A. S. G., et al. 2012, MNRAS, 421, 1007

\bibitem[\protect\citeauthoryear{Kennicutt}{1998a}]{Kennicutt1998a}
Kennicutt, R. C. 1998a, ApJ, 498, 541

\bibitem[\protect\citeauthoryear{Kennicutt}{1998b}]{Kennicutt1998b}
Kennicutt, R. C. 1998b, AR\&A, 36, 189

\bibitem[\protect\citeauthoryear{Kennicutt et al.}{2007}]{Kennicutt2007}
Kennicutt, R. C., Calzetti, D., Walter, F., et al. 2007, ApJ, 671, 333

\bibitem[\protect\citeauthoryear{Kennicutt et al.}{2011}]{Kennicutt2011}
Kennicutt, R. C., Calzetti, D., Aniano, G., et al. 2011, PASP, 123, 1347

\bibitem[\protect\citeauthoryear{Krumholz \& McKee}{2005}]{Krumholz2005}
Krumholz, M. R., \& McKee, C. F. 2005, ApJ, 630, 250

\bibitem[\protect\citeauthoryear{Krumholz \& Tan}{2007}]{KrumholzTan2007}
Krumholz, M. R. \& Tan, J. 2007, ApJ, 654, 304

\bibitem[\protect\citeauthoryear{Krumholz, Dekel \& McKee}{2012}]{Krumholz2012}
Krumholz, M. R., Dekel, A., \& McKee, C. F. 2012a, ApJ, 745, 69

\bibitem[\protect\citeauthoryear{Krumholz \& Burkhart}{2016}]{Krumholz2016}
Krumholz, M. R. \& Burkhart, B. 2016, MNRAS, 468, 1671

\bibitem[\protect\citeauthoryear{Larson}{1981}]{Larson1981}
Larson, R. B. 1981, MNRAS, 194, 809

\bibitem[\protect\citeauthoryear{Lang et al.}{2014}]{Lang2014}
Lang, P., Wuyts, S., Somerville, R. S., et al. 2014, ApJ, 788, 11

\bibitem[\protect\citeauthoryear{Law et al.}{2012}]{Law2012a}
Law, D. R., Steidel, C. C., Shapley, A. E., et al. 2012, ApJ, 85, 85

\bibitem[\protect\citeauthoryear{Lehnert et al.}{2009}]{Lehnert2009}
Lehnert, M., Nesvadba, N., Le Tiran, L., et al. 2009, ApJ, 699, 1660

\bibitem[\protect\citeauthoryear{Leroy et al.}{2008}]{Leroy2008}
Leroy, A., Walter, F., Brinks, E., et al. 2008, AJ, 136, 2782

\bibitem[\protect\citeauthoryear{Leroy et al.}{2009}]{Leroy2009}
Leroy, A., Walter, F., Bigiel, F., et al. 2009, AJ, 137, 4670

\bibitem[\protect\citeauthoryear{Leroy et al.}{2013}]{Leroy2013}
Leroy, A., Walter, F., Sandstrom, K., et al. 2013, AJ, 146, 19

\bibitem[\protect\citeauthoryear{Liske et al.}{2015}]{Liske2015}
Liske, J., Baldry, I. K., Driver, S. P., et al. 2015, MNRAS, 452, 2087

\bibitem[\protect\citeauthoryear{Lowe et al.}{1994}]{Lowe1994}
Lowe S., Roberts W., Yang J., et al. 1994, ApJ, 427, 184

\bibitem[\protect\citeauthoryear{Madden et al.}{1993}]{Madden1993}
Madden, S. C., Geis, N., Genzel, R., et al. 1993, ApJ, 407, 579

\bibitem[\protect\citeauthoryear{Malhotra et al.}{1997}]{Malhotra1997}
Malhotra, S., Helou, G., Stacey, G., et al. 1997, ApJ, 491, 27

\bibitem[\protect\citeauthoryear{Malhotra et al.}{2001}]{Malhotra2001}
Malhotra, S., Kaufman, M. J., Hollenbach, D., et al. 2001, ApJ, 561, 766

\bibitem[\protect\citeauthoryear{Mogotsi et al.}{2016}]{Mogotsi2016}
Mogotsi, K., de Blok, W., Cald\'u-Primo, A., et al. 2016, AJ, 151, 15

\bibitem[\protect\citeauthoryear{Molina et al.}{2017}]{Molina2017}
Molina, J., Ibar, E., Swinbank, A. M., et al. 2017, MNRAS, 466, 892

\bibitem[\protect\citeauthoryear{Mosenkov et al.}{2015}]{Mosenkov2015}
Mosenkov, A., Sotnikova, N., Reshetnikov, V., et al. 2015, MNRAS, 451, 2376

\bibitem[\protect\citeauthoryear{Narayanan et al.}{2012}]{Narayanan2012}
Narayanan, D., Krumholz, M., Ostriker, E., \& Hernquist, L. 2012, MNRAS, 421, 3127

\bibitem[\protect\citeauthoryear{Neugebauer et al.}{1984}]{Neugebauer1984}
Neugebauer, G., Habing, H. J., van Duinen, R., et al. 1984, ApJ, 278, 1

\bibitem[\protect\citeauthoryear{Nozawa \& Kozasa}{2013}]{Nozawa2013}
Nozawa, T. \& Kozasa, T. 2013, ApJ, 776, 24 

\bibitem[\protect\citeauthoryear{Oberst et al.}{2006}]{Oberst2006}
Oberst, T. E., Parshley, S. C., Stacey, G. J., et al. 2006, ApJL, 652, 125

\bibitem[\protect\citeauthoryear{Papadopoulos \& Seaquist}{1999}]{PapaSea1999}
Papadopoulos, P. P., Seaquist, E. R. 1999, ApJ, 516, 114

\bibitem[\protect\citeauthoryear{Papadopoulos et al.}{2012}]{Papa2012}
Papadopoulos, P., van der Werf, P., Xilouris, E., et al. 2012, ApJ, 751, 10 

\bibitem[\protect\citeauthoryear{Peng et al.}{2010}]{Peng2010}
Peng, C. Y., Ho, L. C., Impey, C. D. \& Rix, H. 2010, AJ, 139, 2097

\bibitem[\protect\citeauthoryear{Pilbratt et al.}{2010}]{Pilbratt2010}
Pilbratt, G., Riedinger, J., Passvogel, T., et al. 2010, A\&A, 518, 1

\bibitem[\protect\citeauthoryear{Pineda et al.}{2013}]{Pineda2013}
Pineda, J. L., Langer, W. D., Velusamy, T., et al.2013, A\&A, 554, 103

\bibitem[\protect\citeauthoryear{Pineda, Langer \& Goldsmith}{2014}]{Pineda2014}
Pineda, J. E., Langer, W. D. \& Goldsmith, P. F.  2014 A\&A, 570, 121

\bibitem[\protect\citeauthoryear{Poglitsch et al.}{2010}]{Poglitsch2010}
Poglitsch, Waelkens, C., Geis, N., et al. 2010, A\&A, 518, 2

\bibitem[\protect\citeauthoryear{Rafikov}{2001}]{Rafikov2001}
Rafikov, R. R. 2001, MNRAS, 323, 445

\bibitem[\protect\citeauthoryear{R\"ollig et al.}{2007}]{Rollig2007}
R\"ollig, M., Abel, N. P., Bell, T., et al. 2007, A\&A, 467, 187

\bibitem[\protect\citeauthoryear{Romeo}{1992}]{Romeo1992}
Romeo, A, B. 1992, MNRAS, 256, 307

\bibitem[\protect\citeauthoryear{Romeo, Burkert \& Agertz}{2010}]{Romeo2010}
Romeo, A, B., Burkert, A. \& Agertz, O. 2010, MNRAS, 407, 1223

\bibitem[\protect\citeauthoryear{Romeo \& Wiegert}{2011}]{RW2011}
Romeo, A, B. \& Wiegert, J. 2011, MNRAS, 416, 1191

\bibitem[\protect\citeauthoryear{Romeo \& Falstad}{2013}]{Romeo2013}
Romeo, A, B. \& Falstad, N. 2013, MNRAS, 433, 1389

\bibitem[\protect\citeauthoryear{Saintonge et al.}{2013}]{Saintonge2013}
Saintonge, A., Lutz, D., Genzel, R., et al. 2013, ApJ, 778, 2

\bibitem[\protect\citeauthoryear{Schaye et al.}{2015}]{Schaye2015}
Schaye, J., Crain, R. A., Bower, R. G., et al. 2015, MNRAS, 446, 521

\bibitem[\protect\citeauthoryear{Schmidt}{1959}]{Schmidt1959}
Schmidt, M. 1959, ApJ, 129, 243

\bibitem[\protect\citeauthoryear{S\'ersic}{1963}]{Sersic1963}
S\'ersic, J. L., BAAA, 6, 41

\bibitem[\protect\citeauthoryear{Shapiro et al. }{2008}]{Shapiro2008}
Shapiro, K., Genzel, R., F\"{o}rster Schreiber, N., et al. 2008, ApJ, 193, 335

\bibitem[\protect\citeauthoryear{Shetty et al.}{2011b}]{Shetty2011b}
Shetty, R., Glover, S., Dullemond, C., et al. 2011b, MNRAS, 415, 3253 

\bibitem[\protect\citeauthoryear{Silk}{1997}]{Silk1997}
Silk, J. 1997, ApJ, 481, 703

\bibitem[\protect\citeauthoryear{Smith et al.}{2017}]{Smith2017}
Smith, J. D. T., Croxall, K., Draine, B., et al. 2017, ApJ, 834, 5

\bibitem[\protect\citeauthoryear{Solomon et al.}{1987}]{Solomon1987}
Solomon, P. M., Rivolo, A. R., Barrett, J. \& Yahil, A. 1987, ApJ, 319, 730

\bibitem[\protect\citeauthoryear{Solomon et al.}{1997}]{Solomon1997}
Solomon, P. M., Downes, D., Radford, S. J. E. \& Barrett J. W. 1997, ApJ, 478, 144

\bibitem[\protect\citeauthoryear{Solomon \& Vanden Bout}{2005}]{SolomonVandenBout2005}
Solomon, P. M., \& Vanden Bout, P. A., 2005, ARA\&A, 43, 677

\bibitem[\protect\citeauthoryear{Springel \& Hernquist}{2003}]{Springel2003}
Springel, V., \& Hernquist, L., 2003, MNRAS, 339, 289

\bibitem[\protect\citeauthoryear{Stacey et al.}{1991}]{Stacey1991}
Stacey, G., Geis, N., Genzel, R., et al. 1991, ApJ, 373, 423

\bibitem[\protect\citeauthoryear{Stacey et al.}{2010a}]{Stacey2010a}
Stacey, G. J., Hailey-Dunsheath, S., Ferkinhoff, C., et al. 2010a, ApJ, 724, 957

\bibitem[\protect\citeauthoryear{Stott et al.}{2016}]{Stott2016}
Stott, J. P., Swinbank, A. M., Johnson, H. L., et al. 2016, MNRAS, 457, 1888

\bibitem[\protect\citeauthoryear{Swinbank et al.}{2012a}]{Swinbank2012a}
Swinbank, A. M., Sobral., D., Smail, I., et al. 2012a, MNRAS, 426, 935

\bibitem[\protect\citeauthoryear{Swinbank et al.}{2012b}]{Swinbank2012b}
Swinbank, A. M., Smail, I., Sobral., D., et al. 2012b, ApJ,  760, 130

\bibitem[\protect\citeauthoryear{Swinbank et al.}{2017}]{Swinbank2017}
Swinbank, A. M., Harrison, C. M., Trayford, J., et al. 2017, MNRAS, 467, 3140

\bibitem[\protect\citeauthoryear{Tielens \& Hollenbach}{1985}]{Tielens1985}
Tielens, A. \& Hollenbach, D. 1985, ApJ, 291, 722

\bibitem[\protect\citeauthoryear{Toomre}{1964}]{Toomre1964}
Toomre, A., 1964, ApJ, 139, 1217

\bibitem[\protect\citeauthoryear{Turner}{2017}]{Turner2017}
Turner, O., Cirasuolo, M., Harrison, C., et al. 2017, MNRAS, 471, 1280

\bibitem[\protect\citeauthoryear{Utreras, Becerra \& Escala }{2016}]{Utreras2016}
Utreras, J., Becerra, F. \& Escala, A. 2016, ApJ, 833, 13

\bibitem[\protect\citeauthoryear{Valiante et al.}{2016}]{Valiante2016}
Valiante, E., Smith, M. W. L., Eales, S., et al. 2016, MNRAS, 462, 3146

\bibitem[\protect\citeauthoryear{Villanueva et al.}{2017}]{Villanueva2017}
Villanueva, V., Ibar, E., Hughes, T. M., et al. 2017, MNRAS, 470, 3775

\bibitem[\protect\citeauthoryear{Vogelsberger et al.}{2014}]{Vogelsverger2014}
Vogelsberger, M., Genel, S., Springel, V., et al. 2014, MNRAS, 444, 1518

\bibitem[\protect\citeauthoryear{Walter et al.}{2008}]{Walter2008}
Walter, F., Brinks, E., de Blok, W., et al. 2008, 136, 2563 

\bibitem[\protect\citeauthoryear{White et al.}{2017}]{White2017}
White, H., Fisher, D., Murray, N., et al. 2017, ApJ, 846, 35

\bibitem[\protect\citeauthoryear{Wisnioski}{2012}]{Wisnioski2012}
Wisnioski, E., Glazebrook, K., Blake, C., et al. 2012, MNRAS, 422, 3339

\bibitem[\protect\citeauthoryear{Wisnioski}{2015}]{Wisnioski2015}
Wisnioski, E., F\"{o}rster Schreiber, N. M., Wuyts, S., et al. 2015, ApJ, 799, 209

\bibitem[\protect\citeauthoryear{Wong \& Blitz}{2002}]{Wong2002}
Wong, T., \& Blitz, L. 2002, ApJ, 569, 157

\bibitem[\protect\citeauthoryear{Wright}{2010}]{Wright2010}
Wright, E., Eisenhardt, P., Mainzer, A., et al. 2010, AJ, 140, 1868

\bibitem[\protect\citeauthoryear{Wuyts et al.}{2011}]{Wuyts2011b}
Wuyts, S., Förster Schreiber, N. M., van der Wel, A., et al. 2011b, ApJ, 742, 96

\bibitem[\protect\citeauthoryear{Zhou et al.}{2017}]{Zhou2017}
Zhou, L., Federrath, C., Yuan, T., et al. 2017, MNRAS, 470, 4573

\end{thebibliography}
\end{document}